%% file: main.tex
\documentclass[phd, greychapternumbers]{mqthesis}
\usepackage{helvet}
\usepackage{multirow}
\usepackage{etoolbox}
\usepackage{setspace}
\usepackage{array,longtable}

\newtoggle{smallcapschapter}
\togglefalse{smallcapschapter} 
\usepackage{lipsum}
\usepackage{pgfplots}
\usepackage{pgfkeys}
\usepackage{enumerate}
\newenvironment{customlegend}[1][]{%
        \begingroup
        \csname pgfplots@init@cleared@structures\endcsname
        \pgfplotsset{#1}%
    }{%
        \csname pgfplots@createlegend\endcsname
        \endgroup
    }%

    \def\addlegendimage{\csname pgfplots@addlegendimage\endcsname}
\pgfplotsset{
cycle list={%
{draw=black,mark=star,solid},
{draw=black, mark=square,solid},
{draw=black,mark=+,solid},
{black,mark=o},}}

\begin{document}

\frontmatter

\title{Diffusion on dynamic contact networks with indirect transmission links} 
\author{Md Shahzamal} 
\department{Computing}  

\titlepage

\input{acknowledge} 
\input{abstract} 

\tableofcontents 
\listoffigures
\listoftables
\input{listofpublications}

\mainmatter

\makeatletter
\let\old@makechapterhead\@makechapterhead
\makeatletter
\def\thickhrulefill{\leavevmode \leaders \hrule height 1ex \hfill \kern \z@}
\def\@makechapterhead#1{%
  {\parindent \z@ \raggedright
    \reset@font
    \hrule
    \vspace*{10\p@}%
    \par
    \iftoggle{smallcapschapter}{  
   	 \Large \sc \@chapapp{} \Huge\bfseries \thechapter
	}{
	\Large \@chapapp{} \Huge\bfseries \thechapter
	}
    \par\nobreak
    \vspace*{10\p@}%
    \hrule
    \par
    \vspace*{1\p@}%
    \hrule
    \vspace*{20\p@}
    \Huge \bfseries #1\par\nobreak
    \vskip 70\p@
  }}

\input{1_chap_intro.tex}

\input{2_chap_backg.tex}

\input{3_chap_spdtmodel.tex}

\input{4_chap_net.tex}

\input{5_chap_spdtchar.tex}
\input{6_chap_prevac.tex}
\input{7_chap_conclusion}

\makeatletter
\let\@makechapterhead\old@makechapterhead
\backmatter

\input{listofsymbols}
\bibliographystyle{unsrt}
\bibliography{references}

\appendix

\input{chap_appendix}

\end{document}

%% file: acknowledge.tex
\chapter{Acknowledgements}

The research included in this thesis could not have been performed without the assistance and support of many people, whom I would like to thank here. First and foremost, I would like to thank my advisor Prof. Bernard Mans for invaluable guidance I have received from him. Without his encouragement and
support, this thesis would not have been possible.  I am very grateful that he gave me the opportunity to present my works on
international conferences and that he helped me extend my knowledge through discussions and visiting relevant research groups.

I am grateful to Prof. Raja Jurdak for being a supervisor. I am very grateful that he gave me the opportunity to work with Data61, CSIRO. I would also be nowhere without his support and advice. He has always been available and ready to listen and discuss my research problems. His huge supports has helped me to make the best of my research. 

I would like to thank Dr. Frank de Hoog, for always being positive and for many fruitful discussions. It was always a pleasure to discuss my work with him. I would like to extend my thanks to everyone at DSS group at Data61, CSIRO for the warm welcome they gave me. I very much wish that some sort of collaboration continues or at least to visit again.

All my gratitude goes to my family for motivating and helping me through the ups and downs of writing my thesis. They were there for me at any time and never lost hope.

%% file: abstract.tex
\chapter{Abstract}
Modelling diffusion processes on dynamic contact networks is an important research area for epidemiology, marketing, cybersecurity, and ecology. For these diffusion processes, the interactions among individuals build a transmission network of contagious items that spread out over the contact network. The diffusion dynamics of contagious items on dynamic contact networks are strongly determined by the underlying interaction mechanism between individuals. Thus, various research has been conducted to understand the correlation between diffusion dynamics and interaction patterns. However, current diffusion models only assume that contagious items transmit through interactions where both infected and susceptible individuals are present at a physical space together (e.g. visiting a location at the same time) or active in virtual space at the same time (e.g. friendship in online social networks).  

The focus on concurrent presence (real or virtual), however, is not sufficiently representative of a class of diffusion scenarios where transmissions can occur with indirect interactions, i.e. where susceptible individuals receive contagious items even if the infected individuals have left the interaction space. For example, an individual infected by the airborne disease can release infectious particles in the air through coughing or sneezing. The particles are then suspended in the air so that a susceptible individual arriving after the departure of the infected individual can still get infected. In this scenario, current diffusion models can miss significant spreading events during delayed indirect interactions.

In this thesis, a novel diffusion model called the same place different time transmission based diffusion (SPDT) is introduced to take into account the transmissions through indirect interactions. The behaviour of SPDT diffusion is analysed on real dynamic contact networks and a significant amplification in diffusion dynamics is observed. The SPDT model also introduces some novel behaviours different to current diffusion models. In this work, a new SPDT graph model is also developed to generate synthetic traces to explore SPDT diffusion in several scenarios. The analysis shows that the emergence of new diffusion becomes common thanks to the inclusion of indirect transmissions within the SPDT model. This work finally investigates how diffusion can be controlled and develops new methods to hinder diffusion. This study undertakes infectious diseases spreading as a case study as it captures all aspects of SPDT diffusion processes. The real co-location contact networks constructed by users of a location based social networking application are used for this study. All results are compared with the current diffusion models which are based on direct transmission links.

%% file: listofpublications.tex
\chapter{List of Publications}

\textbf{\large Papers accepted and published}
\begin{itemize}
\item Md. Shahzamal, R. Jurdak, B. Mans and F. de Hoog. \textbf{ A Graph Model with Indirect Co-location Links}.  14th Workshop on Mining and Learning with Graphs, as part of KDD 2018, London, UK, August 19-23, 2018

\item Md. Shahzamal, R. Jurdak, B. Mans, A. El Shoghri and F. de Hoog. \textbf{Impacts of Indirect Contacts in Emerging Infectious Diseases on Social Networks}. orkshop on Big Data Analytics for Social Computing, (part of PAKDD), Melbourne, Australia, June, 2018

\item Md Shahzamal,  R. Jurdak,  R. Arablouei,  M. Kim,  K. Thilakarathna,  and B. Mans.\textbf{ Airborne Disease Propagation on Large Scale Social Contact Networks}. The 2017 International Workshop on Social Sensing, Pittsburgh, USA, April 21, 201
\end{itemize} 

\noindent
\textbf{\large Paper under review}
\begin{itemize}
\item Md. Shahzamal, R. Jurdak, B. Mans and F. De Hoog.\textbf{ Indirect interactions influence contact network structure and diffusion dynamics}. Journal of The Royal Society Open Science, 2019
\end{itemize}

%% file: 1_chap_intro.tex

\chapter{Introduction}
\section{Background and Motivations}
Diffusion is an important underlying process for many real-world phenomena on networks. Recently, there has been a substantial amount of research to understand and model various diffusion processes occurring in a broad range of applications: information diffusion, viral marketing, computer virus spreading, technology diffusion and disease spreading etc ~\cite{huang2016insights, keeling2008modeling, gao2015modeling, stummer2015innovation, blonder2011time}. The common mechanism of all these diffusion phenomena is that contagious items appear at one or more nodes of the interacting systems and then transmit from one node to other nodes through the transmission links created due to the inter-node interactions. The underlying interacting systems of diffusion processes are often individual contact networks where individuals interact with each other to create transmission links. For example, in the spreading of infectious diseases, the contagious items are infectious particles generated by infected individuals and are transferred to susceptible individuals when they are exposed to infectious particles~\cite{huang2016insights}. In a similar way, an online post generated by a user is disseminated over an online social network (OSN) where other users learn about it~\cite{gao2015modeling} after reading the post. A new type of behaviour or action, e.g., purchasing a new product, can spread within a population following this approach as well~\cite{stummer2015innovation}. 


The above diffusion processes show that the key underlying operation of a diffusion process is transferring contagious items from one individual to another individual, called local transmission or individual level transmission, through their contacts. These contacts can be created when two individuals are in the same proximity (interactions between two individuals in the physical space), or when a user is reading and learning posts of other users in an online social network (interactions between two individuals in virtual space) etc. In a diffusion process, therefore, an underlying contagious items transmission network is created due to a series of such local transmissions which occurs due to the movements of individuals (e.g. movements of infected individuals to different locations or passing messages from one online blog to other blogs by the users). These local transmissions can occur in two different ways based on the characteristics of the contacts mechanism. In the first way, both individuals are active at the same place same time and contagious items are directly transferred from one individual to another individual. The transmission link for this local transmission is called direct transmission link and the diffusion phenomena based on these transmissions are defined as the same place same time transmission based diffusion (SPST diffusion). The example includes diffusion in Mobile Ad-hoc Network (MANET)~\cite{jacquet2010information,shahzamal2016smartphones} where a sender individual and a receiver individual are present together at the same time at a location to make a transmission of messages and infectious disease spreading through physical touches.

In the second way of local transmission, both individuals are not required to be active at the same place at the same time to make a local transmission as the contagious items have the opportunity to transfer later on. In this mode, an infected individual deposits the contagious items to the medium (physical space, websites or blogs) and other individuals are exposed to that contagious items later on and receive it. The examples include: i) content diffusion by reposting in OSN where posts are seen by friends later on~\cite{lee2015message, gao2015modeling}, and ii) infectious disease spreading where infectious particles generated by the infected individuals deposit on the surface, objects, or suspend in the air and transmit if susceptible individuals come across it further~\cite{fernstrom2013aerobiology,brankston2007transmission}. Therefore, the contagious items can be transmitted with the delayed opportunity and this mode of local transmission is called indirect transmission. In indirect transmissions, the opportunity of transferring contagious item decays as time passes from its generation. For example, the infectious particles generated by an infected individual lose their infectivity over time and is removed from the interaction location, i.e. the probability of disease transmission (contacted susceptible individual gets infected) decays with time for these indirect links~\cite{fernstrom2013aerobiology,han2014risk}. Similarly, an online post in Facebook gradually disappears from the Facebook wall and has less probability to be read and learnt by the other users as time passes. The diffusion phenomena including the indirect transmission opportunities are defined as the same place different time transmission based diffusion (SPDT diffusion). The aim of this thesis is to study the SPDT diffusion dynamics on dynamic contact networks including indirect transmission links.


Diffusion dynamics modelling have been extensively studied for SPST diffusion processes and there have been several promising approaches in the literature~\cite{kermack1927contribution,burke2006individual,meloni2011modeling,koopman2005mass}. However, these methods are not directly applicable to SPDT diffusion processes as they do not consider the indirect transmission opportunities in their modelling. The SPDT diffusion modelling is required to account the indirect transmissions and its decaying properties along with the direct transmissions. In the literature, there have been a few research works to include indirect transmissions and to understand the impacts of decay in transmission opportunities. In online social networks (OSN), the fading of popularity over time to the generated contents are studied in the works of ~\cite{wu2007novelty, gao2015modeling}. These studies have analysed the decay rate of the content popularity in the entire networks. In the SPDT model, however, it is required to understand how popularity (transmission probability) decays over the inter-event time between two consecutive posts. The decay in airborne disease transmission probability through interaction has also been studied in the works of ~\cite{han2014risk, sze2010review}. These works have mainly studied what the temporal dynamics of infection risk is at a location where the infected individual has been. However, they did not consider how these decays in individual level risk impacts on the diffusion dynamics on the dynamic contact networks. 

There are a few research works~\cite{lange2016relevance,sorensen2014impacts,huang2010quantifying} that have accounted for the importance of indirect transmission in infectious diseases spreading. These works have investigated the overall impact of indirect transmissions on disease diffusion, but not how the individual level indirect transmission influence diffusion dynamics. To our knowledge, the work of ~\cite{richardson2015beyond} is the only research work that investigated the impacts of individual level indirect transmissions with decay properties. This study has analysed the diffusion dynamics in a social insect colony of 245 individuals and identifies the contribution of indirect transmission in spreading. However, there is no detailed explanation of diffusion behaviours and how it impacts on other diffusion related activities such as controlling diffusion dynamics. Therefore, it is required to conduct research on the SPDT diffusion process for developing models and tools to study its behaviours with a large population.

The diffusion processes on dynamic contact networks are studied in two ways: analytically and by simulation~\cite{nicas1996analytical,valdano2015analytical,stehle2011simulation}. The analytical approaches require explicit knowledge of the diffusion dynamic equations. However, there are always many assumptions about the diffusion process and underlying contact networks. These assumptions are often fixed and their variation fails the model. Thus, there might be an incomplete understanding of the real diffusion processes. Therefore, a data-driven approach is often the first-hand methodology to understand the diffusion dynamics and relevant applications. In the data-driven approach, the assumptions are varied largely to capture reality. Thus, this easily allows studying the if-else situations (e.g. varying model parameters from a different perspective) to understand the broad scenarios. Recently, data-driven modelling has emerged due to the social data revolution with the flexibility of collecting high-resolution data sets from mobile devices, communication, and prevasive technologies~\cite{zhao2011social,starnini2014time,stehle2011simulation,huang2016insights,tatem2014integrating}. The current widespread of mobile technologies allows sensing human behaviours at a granular level, even with longitudinal dimension. For example, some location-based social networking applications provide geo-tagged locations data about the movements of users~\cite{jurdak2015understanding,chen2013and,abbasi2015utilising}. These digital traces of human activities bring forth new opportunities to transform the ways of modelling socio-technical systems. Thus, understanding and modelling social dynamics using individual level behaviours are on the way to be shifted to the next paradigm. To study the SPDT diffusion process on a dynamic contact network, individual level interaction data is the prerequisite. Hence, there is a lack of data-driven approaches for studying SPDT diffusion. It is necessary to find suitable data sets, understand their characteristics to model SPDT diffusion, and develop a required framework.


Although a large amount of data on individual interactions relevant to diffusion is routinely collected, not all relevant data is available due to privacy and cost considerations. A popular approach, therefore, is to capture the statistical properties from the limited data and incorporate them into a graph model where individuals are presented as nodes and contacts (interactions that can cause transmission of disease) between individuals as links~\cite{porter2016network,zhao2011social, xu2017synthetic, zhang2016modelling,starnini2013modeling}. This graph model allows one to generate various contact networks and study diffusion processes in different scenarios with a view to developing analytical models of diffusion processes. This is helpful to explore diffusion control strategies in various setting. There have been a fair amount of well-defined graph models for studying SPST diffusion processes on individual contact networks~\cite{holme2015modern,laurent2015calls,shahzamal2018graph}. However, current graph models only consider direct transmission links created for concurrent co-location interactions among individuals. Thus, it is required to develop a new graph model integrating indirect links created for delayed interactions.


The diffusion processes are highly influenced by the underlying network structure which is often determined by the contact mechanism~\cite{mossong2008social,mikolajczyk2008social,hens2009mining,iribarren2009impact}. Thus, the impact of the underlying network structure on diffusion process depends on the characteristics of contacts. The distribution of individual degree is one of the important network metrics to characterise the networks~\cite{de2014role,freeman1978centrality,britton2007graphs}. It is found that variation in degree distribution requires various transmission conditions to get the epidemic (disease reaches to a significant number of individuals before dies out) in the network~\cite{bansal2007individual}. The clustering coefficients are studied in the network to understand the connectivity in the network~\cite{wang2003complex,smieszek2009models}. If the network is strongly connected, it has more chance to transfer the contagious item easily between the neighbours. This local connectivity is leveraged to understand the community structure and the importance of individuals in the network~\cite{newman2006finding,geard2008group}. When the indirect transmission link is included, the above discussed network properties may change, thereby changing the diffusion conditions in the networks. As the indirect transmission links are different from the direct transmission links, it is very important to understand the potential and impacts of these links to form network properties and diffusion dynamics.


Controlling diffusion is the ultimate goal of diffusion modelling. Control strategies for a network vary based on the application and context~\cite{al2018analysis,miller2007effective,britton2007graphs}. In addition, the contact patterns in a network often influences developing diffusion control strategies. For online marketing, the aim is to maximise the diffusion while the aim is to minimise the diffusion for containing the spread of disease on social contact networks. However, the key task in both cases is to find a set of individuals who have strong influence in spreading. To maximise diffusion, individuals in these sets are taken as the source or seed of spreading as they spread contagious items to the maximum number of individuals in the network~\cite{zhang2016least,al2018analysis}. On the other hand, to minimise the spread, the selected set is removed from the spreading process to slow down the spreading~\cite{holme2004efficient,britton2007graphs}. The spreading potentiality of an individual depends on his connectivity to the other individuals and their position in the network. When the indirect transmission links are included in the network, the connectivity of an individual to the other individuals is changed. Therefore, individual importance is changed and hence spreading potential. This means that the effectiveness of current diffusion controlling methods may vary for the SPDT model and a new controlling strategy may be required to develop.

\section{Research Objectives}
The aim of this thesis is to investigate the diffusion process on dynamic contact networks including indirect transmission links along with direct transmission links. Accordingly, a new diffusion model called the same place different time transmission based diffusion (SPDT diffusion) is introduced. In the SPDT model, the transmission links are noted as SPDT links and consist of direct and/or indirect transmission link components. The probability of transmitting contagious items through an SPDT link decays over time. The inclusion of indirect transmissions in the SPDT diffusion model introduces challenges for the methods of current SPST diffusion with direct transmission links to study SPDT diffusion dynamics. The investigations are conducted in this thesis for following objectives.

\begin{enumerate}[i)]
\item SPDT model: Current diffusion models (SPST models) only consider direct transmission links to determine diffusion dynamics. Therefore, it is necessary to analyse the SPST models to find ways of adding indirect transmission opportunities with them. The inclusion of indirect transmission links in the SPDT model increases the spreading potential of individuals. It is, therefore, of interesting to quantify the changes in the SPDT diffusion dynamics on the contact networks compared to the SPST diffusion dynamics. However, the changes in SPDT diffusion dynamics will depend on the decay rates of links transmissions probabilities. This may also introduce new diffusion behaviours for the SPDT model due to the dynamic behaviours of SPDT links. Thus, investigations have to be conducted to identify the novel behaviours of SPDT model. These issues are addressed in Chapter 3 of this thesis.

\item SPDT graph: The characterisation of diffusion processes are conducted by exploring diffusion behaviours with various scenarios. The synthetic contact network is a widely used tool for this purpose. However, the current contact graph models can only generate contact networks with direct transmission links. Thus, a graph model is required that can generate the contact networks among nodes considering both direct and indirect interactions. Current graph models are analysed to find an appropriate graph model for supporting SPDT diffusion studies. Chapter 4 focuses on finding an appropriate graph model for SPDT diffusion.

\item Node potential in SPDT: Contact mechanism defines the network properties which in turn change the diffusion dynamics unfolded on the contact network. The contact patterns between nodes and their neighbours form local social contact structures. These local contact structures also define the position of the nodes in the network and define the diffusion phenomena such as spreading potential of nodes (influence of a node to make diffusion of infectious items), super-spreader behaviours and flexibility of emerging diffusion. It is interesting to know what is the potential of indirect links at the node level to shape these diffusion phenomena. A detailed analysis and discussion of these issues are presented in Chapter 5. 

\item Control: Diffusion controlling strategies are often developed based on the contact network properties. However, the contact network properties in the SPDT model are changed for including indirect transmissions. For example, the inclusion of indirect links can add a new neighbour to a node. The effectiveness of the current controlling methods is analysed to understand the effect of indirect links. The other challenge in SPDT model is to identify the neighbour nodes by a host node as the neighbour nodes connected with only indirect links are invisible to them. Thus, controlling strategies might lose effectiveness. Thus, it may require to find the new controlling strategy for SPDT model. Controlling of SPDT diffusion is investigated in Chapter 6.

\end{enumerate}

\section{Thesis Structure}
This thesis contains seven separate chapters and one appendix that includes additional materials and results which are not discussed in the main body text of the thesis. The outline of the chapters is as follows:

Chapter 2, entitled Overview and Relevant Works, discusses the technical background and relevant works on diffusion processes. This chapter analyses diffusion modelling approaches (mainly network-based models) integrating realistic contact properties to understand their applicability to study SPDT diffusion processes. In this thesis, the airborne disease spreading is taken as a case study of SPDT diffusion processes. Therefore, the literature is reviewed around the airborne disease spreading modelling using dynamic contact networks.

In Chapter 3, entitled SPDT Diffusion Model, an SPDT diffusion model is introduced and analysed in details. This chapter also presents the methods for assessing the transmission probability of links with indirect transmissions. Construction of empirical contact networks using GPS locations from the users of a social networking application Momo is presented here. Using different configurations of empirical contact networks, SPDT diffusion behaviours are explored and compared with the SPST diffusion. Finally, the novel diffusion behaviours are identified and verified through various settings of simulations.

Chapter 4, entitled SPDT Graph Model, describes the development of a graph model called SPDT graph that can generate synthetic dynamic contact networks of nodes that create links through both direct and indirect interactions. The graph model is developed following the principals of activity driven time-varying network modelling. Network generation methods are also developed and fitted with real data set. Then, the SPDT graph model are validated by comparing network properties with that of real networks and by analysing the capability of simulating SPDT diffusion processes.

The potential of the indirect transmission links at the node level is discussed in Chapter 5, entitled Indirect Link Potential. The nodes that have only indirect interactions with other nodes during their infectious periods are called \textit{hidden spreaders}. The potential of indirect transmission links is studied through the spreading abilities of hidden spreaders. This chapter also discusses how nodes have the potential of being super-spreaders due to including the indirect transmission links. Finally, the impact of indirect links to emerging diffusion process is studied in this chapter.

Chapter 6, entitled Controlling SPDT Diffusion, discusses controlling of SPDT diffusion through vaccination strategies for the airborne diseases. A new vaccination strategy is developed and studied for both preventive and reactive vaccination scenarios. The proposed strategy is examined for both the SPST and SPDT diffusion models. The effectiveness of vaccination strategies has investigated varying the scale of contact information collection for vaccination procedures where a proportion of nodes provide contact information. 

In Chapter 7, entitled Discussion and Future Works, a conclusion is provided about the research activities conducted and research outcomes for this thesis. This chapter also presents the future research directions for studying SPDT diffusion processes. The limitations of the research works are also discussed here.


%% file: 2_chap_backg.tex
\textbf{\chapter{Overview and Relevant Works}}
Diffusion processes are used to explain dynamic phenomena evolving on networked systems. In a diffusion process, contagious items spread out in the networked system through the inter-node interactions. The evolution behaviours of the diffusion process are strongly influenced by the characteristics of the underlying system and the mechanism of the diffusion process itself. Thus, understanding the characteristics of the underlying system and their impacts on the diffusion process have recently attracted intensive research. Current research on diffusion processes assumes that the sender node and the receiver node are concurrently active to participate in transferring contagious items. However, there are several diffusion processes such as infectious disease spreading and message dissemination in Online Social Networks (OSN) where interacting nodes do not require to be active concurrently to create infectious items transmission links. In these diffusion processes, the contagious items can transmit even with delayed interaction between contagious items and receiver nodes. Thus, these diffusion processes occur through both direct and indirect interactions. This chapter discusses the current literature of diffusion studies and describes the applicability of methods and theories for diffusion dynamics with indirect interactions. 

\input{2.1_chap_diff}
\input{2.2_chap_idis}

\input{2.3_chap_net}

\input{2.4_chap_pro}

\input{2.5_chap_con}

%% file: 2.1_chap_diff.tex
\section{Diffusion processes}
\subsection{Real world diffusion}
A diffusion processes is a dynamic phenomena on a networked system that starts from a node or a set of nodes and spread over the networked system through inter-node interactions. Examples include spreading of infectious diseases, messages, opinions, and belief dissemination etc. in social contact networks. In these diffusion processes, contagious items (infectious particles, a piece of information, innovation and a specific behaviour etc.) initially grow on one or more nodes of the networks and then spread through neighbouring nodes over the network. The interactions between among nodes are responsible for transmitting contagious items from one node to the other nodes. The interactions can happen through physical contact such as being in a common location, or through logical contact such as reading and learning messages from other users in OSN. In this thesis, the nodes of a networked system are individuals and the inter-node interactions causing transmission of infectious items are called contacts, edges, or links. Some of the diffusion scenarios are described below.
 
\textbf{Infectious disease spreading:} The spreading of infectious diseases is a widely studied diffusion process in the literature~\cite{moss2016model,huang2016insights,shahzamal2017airborne}. In these scenarios, infected individuals interact with susceptible individuals and disease is transmitted to susceptible individuals. Infectious diseases are a significant threat to human society. According to the report of the World Health Organisation (WHO), about 4.2 million people die annually due to infectious disease. The infectious diseases can also spread through animal contact networks, insect contact networks and even with plant contact networks~\cite{ keeling2008modeling, richardson2015beyond,jeger2007modelling}.


\textbf{Information diffusion:} Currently, online contact networks are a frequently discussed medium for information or virus spreading. In online social contact networks (OSN), messages are disseminated by sharing and re-sharing activities of users~\cite{gao2015modeling}. A piece of information can spread to a large population through online social blogs where information (message, tweets, news etc.) is transferred from users of one blog to the users of another blog. A belief or behaviour can be spread out in the social networks in similar ways~\cite{stummer2015innovation}. These diffusion platforms have created numerous opportunities for economic and social activities. For example, the social network based viral marketing has brought a new dimension to the traditional televised or roadside-billboard advertising campaigns~\cite{lu2016towards}. Viral marketing exploits the power of "word-of-mouth" to spread product sale through self-replicating transmission processes. Thus, it allows optimising business performance. The diffusion of rumours on online social media is also applied for political campaigning~\cite{shin2018diffusion}.

\textbf{Knowledge diffusion:} Knowledge can also diffuse over social contact networks, business networks and collaboration networks~\cite{stummer2015innovation,kiesling2012agent,young2009innovation,lambiotte2009communities}. The knowledge of new technology and science are distributed within the technical community immediately. It is also received by the collaborating communities. New members joining collaboration groups can receive knowledge from the existing group members. Innovation and technology spreading has been studied by analysing citations and co-authorship, and there is now a strong understanding of how the technology is distributed over the collaboration networks~\cite{abbasi2012egocentric,abbasi2011identifying}.

\textbf{Diffusion in cell biology:} A range of inter-molecular interactions such as protein-protein interactions and protein-DNA interactions etc. occur to function biological processes~\cite{cho2012network}. These interactions can be represented by network models and often explain the spreading of many complex disease in cell biology. The spreading of a disorder in the brain cell can be explained by the diffusion on brain networks~\cite{raj2012network}.

\textbf{Diffusion in ecology:} Diffusion can occur in ecology as well. The spreading of a new social behaviour (e.g. new food search strategy) in animal society can be modelled with diffusion processes~\cite{franz2009network}. In the social insect colonies, insects often exploit the information provided by other insects to take decisions such regarding food location, predator threats and queen instruction ~\cite{blonder2011time,richardson2015beyond,waters2012information}. For example, in ant colonies, the queen's messages are disseminated through the interactions among ants representing the encoded queen's message~\cite{duboscq2016social}. The queen sends messages to fellow ants through a chemical spray which is touched by the fellow ants. The recipient ants regenerate the encoded queen's message and other ants encountered receive the queen messages. 

\subsection{Factors affecting diffusion processes}
The underlying connected systems of diffusion processes in this thesis are the dynamic contact networks where individuals interact with each other. The spreading of contagious items over a contact network can be treated as the coupling of results of three factors, namely individual interactions, characteristics of contagious items, and environments. These three factors make strong contributions to diffusion phenomena on a contact network. The roles and influence of these factors depend on the context of diffusion processes as described below:

\textbf{Contact structure:} The main drivers for spreading contagious items on contact networks are interactions between individuals. The individual interaction patterns provide pathways to spread contagious items. Thus, the interaction patterns play key roles in developing diffusion phenomena on contact networks. For example, how many susceptible individuals infected individuals meets during their infectious period determine the spreading speed of the infectious disease. If an infected individual is connected to many other individuals, there is a high probability to transmit disease to others by him. On the other hand, if an infected individual has no contact with other individuals, the disease is not transmitted. Thus, the final size of the epidemic depends on the contact patterns distribution in the population. Similarly, an online post in OSN can spread quickly if the user generated post and following recipient users have a high number of connections to other users. There are several interaction properties such as how contact happens, contact frequency and contact duration etc. at the individual levels and contact degree distribution, clustering coefficient etc. at the network level that are studied to understand and model diffusion processes.

\textbf{Contagious items:} The contagious items can be infectious particles, a piece of information (Twitter posts), and a novel behaviour (purchasing a new product) etc. The contagious items themselves play strong roles in their spread in a population. The internal spreading potentiality of contagious items varies based on its characteristics. For example, the infectiousness of contagious particles is defined by the disease types and highly infectious diseases usually spread faster in a population. Similarly, political news spreads quickly in OSN compared to the business news. The spreading potential is also affected by the process of news generation, e.g., if the news is generated or endorsed by a celebrity, it may have a high chance to spread quickly. The impacts of contagious items also vary depending on the recipient individual behaviours. For example, the impacts of infectious particles varies according to the susceptibility of the individuals. Therefore, the characteristics of contagious items are often considered in the modelling of diffusion processes. 

\textbf{Environments:} The spreading of contagious items is often influenced by the environment. The environments represent the characteristics of the space where diffusion processes occur. For example, the impacts of infectious particles are determined by weather conditions such as temperature and humidity etc. The infectious particles generally lose their infectiousness over time and may also depend on the weather conditions. Similarly, the spreading behaviours of product purchasers can be influenced by advertisements in the traditional mass media. Another example is that spreading of rumours is affected by political situations. The underlying medium where contagious items spread can also be heterogeneous. For example, if the news is capable of spreading in multiple social networks, the rate of spreading increases. Similarly, a disease can spread through multiple platforms such as proximity contact networks, transportation contact networks, and air-travel contact networks. Thus, diffusion modelling is also required to consider the heterogeneity of diffusion medium.

The roles of properties of contagious items and heterogeneity in the environment for shaping diffusion dynamics are often defined by the underlying contact structure. If an infected individual has no contact with other susceptible individuals, the disease cannot transmit even if the infectious particles have high infectiousness. Thus, the influence of contagious items properties often depends on the interaction properties in the population. Similarly, impacts of the environment also depend on the interaction patterns as the highly favourable environmental conditions have no influence if there are not sufficient contact opportunities to transmit disease. The impacts of contagious properties and environment have been studied for a long time in the literature~\cite{kim2018real}. However, analysing contact structure to understand the spreading of contagious items is comparatively new. The recent exploration of data on individual interactions have fuelled research on unravelling contact patterns affecting contagious spreading~\cite{shahzamal2019indirect, gonzalez2007complex,gonzalez2008understanding,barthelemy2005dynamical}. This thesis has also the same aim. The following subsection discusses on the approaches to integrate impacts of contact patterns with diffusion modelling.

\subsection{Diffusion modelling approaches}
Diffusion modelling is an intensively researched area due to its wide applications. As the area of diffusion is diverse, the models developed are extremely varied in their approaches. Broadly speaking, the models developed can be divided into two groups based on their purposes: i) explanatory models, and ii) predictive models~\cite{meade2006modelling,guille2013information,pastor2015epidemic,barrat2008dynamical}. The explanatory models are usually developed to understand the factors affecting diffusion dynamics on a contact network. This often allows one to answer the questions such as which nodes are influential, what is the underlying reason for the way diffusion occurs, and what is appropriate diffusion controlling strategy? On the other hand, predictive models usually predict the spreading intensity and the final number of individuals received contagious items based on certain factors. It is often the case that the explanatory models find the key influential factors and apply them for developing predictive models. As this thesis focuses on understanding the impacts of contact mechanism on diffusion dynamics occurring on dynamic contact networks, the diffusing approaches for explanatory models are discussed here. There has been a range of approaches for explanatory models of diffusion and this section discusses some of them that are widely used in different fields of diffusion ranging from disease spreading on individual contact networks to information spreading on online social networks (OSN)~\cite{pastor2015epidemic,barrat2008dynamical}.

\subsubsection{Compartment epidemic models}
Compartment epidemic models are frequently applied to study diffusion in many applications such as information diffusion, innovation diffusion and computer virus spreading~\cite{anderson1992infectious,himmelstein1984compartmental,gruhl2004information}. The fundamental concept of these approaches is to divide the system into several compartments or partitions. Each compartment represents a set of individuals having a specific status. Then, the dynamics of the diffusion are determined by the flows between these compartments. Widely used compartments are Susceptible having individuals who are exposed to the contagious items, Infected having individuals who have adopted the contagious items and started forwarding items (infecting) others, and Recovered having individuals who were infected but now recovered. This compartment model is called SIR propagation model and the dynamics of the compartments are given by
\begin{equation*}
\frac{dS}{dt}=-\beta IS , \quad
\frac{dI}{dt}=\beta IS -\tau I , \quad
\frac{dR}{dt}=-\tau I
\end{equation*}
where $S$, $I$ and $R$ are the fractions of the population in the Susceptible, Infected and Recovered compartments respectively, $\beta$ is the transmission rate from Susceptible to Infected compartment and $\tau$ is the transmission rate from Infected to Recovered compartment. Thus, the dynamics of the system is given by $S(t)+I(t)+R(t)=1$. In the diffusion modelling, $\beta$ is given as $\beta=\gamma \beta_{0}$, where $\gamma$ is the number of potential contacts on average individuals has with others through which contagious items transmit with a rate $\beta_{0}$. By changing the value of $\gamma$, $\tau$ and $\beta_{0}$ one, therefore, can study the diffusion dynamics of diffusion processes. 

The compartment model can be analysed easily mathematically in the simple case. It requires no more details than needed to reproduce and explain observed behaviours. It reduces data collection cost and computational cost. Clearly, it can be applied in many situations where high precision is not necessary. However, the assumption of having homogeneous interaction $\gamma$ between individuals is not realistic. Many researchers have pointed out that the interaction between individuals is clearly heterogeneous as individuals do not have the same level of contact with all its neighbours~\cite{mikolajczyk2008social,mossong2008social,hens2009mining,mossong2008social}. Moreover, the constant rate of transmission probability $\beta_{0}$ and constant recovery rate $\tau$ are not realistic in many diffusion processes. This is because individuals have different contact intensity with infected individuals and heterogeneous susceptibilities to the contagious items~\cite{wearing2005appropriate,smieszek2009mechanistic,smieszek2009models}. 

Heterogeneity is often integrated into the compartment model by dividing the main compartments (such as S, I, R) into sub-compartments. These sub-divisions can be constructed based on the age, risk behaviours, or spatial diversities of individuals. Then, the transmission probability can be divided into $k$ sub-classes and the model can be parameterised by means of a $k\times k$ transmission rate matrix instead of a constant transmission rate $\beta$~\cite{rhomberg2017parallelization, yorke1978dynamics}. For example, some disease spread models split individuals spatially (divide population with different regions) and assign heterogeneity for infection risks~\cite{lloyd1996spatial,pitcher2017network,meakin2018metapopulation}. These approaches are called meta-population models. Similarly, different infectious periods can be implemented in the model by dividing population into $k$ sub-classes which resolve the limitations of the constant rate of recovery~\cite{wearing2005appropriate}.

The integration of heterogeneity with sub-compartmentalisation relax some of the most unrealistic assumptions of basic compartmental models. However, many limitations of the compartment model still prevail and new issues arise by doing sub-compartmentalisation. The analysis in ~\cite{koopman2005mass} shows that individual's interaction is still random and transient in these models. Hence, individuals in the divided sub-populations behave homogeneously. In case of meta-population models, dividing a population into the various spatial groups can create asynchronous between these groups, but for time $t \rightarrow \infty$ they become homogeneous~\cite{keeling2005extensions}. Therefore, the real steady state heterogeneity cannot be captured by meta-population models. Thus, high partitioning in compartmental models lose the simplicity.  It also requires more data to fit the model and increases the data collection cost.

\subsubsection{Network models}
To overcome the limitations of the compartment models and capture realistic contact patterns, network science is adopted for modelling diffusion processes on the networked systems~\cite{newman2002spread,keeling2005networks}. The network based diffusion modelling is also empowered by the graph theory where contact networks are often generated by graph models. In addition, graph theory is applied to study the characteristics of contact networks. The core entities in the network-based modelling are nodes, representing individuals, and links connecting one node to other nodes in the network, which represent interactions between individuals. A contact network can be represented by a graph where vertices correspond to nodes and edges to links. The network models provide a range of flexibility for assigning nodes various attributes and defining links with a range of properties. A wide range of efficient network based models has been developed in the literature for studying diffusion processes~\cite{newman2002spread,keeling2005networks,holme2015modern}. For generating contact networks, a fundamental aspect is to build network structure called the network topology based upon which nodes interact with each other. There are four types of network structure namely regular lattice networks, random networks, small-world networks and scale-free networks that are frequently used to study diffusion processes~\cite{seyed2006scale, wang2003complex,hegselmann1998modeling,hegselmann1998understanding}. In addition, data-driven network models are also derived from real-world data and they may assume some properties of the theoretical models.

Regular lattices are the simplest representation of contact network structures where nodes are only connected to their nearest neighbour nodes in a lattice with a regular fashion ~\cite{hegselmann1998modeling,hegselmann1998understanding}. The regular lattice networks assume large path lengths, i.e. the average distance between two nodes is very high and the clustering coefficient is very high as well. Therefore, these are not realistic~\cite{sen2003small,zhou2005maximal}. The random network models improve regular lattice model where nodes contact with each other in a random fashion and each pair of nodes has an equal probability to be connected. Furthermore, the average path lengths in the random networks match with many real-world networks with appropriate contact probabilities~\cite{barbour1990epidemics}. However, the clustering coefficient is too low for these networks. Recently an approach is introduced called the small-world network model based on six degrees of separation phenomena which states that if you choose any two individuals anywhere on Earth, you will find a path of only six acquaintances on average between them. In a small-world network, most of the nodes are not neighbours of each other, but the neighbours of a given node are likely to be neighbours of each other. The Watts-Strogatz model~\cite{watts1998collective} generates such networks where the existing links of a regular lattice are re-wired with a defined probability. The generated networks assume high local clustering co-efficient and short path length~\cite{newman1999renormalization}. The latest approach to generate contact structure is scale-free networks developed by the Barabasi-Albert model~\cite{albert2002statistical} where node's degree follows a power-law distribution. In the Barabasi-Albert model, the scale-free networks are self-organised with growing and preferential attachment processes. The research found that many real-world systems have a power-law degree distribution~\cite{adamic2001search,clauset2009power}. 

The above structured contact networks can be analysed mathematically and numerically. The authors of ~\cite{morris2004network} has presented surveys on the methods of designing contact networks applying real contact data. These networks are, however, static in nature where node attributes and link properties are not changed during the observation period. These contact networks are often represented with adjacency matrices of binary values. This allows the use of algebra to calculate various network properties and corelate it with diffusion dynamics unfolded on it. While having some strong benefits over compartmental models, these network models still have several shortcomings, namely that in these models, the quality of contacts is overlooked. For example, the duration of contacts affects the transmission probability, and frequency of contact etc. There have, however, been some models to overcome these limitations with weighted contacts~\cite{onnela2007analysis,kamp2013epidemic,chu2009epidemic}. However, the weighted contact networks do not capture burstiness of the contact which is found to have an impact on spreading dynamics~\cite{lambiotte2013burstiness}. The other crucial temporal factor is that the contact sequences among individual are completely missing in these network models.

For studying diffusion processes with realistic contacts, there have been several approaches to make the contact network dynamic as well~\cite{holme2015modern,perra2012activity,shahzamal2018graph}. The dynamic networks assume the links are transient in status, i.e., links appear and disappear. However, the relationship between two linked nodes is often permanent. In the dynamic contact network models, the above static network models can be implemented as the underlying structure (capturing permanent social relationship among individuals) and an additional mechanism is added on top of that to maintain the link dynamic. The dynamic contact network models are often difficult to analyse with exact mathematical solutions. Thus, an approximation is often used to characterise the system dynamics~\cite{valdano2015analytical}. There are no analytical solutions for many dynamic contact network models and such models are only used for simulations to explore diffusion dynamics for wide scenarios of developed models. These classes of contact networks are often efficient tools to validate the simulations results of data-driven individual-level diffusion models. Current approaches for developing dynamic contact network models are discussed in details in Section 2.3.

\subsubsection{Individual-based models}
The other trend of diffusion modelling on contact networks is to apply individual-based models~\cite{bansal2007individual,cauchemez2008estimating,stehle2011simulation,toth2015role}. In these models, all operations are executed at the individual-level and thus the integration of many realistic contact properties becomes easier. The other fundamental concept is that individual-based models are implemented upon a community of targeted individuals and that are situated in an environment. In these models, every individual plays its role and interacts with its respective environment. Thus, infectious items are received by an individual according to its behaviour and surrounding conditions, and it transmits contagious items to other individuals by regenerating it. However, it is not so easy to define the boundaries of the model class based on individual compared to compartmental models or network models as the assumptions in the individual based models varies largely. 

The system dynamics in the individual-based model are generated with all individual actions happened simultaneously within the respective simulation environment of the individual~\cite{holland2006studying}. The respective environment depends on the modelling approaches and it may include parts or sometimes all of the other individuals. Thus, all individuals are affected by the state of neighbours in their simulation environment at the same time. Individual response to a specific environment can be deterministic or can be stochastic events. The reactions process to a simulation environment is often implemented with a set of rules (e.g. IF-THEN operations). For disease simulation, such a rule can be IF the individual is susceptible and if there was a contact with an infected individual, THEN switch the status from susceptible to infectious with a certain probability $P$. Some individual-based models implement the process of adaptation, learning or evolution. These models are called agent-based models, which is a subset of individual-based models, and have simulated intelligence~\cite{kiesling2012agent,copren2005individual, xu2017synthetic, stummer2015innovation}. 

The most significant advantage of individual-based models is that they allow for the inclusion of natural mechanisms for every desired aspect of the model to be as realistic as possible. They can offer characterisation even at the link level and environmental conditioning including complex biological mechanisms~\cite{keeling2008modeling,machens2013infectious}. The current exploration of data on social interactions leverages the benefit of this class of model as they easily allow modelling of individual interactions over time. The complexity of implementing higher-level architecture such as clustering and community structure are reduced as the network formation mechanism is implemented at a lower level. The contact networks created by dynamic contact network models can also be simulated with individual-level models. However, the individual based models become difficult with detailed information and require more effort to analyse sensitivity. To achieve stable insights, repeated simulations are conducted with a high number of parameter combinations. Therefore, the individual based model requires more computing resources and computation time. These models often cannot be analysed mathematically due to their stochastic nature and large number of parameters. This thesis studies diffusion processes on data-driven contact networks using individual level models and also applies synthetic contact networks generated by the dynamic contact graph models. 

\subsection{Challenges in diffusion modelling}
The above discussion shows that the diffusion modelling is a wide area for multi-disciplinary research but their approaches vary largely due to their occurrence in varieties of context. Diffusion dynamics on contact networks have been studied for a long time in the field of infectious disease spreading~\cite{hamer1906milroy}. The current development of communication technology has provided many applications of diffusion processes as well as creating business opportunities~\cite{kermack1927contribution,burke2006individual,meloni2011modeling,koopman2005mass}. Many factors affect the spreading dynamics of contagious items on contact networks. However, it is clear that interaction pattern of individuals is one of the key factors in driving diffusion processes on contact networks. There have, therefore, been a wide range of efforts to understand and integrate the impacts of interaction patterns with diffusion modelling~\cite{mossong2008social,mikolajczyk2008social,hens2009mining,iribarren2009impact}. There are, however, still some critical factors to be addressed in constructing proper diffusion models that capture realistic contact patterns. In addition, current opportunities for gathering individual-level contact data have attracted the researchers to deep dive further in this field by looking at contact patterns at the granular level~\cite{starnini2014time,stehle2011simulation,huang2016insights,tatem2014integrating,shahzamal2018impact}. This thesis investigates one of the key challenges from this field by investigating impacts of individual level interactions on diffusion dynamics.

The basic mechanism for developing diffusion phenomena on a contact network is the execution of a series of individual-level transmissions called local transmissions where contagious items transmit from one individual to other individuals through interactions. These local transmissions occur in two ways by 1) direct interactions and 2) indirect interactions. In the direct interactions, the infected individuals and susceptible individuals are present together at a location or in contact as a friend in OSN~\cite{huang2016insights,gao2015modeling}. For example, diffusion in Mobile Ad-hoc Network (MANET), information diffusion in online social blogs, and disease spreading through physical touch~\cite{shahzamal2017airborne,gruhl2004information,shahzamal2015delay,shahzamal2016smartphones,shahzamal2018lightweight}. Most of the above diffusion phenomena discussed consider only direct interactions to execute local transmissions and spread contagious items. On the other hand, the indirect interactions are created when a susceptible individual interacts with the contagious items deposited by the infected individuals in the absence of the depositor. For example, an individual infected with the airborne disease can deposit infectious particles to the environment~\cite{fernstrom2013aerobiology,brankston2007transmission}. These particles can be suspended in the air for a long time and the susceptible individuals can inhale these particles even after the infected individual has left the interaction location. Therefore, both individuals are not required to be present in the same location for a local transmission through indirect interaction. A similar mechanism is observed in message diffusion in online social blogs. A message posted by a current member in an online social blog can be learnt by a newly joined member, even though the new member was not present when the message was posted and indirect transmission links are created~\cite{gruhl2004information,gao2015modeling}. Therefore, it is not required to be a member of the blog at the same time to learn a posted message. In the ecology, the diffusion can happen following this mechanism as well. Queen message dissemination in the social ant colonies and pollen dissemination in the farms are examples of this mechanism ~\cite{richardson2015beyond}. However, there are no comprehensive studies on how these indirect interactions impact on diffusion dynamics on contact networks. In this thesis, this particular gap of current diffusion models is investigated in detail.

Integration of the impacts of indirect transmission is not straight forward. The interaction mechanism in a contact network defines the network properties such as degree distribution, clustering coefficient and path lengths etc. which in turn determine the diffusion dynamics~\cite{de2014role,freeman1978centrality,britton2007graphs}. It is clearly seen that the inclusion of indirect interaction creates more opportunities for transmitting contagious items and hence increases the diffusion dynamics. Thus, there is the question of how much diffusion dynamics are amplified by including indirect interactions? The analysis of diffusion modelling approaches shows that the current approach does not include indirect interactions in diffusion modelling. Therefore, it is required to include transmission opportunities due to indirect interactions with the diffusion model. Indirect interactions not only increase transmission opportunities but it also changes the network properties. For example, the inclusion of indirect links may increase the contact degrees of individuals adding new neighbours through indirect interactions. Thus, local clustering coefficient may also change. Similarly, temporal network properties can be changed due to indirect interactions~\cite{scholtes2016higher}. These network properties play vital roles in developing different diffusion phenomena such as the emergence of diffusion on contact networks, spreading potential of individuals and controlling of diffusion~\cite{scholtes2014causality,masuda2013predicting}. Thus, the inclusion of indirect interactions in diffusion models also needs to investigate these diffusion phenomena. For exploring diffusion phenomena, synthetic contact network generated by network models and graph models are frequently applied. The current contact network generation models are only developed based on contacts created by direct interactions~\cite{holme2015modern,laurent2015calls,shahzamal2018graph}. Therefore, it is also required to develop a suitable contact network generation model (it can be a contact graph model) that support investigation of diffusion processes with indirect interactions. The following sections discuss the current methods and approaches to solving the raised challenges of diffusion modelling due to including indirect interactions. The spreading of infectious disease is taken as the study case of this thesis.

%% file: 2.2_chap_idis.tex
\section{Infectious disease diffusion}
\subsection{Infectious disease transmission}
An infectious disease is transmitted from an infected individual to susceptible individuals via transferring organisms/microbes capable of causing infection~\cite{fernstrom2013aerobiology,boone2007significance}. These organisms/microbes are called pathogens. In this thesis, contagious items or infectious particles refer to these pathogens. The infectious items enter the body of susceptible individuals and deposit on mucus membranes of body parts such as mouth, nose, throat, and lungs where they can cause an infection. Therefore, for an infectious disease to persist within a population, relevant contagious items are required to be transmitted continuously to new bodies. The contagious items are transmitted through two mechanisms: 1) direct transmission and 2) indirect transmission. Direct transmission occurs through individual-to-individual interactions transferring contagious items without any intermediate transmission medium between these two individuals. The example includes physical touches (such as shaking hands, kissing etc.) and contact of blood and body fluids. Direct transmission is found in infectious disease such as common colds, sexually transmitted diseases etc. For many infectious diseases~\cite{rottier2003controlling,brankston2007transmission,fernstrom2013aerobiology,boone2007significance}, infected individuals generate particles containing infectious microbes by their respiratory activities like talking, laughing, coughing or sneezing. These particles are scattered into the environment of the proximity of the infected individuals. The infectious particles then deposit onto objects or surfaces and survive long enough time to transfer to other susceptible individuals who subsequently touch the objects. This creates the indirect transmission of diseases where intermediate medium or objects are required to transmit infectious particles. Examples of diseases with the indirect transmission are Coronavirus, Rhinovirus, and Influenza etc.

The ways indirect transmission occur are not the same in all cases and can be classified into different modes which are based on the roles of the intermediate medium and the properties of the infectious particles when transmitting through the intermediate medium. The respiratory activities of infected individuals generate droplets containing infectious particles and the sizes of the droplets often define the mode of transmission. Droplet whose size is comparatively large, often assumes to be greater than 5$\mu$m, are transmitted through the air to nearby susceptible individuals. This mode of indirect transmission is called droplet transmission. However, the droplet whose size is small, often assumes to be less than 5$\mu$m, evaporates quickly and becomes droplet nuclei. These droplet nuclei are suspended in the air for a long time and can travel large distances. Thus, they can transmit to susceptible individuals with a long time delay after their generation, even to susceptible individuals who are far away up to 100m from the source infected individual. This mode of indirect transmission is called airborne transmission. Indirect transmission of infectious particles can also happen through vectors (mosquitoes, flies and mites etc.) that carry infectious particle from an infected individual to susceptible individuals with delay and at substantial distances. In this situation, the contagious items present in the blood or skin of an infected individual are ingested by vectors. Then, it is developed in the vectors itself. Susceptible individuals are usually infected through the bite of an infectious vector, though other ways of entry are possible. Examples of vector-borne disease transmissions are yellow fever, malaria, plague and dengue etc.~\cite{rottier2003controlling,brankston2007transmission,fernstrom2013aerobiology,stoddard2009role}. Another indirect mode of transmission is to spread disease through contaminated objects spatially. Examples of such diseases includes water-borne diseases, food-borne diseases~\cite{brennan2008direct,lange2016relevance}.

It has been observed that the spreading of some infectious diseases is dominated by only direct transmission and can be modelled by creating direct transmission links for co-presence interaction between infected and susceptible individuals. However, the infectious diseases that spread based on indirect transmission or have additional indirect transmission along with direct transmission cannot be modelled by creating only direct transmission links for co-presence interactions. This thesis considers airborne infectious disease spreading as a case study for understanding and modelling the impacts of indirect transmission links. For airborne diseases, infected individuals generate droplets containing infectious particles through various respiratory activities. The authors of ~\cite{lindsley2012quantity} have found that an infected individual generates on average 75,000 particles/cough but it can be up to 500,000 particles/cough. They have also found that 60\% of these particles can reach the alveolar region of lungs if the particles are inhaled by another individual. It is found that the cough frequency of an infected individual is on average 18/hr~\cite{loudon1967cough}. Thus, an infected individual deposits about $1.36\times10^{6}$ particles during a one hour stay at a location. Up to 50\% of these particles evaporate and become droplet nuclei (airborne particles) which are suspended in the air for a longer time~\cite{thomas2013particle}. Airborne particles are also added to the environment by breathing, talking and laughing. There have been a wide range of studies to understand the viral load of airborne particles. The studies show that most of the influenza virus is contained in the droplets whose sizes are $<5\mu$m. The works of ~\cite{lindsley2010measurements,lindsley2015viable,lindsley2010measurements} have found that up to 75\% virus is contained within droplets with sizes $<5\mu$m. The exhaled breath of an influenza patient can generate on average 0.5 plaque-forming units (PFU) for influenza viruses~\cite{nikitin2014influenza}. The study of ~\cite{lindsley2015viable} found that a cough can generate up to 77 PFU virus. Inhalation of 0.7 - 3.5 PFU of influenza is sufficient to cause infection in 50\% of susceptible individuals~\cite{alford1966human}. Therefore, it can be concluded that the generated airborne particles have sufficient viral load to cause infection if they are inhaled.

The impact of airborne transmission is different to the other model of indirect transmission as airborne particles can travel spatially while large droplets settle nearby. The literature indicates the various range of travel distances for airborne particles. The travel distances depend on the weather conditions and air-flows. The authors of ~\cite{han2014risk} show that airborne particles can travel up to 100m in the direction of air-flow. The travel distance can also be interpreted from the analysis of SARS outbreak occurred in the Amoy Gardens Hong Kong in the year 2003. The study of~\cite{yu2004evidence} has revealed that the infection had reached the Block-E which was at 60m distance from the Block-B where infection had started, although there was no indication of physical interaction among the residences of these buildings. Thus, it was concluded that the infection particles travelled to the Block-E through airborne transmission. A number of studies have also shown that it is also possible to disperse airborne particles between flats in a building~\cite{mao2015airborne,gao2008airborne} and between wards in a hospital~\cite{beggs2003airborne}. The experiment of ~\cite{gao2008airborne} shows that airborne particles can also travel from one building to the nearby buildings. The travel distance of airborne particles is extended in the open area. The authors of ~\cite{corzo2013airborne} have studied the presence of influenza A virus around pigs farms by collecting air samples at different distances from the farms. They have noticed a significant amount of RNA copies of the virus at the distance of 1.5Km from the pig farms that had influenza A infected pigs. Therefore, the airborne indirect transmission mode of infectious disease has strong potential to spread diseases. The airborne infectious diseases spreading is an important application of diffusion process with indirect transmissions.

\subsection{Force of Infection}
An interaction (e.g. being in the same location) between infected and susceptible individuals poses an infection risk for the susceptible individual. Infection risk assessment can be divided into two steps: determining the intake dose of infectious particles and finding the corresponding infection probability~\cite{sze2010review}. The infectious particles that reach the target infection site are called the intake dose. The intake dose is estimated based on the exposure level to the infectious particles, the pulmonary ventilation rate of susceptible individuals, the exposure time interval, and the respiratory deposition of the infectious particles. Then, the infection probability is calculated by a mathematical formula. Two approaches are applied to determine if an infection occurs: deterministic and stochastic. The first approach assumes that each individual has an inherent resistance up to a dose of infectious particles. Thus, a susceptible individual contracts the disease when a target infection site is exposed to a dose equivalent to or exceeding the threshold dose. In the stochastic approach, any amount of intake dose causes disease with a certain probability. The infectious particles are usually randomly distributed in the suspension medium. Thus, the estimated exposure level and intake dose of airborne particles are always expected values rather than exact values. Therefore, the stochastic models are appropriate for studying airborne disease spread. Models that are frequently used for assessing infection risk for airborne diseases are now discussed.

A wide range of models has been developed for the spread of airborne disease. These range from simple models that are easy to apply to complex models that require greater detail of the disease spreading process. Unfortunately, these details is not always available for many diseases. In the literature, the Wells-Riley model or its modification are widely used to estimate infection risks~\cite{sze2010review,fennelly1998relative}. The Wells-Riley equation is given as  
\begin{equation}
P_I=1-exp\left(-\frac{Igpt}{Q}\right)
\end{equation}
where $P_I$ is the probability of causing infection to a susceptible individual for the intake dose $E=Igpt/Q$, $I$ is the number of infected individuals at the interaction room, $p$ is the breathing rate of the susceptible individual (L/s), $g$ is the average quanta generation rate (quanta/s), $t$ is the exposure time interval, and $Q$ is the room ventilation rate (L/s). The $P_I$ is, in fact, the ratio between the number of infections caused for $E$ and the susceptible individuals. This model is based on the concept of quanta which is the number of droplet nuclei required to cause infection for $63\%$ of all exposed susceptible individuals. The ratio $P_I$ provides the reproduction number of the studied diseases which is frequently used to determine disease spreading dynamics for the large population. The model parameter quanta generation rate $\vartheta $ is required to be estimated from the real outbreak cases. This is very difficult for many diseases as it requires data from real outbreak scenarios. The model is also limited due to its assumption that particles are homogeneously distributed in the air, and that every particle reaches to the target infection site. It does not consider the duration of particle generation.

There have been several modifications to overcome these limitations. The authors of ~\cite{fennelly1998relative} incorporated the effect of respiratory protection system that may filter the inhaled infectious particles by multiplying a fraction term with the intake dose as $E=\frac{Igpt \theta}{Q}$, where $\theta$ is the fraction of infectious particles reached to a target infection site. Air disinfection and particle filtration are used in the many buildings that reduce the effective infectious particles to cause infection. These factors are included for the Wells-Riley equation in the work of ~\cite{nazaroff1998framework}. However, collecting such data is difficult and expensive for large scale simulation. The assumption about the homogeneity of particle distribution in the interaction area is addressed by ~\cite{gammaitoni1997using}. They considered the time-weighted average pathogen concentration in the room air to incorporate the non-steady-state conditions in the Wells-Reily equation. This model is given by
\begin{equation*}
P_I=1-exp\left(-\frac{pIg (Qt+e^{-\varphi t}-1)}{VQ^2}\right)
\end{equation*}
where $Q$ is the air change rate or disinfection rate, $\varphi $ is the particle accumulation rate and $V$ is the volume of interaction area. In spite of these improvements, the Wells-Reily model still requires the total exposure during an outbreak to find the quanta generation rate and that is not possible for many diseases. 

Rudnick and Milton~\cite{rudnick2003risk} developed a model where the exhaled air volume fraction is used to estimate the number of quanta that the susceptible individuals are exposed to:
\begin{equation*}
P_I=1-exp\left(-\frac{ g \bar \omega It}{N}\right)
\end{equation*}
where $\bar \omega $ is the average volume fraction of room air that is exhaled breath and $N$ is the total number of people in the premises. To find the quanta generation rate based on the $\bar \omega $, one requires a knowledge of carbon dioxide concentration in the room. These models still follow the well-mixed assumption of particle concentration. Some works~\cite{gao2008airborne,tung2008infection} address this problem by experimenting the dispersion of tracer gas and integrating impacts with model.

In the models discussed above, the quanta generation rates are not well understood for many diseases. However, the infectious particles generation rates, their formation, pathogen loads and their survivable time etc. are now becoming available. The authors of~\cite{nicas1996analytical} first introduce a dose response model based on the infectious particles concentration instead of quanta. The model is 
\begin{equation*}
P_I=1-exp\left(-\frac{Ig\theta pt}{Q}\right)
\end{equation*}
where $g$ is the number of infectious particles released per infected per unit time and $\theta$ is the fraction of infectious particles reaches the target site. In this equation, the quanta generation rate $g$ is replaced by $\theta g$. The authors defined the source strength $g$ with cough frequency, pathogen concentration in the respiratory fluids and the volume of expiatory droplets introduced into the air in a cough. This model also based on the homogeneity. Recently, the authors of ~\cite{issarow2015modelling} have also introduced a model based on the infectious particle concentration considering non-steady-state conditions as
\begin{equation*}
P_I=1-exp\left(-\frac{I(g -r)\theta p t}{Q}\left[ 1-\frac{V}{QT}\left(1-e^{-\frac{Q \tau }{V}}\right)\right]\right)
\end{equation*}
where g is the particles generation rate, $r$ is the mortality rate of the generated particles, $\theta$ is the deposition fraction of the inhaled particles, $\tau$ is the duration of particle generation, and $t$ is the duration susceptible individuals breath in infectious particles.

In the above equations, the temporal variation in the particle concentration is captured using a non-steady-state model. However, this model assumes that all infected individuals arrive at the same time and this may not happen in reality. The variations in the arrival time of infected individuals also introduce the fluctuations in the particles concentration. The current models also do not capture the exposure that susceptible individuals receive after infected individuals leave the interaction locations. Thus, it would be more appropriate to find exposure level due to contact with each infected individual and sum them up to find total exposure. Therefore, the arrival and departure of each infected individual can be tracked independently and hence the exposure during indirect interactions. This also allows one to assign a random value of $Q$ to each contact to capture heterogeneous particle concentrations at different locations.

%% file: 2.3_chap_net.tex
\section{Dynamic contact graphs}
Contact networks are often generated by graph models to study diffusion processes. However, current graph models allow generating contact networks based on the contacts created by direct interactions. Thus, it is required to develop graph model capable of supporting investigation of diffusion processes having transmission of contagious items through indirect interactions. The development of dynamic graph models for generating dynamic contact networks is still at an early stage compared to the static models. There have been limited number of approaches to develop dynamic graph model. This subsection presents a brief details of current dynamic graph modelling approaches. In this thesis, dynamic contact graphs/networks represent temporal or time-varying graphs/networks where edges between a pair of nodes are dynamic as their availability for transmission are not permanent. The addition or removal of nodes to the graph is not considered in this thesis.

\subsection{Dynamic graphs representation}
The evolution of a dynamic contact graph can be captured in many ways. The evolution in the graph can occur due to changes in the status of nodes and status of links. The links in the static graphs represent a relationship between a pair of nodes and is created there is at least one interaction during observation period~\cite{zhang2016modelling,starnini2013modeling}. In dynamic graphs, however, the links are often differentiated from contacts (links and contacts have different meaning)~\cite{holme2015modern,laurent2015calls,shahzamal2018graph}. The contacts indicate interactions between a linked pair of nodes occurring at certain times during an observation period. Dynamic graph modelling is required to incorporate timing information of these contacts with link dynamics. The development of a dynamic graph model often depends on how the graph is represented. The dynamic contact graphs can be represented in the following ways.

\subsubsection{Contact sequences}
Many real world interaction data sets comes with the entries containing identities of interacted nodes and the time when the interaction happened, even with other some meta information such as gender and locations. The interaction time can be a time stamp  or a time interval sequence. For examples, works of ~\cite{mastrandrea2015contact,stehle2011simulation} have collected interactions between two individuals using RFID and wearable sensors. This representation is a straightforward and practical format computationally. However, analysing diffusion processes on the graphs with this format would be difficult as they do not count some properties such as contact duration. It is also difficult to visualise the contact graphs and hence representing it to audience.

\subsubsection{Multi-layer graphs}
The dynamic graph can be visualised well if it is represented with as a sequence of static graphs. In this method, the observation time is divided into discrete time steps and a static graph is constructed for each time step~\cite{lee2015towards,kivela2014multilayer}. Thus, the dynamic graph becomes a multi-layer graph with each sequence of the static graph as a single layer. This allows one to understand and analyse the dynamic graph using static graph theories and then combine the results for the sequence of times to obtain overall results. This method is applicable where the time resolution is high (or continuous) compared to dynamic process on the studied graph. For studying infectious disease, this method has a limitation as the disease cannot be transmitted over a path during a sequence of the graph and thus analysing multi-layer graphs cannot capture real dynamics. The time-lines representation of contacts is one of the extended approaches where nodes are placed in one axis and times in another axis. The advantage with this representation is that the time-respecting paths (sequences of contacts of increasing times) between nodes are easy to identify as these are all paths that do not turn backwards in the time dimension. The structure of time-respecting paths can be represented as a binary matrix as it is in a static graph with an adjacency matrix. Thus, the dynamic graph can be expressed as a binary tensor. The limitation is that the corresponding adjacency tensor, as a data structure, takes a lot of memory and requires high computational overhead to process such graph ~\cite{valdano2015analytical,bach2013visualizing}.

\subsubsection{Dynamic links graph}
In this representation, temporal variations are captured with only one dynamic graph where nodes and links change their status over time. The underlying graph is a static graph with the fixed links among nodes. In fact, the underlying graph captures the fixed topology of the dynamic graph and can be treated as the foot-print of nodes~\cite{holme2013epidemiologically,sarzynska2015null}. Then, the static graph structure evolves over time where contacts can appear and disappear. This means a time dimension is added with the static network. This approach is considered for the class of graphs where the targeted research question is to understand how the structure has evolved and how it affects the diffusion process unfolded on the graph. The dynamic graphs are typically data-oriented where the focus of study is on a data set, its structure, and how something behaves on it e.g. how disease spreading would behave on the graph. In addition, these observations may vary with used data sets and generalisation of results are often difficult. However, it is used widely due to its flexibility to implement and capture properties of real contact networks. 

\subsubsection{Time-node graphs}
The recent trend of dynamic graph modelling is to extend the concept of node into temporal node i.e at each time step the same node is considered as a different node. Then, the graph is built among the temporal nodes. This approach is called the static expansion of a temporal graph~\cite{michail2016introduction}. This type of graph can be practical since it is straightforward to apply static graph methods also over the time dimension. Eventually one usually needs to map the time nodes back to the original nodes. This requires high computational power which is available in the current technology. However, the applicability is limited by the size of the networks.

\subsection{Dynamic graph modelling approaches}
Based on the above discussion, a general representation of the dynamic contact graph is described here. Consider a dynamic contact graph $G_T$ that is built with a set of nodes $Z$, a set of relationships $L$ between these nodes (links, contacts), and a labelling sets $Y$ which represents any property such as links weights, set of node attributes; that is, $L \subseteq Z \times Z \times Y$. The relations between nodes are assumed to take place over a time span $\Gamma \subseteq \mathbb{T}$ denoting the lifetime of the system. The temporal domain $\mathbb{T}$ is generally assumed to be $N+$ for discrete-time systems or $R+$ for continuous-time systems. The dynamics of the system can be
subsequently described by a dynamic contact graph, $G_T = (Z, L, \Gamma, \phi, \psi )$, where\\
\[ \phi  : L \times \Gamma \rightarrow \{0,1\} \] $\phi$ is called presence function, indicates whether a given link is available at a given
time. The status of node can also be varied over time where they can be active or inactive to create links at a certain time. Thus, the model can be extended by adding a node status function \[ \psi : Z \times \Gamma \rightarrow \{0,1\} \]
where the activation function $\psi$ of nodes depend on a given time. Given a  $G_T = (Z, L, \Gamma, \phi, \psi )$, the graph $G = (Z, L)$ is called underlying graph of $G_T$. This static graph $G$ should be seen as a sort of footprint of $G_T$, which flattens the time dimension and indicates only the pairs
of nodes that have relations at some time in $\Gamma$. The connectivity of $G = (Z, L)$ does not imply that $G_T$ is connected at a given
time instant with the connectivity of $G$. Within this general graph definition, various contact networks can be generated by varying details in links and nodes definition. Thus, graph models are often accompanied with a network generation method. This network generation method handles incorporating details to links and nodes while graph defines the relationships at the abstract level. Some network generation methods that have been designed under the frameworks of various graph models are discussed here.

\subsubsection{Static graphs with link dynamics}
The simplest method of generating a dynamic graph is first to generate a static graph with links using a static graph model and then define a sequence of contacts for each link generated. To avoid complexity, contact generation process is often kept independent of the network position
of the links in this approach. The authors of ~\cite{holme2013epidemiologically} have applied the following procedures to generate a dynamic graph.

\begin{quote}
i) a static graph is constructed from a multigraph that is generated using the configuration model~\cite{albano2013matter} and deleting the duplicate links and self-links

ii) an active interval is generated for each link when contacts can occur i.e. nodes are present in the graph. The duration of active intervals is generated using a truncated power-law. For a link, the active interval starts at a uniformly random starting time within a sampling time frame (observation period)

iii) a sequence of contact times is generated following an inter-event time distribution over the observation period. This also generates burstiness in the contact patterns. 

iv) finally the contact time sequence is wrapped with the corresponding active interval of each link. The contact sequence times (step 2) which are within the active time interval are taken and other are deleted for a link. The wrapping is done for all links in the graph and a dynamic contact graph is obtained 
\end{quote}
The authors of ~\cite{sarzynska2015null} used a similar method to generate contact graph. These methods easily integrate contact dynamics and burstiness to the topology of a static graph. However, the inter-event times are not influenced by the topology structure and node properties. 

\subsubsection{Temporal exponential random graphs}
The exponential random graph model (ERGM)~\cite{wang2009exponential} is widely used to generate static graphs in the study of social dynamics. The ERGM model parameters reflect the importance and weight of selected topological elements and sub-graphs such as triangles and stars. The ERGM model generates a class of graphs and the model parameter inferred from an empirical data capture the corresponding graphs. A similar modelling framework is used by the authors~\cite{hanneke2010discrete, krivitsky2012modeling} to generate temporal exponential random graphs (TERGM). Thus, the temporal dynamics of links are influenced by the network topology in TERGM. In TERGM, a set of states of the nodes observed over a time window for an ongoing dynamical process is applied to estimate the model parameters. Thus, the probability of making a contact between a pair of nodes is bounded with the time window of applied data set for modelling. The temporal exponential random graph models are not node-oriented model (nodes connectivity is not build at the node level, but connectivity is defined based on the network topology) that makes possible to change the network with its basic building block such as links. Thus, it is difficult to achieve a good fit unless the successive networks are close to each other~\cite{snijders2011statistical}.

\subsubsection{Coordinated temporal graph models}
The above models cannot implement node and link level operations such as which neighbour nodes a host node contacts frequently, how recent changes in contact sequences affect the future contacts, whether the contact creation and deletion follow any specific social mechanism. There have been several works to incorporate these characteristics of dynamic contact graphs~\cite{cho2013latent, masuda2013self, stadtfeld2017dynamic, starnini2013modeling}. The work of ~\cite{starnini2013modeling} has developed a dynamic graph model for face-to-face interactions. This is a spatio-temporal graph implemented with a two-dimensional random walk. In this model, the propensity of walking closer to a node is proportional to the attractiveness assigned to it. Therefore, the
more attracted a walker is to its neighbour nodes, the slower its
walk becomes. A similar approach is implemented in the dynamic graph model developed by ~\cite{mantzaris2012model}. This model for online setting and their assumption is that some individuals are much more central in a temporal graph than they are in an aggregated static graph. Thus, random communication partners are assigned to a node by a basal rate and a positive feedback mechanism. The authors have applied stochastic point processes to model dynamic contact graphs. In this model, a node creates and breaks links according to a Bernoulli process with memory. The probability of an event between two nodes increases with the number of events recently occured between them.

\subsubsection{Activity-driven graph modelling}
A comparatively simple dynamic graph modelling approaches is proposed by~\cite{perra2012activity}. This model is called activity driven network modelling (ADN). They adapt the graph sequence
framework of dynamic graph modelling and generate a simple graph $G_T$ at
(it is a discrete time system) time $t$. The graph generation procedures are as follows:

\begin{quote}
 i) node $i$ is assigned an activity potential $a_i$. This is usually done with a power-law distribution. The activity potential is assigned to all nodes in the graph

ii) the graph generation process goes through increasing a time counter to $t$ and assume that $G_T$ is empty, i.e. all $N$ nodes have no link and contact memory from previous time step

iii) activate node $i$ with a probability $a_{i} \Delta t$. If node $i$ is activated, it is connected with other $m$ randomly chosen distinct nodes. Repeat the step 2 and step 3
\end{quote}

The distinguishing characteristics of this model are that the activity of the nodes governs the link creation. In contrast, the previous models are connectivity driven where the network's topology is at the core of the model formation. The ADN overcomes the timescale separation assumption and explicitly accounting for the concurrent evolution of the interactions in a graph and the dynamic process evolving on it~\cite{rizzo2016innovation}. The studies of ~\cite{gomez2011nonperturbative,sun2015contrasting} show that many important aspects of the system dynamics can be characterised using a heterogeneous mean-field approach. Interestingly, it is found that some system properties are directly related to the activity potential of nodes. The diffusion dynamics are quite different from that of aggregated static networks~\cite{sun2015contrasting}. This thesis takes the ADN approach for developing an contact graph generative model capturing indirect interactions. However, the basic ADN model has several limitations such as nodes contact with a fixed number of links during each activation, the contacts are not repetitive, and social structure among individual is not maintained. Thefore, it is required to incorporate these properties with the basic model.

\subsection{Realistic activity driven graph models}
When nodes are active in the basic activity driven network, they randomly create connections with other nodes. In other words, at any time, a node may connect with any other nodes in the graphs. That means this method does not apply any prior knowledge of social or geographic relationships that could alter the selection of one link over another. However, this assumption severely challenges the feasibility of the ADN modelling when it is implemented for real graphs. There have been great efforts to incorporate realistic features with the basic ADN. An individual can interact with other individuals arbitrarily or by choice. Thus, some links tend to be persistent in time and such links are generated in the household, at the office and with close friends. This property is integrated introducing memory effects in the link formation. Therefore, social graphs can have two types of links. The first class describes strong ties that identify time repeated and frequent interactions among specific couples of nodes. The second class characterises weak ties among agents that are activated only occasionally. It is natural to assume that strong ties are the first to appear in the system, while weak ties are incrementally added to the contact set of each node. This approach is studied in the work of ~\cite{karsai2014time} where it is assumed that a node will connect to a new node with a probability $P(n+1)=\frac{\eta}{\eta+n}$, where $n$ is the current contact set sizes of the node and $\eta$ is the tendency to broaden their contact set sizes. Therefore, the probability of contacting with a node from previously contacted nodes is $1-P(n)$. This method of repeating with the old contacted nodes and extending contact set size is called the reinforcement process. In the above process, every node has the same tendency to extend the contact graph. In reality, however, individual have heterogeneous tendency to extend the contact set size. This issue is addressed by the work of ~\cite{ubaldi2017burstiness} where they proposed to assign a heterogeneous value of $\eta$. For a node $i$, the probability of contacting a new node is given by
\begin{equation*}
    p(n_i)= \left(1+\frac{n_i}{\eta_i}\right)^{-\alpha _i}
\end{equation*}
where $\alpha_i$ is the reinforcement of node $i$ and $\eta_i$ is the characteristic number indicating the size of contact set size before reinforcement start. The value of $\eta_i$ is often assigned with power-law and the distribution of resultant contact set size will be a power-law. 

In the social graphs, individuals have a tendency to make a close social circle and make a community. Therefore, the underlying social structure of a dynamic contact graph also should maintain community structure. The community structure emerges in a graph by creating triadic closure when making a new connections~\cite{bianconi2014triadic}. One typical mechanism to make triadic closure is to use common neighbours (CN) indices, where two nodes $i$ and $j$ are going to interact if their neighbour nodes set has substantially overlap. This means that the probability of these two nodes interacting is proportional to the number of common neighbours. In other words, triadic closure can be created if the host node chose a new neighbour from its neighbour's neighbour~\cite{bianconi2014triadic,daminelli2015common}. The random new neighbour selection mechanism of basic activity driven graph can not emerge the community saturate. The work of ~\cite{laurent2015calls} has upgraded the basic ADN integrating a triadic closure creation mechanism. If a host node $i$ has not contacted any node yet, it randomly picks another node from the entire graph $j$ and creates a link. Otherwise, the host node tries to make a new link with the triadic closure mechanism. As the first step, it selects randomly one neighbour node $j$ from his contact set with a probability. If node $j$ is not selected or has no other neighbours node except node $i$, node $i$ looks for another random new node and creates a link. If node $j$ is elected and has neighbour nodes, then it selects a random neighbour node $k$ from the neighbour set of $j$. Then node $i$ from a link with $k$ and create triadic closure.

The basic activity-driven graph model assigns heterogeneous potentiality to nodes. This generates a heterogeneous distribution of interactions. However, the research on social interaction shows that individual's interactions have bursty nature, i.e the inter-event time of activation of nodes in ADN is required to be heterogeneous~\cite{stehle2010dynamical,lambiotte2013burstiness}. This bursty activity has a strong influence on graph evolution and diffusion unfolded on it. The inter-event time $t_i$ is directly connected with the activity of node $i$ and can defined as $a_{i}=\frac{1}{<t_i>}$. The inter-event time usually spans over several orders of magnitude. The authors of ~\cite{ubaldi2016burstiness} shows a mechanism to capture this bursty nature of human dynamics using a power-law distribution. The basic ADN can generate heterogeneous contact degree distribution based on the value of $a_i$. However, the growing of contact set size is completely dependent on $\eta$. In addition, the model can capture link heterogeneity of directed graphs where the in-link propensity can be different to out-link density. This issue is addressed by the work of ~\cite{pozzana2017epidemic} where each node is assigned an attractiveness. When a node selects a new neighbour node, a node will be chosen based on its attractiveness. However, any current modification of ADN can generate contact networks with links for indirect interactions. Thus, it is necessary to modify the ADN model to create links for indirect interactions.

%% file: 2.4_chap_pro.tex
\section{Controlling diffusion dynamics}
\subsection{Controlling diffusion in contact networks}
Controlling diffusion dynamics on individual contact networks has a wide range of applications ranging from marketing products to mitigating the spread of infectious diseases. The methods developed for controlling diffusion depends on the context and applications. For example, controlling diffusion for marketing a products focuses on maximising the spreading of items to the largest proportion of populations~\cite{van2007new, peres2010innovation} while diffusion controlling for infectious disease focuses on minimising the number of infections reducing the number of individuals received spreading items~\cite{yang2016optimal}. However, the key task in all controlling methods is to find a set of individuals and change their behaviours to alter spreading rates. These individuals often have high spreading potential and are called super-spreaders. The size of the set should be minimal to reduce vaccination cost as well as achieve the control goals. 

Most of the efforts of developing a control strategy are put on finding the optimal set of individuals. Accordingly, researchers search for the contact properties of individuals and their behaviours relevant to the spreading of contagious items~\cite{al2018analysis,scholtes2016higher,kas2013incremental}. Individual's preference and exposure intensity to contagious items, personal status and their surrounding environment often define the spreading potential. Thus, the influential individuals are often searched based on the individual's behaviours~\cite{ma2016rumor,ma2018rumor}. The network properties such as the number of connection of individuals to others are also key factors to determine one's spreading potential. Understanding personal behaviours and modelling is a complex process. In addition, this thesis focuses on understanding the impacts of contact properties on diffusion dynamics. Therefore, the contact properties based methods are investigated to develop control strategies with indirect interactions. 

Various measures of network properties are applied to find influential individuals for controlling diffusion of contagious items. The widely used measures to find the important individuals are degree centrality, betweenness centrality, k-core score and PageRank centrality etc~\cite{al2018analysis}. Individuals contact degrees defined the number of connections to other individuals is frequently used as topological measures of influence. In the social contact networks with broad degree distribution, it is observed that individuals with high degree determine the diffusion dynamics~\cite{albert2000error}. However, degree based methods sometimes underestimate the low degree individuals that can be influential through connecting high degree individuals. The page ranking algorithm developed to rank the content in the World Wide Web is also adapted to find the pivotal individuals in social contact networks. The ranking mechanism of PageRank is simple and straight forward~\cite{brin1998anatomy} where the importance of page is measured by counting the number and quality of links to that page. The PageRank algorithm is only applicable to directed networks. Betweenness centrality is also a good candidate in many applications as it is a measure of the number of shortest paths passing through one individual ~\cite{freeman1978centrality}. Thus, it is most likely that individuals having high betweenness centrality will play key roles in shaping the diffusion dynamics on the networks. It is efficient but requires high computational resources. Moreover, it is applicable for undirected networks. The K-core score present the positions of individuals in the social networks with the k-index obtained by iteratively removing k-degree nodes~\cite{wuchty2005evolutionary}. These networks measures are based on static networks and may not provide optimal performance in dynamic contact networks. 

The one way of finding influential nodes in dynamic contact networks is to use temporal versions of traditional centrality metrics~\cite{scholtes2016higher,kas2013incremental}. In this work, time respecting paths, paths are created based on the time order of links availability, are used to calculate the betweenness centrality and closeness centrality. Eigenvector centrality is also modified for dynamic networks~\cite{taylor2017eigenvector}. These methods require complete information regarding contact networks. The random walk is applied in ~\cite{rocha2014random} for measuring temporal centrality. This does not require global information. However, all these algorithms require huge computational resources. The temporal centrality measurement is in the early stage and is not still feasible for applying in large social contact networks. 

Applicability of these methods depends on the application scenarios. For example, betweenness centrality can be applied to find the influential individuals in online social contact networks as the contact information often is available~\cite{al2018analysis,newman2005measure}. However, it is difficult to apply this approach for controlling disease spreading as it is quite difficult to collect contact information of a population. The vaccination strategies are required to be developed based on the contact information that can be obtained locally. The following section describes the current approaches for controlling diffusion through vaccination strategies.

\subsection{Diffusion control with vaccination strategies}
Infectious disease spreading is controlled by implementing vaccination strategies. The key task of a vaccination strategy is to choose a set of individuals based on the local contact information. The simplest way of selecting a set of individuals is to choose randomly from the population and is called random vaccination~\cite{madar2004immunization,cohen2003efficient,lelarge2009efficient}. This approach does not consider the disease spreading behaviours of the chosen individuals. Therefore, the information collection cost is minimal. However, it requires a large set of individuals to be vaccinated for achieving hard immunity to disease. Thus, the research is directed to select the individuals who have strong disease spreading potentials~\cite{pastor2002immunization}. These methods are called targeted vaccination. In the targeted vaccination, the number of individuals to be vaccinated is often small and the effectiveness of strategies is substantially high if an appropriate set of individuals are chosen. Therefore, the infection cost can be substantially lower in a targeted vaccination strategy with the reasonable cost of information collection and vaccination cost.  

There has been a wide range of vaccination strategies using obtainable contact information~\cite{deijfen2011epidemics,britton2007graphs,mao2009efficient}. All these methods do not develop vaccination strategies based on local contact information. Sometimes global information is also used and methods are developed to find the global metrics with locally obtainable contact data. There have been several other methods that depend on the movement behaviours of individuals instead of collecting information on interactions between individuals. Vaccination strategies are also varied based on the implementation scenarios. There are two specific vaccination scenarios: preventive vaccination (pre-outbreak) and reactive vaccination (post-outbreak). The vaccination strategies that apply local contact information is now first discussed. Then, the implementation of vaccination strategies is discussed. As examining vaccination strategies in the real-world scenarios are expensive and difficult, empirical contact networks or synthetic contact networks are frequently applied to test and validate developed vaccination strategies. The discussion includes the network model based vaccination strategies.

\subsubsection{Vaccination strategies}
The authors of~\cite{cohen2003efficient, britton2007graphs} present an elegant way of implementing vaccination using local contact information called acquaintance vaccination (AV). According to this strategy, a randomly picked node is asked to name a neighbour node to be vaccinated. Therefore, no knowledge of the node degrees and any other global information of network are required. In fact, it selects the node that has a large number of connection to other nodes. Its efficiency greatly exceeds that of random vaccination. The acquaintance method is also improved in few other ways. Instead of vaccinating random acquaintance, it is more effective to vaccinate the acquaintance who has more frequent contact~\cite{deijfen2011epidemics}. That means the selected individual should be asked to name friends who contact frequently. This method substantially improves the efficiency of AV strategy. The works of ~\cite{holme2004efficient, chen2015improved,gallos2007improving} have shown that if neighbour nodes with many connections are vaccinated then the performances are improved significantly.

The acquaintance based strategies still require a large number of individuals to be vaccinated to achieve control goal. Thus, there have been a fair amount of works to search for other contact properties that can be obtained locally. The analysis of social contact networks shows that individuals are connected to various communities~\cite{girvan2002community,guimera2003self}. These properties of the contact network are exploited by several works where the concept of the bridge nodes is introduced as these nodes provide the pathways for a disease to propagate from one community to another community. Therefore, vaccinating such nodes will be a more effective strategy than of selecting random acquaintance. However, it requires the searching methods that use only the local contact information. The searching algorithm developed by the work of ~\cite{gong2013efficient} can find the bridge nodes using stochastic searching methods that need only local structural information. They found that the developed strategy based on these bridge nodes is more efficient than random strategies. A similar approach is applied in ~\cite{salathe2010dynamics} where bridge hub nodes, nodes bridging between two communities, are chosen for vaccination.

The above vaccination strategies have been developed based on static network properties. However, the real-world social contact networks are dynamic which has a strong impact on disease spreading and hence designing a vaccination strategy. The study~\cite{mastrandrea2015contact} shows that the contact rates between a pair of nodes are broadly distributed. Therefore, the selection of an acquaintance in AV strategy is not sufficient to find the appropriate nodes to be vaccinated. For example, the infection risk for being in a contact with an infected individual is relevant to the contact duration. Moreover, there is a higher risk if one susceptible interacts frequently with the infected individual. The works of ~\cite{lee2012exploiting,starnini2013immunization,masuda2013predicting} consider this information in neighbour selecting instead of selecting random neighbours. The authors of~\cite{lee2012exploiting} use the most recent contact for vaccination and they also apply weight to capture the contact rates with the neighbouring nodes. Theses vaccination strategies outperform acquaintance vaccination.

The collection of contact information is often difficult. Thus, vaccination with detailed local contact information may be infeasible in real-world scenarios and lose the benefit of using the local contact information. There have been several other vaccination models based on individual movement behaviours where individuals contact properties are not consider explicitly~\cite{mao2009efficient}. Beyond the contact properties, the work of ~\cite{mao2010dynamic} consider the individuals who travel long distance for vaccination. The similar approach is taken by ~\cite{miller2007effective} where individuals who visit many locations are vaccinated. 

\subsubsection{Vaccination implementation}
Vaccination is implemented for two types of scenarios: preventive vaccination and post-outbreak vaccination~\cite{takeuchi2006effectiveness}. In the preventive vaccination, a set of individuals are vaccinated to protect the population from future outbreaks. In the vaccination processes, there are three costs: i) information collection cost, ii) vaccination cost, and iii) infection cost~\cite{holme2017cost}. Using a ranking procedure, the implementation target is to achieve the optimal costs of vaccination. Thus, vaccination implementation procedure defines the cost of a specific ranking method. For example, the degree based method can be applied to collecting degrees of all individuals and vaccinating top-ranked individuals. In a second way, a threshold can be set for degree centrality. If an individual has degree greater than the threshold, the individual will be vaccinated. Using threshold information collection cost can be reduced. The other approaches to applying vaccination to the more vulnerable region. 

In the post-outbreak scenario, the vaccination is implemented after noticing disease emergence. The purpose of vaccination is to eradicate the disease as early as possible with causing minimal infection cost. The post-outbreak vaccination can be implemented by mass vaccination called population level vaccination and node level vaccination~\cite{porco2004logistics,keeling2003modelling}. Mass vaccination can be implemented by top-ranked individuals of targeted vaccination strategies. The other popular approach is to find the infected individual and vaccinate their acquaintance randomly. The above section has described several approaches to improve acquaintance vaccination. The acquaintance can be deployed selecting high ranked neighbours. In the post-outbreak scenarios, the strategies should perform well if partial information about the infected nodes is obtained. The node level vaccination may not provide strong protection from disease spreading as the outbreak can start again from the non vaccinated individual. Thus, population vaccination can be applied with the combination of ring vaccination~\cite{kucharski2016effectiveness,keeling2003modelling} where a proportion of neighbours is vaccinated. Similar to preventive vaccination, post-outbreak vaccination can be applied to the endemic region. 

The review of the vaccination strategies shows that node ranking is conducted based on the contact information about neighbouring nodes. However, in the diffusion processes with indirect interactions, it is difficult to identify the neighbours contacted through indirect interactions. Thus, there is a need for an investigation to understand the efficiency of strategies with indirect interaction and find the best strategy.

%% file: 2.5_chap_con.tex
\section{Research gaps and questions}
\subsection{Research gap analysis}
Diffusion process modelling on dynamic contact networks is an attractive research domain due to its wide range of applications. The underlying contact patterns are considered as the key factors for shaping the diffusion on contact networks. The literature review shows that there has been a large body of works to capture real contact properties with diffusion models ranging from epidemic models to dynamic contact network models. This trend of research has been fuelled by the recent mass gathering of individual contact data and researchers deep dive into microscopic contact mechanisms relevant to diffusion. However, there is no comprehensive study in the current literature for a class of diffusion processes where contagious items can be transmitted through indirect interactions. If the indirect interactions are included with the diffusion model, it changes the network properties and hence diffusion dynamics. Therefore, there are several open research issues for including indirect links with the diffusion model. Some of the key research gaps are discussed below.

\subsubsection{Diffusion dynamics}
The literature review shows that indirect transmission is a strong transmission route for airborne disease spreading. The transmission route becomes stronger as the airborne particles stay a long time at the visited places of the infected individual and disperse spatially. There have been few works that indicate the importance of indirect routes for disease spreading~\cite{lange2016relevance,sorensen2014impacts,huang2010quantifying}. However, there is no study on how these routes play roles at the individual-level and their impacts on diffusion dynamics. For studying diffusion dynamics, it is required to determine infection risk for each interaction between infected and susceptible individuals. There have been several models to calculate the infection risk for direct interactions with infected individuals~\cite{rudnick2003risk,fennelly1998relative,nazaroff1998framework}. The model proposed by Wells-Riley and its variations are widely used to assess infection risks~\cite{sze2010review,fennelly1998relative}. However, these models do not consider the stay times of infected and susceptible individuals at the interaction locations separately. It is assumed that all individuals arrive at the same time. Thus, they can not identify the indirect periods and capture the transmission during indirect interactions. The infection risk assessment model, therefore, has to be upgraded to include indirect transmissions. The indirect transmissions create the opportunities to spread disease at different locations in parallel by an infected individual, such as at the locations they recently left as infectious particles may prevail creating indirect transmission, and at the current location for direct transmission. Therefore, disease transmission can occur at multiple locations by the same infected individual at the same time. No model discussed in the literature account for this approach of diffusion. There is, therefore, a need for a new diffusion model.

\subsubsection{Contact graph modelling}
The analysis of graph models has shown that the dynamic graph models capture the realistic contact patterns among individuals~\cite{holme2013epidemiologically,holme2013epidemiologically,holme2015modern}. However, the limitations of these graph models are that they do not consider creating links with indirect transmissions. The other features of contact networks with indirect transmission links are that individuals create links at multiple locations in parallel. This feature is not addressed by current graph models. In the spreading of airborne diseases, infected individuals move at various locations and spread disease. Therefore, the activity driven time-varying network model (ADN) can capture these movements of individuals through the mechanism of activating nodes at a time step with a certain probability~\cite{perra2012activity}. In the case of disease spreading, however, the node is required to be active for consecutive time steps to form a visit to a location. Thus, the current models are required to be extended to stay active for a consecutive time step. During the active states in ADN model, nodes create links with other nodes. However, the arrival of an infected individual at a location may not create links with neighbours as susceptible individuals may arrive at the location later on. It also requires a mechanism to allow creating links after a node leaves the location. Thus, the node can create indirect transmission links. The ADN model deletes all links from the previous time step. But, neighbours can stay at a location for a consecutive time steps. Thus, there is a need to preserve links form the previous time step.

\subsubsection{Diffusion phenomena}
The network architecture defined by the contact patterns has profound impacts on the diffusion phenomena~\cite{perra2012activity}. The inclusion of indirect links may bring changes in the network properties. The direct transmission links between two individuals get stronger with adding indirect transmission links. Thus, the spreading power of the infected individual may increase. Secondly, new neighbours can be connected to the host individual for adding indirect transmission links. This changes the degree distributions, clustering coefficient, and path lengths etc. in the networks~\cite{de2014role,freeman1978centrality,wang2003complex}. This also modifies the node's position in the network. Thus, the spreading behaviours of nodes change when considering indirect transmissions. Then, there is a question of what potential the indirect transmission links have to characterise network properties and diffusion dynamics. Does they bring any new network properties? How do these changes affect the diffusion phenomena such as super-spreaders and the emergence of diffusion? There is no analysis in the literature for these questions on diffusion dynamics with indirect transmissions.

\subsubsection{Diffusion controlling}
The literature review has shown that the contact mechanism is the key features to design the diffusion controlling methods~\cite{holme2004efficient,britton2007graphs}. The controlling strategies of disease spreading are studied through vaccination strategies. The higher order network metrics are not applicable in vaccination strategies as they are very complex and expensive to collect. Thus, local contact information based strategies are developed. The literature review shows that current vaccination strategies frequently use contact information about neighbouring individuals~\cite{cohen2003efficient, britton2007graphs}. However, there might be neighbours who are connected with indirect interactions only. Then, it may not possible by the host individuals to identify these neighbours. Thus, the current vaccination strategies may not be effective in realistic networks with indirect links. There is a need for a vaccination strategy working with both direct and indirect interactions. 

\subsection{Research questions}
Based on the research gaps analysed above, this study formulates four research questions to investigate in this thesis. Each research question below is followed by the general methodology that will be applied to answer that question. The more specific methodologies for individual tasks will be described in the relevant chapter. 

\begin{quote}
\textbf{RQ1: How can diffusion dynamics on dynamic contact networks capture indirect transmission opportunities?} \\
Current diffusion models only consider direct transmission links. Therefore, it is required to integrate the indirect transmission opportunities to the diffusion dynamics. The current SPST diffusion models calculate the transmission probability of a link based on the duration susceptible individual stay together with the infected individuals (direct interactions only). The risk assessment models of airborne disease do not account the arrival and departure time of both susceptible and infected individuals at the interacted locations (Section 2.2.2). Therefore, this model can not capture indirect transmission. The task of Question 1 is to find a way to integrate arrival and departure time with the risk assessment model. Therefore, they would be capable of capturing transmission during indirect interactions. Secondly, each visit for locations is required to distinguish and account transmission separately. 
\end{quote}

\begin{quote}
\textbf{RQ2: How can the underlying dynamic contact network for a diffusion process be modelled including indirect interactions among individuals? }\\
The synthetic contact network is a widely used tool to study the diffusion process. However, the current contact graph models can only generate contact networks with direct transmission links. In the SPDT diffusion model, infected individuals visit various locations and transmit disease. Therefore, the activity driven time-varying network model can be adapted for modelling contact graph with indirect transmission links. The activation of nodes resembling the activation of an infected individual at a location. Activation of a node at a location creates one copy of it. This copy of the node is responsible for creating links with other nodes for that location.

\end{quote}

\begin{quote}
\textbf{RQ3: What is the importance of indirect links in developing diffusion phenomena on dynamic contact networks?}\\
The SPDT model includes indirect transmission links with the existing direct transmissions links and hence transmission probability is increased. Therefore, the probability of emerging diffusion is increased as each individual has stronger transmission opportunities as well as receiving opportunities. In the SPDT model, there are some nodes who do not have direct transmission links during their infectious periods. These nodes are invisible in the SPST model and are called \textit{hidden spreaders}. As the hidden spreaders are connected through direct links only, their potential indicates the potential of indirect links. The potential of indirect links is studied through the behaviour of hidden spreaders.
\end{quote}

\begin{quote}
\textbf{RQ4: How can diffusion dynamics be controlled on dynamic contact networks having indirect transmission links? }\\
The diffusion controlling strategy in a contact network is strongly influenced by the contact properties among individuals. The previous controlling strategies in the literature are developed with contact properties formed with direct transmission links. The disease spreading is controlled through vaccination strategies. The literature review shows that contact information with the neighbour is collected. However, the neighbour contacted through indirect interactions cannot be seen by the host individuals. Thus, the neighbour based method may not be effective. It is also found that several vaccination strategies considering movement based methods that can be adopted for vaccination in the SPDT diffusion model.
\end{quote}
\vfil \break

%% file: 3_chap_spdtmodel.tex
\chapter{SPDT Diffusion Model}
This chapter describes a diffusion model called the same place different time transmission based diffusion (SPDT diffusion) for including the indirect transmission opportunities of contagious items in diffusion phenomena on dynamic contact networks. An infection risk assessment method for links is integrated with the SPDT model. Then, the SPDT diffusion behaviours are investigated simulating airborne disease spreading on data-driven contact networks which are built using GPS location updates of users from a social networking application Momo. Finally, SPDT diffusion dynamics are characterised investigating diffusion on networks where underlying connectivity are varied. This chapter addresses the first research objective of this thesis and answers the question of how can diffusion dynamics on dynamic contact networks capture indirect transmission opportunities? The simulations setup and developed contact networks are often used in the next chapters as well.

\input{3.1_chap_intro.tex}

\input{3.2_chap_spdtpr.tex}

\input{3.3_chap_riskmodel.tex}
\input{3.4_chap_diffusion.tex}
\input{3.5_chap_discussion.tex}

%% file: 3.1_chap_intro.tex
\section{Introduction}
The proposed SPDT diffusion model is studied in the context of the spreading of airborne infectious diseases. The spreading processes of airborne diseases on dynamic contact networks capture elaborately various aspects of SPDT diffusion. Thus, the defined SPDT diffusion model captures disease transmission events occurring for both the direct interactions where susceptible and infected individuals are present at visited locations and indirect interactions where susceptible individuals are present or arrive after the infected individuals have left the locations. In this model, the disease transmission probabilities of links depend on various factors such as the presence of susceptible and infected individuals at the interaction location, removal rates of infectious particles from the interaction area and environmental conditions of interaction area~\cite{issarow2015modelling,sze2010review,fernstrom2013aerobiology}. To assess the infection transmission risk for interaction between infected and susceptible individuals in airborne disease spreading, the Wells-Riley model and its variants are widely used~\cite{issarow2015modelling,sze2010review}. These models determine infection risks based on direct transmission only. Thus, a new SPDT infection risk assessment model is developed based on the Wells-Riley model that accounts for indirect transmission among susceptible and infected individuals as well as the impacts of environmental and structural factors of interaction locations through a model parameter called particles decay rate.

The SPDT and SPST diffusion dynamics are analysed through simulations of airborne disease spreading on real contact networks constructed from location updates of a social networking application called Momo~\cite{thilakarathna2016deep}. About 56 million location updates from 0.6M Momo users of Beijing city are processed to extract all possible direct and indirect disease transmission links. The constructed SPDT contact network has 6.86M SPDT links connecting 364K individuals. Exclusion of indirect links from the above SPDT network provides the corresponding SPST networks with only direct transmission links. These networks are reconfigured to build four other contact networks, controlling link densities, for characterising and comprehensively understanding SPDT diffusion dynamics. To simulate SPDT based airborne disease spreading on these networks, a generic Susceptible-Infected-Recovered (SIR) epidemic model~\cite{allen2008mathematical} is adopted. Realistic disease parameters from the literature are used for the simulations~\cite{huang2016insights,alford1966human,yan2018infectious,lindsley2015viable}.

Integration of indirect transmission links makes the underlying connectivity of SPDT model denser and stronger comparing to SPST model. This is because the individual may connect with the new neighbouring individuals through the indirect transmission links and may have additional links with the already contacted neighbours. Hence, SPDT diffusion dynamics are amplified relative to the SPST diffusion dynamics. However, the underlying contact dynamics of SPDT model vary with the transmission decay rate of contact links as the delayed indirect interaction opportunities shrink at the faster decay rates. The variation in the underlying connectivity affects the diffusion dynamics~\cite{genois2015compensating,lopes2016infection}. Therefore, the amplification of SPDT diffusion is studied for various transmission decay rates of SPDT links. The airborne disease parameters such as infectiousness of particles and infectious period of a disease~\cite{rey2016finding} also influence the spreading dynamics and interact with the effects of transmission decay rates of SPDT links. Thus, the impact of decay rates varying with biological disease parameters is studied in this chapter as well. Then, the SPDT diffusion is compared with the SPST diffusion to identify the novel behaviours that the SPDT model introduces. The identified novel behaviours are verified through various simulations with controlling underlying contact dynamics.  Finally, the reproducibility of the SPDT diffusion dynamics by the current SPST models is studied. The key contributions of this chapter are:
\begin{itemize}
\item  Introducing an SPDT diffusion model;
\item Integrating indirect transmission in SPDT diffusion;
\item Understanding SPDT diffusion dynamics in various scenarios;
\item Identifying novel SPDT diffusion behaviours.
\end{itemize}

%% file: 3.2_chap_spdtpr.tex
\section{Proposed diffusion model}
This section first describes the SPDT model and its link creation mechanism. Then, an SPDT infection risk assessment model is developed to include the indirect transmission links in the diffusion dynamics. The SPDT risk assessment model considers the characteristics of SPDT links in determining the transmission probability. 

\subsection{SPDT diffusion processes}
The SPDT diffusion is a representative of a broad range of applications. In the case of respiratory-based infectious diseases spreading, the SPDT diffusion may occur as the infected individuals leave their organisms in the environment which can transfer to the susceptible individuals and get contrast with the disease~\cite{li2009dynamics}. In online social networks, information diffusion within virtual spaces such as groups can include indirect spreading as the influence of a post in a group affect a newly joined members after the original poster has left the group. Similar processes occur in the communication of ant colonies~\cite{richardson2015beyond}. Without loss of generality, the SPDT diffusion mechanism can be described by considering the spread of airborne diseases where infected individuals deposit infectious particles at locations they visit. These particles persist in the environment and are transferred to susceptible individuals who are currently present nearby (direct transmission) as well as to individuals who visit the location later on (indirect transmission)~\cite{fernstrom2013aerobiology,li2009dynamics}. Therefore, the contagious items transmission network in the SPDT diffusion model is built at the individual level with the direct and (delayed) indirect transmission links.
\begin{figure}[h!]
    \centering
    \includegraphics[width=0.88\linewidth, height=9.0 cm]{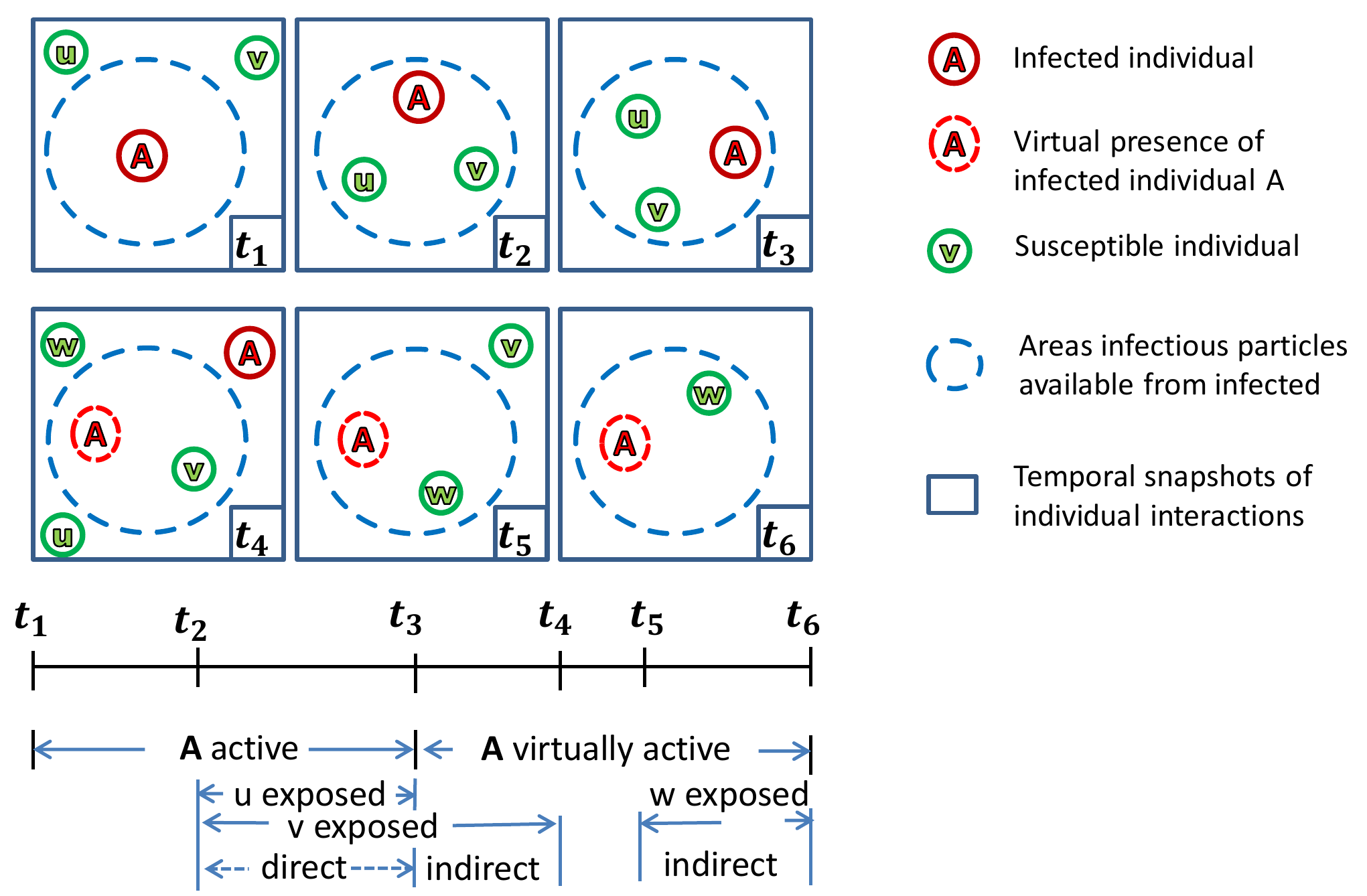}
    \caption{Disease transmission links creation for co-located interactions among individuals in SPDT model. The upper part shows the six snapshots of interactions over time and the lower part shows the periods of exposure through direct and indirect interactions. Susceptible individuals are linked with the infected individual if they enter the blue dashed circle areas within which infectious particles are available to cause infection}
    \label{fig:spdtp}
\end{figure}

The link creation procedure for this model is explained by airborne disease spreading phenomena as shown in Fig.~\ref{fig:spdtp}. In this particular scenario, an infected individual A (host individual) arrives at location L at time $t_{1}$ followed by the arrival of susceptible individuals u and v at time $t_{2}$. The appearance of v at L creates a directed link for transmitting infectious particles from A to v and lasts until time $t_{4}$ making direct contact during $[t_{2},t_{3}]$ and indirect contact during $[t_{3},t_{4}]$. The indirect contact is created as the impact of A still persists (as the virtual presence of A is shown by the dashed circle surrounding A) after it left L at time $t_3$, due to the survival of the airborne infectious particles in the air of L. But, the appearance of u has only created direct links from A to u during $[t_{2},t_{3}]$. Another susceptible individual w arrives at location L at time $t_{5}$ and a link is created from A to w through the indirect contact due to A's infectious particles still being active at L. However, the time difference between $t_5$ (arrival time of w) and $t_4$ (departure time of A) should be the maximum $\delta$ such that a significant particles concentration is still present after A left at $t_4$.

The infected individual makes several such visits, termed as active visits, to different locations and transmits disease to susceptible individuals. However, visits of infected individuals at locations where no susceptible individuals are present or visit after the time period $\delta$ do not lead to transmissions and hence are considered as invalid visits. During the active visits, directed transmission links between infected individuals and susceptible individuals are created through location and time. These links are called SPDT links that can have components: direct transmission links where both susceptible and infected individuals are present and/or indirect links where susceptible individuals are only present or arrive after the infected individual has left the location $L$. Unlike SPST diffusion, that only considers the concurrent presence of interacting individuals for contacts, defining the link duration and their impacts on diffusion in the SPDT model is not straight forward. The disease transmission probability over an SPDT link is influenced by the indirect link duration along with the direct link duration, the time delay between neighbour and host appearance at the interaction location, removal rates of infectious particles from the location, and spatial features of the location. The removal rates of infectious particles depend on air exchange rates at the interacted areas, temperature, humidity and wind flow directions etc. ~\cite{shahzamal2017airborne,fernstrom2013aerobiology}.

\subsection{Infection risk for SPDT links}
A new infection risk assessment model is developed to determine the transmission probability of the SPDT links. The proposed model incorporates the above characteristics of SPDT link in determining infection risk. The Wells-Riley model and its variation with homogeneous infectious particles distributions in the interaction area are simple and are widely used~\cite{sze2010review, tung2008infection}. On the other hand, the models capturing heterogeneous distributions of particles can provide more accurate results. However, heterogeneous models are inconvenient for simulating airborne diseases on large scale contact networks as finding the appropriate parameters is complex and expensive. An infection risk assessment model is developed considering the arrival and departure times of infected individual and susceptible individuals, impacts of suspended infectious particles and environmental conditions. Thus, it can determine the infection risk for a susceptible individual that has SPDT links with the infected individuals. For airborne diseases such as Influenza A, the infectious particles containing viruses take time ranging from minutes to hours~\cite{fernstrom2013aerobiology,han2014risk} to settle down to the ground. The settle periods depend on the temperature, humidity and airflow etc. at the interaction areas. The existence of the particles in the generated proximity is highly dependent on the air exchange rates. In addition, the generated particles also lose their infectivity as time goes~\cite{han2014risk}. Therefore, the generated active virus particles (concentration of active infectious particles in the air of interaction location) are reduced since they were generated. For large scale simulations, it is hard to distinguish the architecture of each interaction area. Thus, in the proposed model definition it is considered that the infectious particles are continuously added to the air at the interaction area because the presence of infected individuals and homogeneously distributed into the interaction area. Concurrently, the generated particles are removed from the proximity of infected individual with a rate representing reduction for all factors. The formulation of the proposed model is explained as follows.

The disease diffusion system is analysed over the discrete time of time step $\Delta t$. Suppose that an infected individual A arrive at a location L at time $t_s$ and deposits airborne infectious particles into his proximity with a rate $g$ (particles/s) until he leaves the location at $t_l$. These particles are homogeneously distributed into the air volume $V$ of proximity and the particle concentration keeps increasing until it reaches  steady state. Simultaneously, active particles decay at a rate $b$ (proportion/s) from the proximity due to various reduction processes such as air conditioning, settling down of particles to the ground and losing infectivity of particles. Thus, the accumulation rate of particles in the proximity can be given by 
\begin{equation}
V\frac{\mathrm{d}C }{\mathrm{d} t}=g-bVC
\end{equation}
where $C$ is the current number of particles in one $m^{3}$ of air at L.Thus, the particle concentration $C_t$ at time $t$ after the infected individual A arrives at L is given by~\cite{he2005particle,issarow2015modelling}
\begin{equation*}
\int_{0}^{C_t}\frac{dC}{g-bVC}=\frac{1}{V}\int_{t_s}^{t}dt
\end{equation*}
This leads to
\begin{equation} \label{eq:dpc}
C_t=\frac{g}{bV}\left(1-e^{-b(t-t_s)}\right)
\end{equation}

If a susceptible individual u (as Fig.~\ref{fig:spdtp}) arrives at location L at time $t_{s}^{\prime}\geq t_s$ and continues staying with A up to $t_l^{\prime}<t_l$, the number of particles $E_d$ inhaled by u through this direct link is
\begin{equation*}
E_d=\frac{gp}{bV}\int_{t_{s}^{\prime}}^{t_l^{\prime}}\left(1-e^{-b(t-t_s)}\right) dt
\end{equation*}
where $p$ is the pulmonary rate of u. Thus, the integration gives
\begin{equation*}
E_d=\frac{gp}{bV}\left[\left(t_l^{\prime} + \frac{1}{b} e^{-b(t_l^{\prime}-t_s)}\right)-\left(t_s^{\prime}+\frac{1}{b}e^{-b(t_s^{\prime}-t_s)}\right)\right]
\end{equation*}

\begin{equation}\label{eq:dexp}
E_d=\frac{gp}{bV}\left[t_l^{\prime}-t_s^{\prime} + \frac{1}{b} e^{-b(t_l^{\prime}-t_s)}-\frac{1}{b}e^{-b(t_s^{\prime}-t_s)}\right]
\end{equation}

If a susceptible individual v (as Fig.~\ref{fig:spdtp}) stays with the infected individual A up to $t_l$ when A leaves L,where $ (t_l^{\prime}\geq t_l)$, and continues staying at L after A departs, he will have both direct and indirect transmission links. The value of $E_d$ for v within the time $t_s^{\prime}$ and $t_l$ is given by
\begin{equation}
E_d=\frac{gp}{bV}\left[t_l-t_s^{\prime} + \frac{1}{b} e^{-b(t_l-t_s)}-\frac{1}{b}e^{-b(t_s^{\prime}-t_s)}\right]
\end{equation}
For the indirect link from time $t_l$ to $t_l^{\prime}$, the particle concentration are first computed during this period which decreases after A leaves at $t_l$. The particle concentration $C_t$ at time $t$ is given by
\begin{equation*}
C_{t_l}=\frac{g}{bV}\left(1-e^{-b(t_l-t_s)}\right)
\end{equation*}
The particle concentration decreases after the infected individual leaves the proximity at time $t_l$ and is given by
\begin{equation*}
\frac{dC}{dt}=-bC
\end{equation*}
Thus, the concentration at time $t$ will be
\begin{equation*}
\int_{C_{t_l}}^{C_t}\frac{dC}{C}=-b\int_{t_l}^{t} dt
\end{equation*}
This leads to
\begin{equation*}
C_t=C_{t_l} e^{-b(t-t_l)}=\frac{g}{bV}\left(1-e^{-b(t_l-t_s)}\right)e^{-b(t-t_l)}
\end{equation*}
The individual v inhales particles $E_i$ during the indirect period from $t_l$ to $t_l^{\prime}$ is
\begin{equation*}
E_i=\int_{t_l}^{t_l^{\prime}}pC_t dt 
\end{equation*}

\begin{equation*}
=\frac{gp}{bV}\left(1-e^{-b(t_l-t_s)}\right)\int_{t_l}^{t_l^{\prime}} e^{-b(t-t_l)} dt 
\end{equation*}

\begin{equation}
E_i=\frac{gp}{Vb^2}\left(1-e^{-b(t_l-t_s)}\right)\left[1-e^{-b(t_l^{\prime}-t_l)}\right]
\end{equation}
If a susceptible individual w is only present for the indirect period at the proximity (as Fig.~\ref{fig:spdtp}), the number of inhaled particles for the indirect period from $t_s^{\prime}$ to $t_l^{\prime}$ is given by
\begin{equation*}
\frac{gp}{bV}\left(1-e^{-b(t_l-t_s)}\right)\int_{t_s^{\prime}}^{t_t^{\prime}} e^{-b(t-t_l)} dt 
\end{equation*}
\begin{equation}
=\frac{gp}{Vb^2}\left(1-e^{-b(t_l-t_s)}\right)\left[e^{-b(t_s^{\prime}-t_l)}-e^{-b(t_l^{\prime}-t_l)}\right]
\end{equation}
To generalise the intake dose equations into one equation that counts any configuration of SPDT link: direct and/or indirect transmission components, a link characterising time $t_i$ is introduced. Based on the $t_i$, the above equations for direct and indirect transmission links can be written as
\begin{equation*}
E_d=\frac{gp}{bV}\left[t_i-t_s^{\prime} + \frac{1}{b} e^{-b(t_i-t_s)}-\frac{1}{b}e^{-b(t_s^{\prime}-t_s)}\right]
\end{equation*}
\begin{equation*}
E_i=\frac{gp}{Vb^2}\left(1-e^{-b(t_l-t_s)}\right)\left[e^{-b(t_i-t_l)}-e^{-b(t_l^{\prime}-t_l)}\right]
\end{equation*}
Value of $t_i$ is given as follows: $t_i=t_l^{\prime}$ for the SPDT links with only direct component, $t_i=t_l$ if SPDT link has both direct and indirect components, and otherwise $t_i=t_s^{\prime}$. Thus, the total inhaled particles can be given for a susceptible individual due to a SPDT link by
\begin{equation*}
E_l=E_d + E_i
\end{equation*}
\begin{equation}\label{eq:expo}
E_l=\frac{gp}{Vb^2}\left[b\left(t_i-t_s^{\prime}\right)+ e^{bt_{l}}\left(e^{-bt_i}-e^{-bt_l^{\prime}} \right)+e^{bt_{s}}\left(e^{-bt_l^{\prime}}-e^{-bt_s^{\prime}} \right)\right]
\end{equation}
The equations determine the received exposure for one SPDT link with an infected individual, comprising both direct and indirect links. In this equation, $t_s$ can not be greater than $t_s^{\prime}$ as SPDT link is created after infected individual arrive at a location. If $t_s>t_s^{\prime}$, it is required to set $t_s=t_s^{\prime}$ for calculating appropriate exposure. If a susceptible individual has $m$ SPDT links during an observation period, the total exposure is 
\begin{equation*}
E=\sum_{k=0}^{m}E_{l}^{k}
\end{equation*}
where $E_{l}^{k}$ is the received exposure for k$^{th}$ link. The probability of infection for causing disease can be determined by the dose-response relationship defined as 
\begin{equation}\label{eq:prob}
P_I=1-e^{-\sigma E}
\end{equation}
where $\sigma$ is the infectiousness of the virus to cause infection. The infectiousness depends on the disease types and even virus types~\cite{fernstrom2013aerobiology, riley1978airborne}.

%% file: 3.3_chap_riskmodel.tex
\section{Empirical networks}
In this study, empirical contact networks are applied to studying SPDT diffusion. These networks are constructed using the location updates of users from a social networking application called Momo. These location updates are processed to find the possible direct and indirect transmission links. Then, the obtained links are used to build various contact networks. The used data set and the network construction procedures are explained in this section.

\subsection{Data set}
This study applies location update information from users of a social discovery network \emph{Momo}\footnote{https://www.immomo.com}. Momo App enables users to interact with nearby users by sharing their current locations. Whenever a user launches the Momo app, the current location is forwarded to the Momo server. The server sends back the latest location updates of all users from the close geographical area. These location updates have been previously collected by the authors of~\cite{thilakarathna2016deep} using a set of network API communicating with Momo server. The API retrieved location updates every 15 minutes over a period of 71 days (from May to October 2012). The data set contains 356 million location updates from about 6 million Momo users around the world, but primarily in China. Each database entry includes GPS coordinates of the location, time of update and user ID. For this study, the updates from Beijing are separated as it is the city with the highest number of updates for the period of 32 days from 17 September, 2012 to 19 October, 2012. The 32 days period is sufficient for studying the disease spreading of influenza-like diseases~\cite{huang2016insights}. This data contains almost 56 million location updates from 0.6 million users. 

\subsection{Disease transmission links}
All possible disease transmission links according to SPDT diffusion model definition are extracted from the location updates of Momo users. To create an SPDT link between a host user (assumed infected with disease) $v$ and a neighbour user $u$ (susceptible user), it is required to find the arrival times ($t_s, t_s^{\prime}$) and departure times ($t_l, t_l^{\prime}$) of two users. As the first step, it is identified that an infected host user $v$ is staying at a location. Consecutive updates, $X=\{(x_{1},t_{1}),(x_{2},t_{2}),\ldots (x_{k},t_{k})\}$ where $x_{i}$ are the co-ordinate values and $t_{i}$ are the update times, from a user $v$ within a radius of 20m (travel distance of airborne infection particles~\cite{han2014risk}) of the initial update's location $x_{1}$ are indicative of the user staying within the same proximity of $x_1$. A threshold is set for time difference of any two updates to 30 minutes to make sure infected host remain within the same proximity, as longer gaps may indicate a data gap in the user pattern. Then, the central co-ordinate in the update $X$ is searched where the distances from each update to all other updates are added together and the update $x_c$ with the minimum sum is selected as central co-ordinate. For the host user $v$, its visit to the proximity of $x_{c}$ will represent a valid visit if a susceptible user $u$ has location updates starting at $t^{'}_{1}$ while $v$ is present, or within $\delta$ seconds after $v$ leaves the area. The user $u$ should have at least two updates within 20m of $x_{c}$ to be valid to ensure that it is in fact staying at the same proximity, and therefore can be exposed to the infected particles, rather than simply passing by. The stay period of host user $v$ at the proximity of $x_c$ is ($t_{l}=t_{k}, t_s=t_{1}$), where $t_k$ represents the end of the current stay period. If $u$'s last update within 20m around $x_{c}$ is $(x^{'}_{j},t^{'}_{j})$, the created SPDT link has a link duration ($t_l^{\prime}=t^{\prime}_{j},t_s^{\prime}=t^{'}_{1}$) due to active visit ($t_{l}=t_{k}, t_s=t_{1}$). All links to other users for this active visit ($t_{l}=t_{k}, t_s=t_{1}$) are computed. In the similar way, all visits made by $v$ are searched over the updates of 32 days and SPDT links are extracted. This process is executed for all users present in the data set to find the all possible SPDT disease transmission links and provides a real contact network with SPDT links among users. A SPDT link is noted as $e_{vu}=\left(v(t_s,t_l),t_s^{\prime},t_l^{\prime}\right)$ which means that a user $v$ visit a location during $(t_s,t_l)$ where another user $u$ is present for the duration $(t_s^{\prime},t_l^{\prime})$ where $t_s^{\prime} \geq t_s$. Each link between the two same users are distinguished by the time intervals $(t_s,t_l)$ and $(t_s^{\prime},t_l^{\prime}
)$ as there can be multiple links to a neighbour for the same stay interval $(t_s,t_l)$ of the host user.

\subsection{Contact networks}
Real individual SPDT contact network is obtained from the above procedure. This network connects 364K individuals through 6.86M SPDT links. A corresponding SPST network which excludes the indirect transmission links from the SPDT network is built. The SPST network can connect 264K individuals and has 2.10M direct links. The number of connected individuals reduces because many individuals are isolated when the indirect transmission links are removed. The analyses show that the users stay in the system only 3-4 days on average and then disappear for the remainder of the simulation period. Thus, these networks have low links density and are called Sparse SPDT network (SDT network) and Sparse SPST network (SST network). It can be assumed that links captured in the SDT network are the typical behaviours of individual. Thus, a Dense SPDT network (DDT network) is constructed from the SDT network where the links from the available days of a user are repeated for the missing days for that user. The corresponding Dense SPST network (DST network) is built excluding indirect links from the DDT network. Now, the DDT network has 47.11M links connecting 364K users while DST network has 13.24M links connecting 264K users. Thus, the link densities increase in these dense networks. As the users are present every day in these networks, the diffusion dynamics obtained through these networks will be more realistic. 

\begin{table}[h!] \label{netst}
\caption{Network properties for the constructed networks}
\vspace{1em}
\centering
\begin{tabular}{|c|c|c|c|c|c|c|}
\hline
Networks &  SDT & SST & DDT & DST & LDT & LST             \\ \hline
Total links (M) &  6.86  & 2.10 & 47.11 & 13.24 & 47.11 & 47.11              \\ \hline
Connected users (K) & 364 & 264 & 364 & 264 & 364 & 364    \\ \hline
Link density per user & 18.86 & 7.95 & 129.42 & 50.15 & 129.42 & 129.42 \\    
\hline
\end{tabular}
\end{table}

The users who are connected with other users through only indirect links in the above SPDT networks become isolated in the SPST networks as indirect links do not exist in SPST networks. In addition, link density reduces in SST network to 7.95 links from 17.86 links of SDT network and 50.15 links for DST network from 129.42 links of DDT network. The underlying social structure is also reshaped since connected users reduce in the SST and DST networks. Thus, two other networks termed as LDT and LST networks are reconstructed that maintain the same link densities as that of the DDT networks. In this format, neighbour user's arrival time $t_s^{\prime}$ of the SPDT link that has only indirect components in DDT networks is set to $t_s$ of host user and an LDT network is obtained. Then, indirect components of links are removed from the LDT network to built the LST network which now has the same link density of 129.42 and no isolated user. The LDT network can be considered as the extension of the LST network where some of the direct links of LST network is appended with the indirect links. Thus, the underlying contact structures are the same in both LST and LDT networks. A summary of the constructed networks is presented in Table 3.1.

%% file: 3.4_chap_diffusion.tex
\section{SPDT diffusion analysis}
Now the spreading dynamics of airborne infectious disease based on the SPDT diffusion model are explored on real contact networks. Intensive simulations are conducted on the both SPDT and SPST networks with various configurations to characterise spreading dynamics and reveal interactions between diffusion dynamics and disease modelling parameters: particle decay rates, the infectiousness of inhaled particles, and infectious periods of infected individuals.

\subsection{Disease simulation setup}
The diffusion dynamics of airborne disease are studied using the SPDT model Equation~\ref{eq:prob}. The disease propagation model, SPDT infection risk assessment model parameters and diffusion characterising metrics are described here.

\subsubsection{Propagation model}
A simplified Susceptible-Infected-Recovered (SIR) epidemic model is adapted to emulate airborne disease propagation on the constructed contact network. Individuals are in one of the three states any time and change the states as in a stochastic process through the three possible disease states, namely, susceptible, infectious and recovered. If an individual in the susceptible state comes within close proximity where an infected individual have been, the former will be exposed to the infectious particles and may contract with the disease i.e individual may move to the infectious state. The probability of getting infected due to be in contact with an infected individual is derived by Eqn.~\ref{eq:prob}. Then, the infectious individual continues to produce infectious particles over its infectious period $\tau$ days until they enter the recovered state, where $\tau^{-1}$ is the rate of recovering from the disease. Recovered individuals are immune to all possible strains of the disease. As the values $\tau$ can vary for each individual even for the same disease, the parameter $\tau$ is derived from a uniform random distribution within the observed empirical ranges. The uniform distribution allow to maintain a constant recovery rate in the networks and allow to quantify the reproduction number easily. In this model, no event of births, deaths or entries of a new individual is considered. 

The user's disease states change stochastically in the conducted simulations. There are two approaches to simulate such stochastic process~\cite{kiss2017mathematics}. In the first approach, execution times of next events are defined by knowing the combined rate of all possible events at the current time and are selected from an exponential distribution with that rate. Then, the transition events are occurred randomly and assigned time from the above event time set. For the large scale simulation, it becomes computationally expensive and complex. In the second approach, the simulation is step forwarded in time over time intervals (the observation period is divided into several time intervals). At each time step, the transition probability for changing from one disease state to another possible state in the next interval is calculated. Then, the state transition events that will occur in the next interval is generated using a random number generator (e.g. using a Bernoulli process with transition probability). The limitation of this approach is that multiple events can happen during a large interval and the new events could bring another event. In the conducted simulations, the second approach is selected with the step forward interval of one day. The authors of ~\cite{stehle2011simulation,toth2015role} have studied that aggregating contact information in one day interval provides similar disease spreading dynamics. Besides, newly infected individuals in influenza-like disease do not start infecting susceptible individuals immediately~\cite{toth2015role}. Thus, the large interval would function as a latent period that susceptible individuals require to be infectious.

\subsubsection{Parameters setting}
Various practical experiments have been conducted to understand the viral load of influenza A disease in the literature. The experiment of~\cite{yan2018infectious} have analysed breath for the influenza A patients and found that $65\%$ virus is contained within the droplets of sizes $<5\mu m$. These droplets can be airborne and suspended in the air for a long time. During the 30 minutes of breathing, patients have generated about 140 PFU (plaque-forming unit) virus contained particles. In the experiment of ~\cite{lindsley2015viable}, 5-538 PFU viable influenza A virus is found in the samples of six coughs from each patient. Thus, the average generation of virus for each cough is $269/6=44.8$ PFU per cough. The mean cough frequency for a human is 18 per hour. Therefore, virus generation rate per second by an influenza A patient through breathing and coughing will be
\[g=(140/1800)+(44.8*18/3600)=0.304 PFU/s\]

These generated particles are removed from the interaction location in various ways. The main particles removal mechanism often comes from the air change rates (ACR) in the proximity. In the residential areas, air change rates are measured below 1h$^{-1}$. However, it becomes higher up to 3h$^{-1}$ with opening the windows and doors~\cite{howard2002effect,shi2015air}. In the classroom setting, ACR is found from 2 to 6h$^{-1}$ with the median 3h$^{-1}$. The air exchange rates in the office buildings are 3-4h$^{-1}$. In the open public places, ACR is varied for a wide range of 0.5 to 6h$^{-1}$. The present particles of proximity also lose their infectivity over time. The other removal mechanism is to settle down on the surfaces. Considering all removal mechanism, it is assumed that, in this study, the upper bound of particle decay rates (combining all removal rates) from interacted proximity can be up to 8h$^{-1}$. As many studies found the suspended particles remain in the proximity up to several hours. Thus, the lower limit of particles decay rates is considered to be 0.2h$^{-1}$ after that there will not active infectious particles in the proximity. Therefore, the deposited particles may require 7.5 min to 300 min to be removed from interaction areas after their generation. In the simulations. the particle decay rates $b$ for SPDT links are selected as $b=\frac{1}{60 r}$ where $r$ are particle decay time (in minutes) randomly chosen from [7.5-300] min given a median particle decay time $r_t$. The particle decay rates $b$ in this thesis, for convenience, are discussed with $r$ and $r_t$ where particle decay rate $r$ means the corresponding $b$ and particle decay rates $r_t$ mean the corresponding particle removal rates $b$ drawn from the above process for all links in the network. The volume $V$ of proximity in Eqn.~\ref{eq:expo} is fixed to 2512 m$^{3}$ assuming that the distance, within which a susceptible individual can inhale the infectious particles from an infected individual, is 20m and the particles will be available up to the height of 2m~\cite{han2014risk,fernstrom2013aerobiology}. The other parameters are assigned as follows: the pulmonary rate of susceptible individual $p=7.5$ litre/min ~\cite{yan2018infectious,lindsley2015viable,han2014risk}. and infectiousness of particle $\sigma$ is set to 0.33 as the median value of required exposures for influenza to induce disease in 50$\%$ susceptible individuals is 2.1 PFU~\cite{alford1966human}.

\subsubsection{Simulation progression}
The simulations in each experiment start with randomly selected 500 seed individuals (initially infected). All simulations are run for a period of 32 days. The simulations are step-forwarded with the one-day interval. During each day of disease simulation, the received SPDT links for each susceptible individual from infected individuals are separated and infection probabilities are calculated by Eqn.~\ref{eq:prob}. Susceptible individuals stochastically switch to the infected states in the next day of simulation according to the Bernoulli process with the infection probability $P_I$ (Eqn~\ref{eq:prob}). Individual stays infected up to $\tau$ days randomly picked up from 3-5 days maintaining $\bar{\tau}=4$ days (except when other ranges are mentioned explicitly)~\cite{huang2016insights}. The daily simulation outcomes are obtained for the epidemic parameters: the number of new infections $I_n$, the disease prevalence $I_p$ as the number of current infections in the system and the cumulative number of infections $I_a$ which is the total number of infection up to the current simulation day.

\subsubsection{Diffusion characterisation metric}
The obtained simulation outputs are characterised with the disease reproduction rate in the networks. The disease reproduction rate $R$ is derived in the following way. Assuming that the average infection rate by the current infected individuals at a time $t$ is $\beta(t)$ (as $\beta$ changes in the heterogeneous network over time) and the recovery rate $\tau^{-1}$ which is constant over the simulation period, the dynamics of disease prevalence $I_p$ is given by 
\begin{equation}
\frac{\mathrm{d }I_p(t) }{\mathrm{d} t}=\left( \beta(t) S(t) -\tau^{-1} \right) I_p(t)
\end{equation}
where $S(t)$ is the number of susceptible individuals at $t$. If $\beta(t) S(t)>\tau^{-1}$, the disease prevalence get stronger. The ratio $\beta(0) S(0)/\tau^{-1}$ is called basic reproduction number and termed as $R_0$ where $S(0)$ is the total susceptible individual in the network at $t=0$. As it is difficult to quantify the reproduction number for the dynamic heterogeneous networks, every day's $R_0$ is computed and the average reproduction rate $R$ is taken as the measure of infection reproduction ability by a network. Thus, the reproduction rate $R(t)$ is given by
\begin{equation*}
\frac{\mathrm{d }I_p(t) }{\mathrm{d} t}=\tau^{-1} \left(R(t)-1\right) I_p(t)
\end{equation*}
\begin{equation*}
I_p^{-1}(t)\frac{\mathrm{d }I_p(t) }{\mathrm{d} t}=\frac{\mathrm{d }\ln \left( I_p(t)\right) }{\mathrm{d} t}=\tau^{-1} \left(R(t)-1\right) 
\end{equation*}
Now integrating over the $\Delta t$ time leads to
\begin{equation*}
\int_{t}^{t+\Delta t} \frac{\mathrm{d }\ln \left( I_p(t)\right) }{\mathrm{d} t}=\int_{t}^{t+\Delta t} \tau^{-1} \left(R(t)-1\right) 
\end{equation*}
\begin{equation*}
\ln \left( \frac{I_p(t+\Delta t)}{I_p(t)}\right)\approx \tau^{-1} \Delta t \left(R(t)-1\right)
\end{equation*}
Therefore, the disease reproduction rate for a period $\Delta t$ can be given by
\begin{equation}
R(t)\approx 1+\frac{\tau}{\Delta} \ln \left( \frac{I_p(t+\Delta t)}{I_p(t)}\right)
\end{equation}
If $\Delta t =1$ day, the $R(t)$ is the average reproduction rate over a day. In this study, $R(t)$ is calculated for every day of simulation periods and find the average $R$ as the measure of disease reproduction abilities of infected individuals for a network.

\subsection{SPDT contact dynamics}
\begin{figure}[h!]
    \centering
    \includegraphics[width=0.48\linewidth, height=5.0 cm]{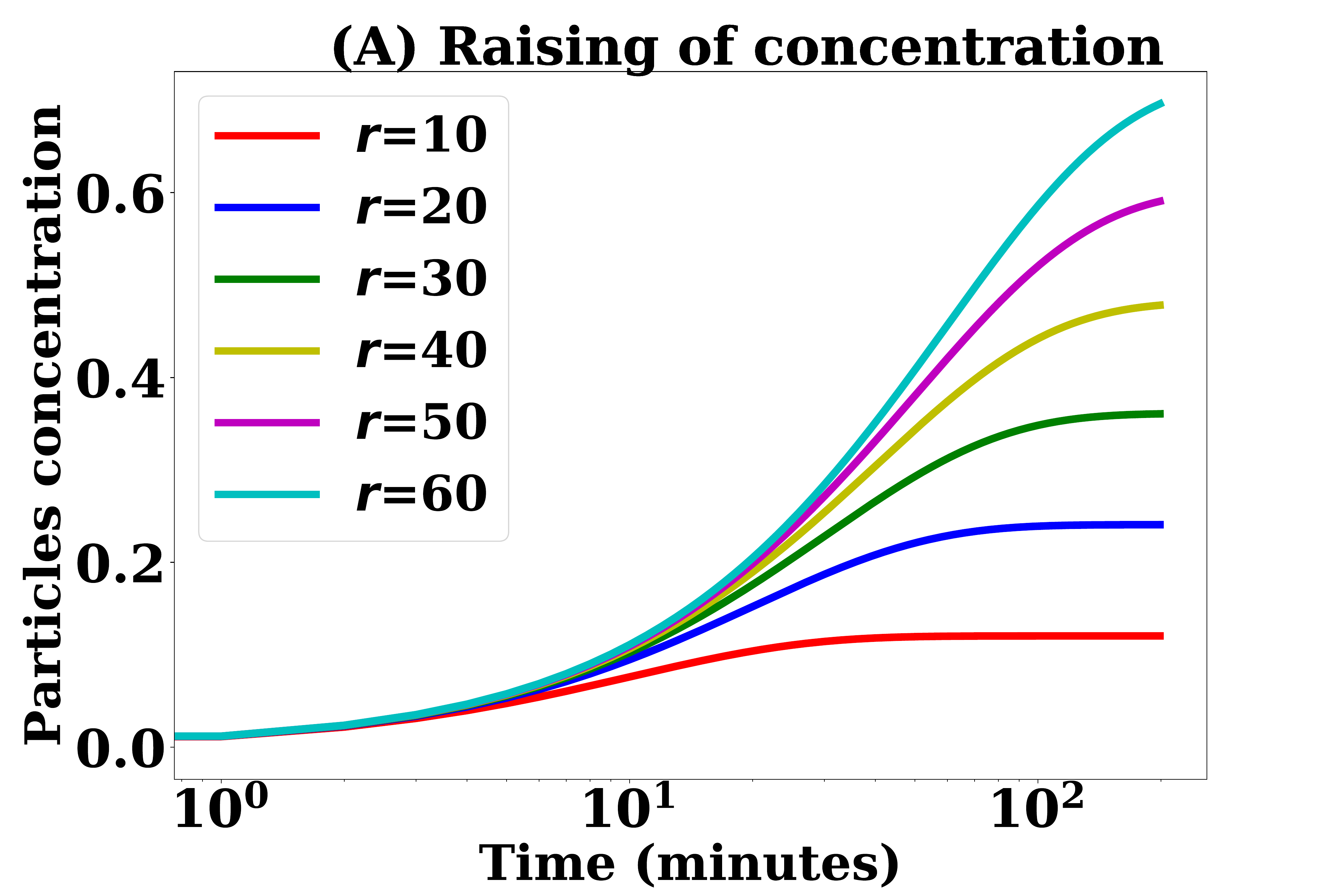}
      \includegraphics[width=0.48\linewidth, height=5.0 cm]{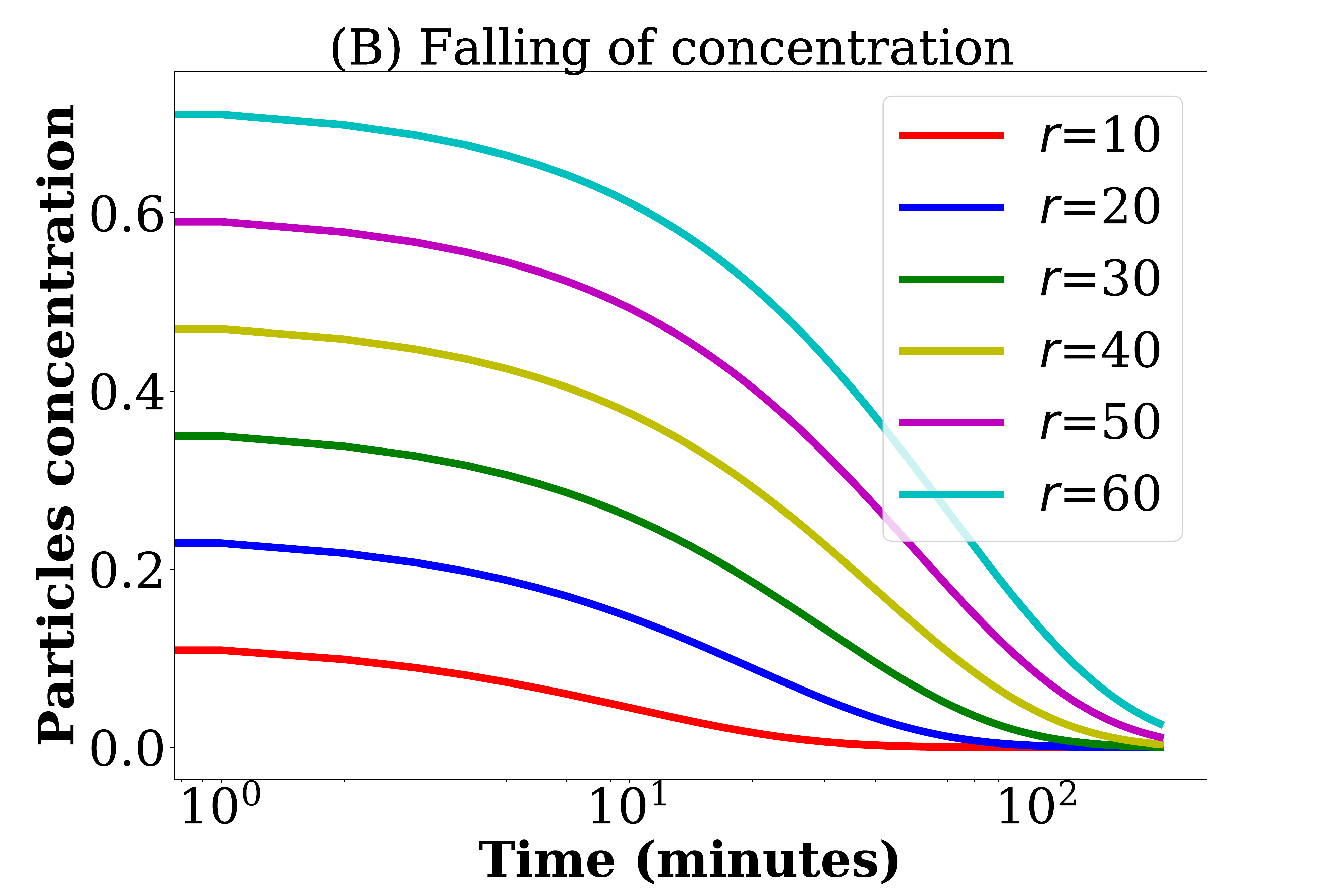}\\
          \includegraphics[width=0.48\linewidth, height=5.0 cm]{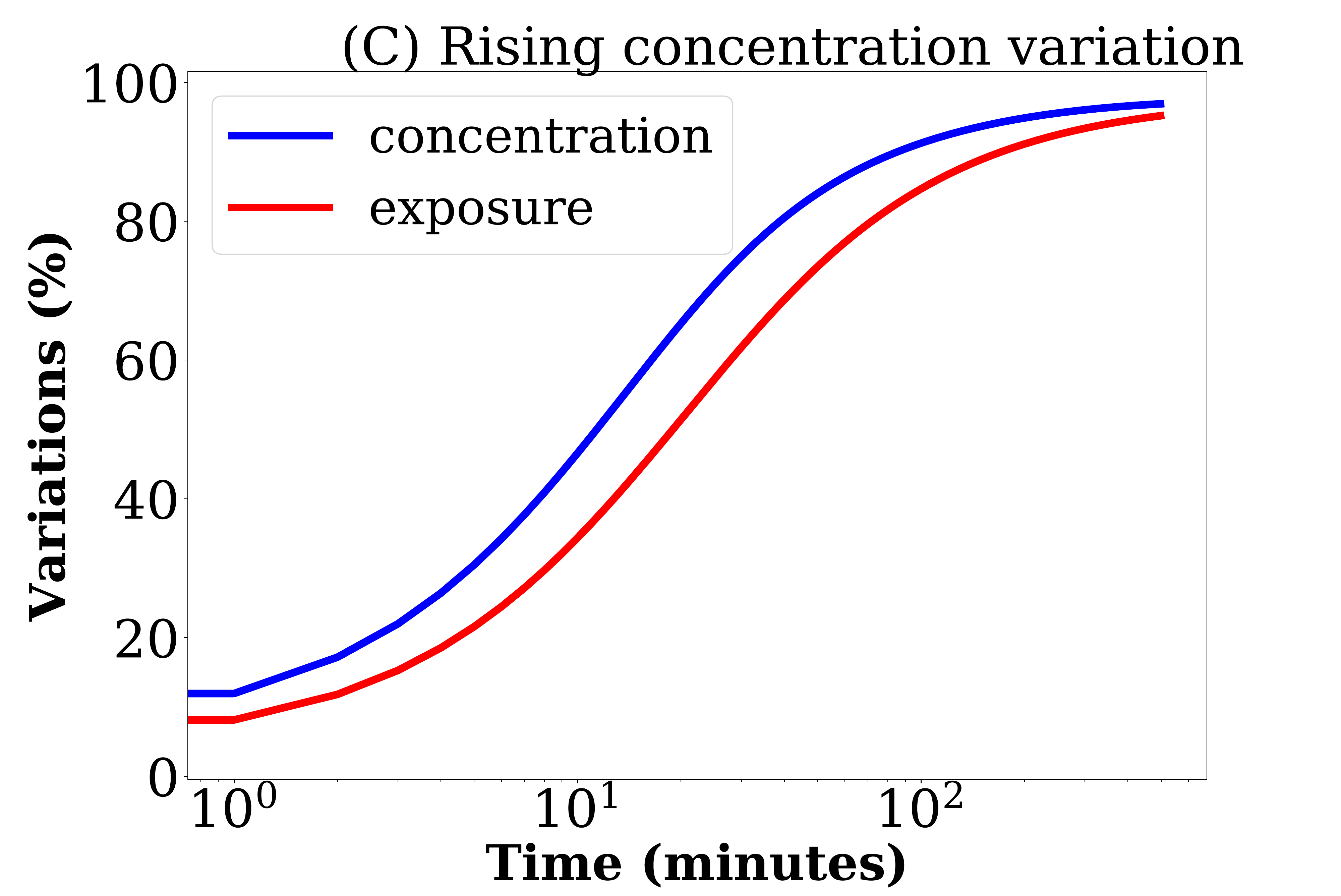}
      \includegraphics[width=0.48\linewidth, height=5.0 cm]{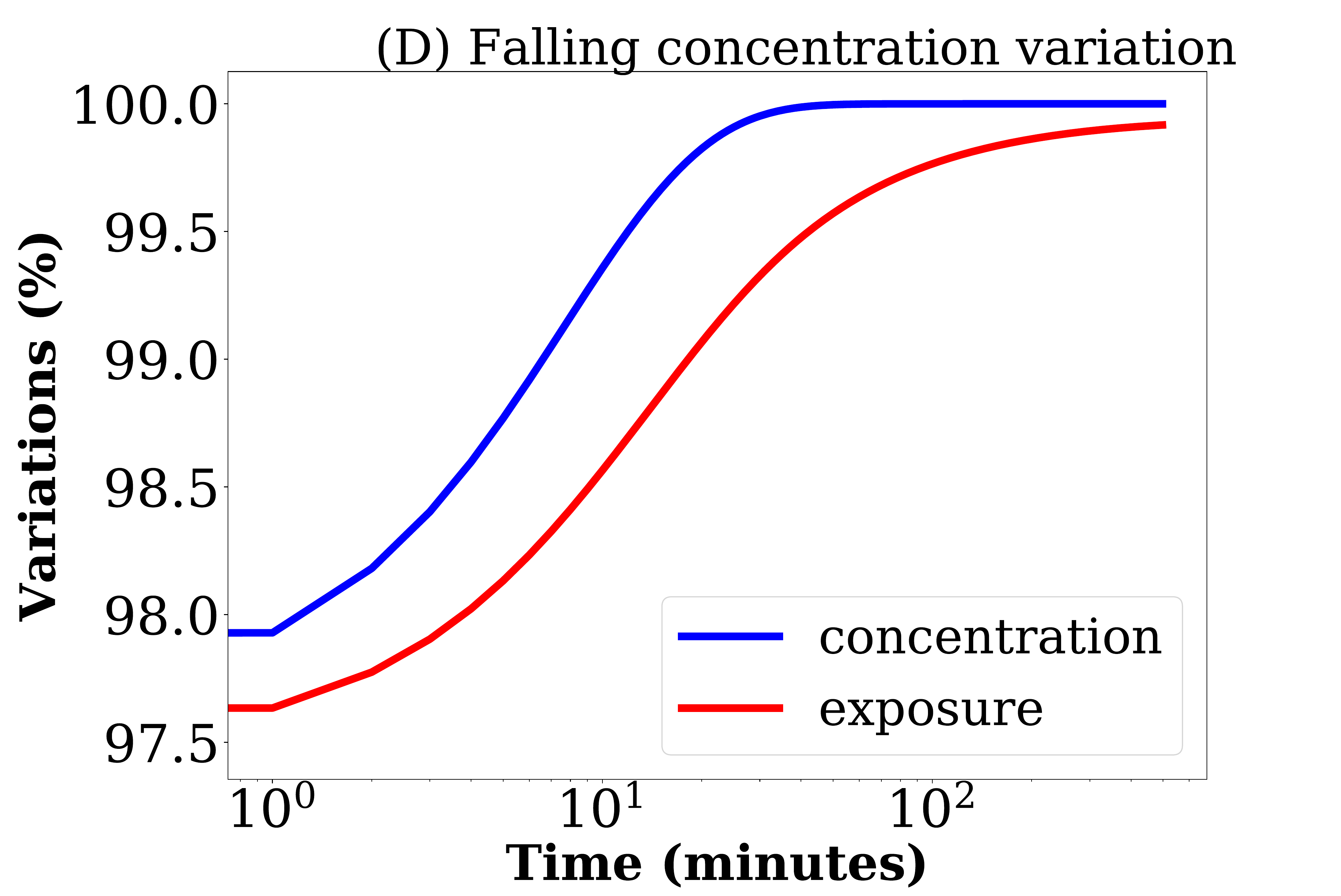}
    \caption{A) increase of the virus concentration when infected individual is present at the proximity and B) decrease of concentration of particles after he leaves the proximity having stayed 200 minutes, C) variation in exposure for changing stay time, and D) variation in concentration for changing stay}
    \label{fig:pcon}
    \vspace{-0.5 em}
\end{figure}

The inclusion of indirect transmission links makes the underlying transmission network of SPDT model strongly connected compared to the SPST model. Individuals who are not connected in the SPST model may get connected in the SPDT model while adding indirect transmission links. Hence, the underlying connectivity becomes denser in SPDT model. Secondly, the number of links between two connected individuals may increase in the SPDT model due to the inclusion of indirect links. Besides, a direct link connecting two individuals can be extended by appending an indirect transmission link. These enhancements increase the disease transmission probabilities among individuals in the SPDT model. Therefore, the diffusion dynamics will be amplified in the SPDT model compared to the SPST model. However, the extent of these enhancements and their impacts on diffusion dynamics are subject to the particle decay rates at the interaction locations. The underlying reasons are analysed here.

\subsubsection{Links dynamics}
The range of influence of particle decay rate $r$ is analysed through the variations of particle concentrations at a location where an infected individual has been (Fig~\ref{fig:pcon}). The Fig~\ref{fig:pcon}A shows the increase of particle concentrations over the stay time due to the visit of the infected individual for $r=\{10,20,30,40,50,60\}$ min. These values are estimated using the Equation~\ref{eq:dpc}. If the stay time is below 10 min, the particle concentrations are similar for all $r$. However, the variations in particle concentrations are increased as stay times are increased. At $r=10$ min, the particle concentration quickly reaches to the saturated value 0.12 particles/m$^{3}$. This steady state value is very low comparing to the 0.72 particles/m$^{3}$ at $r=60$ min. Therefore, the particle concentration varies significantly with $r$ and the maximum variation is found to be 3.35 particles/m$^{3}$ for changing the lower limit $r=7.5$ min to upper limit $r=300$ min in the defined limits. The upper bound of variations in particle concentrations is shown in the Fig~\ref{fig:pcon}C where the variation is found between $r=7.5$ min and $r=300$ min. Therefore, the maximum 97\% variation can happen in particle concentrations for changing $r$ if the stay time of an infected individual is long. Simultaneously, the variations in inhaled exposure by a susceptible individual are estimated assuming that the susceptible individual is staying with the infected individual concurrently. Large variation is found in inhaled exposure for changing $r$.

Now, the variation in particle concentrations during the indirect transmissions period are shown in the Fig~\ref{fig:pcon}B and Fig~\ref{fig:pcon}D. The variations at the different $r$ are shown in the Fig~\ref{fig:pcon}B where it is assumed that the infected individual has been at the visited location for a long time so that the particle concentration reaches the steady state. There are also large variations in the particle concentrations for varying $r=10$ min to $r=60$ min. At $r=10$ min, the particle concentration reduces to 0.01 particles/m$^{3}$ (that can cause infection found empirically) at 24 min. But, the particle concentration reduces to 0.01 particles/m$^{3}$ at 156 min at $r=60$ min. The time difference for reducing the particle concentration to 0.01 particles/m$^{3}$ is about 350 min for changing $r=10$ min to $r=300$ min. The variations in the inhaled exposures and particle concentrations are very high during the indirect transmission period (Fig~\ref{fig:pcon}D). The initial variation is about 98\% for changing $r=7.5$ min to $r=300$ min.

\begin{figure}[h!]
\centering    
 \includegraphics[width=0.45\linewidth, height=5.5 cm]{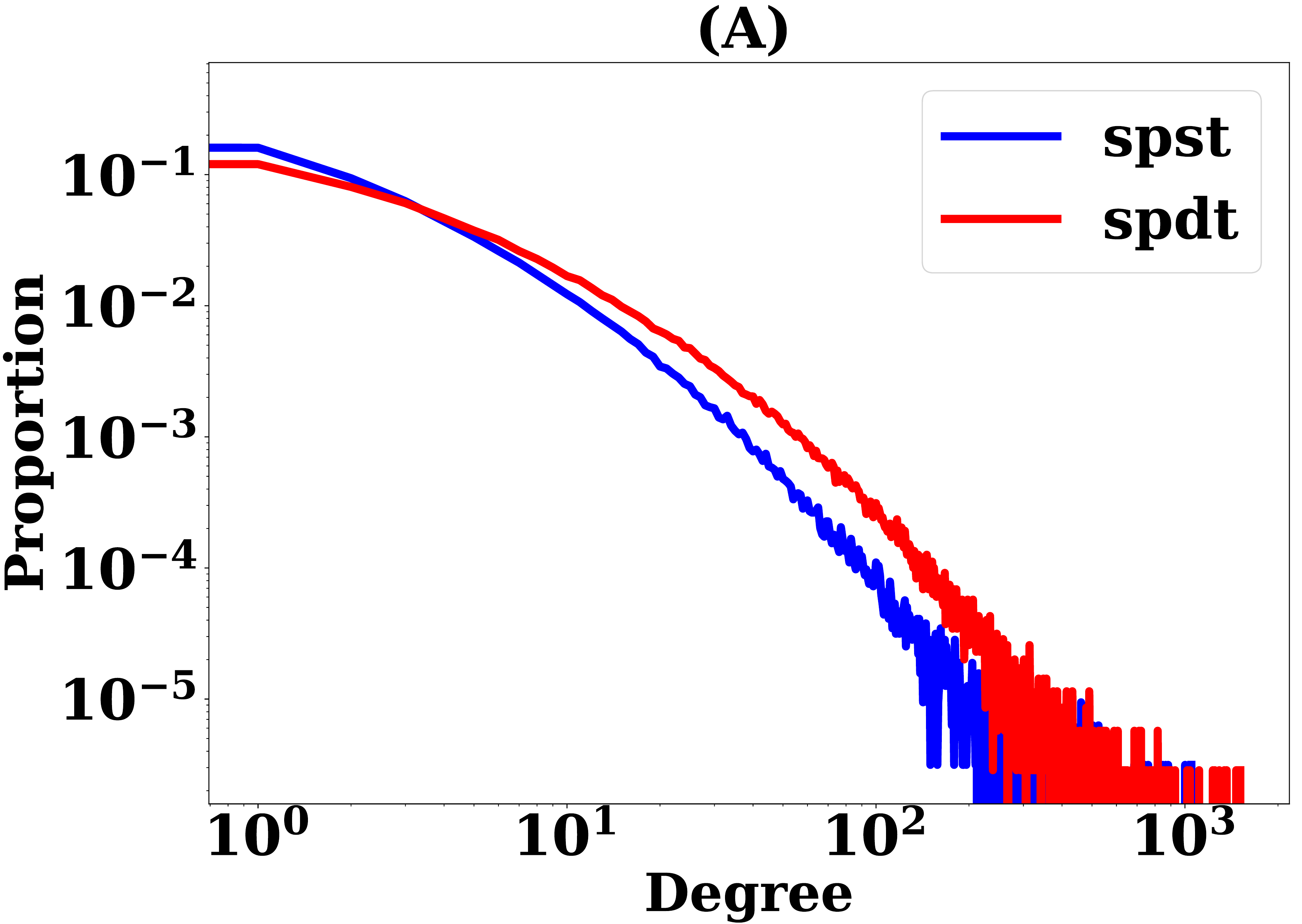} \quad
\includegraphics[width=0.45\linewidth, height=5.5 cm]{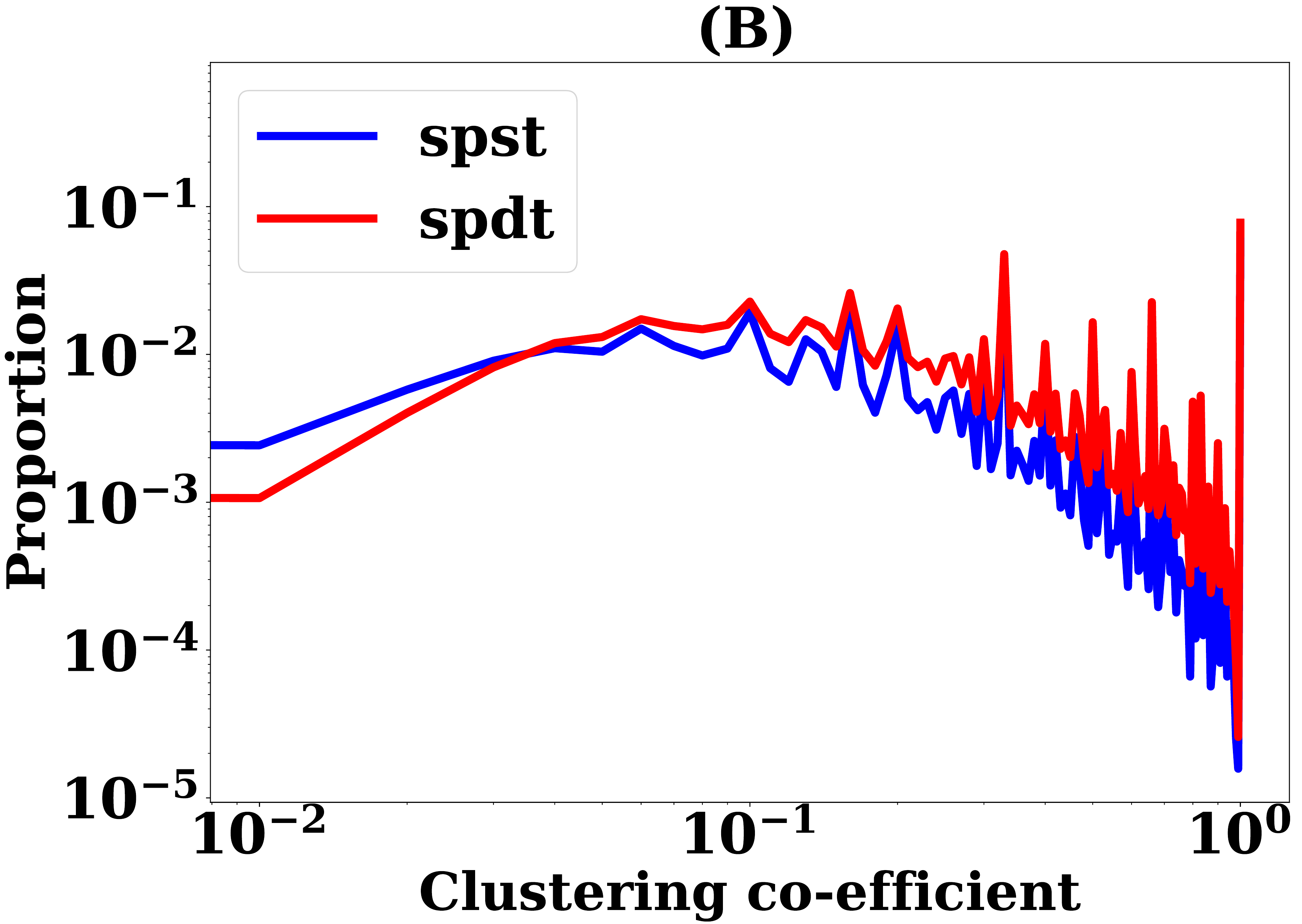}\\ \vspace{0.5em}
\includegraphics[width=0.45\linewidth, height=6.3 cm]{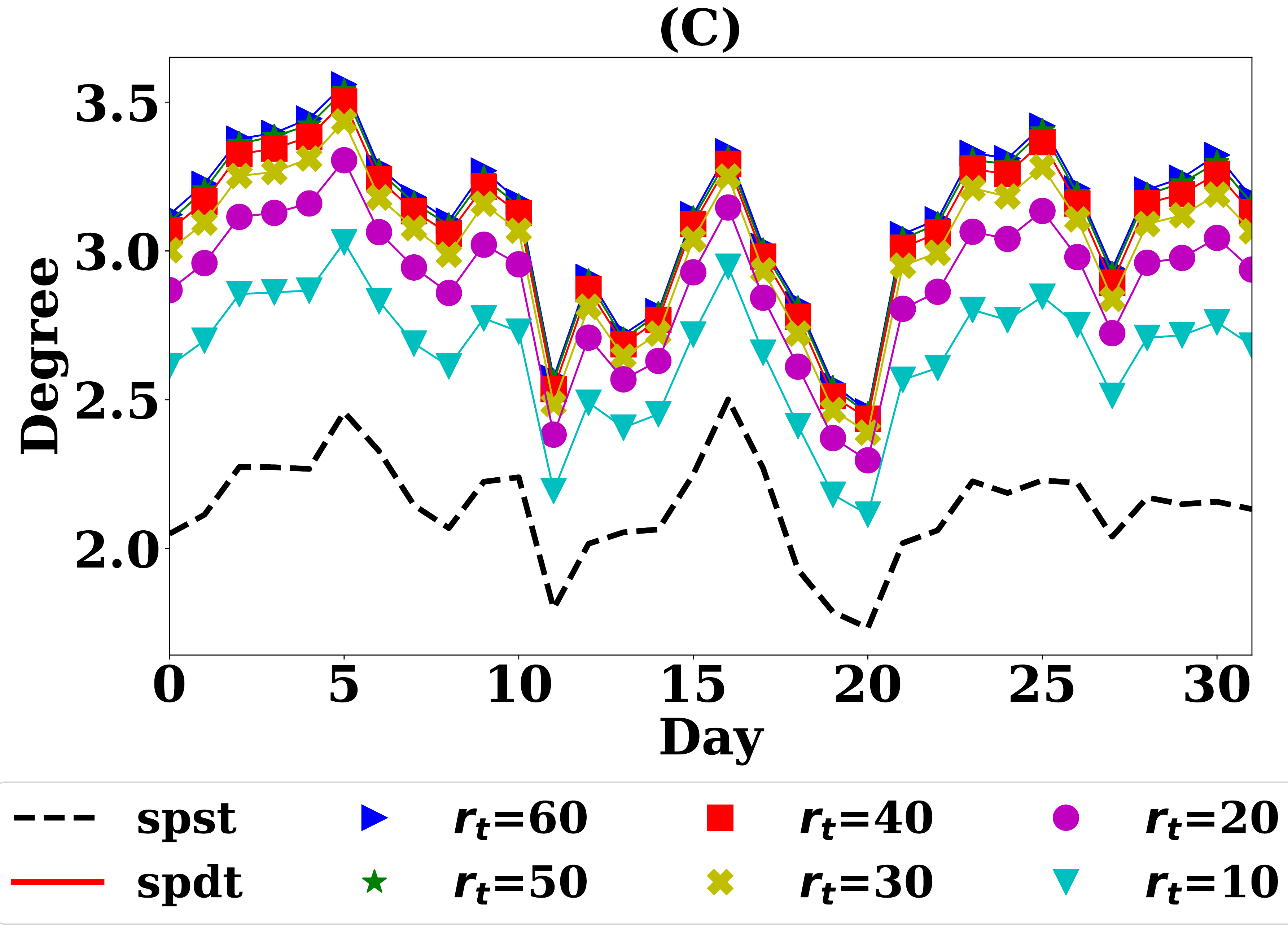} \quad
\includegraphics[width=0.45\linewidth, height=6.3 cm]{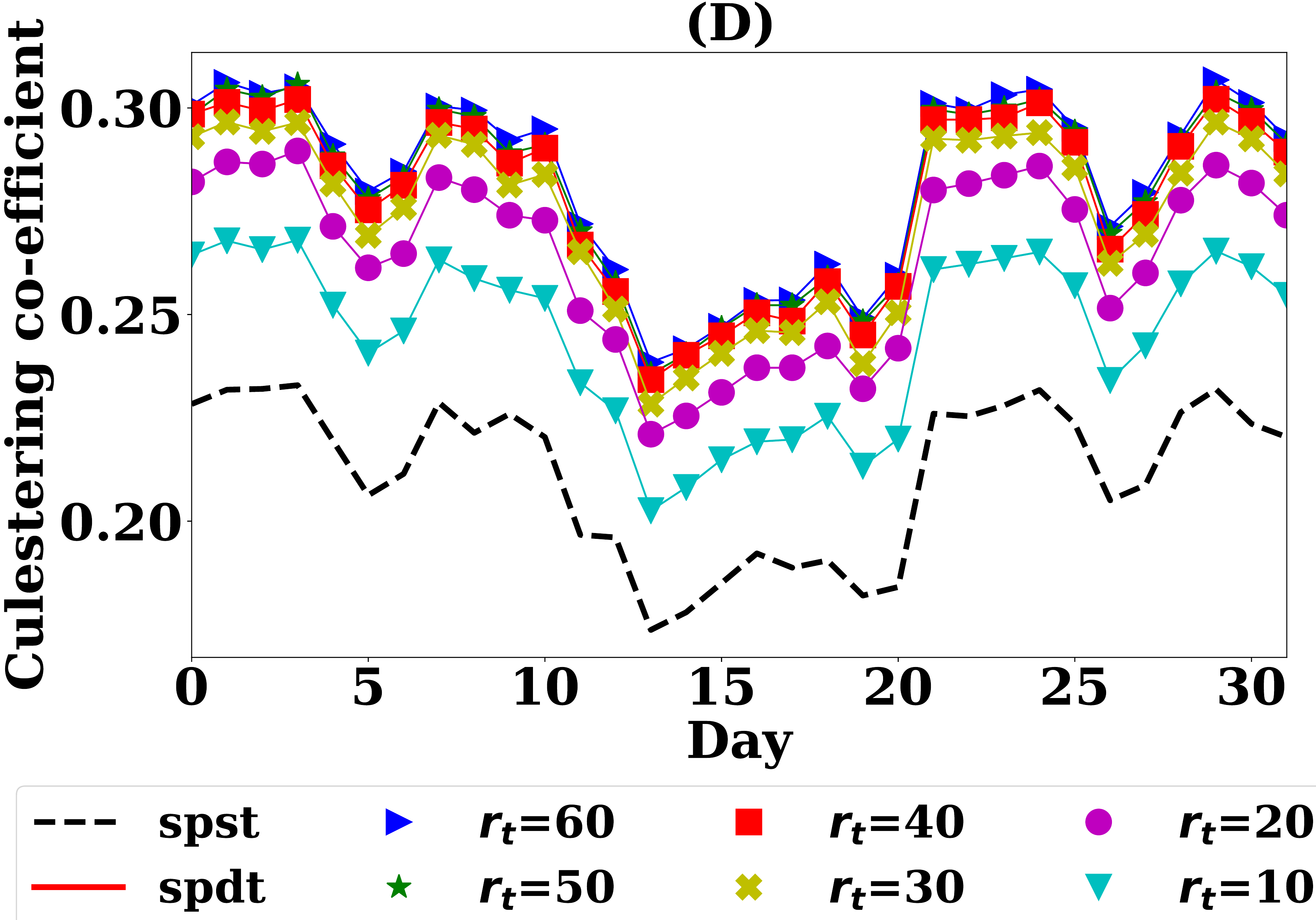}
\caption{Comparing SPST and SPDT networks properties: A) degree distribution in static networks, B) clustering co-efficient distribution in static networks, C) daily average degree in dynamic networks, and D) daily average clustering co-efficient in dynamic networks}
    \label{fig:netan}
\end{figure}

\subsubsection{Connectivity dynamics}
It is seen that the link creation window $\delta$ after infected individual leave interaction proximity is strongly varied with particle decay rate $r$. Thus, the number of neighbour individuals visiting the locations increases as $r$ increases and hence the underlying network connectivity of SPDT get denser. The enhancements in the underlying connectivity of the SPDT model are analysed exploring two network metrics: degree centrality and local clustering coefficient. These metrics are studied for static and dynamic representations of sparse networks (SST and SDT) over 32 days. In static networks, an edge between two individuals is created once they have a link. A SPDT link will be an edge when the inhaled exposure $E_l$ by the susceptible individual is $E_l\geq 0.01$ particles (as Eqn.\ref{eq:prob} shows $P_I$ negligible at $E_l=0.01$). The particle decay rates $r_t$ is set to 60 min for understanding the maximum enhancements by the SPDT model. The degree distribution, number of individual neighbours an individual has contacted over 32 days, and the local clustering coefficient, which is the ratio of the number of triangles present among the neighbours and the possible maximum triangles among neighbours are calculated. To compute the clustering coefficient, the directions of links are neglected as the main goal is to understand the changes in the network connectivity. The results are obtained using NetworkX~\cite{hagberg2008exploring} and are shown in Fig.~\ref{fig:netan}A and Fig.~\ref{fig:netan}B. In the SPDT network, the number of individuals with low degree decreases while the number of individuals with high degree increases compared to that of SPST network (Fig.~\ref{fig:netan}A). This is because the individuals are connected with more neighbour individuals for adding indirect links. The same changes are found for the clustering coefficient as well (Fig.~\ref{fig:netan}B). 

The dynamic representations are created by aggregating networks over each day where an edge between two individuals is created once they have a link on that particular day. Then, the daily average degree and clustering coefficient are measured for the SDT network changing the values of $r_t=\{10,20,30,40,50,60\}$ min and also for the SST network, $r_t$ is changed for SST. The results are shown in Fig.~\ref{fig:netan}C and Fig.~\ref{fig:netan}D. The daily average individual degree and clustering coefficient in the SPDT model is significantly larger than that in the SPST model and the difference increases as $r_t$ increases. The inclusion of indirect links increases the degree and clustering coefficient of individuals. As higher $r_t$ increase opportunities to create SPDT links, these metrics values are increased. However, the growth in average degree and clustering coefficient decrease at the higher values of $r_t$ as the particle concentration reaches decay rates do not have the impact for the short stays at certain $r_t$. Both the static and dynamic networks show the stronger connectivity in the SPDT model than in the SPST model and vary with $r_t$.

\subsection{SPDT diffusion dynamics}
The underlying network connectivity of SPDT model becomes denser with increasing $r$. In addition, the exposure through indirect links become stronger as $r$ increases at the interaction area. These variations influence diffusion dynamics on the SPDT networks. The first experiment, therefore, study the variations in the diffusion dynamics with particle decay rates $r_t$. The disease parameters are known to influence the spreading dynamics as well~\cite{rey2016finding}. The SPDT model definition shows the interaction between the infectiousness of disease and the particle decay rates (Eqn~\ref{eq:prob}). For example, the higher value of infectiousness $\sigma$ will increase infection risk of SPDT links. Thus, the required threshold of the particle concentration to cause infection through an SPDT link reduces and the indirect link creation window $\delta$ will be longer. This means that more susceptible individuals will create links with the infected individual. Therefore, the underlying connectivity gets stronger with high $\sigma$ for a given particle decay rate $r_t$. Besides, the other disease parameter infectious period $\tau$ also affects disease spreading varying the recovery rates of infected individuals. Thus, the second experiment studies how the SPDT model behaves with stronger disease parameters and various $r_t$. The final experiment investigates the novel SPDT diffusion behaviours and studies the reproducibility of SPDT diffusion with the SPST model.

\subsubsection{Diffusion for various particle decay rates}
\begin{figure}[h!]
\centering    
 \includegraphics[width=0.45\linewidth, height=5.5 cm]{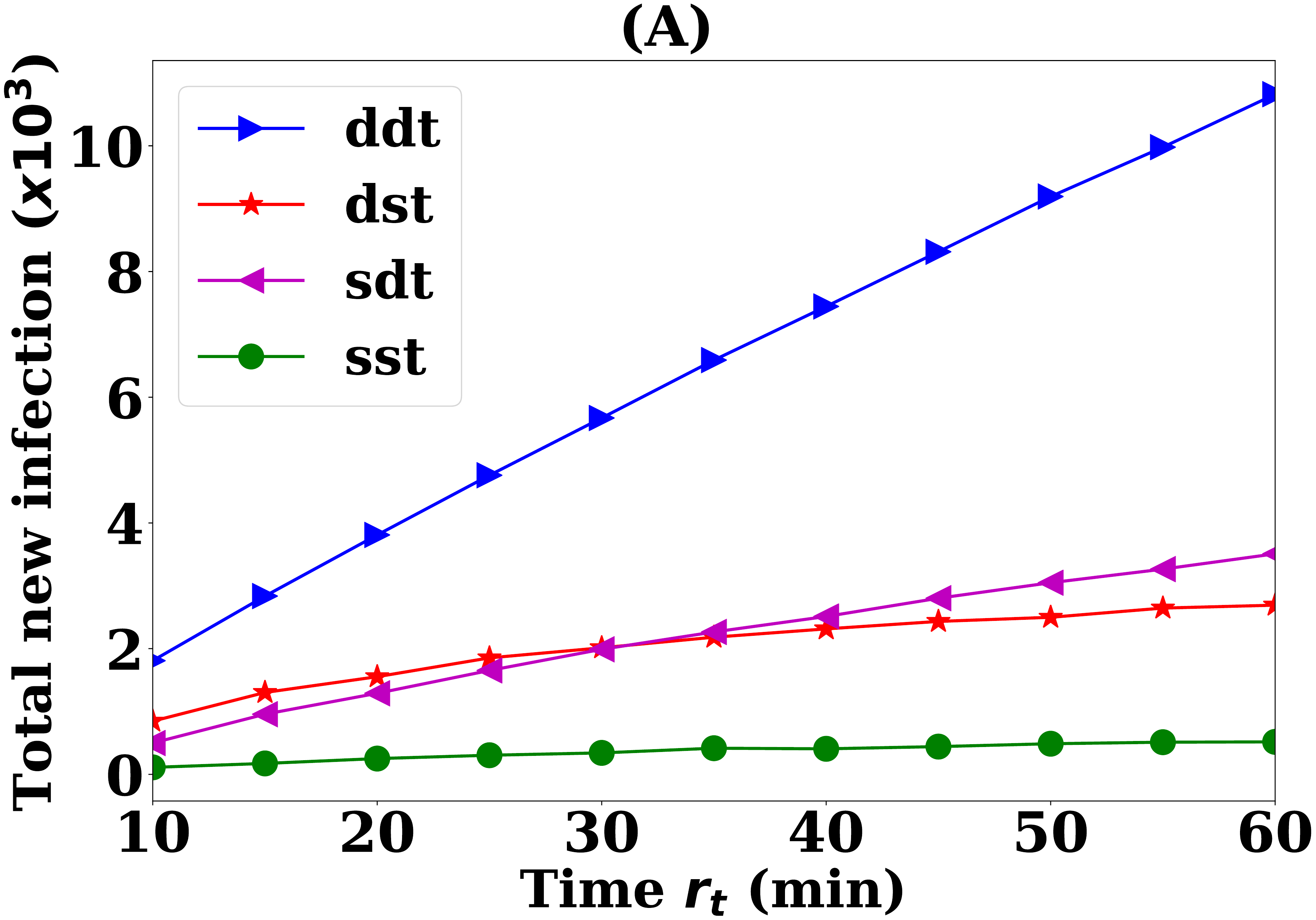}~
 \includegraphics[width=0.45\linewidth, height=5.5 cm]{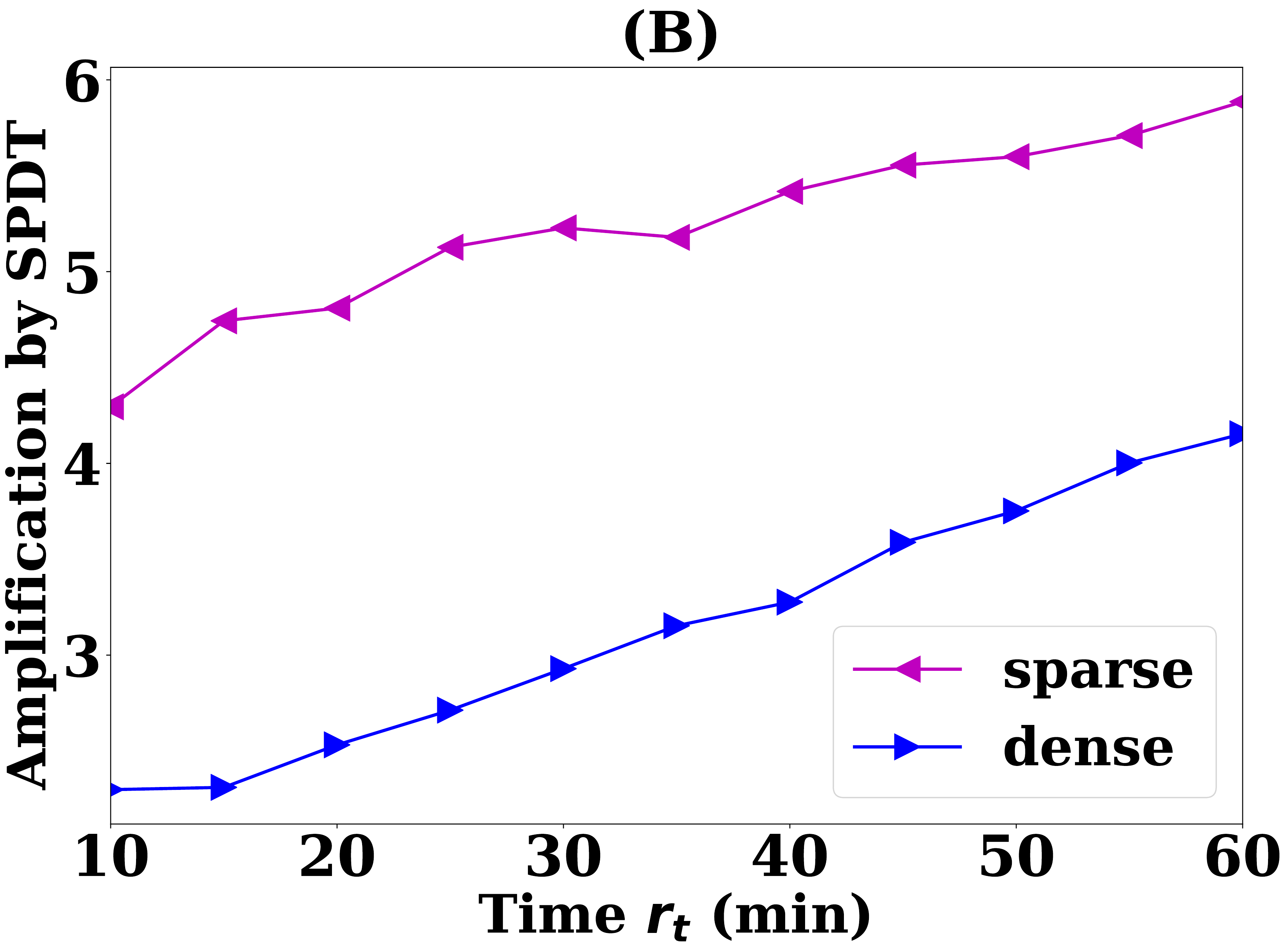}\\ \vspace{0.5 em}
\includegraphics[width=0.45\linewidth, height=5.5 cm]{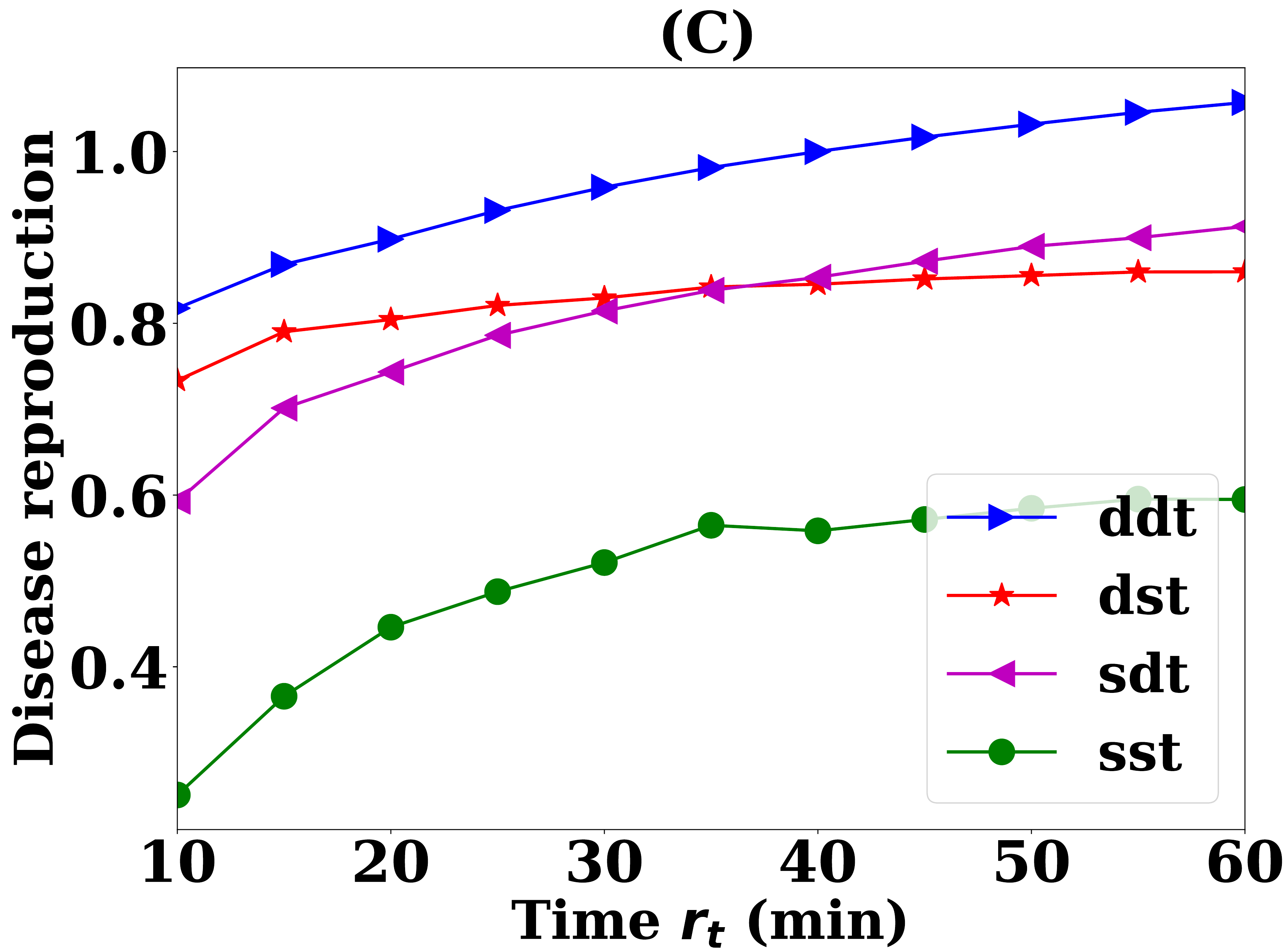}~
\includegraphics[width=0.45\linewidth, height=5.5 cm]{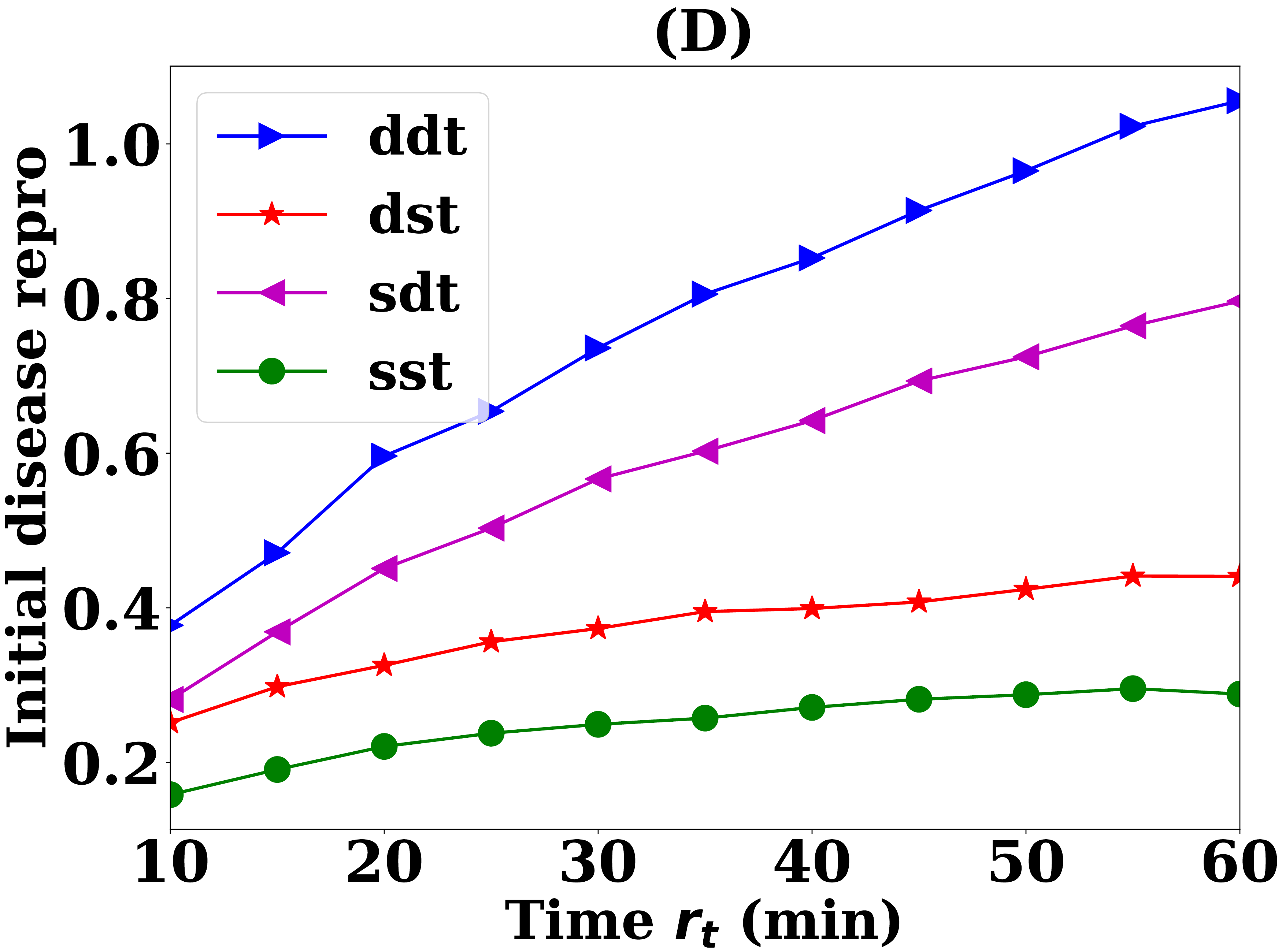}~
\caption{Diffusion dynamics including interquartile ranges on SPST and SPDT networks with particles decay rates $r_t$ : A) total number of infections - outbreak sizes, B)  amplification by SPDT model, C) disease reproduction number, and D) initial disease reproduction number}
    \label{fig:ovldif}
\end{figure}

This  experiment explores the influence of particle decay rates on diffusion varying $r_t$ in the range [10, 60] minutes with step of 5 minutes. 1000 simulations are run for each $r_t$ on the both SPDT and SPST networks with sparse and dense configurations. Fig.~\ref{fig:ovldif} shows the overall diffusion characteristics for all $r_t$. In total, the new infections in SPDT model increases linearly with $r_t$ (Fig.~\ref{fig:ovldif}A). The amplification with the SPDT model is up to 5.6 times for sparse network and 4.3 times for dense network at $r_t=60$ min (Fig.~\ref{fig:ovldif}B). The individuals in the SPDT model achieve strong disease reproduction abilities relative to SPST (Fig.~\ref{fig:ovldif}C). Thus, outbreak sizes are amplified in the SPDT model. The initial disease reproduction abilities (Fig.~\ref{fig:ovldif}D), which is calculated at the first simulation day, shows how the contact networks become favourable for diffusion with increasing $r_t$ in both SPDT and SPST models. The initial disease reproduction abilities in SPDT model are strongly influenced by $r_t$.

\begin{figure}[h!]
\centering    \includegraphics[width=0.45\linewidth, height=5.3 cm]{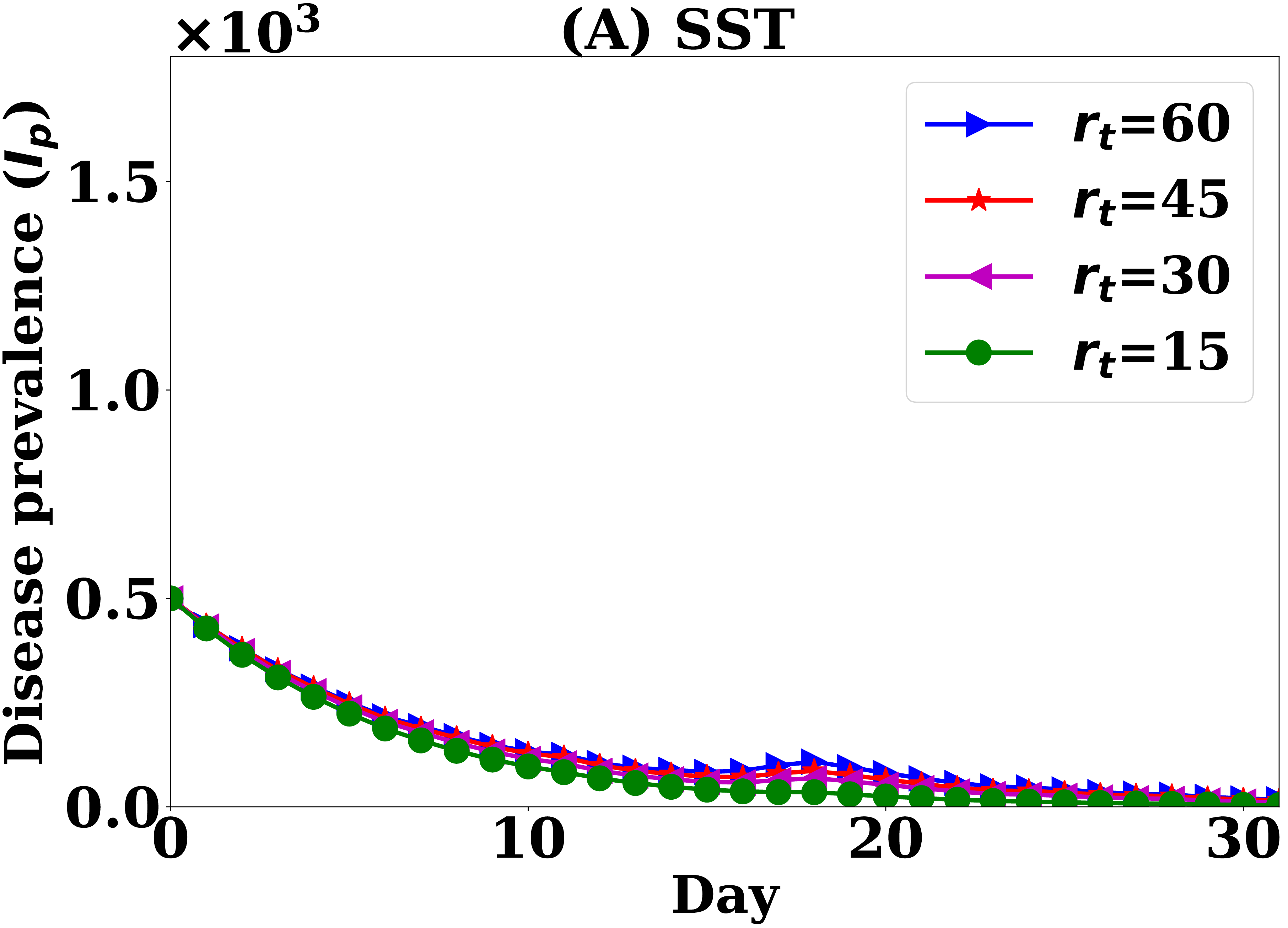}~
\includegraphics[width=0.45\linewidth, height=5.3 cm]{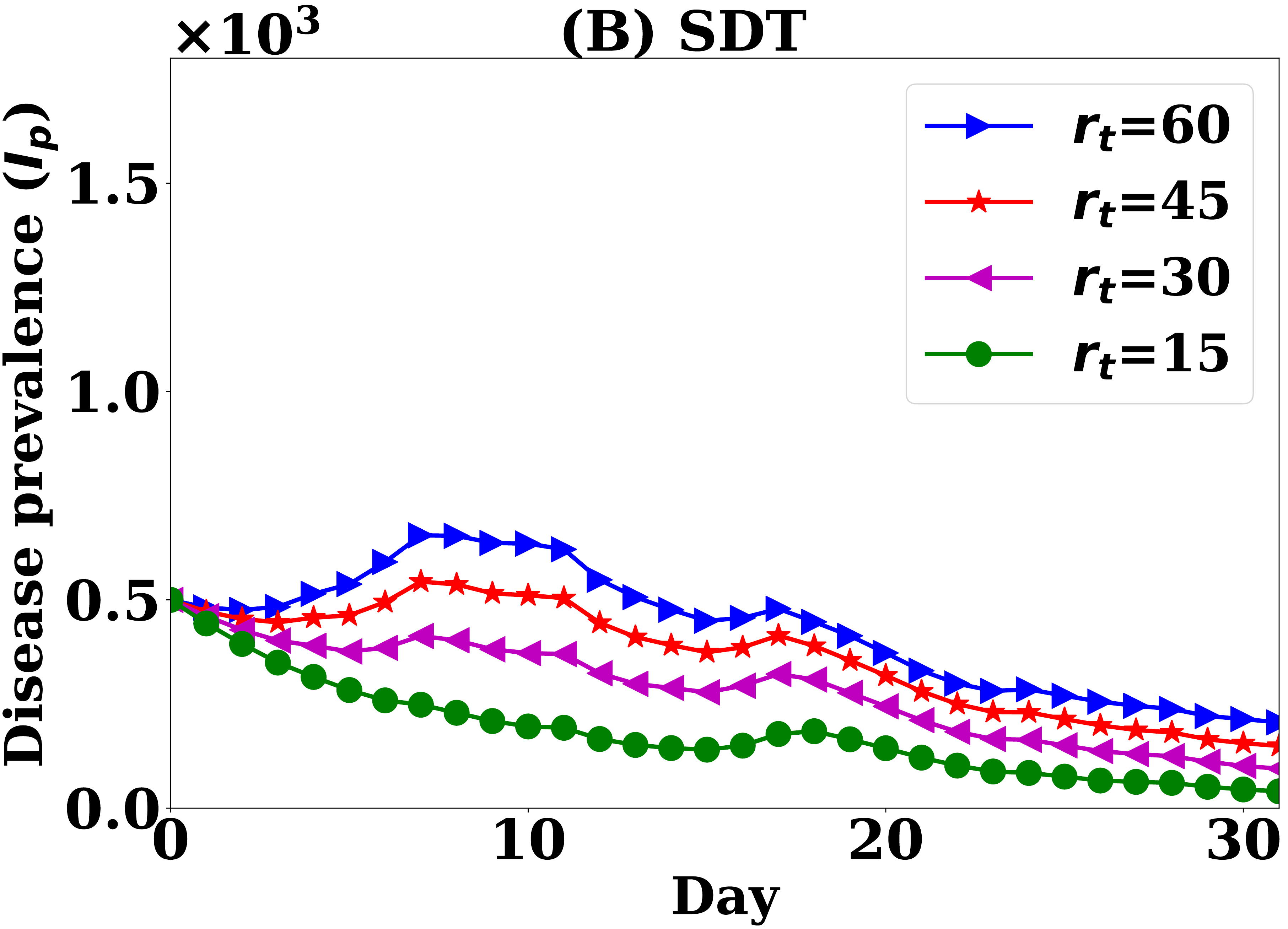}\\ \vspace{0.5em}
\includegraphics[width=0.45\linewidth, height=5.3 cm]{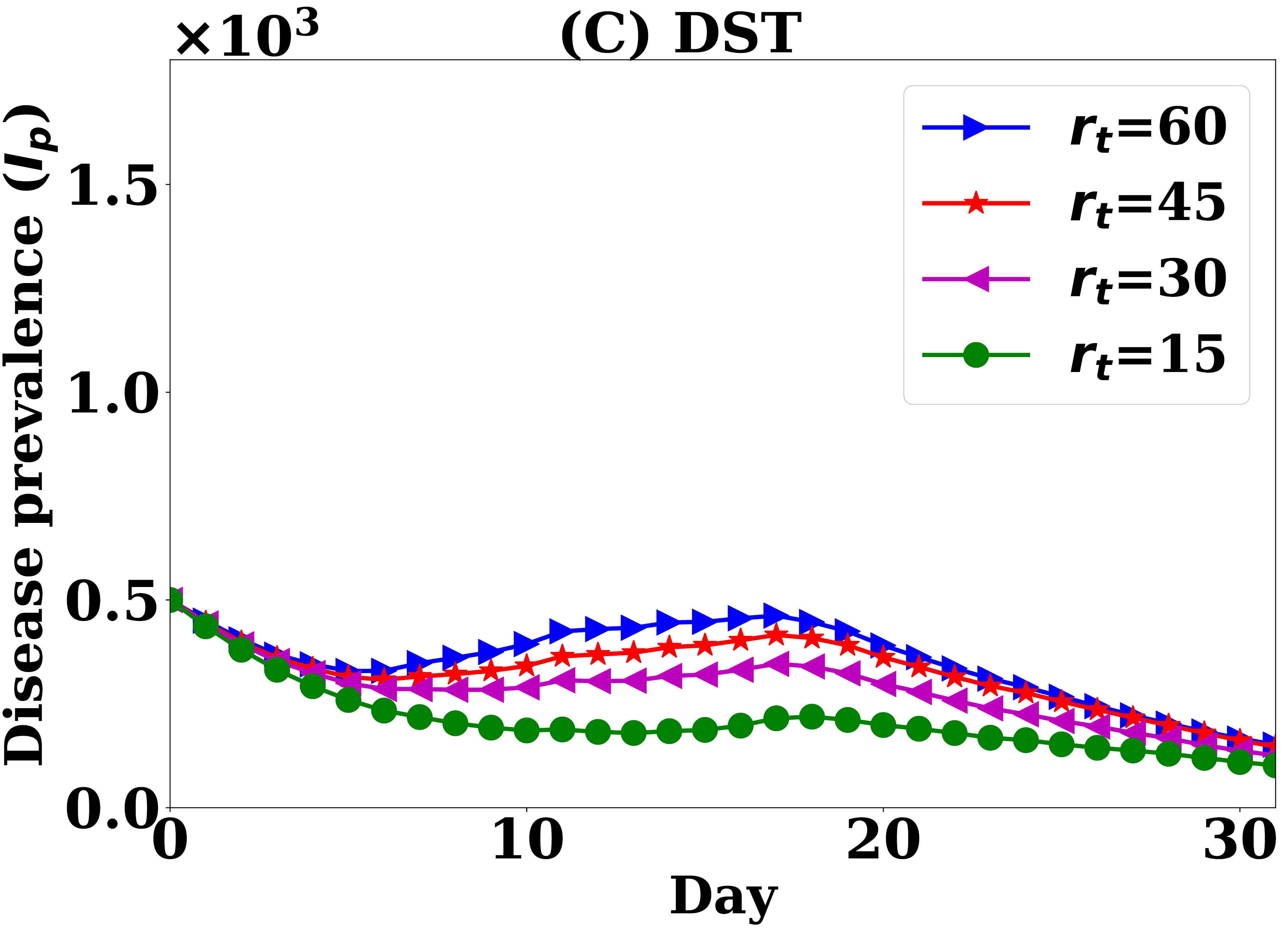}~
\includegraphics[width=0.45\linewidth, height=5.3 cm]{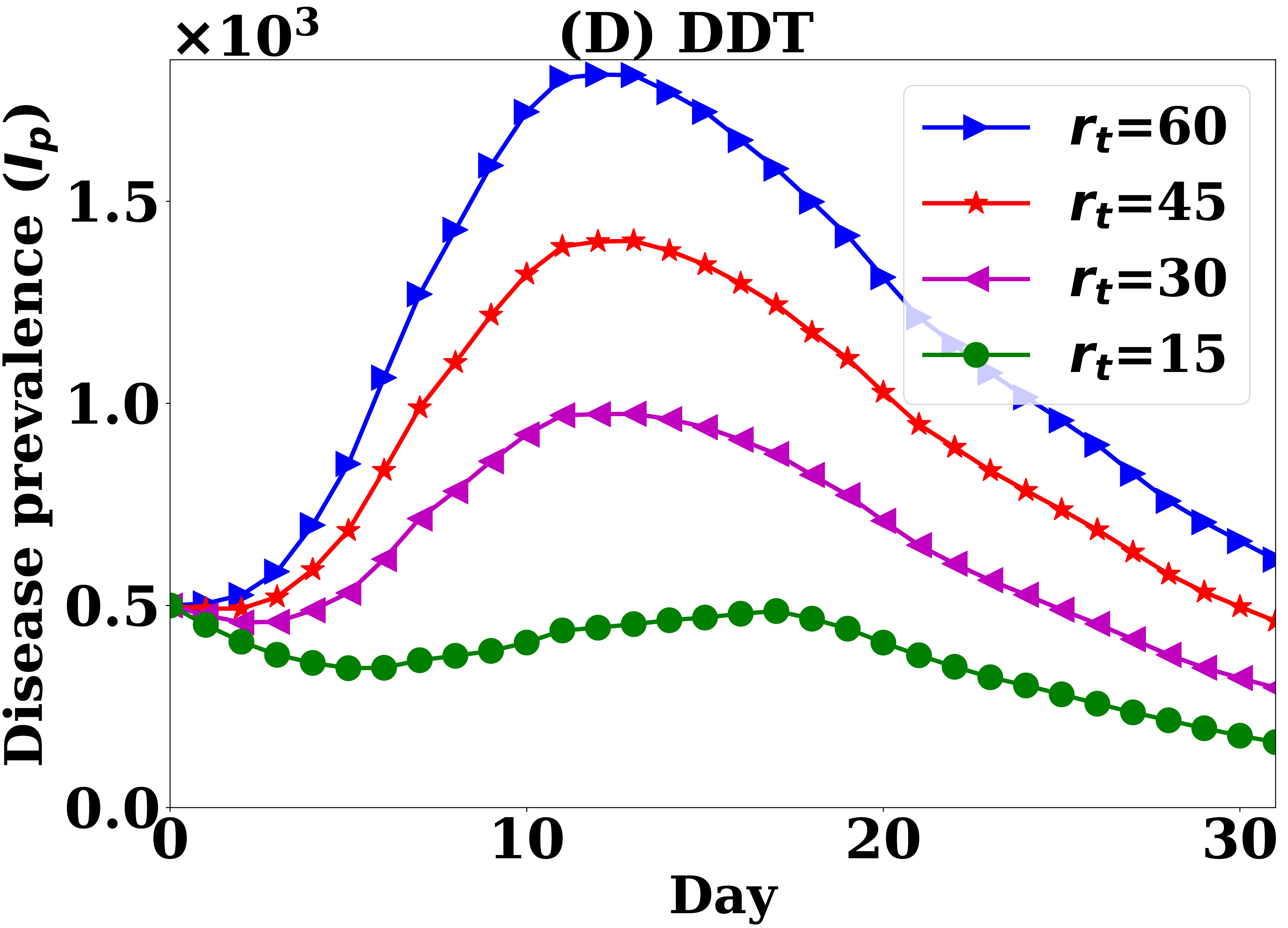}
\caption{Daily variations in disease reproduction rate $R_t$ with particle decay rates $r_t$: A) sparse SPST network (SST), B) sparse SPDT network (SDT), C) dense SPST network (DST), and D) dense SPDT network (DDT)}
    \label{fig:disprv}
\vspace{-1.5 em}
\end{figure}

\begin{figure}[h!]
    \centering    
 \includegraphics[width=0.45\linewidth, height=5.3 cm]{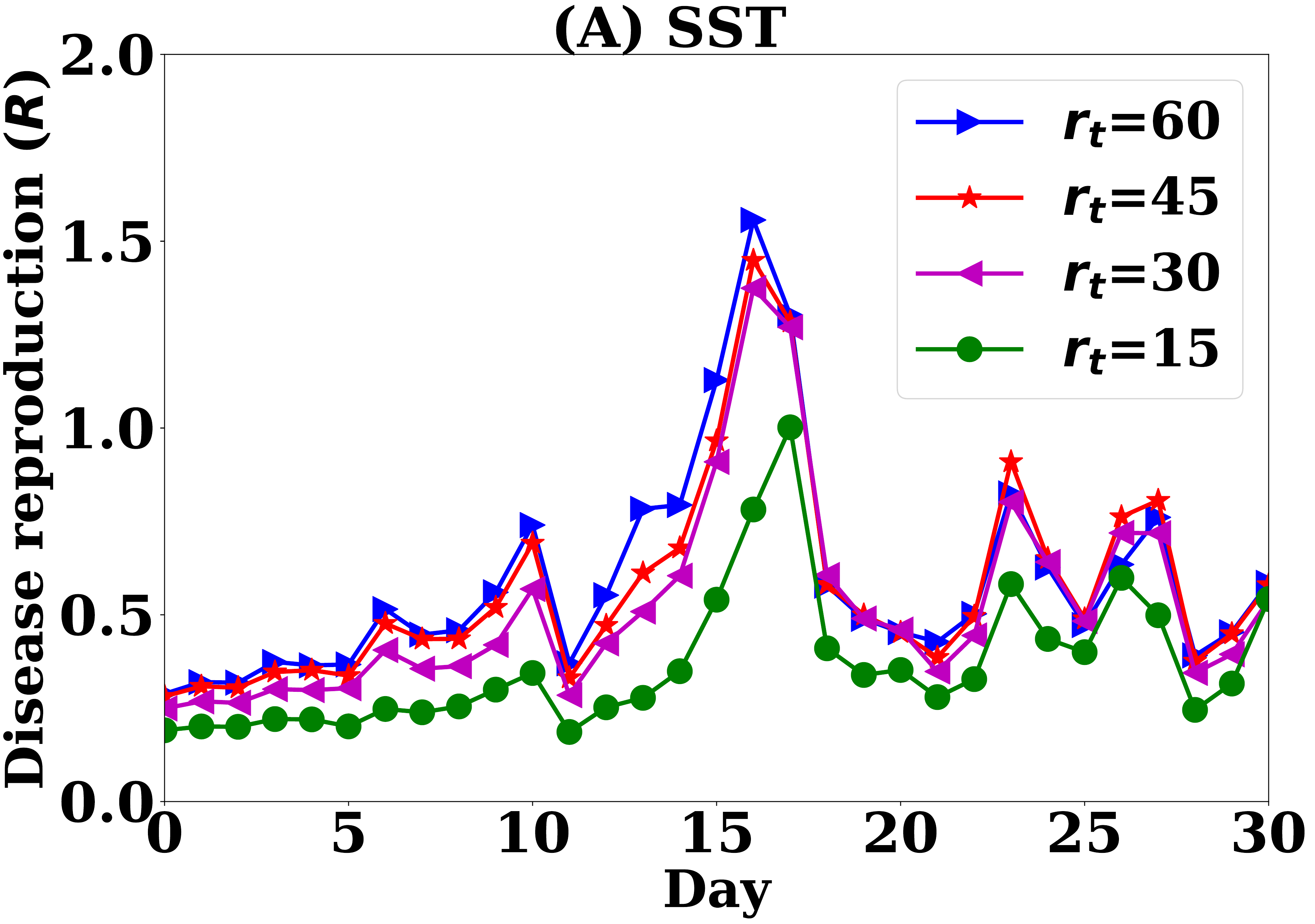}~
\includegraphics[width=0.45\linewidth, height=5.3 cm]{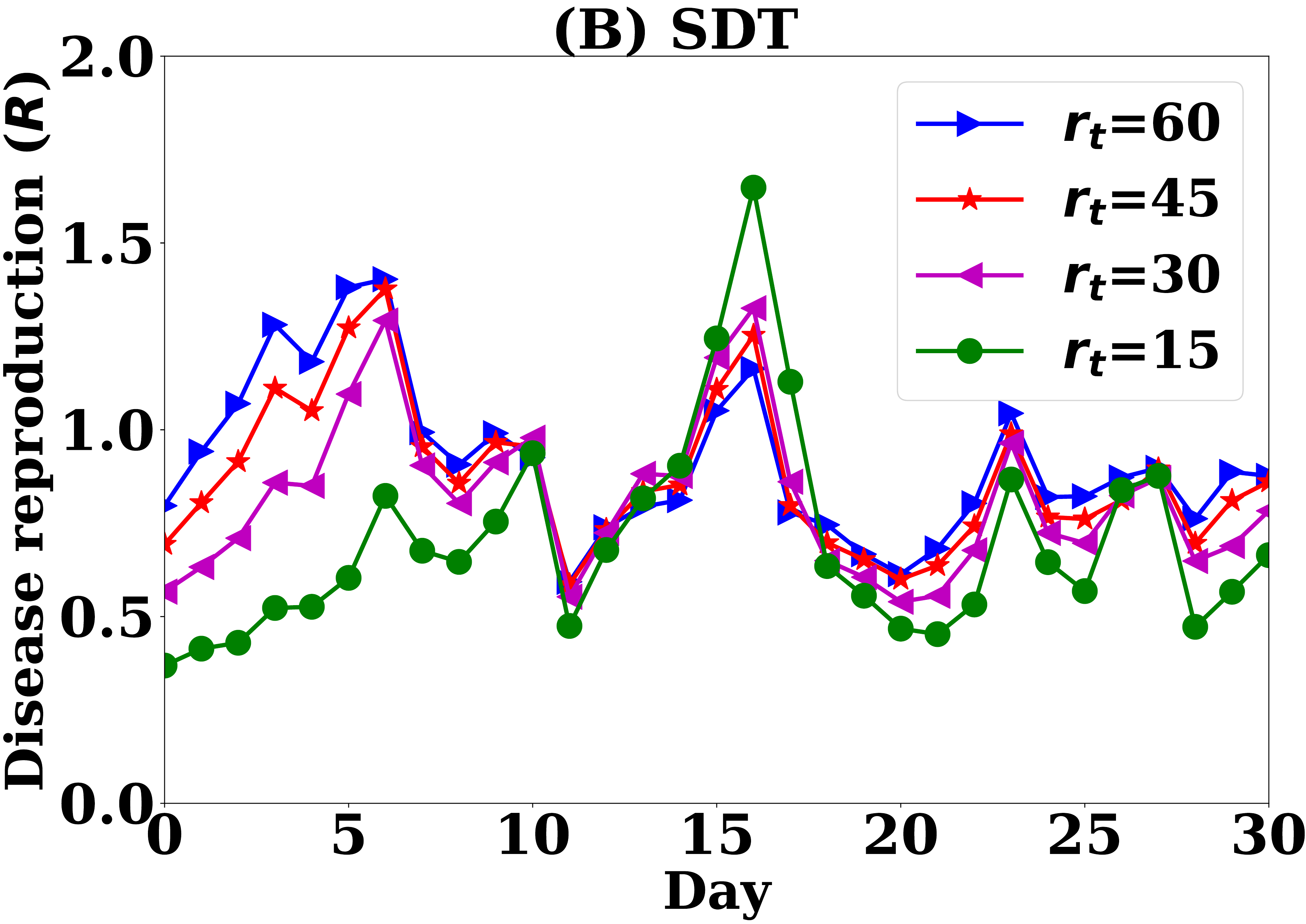}\\ \vspace{0.5em}
\includegraphics[width=0.45\linewidth, height=5.3 cm]{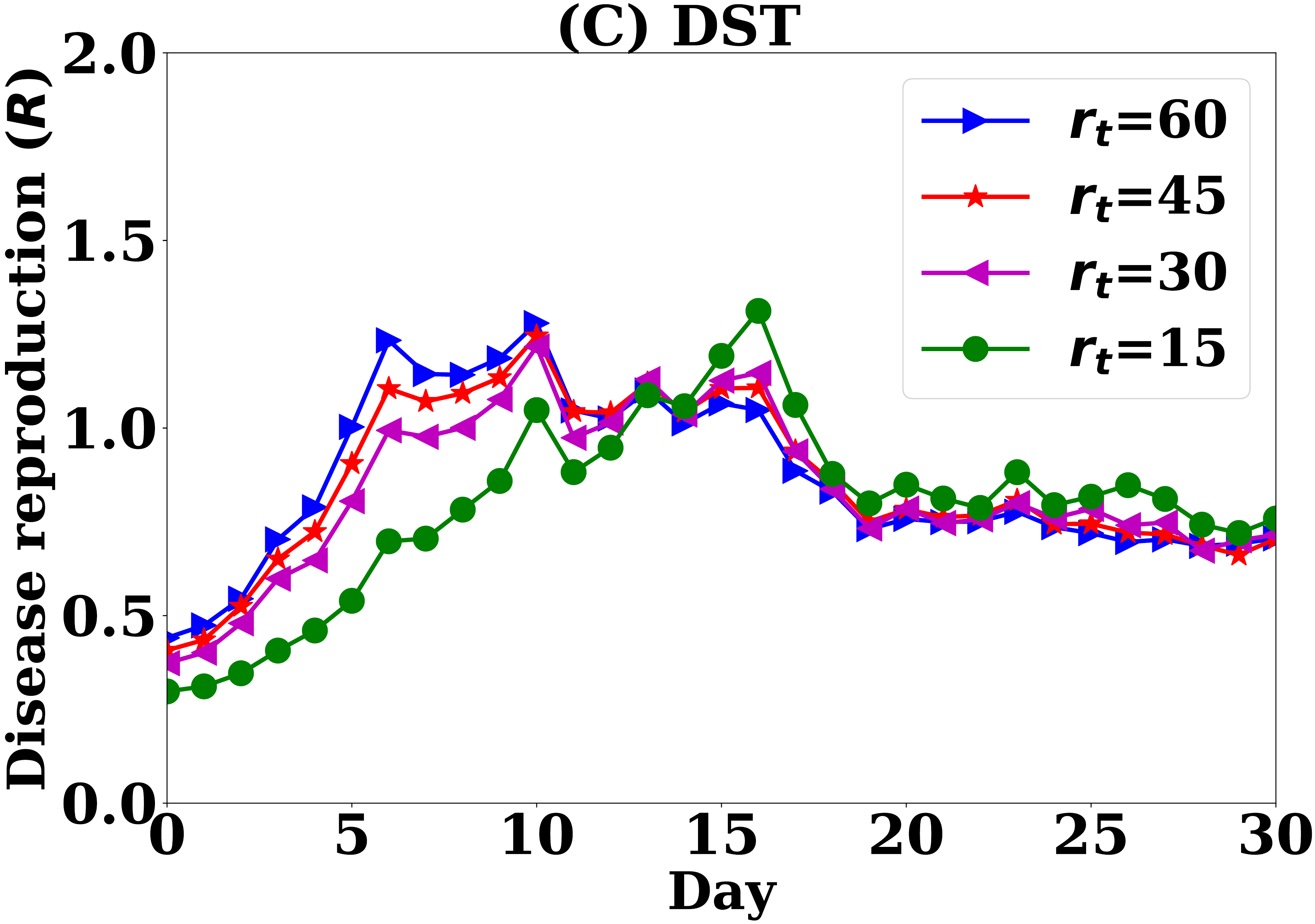}~
\includegraphics[width=0.45\linewidth, height=5.3 cm]{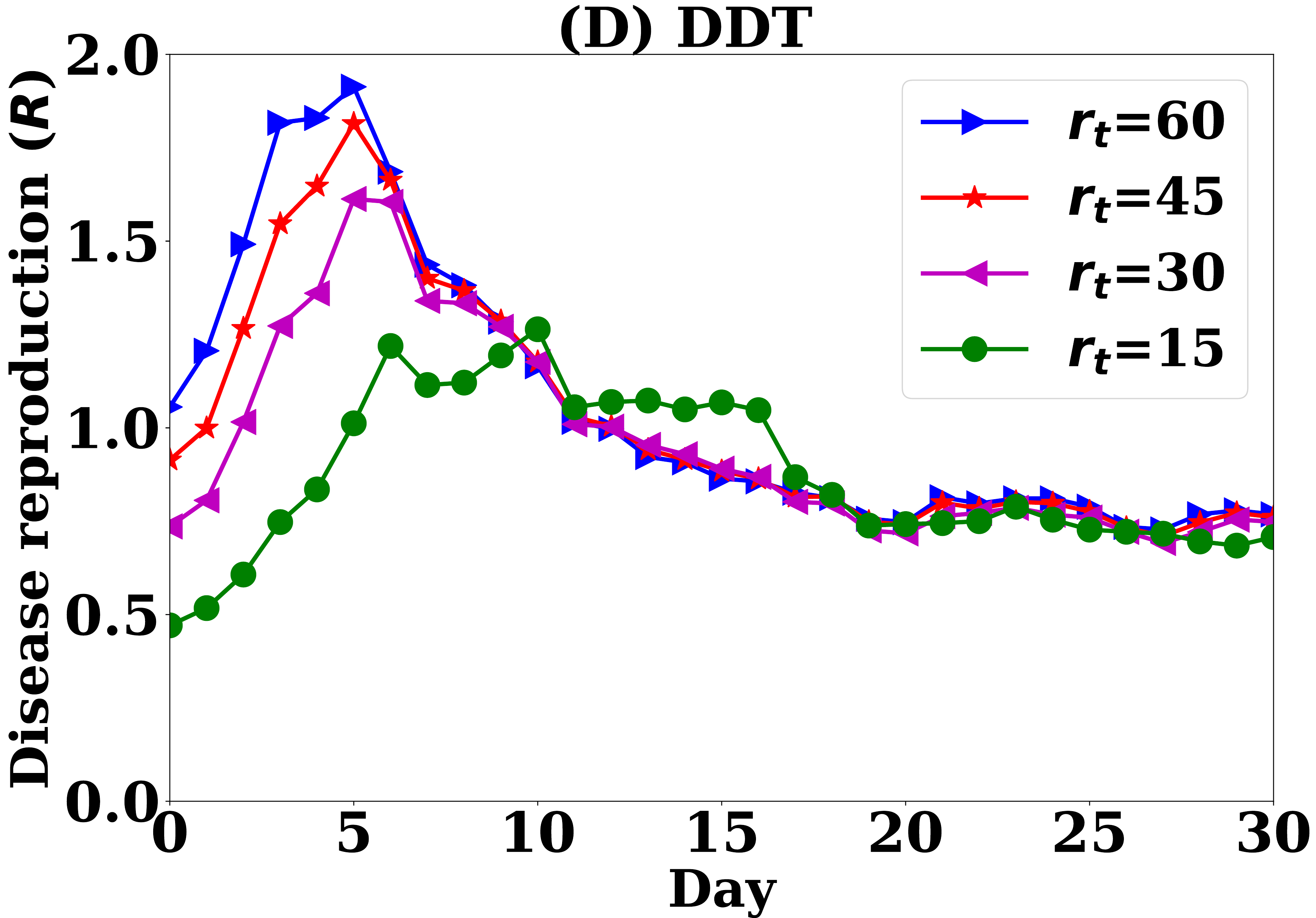}
\caption{Daily variations in disease reproduction rate $R_t$ with particle decay rates $r_t$: A) sparse SPST network (SST), B) sparse SPDT network (SDT), C) dense SPST network (DST), and D) dense SPDT network (DDT)}
    \label{fig:difp}
\end{figure}

The temporal variation of disease prevalence for $r_t=\{15,30,45,60\}$ min are presented in Fig.~\ref{fig:disprv} while Fig.~\ref{fig:difp} shows the variations in disease reproduction rate $R$. The results show that the diffusion dynamics are strongly governed by $r_t$. The SDT network shows growing of disease prevalence $I_p$ from the initial 500 infected individuals for values of $r_t \geq 45$ min while $I_p$ drops for all other values (Fig.~\ref{fig:disprv}B).
In the heterogeneous contact networks, disease gradually reaches to the individuals, called higher degree individuals, who have high contact rates and hence the value of $R$ gradually increases~\cite{may2001infection,shirley2005impacts}.
 This growth of $R$ is faster at high $r_t$ due to strong underlying connectivity (Fig.~\ref{fig:difp}B). However, individuals with a high degree get infected earlier and the number of susceptible individuals reduces through time.
 Hence, an infection resistance force grows in the network and the rate $R$ of infected individuals decreases. Therefore, an initial small $R$ for $r_t\geq 45$ min quickly increases above one which grows $I_p$ as long as $R$ remains above one and then decreases (Fig.~\ref{fig:difp}B). For $r_t<45$ min, $R$ slowly grows above one due to the weak underlying network connectivity with low $r_t$ and $I_p$ decreases significantly with time.
 As a result, $I_p$ increases slightly and then start dropping. In the SST network, $I_p$ could not grow for any value of $r_t$ due to very small initial $R$ and lack of connectivity for considering only direct links (Fig.~\ref{fig:disprv}A and  Fig.~\ref{fig:difp}A).
The SDT network at $r_t=10$ min shows similar trend for the SST network (Fig.~\ref{fig:difp}A and Fig.~\ref{fig:difp}B). This is because the SDT network becomes similar to the SST network due to weak underlying connectivity for this low $r_t$

The impact of SPDT model becomes stronger in the dense DDT network where infected individuals apply their full infection potential staying every day in the network. Hence, the total new infections  and disease prevalence $I_p$ increase significantly (Fig.~\ref{fig:ovldif}A and Fig.~\ref{fig:disprv}D). The DDT network is capable of increasing $I_p$ even at lower $r_t\geq 20 $ min. Due to high link density, disease reproduction rate $R$ in the DDT network reaches one very quickly and then increases faster as time goes (Fig.~\ref{fig:difp}D). The rate $R$ has multiple effects on the disease prevalence $I_p$. At the higher value of $I_p$ and $R>1$, for small changes in $R$ have a large increase in $I_p$. Thus, small variations in $r_t$ change $R$ which in turn significantly changes $I_p$ in the DDT network. On the other hand, initially, $I_p$ drops for all values of $r_t$ in the DST network and then starts increasing after some days as the high degree individuals are infected. However, this increase is not within the same range than the DDT network due to a weak $R$ and a lack of underlying connectivity. Similarly, the DDT network with low $r_t<20$ min behaves comparatively to the DST network as the underlying connectivity becomes weak.  
\begin{figure}[h!]
    \centering    \includegraphics[width=0.45\linewidth, height=5.5 cm]{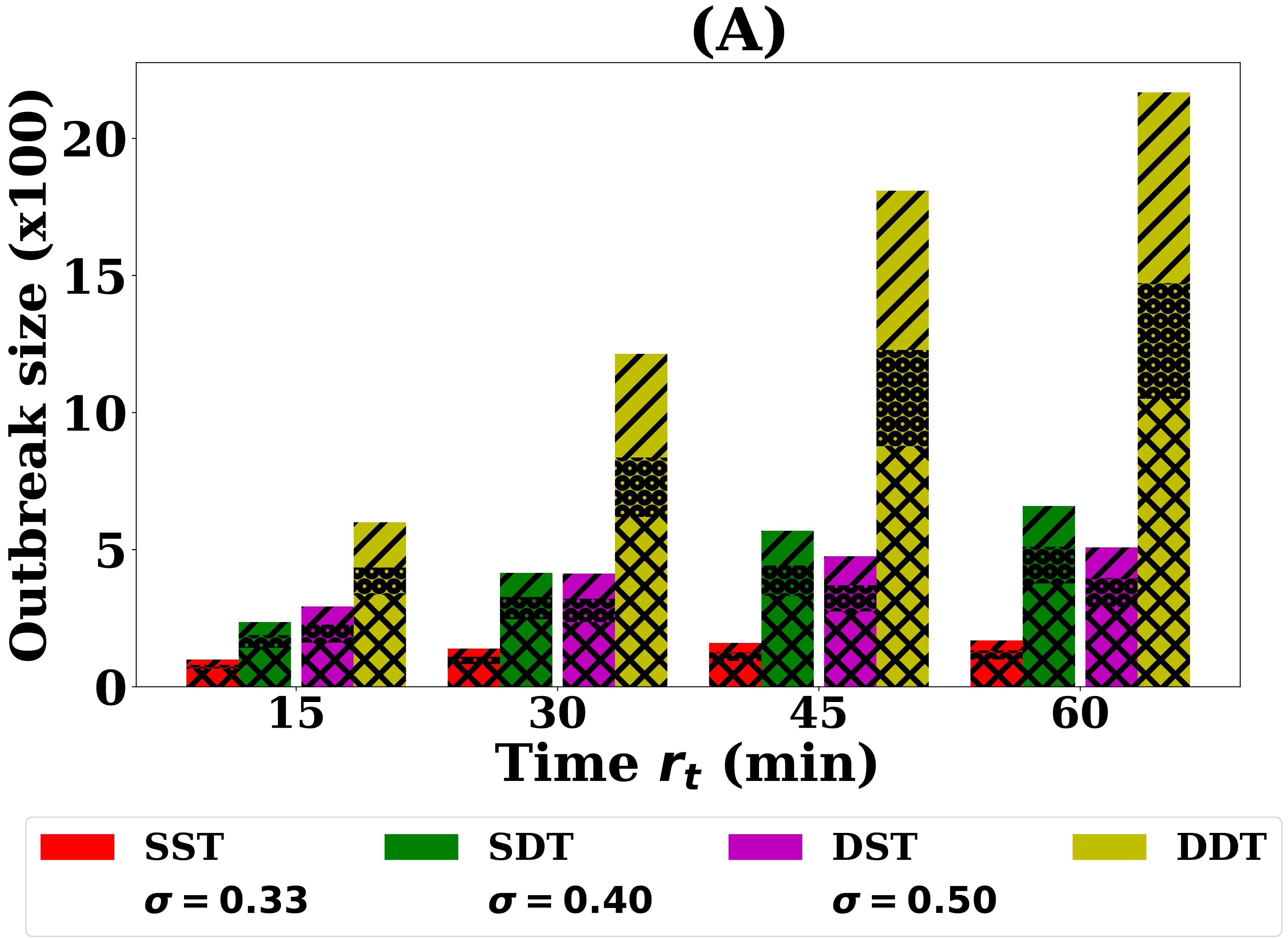}~
\includegraphics[width=0.45\linewidth, height=5.5 cm]{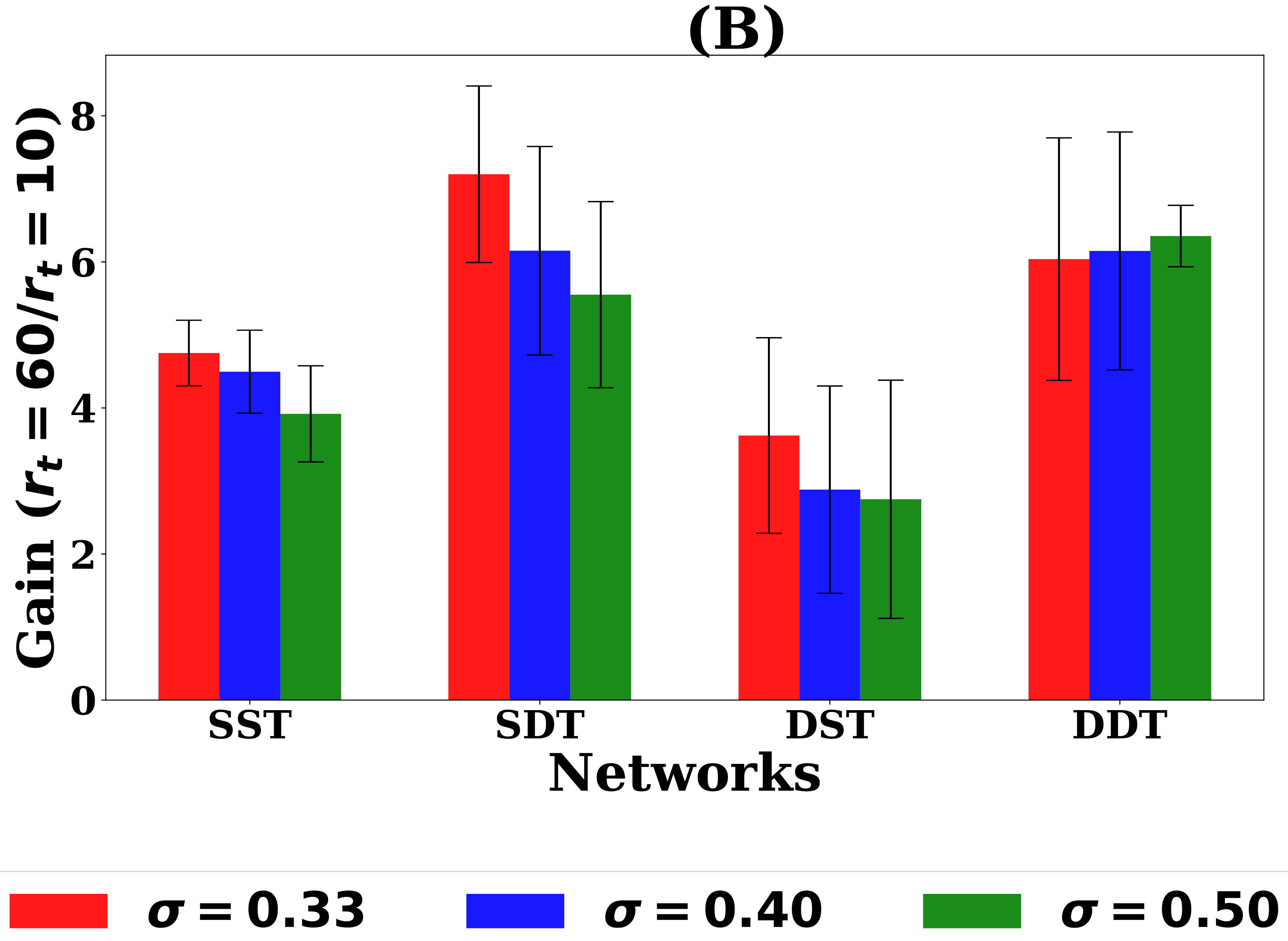}\\[1ex]
\includegraphics[width=0.45\linewidth, height=5.5 cm]{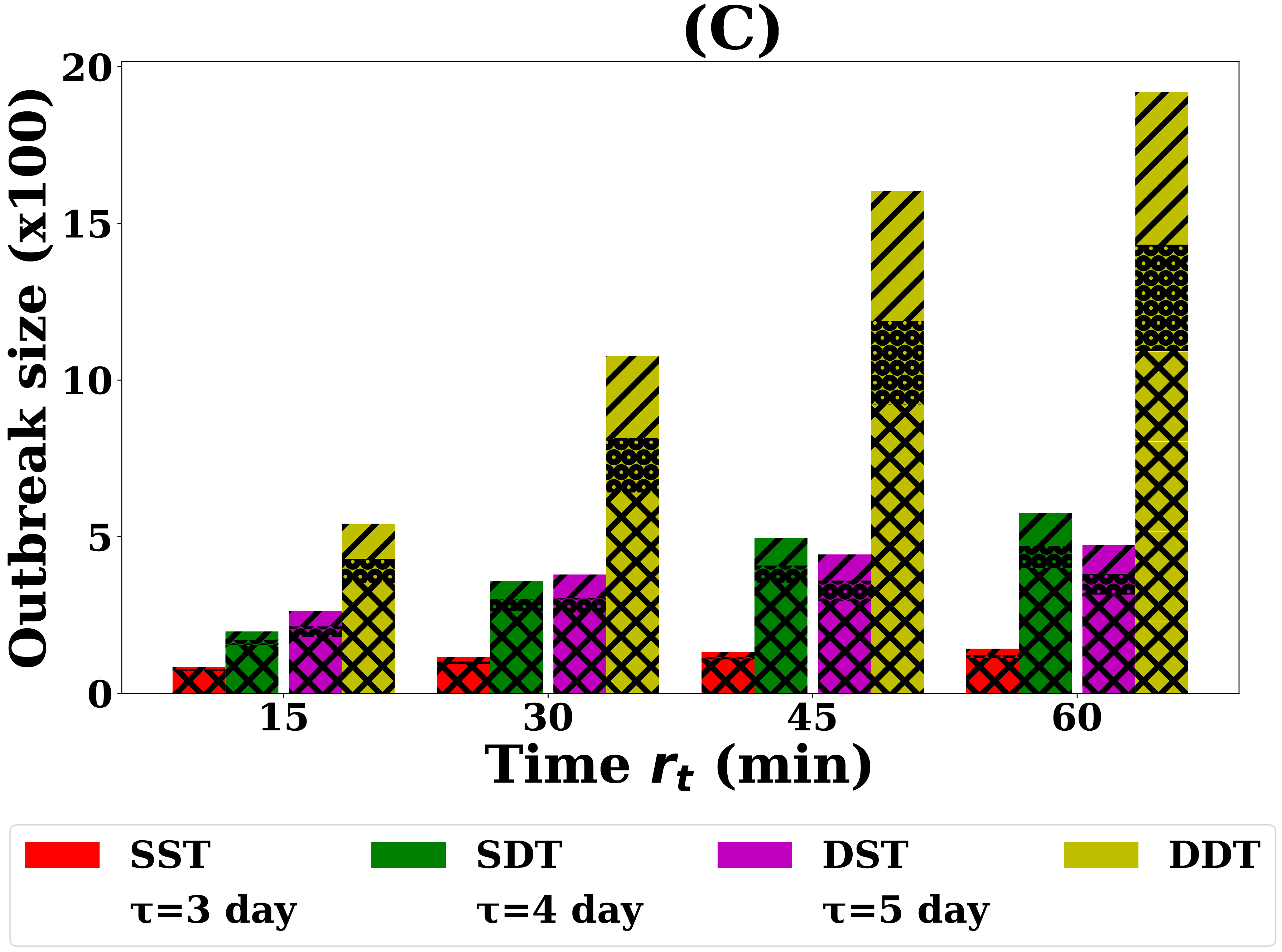}~
\includegraphics[width=0.45\linewidth, height=5.5 cm]{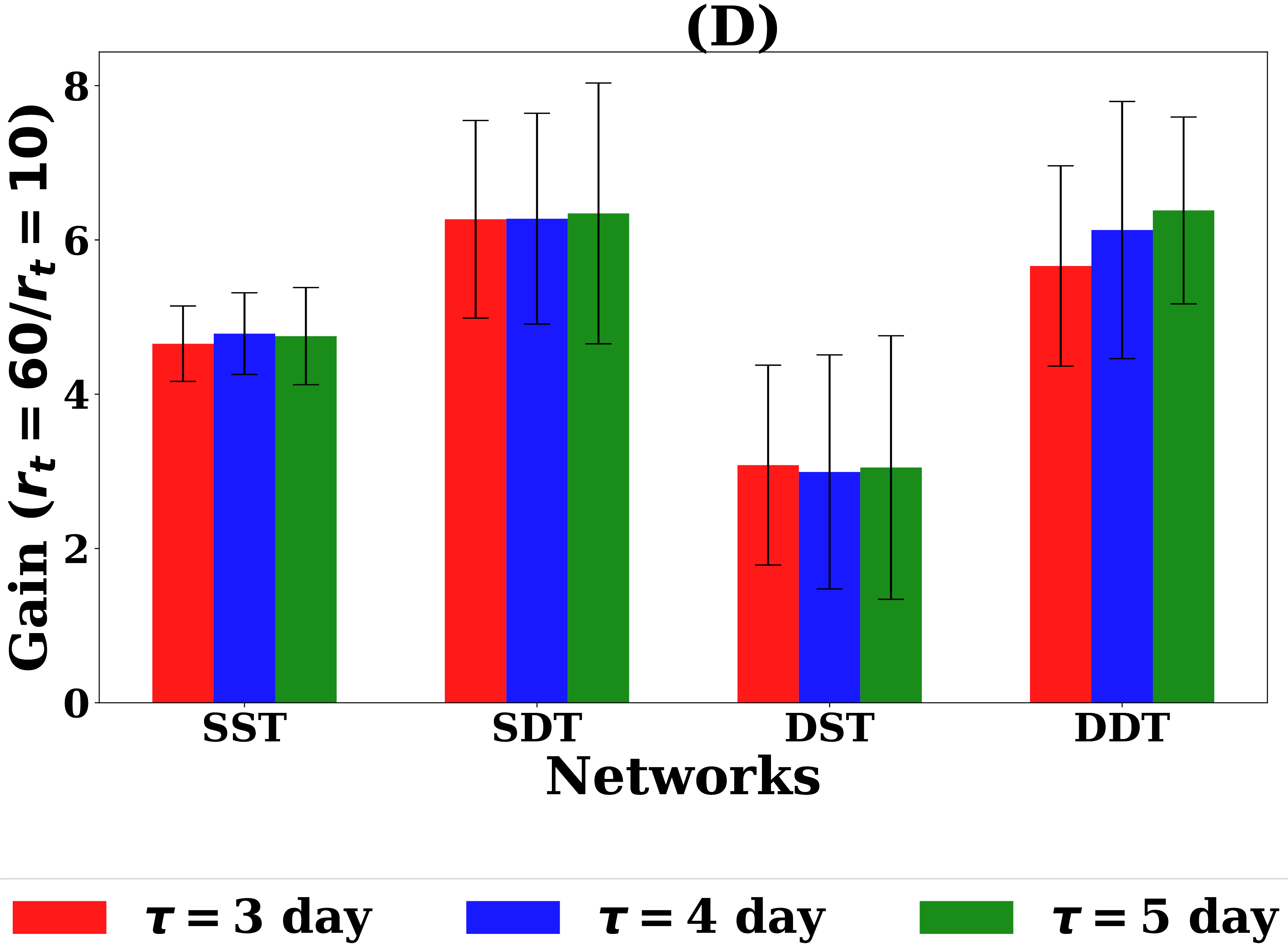}
	\caption{Diffusion dynamics for various disease parameters in the different networks: A) outbreak sizes for various infectiousness $\sigma$ - different pattern showing amplification for increasing $\sigma$, B) gains in outbreak sizes for changing $r_t$ from 10 min to 60 min for various $\sigma$, C) outbreak sizes for various infectious period $\tau$ - different pattern showing amplification for increasing $\tau$, and D) gains in outbreak sizes for changing $r_t$ from 10 min to 60 min for various $\tau$}
    \label{fig:dsp}

\end{figure}

\subsubsection{Diffusion for various disease parameters}
The impact of $r_t$ on SPDT diffusion increases with $\sigma$. For analysis, simulations are run for $\sigma=\{0.33, 0.4, 0.5\}$ on both SPST and SPDT networks. The results are presented in Fig.~\ref{fig:dsp}. The amplification by SPDT model increases as $\sigma$ increases (Fig.~\ref{fig:dsp}A). The required $r_t$ to grow disease prevalence $I_p$ in SPDT model reduces with high $\sigma$. 
It is also noticed that the SPDT models a longer linear amplification by causing large infection with $r_t$ in dense networks. Except for DDT network, the growth in the total infection gain at low $r_t$ due to increasing in $\sigma$ is higher compared to that at high $r_t$. This is shown in the Fig.~\ref{fig:dsp}B through the ratios of total infections at $r_t=60$ min and $r_t=10$ min. The DDT network shows opposite behaviour with the growth in total infection gain at high $r_t$ still increases as $\sigma$ increases. Having high link density and more high degree individuals, the DDT network can achieve stronger infection force to override the infection resistance force coming from the reduction in susceptible individuals. However, the other three networks are affected by the infection resistance force at high $r_t$ more strongly than the low $r_t$ as they have lack of underlying connectivity (the sparse networks SST and DST only account indirect links and SDT has low link density).

Longer infectious period $\tau$ also increases disease reproduction abilities $R$ of infected individuals as recovery forces from infection reduce. The diffusion behaviours for $\bar \tau=\{3,4,5\}$ days with constant $\sigma =0.33$ are shown in the Fig~\ref{fig:dsp}C and Fig~\ref{fig:dsp}D. This also increases total infection and reduces $r_t$ to grow disease prevalence $I_p$. In this case, the DDT network support longer linear amplification as well while for other networks it reaches to the steady state. Besides, the delays to reach peak $I_p$ become longer as $\bar \tau$ increases. This is because $R$ is maintained over one for a long time which grows $I_p$ for a longer time. As a result, the disease persists within the population for a long time in SPDT model than SPST model. 

\begin{figure}[h!]
\centering    
\includegraphics[width=0.42\linewidth, height=5.5 cm]{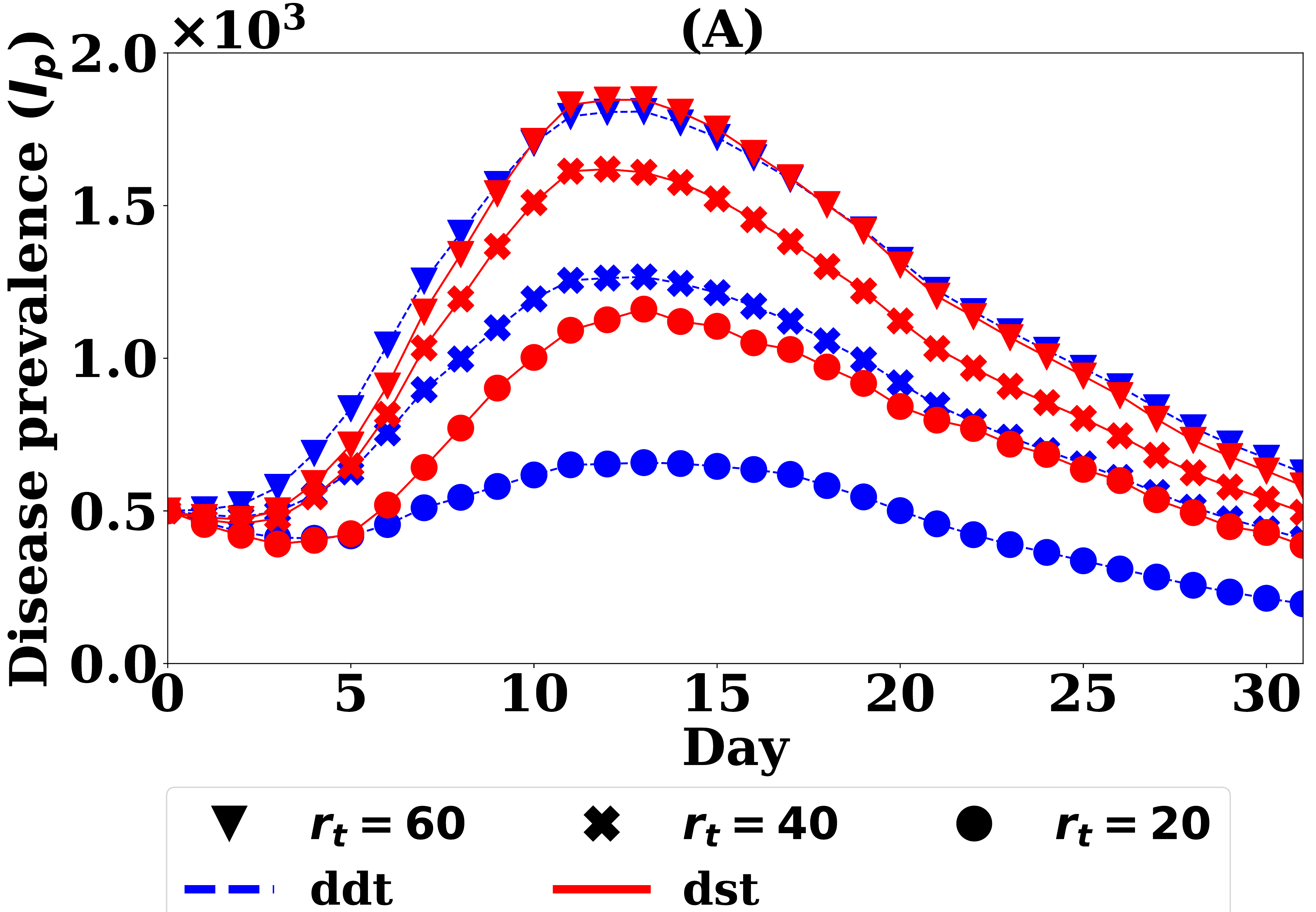}~\quad
\includegraphics[width=0.42\linewidth, height=5.5 cm]{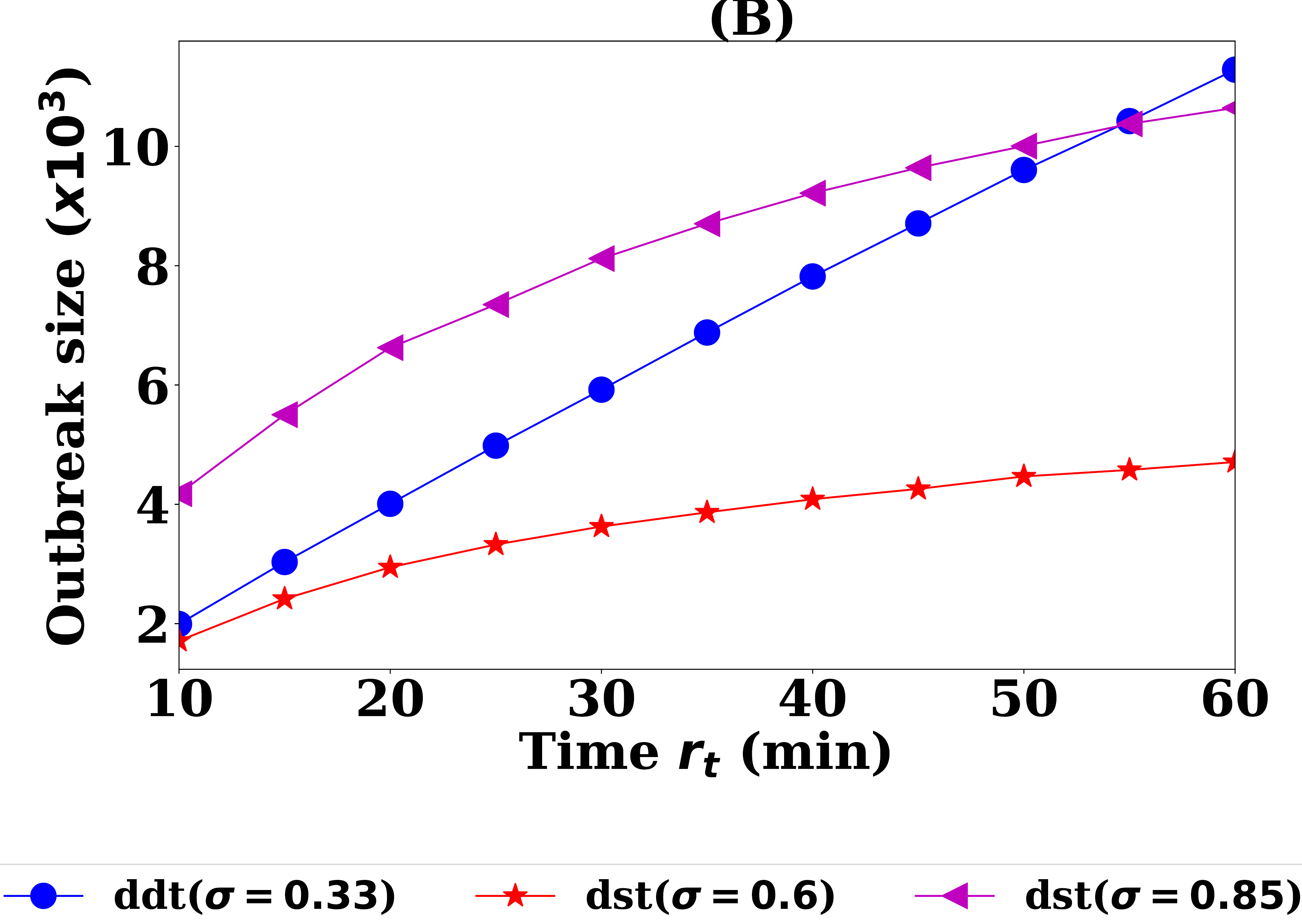}\\ \vspace{0.8em}
\includegraphics[width=0.42\linewidth, height=5.5 cm]{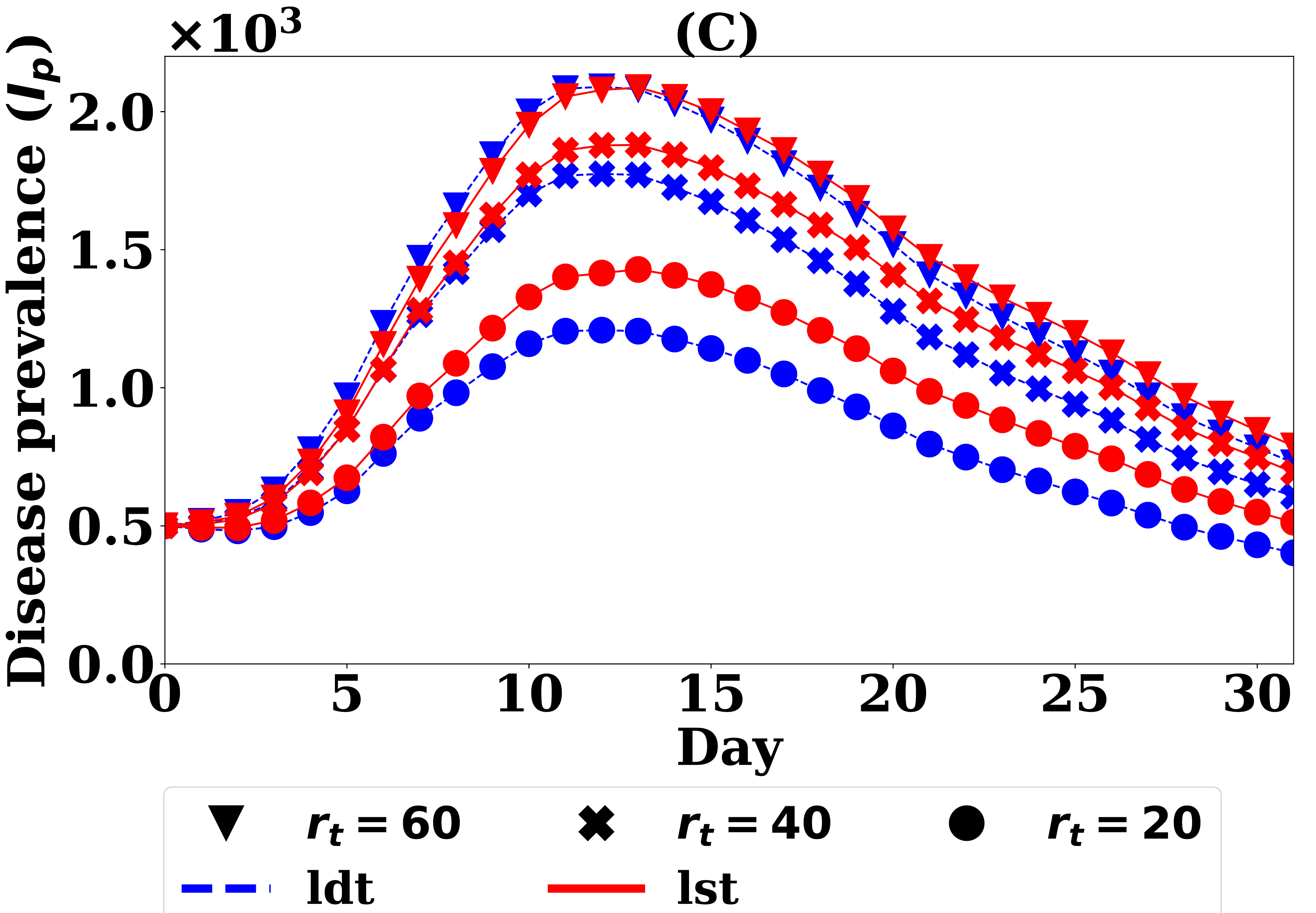}~ \quad
\includegraphics[width=0.42\linewidth, height=5.5 cm]{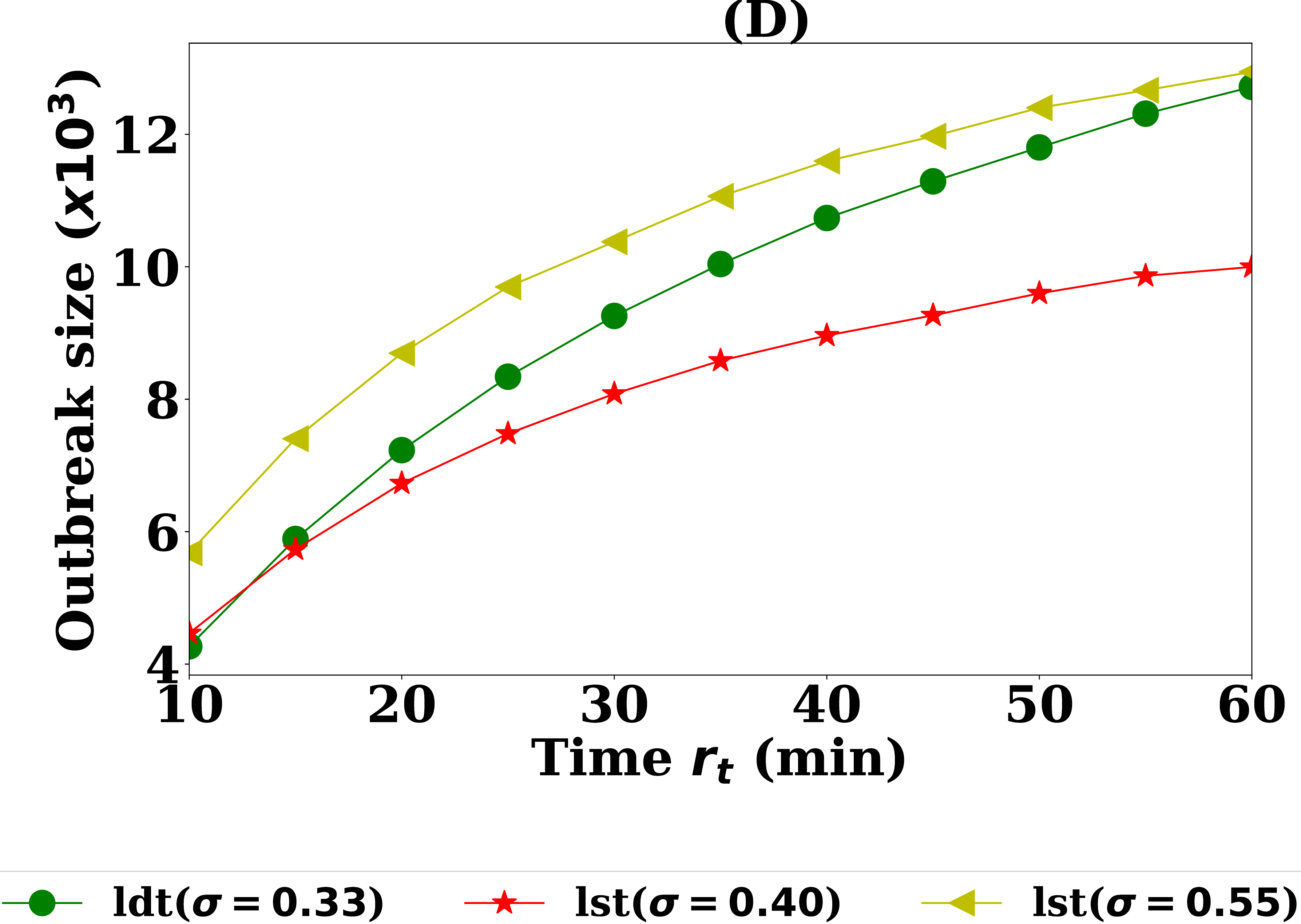}
    \caption{Reconstructing SPDT diffusion dynamics by SPST model and their differences over various particle decay rates $r_t$: A) comparing SPDT diffusion with $\sigma=0.33$ and SPST diffusion with $\sigma=0.85$, B) comparing outbreak sizes of SPDT model with $\sigma=0.33$ and that of SPST model with $\sigma=0.60$ and $\sigma=0.85$, C) comparing LDT diffusion with $\sigma=0.33$ and LST diffusion with $\sigma=0.55$, and D) comparing outbreak sizes of LDT model with $\sigma=0.33$ and that of LST model with $\sigma=0.55$ and $\sigma=0.40$}
    \label{fig:ldfgn}
\end{figure}

\subsubsection{New SPDT diffusion dynamics}
The SPDT model introduces diffusion behaviours that are not observed in the SPST diffusion. The underlying network connectivity in the SPDT model is changed with the particle decay rate $r_t$. The impacts of this property on diffusion dynamics are not reproducible by the SPST model. This is observed by reconstructing equivalent SPDT diffusion dynamics on the corresponding SPST network with strong $\sigma$ (Fig.~\ref{fig:ldfgn}A and Fig.~\ref{fig:ldfgn}B). With $\sigma=0.85$, the SPST diffusion dynamics become similar to that of SPDT model at $r_t=60$ min and total infections in both models are close. However, diffusion dynamics and total infections in SPST model are overestimated at low $r_t$. On the other hand, the SPST model with $\sigma=0.6$ shows same total infection of SPDT model at $r_t$=10 min but underestimates at higher values of $r_t$. This is because the underlying connectivity in SPST model does not vary with $r_t$ and spreading dynamic variations are limited.  

\begin{figure}[h!]
\centering    
\includegraphics[width=0.42\linewidth, height=5.0 cm]{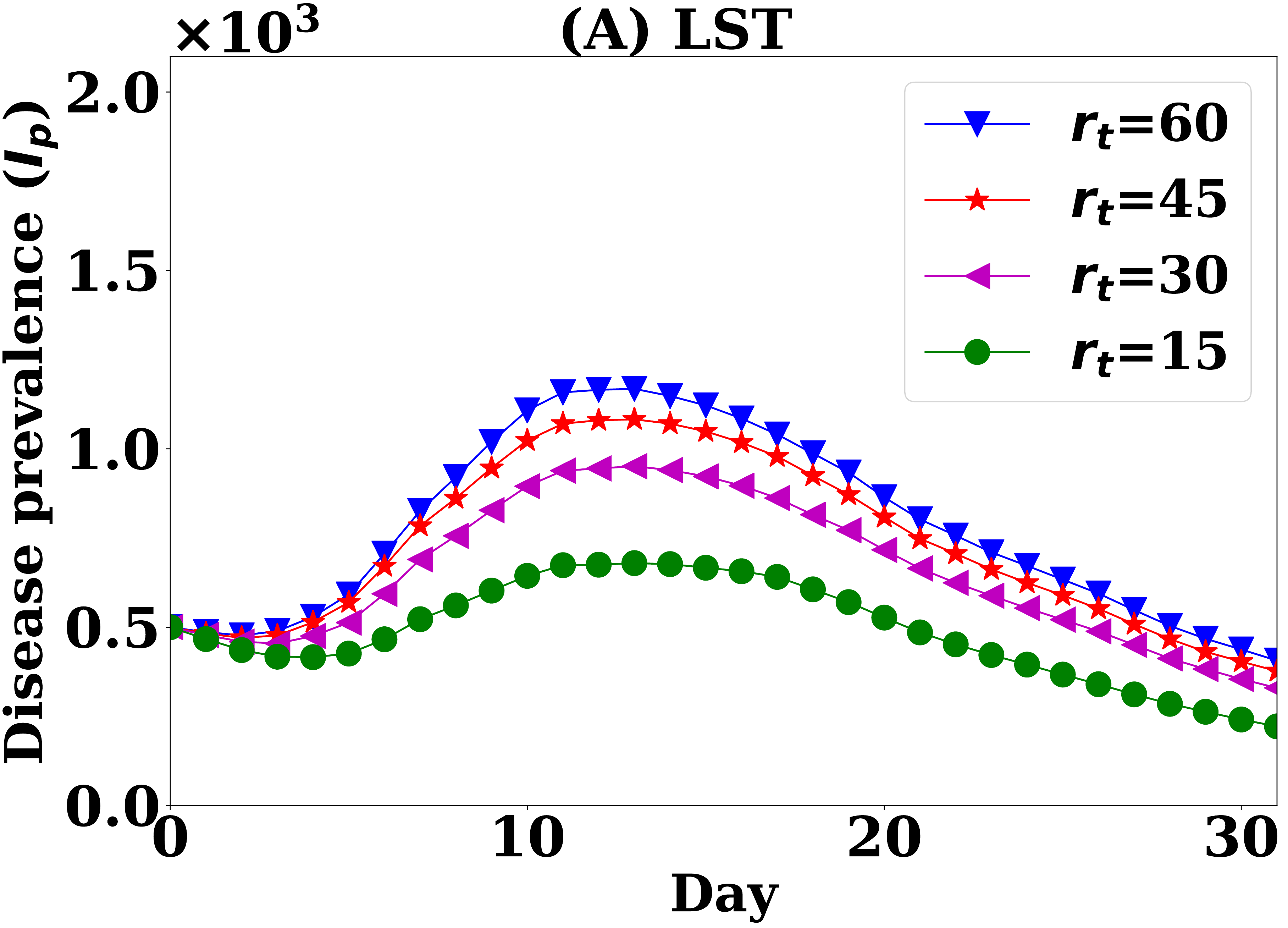}~\quad
\includegraphics[width=0.42\linewidth, height=5.0 cm]{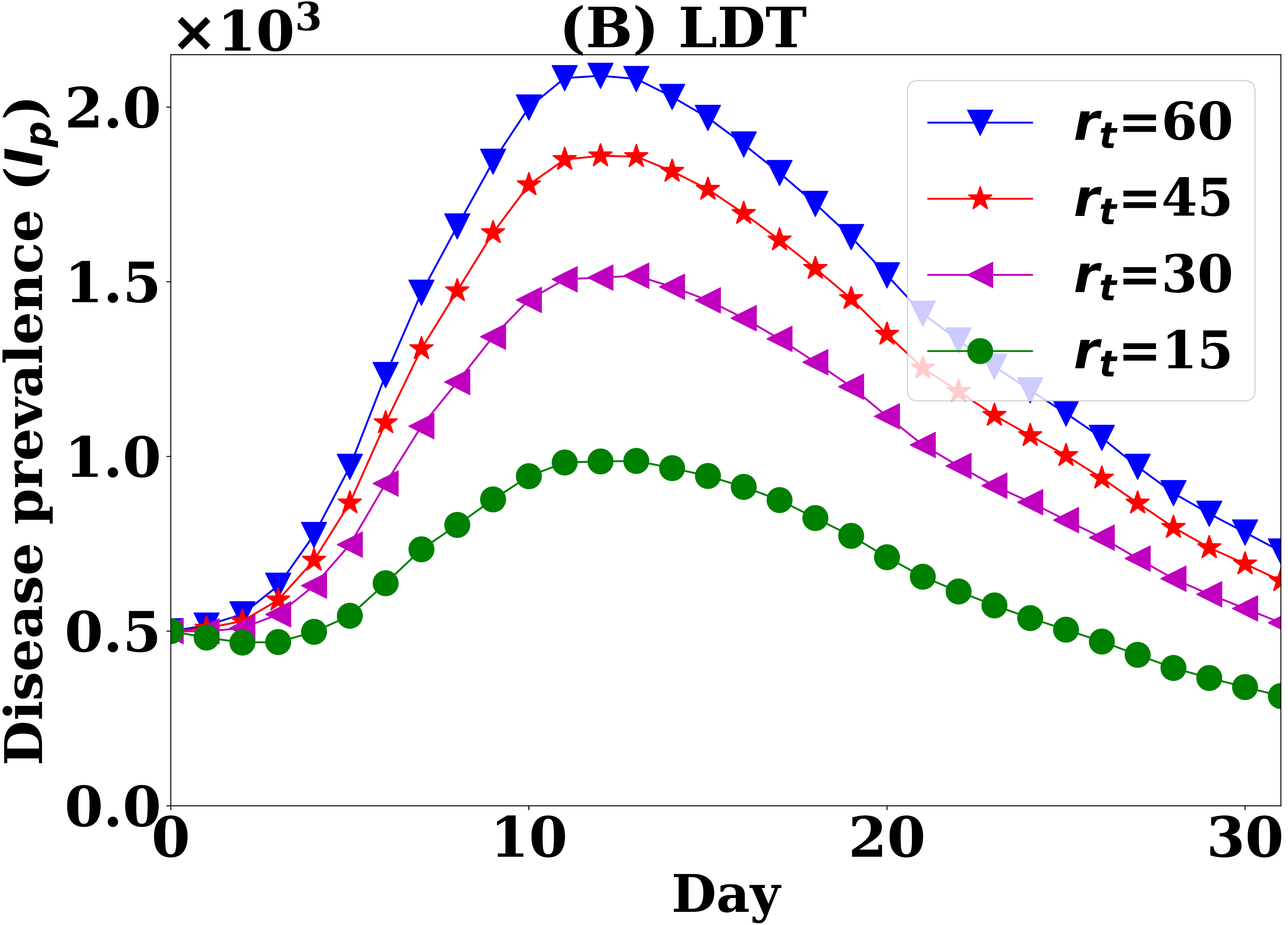}\\ \vspace{0.6em}
\includegraphics[width=0.42\linewidth, height=5.0 cm]{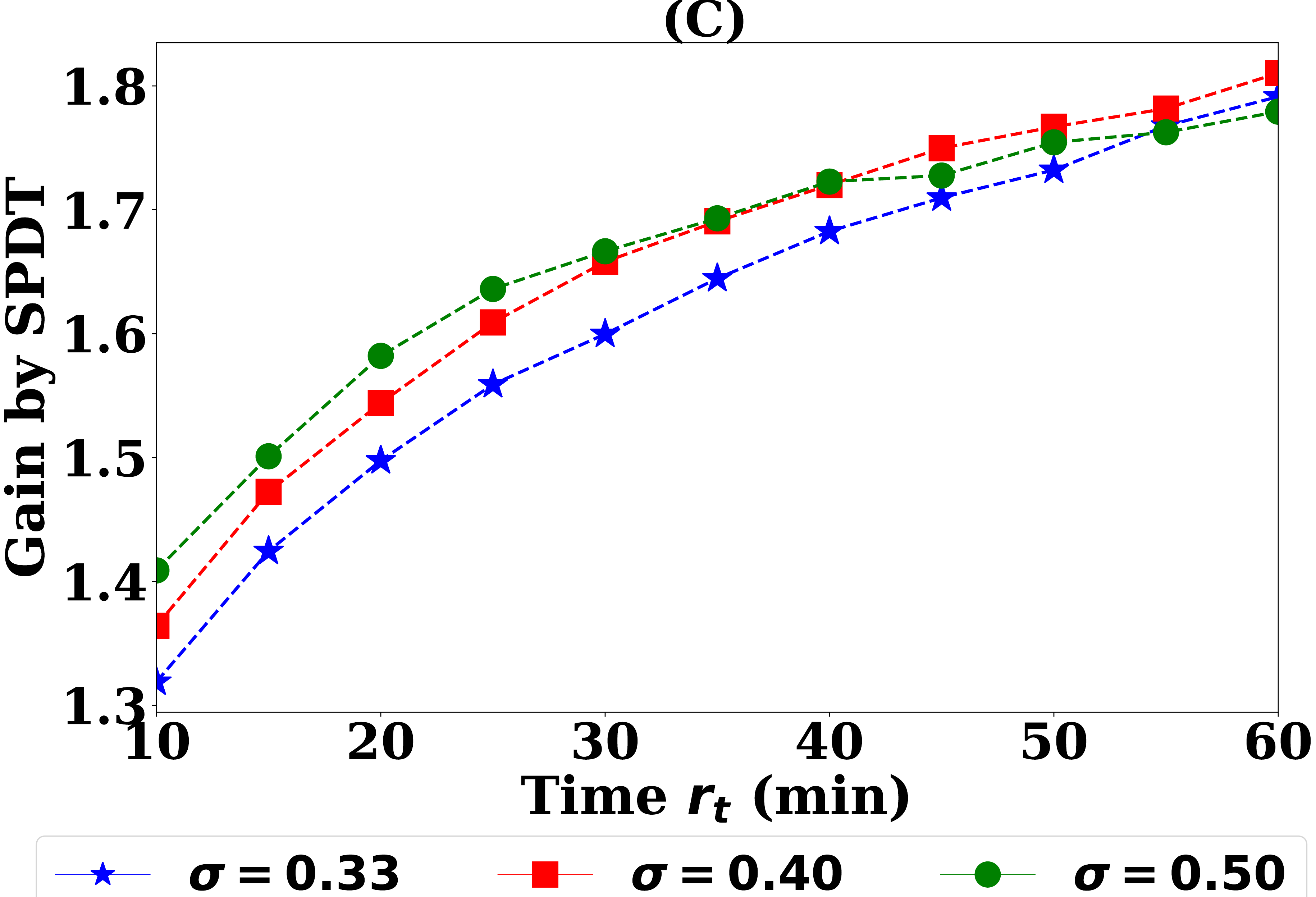}~\quad
\includegraphics[width=0.42\linewidth, height=5.0 cm]{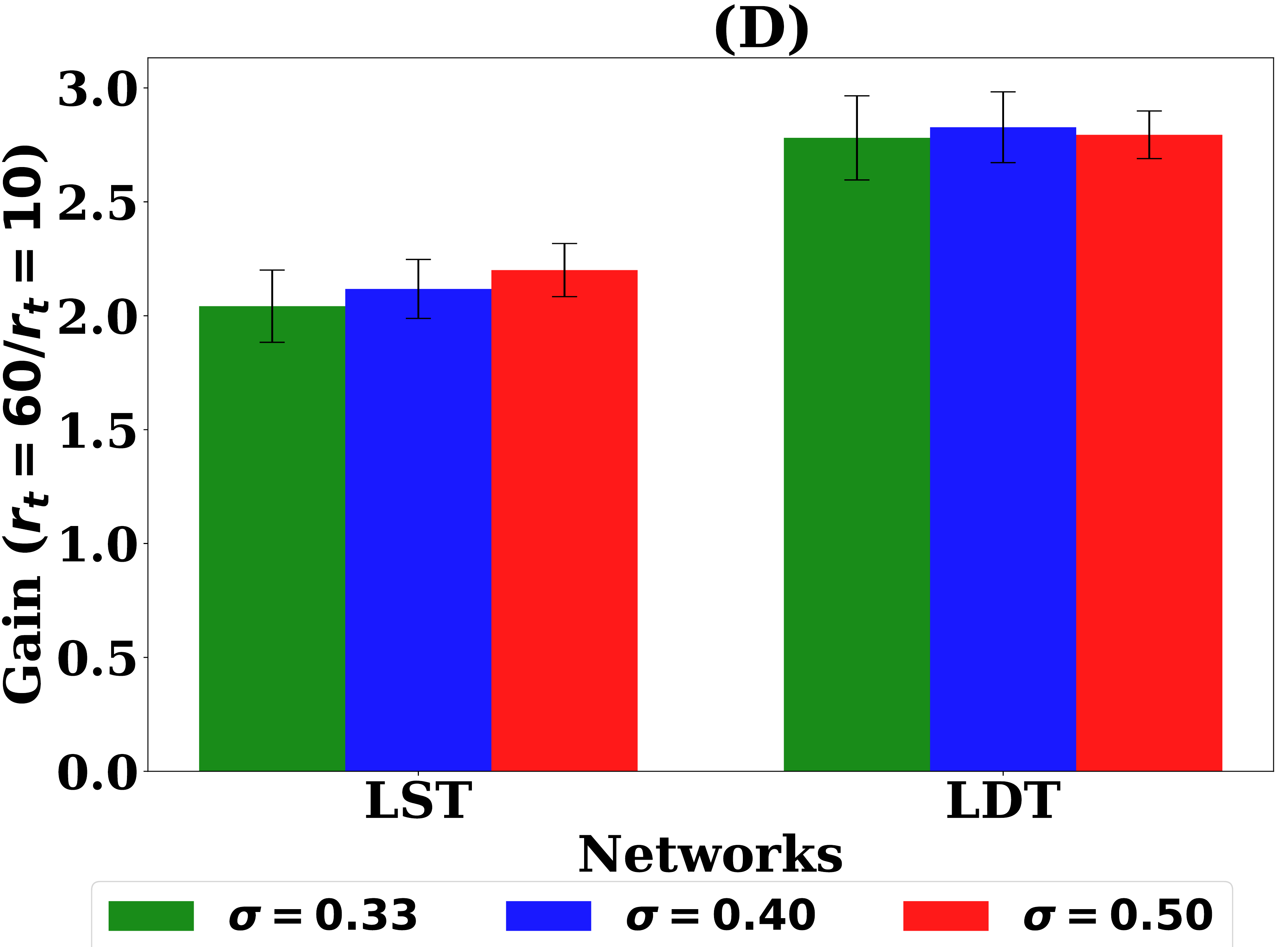}
	\caption{Diffusion on SPST and SPDT networks with controlled links densities: A) diffusion dynamics on LST network at $\sigma=0.33$ varying $r_t$, B) diffusion dynamics on LDT network at $\sigma=0.33$ varying $r_t$, C) gain by SPDT model over SPST model (LDT/LST) at various $\sigma$ when both have same link densities, and D) infection gains for changing $r_t=10$ min to $r_t=60$ min  at various $\sigma$}
    \label{fig:dns}
\end{figure}

The inclusion of indirect transmission links also significantly increases diffusion dynamics even if the underlying connectivity in the SPST and SPDT model is the same. This is observed from the diffusion dynamics on the LST and LDT networks which maintain the same number of active users (having links every day) and the same link densities. Besides, the LDT network can be described such that some direct links of LST network are appended with indirect links. Therefore, the underlying connectivity is not changed with $r_t$, but the links get stronger with increasing $r_t$. The LDT network still shows stronger disease prevalence relative to LST (Fig.~\ref{fig:dns}A and Fig.~\ref{fig:dns}B). Due to including indirect links, the LDT network achieves strong disease reproduction rate $R$ and produces high disease prevalence $I_p$. 

Our experiments also show that indirect transmission links are more sensitive to $r_t$ comparing to direct links. Its effect is also not reproducible by SPST model which is observed reconstructing LDT diffusion dynamics on the LST network with increasing $\sigma$ (Fig.~\ref{fig:ldfgn}C and Fig.~\ref{fig:ldfgn}D). With $\sigma=0.55$, the LST diffusion dynamic becomes similar to the LDT diffusion and causes same number of infections at $r_t=60$ min. However, the diffusion dynamics and total infections are overestimated at low $r_t$. In contrast, LST network with $\sigma=4$ shows same total infection of LDT network at $r_t$=10 min but underestimates at higher values of $r_t$. This difference is happening because the indirect link propensity is strongly sensitive to $r_t$ than direct links. However, the variations in the reconstructed LST diffusion is small comparing to D-SPST network as the underlying connectivity are diminished in LDT network.

The SPDT model assumes longer linear amplifications of infections with $r_t$. It is seen that the growth in SPDT amplification increases at high $r_t$ as $\sigma$ increases while it drops in SPST network (Fig.~\ref{fig:dsp}). The underlying cause is found studying diffusion on LST and LDT networks for $\sigma =\{0.33,0.4,0.5\}$ (Fig.~\ref{fig:dns}C and Fig.~\ref{fig:dns}D). Unlike the previous diffusion dynamics (Fig.~\ref{fig:dsp}), the gain across $r_t$ (total infections at $r_t=60$/total infections at $r_t=10$) slightly drops in LDT network which is the behaviour of SPST model where the growth in amplification at low $r_t$ is more than that at high $r_t$ when $\sigma$ increases. Due to having strong $R$ for indirect links, the infection force in LDT network over rules infection resistance force at $\sigma=0.4$, but it is affected at $\sigma=0.5$. The LDT network gain reaches to steady state which did not happen in DDT network. Thus, the SPDT model with underlying network dynamics supports extended spreading opportunities.
 

%% file: 3.5_chap_discussion.tex
\section{Discussion}
This chapter has introduced a SPDT diffusion model which account for the indirect transmissions created for visiting the same locations where infected individuals have been. A new risk assessment model has been added with the SPDT model to account for indirect links. The developed risk assessment model is simple to implement and capture the heterogeneities of real scenarios by selecting random values of the parameter. Using the developed SPDT model, diffusion dynamics for an influenza-like disease has been studied on real contact networks constructed by the Momo users. The simulation has shown that the inclusion of indirect transmission amplifies the spreading dynamics compared to that of the SPST models. The SPDT model has up to 4 times amplification in the total infections over the SPST model. The SPDT model has shown strong amplification with the dense networks where outbreaks can occur for a value of $r_t$ as low as 10 min. The amplification is also varied with disease parameters. The SPDT model has introduced new network dynamics to the underlying contact network. The underlying connectivity is changed with particle decay rates. This is because the stay time of infectious particles at a location is modified by the particles decay rates. The stay time, in turn, defines the number of individuals connected to the infected individual through this visit. Thus, the connectivity in the network is defined with the decay rates. The impact of the network dynamics is not reproducible by the SPST model as the direct links cannot capture the dynamic of indirect links.

%% file: 4_chap_net.tex
\chapter{SPDT Graph Model}
Graph models are widely used to study the diffusion processes on the contact networks. In the graph models, individuals are represented as nodes and relationships between the individuals as edges or links. Most of the current graph models create links for the concurrent presence of two nodes in physical or virtual space. Thus, these graph models do not support the study of SPDT processes where individuals have opportunities to create links with the delayed interactions. A dynamic contact graph model called the SPDT graph is developed in this chapter. The graph model developed can generate individual dynamic contact networks with SPDT links and is capable of capturing the socio-temporal dynamics of observed empirical SPDT graphs from the Momo data set. The SPDT graph model is developed adopting an activity driven time-varying network modelling approach and is able to generate dynamic contact networks using simple statistical distributions. This model is tunable and fitted with empirical data using maximum likelihood estimation methods. The SPDT graph model allows one to study SPDT diffusion with various scenarios of diffusion. The SDPT diffusion study based on the developed graph model can also avoid the limitations of the empirical contact networks applied in the previous chapter where the available SPDT links of original SDT network are repeated for the missing days for the same users to construct dense networks. However, the links repetition to the missing days of links cannot grow the contact set sizes, number of neighbours, of a user over the observation period. In addition, the time of interactions is the same for repeated links every day. These empirical contact networks cannot be reconfigured to study the diffusion dynamics with various network properties. Thus, the graph model developed provides greater opportunities to study SPDT diffusion in various scenarios. This chapter answers the question of how indirect links can be included with the dynamic contact network underlying a diffusion process.

\section{Introduction}
The study of SPDT diffusion processes requires detailed information about co-located interaction patterns among individuals and their temporal behaviours. The co-located interaction patterns are extracted from the GPS locations or RFID data~\cite{thomas2016diffusion,jurdak2015understanding}. However, these data are often incomplete and limited due to privacy. For example, the location updates from social networking application Momo provides sparse contact information where users did not update their location regularly~\cite{thilakarathna2016deep,shahzamal2018graph}. This data might miss some disease transmission opportunities and might vary the outbreak sizes. Thus, synthetic contact traces are often an alternative to validate the results of empirical data and explore the diffusion in wider scenarios. The synthetic contact network is generated by a graph model where individuals are represented as actors (i.e., nodes) and relationships among them as edges (i.e., links). In the traditional approach, the links between nodes are static. However, links between two individuals are not permanent over time for SPDT interactions. There have been various attempts to construct dynamic contact graphs to capture the temporal dynamics of interactions ~\cite{holme2015modern,casteigts2012time,zhang2016modelling,shahzamal2014mobility}. These dynamic contact graphs are frequently generated using statistical methods. The resulting models provide realistic dynamic contact networks for hypothesis testing, "what-if" scenarios, and simulations, but are often mathematically intractable. 

Dynamic models make graph entities and links evolve over time and simultaneously maintain the underlying social structure~\cite{holme2015modern,masuda2013self}. There have been several approaches to maintain the dynamics of nodes and links in the graph. The simplest method is to construct the static graph and make the links dynamics~\cite{holme2013epidemiologically}. The authors of ~\cite{hanneke2010discrete} also show the methods to produce dynamic networks from the exponential random graph model. The random graph model is fitted with a real data set. However, all these models are connection oriented. The authors of ~\cite{perra2012activity} have shown that the dynamic graph model is required to account for the status of nodes as well. They proposed an activity driven temporal network (ADN) model to generate dynamic contact graphs. In this model, nodes are activated with a probability at a time step and links to other nodes are created. The basic ADN model cannot capture many real-world network properties and it is updated adding memory and clustering co-efficient etc~\cite{karsai2014time,laurent2015calls}. 

Current co-located interaction graph models, however, assume that links between two nodes are created when they are both present at the same location. Thus, the contagious items can be transmitted through a link when both infected and susceptible nodes are only present~\cite{holme2015modern,casteigts2012time}. The diffusion on the contact graph based on these node to node level transmissions is known as SPST diffusion. The focus on concurrent presence, however, is not sufficiently representative of the class of SPDT diffusion processes. There is, therefore, a need for a novel dynamic contact graph model to study SPDT diffusion processes. This chapter proposes a dynamic contact graph model that considers both concurrent (direct) interactions and delayed (indirect) interactions among nodes in forming transmission links. This graph is called the same place different time interactions graph (SPDT graph) and can be represented with the time-node framework.

An SPDT graph generation method is also developed based on the principle of activity driven temporal network model. In the SPDT graph, the activation of a node means its arrival at a location where at least one other node is present. As individuals stay at a location for some time, the node's activation is done in such a way that a node activates for a consecutive of time steps. The node's stay duration (number of time steps) at a location with a potential to spread contagious items is termed as an active period. Therefore, the concept of activating nodes at a time step in ADN is extended to activating nodes for a period of time. In order to represent both direct and indirect transmission opportunities, an active copy of a node is created for each active period of the node. In the proposed model, links are created between the active copy of the host node and neighbour nodes. The active copy survives for the active period, when the host is present at the interaction location, in addition to an indirect transmission period, when the host leaves the location yet the contagious items persist at the location. In the SPDT graph, an active copy creates links at a time step within the active period and indirect transmission period. Thus, the SPDT graph evolves according to temporal changes of link and node's status. SPDT graph generation methods are developed using simple statistical distributions and are fitted with the real graph constructed using Momo data. The model is validated analysing generated network properties and by the capability of simulating SPDT diffusion processes. The results are compared with the ADN models and SPST diffusion. The contributions of this chapter are as follows:
\begin{itemize}
\item Developing a graph model to include indirect transmission links
\item Fitting the graph model with the real dynamic contact graph
\item Analysing the network properties generated by the developed graph model
\item Validating model and analysing sensitivity of the model's parameters
\end{itemize}
\input{4.1_chap_netmdl}
\input{4.2_chap_mdlfit}
\input{4.3_chap_netpro}
\input{4.4_chap_netdif}
\input{4.5_chap_disc}

%% file: 4.1_chap_netmdl.tex
\section{SPDT graph model}
According to the definition of SPDT diffusion model, visits of an infected individual to the locations where at least one susceptible individual is present create SPDT links. The SPDT links are created through space and time and may have components: a direct transmission link and/or an indirect transmission link. However, visits of an infected individual to the locations where no susceptible individuals are present do not lead to transmission of disease. The SPDT graph model is based on these contact opportunities and is defined as follows.

\subsection{Graph definition}
\begin{figure}
\centering
\includegraphics[width=0.9\linewidth, height=2.3cm]{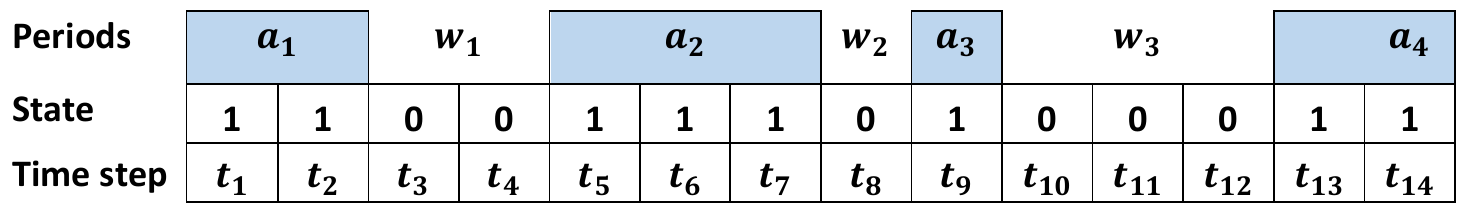}
\caption{Status of a node at each time step and the consecutive same states form the active periods ($a_1,a_2$) and inactive periods $(w_1, w_2)$. The active periods can represent the stay of a node in a location}
\label{fig:states}
\vspace{1 em}
\includegraphics[width=0.87\linewidth, height=2.5cm]{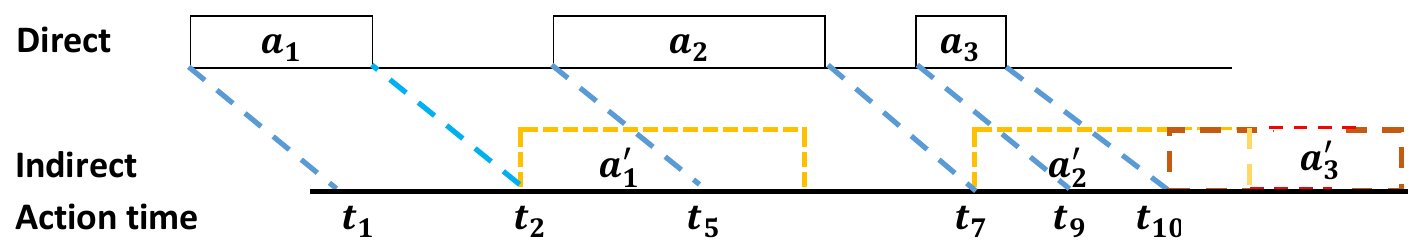}
\caption{Active periods with the corresponding indirect periods and their parallel presence. Indirect links during the inactive periods can transmit disease at the previous location and  direct links transmit disease at the current location}
\label{fig:timing}
\end{figure}

Individuals may move to various places during their infectious periods while travelling to public places such as offices, schools, shopping malls and bus stations and tourist destinations. The main goal is to develop a graph model that is capable of capturing the indirect transmission links along with direct links for co-located interactions among individuals. Thus, the proposed model focuses on link creations in the time domain to ensure scalability and abstracts spatial aspects implicitly. Temporal modelling is sufficient to identify the nodes participating in possible disease transmission links. Accordingly, the link creation events in the proposed scenario can be represented as a process where an infected node activates for a period of time (staying at a location where susceptible nodes are present) and creates SPDT links. Then, the infected nodes become inactive for a period during which it does not create SPDT links. Inactive periods represent the waiting time between two active periods. Thus, the co-located interaction status of an infected node during an observation period can be summarised by a set $\{a_1,w_1,a_2,w_2,..\}$ where $a$ is active period and $w$ is inactive period (see Fig.~\ref{fig:states}).

The proposed SPDT graph is defined as $G_T=(Z, A, L, T)$ to represent all possible disease transmission links among nodes, where $Z$ is the set of nodes. The number of nodes in the graph is constant; however, nodes may have one or more active copies in the graph which captures their ability to spread diseases both at locations they are present and at locations from which they recently departed. The set of active copies for all nodes is represented by $A$. $L$ is the set of links in the graph. The graph is represented over a discrete time set T=$\{t_1,t_2,\ldots t_z \}$. Each node in the graph creates a set of active and inactive periods $\{a_1,w_1,a_2,w_2 \ldots\}$. An active copy $v_{i}=v(t_s^{i},t_l^{i})$ is defined for an active period $a_i$ of node $v$, where $a_i$ starts at time step $t_s$ and finishes at $t_l$. Thus, each node will have several such temporal copies for the observation period. For an active copy $v_{i+1}$ of a node $v$, $t_{s}^{i+1}$ should be greater than $t_{l}^{i}$ of $v_i$ to capture the requirement that a node should have left the first location before arriving in another location. In this graph, a link $e_{vu}\in L$ is defined between an active copy $v_i$ of host node $v$ and neighbour node $u$ (node $u$ visits the current or recent location of node $v$) as $e_{vu}=(v_i,u,t_{s}^{\prime},t_{l}^{\prime})$ where $t_s^{\prime}$ is the joining time and $t_l^{\prime}$ is departure time of $u$ from the interacted location. The value of $t_s^{\prime}$ should be within $t_s^{i}$ and $t_l^{i}+\delta$ where $\delta$ is the time period allowed to create indirect transmission links. Thus, an active copy $v_i$ of a node v expires after $t_l^{i}+\delta$, where $\delta$ captures the decaying probability of infection after $v$ departs. During the indirect transmission period $\delta$, node $v$ can start another active period at another location (see Fig.\ref{fig:timing}). However, if the infected node $v$ leaves and returns to the location of $u$ within a time period $\delta$, then there will be two active copies of $v$, each with a link to the susceptible node $u$. The first copy is due to the persistent of particles from $v$'s last visit, while the second copy is due to $v$'s current visit. 

The evolution of the proposed graph is governed by two dynamic processes: 1) switching of nodes between active and inactive states, as in Figure~\ref{fig:states}; and 2) link creation and deletion for active copies of nodes. The total number of time steps a node remains in one state determines the current active or inactive period, leading to a set of alternating active and inactive periods $\{a_1,w_1,a_2,w_2 \ldots\}$ for an observation period (see Fig.\ref{fig:states}). As stay times at locations are not fixed, a transition probability $\rho$ is defined to determine switching from active state to inactive state (modelling stay and departure events of a node at a location). This induces variable lengths of active periods. Similarly, another transition probability $q$ is defined to determine when a node switches from inactive to an active state (modelling arrival of a node at location). A similar approach is taken to define link update dynamics in the graph. An active copy of a node creates a link to a newly arriving neighbour node with probability $p_{c}$ at each time step until it expires. During an activation period, multiple neighbours can arrive at interaction locations. Thus, an activation degree probability $P(d)$ is defined to model the arrival of multiple new neighbours for an active copy. The created links break (neighbour node leaves the interaction area) with probability $p_{b}$ at each time step. 

\subsection{Network generation}
Based on the above SPDT graph model definition, dynamic contact networks among a set of nodes can be generated in various ways ranging from homogeneous to complex heterogeneous scenarios. The simplest homogeneous version (where node properties are the same) of a model often provides the significant insight about the studied processes, but cannot capture many properties of the actual system. On the other hand, the heterogeneous system (where node properties are different) often captures broad properties with some complexities. The proposed network generation procedures consist of simple to complex network generations. To generate the SPDT contact network, first geometric distributions are used, as they are easy to implement in the computer systems, and then a power-law distribution is added to bring the heterogeneity in the generated networks. The SPDT graph contains a fixed number $N$ of nodes interacting with each other over the observation period $[0,T]$. The graph is implemented over discrete time with the step of $\Delta t$ time. The links generation procedure in SPDT graph model is implemented through a set of defined co-location interaction parameters (CIP) namely: active periods($t_a$), waiting periods($t_w$), activation degree($d$), link creation delay ($t_c$) reflecting time difference between the arrival of infected and susceptible individual, and link duration($t_d$). The generation procedures are as follows:

\subsubsection{Node activation}
Active copies of nodes are the building blocks for constructing an SPDT graph. The active copies of nodes are created over time according to the active periods. Thus, it is first required to generate active periods and intervening inactive periods. In the proposed model, determining whether a node will stay in the current state or transit into the other state at the next time step resembles a Bernoulli process with two outcomes. Thus, the number of time steps a node stays in a state can be obtained from a geometric distribution. With the transitional probability $\rho$ of switching from active to inactive state, the active period duration $t_a$ can be drawn from the following distribution as:
\begin{equation}\label{aprds}
Pr(t_{a}=t)=\rho (1-\rho)^{t-1}
\end{equation}
where $t=\{1,2, \ldots \}$ are the number of time steps. Similarly, the inactive period duration, $t_w$, with the transition probability $q$ can be drawn from the following distribution as:
\begin{equation}\label{iaprds}
Pr(t_{w}=t)=q (1-q)^{t-1}
\end{equation}
where $t=\{1,2, \ldots \}$ are the number of time steps. 

Knowing the active and inactive period duration, the graph generation process requires the initial states of nodes to generate the active copies of nodes. The model's state dynamics can be described with a two state Markov-process with transition matrix
\begin{equation*}
P=\begin{bmatrix}
q & 1-q\\
\rho & 1-\rho
\end{bmatrix}
\end{equation*}
for which the equilibrium probabilities that the node is in inactive state and active state are $\pi_0$ and $\pi_1$ respectively, where
\begin{equation}\label{son}
\pi_{0}=\frac{\rho }{q+\rho}
\mbox{\ \ \ \ \  and \ \ \ \ \ }
\pi_{1}=\frac{q}{q+\rho}
\end{equation}
If the initial state of node $v$ is active, the first active copy $v_1$ is created for the time interval $(t_s^{1}=0,t_l^{1}=t_a)$. Otherwise, $v_1$ will be created for the interval $(t_s^{1}=t_w,t_l^{1}=t_w+t_a)$. Active copy creation continues over the observation period and the corresponding interval $(t_s,t_l)$ is defined according to the drawn $t_a$ and $t_w$. Active copies are generated for each node independently. The values of $\rho$ and $q$ are fitted with real data.

\subsubsection{Activation degree}
The next step in the graph generation process is to define interactions of neighbour nodes with an active copy. Multiple neighbour nodes can be connected with an active copy. The number of neighbour nodes interacting with an active copy is noted as activation degree $d$. The value of $d$ can be used to capture the spatio-temporal dynamics of the real-world social contact networks. The activation degree $d$ are drawn from a geometric distribution (Eq.~\ref{eq:actdgr}) instead of finding the arrival times of neighbour nodes separately. 

\begin{equation} \label{eq:actdgr}
Pr\left (d=k\right )=\left(1-\lambda\right)\lambda^{k-1}
\end{equation}

where $k=\{1,2,\ldots \}$ are the number of neighbour nodes of an active copy and $\lambda$ is a scaling parameter. If the same parameter $\lambda $ is assigned to all nodes in the network, it is a homogeneous contact network where contact set sizes of all nodes are same. However, individuals in the real world contact scenarios have heterogeneous propensity to interact with other individuals. This heterogeneity is incorporated into the model by drawing heterogeneous $\lambda$ from a power law distribution of Equation~\ref{accesb}:

\begin{equation} \label{accesb}
f\left( \lambda_{i}=x\right)= \frac{\alpha x^{-(\alpha+1)}}{\xi^{-\alpha}-\psi^{-\alpha}}
\end{equation}

where $\lambda_{i}$ of a node $i$ can be applied to draw heterogeneous degree from Equation~\ref{eq:actdgr}, $\alpha$ is the scaling parameter of power law distribution, $\xi$ is the lower limit of $\lambda_i$ and $\psi$ is the upper limit which is approximately 1. The value of $\lambda_{i}$ defines the range of variations of $d$ for active copies of a node $i$ and Equation~\ref{eq:actdgr} ensures wide ranges for large values of $\lambda$. Combining geometric and power law distributions is known to generate more realistic degree distribution~\cite{chattopadhyay2014fitting} as is demonstrated in the model fitting section. Given the degree distribution, a link creation delay $t_c$ indicating time gap $(t_s-t_{s}^{\prime})$ between arrivals of host node and neighbour node, and a link duration $t_d$ indicating the stay time $(t_s^{\prime}-t_{l}^{\prime})$ of neighbour node at the interacted location are assigned to each link. 

\subsubsection{Link creation}
With the activation neighbour set, the graph generation process defines the arrival and departure dynamics of neighbour nodes for each link created with an active copy. A similar approach to the definition of active and inactive periods creation with transition probabilities is adapted. Here the assumption is that each link is created with probability $p_{c}$ at each time step during the life period $(t_s,t_{l}+\delta)$ of an active copy and is broken with probability $p_{b}$ after creation. For drawing link creation delay $t_c$, the time gap $(t_s-t_{s}^{\prime})$ between arrivals of host node and neighbour node, the truncated geometric distribution is used: 
\begin{equation} \label{ldelay}
P\left (t_{c}=t\right )=\frac{p_{c} \left(1-p_{c}\right)^{t}}{1 -(1-p_{c})^{t_{a}+\delta}} 
\end{equation}
where $t=\{0, 1,2, \ldots, t_a+\delta \}$ are the number of time steps and $t_a$ is the active period duration of corresponding active copy. Truncation ensures that links are created within $t_{l}+\delta$, i.e. before the active copy expires. In contrast, the link duration $t_{d}$, the stay time of neighbour at the interacted location, does not have a specific upper bound and is generated for each link upon creation through a geometric distribution: 
\begin{equation} \label{ldur}
P\left(t_{d}=t\right)=p_{b} \left(1-p_{b} \right)^{t-1}
\end{equation}
where $t=\{1,2, \ldots \}$ are the number of time steps. For simplicity, $p_b$ is set to $\rho$ as both probabilities relate to how long nodes stay at a location. Each link with an active copy, thus, has timing characteristics as $t_{s}^{\prime}=t_s+t_c$ and $t_{l}^{\prime}=t_s^{\prime}+t_d$. A link with $t_s^{\prime}\geq t_l$ is an indirect transmission only component. Link can also have indirect component if $t_s^{\prime} < t_l$ and $t_l > t_l^{\prime}$. The above graph generation steps capture the temporal behaviour of SPDT links. The social mixing patterns are integrated by selecting the neighbouring node, as it is described below.

\subsubsection{Social structure}
The underlying social structure of the generated graph depends on the selection of a neighbour node for each link. The simplest way of selecting the neighbour node is to pick a node randomly. However, social network analysis has shown that the neighbour selection for creating a link follows a memory-based process. Thus, the reinforcement process~\cite{karsai2014time} is applied to realistically capture the repeated interactions between individuals. In this process, a neighbour node from the set of already contacted nodes is selected with probability $P(n_{t}+1)=n_{t}/(n_{t}+\eta)$ where $n_{t}$ is the number of nodes the host node already contacted up to this time $t$. On the other hand, a new neighbour node is selected with the probability $1-P(n_t+\eta)$. Here, the size of the contact set, the number of nodes a node contact over the observation period, depends on the $\lambda$. This is because the system forces a host node to select a new neighbour when neighbours from the current contact set are already selected during an active period even if $P(n_t+1)$ is true. On the other hand, a new node is selected as a neighbour with the probability $1-P(n_t+1)$. To maintain local clustering among nodes, proportion $\mu$ of new neighbours are selected from the neighbours of neighbour nodes. In addition, when a node $j$ is chosen as a new neighbour by node $i$ in the heterogeneous networks, it is selected with the probability proportional to its $\lambda_j$ as nodes with higher $\lambda$ will be neighbours to more nodes~\cite{alessandretti2017random}. This ensures nodes with higher potential to create links also have a higher potential to receive links.

%% file: 4.2_chap_mdlfit.tex
\section{Estimating model parameters}
The graph generation methods require appropriate values for the distribution parameters that will make the graph representative of real contact dynamics. The SPDT graph generation methods are based on five co-located interaction parameters (CIP) namely active period $(t_{a})$, waiting period $(t_{w})$, activation degree $(d)$, link creation delay $(t_{c})$ and link duration $(t_{d})$ to construct dynamic SPDT graph $G_T=(Z,A,L,T)$. The CIP parameters are collected from real contact networks which are created using locations updates of Momo users. Then, maximum likelihood estimation methods are applied to find the model parameters. 

\subsubsection{Data set and real SPDT graphs}
The location updates from Shanghai city are applied to estimate the model parameters and updates from Beijing city are used to validate the model. The location updates of 7 days from Shanghai city are analyzed to find the co-location interaction parameters. The link extraction procedure, described in Section 3.2, is applied here to collect CIPs. The active period durations $t_a$ are collected over the selected seven days. Thus, every possible length of the active period made by the Momo users is captured. The seven days data contains 2.53 million active period durations from 126K users. The distribution of $t_a$ is presented in the Figure~\ref{fig:fita}A where the periods are presented with the number of time steps of 5 minutes. The users have heterogeneous behaviours of staying which are captured by long tail distributions. Then, the waiting periods $t_w$ are collected. However, the waiting period collection is not straight forward like $t_a$. This is because Momo users do not use the App regularly and many potential visits are not captured. Assuming that individuals are activated at least once a day, the waiting period will be less than 2 days if one active at the beginning of previous days and again active at the end of the current day. The distribution of waiting periods $t_w$ is shown in the Figure~\ref{fig:fita}B. A new co-location interaction parameter called activation frequency $h$ is defined which measures the rate of activation of nodes to avoid the irregularity in the waiting period. This is measured with the number of active periods created by a Momo user in a day. The activation frequencies $h$ are collected for all users over each day during the selected seven days. This provides 0.5M activation frequencies for the 126K users. The distribution of activation frequencies is presented in the Figure~\ref{fig:fita}C. The activation frequencies with active period duration model the inactive periods. As the activation frequencies are daily behaviours of Momo users, modelling active period and waiting period based on $h$ avoids the impact of missing updates. Next, the activation degrees are collected based on the number of other Momo users visits a location where host user has been for the corresponding active period. Here, the users are required to visit a location within $\delta=3$ hours as it is found in the previous chapter that particles can stay up to the four hours. The activation degree distributions are presented in the Figure~\ref{fig:fitd}. This follows the long-tailed distributions which are found in the degree distributions of many real systems. Then, the link creation delays $t_c$ and link duration $t_d$ are collected for all links created over the seven days. Theses also capture all possible length of $t_c$ and $t_d$. The distribution of $t_c$ and $t_d$ are presented in the Figure~\ref{fig:fitl}. These also can be captured by the long tail distributions.

\subsubsection{Goodness of fit}
Using the developed graph model, a synthetic SPDT contact network (GDT network) is created with 126K nodes for 7 days applying the estimated model parameters. Then, the CIP parameters of GDT are compared with the parameters of real SPDT network (SDT) constructed by 7 days updates of 126K users from Beijing city. To understand the model's response across network sizes, another large synthetic SPDT contact network (large GDT network) of 0.5M nodes for 7 days is generated. Results of 100 runs for each network are presented in Figures (~\ref{fig:fita},~\ref{fig:fitd},~\ref{fig:fitl}) where periods are based on time steps of $\Delta t =5$ minutes. The root squared error (RSE) is calculated between the generated and real data distributions of CIP parameters as:
\begin{equation}\label{eq:error}
RSE=\sqrt{\sum_{i=1}^{m}(x_i-y_i)^{2}}
\end{equation}
where observed values are grouped in the $m$ bins as they are discrete, $x_i$ is the proportion of observations for the i$^{th}$ bin, $y_i$ is the proportion of empirical dataset values in the i$^{th}$ bin. As the RSE is computed from the proportion values, bins are naturally weighted so that bins representing larger proportions of events have higher contributions in error. Note that the bins up to the last bin of $x_i$ are considered assuming that the discarded values of $y_i$ provide very small error as the selected distributions are long-tailed. The distributions of network CIP parameters are presented in log scale with RSE error bars for SDT, GDT and large GDT in Figure ~\ref{fig:fita}, Figure~\ref{fig:fitd} and Figure~\ref{fig:fitl}.

\subsubsection{Fitting activation parameters}
\begin{figure}[h!]
\centering
\includegraphics[width=0.33\linewidth, height=4.0cm]{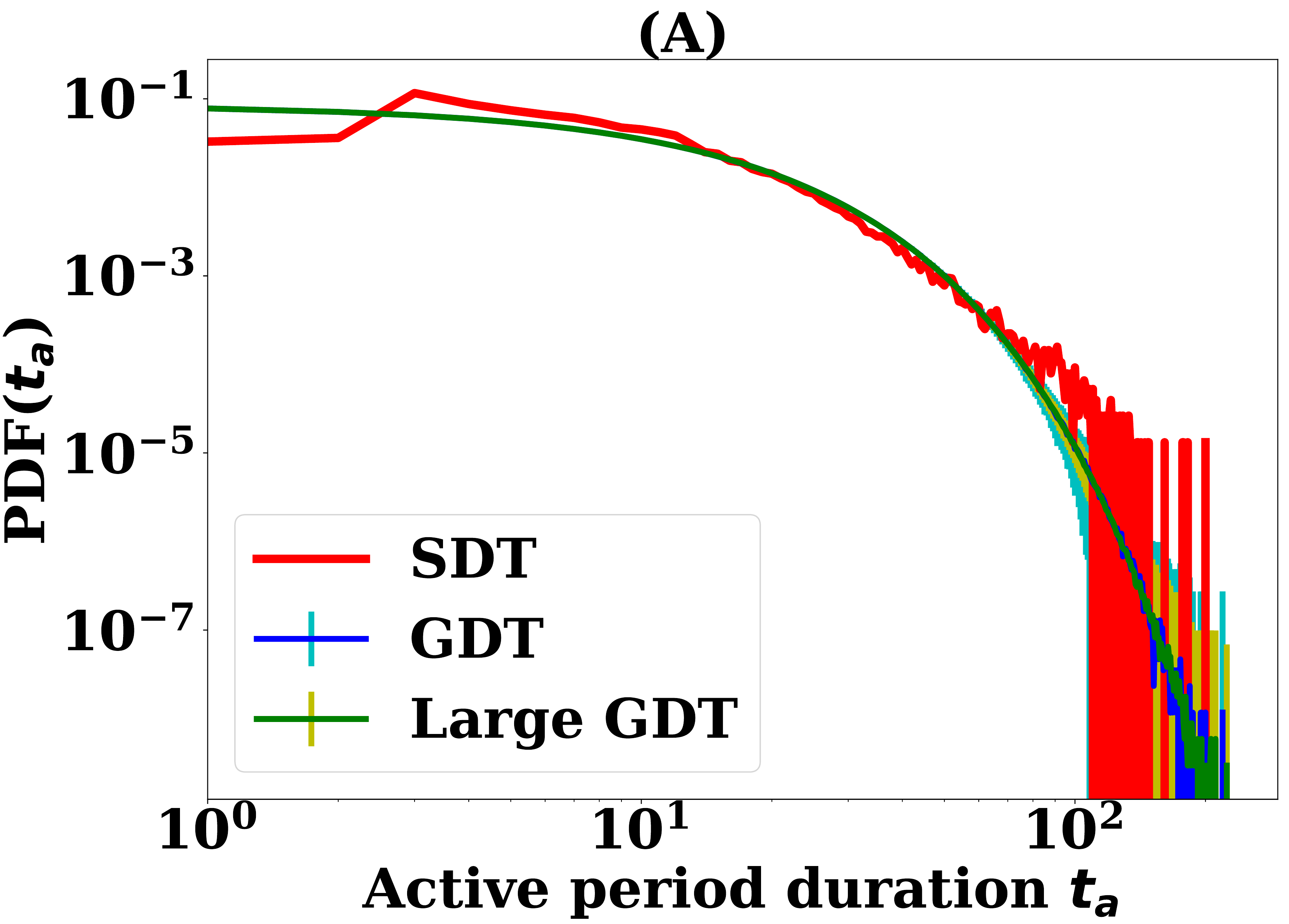}~
\includegraphics[width=0.33\linewidth, height=4.0cm]{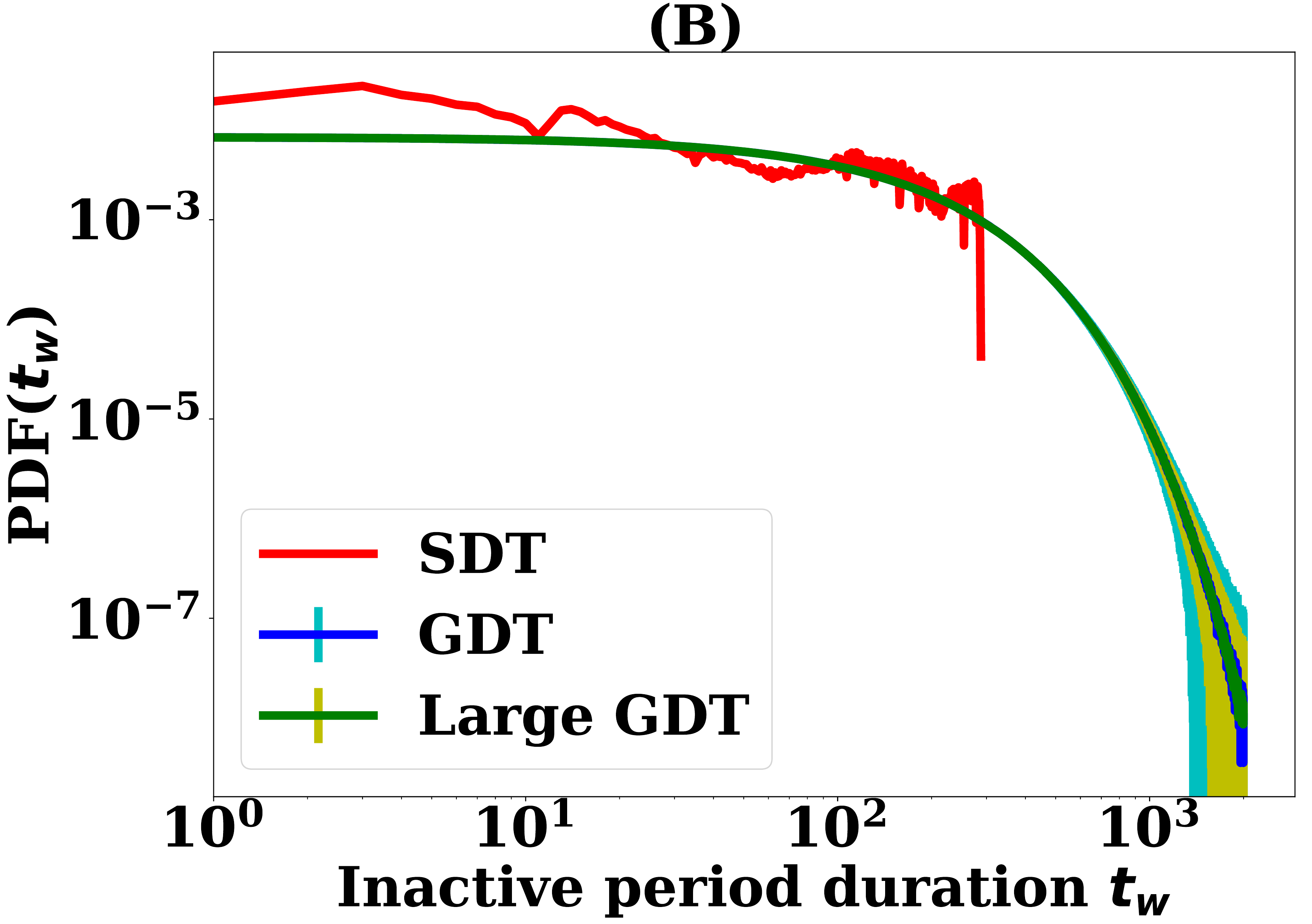}
\vspace{1em}
\includegraphics[width=0.32\linewidth, height=4.0cm]{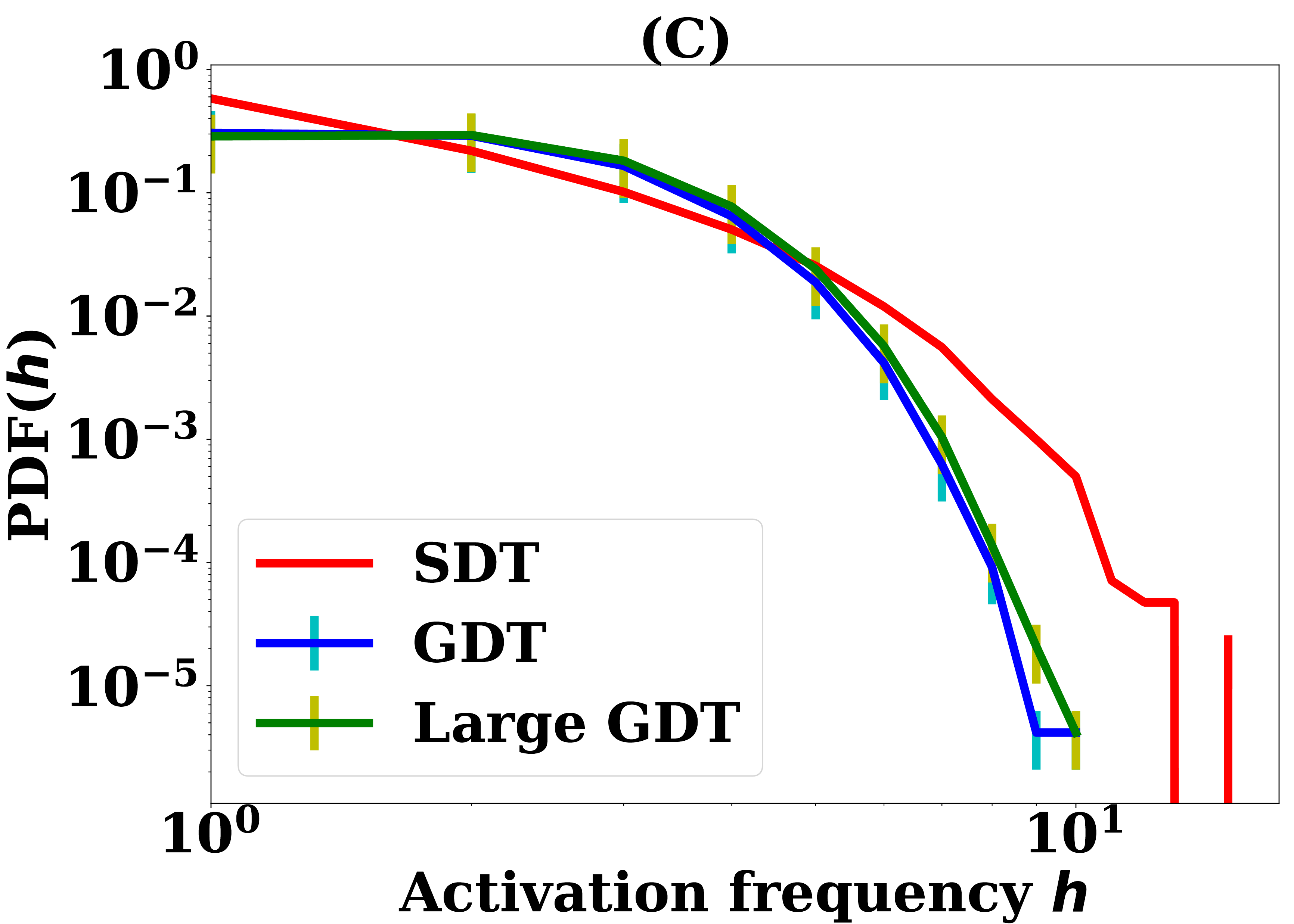}
\caption{Fitting co-located activation parameters (CIP) for activation of nodes and comparing with the real world network (SDT) parameters. A synthetic network (GDT) equivalent to SDT and another large synthetic network (large GDT) are used to understand stability of fitting: A) active period $t_a$ in the number of time steps of 5 minutes, B) waiting period $t_w$ in the number of time steps of 5 minutes and C) activation frequency $h$ in the number of activations per day}
\label{fig:fita}
\end{figure}

In this model, the active periods $t_a$ are drawn from a geometric distribution with the scaling parameter $\rho$. The value of $\rho$ can be estimated from the following MLE condition of geometric distribution applying a sample data set of $t_a$:
\begin{equation}\label{geomle}
\hat \rho =\frac{m}{\sum_{l=0}^{m}s_{l}}
\end{equation}
where $m$ is the size of the selected sample set. About $m=2.53 M$ real activation periods made by $126K$ Momo users over the 7 days are used here. This large set of active periods contain the periods happening from workdays to the weekend and captures all sizes of periods. The above MLE Equation~\ref{geomle} estimates $\hat \rho=0.085$. The distributions of generated $t_a$ for both networks GDT and large GDT are shown in Figure~\ref{fig:fita} A. The RSE error for GDT is 0.0876 with standard deviation (std.dev.) of 0.00041 while RSE for large GDT is very similar giving 0.08758 with std.dev. of 0.00018. The model with fitted parameters consistently generate the active periods $t_a$ for nodes of both network sizes. The distribution of active periods is also supported by the findings of other studies where they found the contact duration is broadly distributed~\cite{scherrer2008description,hui2005pocket}. In these studies, contact duration means staying at a location.

The activation periods $t_a$ have similar patterns for all individuals. However, the inactivation periods, $t_w$ waiting time between two active periods, depends on how frequently individuals visit public places. Thus, the waiting periods $t_w$ are fitted based on the fitting of activation frequencies $h$. The number of transition events from inactive to active states or active to inactive states will be the activation frequency during an observation periods. According to Equation~\ref{son}, the probability of transition event $0 \rightarrow 1$ at a time step is: 
\[p_{01}=\frac{\rho q}{q+\rho}\]
Thus, the number of transition events $0 \rightarrow 1$ during $z$ time steps (here number of time steps in a day) represents the number of activation events $h$. The probability of $h$ activation for a node is given by the binomial distribution as: 
\begin{equation*}
Pr(h\mid q)=\begin{pmatrix}
z\\ 
h
\end{pmatrix}\left(\frac{\rho q}{q+\rho}\right)^{h}
\end{equation*}
The term $\frac{\rho q}{q+\rho}$ becomes small as $\rho=0.083$, as $z$ is usually large and $q<1$. Thus, the above equation can be approximated by a Poisson distribution as:
\begin{equation}\label{eq:wait}
Pr(h\mid q)=\frac{\left(\frac{z \rho q}{q+\rho}\right)^ {h}e^{-\frac{z \rho q}{q+\rho}}}{h!}
\end{equation}
The MLE condition for the Equation~\ref{eq:wait} is derived in the Appendix as 
\begin{equation}\label{actf}
\frac{qz\rho}{q+\rho}=\frac{1}{l}\sum_{i=1}^{l}h_i
\end{equation}

In the selected Momo data set, the users are not present every day. Thus, it is assumed that the number of activations made during a day is representative of the activation frequency. As the focus is to model the waiting time distribution with a geometric distribution, it makes nodes activate with variable frequency over the observation days. The day activation frequencies of $126K$ users are collected over the seven days. Applying the activation sample set $h=\{h_1,h_2,\ldots h_r\}$ of size r=300K to MLE Equation~\ref{actf} gives $\hat q =0.0048$. The generated activation frequencies for GDT and large GDT are presented in Figure~\ref{fig:fita} B, with RSE of 0.100 and std.dev. of 0.005 for both GDT and large GDT compared to real activation frequencies. The plotted waiting periods in Figure~\ref{fig:fita} C also follows the distribution of real $t_w$ with RSE around 0.102 in both networks. The obtained std.dev. of error is 0.01. The $t_w$ is characterized by the irregularity of using Momo Apps by users.

\subsubsection{Fitting activation degree}
\begin{figure}[h!]
\centering
\includegraphics[width=0.42\linewidth, height=5.0cm]{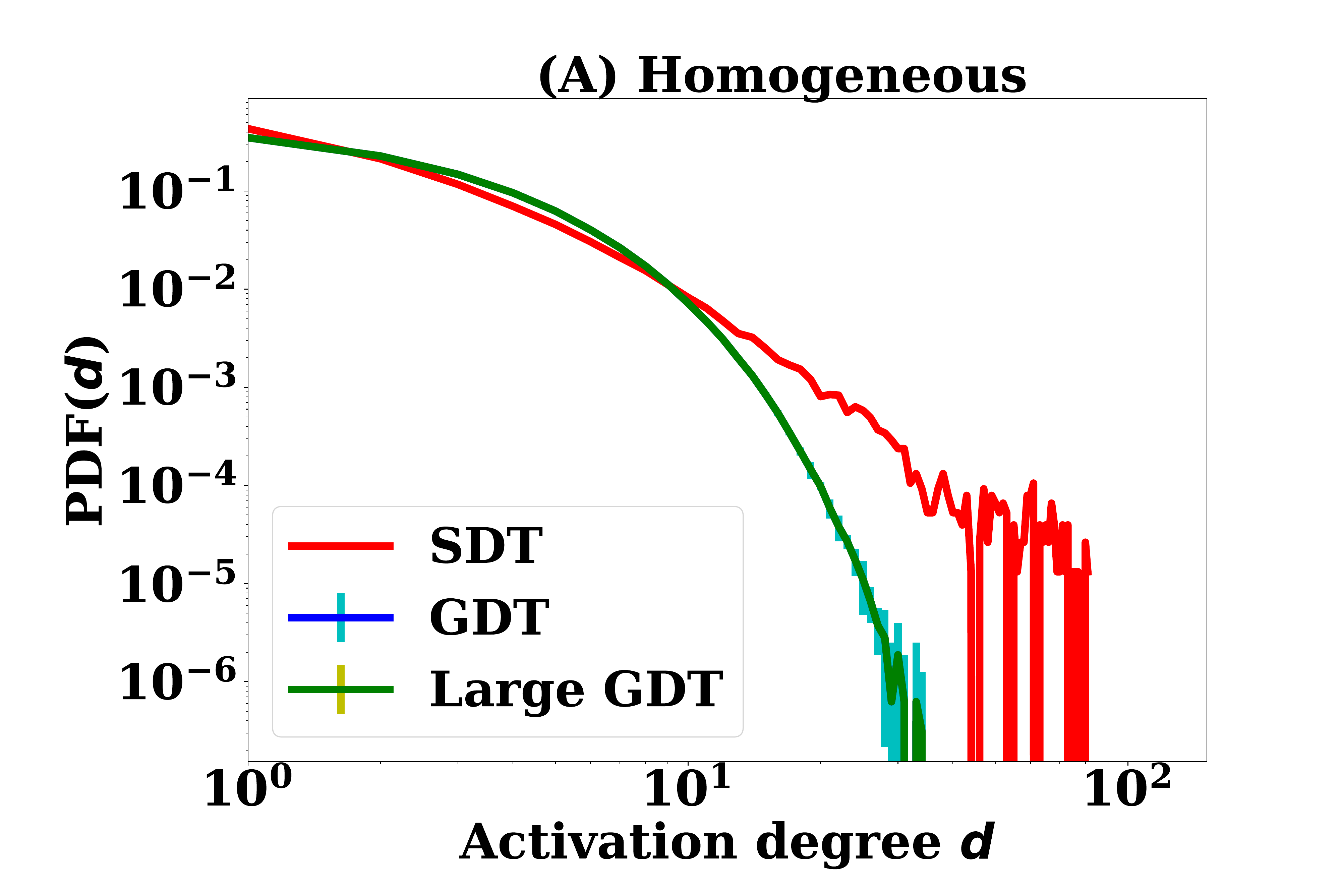}~
\includegraphics[width=0.42\linewidth, height=5.0cm]{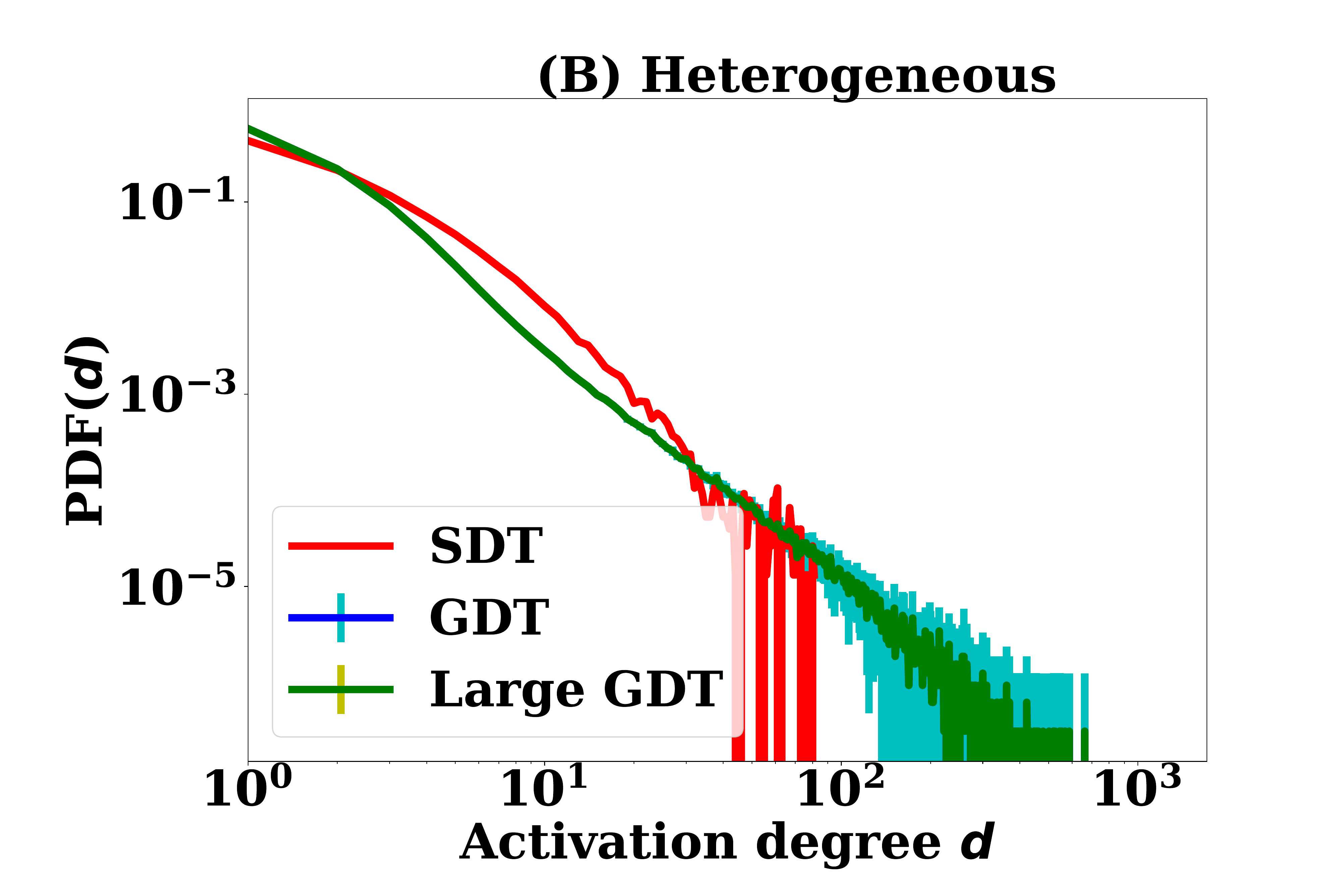}
\caption{Fitting activation degree CIP for both GDT and large GDT networks and comparison with SDT: A) nodes in the GDT networks are homogeneous and B) nodes in the GDT networks are heterogeneous}
\label{fig:fitd}
\end{figure}

The activation degree is drawn from a geometric distribution with the scaling parameter $\lambda$. If it is assumed that nodes in the network are homogeneous, the value of $\lambda$ can be estimated using the following equation 
\begin{equation}\label{geomleac}
\hat \rho =\frac{m}{\sum_{l=0}^{m}d_{l}}
\end{equation}
The activation degree sample set $d=\{d_{1},\ldots,d_{m}\}$ of size m=518K are used to the MLE equation. The estimated value for $\hat \lambda $ is 0.32. The generated activation degree distributions for 
GDT and large GDT and comparison with real network SDT are presented in Figure~\ref{fig:fitd} A. The RSE error is 0.017 with std.dev. of 0.004. The fluctuating error at the tail is due to data sparsity and has little effect on the RSE. 

When nodes in the network have heterogeneous propensity to contact the other individuals, the distribution of $d$ in the network will be given for any $\lambda$ as:
\begin{equation*}
Pr(d)=\frac{\beta }{\xi^{\beta}-1}\int_{\xi}^{1}( \lambda^{d-\beta-2} -\lambda^{d-\beta-1}) d\lambda
\end{equation*}
\begin{equation}\label{eq:dmle}
=\frac{\beta}{\xi^{-\beta} - 1}\left(\frac{1-\xi^{d-\beta-1}}{d-\beta-1}-\frac{1-\xi^{d-\beta}}{d-\beta}\right)
\end{equation}
The MLE condition for the Equation ~\ref{eq:dmle} is derived in the Appendix which provide the following two equations. The MLE condition for $\beta$ assuming $\psi \approx 1 $ is given as 
\begin{equation}\label{eq:dp}
0=\frac{m}{\beta}-\frac{m \xi^\beta \ln\xi}{\xi^\beta-1}+\sum_{k=1}^{n}\frac{\frac{\xi^{d_k-\beta -1} \ln\xi}{d_k-\beta-1}-\frac{1-\xi^{d_k-\beta -1}}{(d_k-\beta -1)^2}-\frac{\xi^{d_k-\beta} \ln\xi}{d_k-\beta}-\frac{1-\xi^{d_k-\beta}}{(d_k-\beta)^2}}{\frac{1-\xi^{d_k-\beta-1}}{d_k-\beta-1}-\frac{1-\xi^{d_k-\beta}}{d_k-\beta}}
\end{equation}
where $m$ is the length of the data set $d=\{d_{1},\ldots,d_{m}\}$. Then the lower limit of power law distribution is estimated with the following MLE equation.
\begin{equation}\label{eq:dl}
0=\frac{m \beta\xi^{-\beta -1}}{\xi^{-\beta}-1}+\sum_{k=1}^{m}\frac{\frac{(d_k-\beta)\xi^{d_k-\beta -1}}{d_k-\beta}-\frac{(d_k-\beta-1)\xi^{d_k-\beta -2}}{d_k-\beta-1}}{\frac{1-\xi^{d_k-\beta-1}}{d_k-\beta-1}-\frac{1-\xi^{d_k-\beta}}{d_k-\beta}}
\end{equation}
As the MLE equations do not have closed form solutions, the Python module fsolve is used to estimate the best value of $\xi$ and $\beta$. To estimate the values, it is assumed that $\beta=2.5$ since it is known that it is in the range 2-3 in the real world network. Then, the value of $\xi=0.26$ is estimated using the MLE Equation~\ref{eq:dl}. Using $\xi=0.26$ in Equation~\ref{eq:dp}, the value of $\beta$ is estimated as 2.963. The generated activation degree distributions with the heterogeneous potential are presented in the Figure~\ref{fig:fitd}. The RSE error is 0.0152 with std.dev. of 0.004. There are few values with high degree in GDT network and that can also happen in real contact networks.

\begin{figure}[h!]
\centering
\includegraphics[width=0.44\linewidth, height=5.0cm]{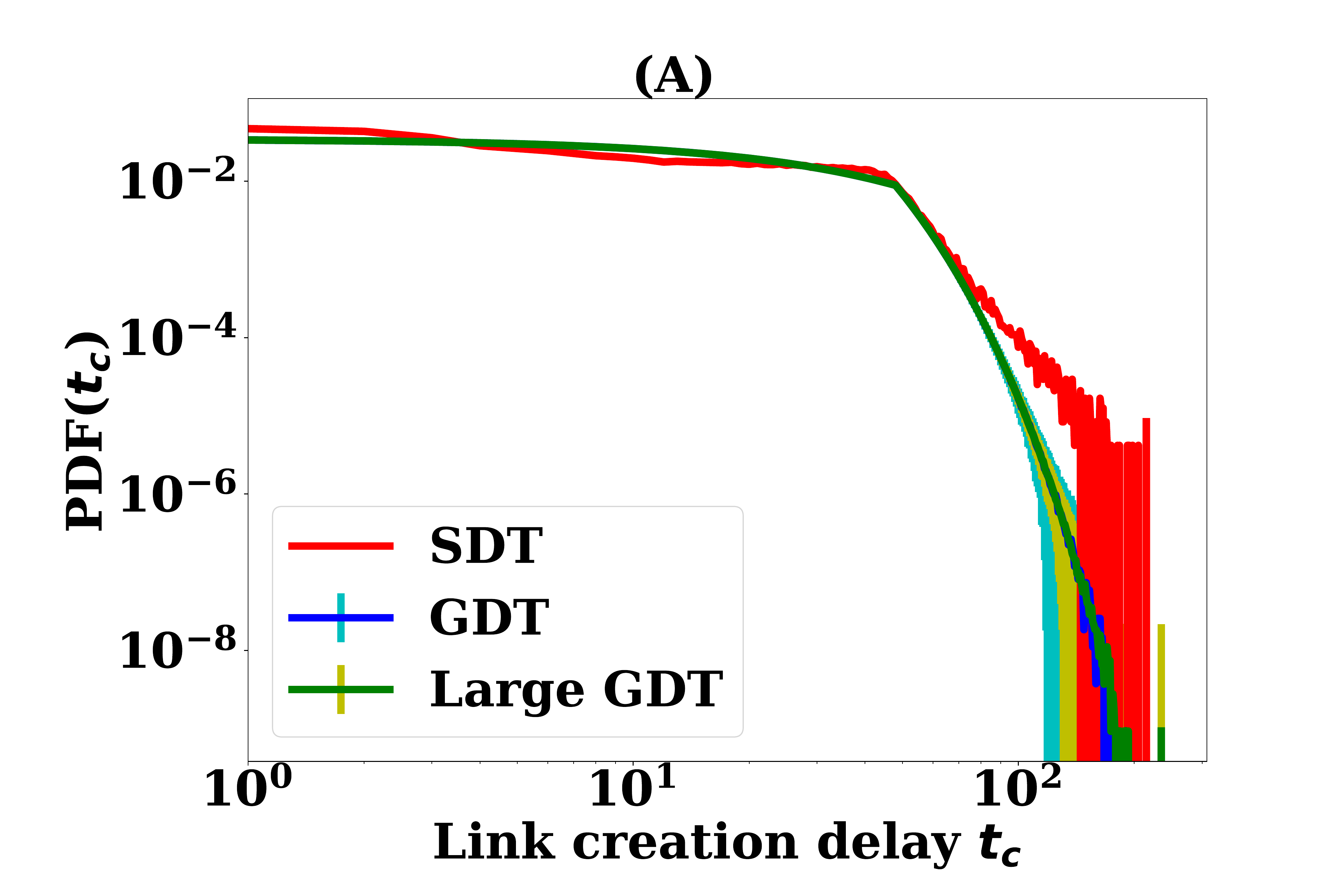}~
\includegraphics[width=0.44\linewidth, height=5.0cm]{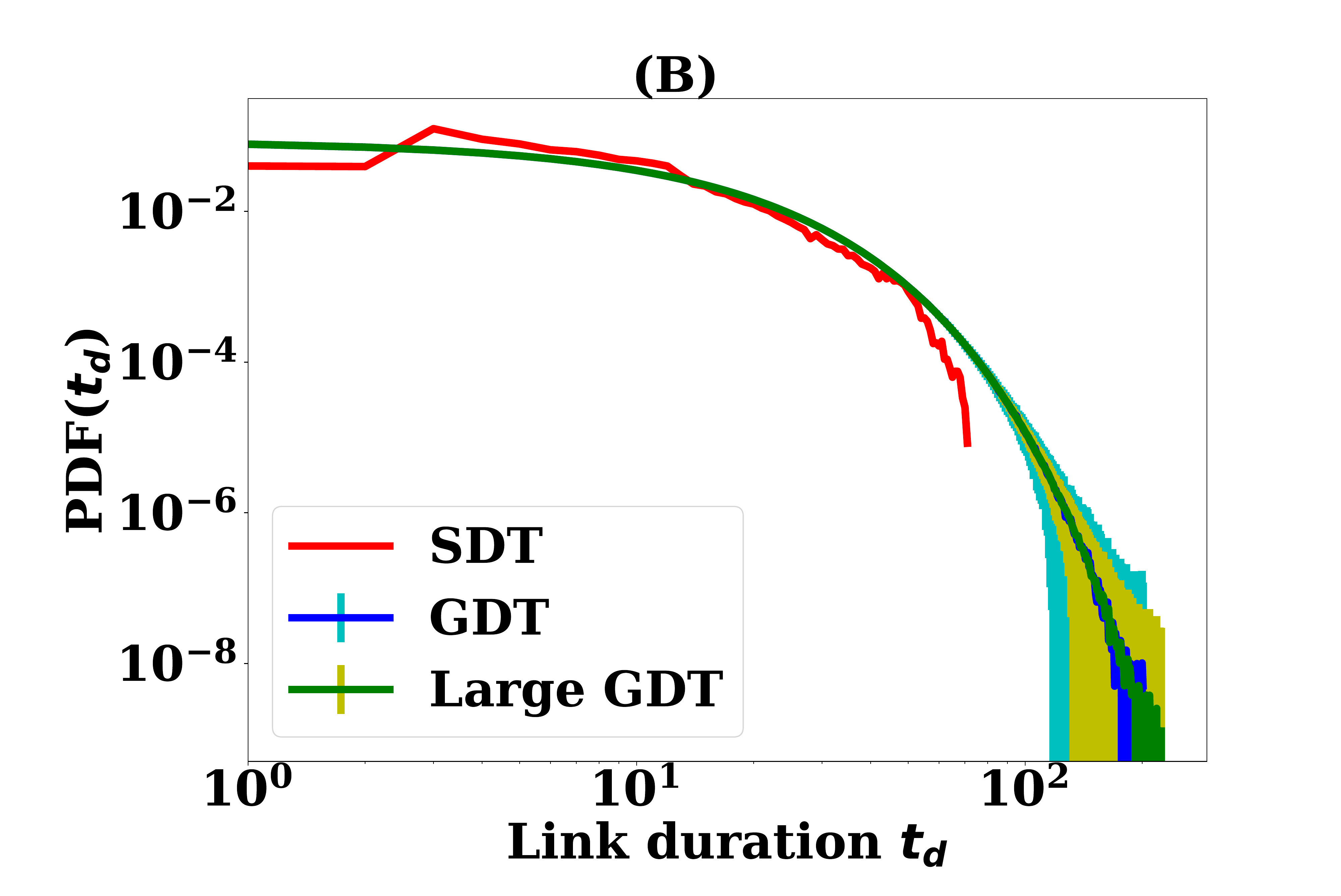}
\caption{Fitting link creation CIP for both GDT and large GDT networks and comparison with SDT:: A) link creation delay $t_c$ in the number of time steps of 5 minutes and B) link duration $t_d$ in the number of time steps of 5 minutes}
\label{fig:fitl}
\end{figure}

\subsubsection{Fitting link creation parameters}
The link creation delay $t_c$ is drawn with the truncated geometric equation. The MLE condition is not a straight forward as the MLE of geometric equation. The MLE for the truncated equation is derived as follows:
\begin{equation*}
0=p_c \sum_{k=1}^{n} \frac{\sum_{l=1}^{m}\frac{(t_{c}^{k}-1)(1-(1-p_c)^{t_a^{l}+\delta})+(t_{a}^{l}+\delta)(1-p_c)^{t_a^l +\delta}}{(1-p_c)(1-(1-p_c)^{t^{l}_{t_a}+\delta})^2}}{\sum_{l=1}^{m} ((1-(1-p_{c})^{t^{l}_{a}+\delta})^{-1}}
\end{equation*}
where $t^{1}_{c},t^{2}_{c},\ldots,t^{n}_{c}$ are sample set of size $n=1.2M$ and $t_a=\{t_a^{1},\ldots,t_a^{m}\}$ with $m=518$K. The estimated value of $\hat p_c$ is 0.02. The generated link creation delays are presented in the Figure~\ref{fig:fitl}A, where the generated $t_c$ have RSE of 0.009 in comparison with the real distribution. The errors were consistent in both GDT and large GDT. The value of $\rho=0.084$, which is fitted with the active periods, is used to generate the link duration $t_d$. Figure~\ref{fig:fitl}B presents the comparison of generated link duration with real duration which has RSE error of 0.022 with std.dev. of 0.004. All the six co-location interactions parameters (CIP) of the generated networks (GDT and large GDT) are a good fit to that of the real network (SDT). The variations for generated networks are very low and consistent in both GDT and large GDT networks.

%% file: 4.3_chap_netpro.tex
\section{Model validation}
The model developed is validated by analyzing the generated network properties and by the capability of simulating SPDT diffusion process on it. First, network properties are analyzed to understand the capability of the graph model to generate properties of the real graph. Then, the diffusion dynamics on the generated networks are analyzed to assess its ability to replicate the diffusion dynamics of the real contact networks. In the previous section, the model's parameters have been tuned using the location updates from Shanghai. Now, location updates from Beijing are used to compare the network properties and diffusion dynamics. The SPDT graph properties and diffusion dynamics unfolded on it are studied generating synthetic contact networks with both homogeneous and heterogeneous degree distributions. In this section, the DDT contact networks constructed in Chapter 3 are selected instead of SDT network since this network has shown realistic diffusion dynamics and the developed graph model are expected to generate realistic diffusion dynamics. It is mentioned earlier that the underlying social structure in both SDT and DDT networks are the same. However, the contact frequencies over the observation days are varied in DDT network comparing to the SDT network. Thus, investigation and comparison of network properties of GDT network with the DDT network is more appropriate.   

\subsection{Network Properties}
The previous section has focused on fitting co-location interaction parameters (CIP) of SPDT graph to empirical data. This section explores the fitted model's ability to reproduce the static and temporal properties of empirical contact networks. 
\begin{figure}[h!]
\centering
\includegraphics[width=0.67\linewidth, height=3.6cm]{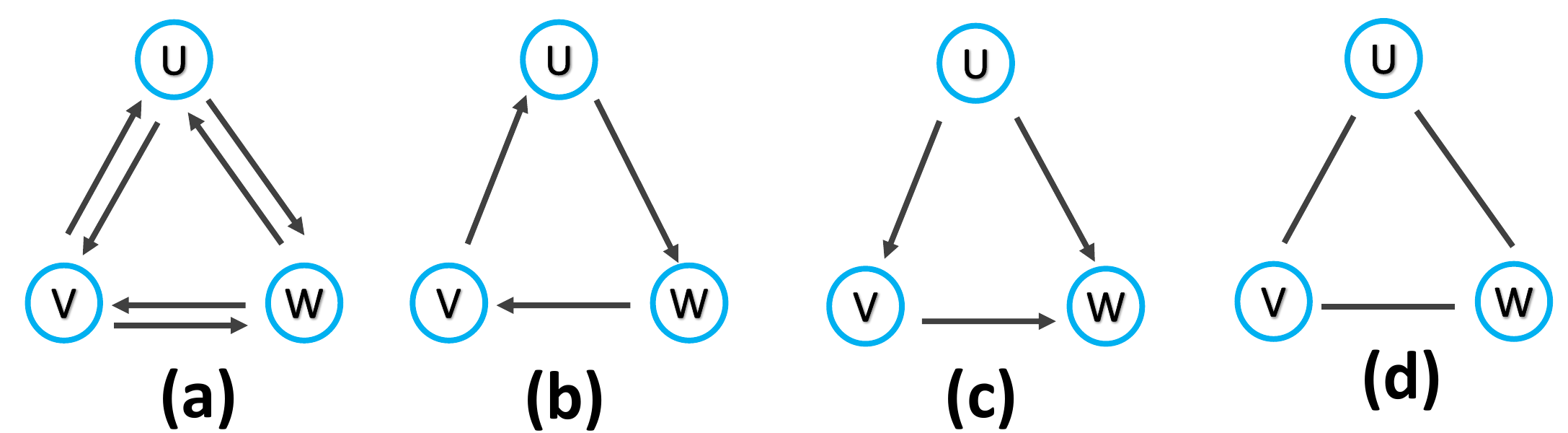}
 \caption{Triangles for various SPDT link configuration among nodes and showing the underlying social structures are same regardless of link direction}
      \label{fig:triangle}
\end{figure}

\subsubsection{Static Properties}
The SPDT contact networks generated by the developed graph model are converted to static contact networks where a directed edge between two nodes is created if they have at least one SPDT link from the host node to neighbour node at any time over the observation period. Then, two static network metrics: degree centrality and clustering coefficient are studied. The degree centrality quantifies the extent of a node's connectedness to other nodes~\cite{freeman1978centrality} how many nodes a host node contacts during the observation period. In a disease spread context, nodes with higher degree centrality get infected quickly as well as infect a higher number of other nodes~\cite{de2014role}. While degree centrality highlights the node connectivity, the clustering coefficient represents how a node is locally connected with the neighbour, i.e. the local social contact structure of nodes in the network~\cite{laurent2015calls}. The local clustering coefficient is defined as the ratio of the number of triangles present among the neighbours and the possible maximum triangles among neighbours. If a node is already infected in a cluster, it is not infected further by other nodes of the same cluster in the SIR epidemic model. Thus, the spreading is varied based on the clustering coefficient and the realistic clustering coefficient is required in the generated network. To compute the clustering coefficient in a simple way, the directions of links are neglected.  However, the triangles still preserve the social structure formed by friends of friends as shown in Figure~\ref{fig:triangle}. All the triangles formed with directional links are the same as the triangle made by Figure~\ref{fig:triangle}. Thus, the generated directed network is converted into an undirected network for computing clustering coefficients.

\begin{figure}[h!]
\includegraphics[width=0.47\linewidth, height=4.6cm]{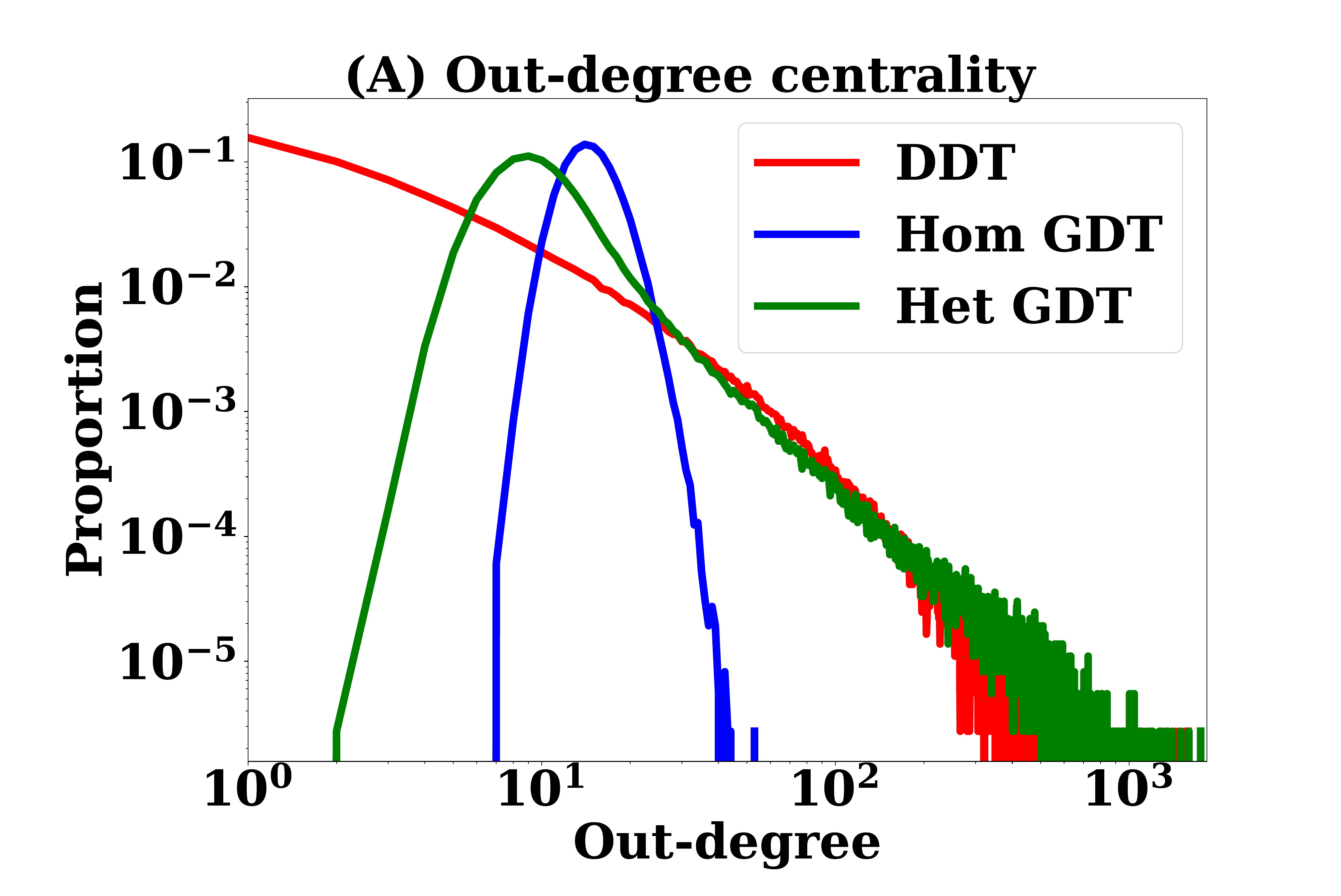}~
\includegraphics[width=0.47\linewidth, height=4.6cm]{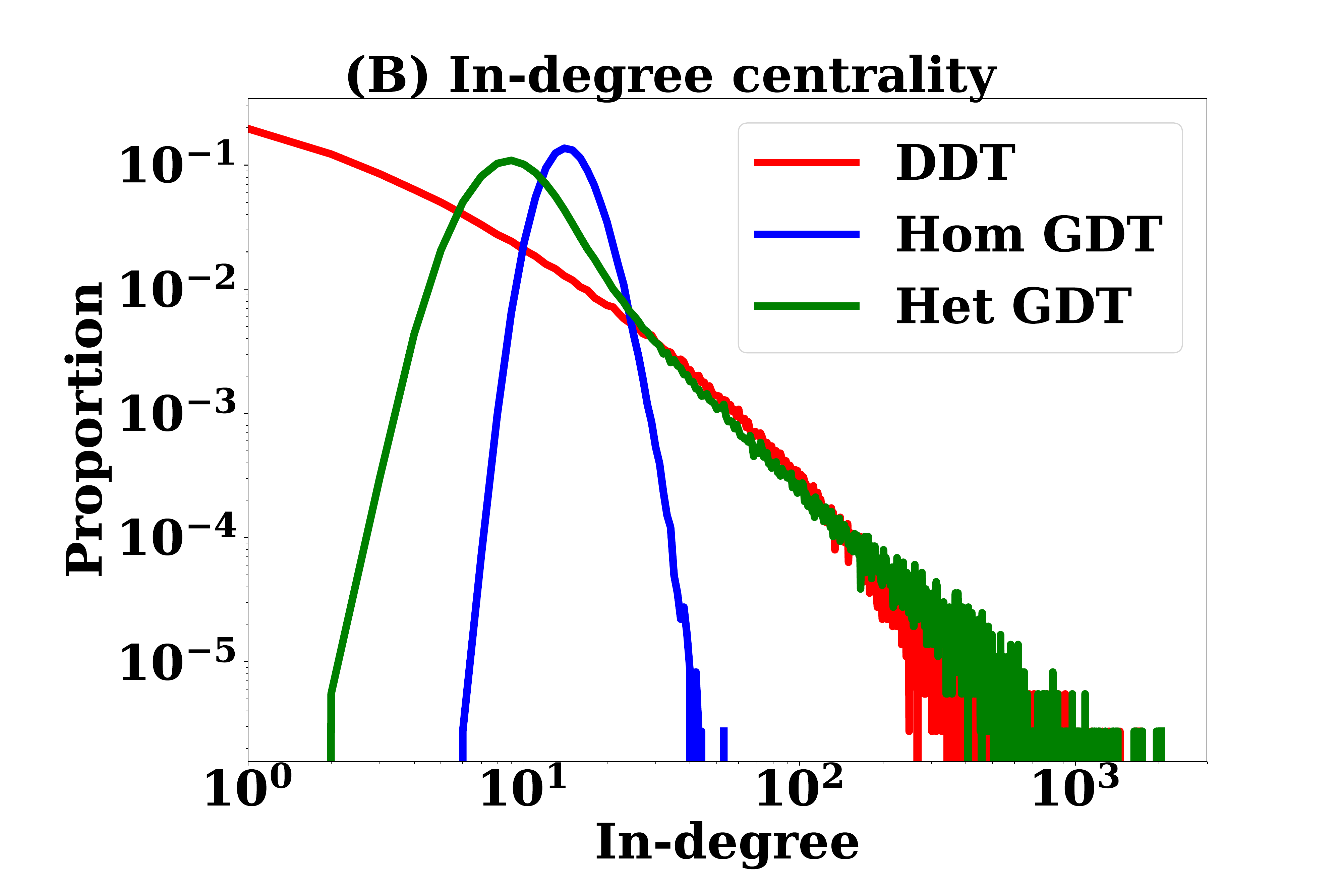}\\

\includegraphics[width=0.47\linewidth, height=4.6cm]{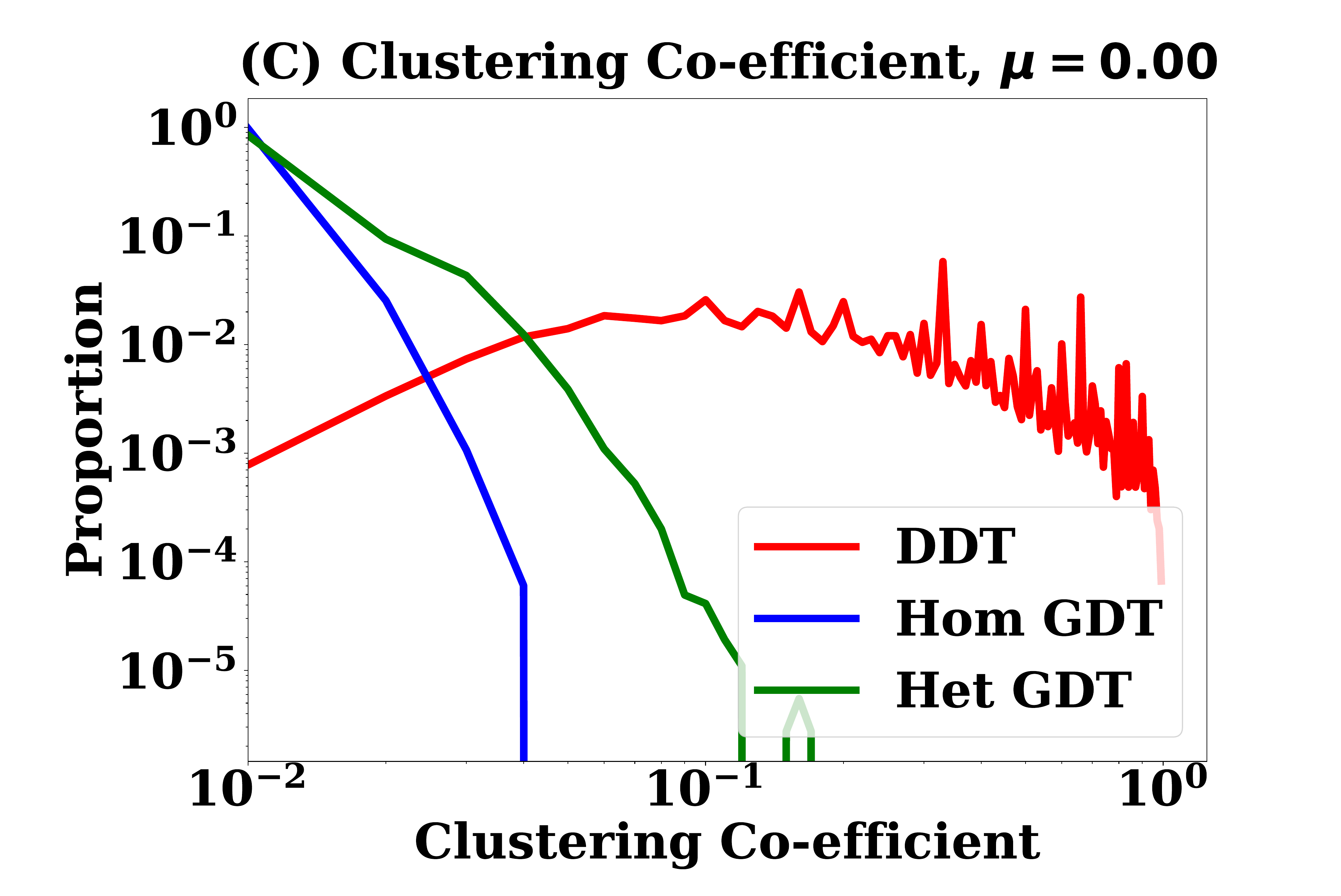}~
\includegraphics[width=0.47\linewidth, height=4.6cm]{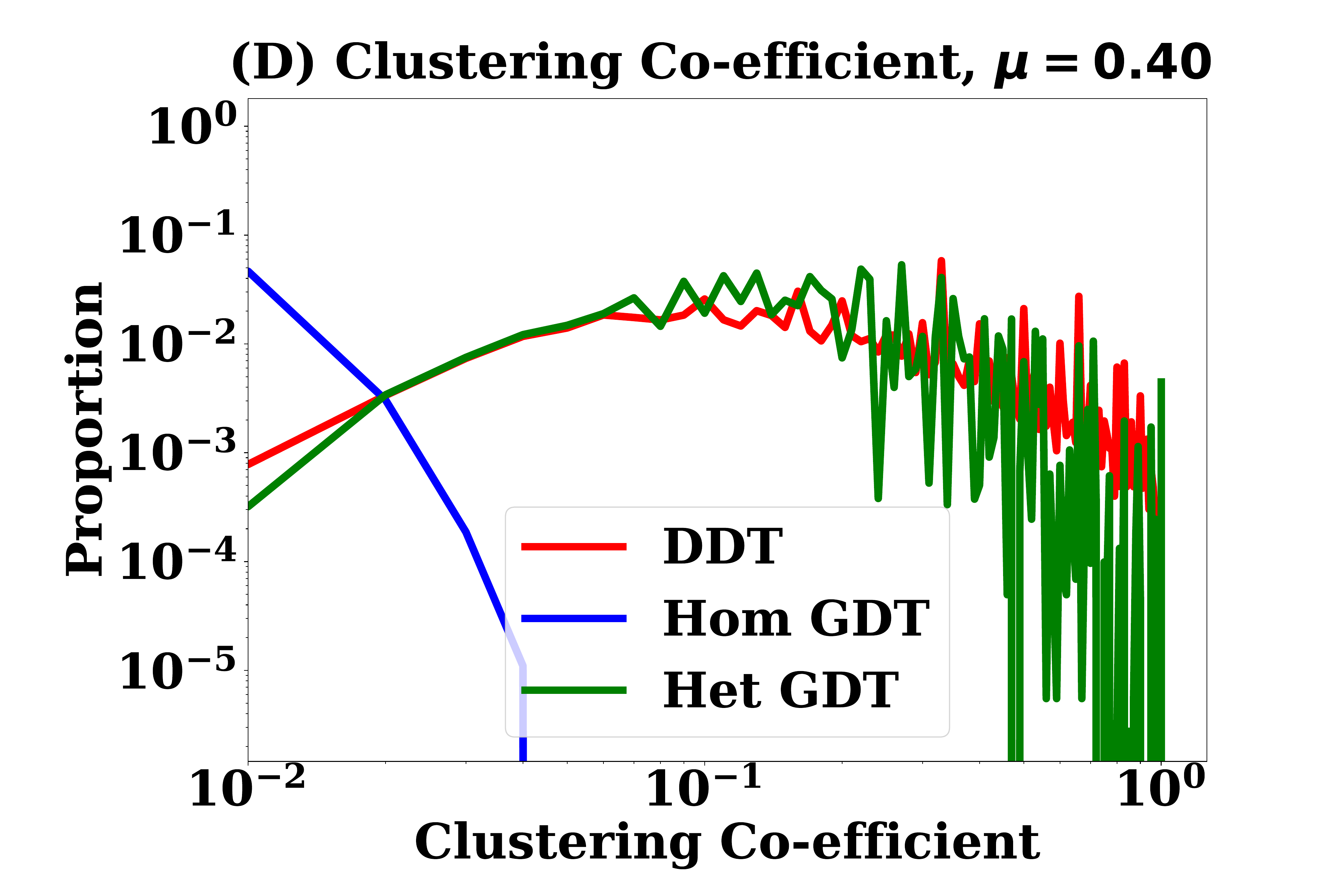}\\

\includegraphics[width=0.47\linewidth, height=4.6cm]{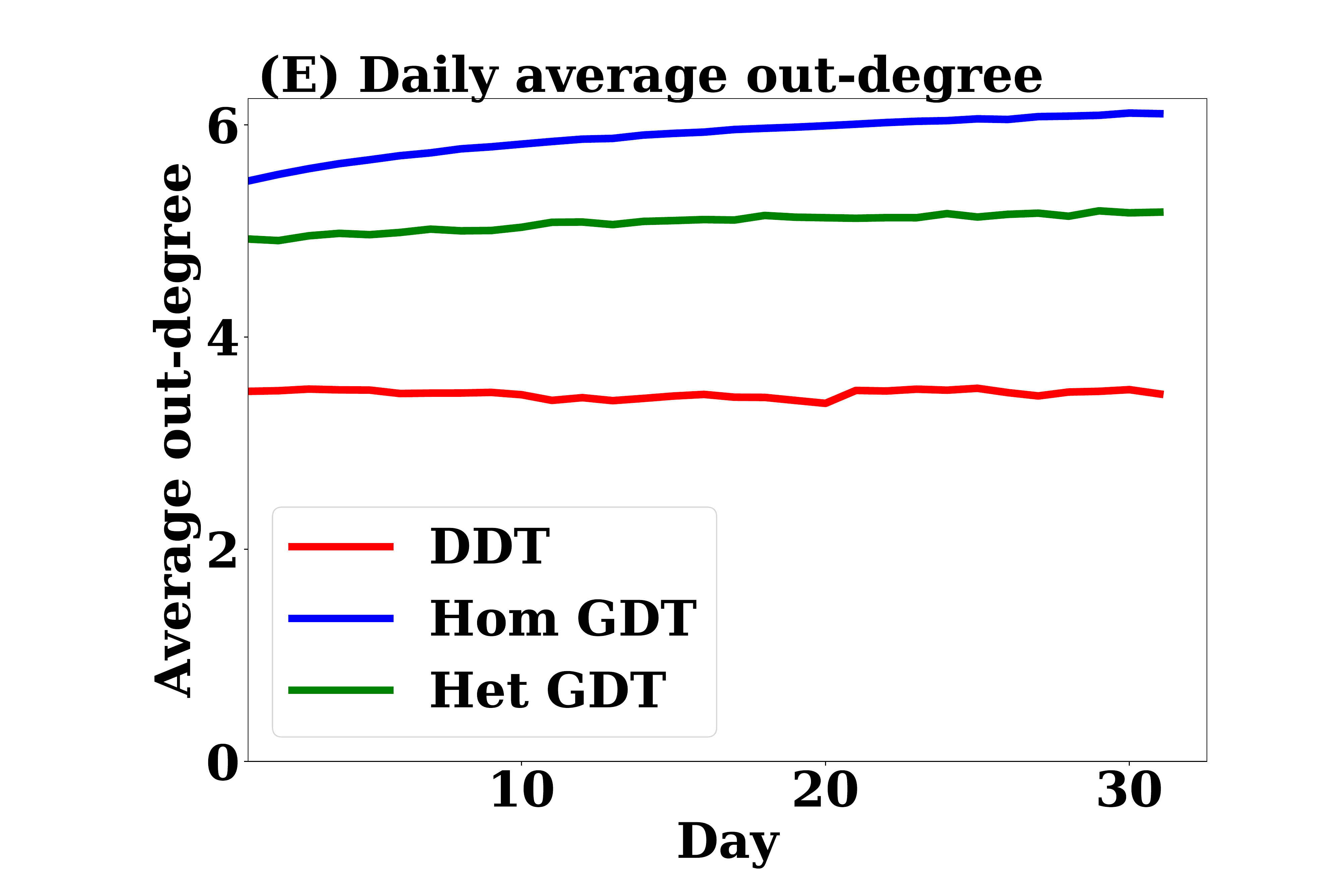}~
\includegraphics[width=0.47\linewidth, height=4.6cm]{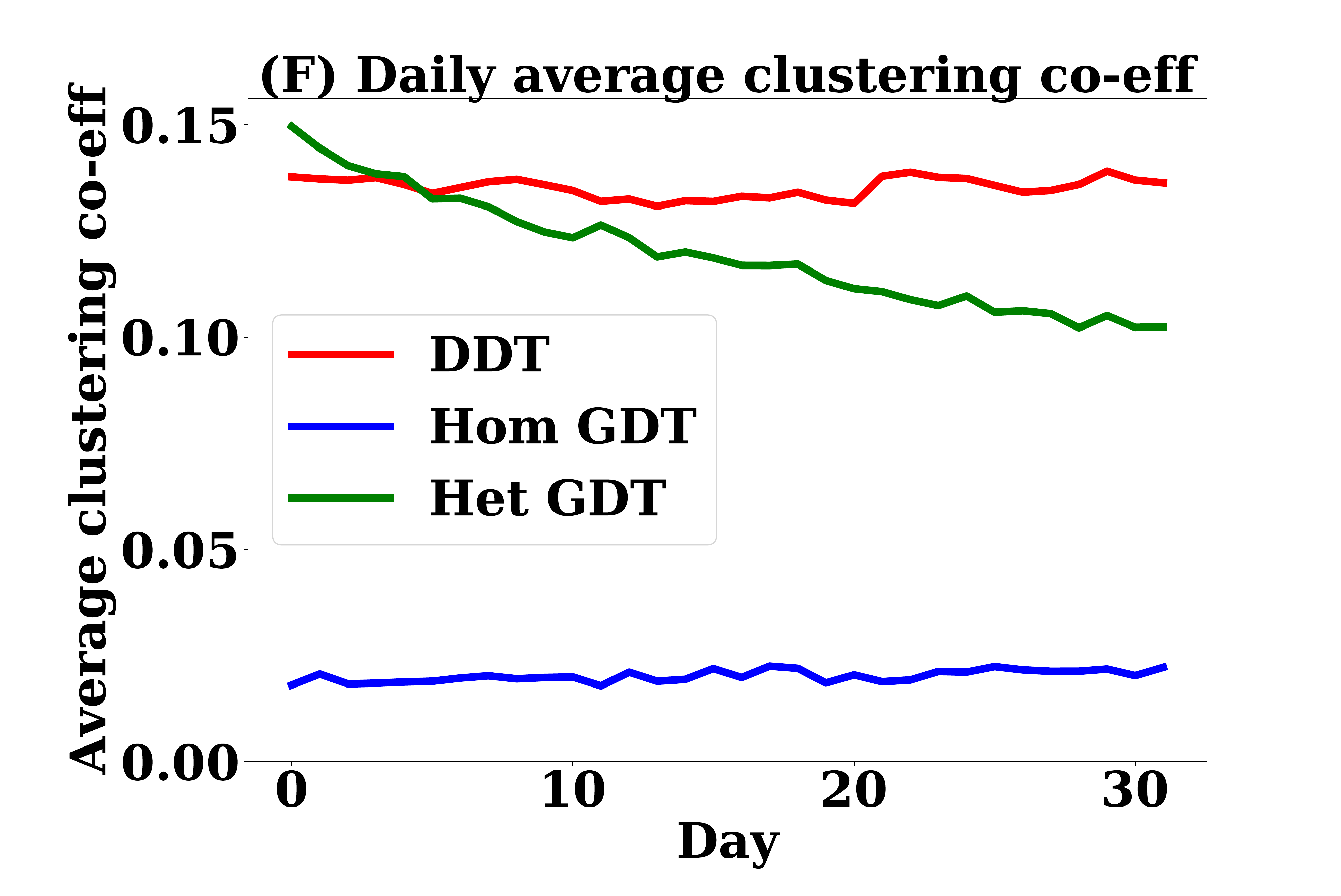}

 \caption{Static network properties for generated homogeneous network (Hom GDT), generated heterogeneous network (Het GDT) and real contact network (DDT): A) out-degree centrality, B) in-degree centrality, C) clustering co-efficient with $\mu=0$, D) clustering co-efficient with $\mu=0.4$, E) daily average out-degree, and F) daily average clustering co-efficient}
      \label{fig:staticpro}
\end{figure}

In the developed graph model, the growth of the contact set of a node is determined by $\lambda$ and the neighbour selection process defined by $P(n_t+1)=n_t/(n_t+\eta)$. The value of $\eta$ controls the degree to which nodes expand their contact set size. In this study, $\eta$ is set to 0.1 and the selection of an optimal value of $\eta$ is beyond the scope of this paper. The growing of contact set size depends on $\lambda$ which is the same for all nodes in the homogeneous network and varies across nodes in the heterogeneous nodes. However, setting $\eta=0.1$ creates opportunities to receive new neighbour nodes. The computed degree centrality and clustering coefficients for both homogeneous and heterogeneous networks are presented in the Figure~\ref{fig:staticpro}. In degree centrality, another desirable feature is that nodes which have more directed links to other nodes also receive more links. This is because individuals who visit crowded places can sent and receive large number of links to others. Thus, the degree centrality is analysed for both out-degree and in-degree and which have the same trends (Fig.~\ref{fig:staticpro}A and Fig.~\ref{fig:staticpro}B). 

The contact set size distribution in DDT network follows the long-tailed distribution. The developed GDT network have a long-tailed at the end, but the distribution of lower degree nodes deviate from that of DDT network. This is more practical compared to DDT network where many individuals have contact set sizes of one. In reality one individual will contact with more than one individuals during 32 days period. Thus, the developed model overcomes the limitation of DDT network for not growing contact set sizes. If $\eta$ is increased, the contact set sizes can be larger. The degree distribution in the homogeneous GDT network follows the poison distribution as the activation degree is generated with a geometric distribution and with the small parameter value. Therefore, the same intensity to involve in interaction with others can not capture the real contact set size distribution. 

In the GDT network, local clustering is formed with selecting $\mu$ proportion neighbours from the neighbours of neighbour nodes. The clustering coefficient for two values $\mu=0$ and $\mu=0.4$ are studied in both heterogeneous and homogeneous networks. The results are presented in Figure~\ref{fig:staticpro}C and Figure~\ref{fig:staticpro}D where nodes are binned based on their clustering coefficient with a step of 0.01. If $\mu$ is set to zero, the clustering coefficient is very low. The clustering coefficient in homogeneous GDT network is 0.005 and 353K nodes are in the first bin, i.e 353k nodes having clustering coefficient below 0.01 while 93K nodes have clustering coefficient below 0.01. The clustering coefficient improves a bit in the heterogeneous GDT network with long-tailed activation degree distribution where 307k nodes having a clustering coefficient below 0.01 and average clustering coefficient in heterogeneous GDT network with $\mu=0$ is 0.008. Then, $\mu$ is set to 0.4 which is found analysing Momo data set. With this value, the heterogeneous GDT network improves significantly to form realistic clustering coefficient and the number of nodes in the first bin is 45K. However, the clustering coefficient is still not realistic in the homogeneous GDT networks and have similar trends in clustering coefficient with $\mu=0$. The heterogeneous degree distribution contributes to making the local clustering as many lower degree nodes can create local cluster through the higher degree nodes connect. In the GDT networks, the clustering coefficients are defined by $\mu$ and contact set sizes of nodes are defined by $\lambda$. However, the value of $\lambda$ affects the required value of $\mu$ and finding their relationship is the out of scope of this thesis.

This experiment also studies the average daily degree and clustering coefficient of nodes. This is measured converting everyday dynamic graph as a static graph where a pair of nodes is linked if once they have contacted. This can indicate how many other nodes a node contact during each day and what is their clustering coefficient. The results are presented in Figure~\ref{fig:staticpro}E and Figure~\ref{fig:staticpro}F. It is observed that nodes in DDT network interact with on average 4 other nodes which are around 5 nodes in the heterogeneous GDT network. As the contact set size grows in the heterogeneous GDT network, it has a high average contact degree over the day. The homogeneous GDT network has the highest average daily degree. This is because nodes in the homogeneous GDT network have moderate activation degree and do not have many nodes with the low degree as it is in the DDT and heterogeneous GDT networks. The daily average clustering coefficient is very low in the homogeneous GDT network. The average daily clustering coefficient in the heterogeneous GDT network is the same to that of DDT network at the early days. However, it decreases as simulation time goes. This is because the contact set sizes grow in the heterogeneous GDT network and nodes have more options to select neighbour nodes. Therefore, common neighbours among nodes decrease and hence clustering coefficients are low. However, it can gradually becomes stable at later days of simulation if simulation is run for longer days.

\subsubsection{Temporal Properties}
The proposed graph model generates the temporal activities of nodes through SPDT link creation. The temporal properties should match those of real networks. Two commonly used temporal network central properties to characterise the network: closeness and betweenness are studied. The closeness of nodes measures the geodesic distance between all pairs of nodes in the network. The node betweenness measures the fraction of geodesic distances which a node is connected between all node pairs in the network~\cite{holme2015modern}. The generated network is converted to the temporal network $G_{T}=(Z,L_{T})$ where $Z$ is the set of nodes and set $L^{T}\subseteq V\times Z\times [0,T]$ of time-stamped links $(u,v;t)\in L_{T}$. In the airborne disease propagation, susceptible individuals who get infected do not start infecting immediately and go through an incubation period before they can infect others. In this study, the incubation period is set to one day, i.e, individuals that become infected on a simulation day can start spreading the disease the next day after infection. Thus, all links of a day between a pair of nodes are aggregated over the day and the timestamp $t_j$ represents the interaction day $j$. If a disease is transmitted from node $u$ to node $v$ through a link $(u,v;t_{j})$, the disease will be transmitted from node $v$ to another node $w$ with the link $(v,w;t_{j+k})$ where $1\leq k\leq 5$ as an node is considered to be infectious for the next five days.The temporal properties are measured based on the time-respecting paths among nodes. A time-respecting path between an infected node $u$ to a susceptible node $v$ may create through a sequence of time-stamped links as follows:
\[(w_{0},w_{1};t_{1}),(w_{1},w_{2};t_{2}),\ldots,(w_{l-1},w_{l};t_{l})\]
where $w_{0}=u,w_{l}=v$ and the sequence of time-stamped links should be $t_{1}\leq t_{2}\leq \ldots \leq t_{l}$. The time difference between two consecutive links of a time-respecting path can be at most 5.

\begin{figure}[h!]
\includegraphics[width=0.49\linewidth, height=4.8cm]{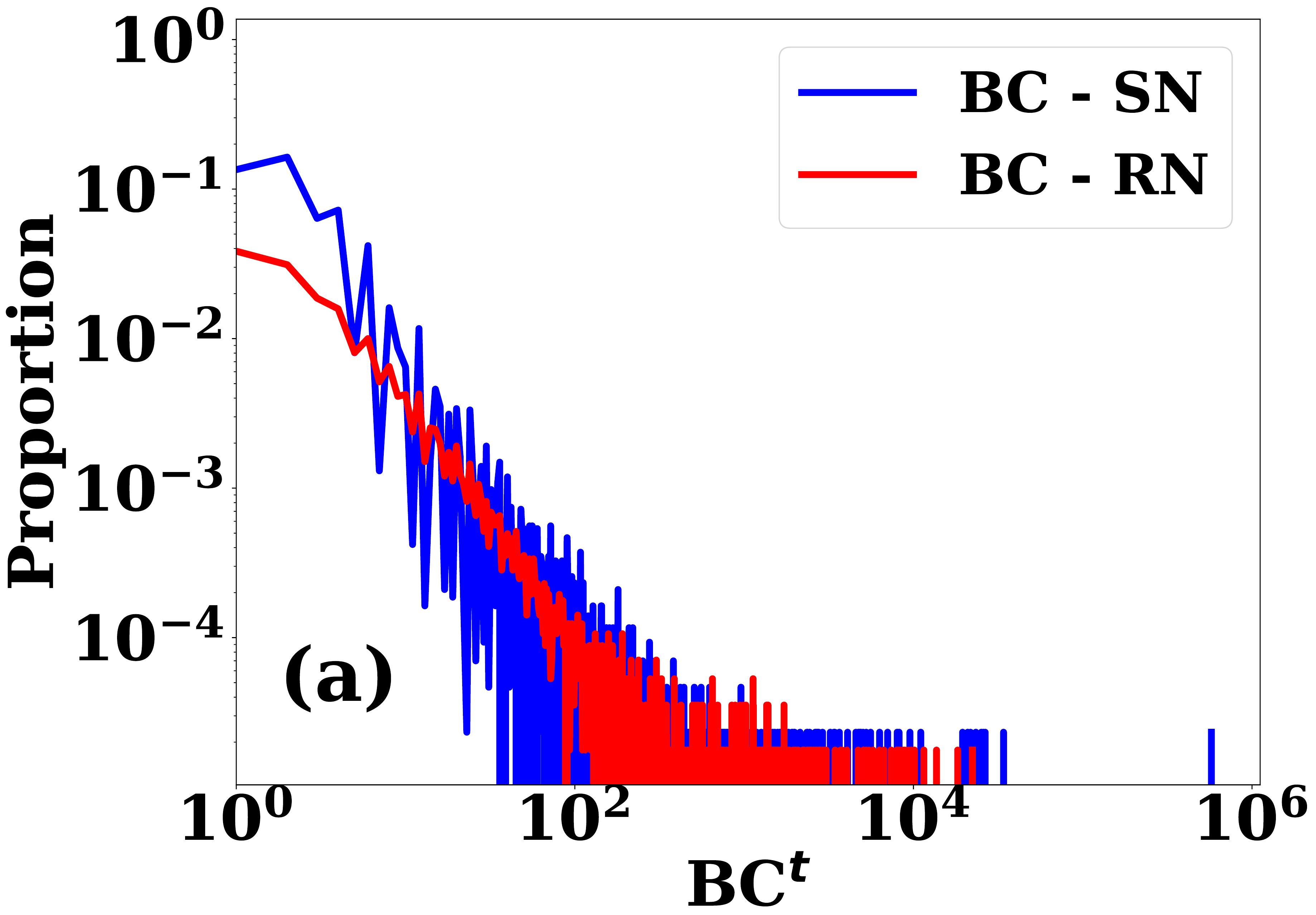}~    \includegraphics[width=0.49\linewidth, height=4.8cm]{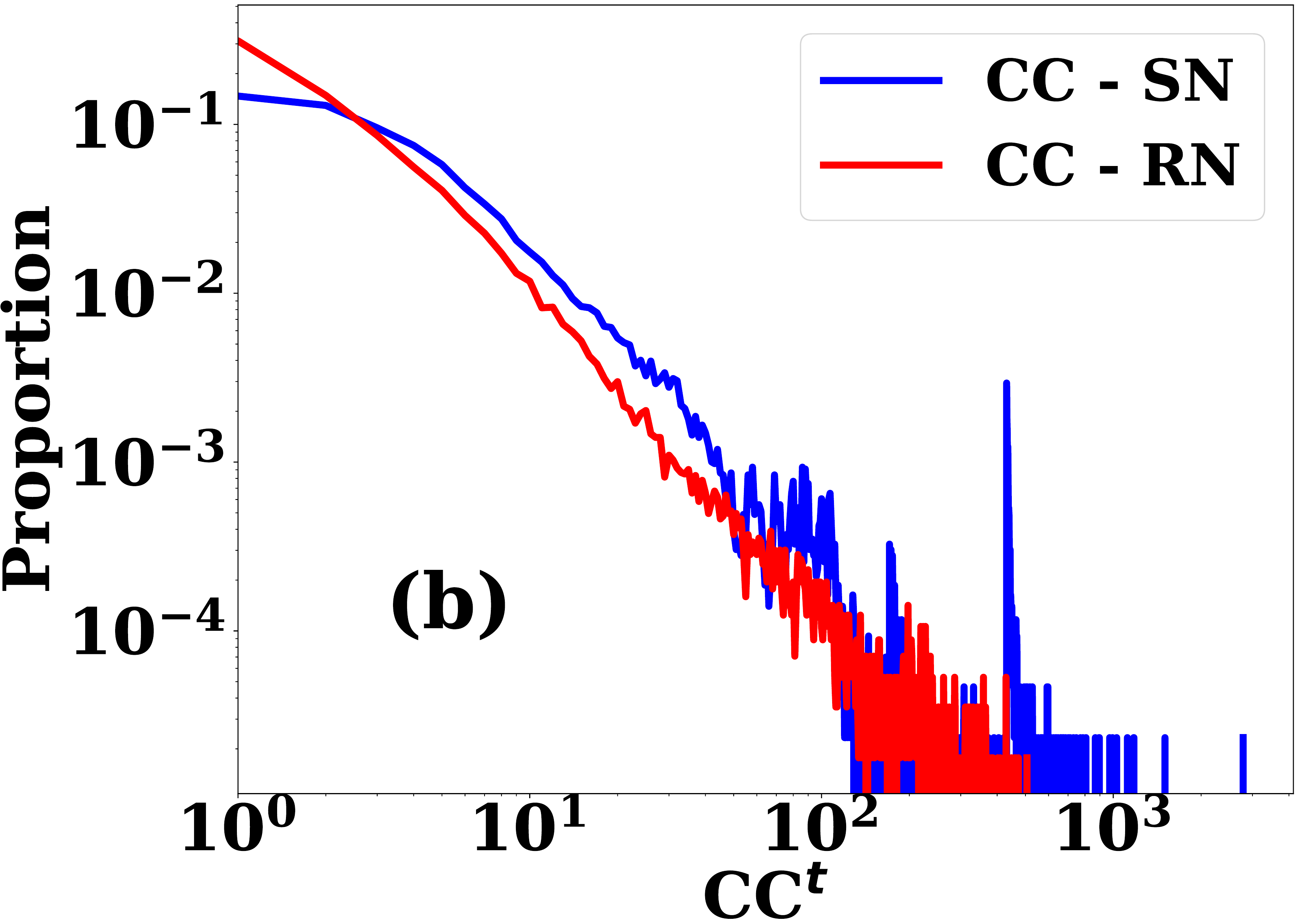} 
\caption{Temporal network properties in the generated contact network and comparison with that of real contact network : A) Betweenness centralities and B) Closeness centralities}
      \label{fig:tempro}
\end{figure}

In the temporal network, the betweenness of a node is defined as the number of shortest time-respecting paths pass through it~\cite{taylor2017eigenvector,scholtes2016higher}. However, the temporal betweenness depends on the starting time $t_{1}$ of time-respecting paths as the path length between the same pair of nodes may change if the starting time is shifted. Thus, shortest paths between a pair of nodes are computed at all starting times and the shortest paths regardless of starting time are counted as the set to measure betweenness. The temporal betweenness centrality $BC^{t}$ of an node $v$ is measured as:
\[BC^{t}(v)=\sum_{u\not= v \not= u}^{}\mid Q^{t}\left(u,w; v\right)\mid\]
where $Q^{t}$ is the set of shortest time-respecting paths between $u$ and $w$ passing through $v$ starting any time during observation period~\cite{scholtes2016higher}. In a similar way, the temporal closeness centrality $CC^{t}$ based on the shortest time respecting paths is defined as: 
\[CC^{t}(v)=\sum_{u\not= v}^{}\frac{1}{dist^{t}\left(u,v\right)}\]
where $dist^{t}$ is the distance of the shortest time-respecting path from $u$ to $v$ starting at any time during the observation period. The temporal network analysis methods proposed by the authors of~\cite{scholtes2016higher} are used in this experiment. The generated networks of 32 days are used and the results obtained are presented in Figure~\ref{fig:tempro} for both the real network (RN) and synthetic network (SN). The betweenness distributions of $BC^{t}$ and $CC^{t}$ have almost similar trendiness. However, the synthetic network has a higher number of nodes with a lower $BC^{t}$. This is because the synthetic network is more random than the real network. In the synthetic network, some nodes may have higher activation degree as it depends on the power law distributed $\lambda_{i}$ and hence have higher $CC^{t}$. The synthetic network overall produces similar network properties of the real network.

%% file: 4.4_chap_netdif.tex
\subsection{Diffusion dynamics}
Now, the capability of simulating SPDT process by the generated networks are studied. Accordingly, airborne disease spreading is simulated on the generated synthetic contact networks (GDT) and the real contact networks (DDT). First, the similarities in the diffusion dynamics are studied on the various networks and then the sensitivity of the graph model to various parameters are studied as the model is required to respond with model parameters realistically. The experimental setup and diffusion analysis are presented below.

\subsubsection{Experimental setup}
For this experiment, homogeneous GDT networks and heterogeneous GDT networks with 364K nodes are generated for 32 days as the DDT network contains 364K nodes and 32 days period. The simulations are also conducted on another synthetic contact networks constructed according to the basic activity driven networks (ADN) model~\cite{perra2012activity} to understand how well the proposed model capture diffusion dynamics comparing to the current graph models. The contact networks generated by the ADN model are only based on the direct transmission links. In these network, nodes activate at each time step $\Delta t$ with a probability $\phi$ and generate $m$ links to others nodes. The probabilities $\vartheta$ of activation of nodes are assigned with a power law degree distribution. The number of links $m$ created during activation can be constant for all nodes or heterogeneous. In this study, both types of $m$ are used and homogeneous ADN networks and heterogeneous ADN networks with homogeneous and heterogeneous degree distribution are generated. For fitting activation potential $\vartheta$, the daily activation frequencies $h$ of Momo users are used. The analysis of Momo users shows that their average stay periods $\Delta t$ are 50 minutes. Therefore, the number of time steps in one day is 28.8 and the potential is $\vartheta=28.8/h$. Thus, the activation frequencies are converted to activation potential $\vartheta$ and fit with a power law degree distribution with a lower limit of $\vartheta$ is 0.02 and an upper limit is 0.18 while power law exponent is 2.95. For the activation degree, it is found that the average activation degree is $m=3$. For heterogeneous degree distribution, the activation degree distribution from the developed SPDT graph model are being used. With these fitted parameters homogeneous ADN and heterogeneous ADN networks are generated for 364K nodes over 32 days.

On these contact networks, disease propagates according to the Susceptible-Infected-Recovered (SIR) epidemic model. In the simulations, the infection probability $P_I$ for inhaling $E$ dose of infectious particles by a node is given by the Equation~\ref{eq:prob}. The previous parameter setting of Equation~\ref{eq:expo} are applied here. Each simulation is started with an intial set of 500 seed nodes. The epidemic parameters: disease prevalence $I_p$, number of infected nodes in the network, and cumulative infection $I_a$, number of the infected individuals up to a day are calculated. To characterize the diffusion reproduction ability, the absolute percentage variation (APV) for infection events are calculated for each network with the reference of real contact networks as: 
\[ APV=100\times \frac{I_r-I_o}{I_r}\]
where $I_r$ is the number of infection events in the real contact network (DDT) and $I_o$ is the infection event in the corresponding synthetic contact networks (GDT or ADN). For the both $I_p$ and $I_a$ are measured for each network.

\begin{table}
\caption{Summary of diffusion dynamics on contact networks}
 \label{difsum}
\vspace{1em}
\centering
\begin{tabular}{|c|c|c|c|c|c|}
\hline
Networks & $r_t$ & $I_p$(max) & $I_a$(max) & average daily var (\%) & average acc. var (\%)  \\ \hline
\multirow{3}*{DDT} & 60 & 1731  & 11221 & 0 & 0 \\ 
\cline{2-6}
 & 40 & 1332  & 8671 & 0 & 0 \\  \cline{2-6}
 & 20 & 966  & 6460 & 0 & 0 \\ \hline
 
\multirow{3}*{Het GDT} & 60 & 2094  & 11929 & 15 & 10 \\ 
\cline{2-6}
 & 40 & 1531  & 9584 & 18 & 12 \\ \cline{2-6}
 & 20 & 987  & 6853 & 22 & 13 \\ \hline
 
\multirow{3}*{Het ADN} & 60 & 1007 & 7809  & 29 & 37 \\ 
\cline{2-6}
 & 40 & 673  & 5666 & 30 & 38 \\ \cline{2-6}
 & 20 & 335  & 3170 & 45 & 44 \\ \hline
 
\multirow{3}*{Hom GDT} & 60 & 698  & 3600 & 61 & 50 \\ 
\cline{2-6}
 & 40 & 626  & 2110 & 69 & 53 \\ \cline{2-6}
 & 20 & 586  & 1592 & 70 & 51 \\ \hline
 
 \multirow{3}*{Hom ADN} & 60 & 600  & 1892 & 73 & 62 \\ 
\cline{2-6}
 & 40 & 581  & 1645 & 73 & 59 \\ \cline{2-6}
 & 20 & 562  & 1386 & 72 & 55 \\ \hline

\end{tabular}
\end{table}

\begin{figure}[h!]
\begin{tikzpicture}
    \begin{customlegend}[legend columns=5,legend style={at={(0.12,1.02)},draw=none,column sep=2ex,line width=2 pt }, legend entries={DDT, Het GDT, Het ADN, Hom GDT, Hom ADN}]
    \addlegendimage{mark=triangle,solid,line legend, color=blue}
    \addlegendimage{mark=x,solid, color=red}   
    \addlegendimage{mark=triangle,solid, color=magenta}
    \addlegendimage{mark=o, color=cyan}
    \addlegendimage{mark=star, color=yellow}
    \end{customlegend}
 \end{tikzpicture}
\vspace{4ex}
\centering
\includegraphics[width=0.45\linewidth, height=5cm]{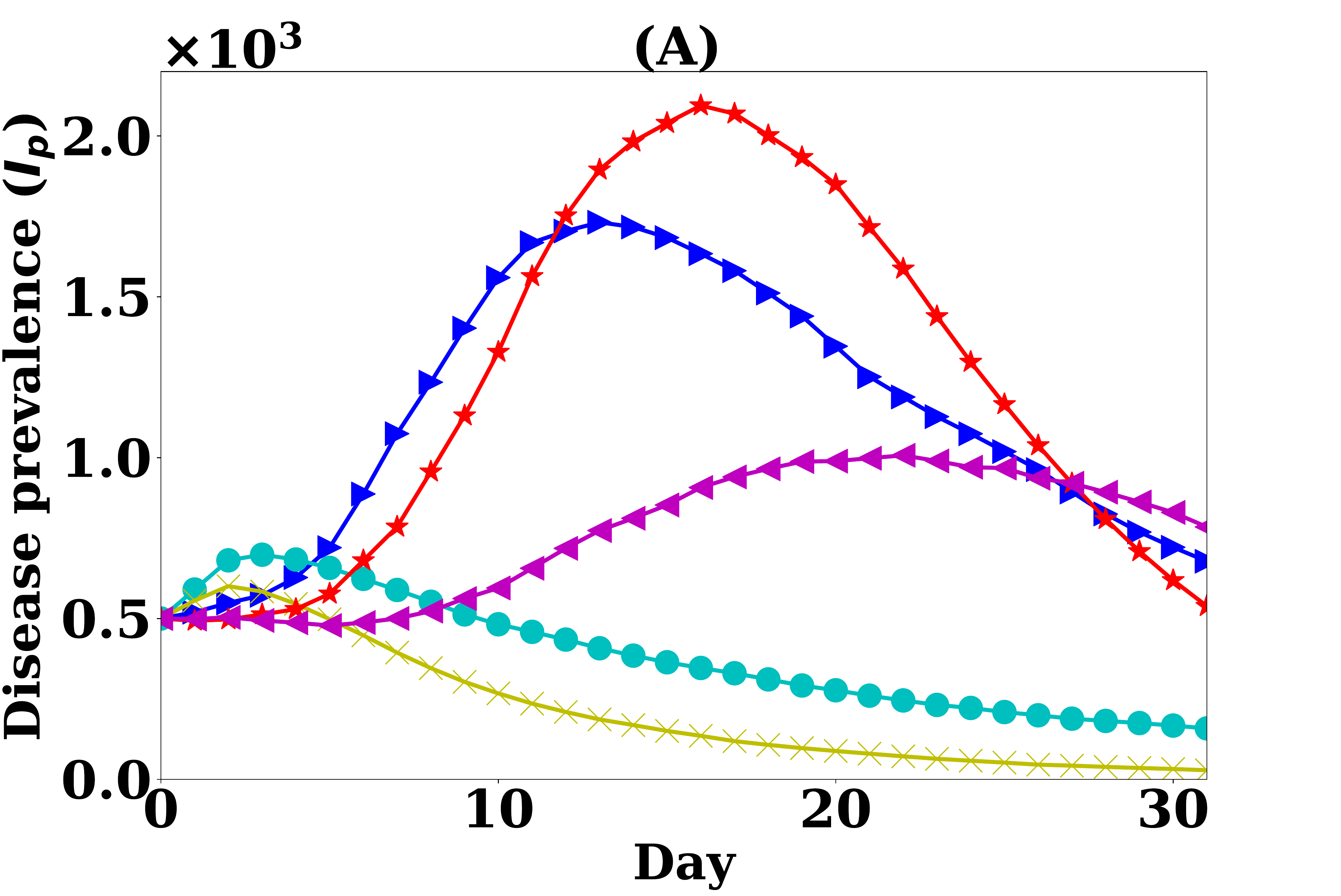}~
\includegraphics[width=0.45\linewidth, height=5cm]{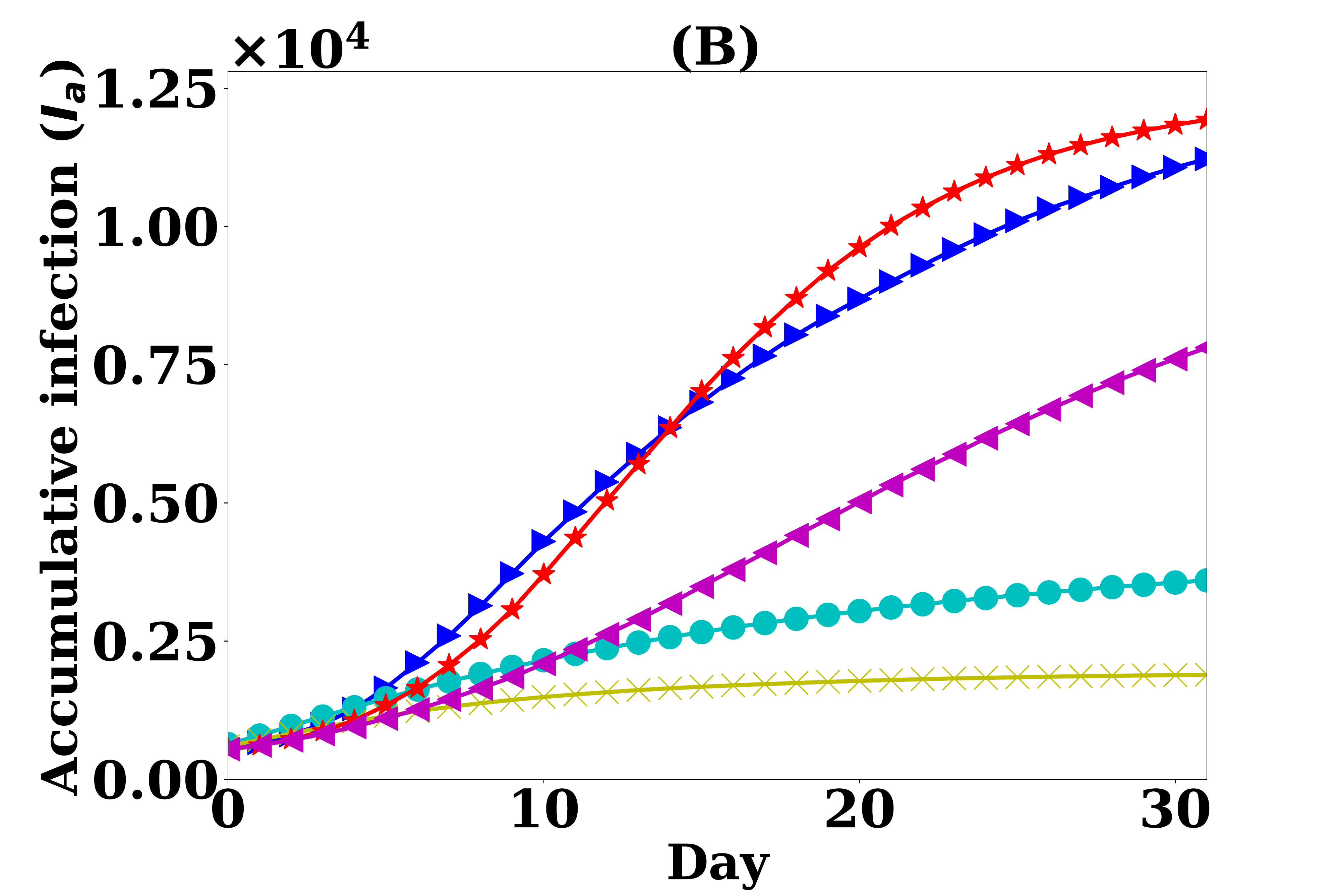}\\
\vspace{1em}
\includegraphics[width=0.45\linewidth, height=5cm]{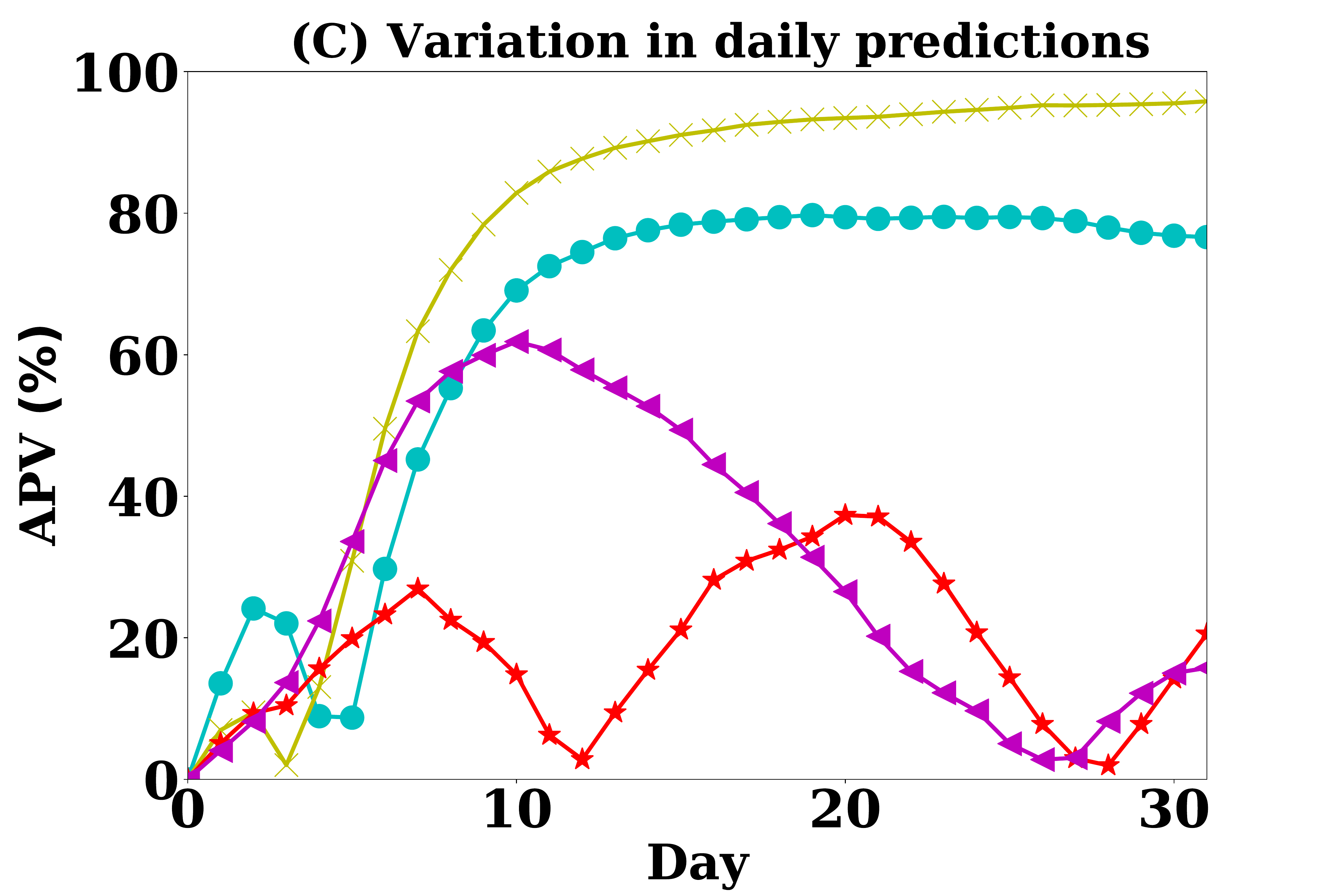}~
\includegraphics[width=0.45\linewidth, height=5cm]{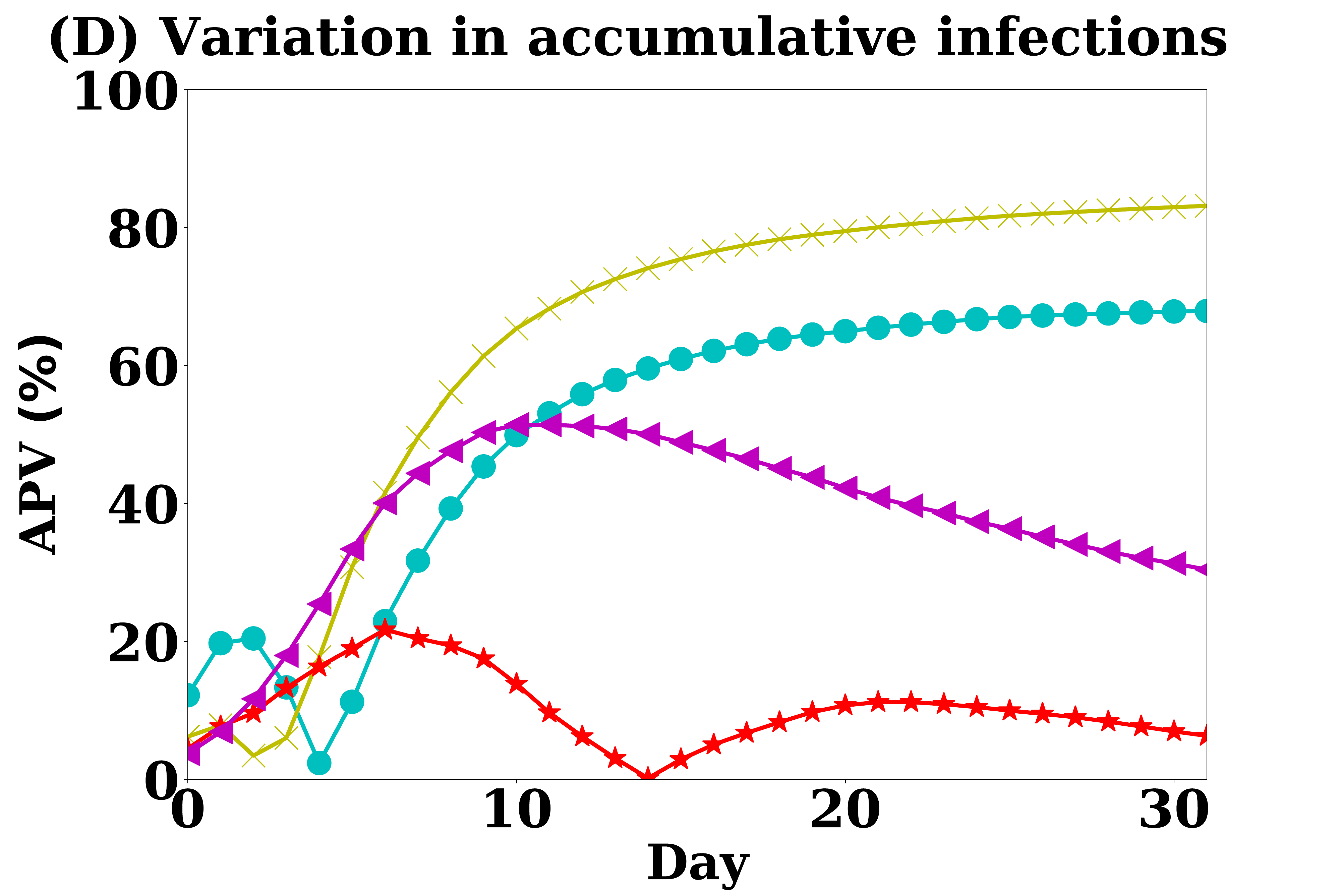}
\caption{Diffusion dynamics on various contact networks -  real contact network (DDT), heterogeneous GDT (Het GDT), homogeneous GDT (Hom GDT), heterogeneous ADN (Het ADN), and homogeneous ADN (Hom ADN) : A) disease prevalence dynamics, B) accumulative infection over simulation days, C) variations in the daily prediction comparing to real contact network and D) variation in the accumulative prediction}
\label{fig:nethmdf}
\end{figure}

\subsubsection{Diffusion analysis}
The selected four synthetic contact networks (Hom ADN, Het ADN, Hom GDT and Het GDT) are generated with 364K nodes for 32 days and simulations are run with the selected disease parameters. The infection probability $P_I$ in Equation~\ref{eq:expo} is calculated with particle decay rates $r_t=60$ min. The average results for 1000 realisation are shown in Figure~\ref{fig:nethmdf}. The proposed graph model with heterogeneous degree distributions (Het GDT network) has diffusion dynamics that are close to real contact networks (DDT network). The heterogeneous GDT network has higher disease prevalence $I_p$ than the DDT network (Fig.~\ref{fig:nethmdf}A). This is because nodes in the heterogeneous GDT network grow their contact sizes substantially compared to DDT network where links between two nodes are repeated. This is observed in the Figure~\ref{fig:staticpro} where nodes with small contact set sizes in GDT network is fewer than that of the DDT network. Thus, infected nodes in the DDT network repeatedly forward SPDT links to the same susceptible nodes and disease propagates faster in the DDT network at the early days of simulations. On the other hand, infected nodes in the heterogeneous GDT networks forward links to more susceptible nodes and disease transmission happen slowly in the heterogeneous GDT network. Thus, the initial spreading rates in the DDT network is faster than the heterogeneous GDT network. For the same reason, the heterogeneous GDT network carries growth of $I_p$ for a longer time and to a higher value. The total infection in heterogeneous GDT network is also close to the total of the real network (Figure~\ref{fig:nethmdf}B). On the other hand, the heterogeneous ADN networks show increasing of $I_p$. But, the diffusion dynamics are underestimated significantly and total infection is comparatively low. This is because the heterogeneous ADN network does not have indirect transmission links. Thus, the growth of $I_p$ is not carried out in this network as it is in the DDT and heterogeneous GDT networks. On the other hand, homogeneous networks cannot replicate the diffusion dynamics of real contact networks. Initially, the disease prevalence $I_p$ increases for a few days in the homogeneous GDT network and declines. A similar trend is found for the homogeneous ADN network. This is because the homogeneous networks do not have higher degree nodes that have strong disease spreading potential and forward disease to different points of the network. In the homogeneous networks, the neighbour nodes quickly get infected and infection resistance force, which is created when the susceptible nodes reduce in the networks, grows due to reducing susceptible neighbour nodes. The prediction variations in APV for all networks are shown in the Figure~\ref{fig:nethmdf} C and D. The heterogeneous GDT network shows the lowest variation in daily predictions among all networks comparing to DDT network. This has on average 15\% variation while heterogeneous ADN network shows on average 29\% variation. However, the variations are very high for the heterogeneous ADN network with the maximum variation 60\%. The accumulative prediction variation in heterogeneous GDT network is averaged at 10\% while this is very high for the heterogeneous ADN networks with 37\%. For the other networks, the homogeneous GDT network has on average 61\% daily variation and homogeneous ADN network has 70\% variation. Having higher variation in daily prediction, proposed model still maintains a cumulative prediction variation around 10\%. The proposed graph model improves up to 31\% in predicting disease spreading dynamics, that is done by the real contact network (DDT), over the current models (see Table~\ref{difsum}).

\begin{figure}[h!]
\begin{tikzpicture}
    \begin{customlegend}[legend columns=4,legend style={at={(0.12,1.02)},draw=none,column sep=2ex,line width=2 pt }, legend entries={DDT, Hom GDT, Hom ADN, Het GDT, Het ADN, $r_t=60$, $r_t=40$, $r_t=20$}]
    \addlegendimage{solid,line legend, color=blue}
    \addlegendimage{solid, color=red}   
    \addlegendimage{solid, color=cyan}
    \addlegendimage{color=yellow}
    \addlegendimage{color=black}
    \addlegendimage{only marks, mark=x}
    \addlegendimage{only marks, mark=o}
    \addlegendimage{only marks, mark= triangle}
    \end{customlegend}
 \end{tikzpicture}
\vspace{4ex}
\centering
\includegraphics[width=0.45\linewidth, height=5cm]{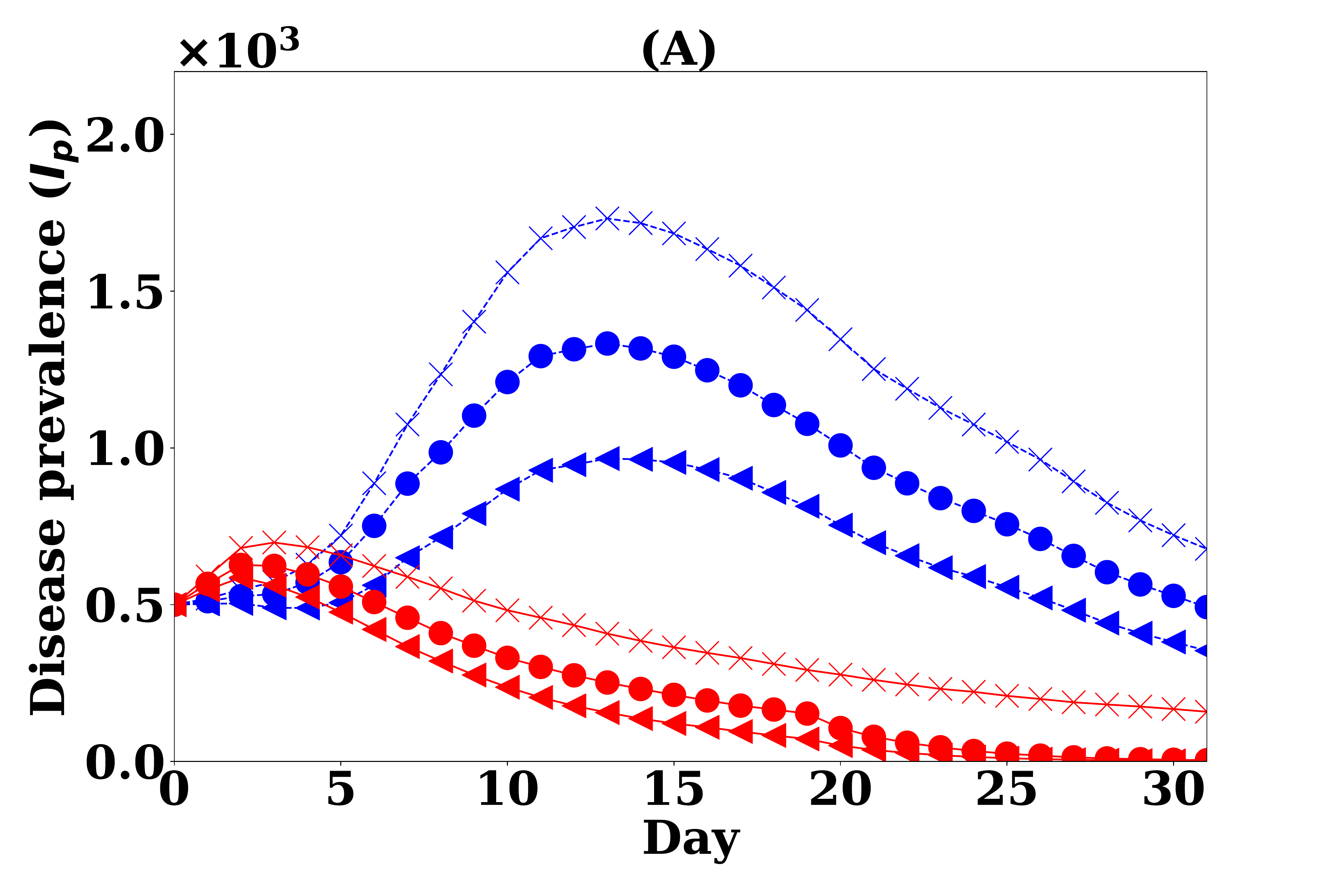}~
\includegraphics[width=0.45\linewidth, height=5cm]{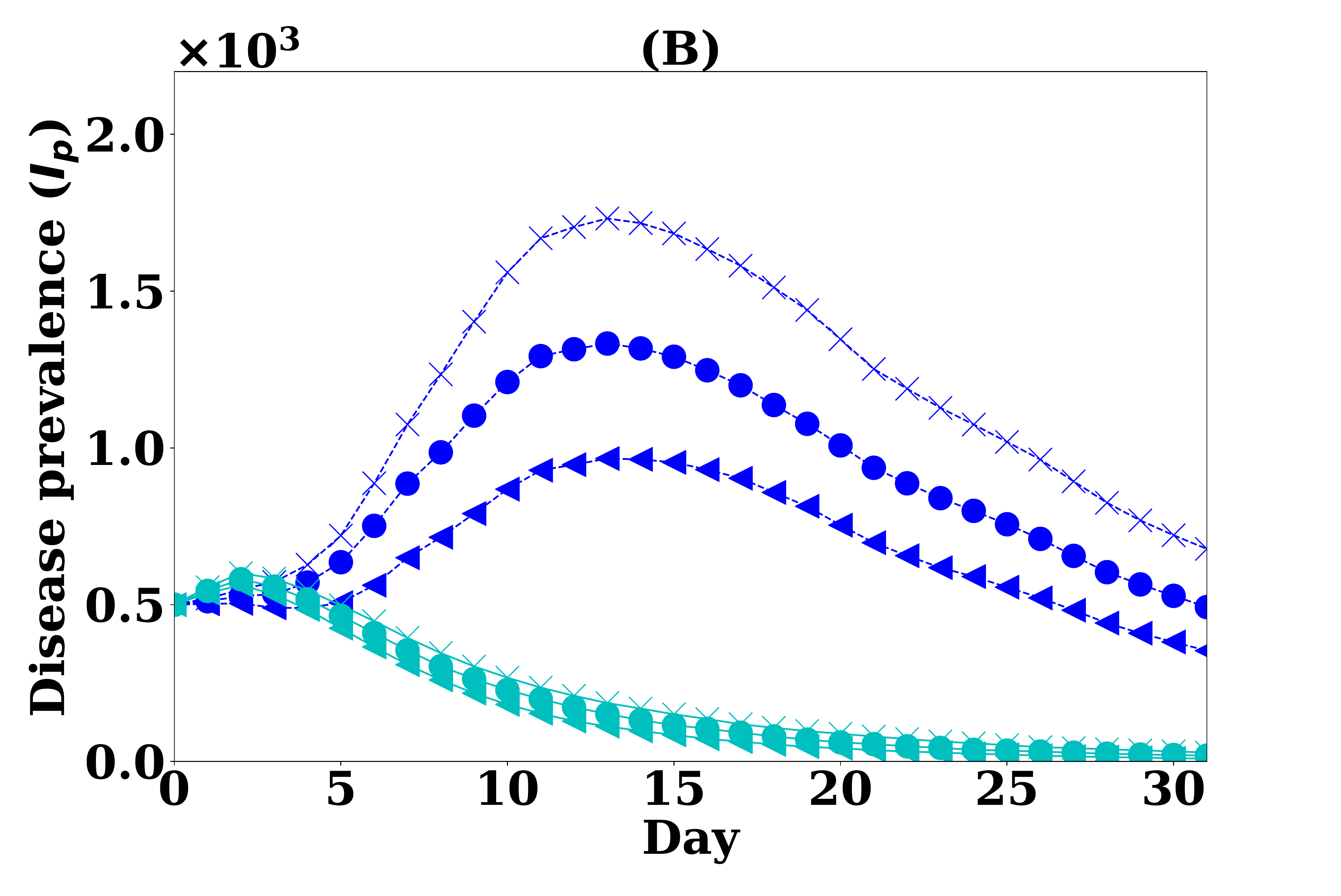}\\
\vspace{1em}
\includegraphics[width=0.45\linewidth, height=5cm]{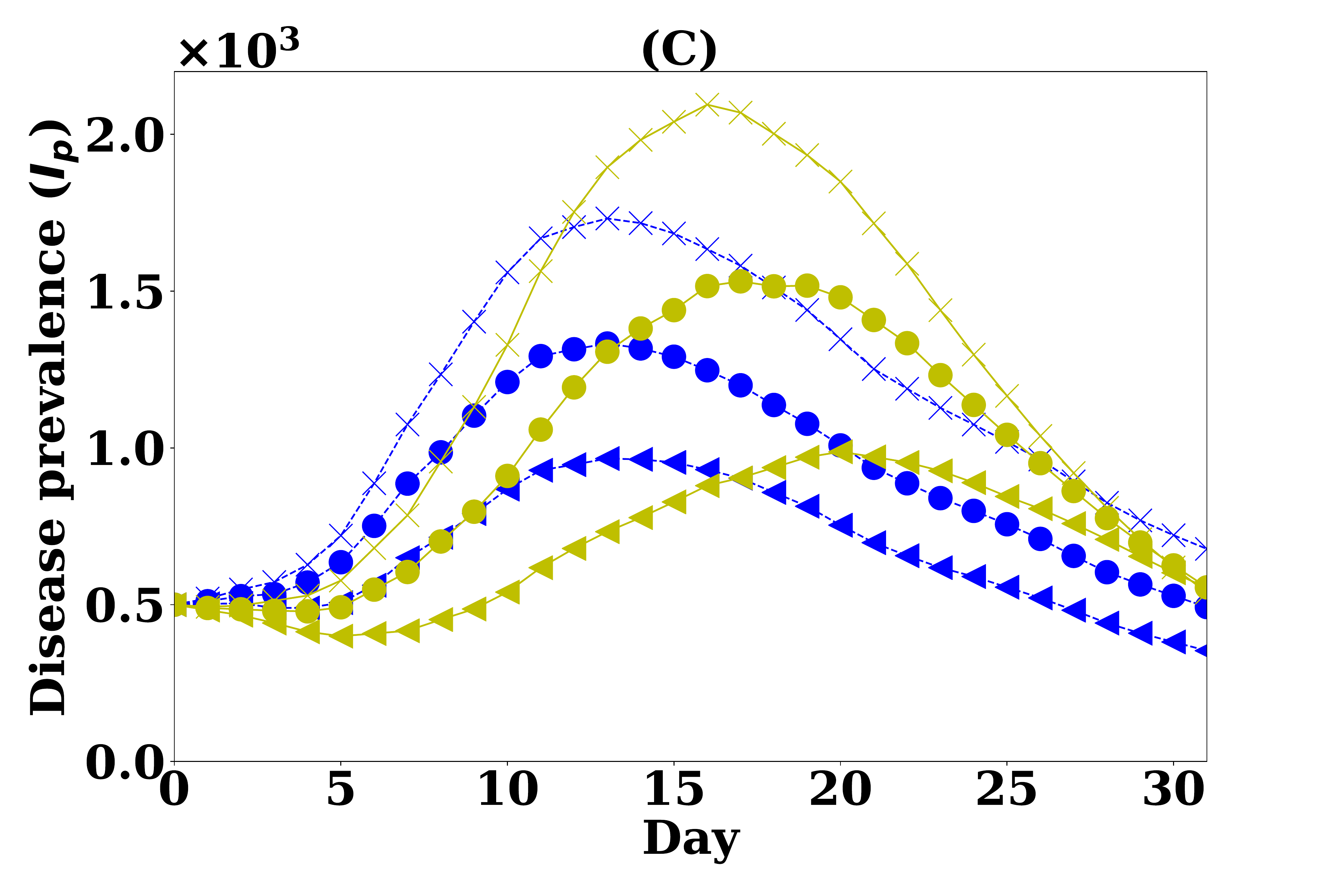}~
\includegraphics[width=0.45\linewidth, height=5cm]{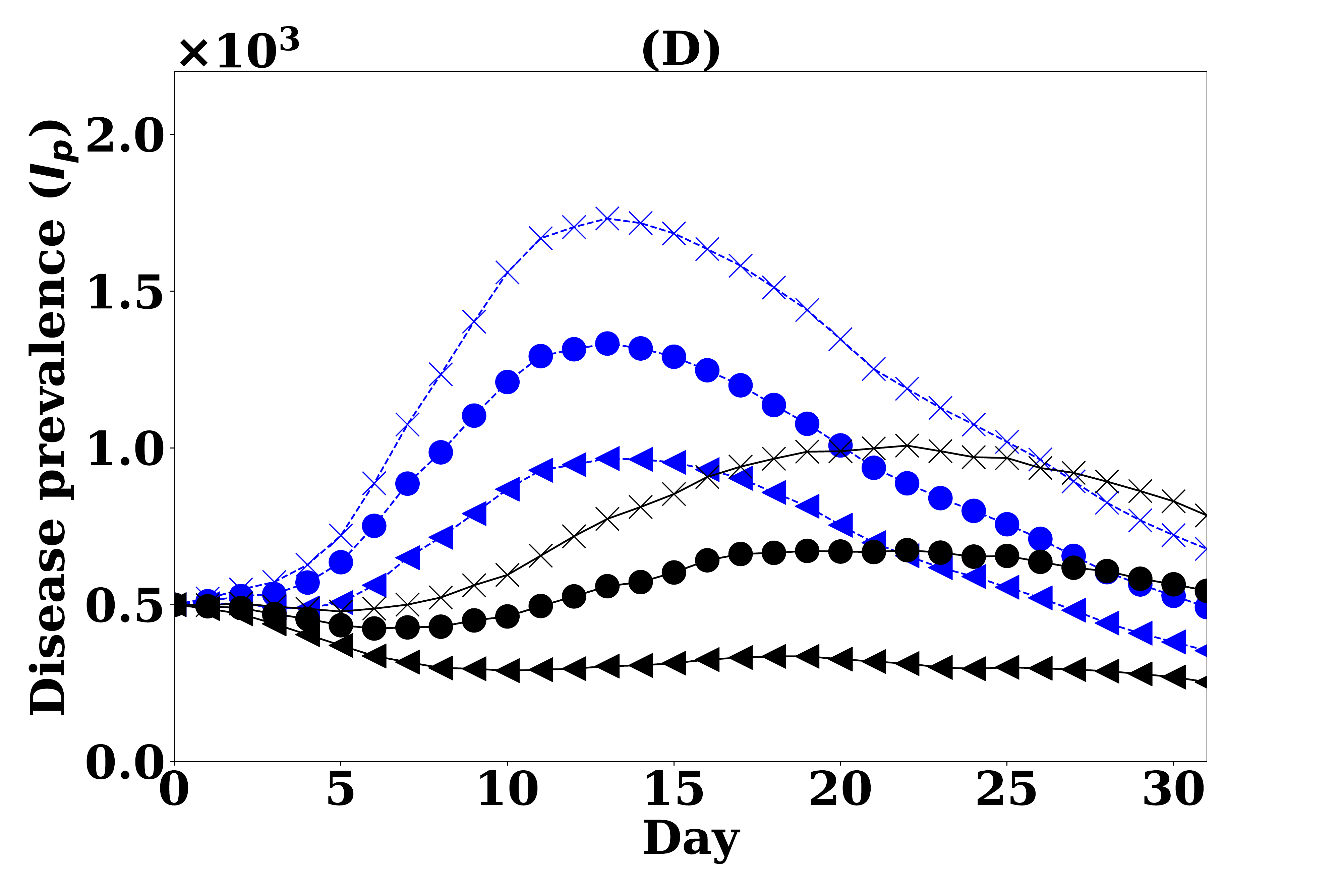}
\caption{Sensitivity analysis of the contact networks with particle decay rates $r_t$: A) homogeneous GDT network, B) homogeneous ADN network, C) heterogeneous GDT network and D) heterogeneous ADN network}
\label{fig:netsens}
\end{figure}

\begin{figure}[h!]
\begin{tikzpicture}
    \begin{customlegend}[legend columns=4,legend style={at={(0.12,1.02)},draw=none,column sep=2ex ,line width=2 pt}, legend entries={DDT, Hom GDT, Hom ADN, Het GDT, Het ADN, $\sigma=0.33$, $\sigma=0.40$, $\tau=4$ days}]
    \addlegendimage{solid,line legend, color=blue}
    \addlegendimage{solid, color=red}   
    \addlegendimage{solid, color=cyan}
    \addlegendimage{color=yellow}
    \addlegendimage{color=black}
    \addlegendimage{only marks,mark=triangle}
    \addlegendimage{only marks,mark=o}
    \addlegendimage{only marks,mark= x }
    \end{customlegend}
 \end{tikzpicture}
\vspace{4ex}
\centering
\includegraphics[width=0.45\linewidth, height=5cm]{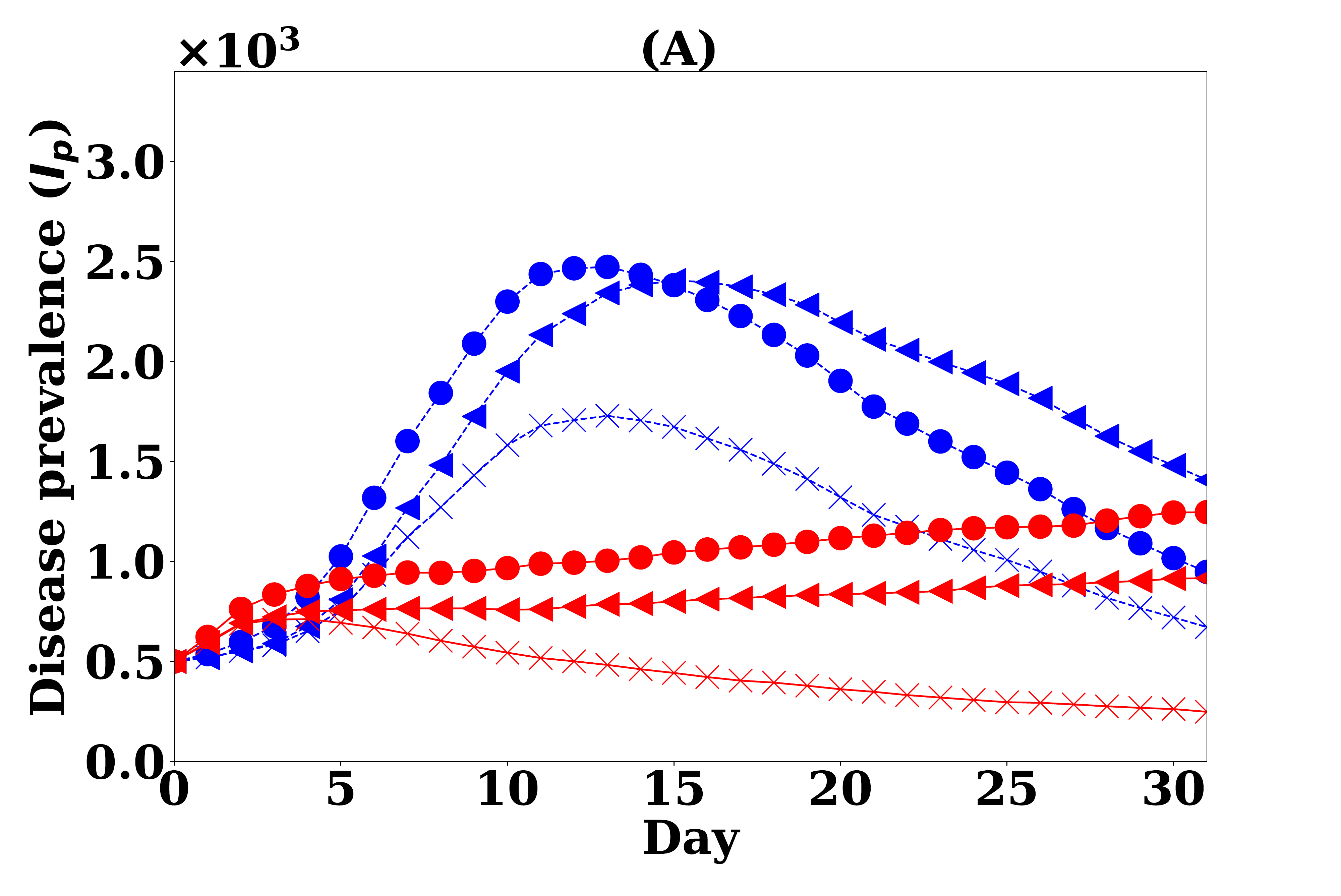}~
\includegraphics[width=0.45\linewidth, height=5cm]{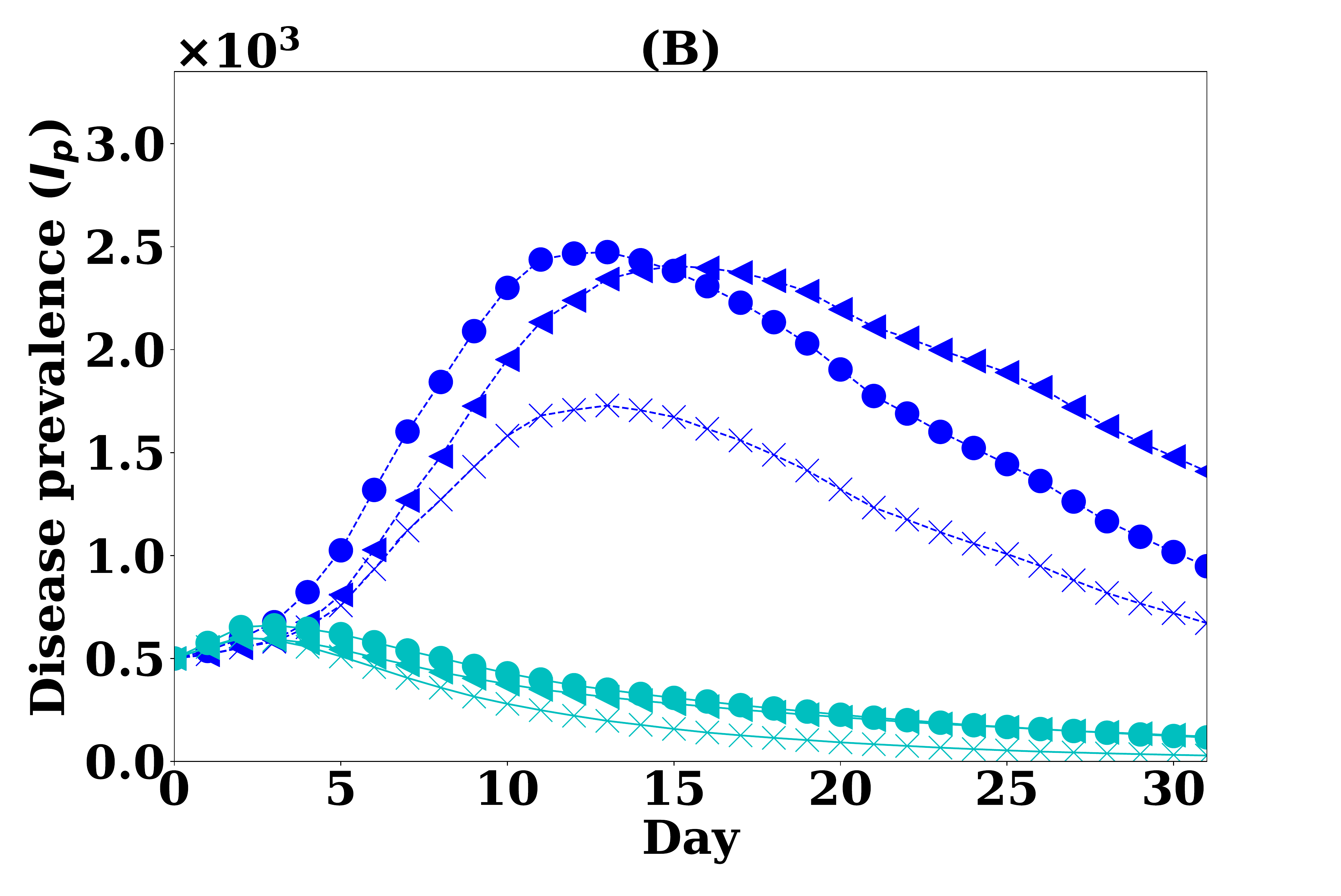}\\
\vspace{1em}
\includegraphics[width=0.45\linewidth, height=5cm]{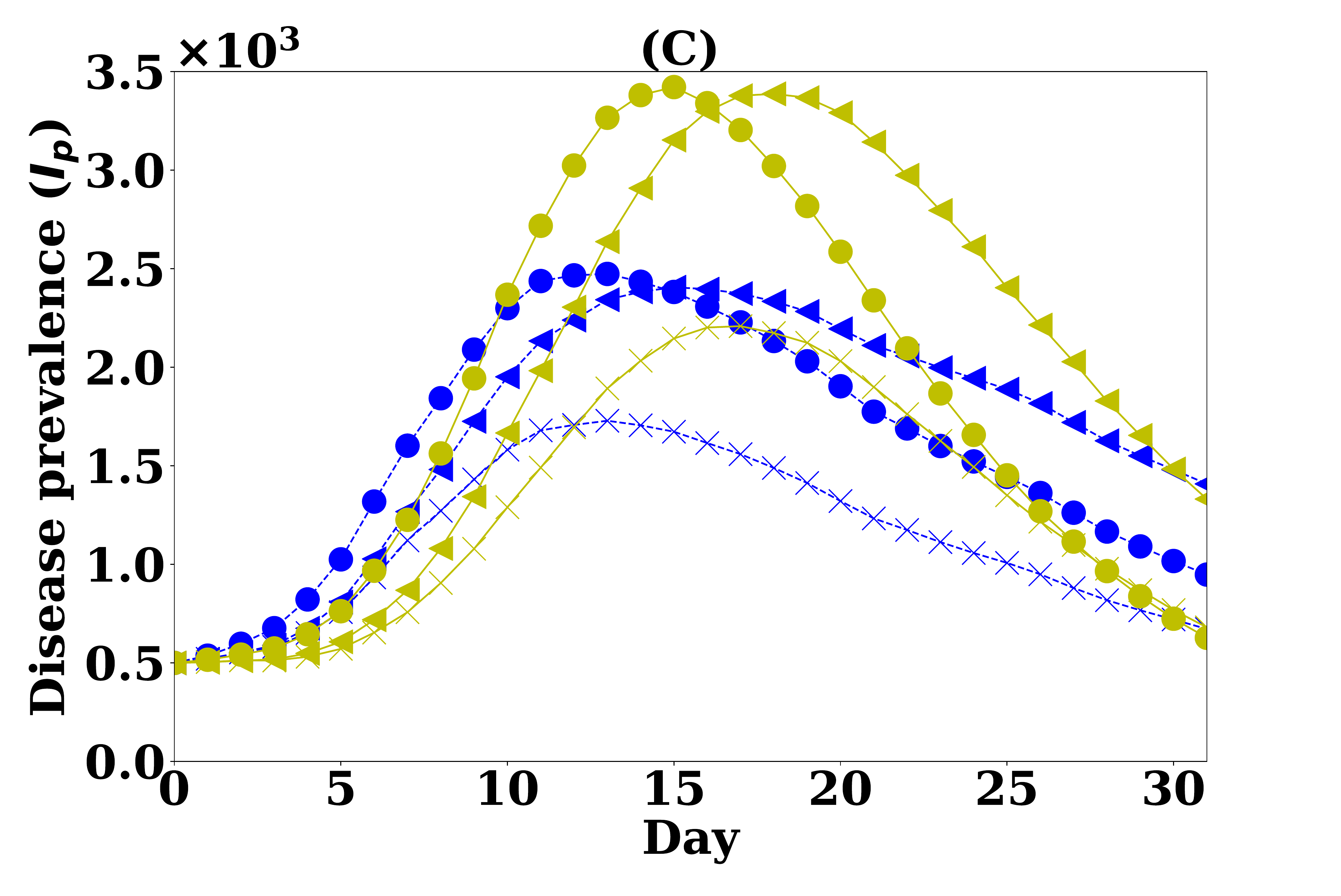}~
\includegraphics[width=0.45\linewidth, height=5cm]{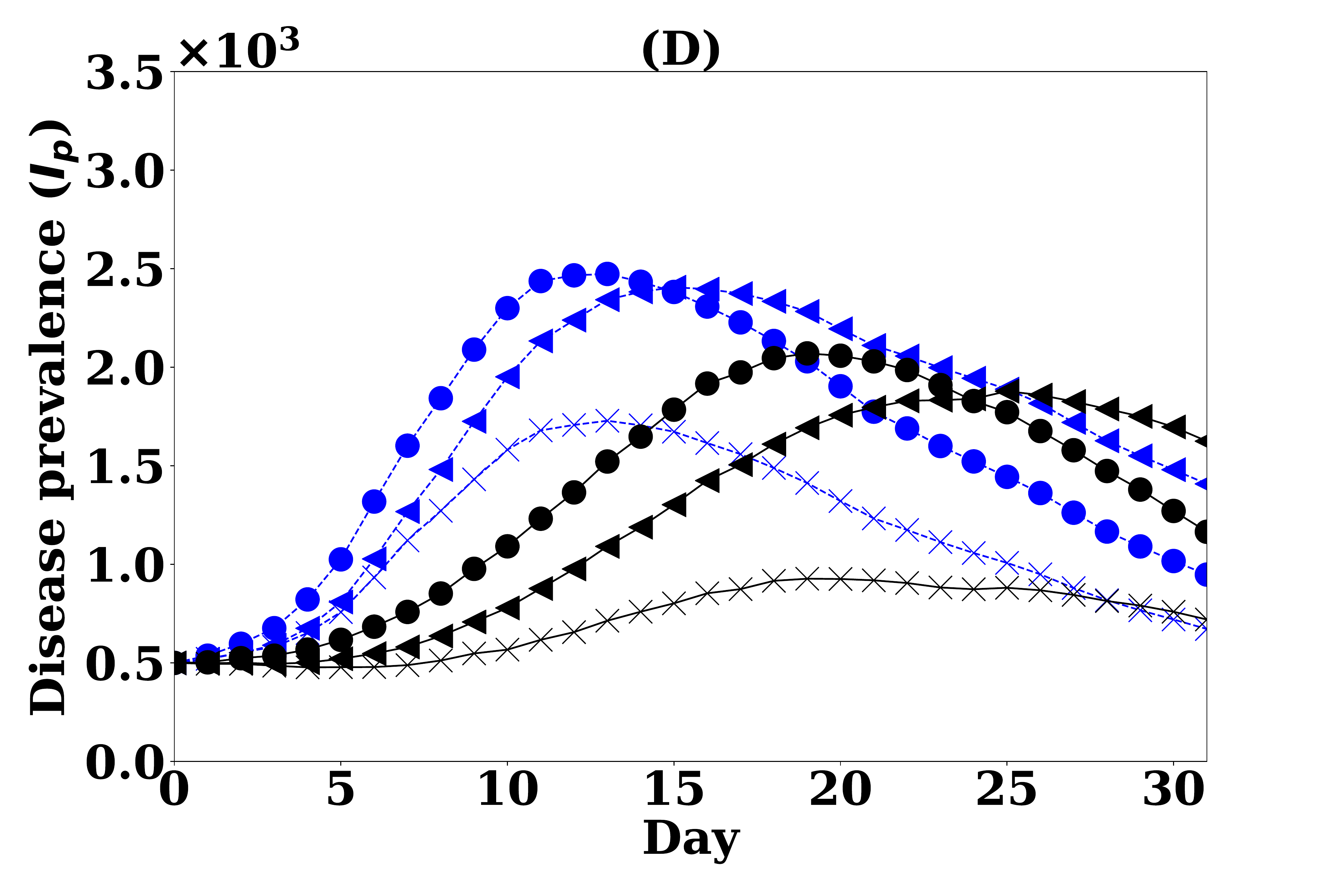}
\caption{Sensitivity analysis of the contact networks with disease parameters infectiousness $\sigma$ and infectious period $\tau$: A) homogeneous GDT network, B) homogeneous ADN network, C) heterogeneous GDT network and D) heterogeneous ADN network}
\label{fig:netdis}
\end{figure}

Now, the model's sensitivity to the various diffusion parameters is studied. The SPDT model is highly sensitive to the particle decay rates $r_t$. Thus, disease spreading is simulated for three different values of $r_t=\{20,40,60\}$ min for all networks and results are presented in Figure~\ref{fig:netsens}. The homogeneous GDT network can model the variation in diffusion dynamics with $r_t$. However, it underestimates the diffusion at all $r_t$. The prediction variations are increased with $r_t$ where the average prediction variation reaches to 70\% with the average accumulative variation of 51\%. The other homogeneous ADN network cannot vary the diffusion with $r_t$. As this network has only direct transmission links. The homogeneous ADN network always has more than 72\% variation in daily prediction for any value of $r_t$. The proposed heterogeneous GDT network shows increasing in daily prediction variation. However, it is still 23\% better than the heterogeneous ADN networks at particle decay rates $r_t=20$ min. The model also maintains the same 32\% improvements in the accumulative prediction variations at $r_t=20$ min comparing to heterogeneous GDT model. Although the prediction variation increases with the raising of particle decay rates, the model still predicts well comparing to other models.

The diffusion dynamics are also influenced by biological factors. This is the properties of contagious items. The model should respond properly for varying the properties in the contagious items. Now, the infectiousness $\sigma$ and infectious period $\tau$ are varied keeping $r_t=60$ min. The results are presented in Figure~\ref{fig:netdis}. For increasing infectiousness $\sigma=0.44$, the disease prevalence gets stronger in all networks. However, the variation in the heterogeneous GDT network is smallest than other networks. The average daily variation is 13\% in the heterogeneous GDT network that less than that is with $\sigma=0.40$. For the other networks, the average daily variations increases with $\sigma=0.40$ and the maximum are 70\% in the homogeneous GDT networks. The similar trends are found for the increase in infectious period $\tau=4$ days. The average daily variation in heterogeneous GDT is 12\%. All others networks show higher average daily variation with a maximum in homogeneous ADN network. The overall responses in heterogeneous GDT follow the trend of DDT network while homogeneous ADN does not have much variation changing biological parameters.

\begin{figure}[h!]
\centering
\includegraphics[width=0.45\linewidth, height=5cm]{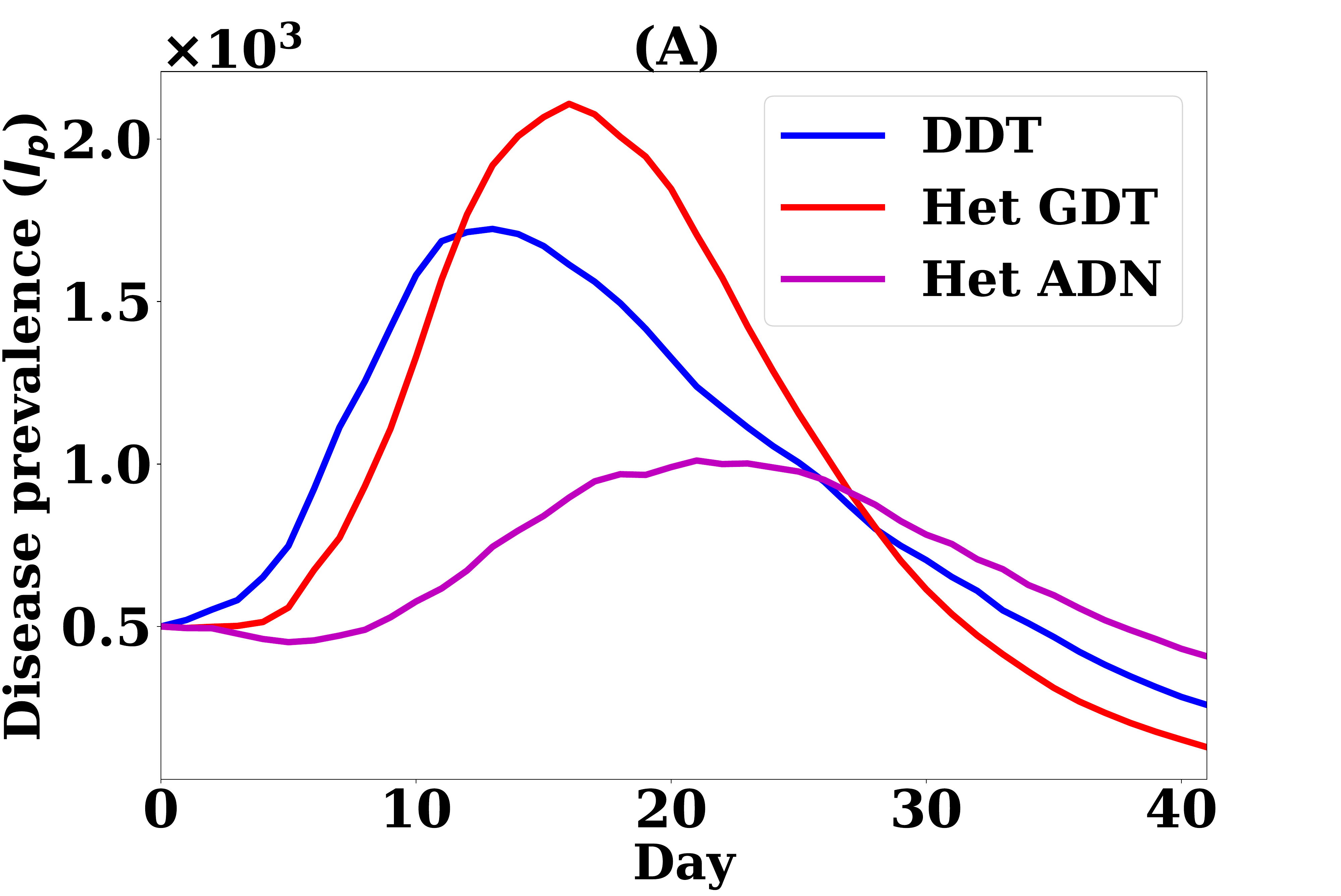}~
\includegraphics[width=0.45\linewidth, height=5cm]{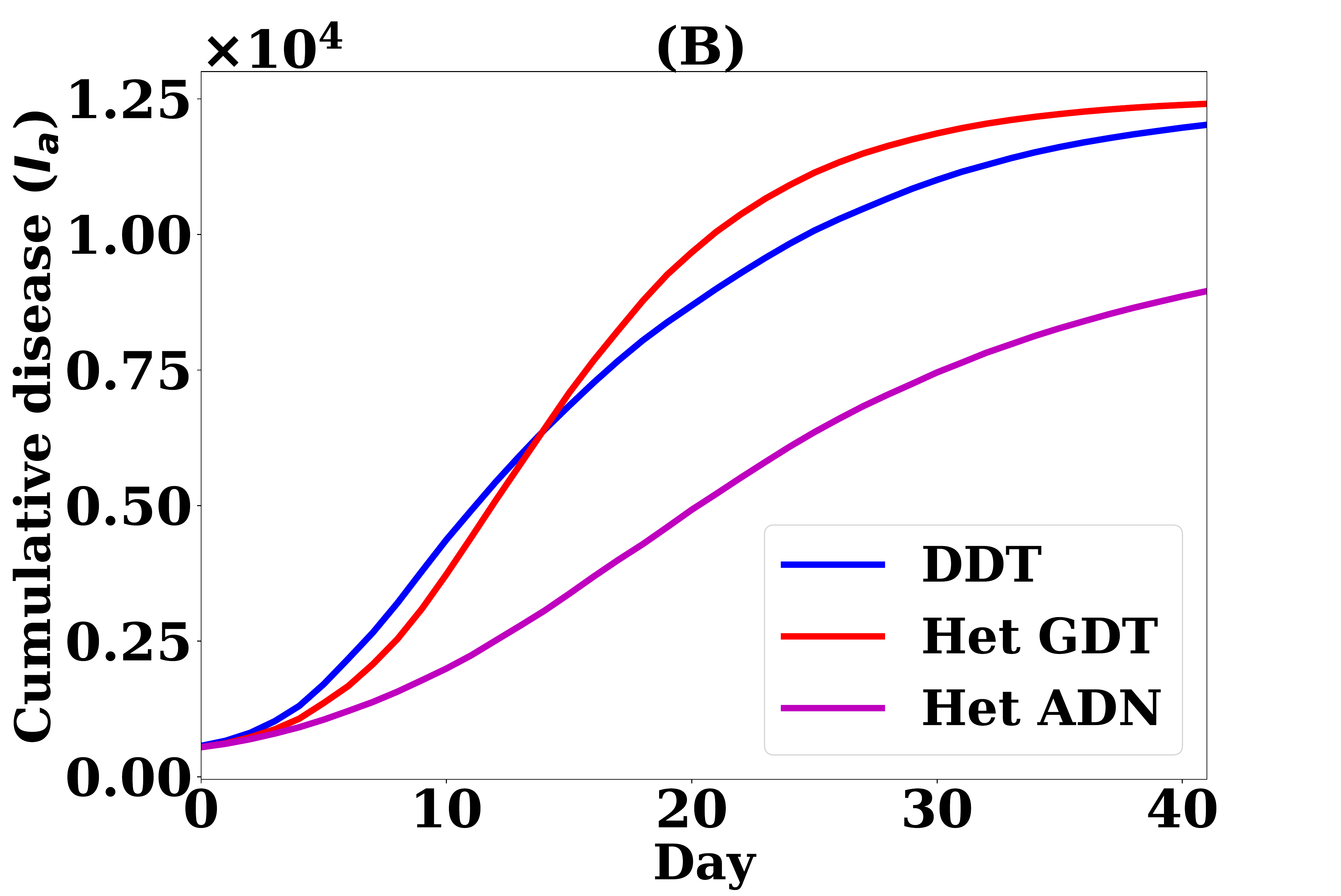}
\caption{Sensitivity of the model with longer observation period: A) disease prevalence dynamics and B) accumulative infection over simulation days}
\label{fig:netszl}
\end{figure}

\begin{figure}[h!]
\begin{tikzpicture}
    \begin{customlegend}[legend columns=5,legend style={at={(0.12,1.02)},draw=none,column sep=2ex,line width=2 pt }, legend entries={Het GDT, Het ADN, 0.364M, 0.50M, 1.0M}]
    \addlegendimage{dash dot,line legend, color=black}
    \addlegendimage{solid, color=black}   
    \addlegendimage{solid, color=red}
    \addlegendimage{solid, color=blue}
    \addlegendimage{solid, color=green}
    \end{customlegend}
 \end{tikzpicture}
\vspace{4ex}
\centering
\includegraphics[width=0.45\linewidth, height=5cm]{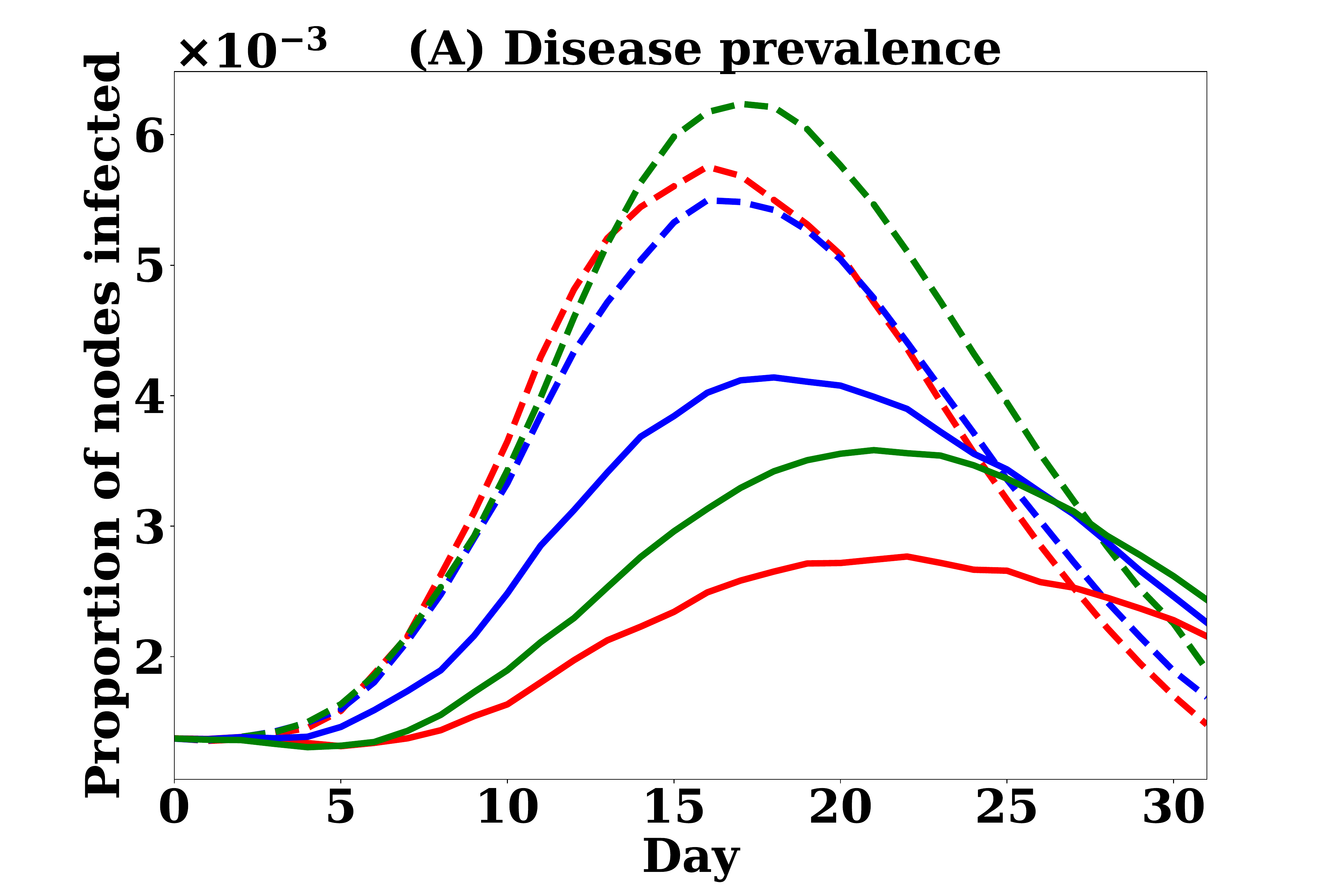}~
\includegraphics[width=0.45\linewidth, height=5cm]{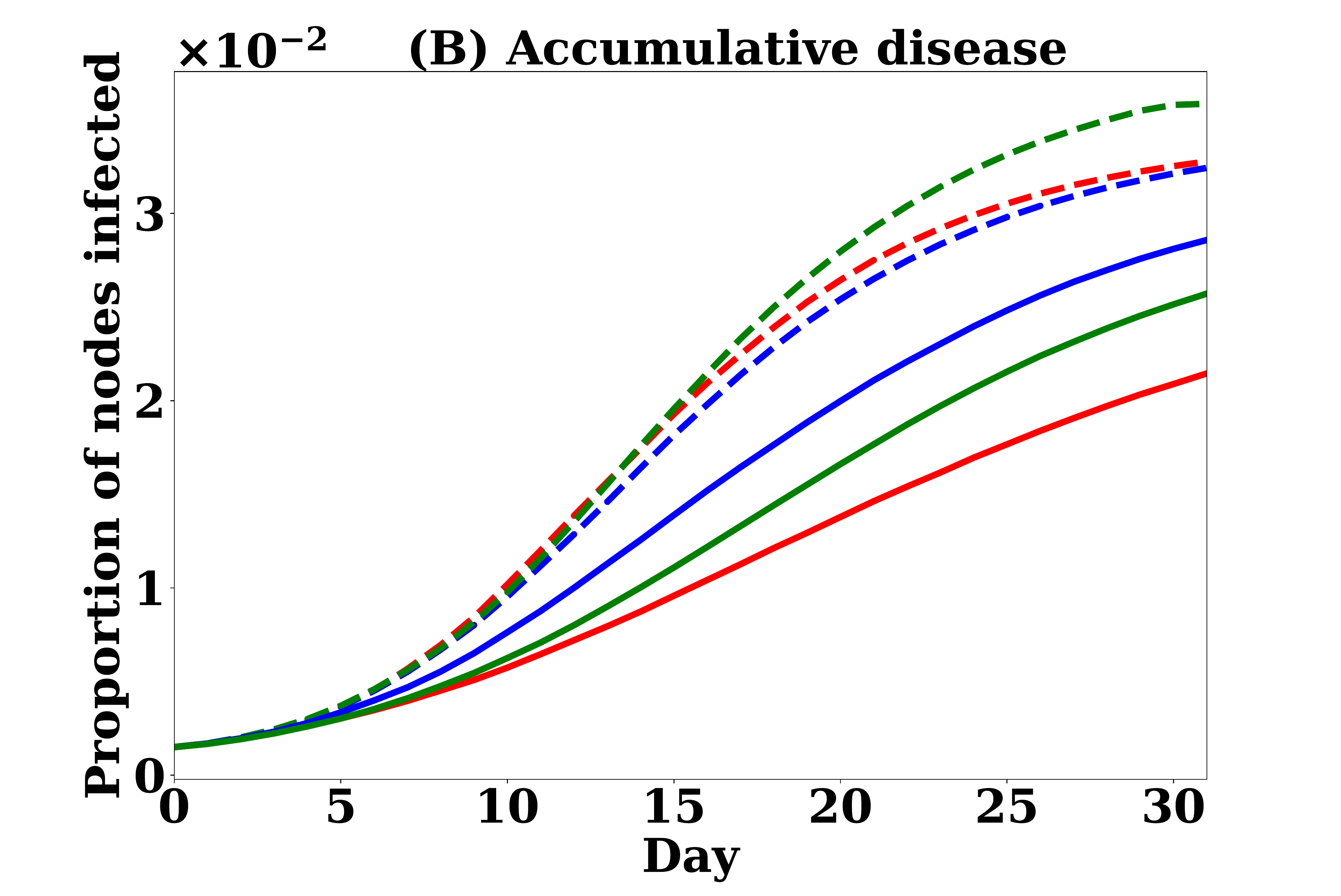}
\caption{Sensitivity of the graph model with large network sizes: A) disease prevalence dynamics and B) cumulative infection over simulation days}
\label{fig:netszn}
\end{figure}

To understand the response of the graph model for larger scale simulation with large network sizes, the simulations are also run on various sizes of the networks. Two dimensions are considered here: increasing the length of the simulation period and number of nodes. In this experiment, only heterogeneous networks are considered as they assume more realistic diffusion dynamics. For this study, a heterogeneous GDT network and a heterogeneous ADN network with 364K nodes are generated for 42 days period. Then, the real contact network DDT is extended to a network of 42 days repeating all links of a day randomly picked up from 32 days and placing to a random day within 33 to 42. The result is presented in the Figure~\ref{fig:netszl}. The diffusion dynamic is consistent in the heterogeneous GDT network. The total infection in the heterogeneous GDT network becomes almost the same at the end of the simulation period. However, the heterogeneous ADN network underestimates the diffusion dynamic and total infection as it is in the previous experiments. The model response is analyzed generating the network with a large number of nodes. For these experiments, the heterogeneous GDT and heterogeneous ADN networks are selected and generated with 0.364M, 0.5M and 1M nodes for 32 days. The simulations are run with the same proportion of seed nodes and results are presented in the Figure~\ref{fig:netszn}. The results show that the heterogeneous GDT network behaves consistently for diffusion dynamic and cumulative diffusion while heterogeneous ADN network shows some variations. This characterizes the limitation of underlying connectivity of GDT network, having the same underlying connectivity to DDT network, comparing to ADN network.   

%% file: 4.5_chap_disc.tex
\section{Discussion}
In this chapter, a novel graph model is developed and analysed to support the study of SPDT diffusion dynamics. The developed SPDT graph can capture both direct and indirect interactions for modelling the spreading of contagious items. The synthetic dynamic contact networks generated by this graph model is capable of capturing SPDT dynamics, applying reinforcement for repetitive interactions and heterogeneous propensity to engage in interactions. This chapter has demonstrated how the model can be fitted to empirical data from a large social networking application and its ability to reproduce both network properties and diffusion dynamics of empirical contact networks. The developed model differs from the real contact graph DDT where the available links have been repeated to the missing days to fill up the data gap. Thus, nodes in the DDT network repeatedly contact the same neighbour nodes and could not grow their contact set sizes. This is solved in the developed model. The developed graph model is up to 30\% more efficient to generate the diffusion dynamics of the real SPDT contact networks. Analysis of sensitivity to the SPDT diffusion model parameters has shown that the SPDT graph model responds similarly to that of the real SPDT graph. The stability of the model is analysed by various simulations settings. This has shown that the graph model developed are consistent to the  model parameters under extreme conditions.

%% file: 5_chap_spdtchar.tex
\chapter{Indirect Link Potential}

The interaction mechanism among individuals defines the network properties. These properties, in turn, influence the diffusion dynamics. In the current diffusion dynamic studies, the impacts of network properties on diffusion dynamics are investigated based on the direct transmission links. However, transmission links in the SPDT model are created for delayed indirect interactions along with the direct interactions. In Chapter 3, it has been observed that the inclusion of indirect links significantly amplifies the diffusion dynamics in SPDT model compared to that in SPST model. Chapter 3 also studies the characteristics of SPDT diffusion dynamics at the network level. This chapter now investigates the enhancements in the diffusion by the SPDT model at the node level. In networks, a local contact structure around a node is created based on how the node is connected with its neighbouring nodes. This shapes some diffusion phenomena such as whether a node can be super-spreader, how easier emergence of a diffusion process is in a contact network etc. It would be interesting to know what is the potential of indirect link for creating diffusion phenomena. In the SPDT model, some nodes do not have direct transmission links when they are infectious. These nodes are not visible disease spreaders in the SPST model and are called as \textit{hidden spreaders} in the SPDT model. The hidden spreaders highlight the inherent properties of the SPDT model. It is critical to understand the behaviour of these nodes as they shape the diffusion on contact networks. In addition, the inclusion of indirect links changes the position of the nodes in the networks and the spreading potentialities of nodes get stronger. As a result, the emergence of disease from a node becomes easier and a disease easily invades a contact network. This chapter examines to what extent the spreading potentialities of nodes increase and invading the networks become easier by investigating diffusion at node level.

\input{5.1_chap_intro.tex}
\input{5_2_chap_method.tex}

\input{5_3_chap_results.tex}

\section{Discussion}
This chapter analyses the importance of indirect transmission links in various diffusion phenomena. Clearly, indirect transmission links have at least the same potential in causing spreading phenomena to direct transmission links. In addition, the individual level interactions patterns have a strong influence on diffusion dynamics. The inclusion of indirect links bring changes in the local contact structures and changes in the position of the nodes in the networks and importance in terms of spreading potential. The analysis shows that the spreading potential of nodes in the SPST network is boasted in the SPDT networks. They create large outbreak sizes in the SPDT model. Interestingly, it is found that the hidden spreaders can also be super-spreaders although they are assumed to have zero spreading potential in the SPST models. Moreover, the networks with only hidden spreaders can also cause outbreaks with a significant number of infections. The indirect links have also strong impacts on controlling diffusion. When the indirect links are considered, the diffusion is easily stopped with a small number of nodes to be vaccinated. However, neglecting indirect transmission links creates a huge additional vaccination cost. 

%% file: 5.1_chap_intro.tex
 \section{Introduction}
The diffusion dynamics on contact networks are influenced by the underlying contact properties. Thus, a wide range of investigation has conducted to understand the correlation among various contact properties: e.g., repetitive behaviours of contacts, clustering, burstiness, community structures, temporal properties, and the dynamics of diffusion processes. Most of these works assume that the interacting individuals are co-located in the same physical or virtual space at the same time to create a disease transmission link~\cite{holme2015modern,machens2013infectious,huang2016insights}. In the SPDT model, however, the transmission links are created between two individuals for direct interactions when both individuals are present at a location; and/or indirect links when the infected individual has left, but the susceptible individual is still present in the location or arrives later on. In Chapter 3, the overall enhancement in diffusion dynamics for including the indirect transmission links are studied. This chapter investigates the impacts of indirect links at the individual levels and to emerging disease in the contact networks.

The spreading potential of a node in a network depends on how it is connected with its neighbours (local social structure) and its position within the network. Based on the direct transmission links, the impacts of the local structure are frequently studied in the literature. However, nodes in the SPDT model are connected through the indirect links along with other direct links. Thus, the local contact structure properties are different due to forming with direct links and indirect links. It is clearly of interest to know the impacts of the addition of these indirect links and characterise these impacts. The spreading potential of a node is quantified by the outbreak size when a diffusion process starts from the node and, alternatively, it can be said that it is the measurement of node's position importance and its local contact structure's influences. In the SPDT diffusion model, there are some nodes that do not have any direct transmission links to the neighbouring nodes when they are infectious. These nodes are invisible in the SPST networks as they do not have direct links and are called \textit{hidden spreaders} in the SPDT model. As the hidden spreaders are connected with other nodes only through the indirect links, the spreading potential of these nodes can indicate the potential of indirect links and local contact structure made by indirect links. Thus, in the first experiment, the outbreak sizes are collected from a set of seed nodes with different proportion of hidden spreader nodes and non-hidden spreader nodes and are analysed. The variations in the outbreak sizes for increasing hidden spreader nodes indicate the ability of indirect links to sustain the diffusion dynamics and hence the potential of indirect links.

The inclusion of indirect links also increases the importance of nodes to drive diffusion on the contact network. Hence, the opportunities for emerging disease on the network increase, i.e. the networks become more vulnerable to an invasion by a disease. Therefore, understanding the vulnerability of a social contact network against disease is incomplete and may be misleading when direct links only are considered. The enhancement in the importance of nodes in causing diffusion is studied through analysing the spreading potential. To do this, nodes are first classified according to contact set sizes, the number of neighbour nodes they have contacted during an observation period through direct links, into groups. Then, the spreading potentials are studied for nodes of each group on both SPST and SPDT networks to analyse the enhancement of node's importance. The spreading potential of hidden spreaders is also investigated. Finally, the increase in invasion strength in SPDT model due to including indirect transmission links is analysed applying diffusion control strategies, i.e. vaccination strategies. The increased rates of vaccination in the SPDT model compared to the SPST model for stopping disease indicates the strength of the invasion in the networks. The above issues are studied through simulating airborne disease spreading on real dynamic contact networks (DDT and DST) and synthetic contact networks (GDT and GST) generated by the developed SPDT graph model. This chapter's contributions are as follows:
\begin{itemize}
\item Analysing the potential of indirect transmission links
\item Investigating the spreading potential of hidden spreaders
\item Understanding the spreading potential of super-spreaders with indirect links 
\item Analysing the diffusion emergence potentiality in the SPDT model
\end{itemize}

%% file: 5_2_chap_method.tex
\section{Methodology}
Intensive simulations of airborne disease spreading are conducted on dynamic contact networks to explore the impacts of indirect transmission links in SPDT diffusion. The dense SPDT contact network (DDT network introduced in Chapter 3) constructed with empirical location updates of Momo users is re-used in this chapter. The DDT network is considered to be realistic since it includes links of users every day. The DDT network contains 364K nodes which interact with each other over the 32 days and creates 6.46M SPDT links. The corresponding SPST network (DST network) is obtained excluding the indirect transmission links from the DDT network. These networks have been created from the location updates of Momo users from Beijing city. To verify the results, another synthetic SPDT network (GDT network) is used which is generated using the developed SPDT graph model in Chapter 4. The GDT network also contains 364K nodes that interact with each other over 32 days. The corresponding SPST network (GST network) are obtained excluding the indirect transmission links from the GDT network.

For propagating disease on these selected contact networks, this chapter also adopts the generic Susceptible-Infected-Recovered (SIR) epidemic model. If a node in the susceptible compartment receives a SPDT link from a node in the infectious compartment, the former will receives the exposure $E_l$ of infectious pathogens for both direct and indirect transmission links according to the following equation developed in the Section 3.3.2
\begin{equation}
E_l =\frac{gp}{Vb^2}\left[b\left(t_i-t_s^{\prime}\right)+ e^{bt_{l}}\left(e^{-bt_i}-e^{-bt_l^{\prime}} \right)+e^{bt_{s}} \left(e^{-bt_l^{\prime}}-e^{-bt_s^{\prime}} \right) \right]
\end{equation}
If the susceptible node receives $m$ SPDT links from infected individuals during an observation period, the total exposure $E$ is 
\begin{equation}\label{eq:liexpo}
E=\sum_{k=0}^{m}E_{l}^{k}
\end{equation}
where $E_{l}^{k}$ is the received exposure for $k^{th}$ link. The probability of infection of causing disease can be determined by the dose-response relationship defined as 
\begin{equation}\label{eq:liprob}
P_I=1-e^{-\sigma E}
\end{equation}
where $\sigma$ is the infectiousness of the particles generated by the infected individuals~\cite{fernstrom2013aerobiology}. If the exposed susceptible individual moves to the infected compartment with the probability $P_I$, he continues to produce infectious particles over its infectious period $\tau$ days until they enter the recovered state. The value of $\tau$ is randomly chosen in the range $3-5$ days maintaining a mean value of 3 days. The parameters of Equation~\ref{eq:liexpo} are the same of previous chapters. The value of $b$ in Equation~\ref{eq:liexpo} is determined as $b=\frac{1}{60 r}$ where $r$ is particles removal time randomly chosen from [5-240] minute given a median particles removal time $r_t$.

The infected nodes in the SPDT model are divided into two groups based on the SPDT links they form during their infectious periods: 1) nodes that have direct and may have indirect transmission links; and 2) nodes that have only indirect transmission links, called \textit{hidden spreaders}. A hidden spreader has zero disease transmission capability in the SPST network but can spread disease in the SPDT network. On the other hand, non-hidden spreaders can transmit disease in both networks. These two types of nodes are applied to characterise the potential of indirect transmission links. For finding a set of hidden spreaders, the SPDT links of first 5 days in both the DDT and GDT networks are analysed and separated the nodes that has no direct links in the first 5 days, where it is assumed that the infected nodes stay infectious for 5 days. This provides a hidden spreader set of 6.12K nodes for GDT network and a set of 10.23K hidden spreader nodes for DDT network. The other nodes outsides of these sets are non-hidden spreaders.

Nodes of the selected networks are also classified based on their contact set sizes which are built by the number of neighbour nodes they have contacted through direct links during the first five days of traces. The purpose of this classification is to understand how the spreading potential of nodes from each group is increased by the inclusion of indirect transmission links and hence the emergence of diffusion is facilitated. Thus, the general sense is used to set the class range. The defined classes are 1) low degree nodes - having 1 to 2 neighbours, 2) average degree nodes - having 3 to 10 neighbours, 3) high degree nodes - having 11 to 20 neighbours and  4) hub nodes - having more than 20 neighbours. The analysis of DDT and GDT networks provide the number of nodes in each class presented in Table~\ref{tb:ndclass}.

\begin{table} 
\caption{Number of nodes in various nodes classes}
\label{tb:ndclass}
\vspace{1em}
\centering

\begin{tabular}{|c|c|c|c|c|}
\hline
Networks &  Low degree & Average degree & High degree & Hub nodes  \\ \hline
DDT &  23K  & 210K & 24K & 7K              \\ \hline
GDT & 16K & 196K & 35K & 12K \\ \hline
\end{tabular}

\end{table}

Vaccination strategies are applied to understand the strength of invasion of a network by the disease due to the inclusion of indirect links. The implications of vaccinating nodes in a diffusion network are to hinder the spreading paths of diffusion. Therefore, the required vaccination rate can indicate the strength of indirect links to emerging disease and invade a network. This also reveals the potency of the current vaccination strategies for SPDT diffusion model. The vaccination strategies are applied for two types of disease spreading scenarios: preventive vaccination, hindering future outbreak of a disease, and post-outbreak vaccination, containing disease outbreaks. For this study, two popular vaccination strategies are considered: mass vaccination and ring vaccination~\cite{takeuchi2006effectiveness,kucharski2016effectiveness}. In the mass vaccination, a proportion of nodes is vaccinated to stop possible disease outbreaks. Nodes are randomly chosen to be vaccinated in this strategy. On the other hand, the ring vaccination is implemented by vaccinating a proportion of susceptible neighbour nodes of infected nodes to stop further spreading of infectious disease. Both vaccination strategies are studied to understand the enhancement in diffusion emergence and invasion strength. The nodes form both DST and GST networks are selected based on the direct transmission links for vaccination and then the diffusion dynamics with these selected nodes are studied in both the SPST and SPDT networks. Then, the differences in outbreak sizes between SPDT and SPST model for same vaccination rate reflect the contribution of indirect links.

%% file: 5_3_chap_results.tex
\section{Indirect link potential analysis}
This section explores and analyses the potential of indirect transmission links through simulations of various disease spreading scenarios. 

\subsection{Transmission ability of indirect link}
This experiment studies the potential of indirect transmission links by analysing the spreading potential of a set of nodes. First, final outbreak sizes are obtained simulating airborne disease spreading from a set of 500 seed nodes on both the real networks (DDT and DST) and synthetic networks (GDT and GST). The selected seed nodes have direct transmission links, i.e. these nodes are selected from the non-hidden spreader nodes set. The outbreak sizes obtained represent the spreading potential of 500 seed nodes without hidden spreaders. Then, seed nodes are selected in such a way that $P$ proportion of seed nodes are hidden spreaders and $1-P$ proportion are non-hidden spreaders. By increasing $P$, the spreading of disease will be reduced in the SPST networks as hidden spreaders have zero disease transmission ability. On the other hand, the spread of disease can be sustained for SPDT networks depending on the potential of the indirect transmission links. Comparison of the spreading speed at different values of $P$ and the speed at $P=0$ indicates the transmission ability of indirect links. In addition, the disease spreading in SPST networks will be zero at $P=1$ but SPDT networks still show outbreaks and the outbreak sizes represent the maximum strength of indirect transmission links.

\begin{figure}[h!]
\begin{tikzpicture}
    \begin{customlegend}[legend columns=6,legend style={at={(0.12,1.02)},draw=none,column sep=2ex,line width=1.5pt,font=\small }, legend entries={ P=0.0, P=0.2, P=0.4, P=0.6, P=0.8, P=1.0}]
    \addlegendimage{solid, color=blue}
    \addlegendimage{color=red}
    \addlegendimage{color=green}
    \addlegendimage{color=cyan}
    \addlegendimage{color=yellow}
    \addlegendimage{color=black}
    \end{customlegend}
 \end{tikzpicture}
\centering
\includegraphics[width=0.48\linewidth, height=5.5cm]{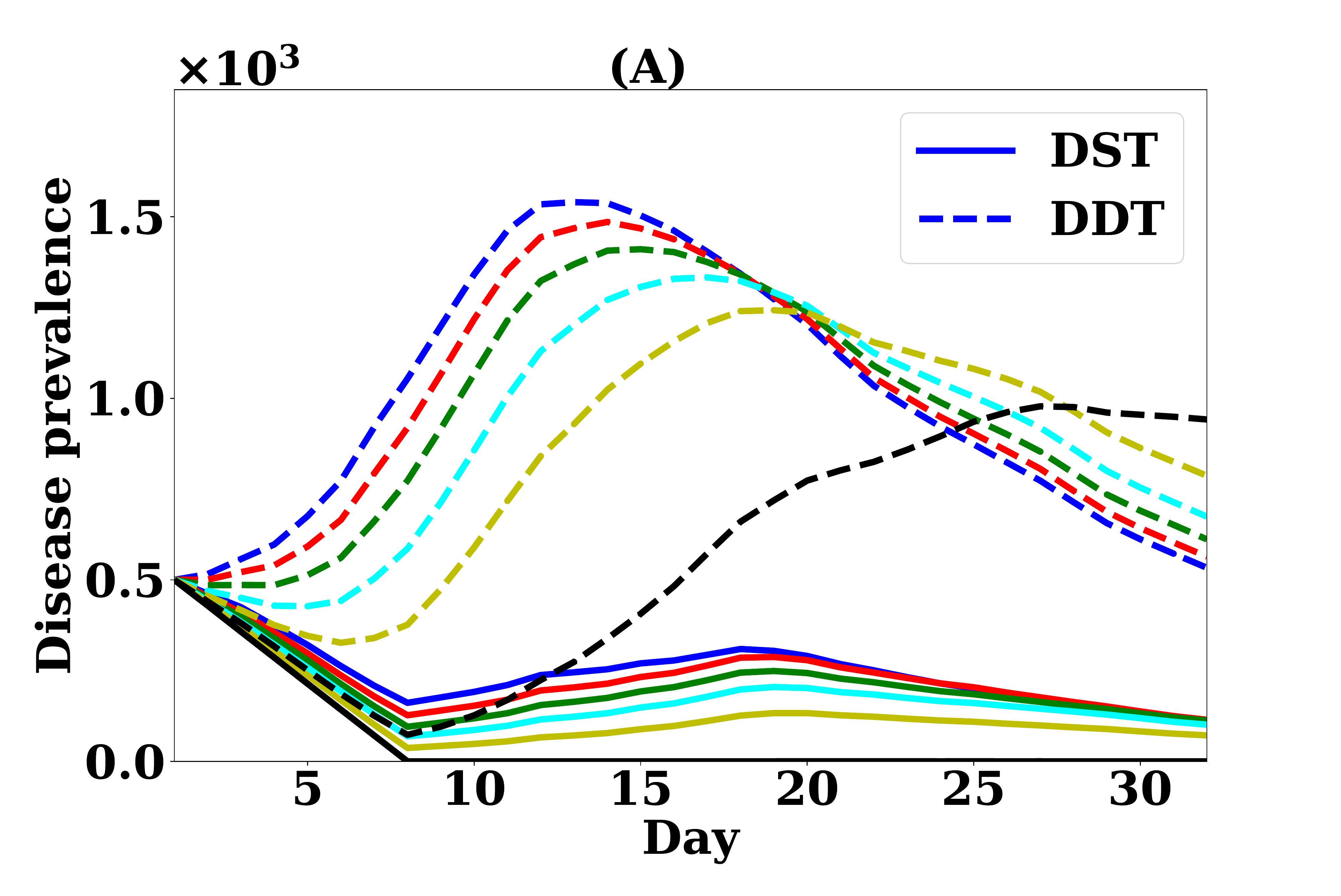}~
\includegraphics[width=0.48\linewidth, height=5.5 cm]{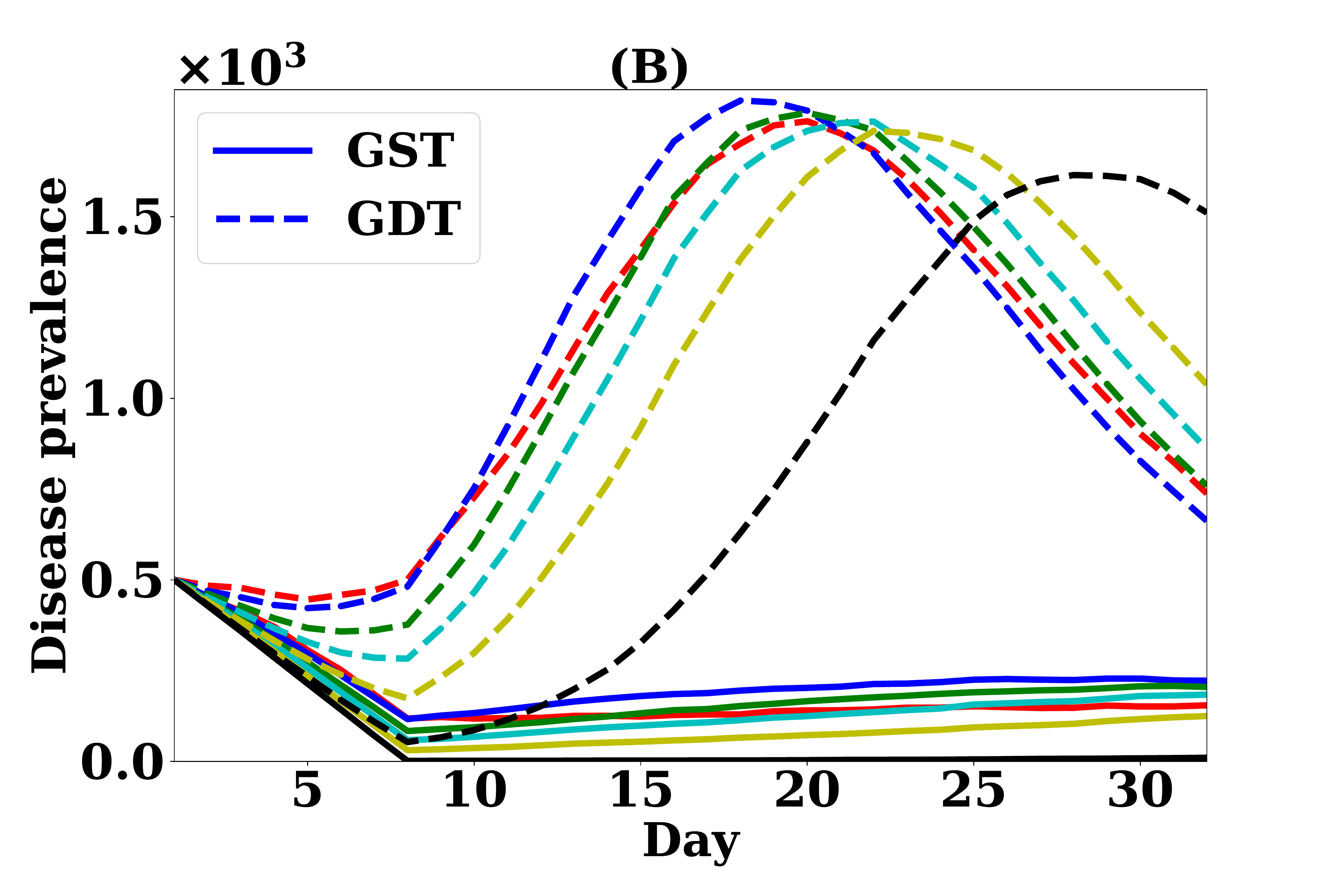}\\ 
\includegraphics[width=0.48\linewidth, height=5.5 cm]{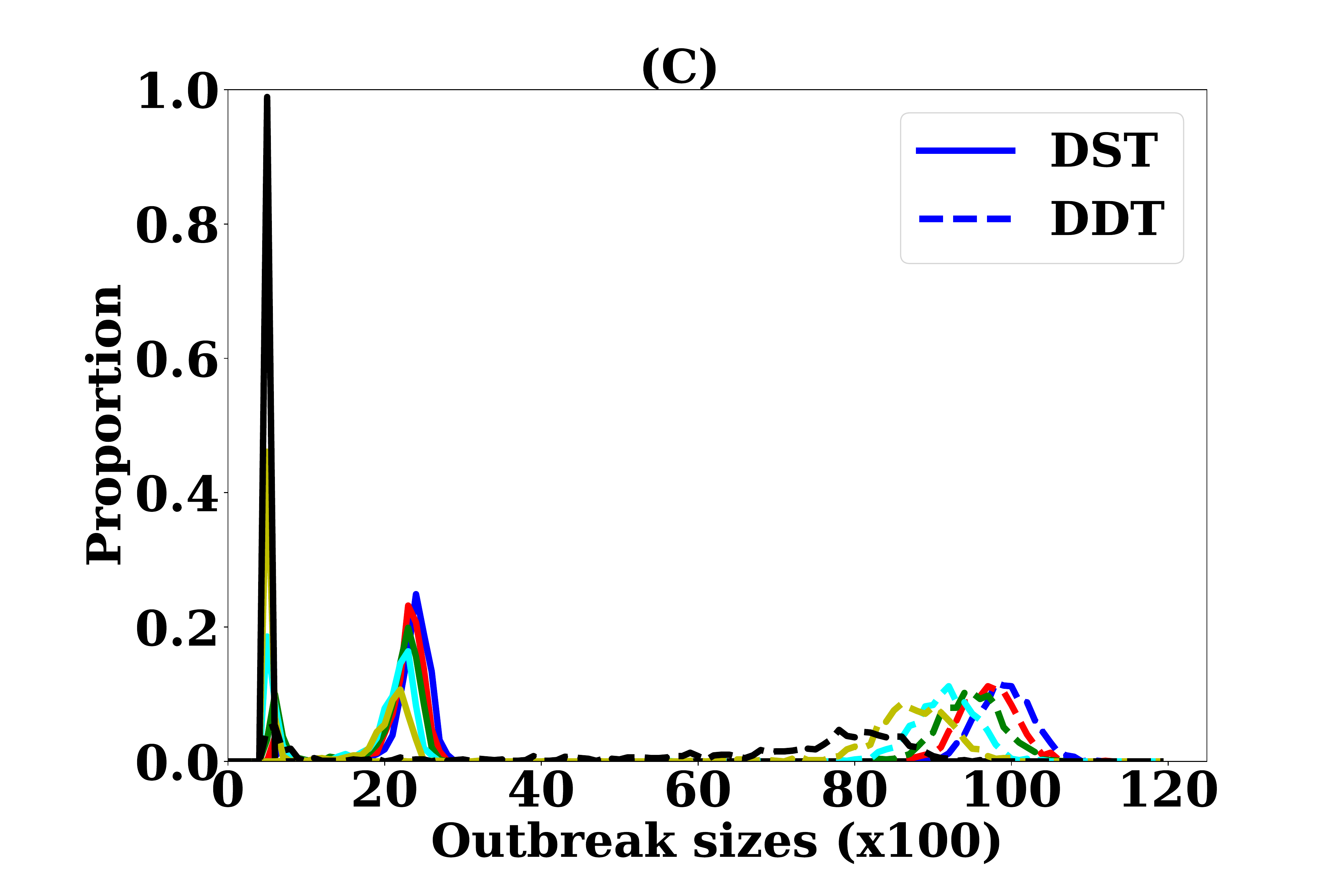}~
\includegraphics[width=0.48\linewidth, height=5.5 cm]{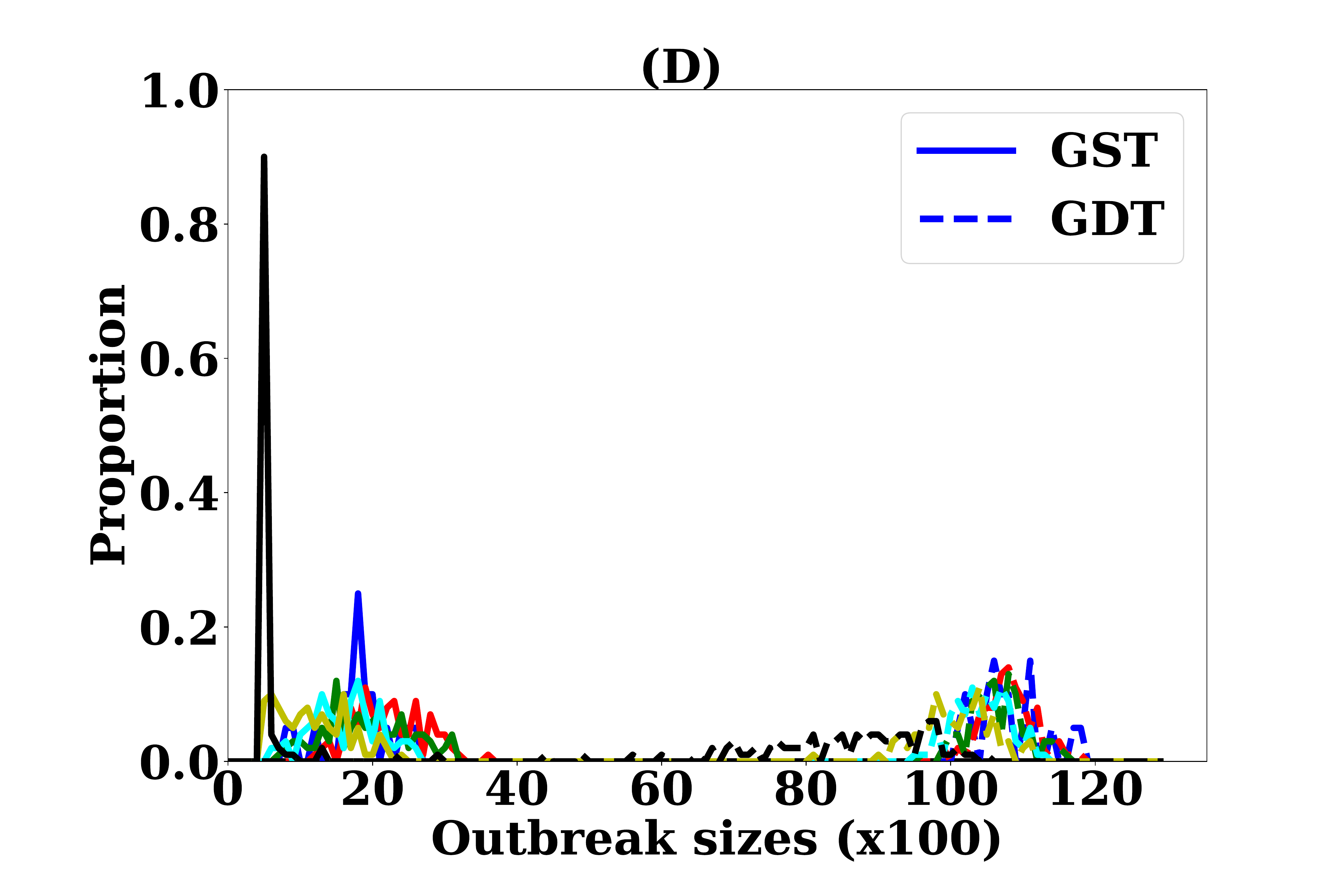}
\caption{Diffusion dynamics for various percentage of hidden spreaders $P$ in the seed nodes on real contact networks (left column) and synthetic contact networks (right column): (A, B) disease prevalence dynamics and (C, D) distribution of generated outbreak sizes}
\label{fig:ipdp}
\vspace{-0.5em}
\end{figure}

\begin{figure}[h!]
\begin{tikzpicture}
    \begin{customlegend}[legend columns=6,legend style={at={(0.12,1.02)},draw=none,column sep=2ex,line width=1.5pt,font=\small }, legend entries={P=0.0, P=0.2, P=0.4, P=0.6, P=0.8, P=1.0}]
    \addlegendimage{solid, color=blue}
    \addlegendimage{color=red}
    \addlegendimage{color=green}
    \addlegendimage{color=cyan}
    \addlegendimage{color=yellow}
    \addlegendimage{color=black}
    \end{customlegend}
 \end{tikzpicture}
\centering
\includegraphics[width=0.48\linewidth, height=5.5cm]{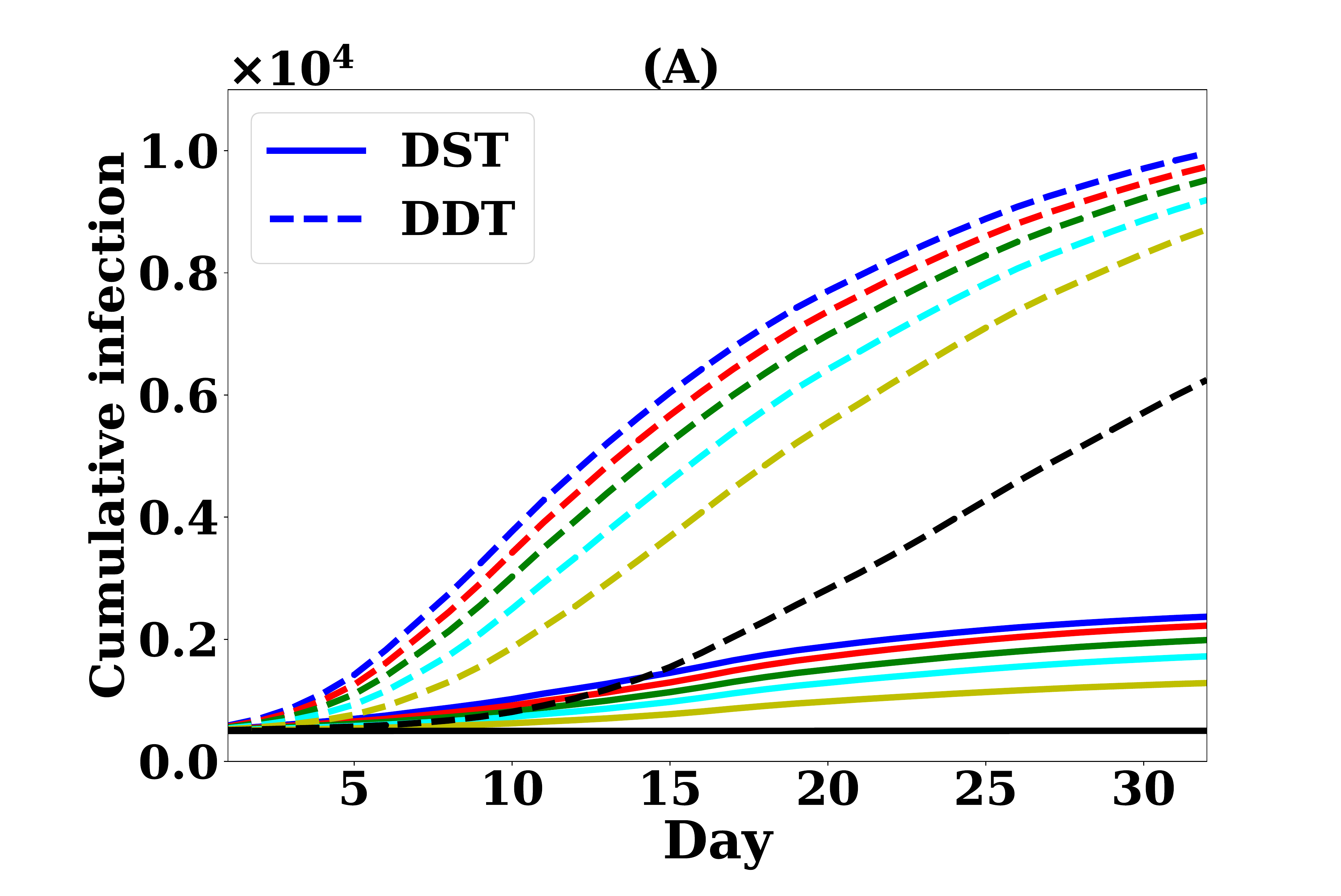}~
\includegraphics[width=0.48\linewidth, height=5.5 cm]{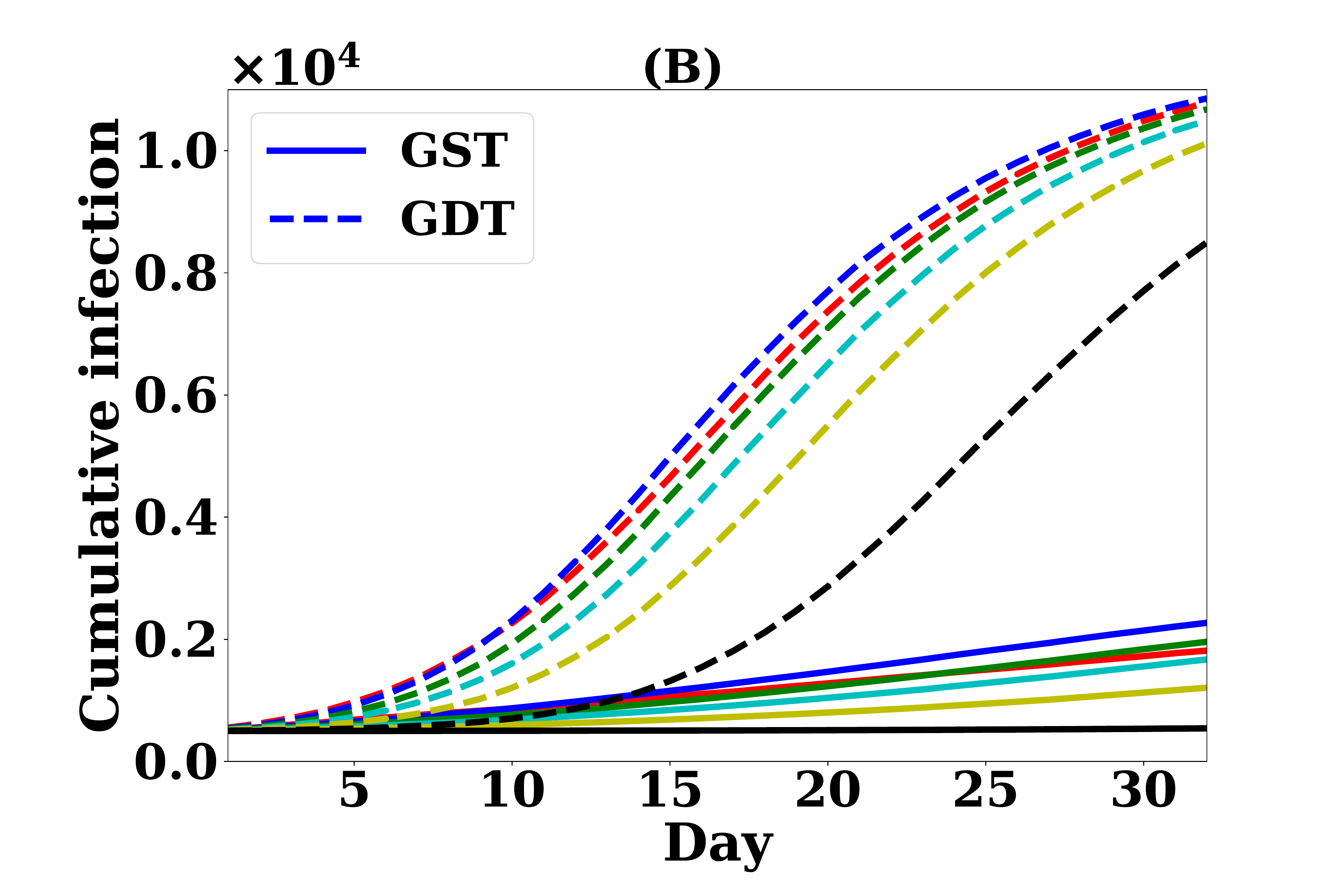}\\
\includegraphics[width=0.48\linewidth, height=5.5 cm]{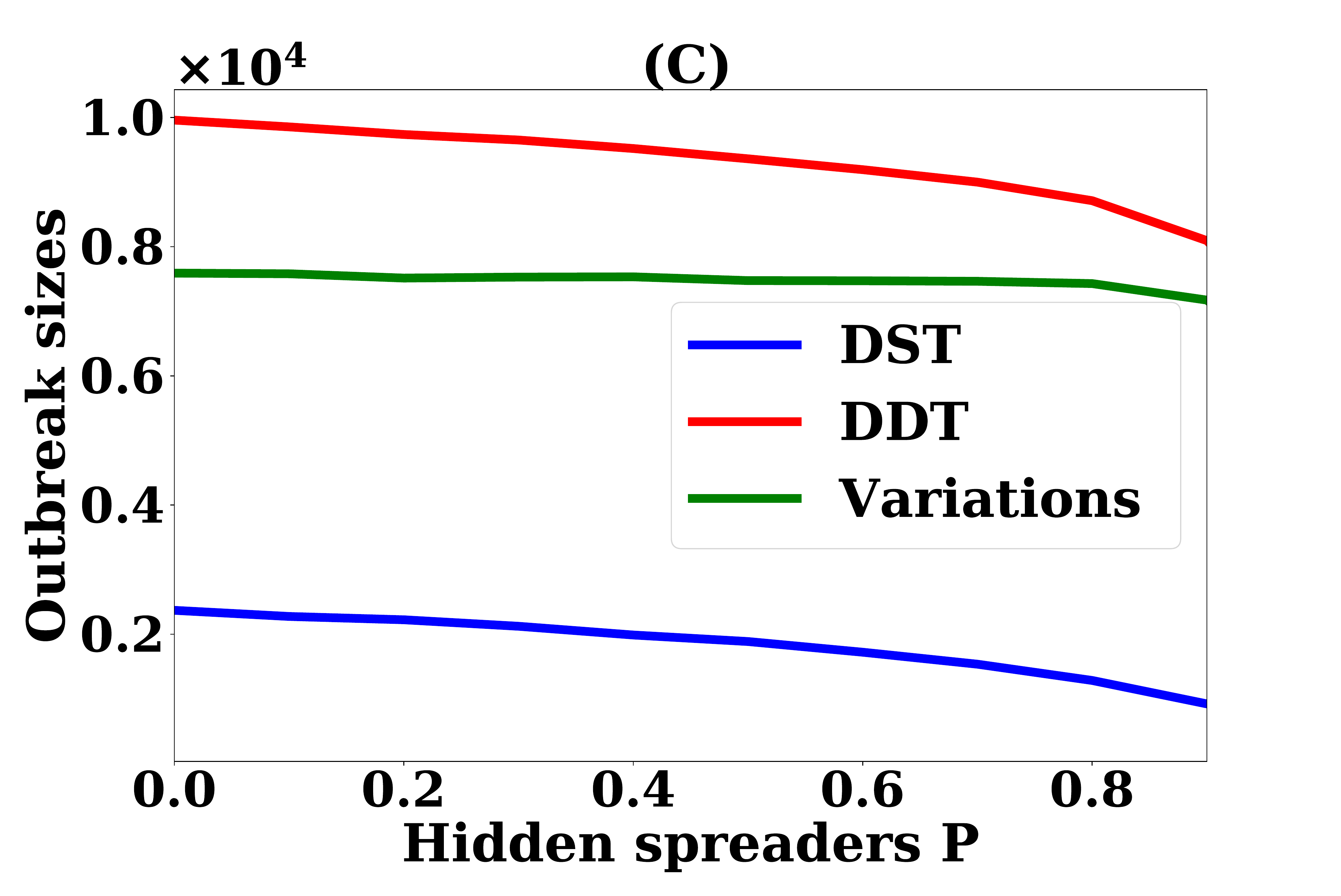}~
\includegraphics[width=0.48\linewidth, height=5.5 cm]{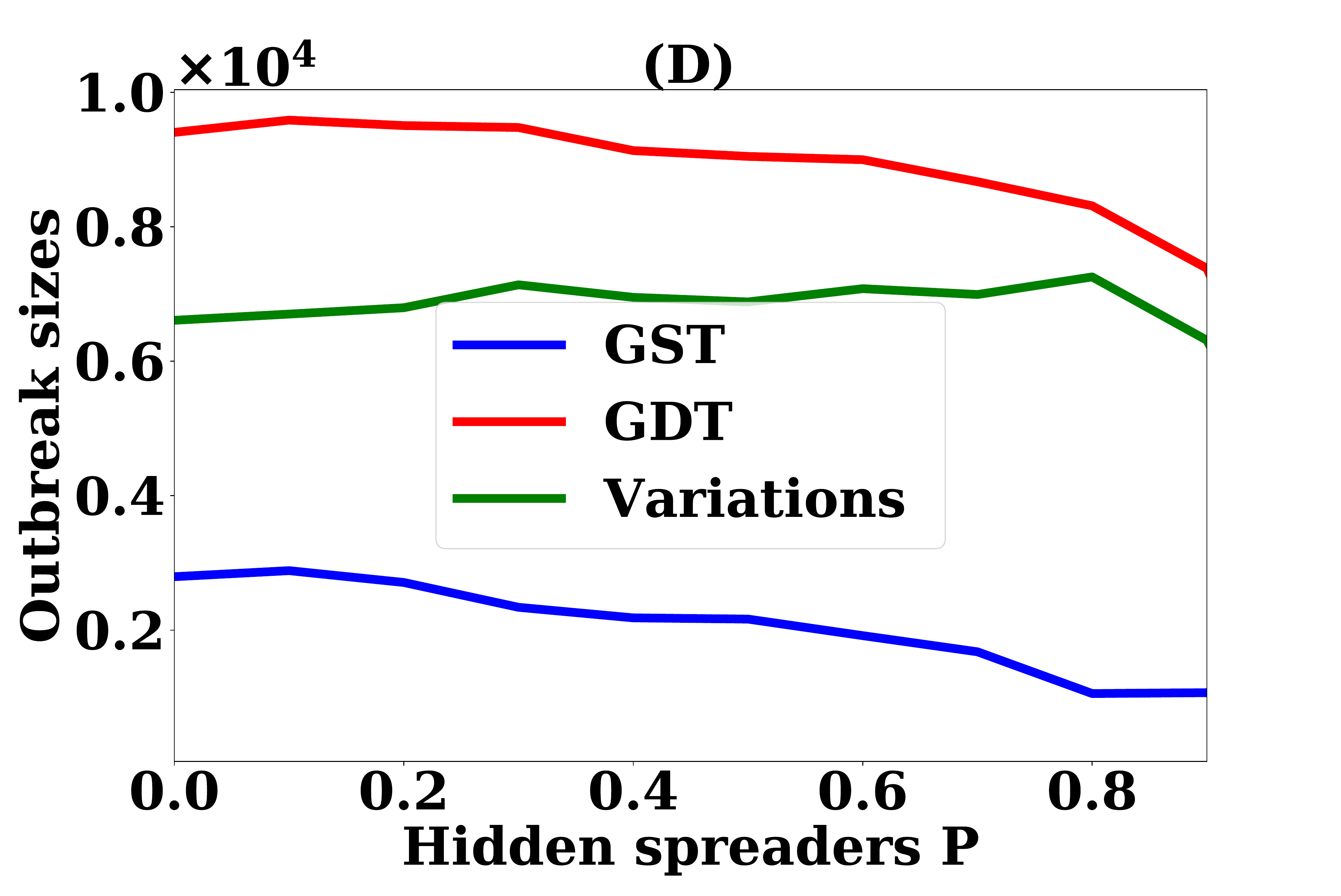}
\caption{Variations in diffusion dynamic predictions with varying $P$ on both the real contact networks (left column) and synthetic contact networks (right column): (A, B) delay in predicting cumulative number of infections and (C, D) variations in the outbreak sizes for SPST and SPDT model}
\label{fig:ipdpdis}
\end{figure}

The seed nodes start infecting susceptible nodes at $T=0$ and continue infecting for the period of days picked up randomly from the range [1, 5] days with a uniform distribution. Simulations are conducted for varying $P$ in the range [0,1] with a step of 0.1 proportion. For each value of $P$, 1000 simulations were run on both the real and synthetic contact networks. Figure~\ref{fig:ipdp} shows the changes in disease spreading dynamics and outbreak sizes, averaged over the 1000 simulation runs, for some $P=\{0.0, 0.2, 0.4, 0.6, 0.8, 1.0\}$. The disease prevalence $I_p$, number of infected individuals in the network on a simulation day, reduces with increasing $P$ for both the SPST and SPDT networks ~(see Fig.~\ref{fig:ipdp} A and Fig.~\ref{fig:ipdp}B) because the increased proportion of hidden spreaders as seed nodes reduces the likelihood of seed nodes to trigger spreading of diseases. The disease prevalence in the SPST networks (DST and GST) vanishes at $P=1$. In the SPDT networks (DDT and GDT), $I_p$ drops gradually with increasing $P$ and the minimum disease prevalence is obtained at $P=1$. With increasing $P$, the disease prevalence initially drops as the potentialities of SPDT links decrease as the direct component reduces in the SPDT links. But, the SPDT networks at $P=1$ are still capable of increasing disease prevalence with the spreading power of indirect transmission links in both the real and synthetic contact networks. However, the peak of disease prevalence is delayed as $P$ increases. The outbreak sizes in the SPDT networks decrease with $P$ ~(see Fig.~\ref{fig:ipdp}C and Fig.~\ref{fig:ipdp}D). At $P=1$, the outbreak size ranges in 6-8K infections in DDT network and 6-11K infections in GDT network while it was in the range of 8-10K infections and 9-12K infections at $P=0.0$ respectively in the DDT and GDT networks. The average total infections over 32 days in the DDT network at $P=1$ is about 5900 which is the 62\% of total infections 9600 at $P=0$. As the GDT network favours diffusion with the network properties of extending contact set sizes (nodes higher contact set sizes in GDT network as in Table~\ref{tb:ndclass}), the indirect links show stronger potential in the GDT compared to that in the DDT network. The generated epidemics sizes are still significant at $P=1$ in both networks and indicate the potential of indirect links. 

The disease prediction performance is now analysed with changing $P$ and is presented in Figure~\ref{fig:ipdpdis} for delving deeper into the contribution of indirect transmission links. Figure~\ref{fig:ipdpdis}A and Figure~\ref{fig:ipdpdis}B show the number of days required to cause a specific number of infections in the networks and how it is delayed with changing $P$. The SPST networks (DST and GST) substantially fail to predict infection dynamics causing more delays to reach a specific number of cumulative infections as $P$ increases (see Figure~\ref{fig:ipdpdis}A). The cumulative infections reach 1000 infections by the day 10 at $P=0$ but it is delayed to day 30 when $P$ is set to 0.8. However, the required days to reach a number of cumulative infections changes slightly for the SPDT model in both real and synthetic networks up to $P=0.8$ as the indirect links have significant potential. However, it is deviated significantly at $P=1$ when all seed nodes are hidden spreaders. The differences in outbreak sizes and disease prevalence between SPST and SPDT networks with the changes in $P$ are shown in Figure~\ref{fig:ipdpdis}C and Figure~\ref{fig:ipdpdis}D. As $P$ changes from 0 to 1, the average outbreak sizes in SPST networks drop by 100\% from 2257 to 500 (number of seed nodes) infections in DST network and from 2795 to 500 infections in GST network. By comparison, the average outbreak sizes drop by 40\% in DDT network changing from 9688 to 5916 infections and 28\% in GDT network changing from 11000 to 8000 infections. Thus, the differences in outbreak sizes between SPST and SPDT networks vary from 4916 infections for $P=0$ to 7000 infections for $P=1$ in the real network while the variations in synthetic networks are from 7000 infections for $P=0$ to 8000 infections for $P=1$. This indicates that the underestimation of disease dynamics by SPST model increases with $P$, i.e. if the disease starts with all hidden spreaders, it shows no emergence of disease. The indirect transmission links carry strong influence for shaping SPDT diffusion dynamics and prove its potentiality to be considered in future modelling. 

\subsection{Hidden spreaders}
\begin{figure}[h!]
\begin{tikzpicture}
    \begin{customlegend}[legend columns=4,legend style={at={(0.12,1.02)},draw=none,column sep=5ex,line width=3.5pt, font=\small }, legend entries={C1, C2, C3, C4}]
    \addlegendimage{only marks,mark=o,color=red}
    \addlegendimage{only marks,mark=o,color=blue}
    \addlegendimage{only marks,mark=o,color=green}
    \addlegendimage{only marks,mark=o,color=yellow}
    \end{customlegend}
 \end{tikzpicture}
\centering
\includegraphics[width=0.48\linewidth, height=5.0cm]{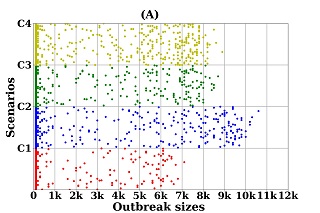}~
\includegraphics[width=0.48\linewidth, height=5.0 cm]{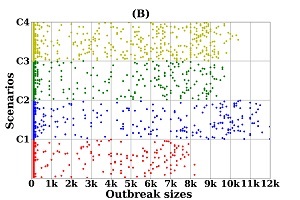}\\
\includegraphics[width=0.48\linewidth, height=5.0 cm]{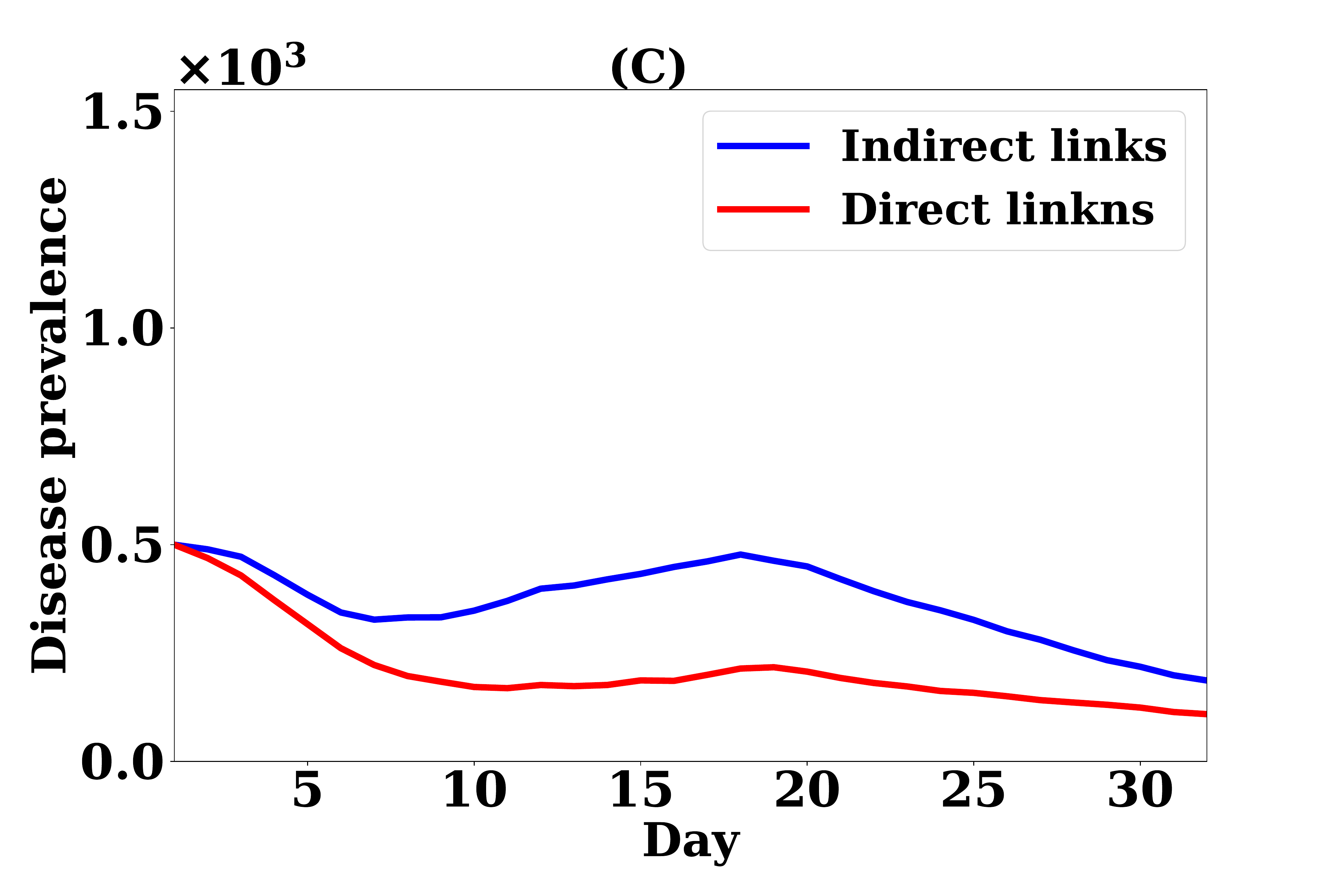}~
\includegraphics[width=0.48\linewidth, height=5.0 cm]{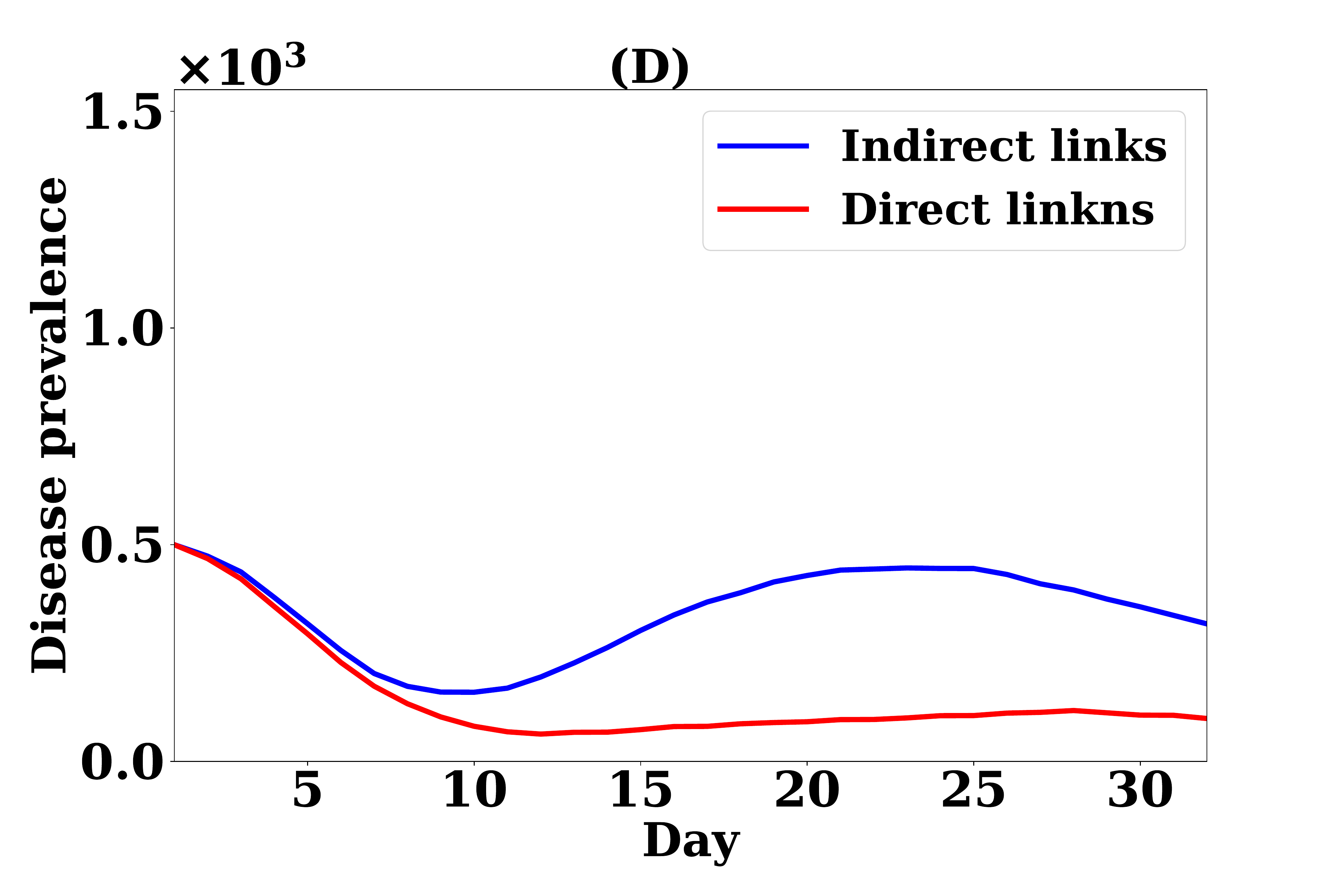}\\
\includegraphics[width=0.48\linewidth, height=5.0 cm]{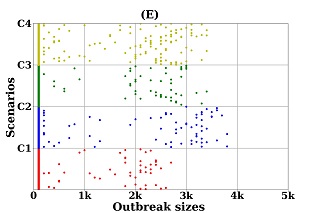}~
\includegraphics[width=0.48\linewidth, height=5.0 cm]{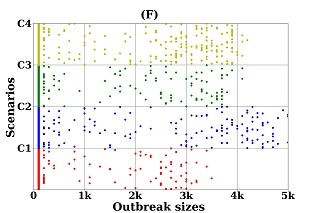}
\caption{Diffusion behaviour of hidden spreaders - left column for real networks and right column for synthetic networks. The dots in the figures represent nodes that are placed into the bins of outbreak sizes where bins are formed with the step of 1000 infections: (A, B) hidden spreaders in the DDT and GDT networks, (C, D) diffusion in the networks with only direct links and in the networks with only indirect links, and (E, F) hidden spreaders in the networks with indirect links only}
\label{fig:hspd}
\end{figure}

It is observed in the previous experiment that the indirect transmission links assume sufficient spreading potential to sustain the disease prevalence in the SPDT networks. Simulations are now conducted to explore whether hidden spreaders can be super-spreaders although they do not have any spreading potential in the SPST networks. For this experiment, two sets of 5000 hidden spreader nodes are selected from both the DDT and GDT networks. Simulations are run for each node picked up from the selected sets of nodes in the various configurations of real and synthetic contacts networks. Each simulation is triggered for a maximum of 10 times if the outbreak sizes are below 100 infections. As the environments of interactions affect the spreading potential in the SPDT model, various scenarios are used to understand the full potential of hidden spreaders. The results are presented in Figure~\ref{fig:hspd}. The previous experiments are conducted with particle decay rates $r_t=45$ min and $\sigma=0.33$ which is assumed to be the typical disease spreading scenario (C1). Simulations are run as well for scenario (C2) increasing $\sigma=0.4$ along with the scenario (C3) increasing $r_t=60$ to the typical scenario (C1). These reconfigured scenarios affect all SPDT links available in the networks. Another scenario (C4) considered is where it is assumed that the particles decay rates at the locations of the infected node visited are very slow. The C4 represents the extreme scenarios for particles decay rates. Thus, the upper bound of particles removal time 240 minutes is taken as the disease spreading scenario $C4$. The scenario C4 shows the impact of indirect links if infected individual visit such locations where infectious particles stay a long time than usual time and how a single individual with critical situation influences disease spreading. The simulations are run on both DDT and GDT networks and results are presented in Figure~\ref{fig:hspd}A and Figure~\ref{fig:hspd}B. The simulations are not run on SPST networks as hidden spreaders cannot spread disease in SPST networks. In these figures, the nodes are placed into the bins of outbreak sizes where bins are formed with the step of 1000 infections. Number of nodes that generates outbreak sizes grater than 100 infections over 32 days of simulations for both networks are presented in Table~\ref{tb:hdnspd}.

\begin{table} 
\caption{Number of nodes generated outbreak sizes grater than 100 infections in both networks}
\label{tb:hdnspd}
\vspace{1em}
\centering

\begin{tabular}{|c|c|c|c|c|}
\hline

Diffusion scenarios &  C1 & C2 & C3 & C4  \\ \hline
Super-spreaders in DDT network &  114  & 240 & 230 & 381 \\ \hline
Super-spreaders in GDT network & 145 & 231 & 213 & 346 \\ \hline
\end{tabular}

\end{table} 

The simulation results show that the hidden spreaders can be super-spreaders and the number of hidden super-spreaders in the networks varies depending on the scenarios. With the scenario C1, about 114 hidden spreader nodes out of 5000 can cause outbreaks greater than 100 infections in DDT network and GDT network has 145 hidden super-spreaders out of 5000 (see Figure~\ref{fig:hspd}A and Figure~\ref{fig:hspd}B). The distribution of hidden spreaders shows that the outbreak sizes can be up to 7K infections in scenario C1. The average infection for 5000 hidden spreaders is about 231 infections in DDT network and 245 infections in GDT network. When the scenario is changed to C2 $\sigma=0.4$, the number of hidden super-spreaders increases to 240 in DDT network and 231 in GDT network. The outbreak sizes are also increased and can be up to 12K infections. The average infection also increases significantly which is 590 infections. Thus, the hidden super-spreaders are influential in the scenario C2. The number of hidden super-spreaders also increase in the scenario C3 ($r_t=60$ min). In this case, 230 nodes are able to make outbreaks greater than 100 in DDT network and the outbreak sizes increase to 9K infections. The GDT network shows that 213 nodes can be super-spreaders and outbreak sizes can be up to 10K infections. At the extreme scenarios created for the visited locations of infected hidden spreaders can also trigger more nodes to be super-spreaders. In the scenario C4, about 381 nodes become supers-readers with outbreak sizes up to 9K in DDT network while 346 hidden spreaders become super-spreaders in the GDT network with the outbreak sizes up to 11K infections. The number of hidden spreaders is a bit higher in the DDT network than in the GDT network. This is because nodes repeatedly send links to the same neighbours in DDT network (due to reconstruction of networks which was discussed in the Section 3.3.3) while links are forward to a broader set of neighbours in GDT network. The above results show that the hidden spreaders have a strong potential similar to non-hidden spreaders with direct links to be prominent spreaders and the potentiality get stronger with the changes in scenarios. 

To increase the understanding of spreading potential of hidden spreaders, simulations are conducted for a special configuration of contact networks. First, diffusion dynamics on the contact networks are investigated when all nodes are hidden spreaders. For these simulations, direct links are excluded from the DDT and GDT networks and contact networks of nodes with only indirect links are obtained. Therefore, the networks obtained have only hidden spreader nodes. Then, disease simulations are run from 500 seed nodes at time $T=0$ and for a period of 32 days. The value of $r_t$ is set to 45 min and $\sigma$ is set to 0.33. The average results for 1000 realisations are shown in Figure~\ref{fig:hspd}C and Figure~\ref{fig:hspd}D. This result is compared with the results of DST and GST networks where only direct links are available. The networks with indirect links show the higher number of infections and strong disease prevalence. This is because the indirect transmission links assume a spreading potential similar to direct links and the 63\% links in the DDT networks and 57\% links in GDT networks are indirect transmission links. The DST network can cause 2670 infections and the GST network can cause 2712 infections while the corresponding hidden spreaders networks (both real and synthetic networks with indirect links only) can cause about 4370 infections and 4731 infections respectively. 

\begin{table} 
\caption{Number of nodes generated outbreak sizes grater than 100 infections in the networks with only indirect links}
\label{tb:hdnspd2}
\vspace{1em}
\centering

\begin{tabular}{|c|c|c|c|c|}
\hline

Diffusion scenarios &  C1 & C2 & C3 & C4  \\ \hline
Super-spreaders in DDT network with indirect links &  48  & 72 & 53 & 118 \\ \hline
Super-spreaders in GDT network with indirect links & 81 & 122 & 96 & 186 \\ \hline
\end{tabular}
\end{table} 

Finally, simulations are also carried out from the single seed nodes in the networks of hidden spreader only, where no node has a direct link, for all previous scenarios (C1-C4). The hidden spreader nodes in the networks of hidden spreader are also capable of triggering outbreaks with a significant number of total infections (see Figure~\ref{fig:hspd}E and Figure~\ref{fig:hspd}F). The number of nodes having outbreak sizes greater than 100 infections are presented in Table~\ref{tb:hdnspd2}. The number of hidden spreaders in DDT network are 48,72,53 and 118 respectively for scenarios C1, C2, C3 and C4 while they are 81,122,96 and 186 in GDT network. The maximum outbreaks sizes are up to 4K infections in the scenario C3 in the DDT network and 5K infections in the GDT network. These results demonstrate that the indirect links play a similar role in emerging SPDT diffusion. The interesting outcome is that there are outbreaks in the SPDT networks which are not identified by the current SPST models.

\section{Emergence of diffusion in SPDT model}
The previous experiments show that the indirect links have significant transmission potential and impacts on the spreading behaviours of a diffusion process. This experiment will examine how these indirect links play a vital roles at the individual level and increase the spreading potential of nodes. Therefore, the emergence of diffusion from a node becomes easier and a disease easily invades the network, this is not estimated by the current SPST diffusion model. First, the enhancement in node's spreading potential due to including the indirect links is studied and then the increased in invading power due to indirect links is examined.

\subsection{Spreading potential with indirect links}
The position of nodes in the network is crucial to achieve high spreading potential (become super-spreaders). The contact set size of a node, built by the number of neighbour nodes the node interacts and represent the static degree for the observation period, has strong influence on its spreading potential as well. To analyse enhancement in spreading potential across different connectivity of nodes, nodes of both real and synthetic networks are classified into four classes according to the contact set sizes of the nodes. The number of nodes in each class for both the SPDT and SPST networks is presented in the Table~\ref{tb:ndclass}. In the SPST networks, various proportions of these nodes from each class can be super-spreaders and large outbreaks are generated if the disease starts from them. If the indirect transmission links are added with the direct transmission links, the importance of nodes can be escalated and new super-spreaders are created. Therefore, the spreading potential of nodes is characterised by the addition of indirect transmission links and changing local contact structure surrounding this node in the SPDT model.
 
\begin{figure}[h!]
\centering
\begin{tikzpicture}
    \begin{customlegend}[legend columns=2,legend style={at={(0.12,1.02)},draw=none,column sep=6ex,line width=3.5pt, font=\small }, legend entries={SPDT, SPST}]
    \addlegendimage{only marks,mark=o,color=blue}
    \addlegendimage{only marks,mark=o,color=red}
      \end{customlegend}
 \end{tikzpicture}
 
\includegraphics[width=0.48\linewidth, height=5.0 cm]{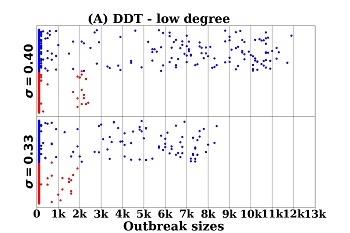}~
\includegraphics[width=0.48\linewidth, height=5.0 cm]{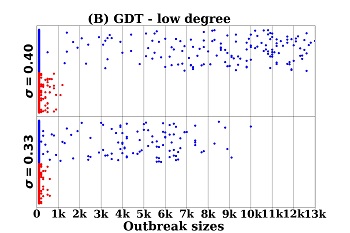}\\
\includegraphics[width=0.48\linewidth, height=5.0 cm]{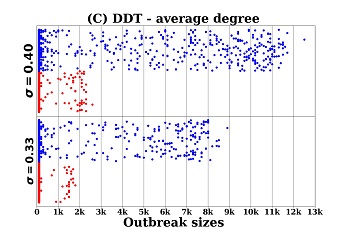}~
\includegraphics[width=0.48\linewidth, height=5.0 cm]{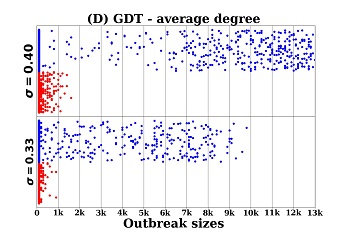}
\includegraphics[width=0.48\linewidth, height=5.0 cm]{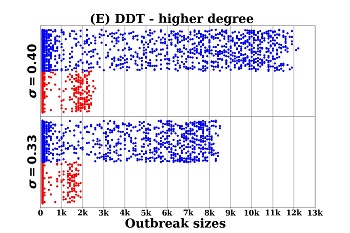}~
\includegraphics[width=0.48\linewidth, height=5.0 cm]{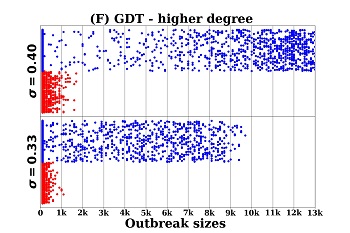}\\
\includegraphics[width=0.48\linewidth, height=5.0 cm]{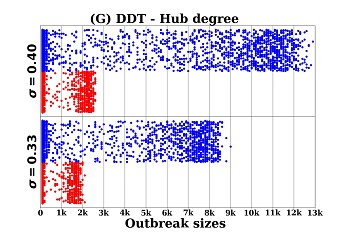}~
\includegraphics[width=0.48\linewidth, height=5.0 cm]{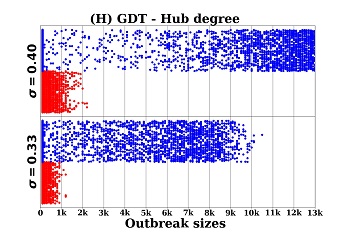}
\caption{Distributions of super-spreaders according to their potential: left column for real networks and right column for synthetic networks. The nodes are binned with their outbreak sizes. Red dots are for SPST networks and blue dots are for SPDT networks}
\label{fig:sprds}
\end{figure}

\begin{figure}[h!]
\centering
\includegraphics[width=0.45\linewidth, height=5.5 cm]{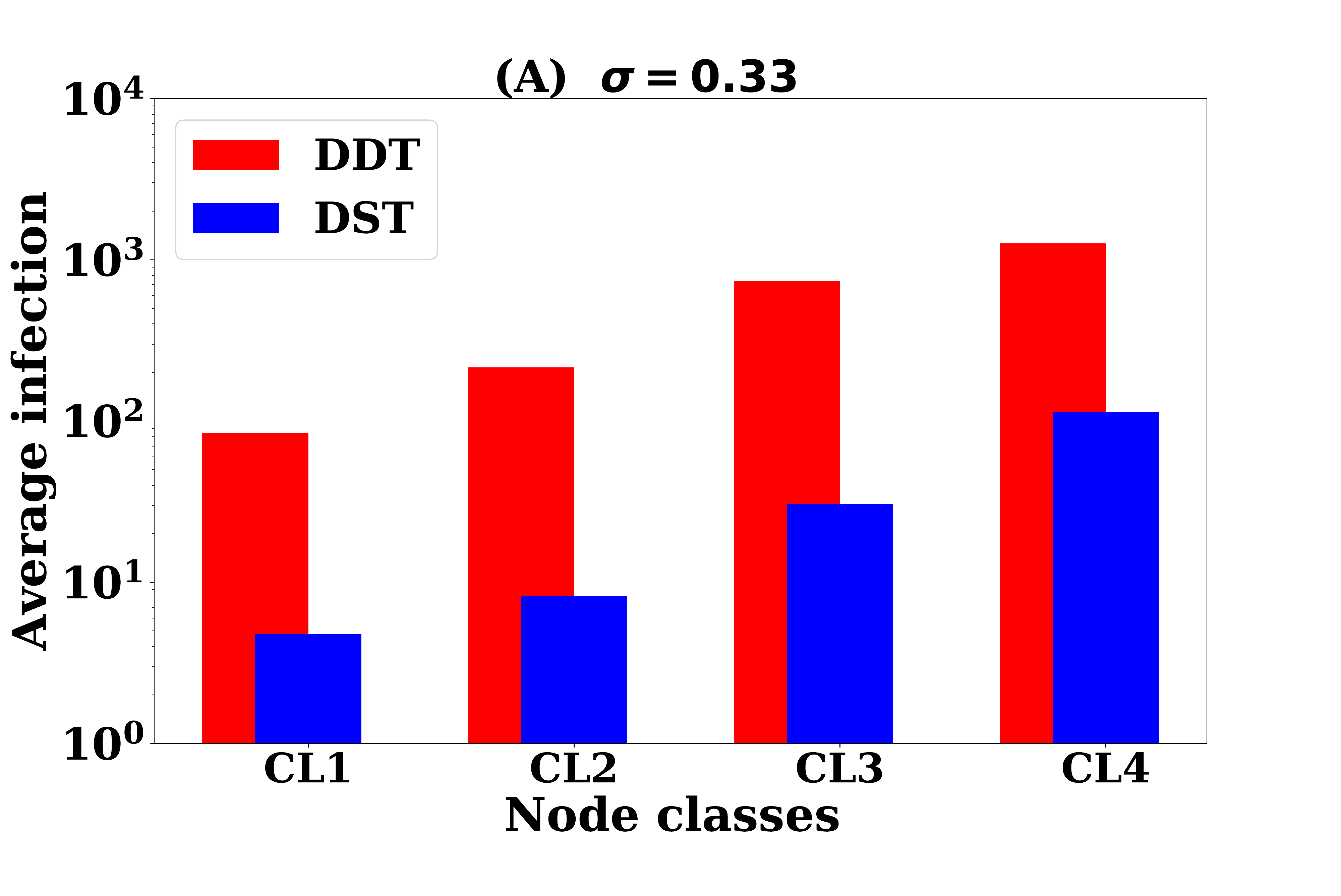}~
\includegraphics[width=0.45\linewidth, height=5.5 cm]{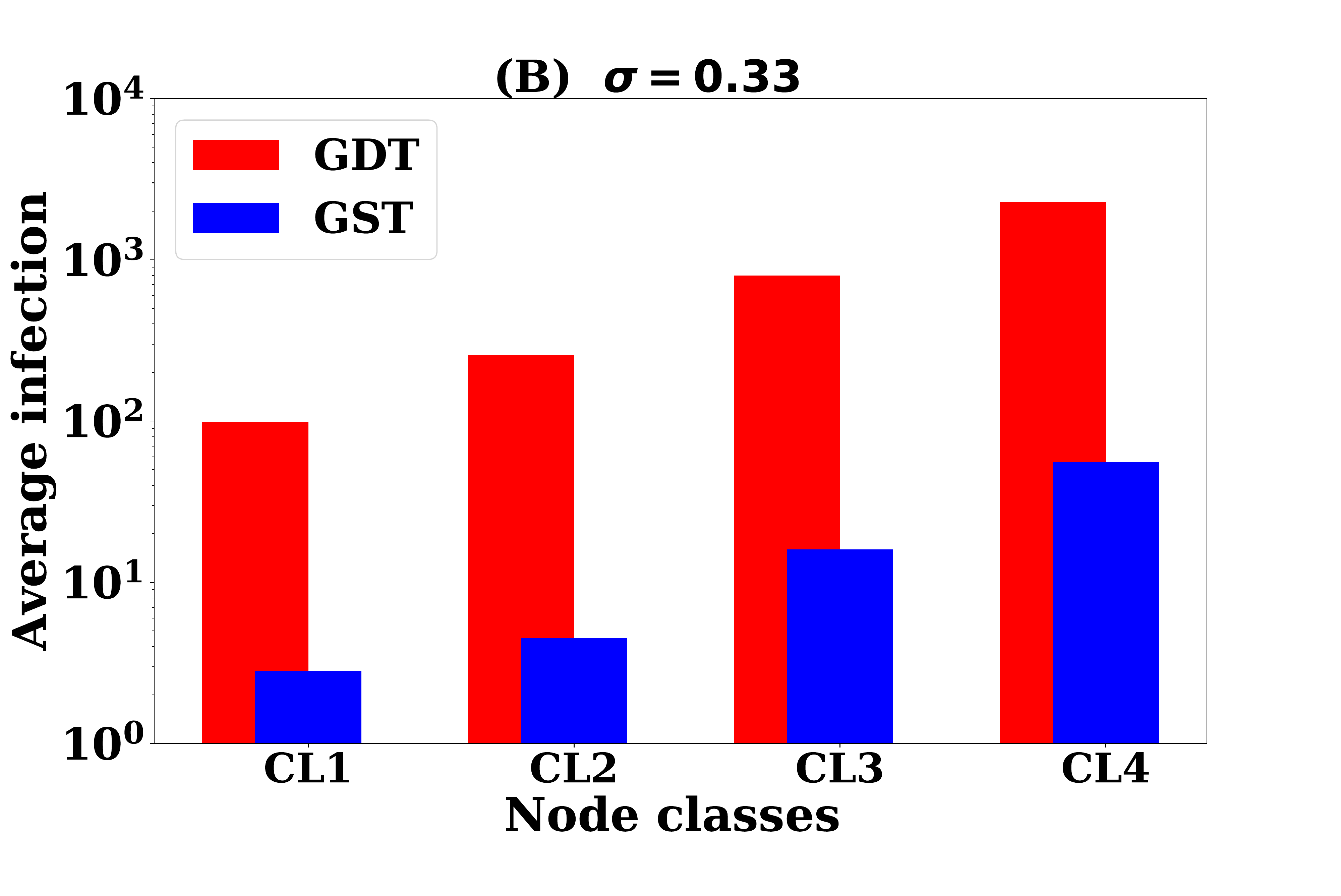}\\
\includegraphics[width=0.45\linewidth, height=5.5 cm]{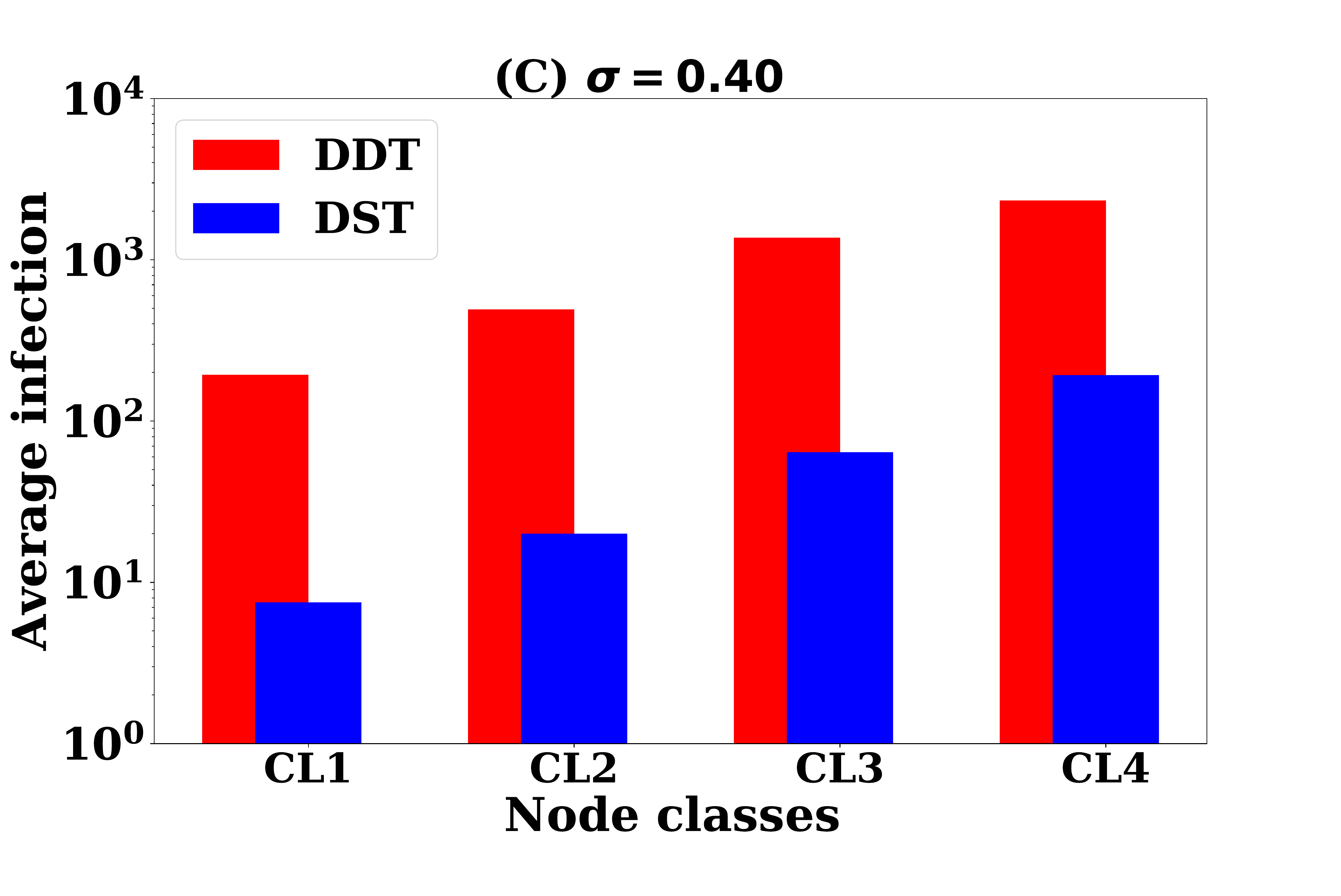}~
\includegraphics[width=0.45\linewidth, height=5.5 cm]{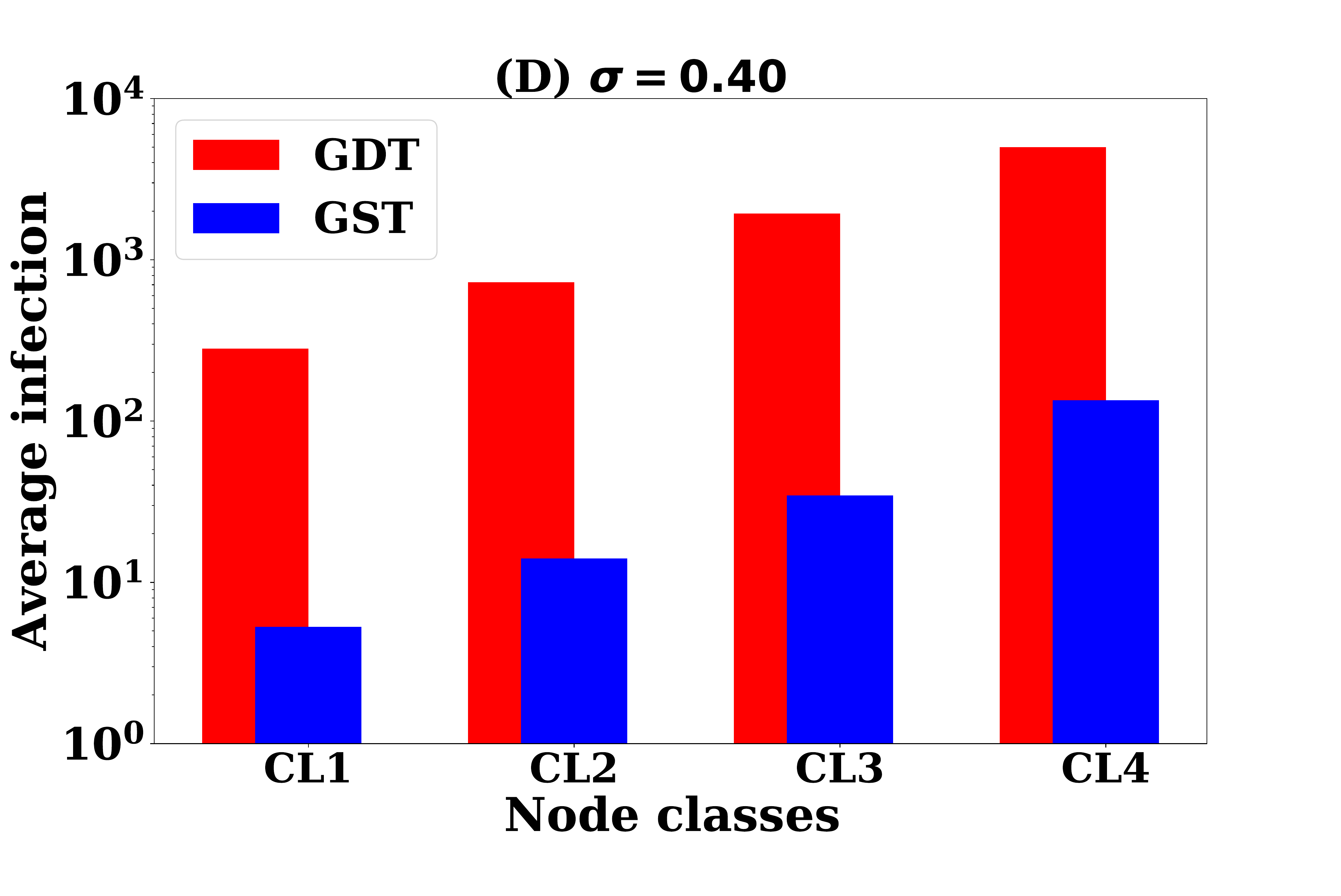}
\vspace{-1.5em}
\caption{Spreading potential for nodes - left column is for real networks (DDT) and right column for synthetic networks (GDT): (A, B) with the infectiousness $\sigma=0.33$ and (C, D) with the infectiousness $\sigma=0.40$}
\label{fig:sprgain}
\vspace{-1.5em}
\end{figure}

Simulations are now conducted from single seed nodes on both the real and synthetic networks and outbreak sizes are collected. From each class defined based on the contact set sizes of nodes, a set of 5000 nodes are randomly selected to study their spreading potential. Simulations iterate through each of the nodes. The seed nodes are able to infect susceptible nodes for 5 days before they recover from the infectious state. Simulations are also run for $\sigma =0.4$ to understand how these nodes play a more significant role under favourable scenarios of disease spreading as it was shown in the previous chapters that the SPDT model behaviours are strongly influenced by disease parameters. From each seed node, simulations are triggered for a maximum of 10 times when the generated outbreak size from the seed node is below 100. This is because the stochastic simulation outcome from a seed node can be varied for infecting different neighbour nodes as some of them may have strong spreading potentials. The simulation results for the selected 5000 single seed nodes are presented in Figure~\ref{fig:sprds} where red dots represent the nodes in SPST networks and blue dots represent the nodes in SPDT networks. The nodes are placed into the bins of outbreak sizes where bins are formed with the step of 1000 infections. The proportion of nodes having outbreak sizes greater than 100 infections is presented in the Table~\ref{sprds}.

\begin{table}
\caption{Number of nodes cause infection above 100 infections in the networks}
\vspace{1em}
\centering

\begin{tabular}{|c|c|c|c|c|c|}
\hline
Networks & $\sigma$ & Low degree & Average degree & High degree & Hub degree  \\ \hline
\multirow{2}*{DDT} & 0.33 & 95  & 227 & 806 & 1190 \\ 
\cline{2-6}
 & 0.40 & 161  & 339 & 1159 & 1581 \\ \hline
 
\multirow{2}*{DST} & 0.33 & 17  & 28 & 115 & 393 \\ 
\cline{2-6}
 & 0.40 & 22  & 61 & 199 & 534 \\ \hline
 
\multirow{2}*{GDT} & 0.33 & 107  & 276 & 793 & 2008 \\ 
\cline{2-6}
 & 0.40 & 162 & 389 & 1004 & 2421 \\ \hline
 
\multirow{2}*{GST} & 0.33 & 28  & 49 & 233 & 797\\ 
\cline{2-6}
 & 0.40 & 53  & 139 & 349 & 1321 \\ \hline

\end{tabular}
\label{sprds}
\end{table}

By including indirect links, nodes spreading behaviours are changed significantly. In the SPST networks (DST and GST), low degree nodes do not have the strong spreading potential to be super-spreaders. Only about 17 nodes out of 5000 nodes are spreaders in DST network and 28 nodes out of 5000 nodes in GST network while this class of nodes have 95 super-spreader nodes in the DDT network and 107 in GDT network (see Figure~\ref{fig:sprds}). When the infectiousness $\sigma$ of disease is changed by 20\% to $\sigma=0.40$, the number of super-spreaders increases to 161 nodes in the DDT networks and 162 nodes in GDT network. However, there are no significant changes in the DST network with $\sigma=0.40$ but it increases to 53 nodes in GST network. The final outbreak sizes in the SPST networks are within 2K infections in DST and 1.5K infections in GST while it can be above 13K infections in the SPDT networks due to including the indirect links. The number of super-spreaders increase substantially for the hub nodes class. In the SPST networks, about 393 nodes become super-spreaders in DST network and 797 nodes in the GST network while 1190 nodes become super-spreader in the DDT network and 2008 nodes in the GDT network. The SPDT model has almost three times more super-spreaders than the SPST model. This is further extended at high $\sigma=0.40$ to 1581 nodes in DDT network and to 2421 nodes in GDT network. For hub nodes, the outbreak sizes generated are distributed in the higher bins significantly and maximum outbreak sizes reach to 13K infections. For the other classes of nodes, the SPDT networks hold a significantly large number of nodes compared to SPST networks and the outbreak sizes increase in the higher bins (see Table~\ref{sprds}). 

The average spreading potential for each node class is presented in the Figure~\ref{fig:sprgain}. This is calculated as the average number of infections caused by the selected 5000 seed nodes. The obtained outbreak size for each seed node are summed and divided by 5000 to get the average infection. The results show that the spreading potential is often gained due to the inclusion of indirect links. The potential gain for low degree nodes in the DDT and GDT networks is about 16 times. However, the spreading potential gain is not changed with increasing $\sigma$ for this node class although a number of super-spreader increases. A greater impact is found for the class of hub degree nodes. The average infection gain in DDT network is 20 at $\sigma=0.33$ which increases to 26 at $\sigma=0.40$ while GDT network shows 50 at $\sigma=0.33$ which increases to 56 at $\sigma=0.40$. The average infection gain for average degree class nodes is 37.8 and 43.1 for the high degree nodes in DDT network. The GDT network shows milder gains which are 12.67 for average degree nodes and 15.1 for high degree nodes. The emergence of diffusion increases for all classes of nodes. However, the hub degree nodes are affected more than the low degree nodes as they have high direct contacts and indirect contacts as well.

\subsection{Strength of diffusion emergence}
The  previous experiment shows that the indirect transmission links boast the emergence of infectious disease in the dynamic contact networks. Therefore, invading the network by an infectious disease becomes easier. This experiment investigates what extent the invasion becomes easier by including indirect links whilst studying the performance of vaccination strategies. If a network requires more vaccination to stop spreading of a disease, it means the invasion strength is stronger in the networks. This experiment also reveals that how much effective the current vaccination strategies for the SPDT model. Two popular vaccination strategies named mass vaccination and ring vaccination are studied on both the SPDT and SPST networks. The performances of vaccination strategies are analysed by conducting disease spreading simulation based on the SIR epidemic model in both real and synthetic contact networks. First, mass vaccination is discussed and then ring vaccination is analysed with various configurations of contact networks.

\subsubsection{Mass vaccination}
In the mass vaccination, a proportion of the population is randomly selected and is vaccinated to prevent future occurring of infectious disease outbreaks. To implement a mass vaccination in this experiment, a proportion $P$ of total nodes is randomly selected and is assigned disease status recovered at the beginning of simulations to represent them as vaccinated nodes in the networks. These nodes are chosen from the first day of traces in both the real and synthetic contact networks. To understand the effectiveness of implemented vaccination strategy, the emergence of disease are simulated from random 5K separate non-vaccinated seed nodes which are also chosen from the first day traces of contact networks. The outbreak sizes for disease spreading over 32 days are collected at the end of the simulations. The average values of all outbreak sizes for these 5K seed nodes indicate the ability of the vaccinated nodes to prevent the emergence of disease. If a network requires more number of vaccination, it concludes that invasion to that network is stronger by the infectious disease. 

\begin{figure}[h!]
\centering
\begin{tikzpicture}
    \begin{customlegend}[legend columns=4,legend style={at={(0.12,1.02)},draw=none,column sep=3ex,line width=1.5pt,font=\small}, legend entries={P=0.0, P=0.1, P=0.2, P=0.3, P=0.4, P=5.0, P=0.6, P=0.7, P=0.8}]
    \addlegendimage{solid, color=blue}
    \addlegendimage{color=red}
    \addlegendimage{color=cyan}
    \addlegendimage{color=magenta}
    \addlegendimage{color=yellow}
    \addlegendimage{color=black}
    \addlegendimage{color=gray}
    \addlegendimage{color=purple}
    \end{customlegend}
 \end{tikzpicture}

\includegraphics[width=0.48\linewidth, height=5cm]{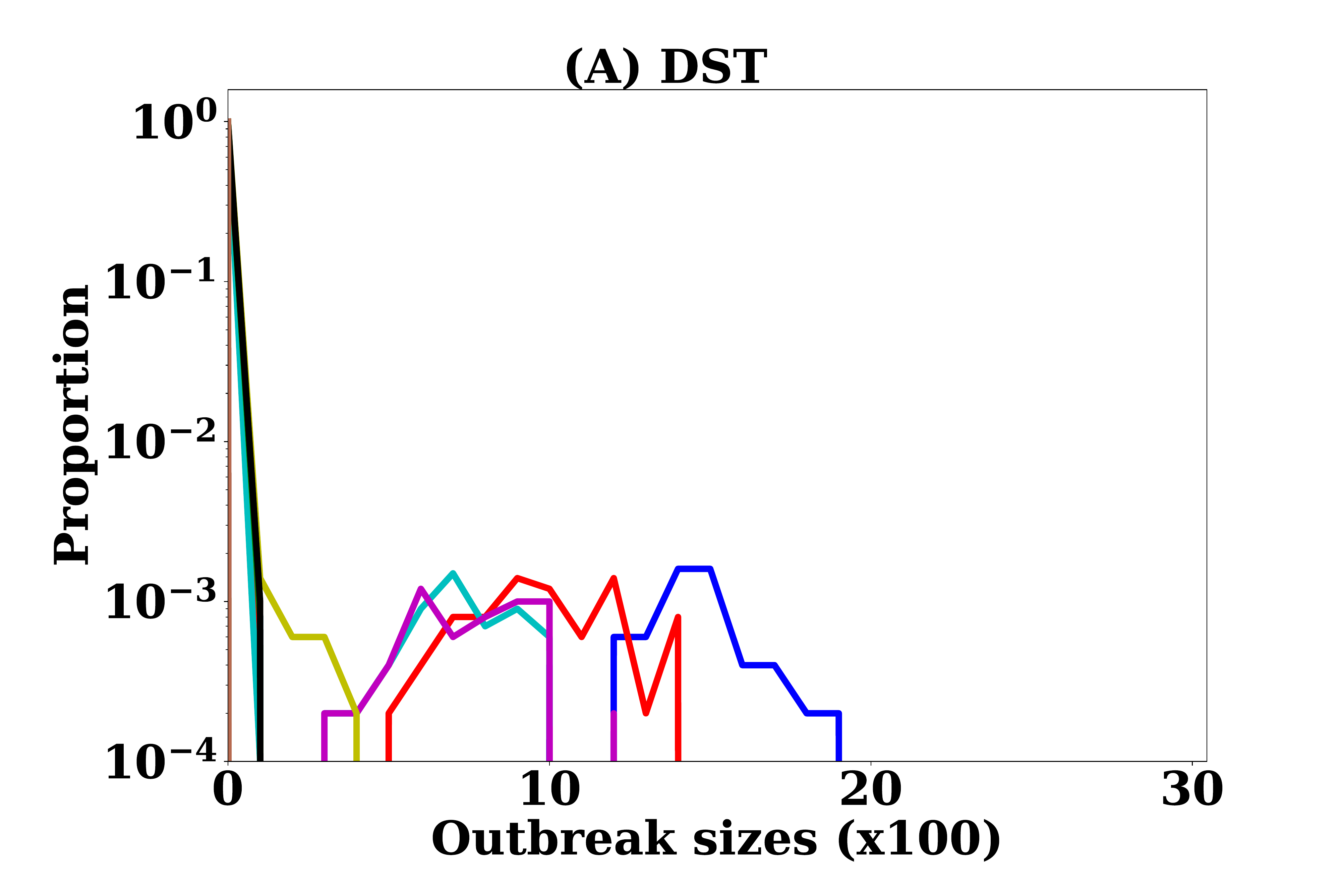}~
\includegraphics[width=0.48\linewidth, height=5cm]{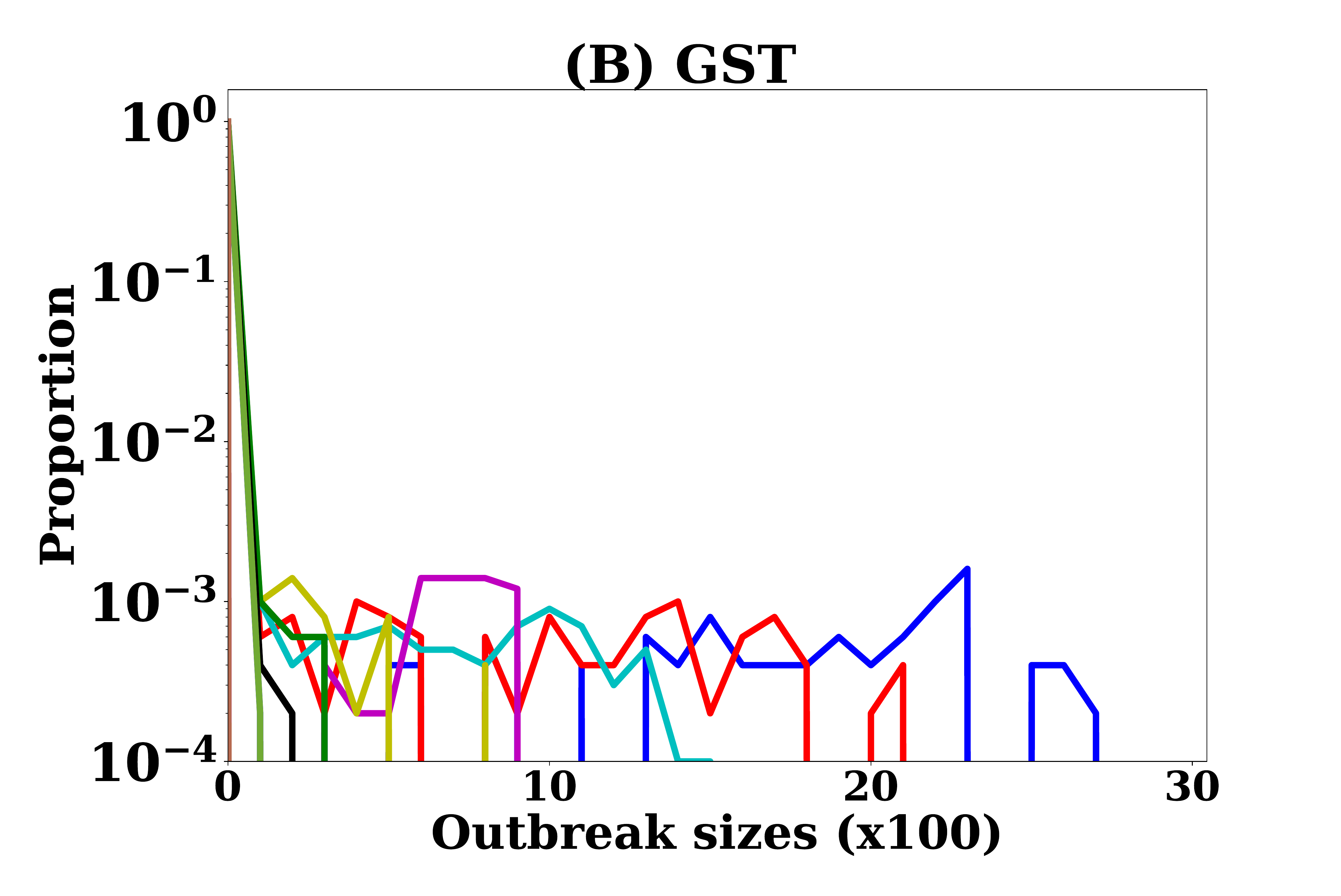}\\
\includegraphics[width=0.48\linewidth, height=5cm]{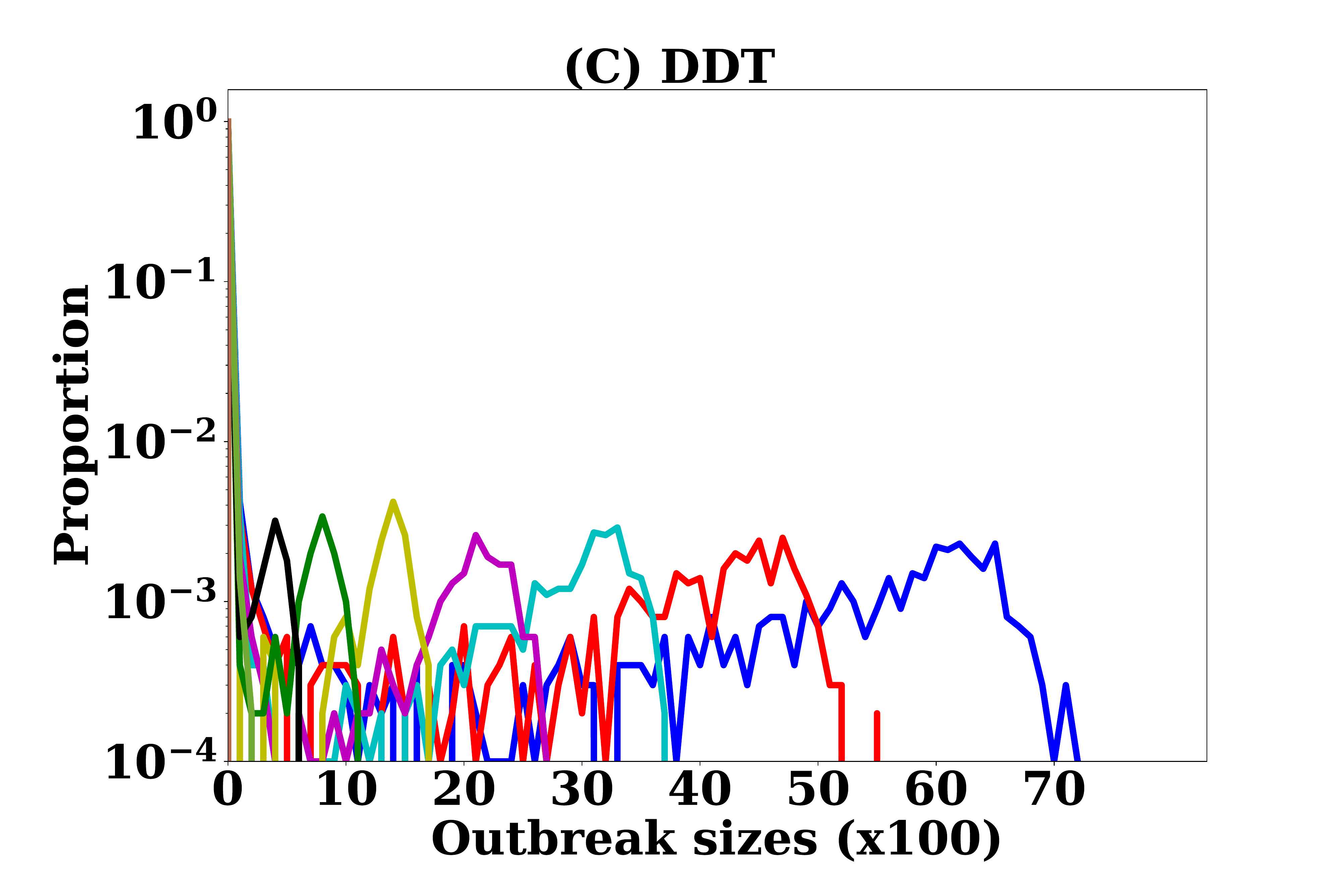}~
\includegraphics[width=0.48\linewidth, height=5cm]{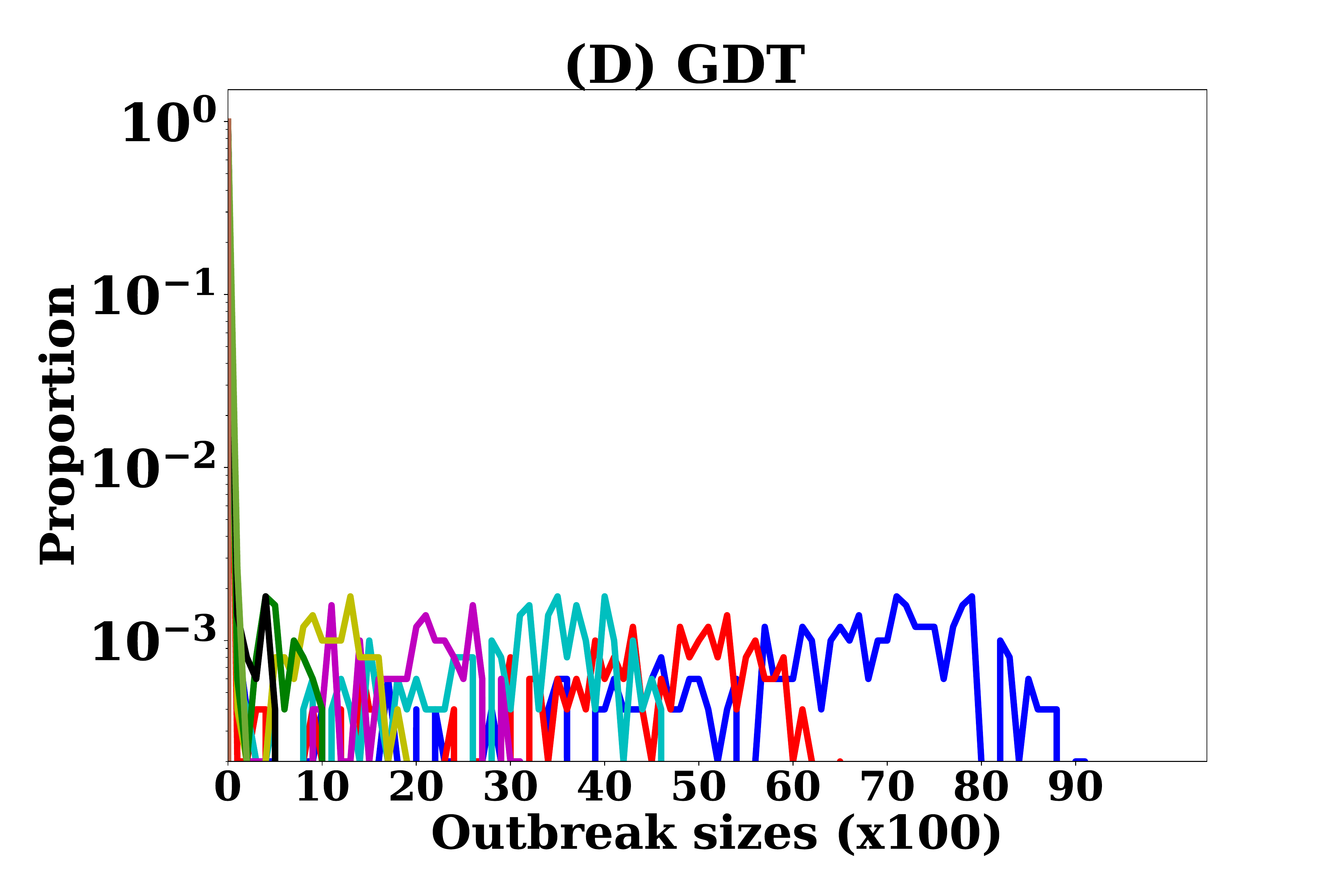}\\
\includegraphics[width=0.48\linewidth, height=5cm]{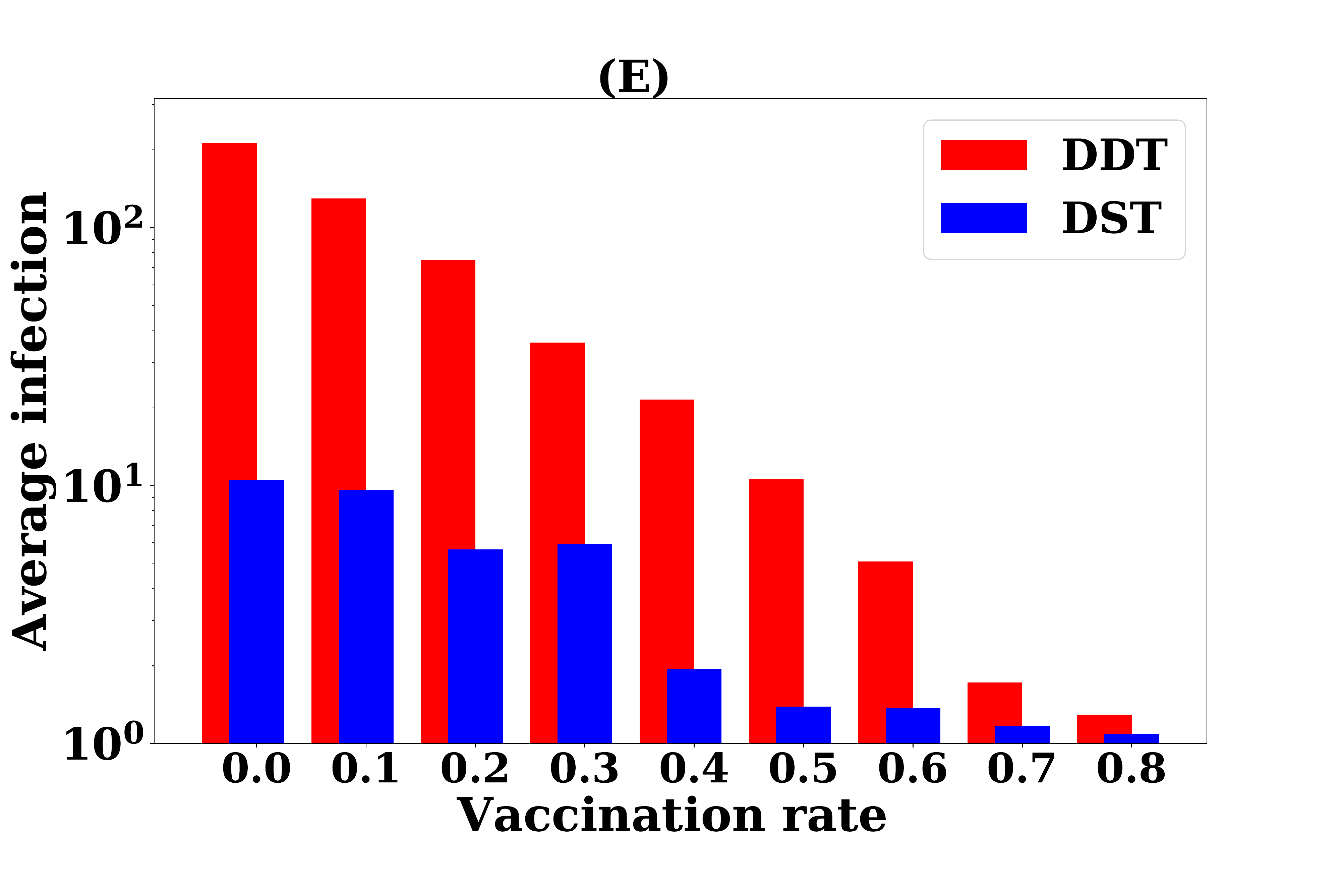}~
\includegraphics[width=0.48\linewidth, height=5cm]{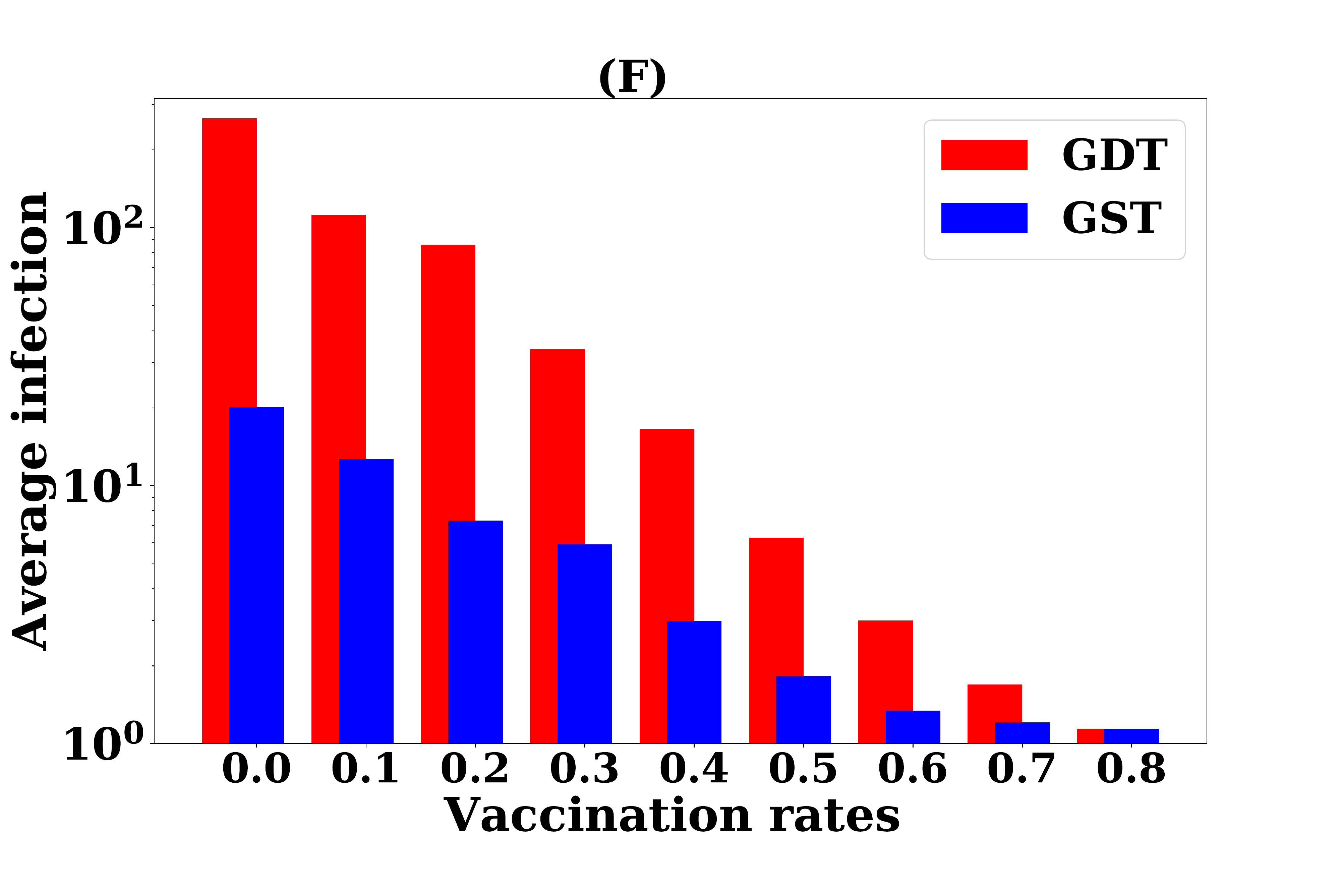}
\vspace{-1.0 em}
\caption{Comparison of effectiveness of mass vaccination in SPDT and SPST networks for various $P$: (A, B) distribution of outbreak sizes in the SPST networks, (C, D) distribution of outbreak sizes in the SPDT networks and (E, F) preventive efficiency, i.e. the reduction in average infection that is obtained without vaccination, for various $P$}
\label{fig:massv}
\vspace{-1.0em}
\end{figure}

The performance of mass vaccination is studied for the vaccination rate $P$ in the range [0,0.8] with a step of 0.1 proportion. In the simulations, the disease starts to spread at time $T=0$ from a randomly chosen nodes that are not vaccinated. The results are shown in the Figure~\ref{fig:massv}. In the SPST networks, the outbreak sizes reduce very quickly as $P$ increases. At $P>0.3$, no node of selected 5K nodes can cause outbreak sizes greater than 1000 infections (Figure~\ref{fig:massv}A and Figure~\ref{fig:massv}B). Thus, the emergence of disease is stopped at $P=0.4$ in the SPST networks (DST and GST) and causes few infections. On the other hand, SPDT networks (DDT and GDT) show the outbreak sizes up to 2K infections in DDT network (Figure~\ref{fig:massv}C) and 3K infections in GDT network (Figure~\ref{fig:massv}D) at $P=0.4$ and the number of nodes that cause outbreaks of more than 100 infections is about 100 nodes in DDT network and 94 nodes in GDT network. The average infection, obtained summing all outbreaks and divided by 5000, at $P=0.4$ is 20 infections in DDT which is 10\% of $P=0$ (Figure~\ref{fig:massv}E). For the GDT network, the spreading control is high where the average infection is 15 and about 80 nodes have outbreaks of more than 100 infections (Figure~\ref{fig:massv}F). In the SPDT networks, however, mass vaccination strategy requires vaccination of 70\% of nodes to eliminate the outbreaks where no outbreak is greater than 100 infections. If the network is constructed with direct transmission links only, one can stop causing future outbreaks with vaccinating 40\% of nodes while it needs an extra 30\% vaccination to mitigate the diffusion contribution of indirect links which indicates the invasion strength due to including indirect links.

\begin{figure}[h!]
\begin{tikzpicture}
    \begin{customlegend}[legend columns=4,legend style={at={(0.12,1.02)},draw=none,column sep=2ex ,line width=1.5pt,font=\small}, legend entries={P=0.0, P=0.1, P=0.2, P=0.3, P=0.4, P=5.0, P=0.6, P=0.7, P=0.8}]
    \addlegendimage{solid, color=blue}
    \addlegendimage{color=red}
    \addlegendimage{color=cyan}
    \addlegendimage{color=magenta}
    \addlegendimage{color=yellow}
    \addlegendimage{color=black}
    \addlegendimage{color=gray}
    \addlegendimage{color=purple}
    \end{customlegend}
 \end{tikzpicture}
\centering
\includegraphics[width=0.48\linewidth, height=5.5 cm]{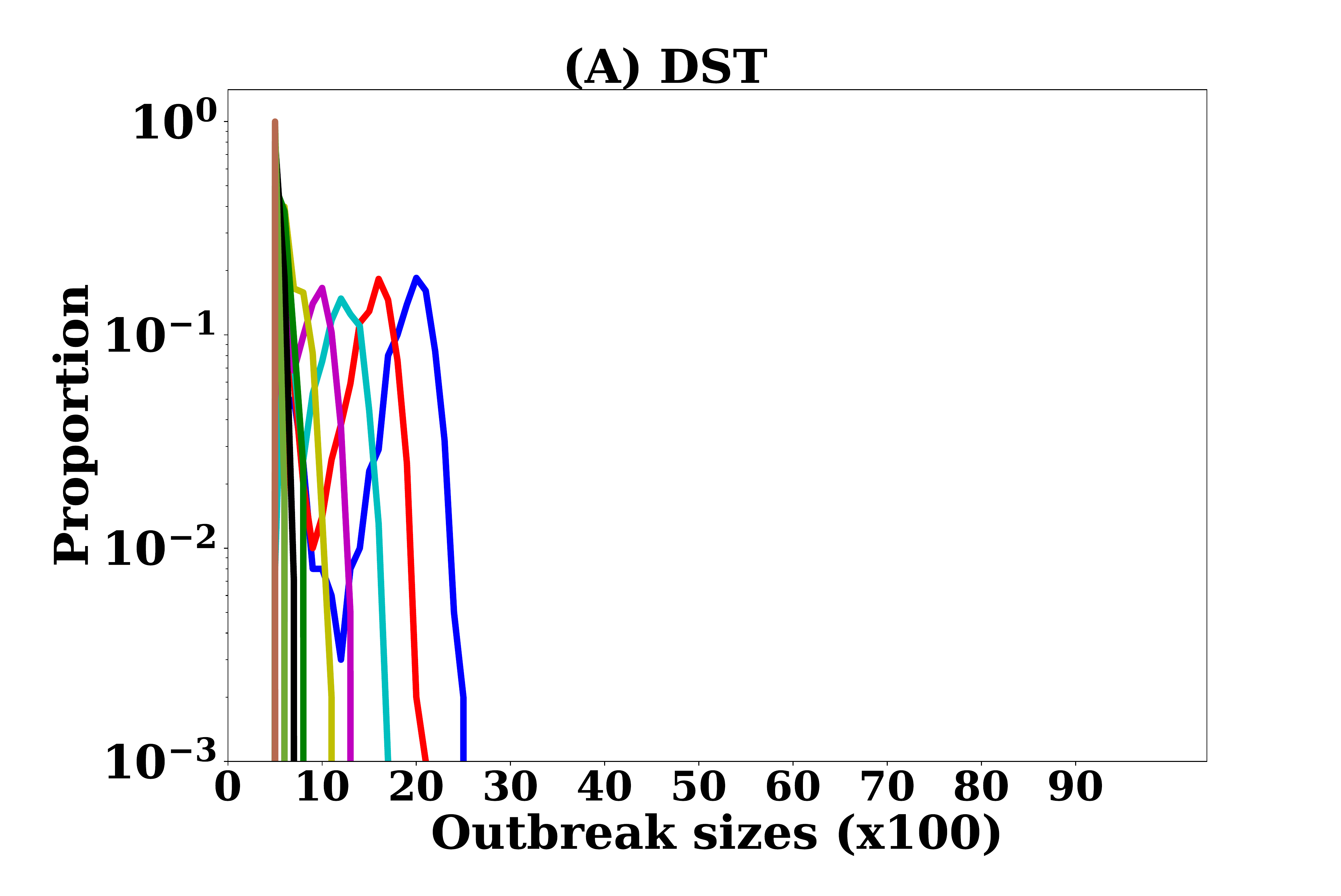}~
\includegraphics[width=0.48\linewidth, height=5.5 cm]{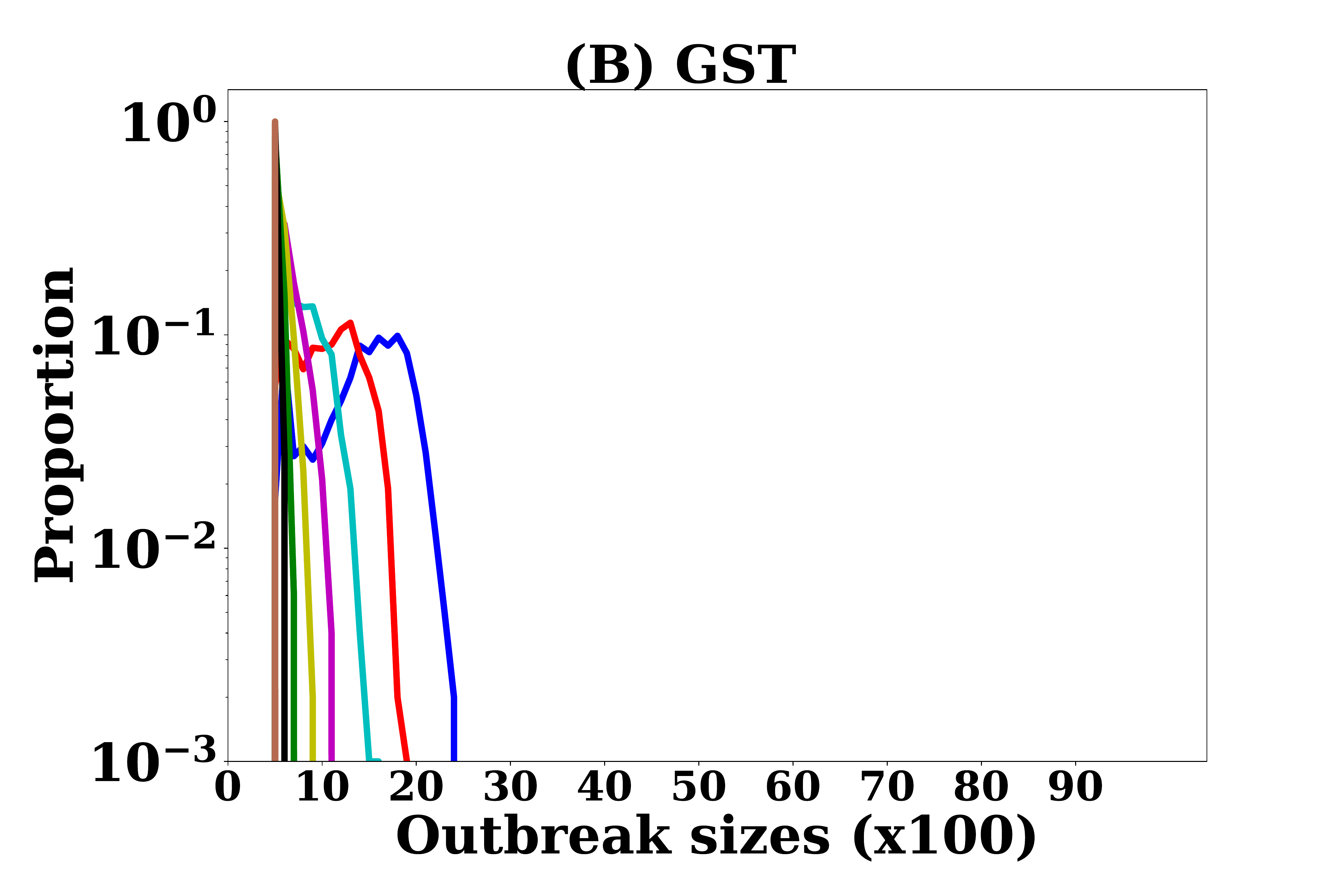}\\
\includegraphics[width=0.48\linewidth, height=5.5 cm]{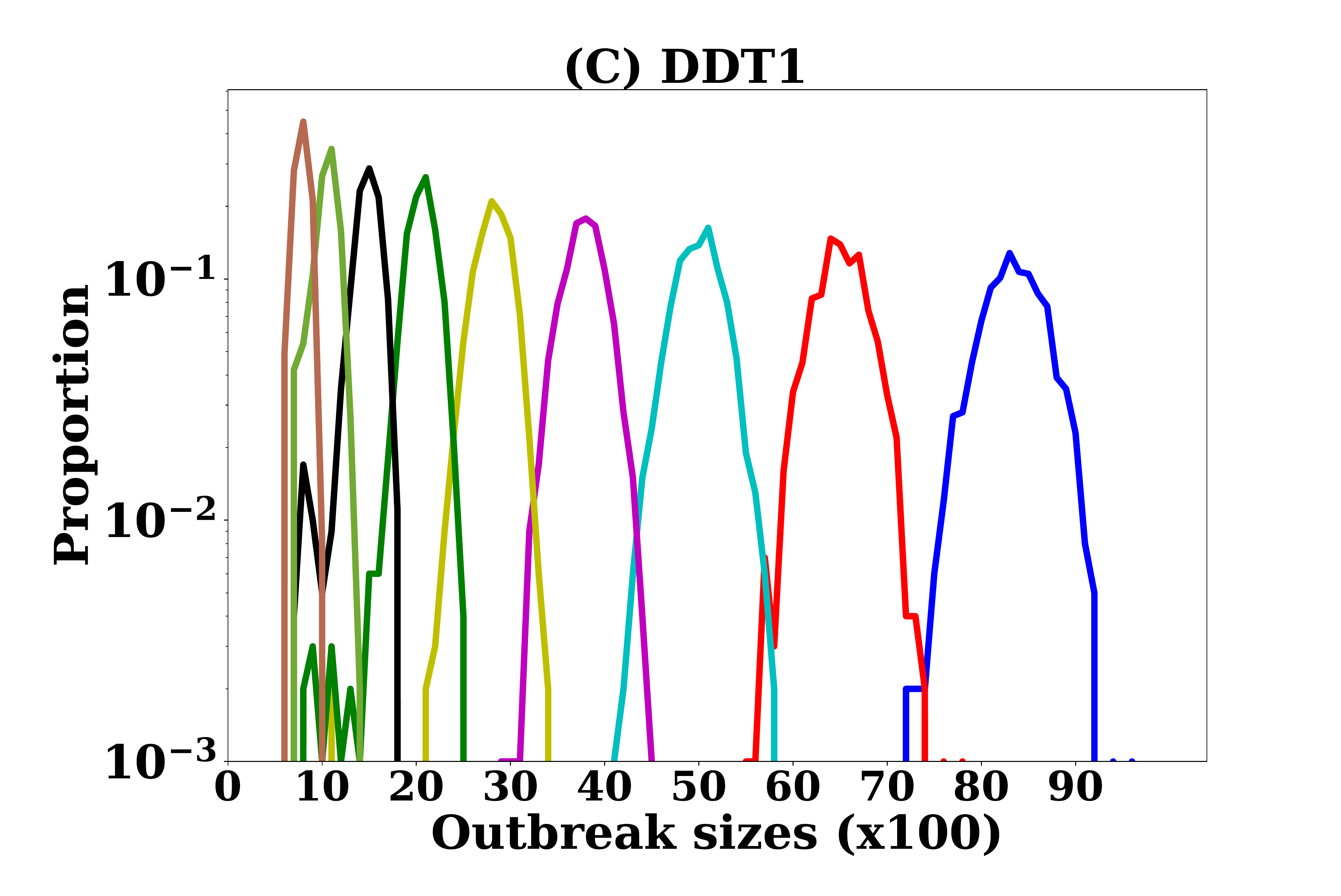}~
\includegraphics[width=0.48\linewidth, height=5.5 cm]{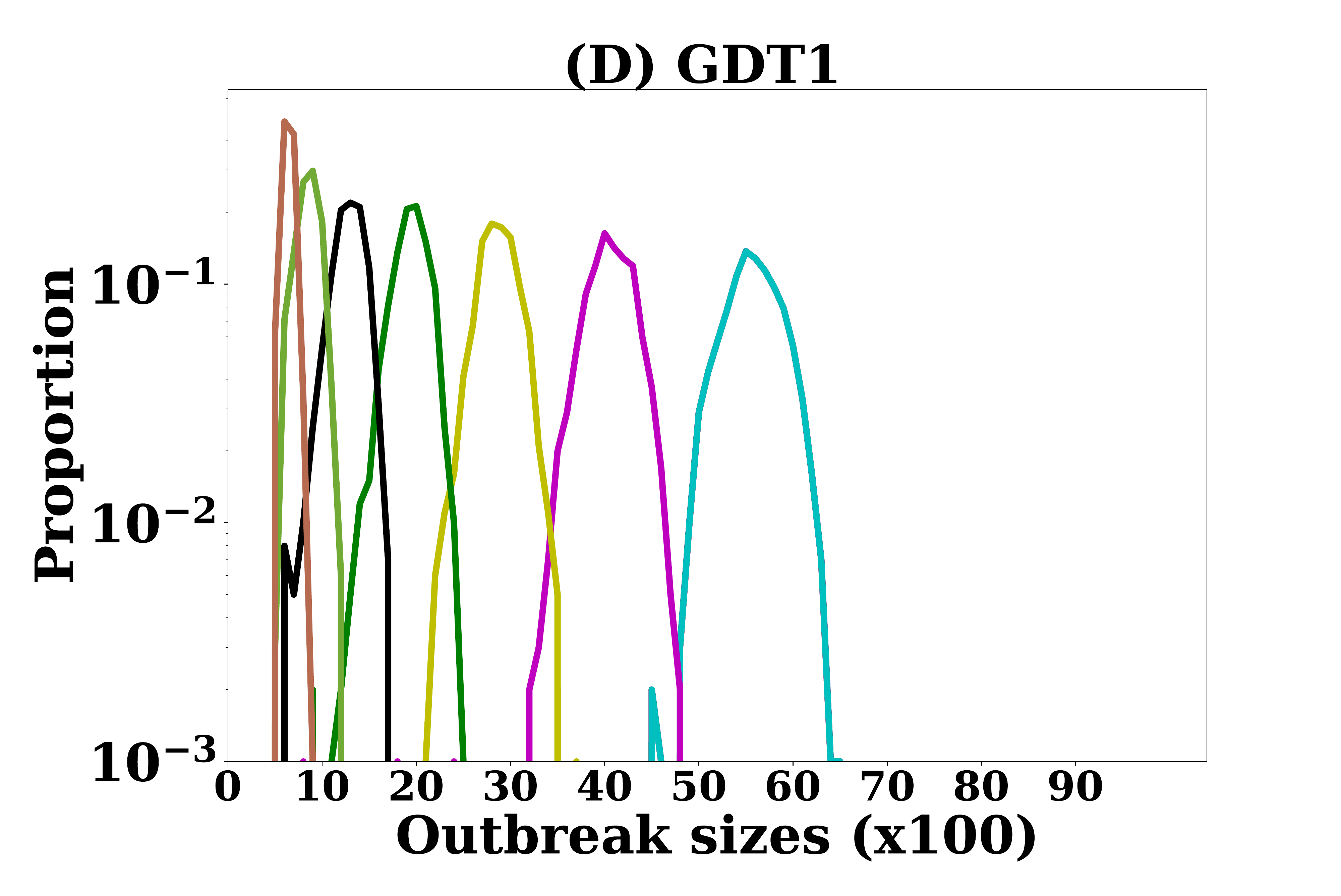}\\
\includegraphics[width=0.48\linewidth, height=5.5 cm]{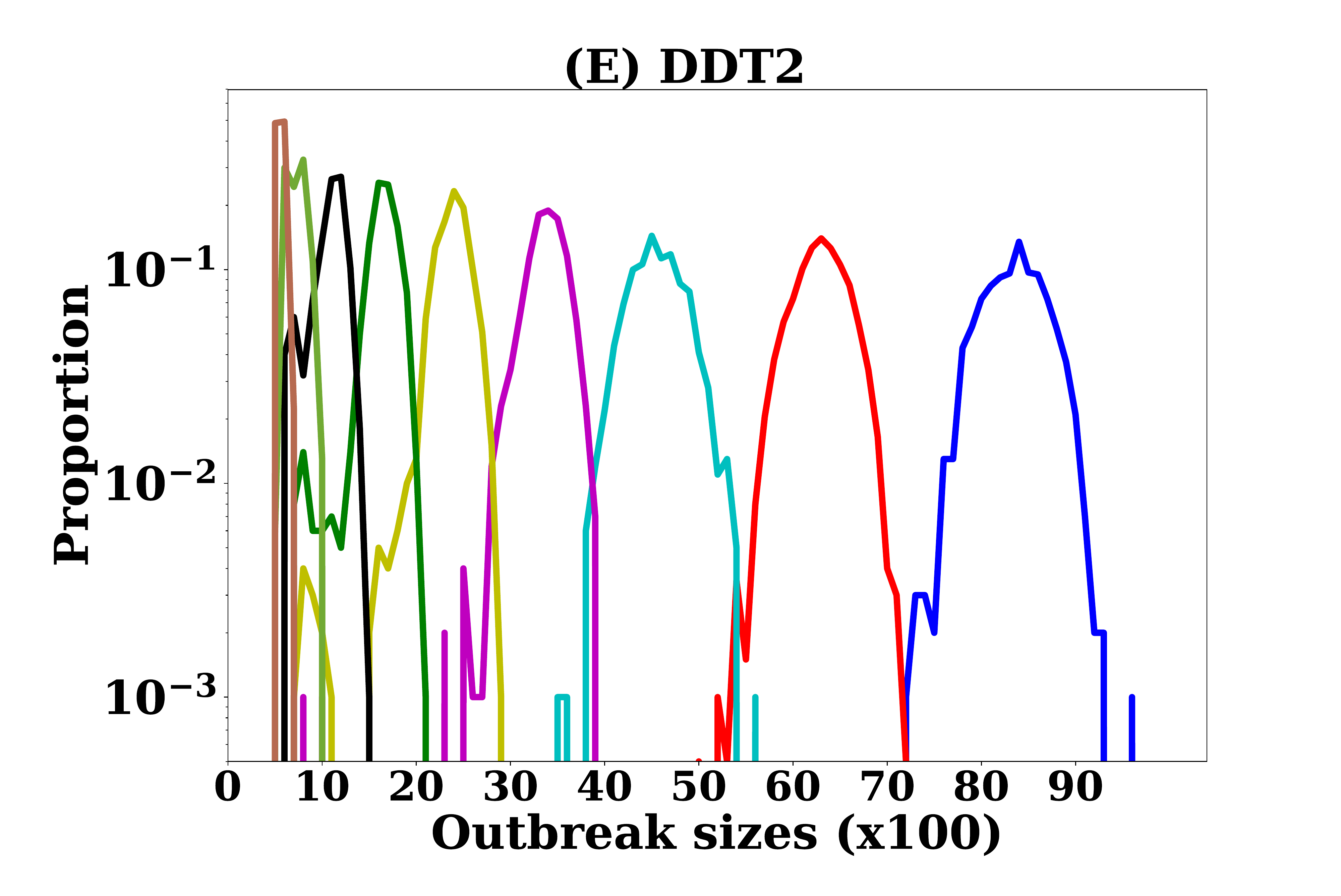}~
\includegraphics[width=0.48\linewidth, height=5.5 cm]{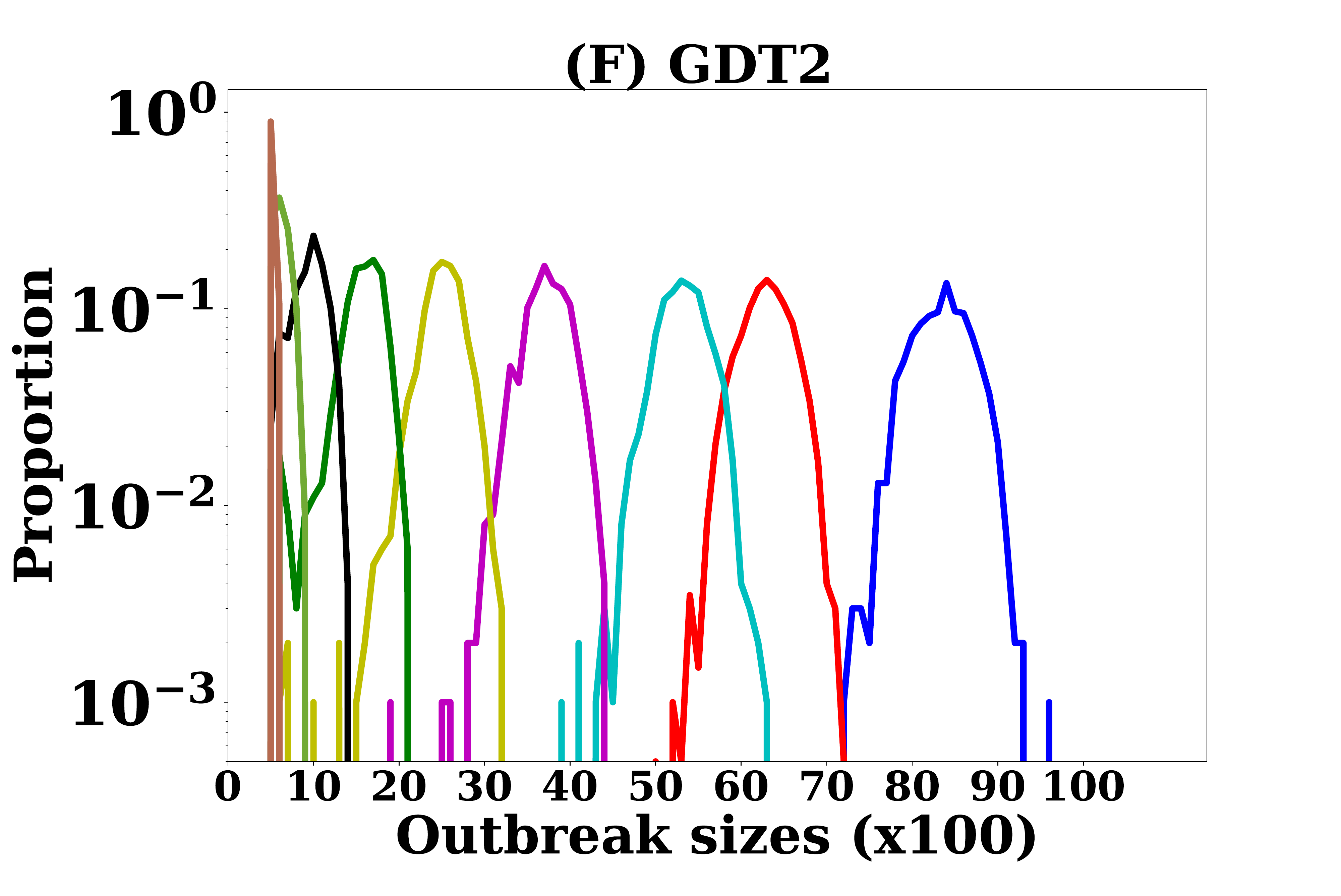}
\vspace{-1.0em}
\caption{Distribution of outbreak sizes in ring vaccination with $R$- left column real networks and right column synthetic networks: (A, B) vaccination with direct links and spreading on the SPST networks, (C, D) vaccination with direct links and spreading on SPDT networks, and (E, F) vaccination with any links and spreading on SPDT networks}
\label{fig:rvc1}
\vspace{-1.0em}
\end{figure}

\begin{figure}[h!]
\centering
\includegraphics[width=0.48\linewidth, height=5.5 cm]{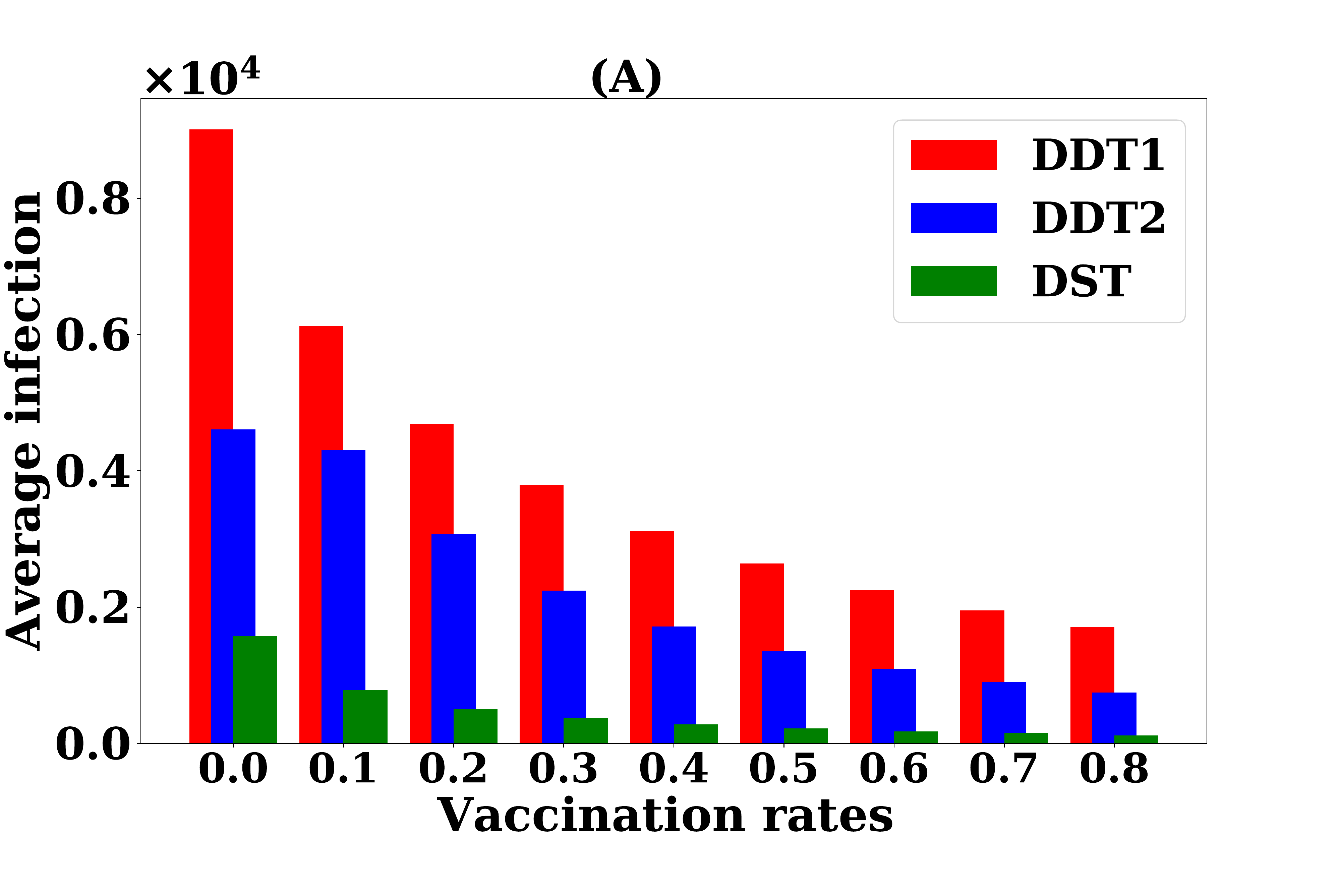}~
\includegraphics[width=0.48\linewidth, height=5.5 cm]{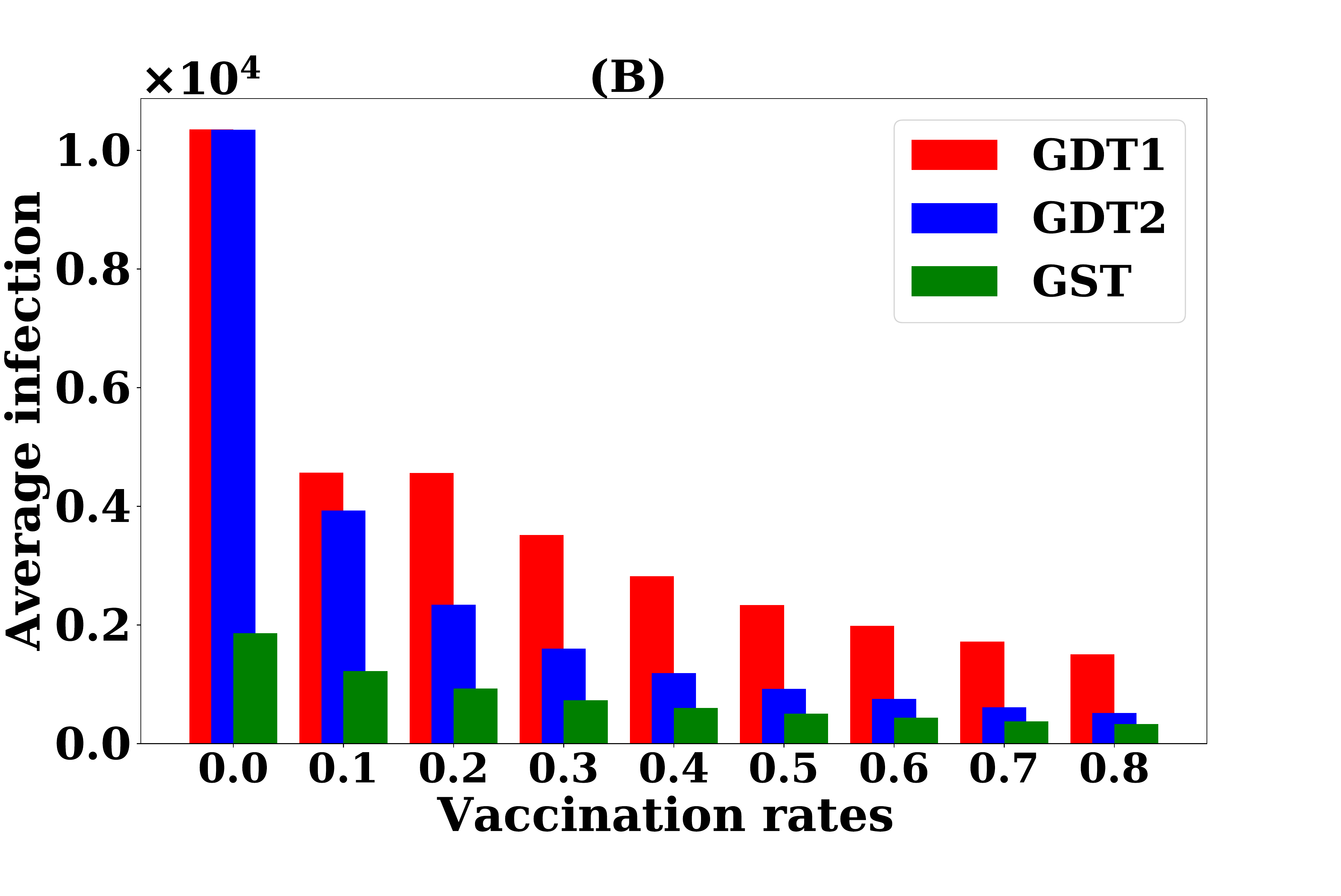}\\ \vspace{-1.0em}
\includegraphics[width=0.48\linewidth, height=5.5 cm]{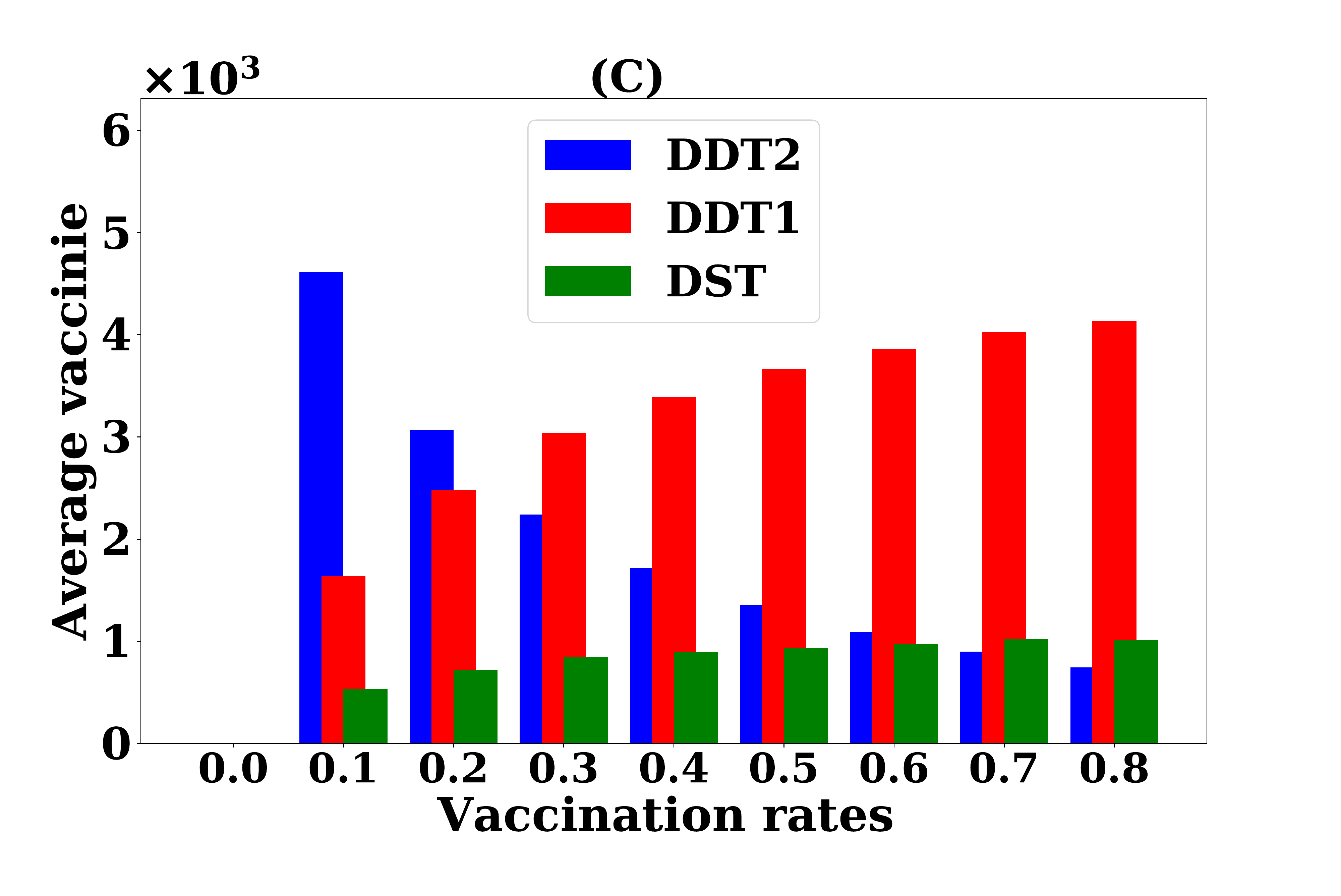}~
\includegraphics[width=0.48\linewidth, height=5.5 cm]{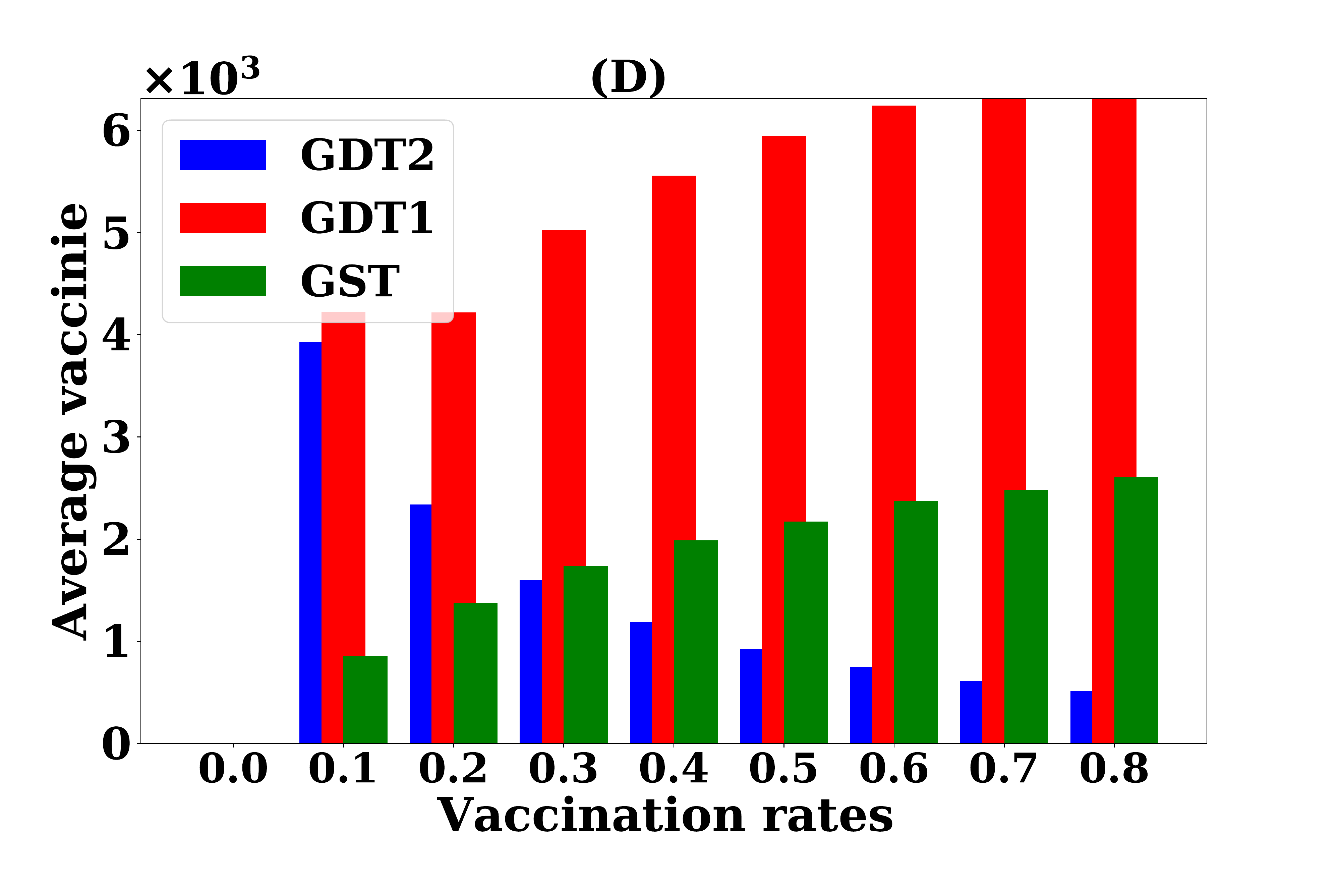}
\vspace{-1.0em}
\caption{Performance of ring vaccination strategy with $B$ : (A, B) average infection obtained summing all outbreak sizes and divided by 5000 and (C, D) average number of nodes vaccinated}
\label{fig:rvc2}
\vspace{-1.0em}
\end{figure}

\subsubsection{Ring vaccination}
In the ring vaccination strategy, a proportion of neighbouring nodes of infected nodes are vaccinated to hinder the spreading of disease. The efficiency of ring vaccination strategy is studied for $P=0$ to $P=0.8$ with a step value of 0.1 on both the real and synthetic networks, where $P$ is the proportion of susceptible neighbour nodes of infected nodes to be vaccinated. Various configuration of node selection is analysed to understand the strength of indirect links to spread disease and invade the networks. In the first experiment, the outbreak sizes on SPST networks are analysed for various $P$. In the SPST networks (DST and GST), users are only connected with direct transmission links. The $P$ proportion of neighbour nodes are vaccinated from the contact set of neighbour nodes whom the infected node has contacted in the past. After selecting a node, its disease status is changed to recover state at the simulation day when an infected node first get infected. Therefore, the selected neighbour nodes to be vaccinated are not infected at the next days of simulations by any infected nodes. Simulations are also run on the SPDT networks (DDT and GDT) for vaccination strategy where neighbour nodes who are connected through only direct links with the infected nodes are vaccinated. Thus, hidden spreaders are not vaccinated in this experiment (DDT1 and GDT1). To understand the impacts of indirect links in the neighbour selection for vaccination, vaccination strategy is implemented with selecting neighbour nodes contacted with a direct or indirect link or both links (DDT2 and GDT2). The simulations start with randomly selected 500 seed nodes, assuming that the outbreak has already started and disease prevalence reaches at 500 infections when it is noticed. Now, disease preventing measure is taken at time $T=0$. The other settings for simulation are taken from the previous experiments. For each value of $P$, 1000 simulations are conducted and the frequencies of outbreak sizes are shown in the Figure~\ref{fig:rvc1}. The average total infections and the average number of nodes vaccinated with vaccination rates $P$ are shown in Figure~\ref{fig:rvc2}.

Ring vaccination also shows a rapid elimination of the disease as $P$ increases in SPST networks (DST and GST). At vaccination rate $P=0.6$, the disease dies out causing infections below 100 in both the DST and GST networks when nodes to be vaccinated are selected based on direct transmission links (see~\ref{fig:rvc2}A and ~\ref{fig:rvc2}B). The number of nodes to be vacated for for vaccination rate $P=0.6$ in DST network is 1000 nodes and  2000 nodes. In addition, he distribution of outbreak sizes shows that no simulation can cause an outbreak larger than 1000 infections in the SPST networks at $P=0.4$ (see~\ref{fig:rvc1}A and ~\ref{fig:rvc1}B).  On the other hand, disease can spread with the outbreak sizes within the 2-4K infections in the SPDT networks at $P=0.4$ and the average total infections is about 2K infections in both the DDT and GDT networks for vaccinating nodes based on direct links (DDT1 and GDT1 plots). Even, when the disease dies out in SPST networks, the average total infection in DDT and GDT networks is about 1K infections. For SPDT networks, it is required to vaccinate 80\% neighbours nodes to keep the outbreak sizes below 100 infections. At this value of $P$, number of nodes to be vaccinated is about 4K nodes in DDT network and 6K nodes in GDT network. However, the hindering of disease spreading is improved significantly when the indirect transmission links are considered in vaccination (DDT2 and GDT2 plots). This mechanism of neighbour selection reduces the outbreak sizes as well decrease the number of nodes to be vaccinated. This requires only 1000 nodes to be vaccinated to contain outbreak sizes below 1K infections in both networks. Clearly, indirect transmission links have strong influence on diffusion control and increase invasion strength in SPDT networks.

%% file: 6_chap_prevac.tex
\chapter{Controlling SPDT Diffusion}
Controlling diffusion has numerous real-life applications and hence there has been a wide range of investigation to develop efficient diffusion controlling strategies for various applications. Current diffusion controlling methods only account for the direct transmission links. The previous chapter has shown that the inclusion of indirect links with the SPDT diffusion model significantly increases the disease spreading through the network and degrades the performance of the current diffusion control methods. Thus, proper diffusion controlling methods are required to control diffusion phenomena that have indirect transmission links.

This chapter considers vaccination strategies as a case study of controlling diffusion and proposes a strategy for efficient vaccination to hinder infectious disease spreading on dynamic contact networks that have direct and indirect transmission links. The proposed vaccination strategy and other popular strategies are analysed for both preventive (pre-outbreak) and reactive (post-outbreak) vaccination scenarios. The primary challenge to implement a vaccination strategy is to collect contact information of individuals. Thus, diffusion control methods that use higher order network metrics such as degree, betweenness and clustering coefficient etc. are not appropriate - collecting these metrics is infeasible in realistic scenarios.
This chapter, therefore, investigates vaccination strategies using local information only that work in both SPST and SPDT models. The research question of controlling diffusion dynamics on dynamic contact networks having indirect transmission links is thus addressed in this chapter.

\input{6.1_chap_intro}
\input{6.2_chap_prevcmdl.tex}
\input{6.3_chap_method.tex}
\input{6.4_chap_preventive.tex}

\input{6.5_chap_postout.tex}

\section{Discussion}
This chapter has investigated vaccination strategies for controlling SPDT diffusion. A simple vaccination strategy called individual movement based vaccination (IMV) is developed using the local contact information. This strategy can work with the coarse-grained contact information instead of exact contact information. The IMV strategy is examined for both preventive and post-outbreak scenarios. The performance is compared with other three vaccination strategies namely random vaccination (RV), acquaintance vaccination (AV) and degree based vaccination (DV). For preventive vaccination scenarios, the proposed IMV strategy shows the performance of DV strategy where 6\% of nodes to be vaccinated for achieving efficiency where no seed nodes have outbreaks of more than 100 infections. However, the RV strategy requires vaccination of 70\% nodes to achieve the same efficiency and the AV requires vaccination of 40\% nodes. The sensitivity of IMV strategy is tested against applying exact contact information and temporal information. There are no significant improvements in applying that information. The vaccination strategies are also studied against the scale of information availability $F$ on nodes. It is found that all strategies have the maximum performance of RV strategy if 50\% of nodes provide contact information for the vaccination procedure. However, IMV and DV strategies show better performance with the higher infection cost for $F<0.5$. Then, the performance of IMV strategy is examined in the post-outbreak scenarios. Post-outbreak vaccination is implemented in two ways population-level vaccination and node-level vaccination. In this case, the IMV strategy shows a similar performance than the DV strategy and requires vaccination of 4\% nodes to contain the outbreak sizes below 1K infections. On the other hand, RV and AV strategies require a higher number of nodes to be vaccinated with 70\% and 40\% respectively. Under the constrained of information availability, the IMV strategy shows better performance in the population level vaccination as well. In the node-level vaccination, the number of nodes to be vaccinated for containing outbreaks about 1K infections with IMV strategy reduces significantly and it is about 2K nodes (0.75\%). This is similar to that of DV strategy. Under the constrained of information availability, the number of nodes to be vaccinated for containing outbreak sizes below 1k infections does not change for $F>0.25$. On the other hand, the RV strategy requires to up 10K (about 3\% of total nodes) nodes at $F=1$ and it is increased if $F$ is reduced. The proposed strategy IMV performs better than RV and AV strategies in both preventive and post-outbreak scenarios.  

%% file: 6.1_chap_intro.tex
\section{Introduction}
One of the main applications of diffusion modelling is controlling diffusion dynamics on contact networks. In the most of the methods, a set of individuals, who have a strong influence in spreading contagious items, is identified and the spreading behaviours of these individuals are then altered to reduce or speed up the rate at which the contagious items can spread~\cite{al2018analysis,scholtes2016higher,kas2013incremental}. An essential task for controlling strategies, therefore, is to identify an appropriate set of individuals and process them to modify the spreading dynamics. The size of the selected set of individuals should be minimal and should require realistic resources to achieve the control goal. Note that finding an optimum set of individuals is relevant to many other challenges such as information collection complexity, allocated budget and required performance of the controlling methods etc~\cite{holme2017cost}. This chapter investigates the development of appropriate diffusion control methods on dynamic contact networks having indirect transmission links. Vaccination strategy is one of the diffusion controlling methods applied to prevent future outbreaks and to slow the ongoing disease outbreaks.  

Infection transmission probability from an infected individual to susceptible individuals depends on the movement and interaction patterns of both susceptible and infected individuals. These interaction patterns are often analysed to find appropriate vaccination strategies~\cite{mao2009efficient,miller2007effective}. Applying detail information of individual interactions to develop a strategy will often result in an effective vaccination. However, the higher order interaction information is quite difficult to obtain for real-world social contact networks due to collection complexity and privacy issues. Thus, the established diffusion control methods such as betweenness centrality, eigenvalue and closeness etc. ~\cite{scholtes2016higher,kas2013incremental,taylor2017eigenvector} are often infeasible for vaccination purposes and vaccination strategies are often developed based on locally obtainable contact information. This chapter, therefore, focuses to study vaccination strategies using local contact information.

There has been a wide range of methods to control diffusion on contact networks using the local contact information. The simplest one is the random vaccination (RV) where a proportion of a population is randomly chosen to be vaccinated~\cite{madar2004immunization,cohen2003efficient,lelarge2009efficient}. This strategy requires a high number of individuals to be vaccinated. The other frequently discussed vaccination approach is the acquaintance vaccination (AV) where a random individual is approached and asked to name a friend~\cite{deijfen2011epidemics}. A name recommended by multiple individuals increases the preference to be vaccinated ~\cite{cohen2003efficient}. This strategy avoids the random selection of RV strategy and provides an opportunity to select individuals who have contact with many other individuals. This approach is called a targeted vaccination where most influential individuals are removed from the disease transmission path. However, AV strategy also requires a large number of individuals to be vaccinated. There have been several modifications of AV strategy to make it more efficient ~\cite{holme2017three}. However, these strategies are only studied on static contact networks. The work of ~\cite{lee2012exploiting} has upgraded AV strategy by prioritising the most recently contacted neighbours and assigning weights to their links to capture contact frequency. These vaccination strategies use the recommendation about a neighbouring individual from another individual. However, such information is often inaccurate and hence an optimal set of individuals may not be selected for vaccination.

This chapter proposes a local contact information based strategy, called the individual's movement based vaccination (IMV) strategy, where individuals are vaccinated based on their movement behaviours. It is very common that individuals do not provide accurate contact information when they are asked about their past interactions. Instead, they give a rough idea about their contact information. In the proposed strategy, individuals are ranked based on their movements to public places. Individuals are asked about the frequencies of visit to the different classes of locations (which are classified based on the intensity of individual visits to those locations such as the popularity of the places). The locations are labelled with a range of values indicating the number of individuals that can be met if the location is visited by an individual. Based on this coarse-grained information, ranking scores are estimated for the sample individuals (picked up as candidates for vaccine) from a population and the highest ranked individuals are vaccinated. The performance of the proposed vaccination strategy is studied in the case of airborne infectious disease spreading. The susceptible-infectious-recovered (SIR) epidemic model is applied to simulate disease spread on dynamic contact networks. First, final outbreak sizes are obtained without vaccination. Then, the effectiveness of the proposed strategy is analysed by reducing the outbreak sizes compared to that of without vaccination. The efficacy of the proposed strategy is compared with the three other strategies: random vaccination(RV), acquaintance vaccination (AV), and degree-based vaccination (DV) where higher degree individuals are vaccinated~\cite{pastor2002immunization}.

Experiments carried out to study vaccination strategies for both preventive and post-outbreak vaccination scenarios. As individuals can only mention their direct interactions, vaccination strategies are first investigated by ranking nodes based on direct interactions. Then, it is examined how the vaccination strategies are affected if both the direct and indirect interactions are accounted for selecting the nodes to be vaccinated. As the proposed strategy depends on the coarse-grained information, the vaccination performance can deviate from that of the exact contact information based vaccination. Thus, the vaccination is analysed by using the exact contact information. The proposed strategy is designed for dynamic contact networks even though the temporal information is not integrated with the node selection process. An investigation is done to understand the impacts of integrating temporal information with the proposed strategy. The other important focus of this chapter is to examine the effectiveness of vaccination strategies with respect to the scale of information collection on the node's contact~\cite{vidondo2012finding}. The contribution of this chapter as follows:

\begin{itemize}
\item Developing a new vaccination strategy using the local contact information
\item Studying preventive vaccination with the developed strategy
\item Analysing post-outbreak vaccination with the developed strategy
\item Studying the performance of vaccinations with the scale of information availability
\end{itemize}

%% file: 6.2_chap_prevcmdl.tex
\section{Method and materials}
This section describes the proposed vaccination strategy and the applied methods for analysing its effectiveness to reduce disease spreading. Simulations of airborne disease spreading are carried out on a real contact network (DDT) and a synthetic contact network (GDT) to investigate vaccination strategies. The description of the methods applied is as follows.

\subsection{Proposed strategy}
The proposed vaccination strategy, called the individual's movement based vaccination (IMV) strategy, is developed based on the individual's movement behaviours and propensity to interact with other individuals. The importance of the individual with respect to the spreading of infectious disease is determined by the individual's past movement behaviours and interaction propensity. This can be reflected by ranking scores to the individuals. Then, the ranking scores are used as the indicators of individuals influence in spreading the disease in the near future. In the IMV strategy, the locations where individuals visit for daily activities such office, restaurants, shopping malls and schools etc. are classified based on the number of neighbour individuals an individual can meet if they visit these locations. The locations are intuitively grouped into 6 classes as in Table~\ref{tab:cls}. This gives an idea of the effectiveness of the model and an accurate classification will the provide exact performance. Then, individuals are asked to tell the number of times they have visited the different classes of locations within a previous time period (as per Table~\ref{tab:aclsf}). The class range is taken assuming that individual usually may not remember or notice the exact number of the individual they have contacted in their past visits to different locations. Instead, they give a rough estimate of their contacts. Individuals also forget to mention their visits for short durations. Thirdly, it easier to gather movement information of individual instead of collecting information for every single visit. 

\begin{table}[h!]
\centering
\vspace{-1.0 em}
\caption{Classification of visits nodes do during their daily activities}
\vspace{2ex}
\label{tab:cls}
\begin{tabular}{|c|c|c|}
\hline
class & contact sizes & locations\\
\hline
class-1 & 1-5 & home,store\\ \hline
class-2 & 6-15 &  coffee shops, bus stops \\ \hline
class-3 & 16-25 & shopping mall, office, local train stations, small parks \\ \hline
class-4 & 26-50 & central train stations, large parks\\ \hline
class-5 & 51-100 &  university, college,  central train stations\\ \hline
class-6 & 101- & university, college, airport\\
\hline
\end{tabular}
\vspace{-1.0 em}
\end{table}

Now a generic method is developed to find the rank of an individual based on the given movement information. Assuming that a susceptible individual $v$ who has been at a location where an infected individual $u$ has visited gets infected with probability $\beta$. The probability of $v$ for not being infected due to this visit is $1-\beta$. If the infected individual $u$ meets $d$ individuals during this visit, the probability of transmitting disease to the neighbours through this visit is given by 
\begin{equation*}
w=1-\left(1-\beta\right)^{d}
\end{equation*}
where $\left(1-\beta\right)^{d}$ is the probability that no neighbour individual is infected from this visit~\cite{alvarez2019dynamic}. Here, the assumption is that the individual $u$ is only the source of infection. All the neighbour individuals are susceptible and contact with $u$ independently. Under these assumptions, $w$ is the spreading potential for a visit of an infected individual based on the number of individuals they meet at the visited location. Now, the spreading potential for visiting a location belonging to a class $i$ is defined as 
\begin{equation*}
w_i=\frac{1}{2}\left(2-\left(1-\beta\right)^{d_{i}^{1}}-\left(1-\beta\right)^{d_{i}^{2}}\right)
\end{equation*}
where $d_{i}^{1}$ is the lower limit of class $i$ and $d_{i}^{2}$ is the upper limit of class $i$. In fact, this is the average spreading potential for the class $i$ locations. Then, the ranking score of an individual for visits to different classes of locations is given as
\begin{equation} \label{rank}
W=\sum_{i=i}^{6} f_i\times w_i 
\end{equation}
where $f_i$ is the frequency of visit to a location belongs to class $i$. In this method, $W$ can be interpreted as the maximum number of disease transmission events during the observation period. As this score is a relative value, it will carry significant information even if the same neighbours are met repeatedly. This is because repeated interaction increases the disease transmission opportunity. In addition, $W$ indicates how easily $u$ get infected due to his movement behaviours and the propensity of interactions. Therefore, individuals are required to provide the following information (see Table~\ref{tab:aclsf}). Then, ranking scores are calculated and a set of a top-ranked individual is chosen to be vaccinated. The IMV strategy can account super-spreaders defined by the degree-based strategy and the intensity of interactions among individuals through $f_i$. It also accounts for the importance of places that 
DV do not consider. It is also easier to remember the visited locations than how many people one has met.

\begin{table}[h!]
\vspace{-0.5em}
\centering
\caption{Information collected from individuals}
\label{tab:aclsf}
\vspace{2ex}
\begin{tabular}{|c|c|c|c|c|c|c|}
\hline
Location classes & class-1 & class-2 & class-3 & class-4 & class-5 & class-6 \\ \hline
Frequency & & & & & & \\ \hline
\end{tabular}
\vspace{-1.0em}
\end{table}

%% file: 6.3_chap_method.tex
\subsection{Experimental setup}
The proposed vaccination strategy is analysed through simulating infectious disease spreading on the dynamic contact networks where individuals are represented as nodes and interactions among individuals as links. This study applies the following settings to simulate disease spreading with the vaccination of nodes. 

\subsubsection{Contact networks}
The dynamic SPDT contact networks capturing direct and indirect transmission links are applied to study the proposed vaccination strategy. The real dense SPDT contact network (DDT network) among Momo users is chosen as this network has shown realistic diffusion dynamics in the previous chapters. The DDT contact network represents the visits of location for a node through its active periods and also provides the number of neighbour nodes the host node contacts during the active period, i.e. for visiting a location. The contact network contains 364K nodes interacting over 32 days. To capture disease spreading for a longer time, the 32 days contact traces are extended to 42 days (6 weeks). In this extension, all links of a randomly selected day are copied to a day within 32 to 42 days. Thus, a DDT network for 42 days is obtained with the same nodes. A synthetic SPDT contact network (GDT) is generated for 42 days with 364K nodes using the developed SPDT graph model. The GDT network can verify the experiment results obtained from the DDT network. The SPDT links from the first week are used to find the ranking scores of nodes based on the applied vaccination strategies and the rest of them are used to simulate disease spreading.

\subsubsection{Disease propagation}
For propagating disease on the contact networks, a generic Susceptible-Infected-Recovered (SIR) epidemic model is adopted that is similar to that described in previous chapters. In this model, nodes remain in one of the three compartments, namely, Susceptible (S), Infectious (I) and Recovered (R). If a node $v$ in the susceptible compartment receives a SPDT link ($e_{uv}=\left(u(t_s,t_l),v, {t}'_s,{t}'_l\right)$) from a node $u$ in the infectious compartment, the former will receives the exposure $E_l$ of infectious pathogens according to the following equation developed in the Chapter 3
\begin{equation}
E =\frac{gp}{Vb^2}\left[b\left(t_i-t_s^{\prime}\right)+ e^{bt_{l}}\left(e^{-bt_i}-e^{-bt_l^{\prime}} \right) +e^{bt_{s}}\left(e^{-bt_l^{\prime}}-e^{-bt_s^{\prime}} \right)\right]
\end{equation}
where $g$ is the particle generation rate of the infected individual, $p$ is the pulmonary rate of the susceptible individual, $V$ is the volume of the interaction area, $b$ is the particles removal rates from the interaction area. The probability of infection is 
\begin{equation}\label{eq:vprob}
P_I=1-e^{-\sigma E}
\end{equation}
where $\sigma$ is the infectiousness of the virus to cause infection~\cite{fernstrom2013aerobiology}. If the susceptible individual moves to the infected compartment with the probability $P_I$, it continues to produce infectious particles over its infectious period $\tau$ days chosen randomly in the range (3,5) and then is deemed to be recovered. The parameters for Equation~\ref{eq:vprob} is the same as described in the previous chapters. Node stays infected up to $\tau$ days picked up from uniform random distribution from the range of 3-5 days maintaining $\bar{\tau}=3$ days~\cite{huang2016insights}. However, the infectious periods of seed nodes are varied depending on the requirements of experiments. The selected nodes for vaccination are moved to the recovered state in the simulations. Thus, they are not infected and do not spread disease.

\subsubsection{Baseline vaccination strategies}
The performance of the proposed vaccination strategy is compared to various vaccination strategies. Some of them are based on local contact information and shows to what extent the proposed strategy improves efficiency for controlling SPDT diffusion. The other strategy is based on the global information and shows to what extent the proposed strategy achieves the performance compared to that of global information-based models.

Random vaccination (RV): This is a straight forward way of vaccination where nodes are chosen randomly to be vaccinated~\cite{rushmore2014network,pastor2002immunization}. To implement this process in preventive vaccination, $P$ proportion of nodes are chosen randomly without knowing their contact behaviours and are vaccinated. The RV strategy is also applied in post-outbreak scenarios where $P$ proportion of neighbour nodes of an infected node are chosen for vaccination. This approach for post-outbreak scenarios is called ring vaccination with random acquaintance selection. 

Acquaintance vaccination (AV): In this strategy, a node is randomly approached and asked to name a neighbour node to be vaccinated~\cite{madar2004immunization,takeuchi2006effectiveness}. To rank the nodes in this strategy, each node present in the network during the first seven days is asked to name a neighbour node and then a list is prepared with each recommended names (the same name can be listed multiple times as they can be recommended by multiple friends). Then, the number of repetitions of each name is counted and ranking scores are obtained for all nodes in the network. The nodes who have contact with a large number of nodes may be named more frequently. Then, $PN$ nodes are chosen from the top-ranked nodes based on the naming score to vaccinate $P$ proportion of nodes, where $N$ is the number of total nodes in the networks. 

Degree based vaccination (DV):
The above strategies require high resources to prevent the spreading of a disease. The best approach to reduce resource cost is to vaccinate a set of target nodes that have strong spreading potential. The widely used targeted vaccination approach is to vaccinate the nodes that have the largest number of contacts (high degree nodes) as they are more prone to get infected and spread disease~\cite{pastor2002immunization,pastor2002immunization}. The contact set sizes for the first seven days of nodes in both networks are separated and the nodes are ranked based on the contact set sizes during this time. Then, $PN$ nodes are chosen from the top-ranked node based on the contact set sizes to vaccinate $P$ proportion of nodes. This strategy requires the information of all contacts a node has. It is often hard in practice to count all other individuals an individual has met. Thus, it is often infeasible in real implementations. However, it is of interest to understand how effective the proposed strategy is compared to DV strategy in reducing disease spread.

\subsubsection{Characterising metric}
For analysing the performance of a vaccination strategy, the first disease spreading is simulated on the selected contact networks without vaccination and the outbreak sizes after 42 days are obtained. These simulations are run from multiple seed node sets and the average outbreak size $z_{r}$ is computed. This indicates the propensity of disease to spread in a network without vaccination and is used as the reference for comparing effectiveness of the applied vaccination strategies. The performance of a vaccination strategy is quantified by how much reduction it can do in the average outbreak sizes comparing to $z_{r}$. Thus, the effectiveness of a vaccination strategy with a vaccination rate $P$ is given by
\[\eta=\frac{z_{r}- z_{c|P}}{z_r}\times 100\]
where $z_{c|P}$ is the average outbreak sizes of the candidate vaccination strategy with the vaccination rate $P$. In the simulation, the outbreak size indicates the number of new infections caused after running simulation over 32 days.

%% file: 6.4_chap_preventive.tex
\section{Preventive vaccination}
The preventive vaccination is implemented to prevent possible future outbreaks of disease. In this section, the effectiveness of the selected vaccination strategies to prevent future outbreaks in the empirical SPDT contact network (DDT network) and the generated synthetic SPDT contact network (GDT network) are investigated. First, simulations are conducted to understand the upper bound of the efficiency of the applied strategies where it is assumed that contact information of all nodes are available and can be collected for vaccinations. Then, the performance is investigated with the simulation scenarios where contact information of only a proportion of nodes are available for ranking and only these are chosen to be vaccinated. However, all nodes participate in propagating disease on the networks.

\subsection{Performance analysis}
\begin{figure}[h!]
\begin{tikzpicture}
    \begin{customlegend}[legend columns=4,legend style={at={(0.12,1.02)},draw=none,column sep=3ex ,line width=2pt,font=\small}, legend entries={RV, AV, IMV, DV, direct, indirect}]
    \addlegendimage{solid, color=blue}
    \addlegendimage{color=red}
    \addlegendimage{color=green}
    \addlegendimage{color=magenta}
    \addlegendimage{color=black}
    \addlegendimage{dashdotted, color=black}
    \end{customlegend}
 \end{tikzpicture}
 \centering
\includegraphics[width=0.48\linewidth, height=5.5 cm]{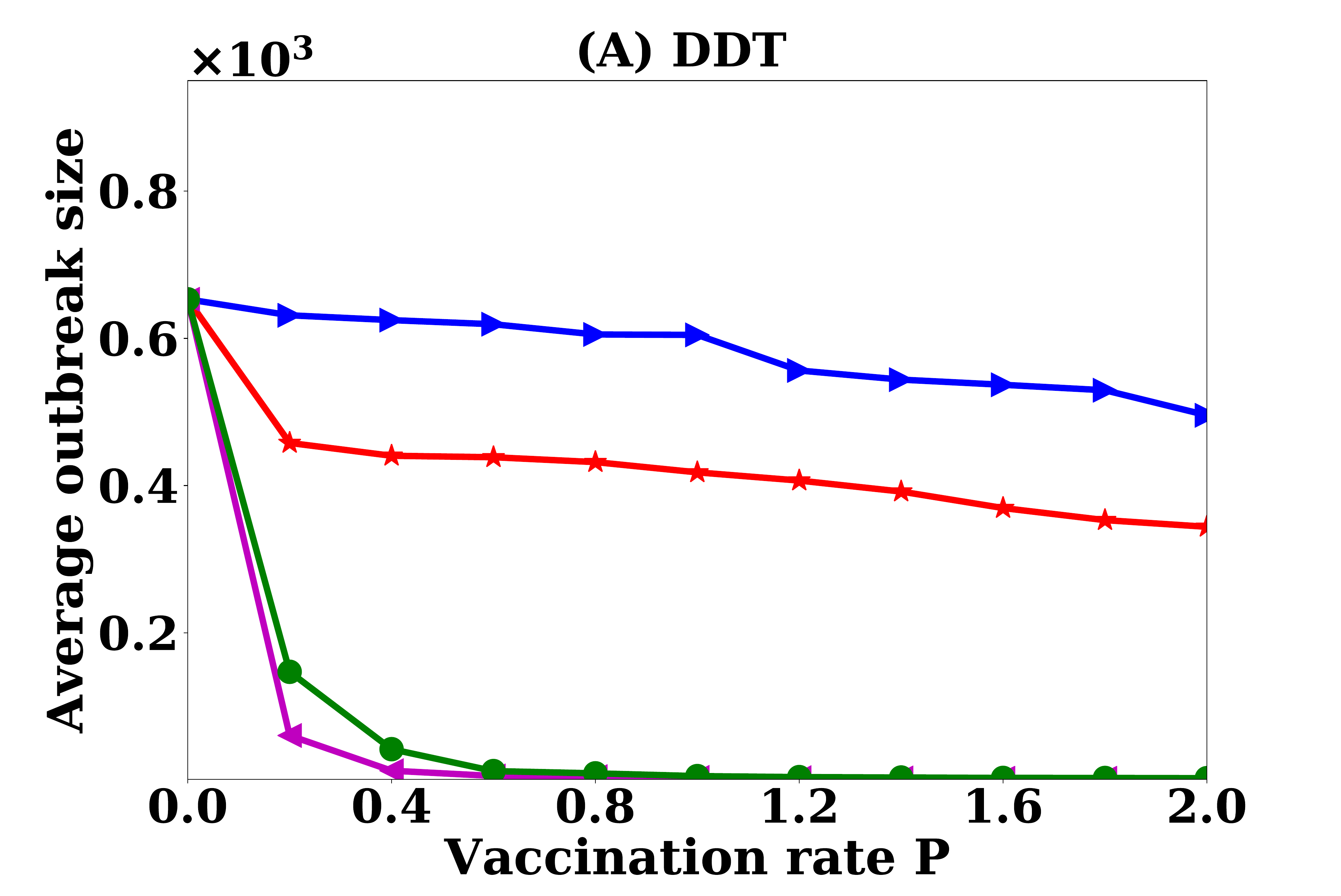}
\includegraphics[width=0.48\linewidth, height=5.5 cm]{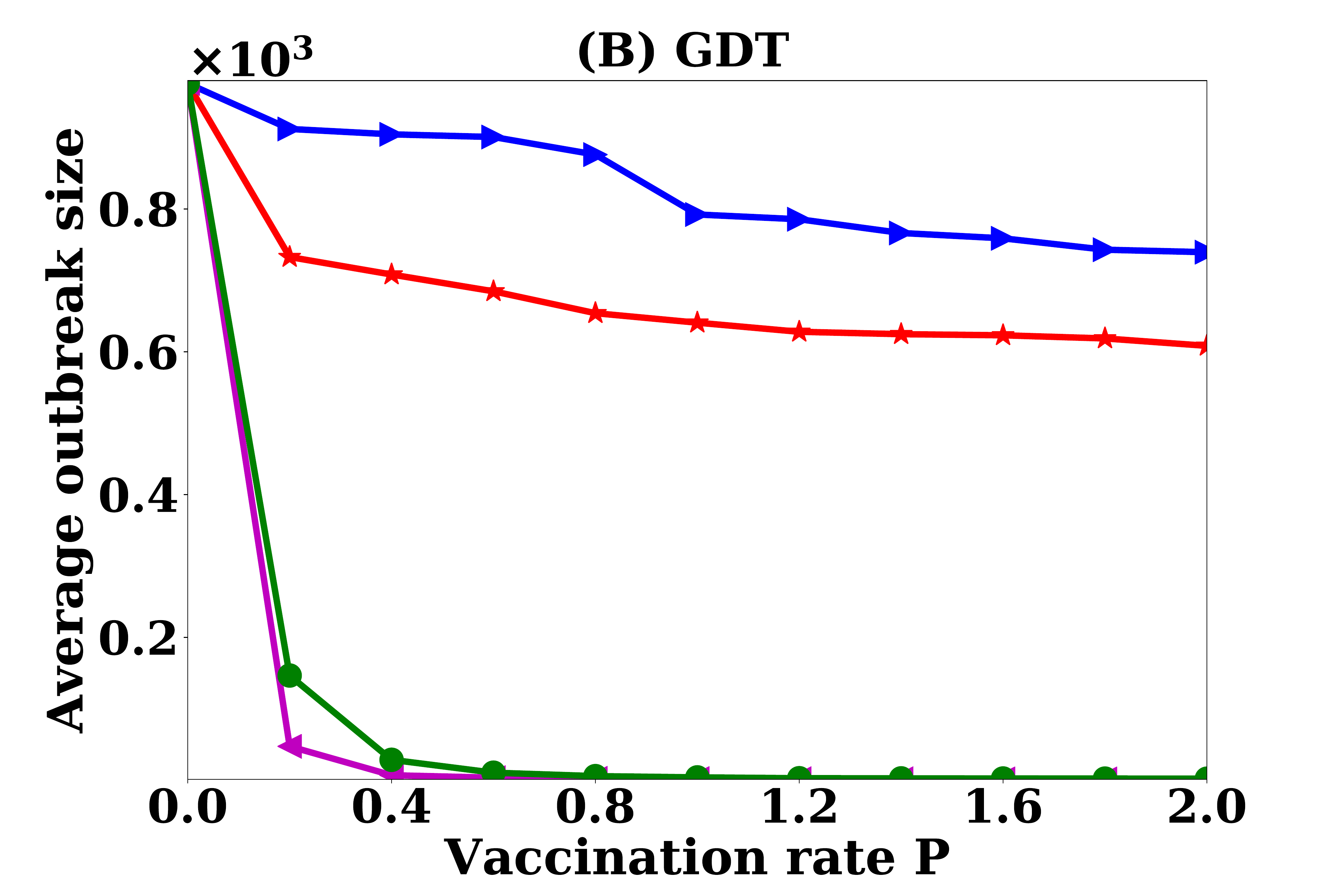}

\includegraphics[width=0.48\linewidth, height=5.5 cm]{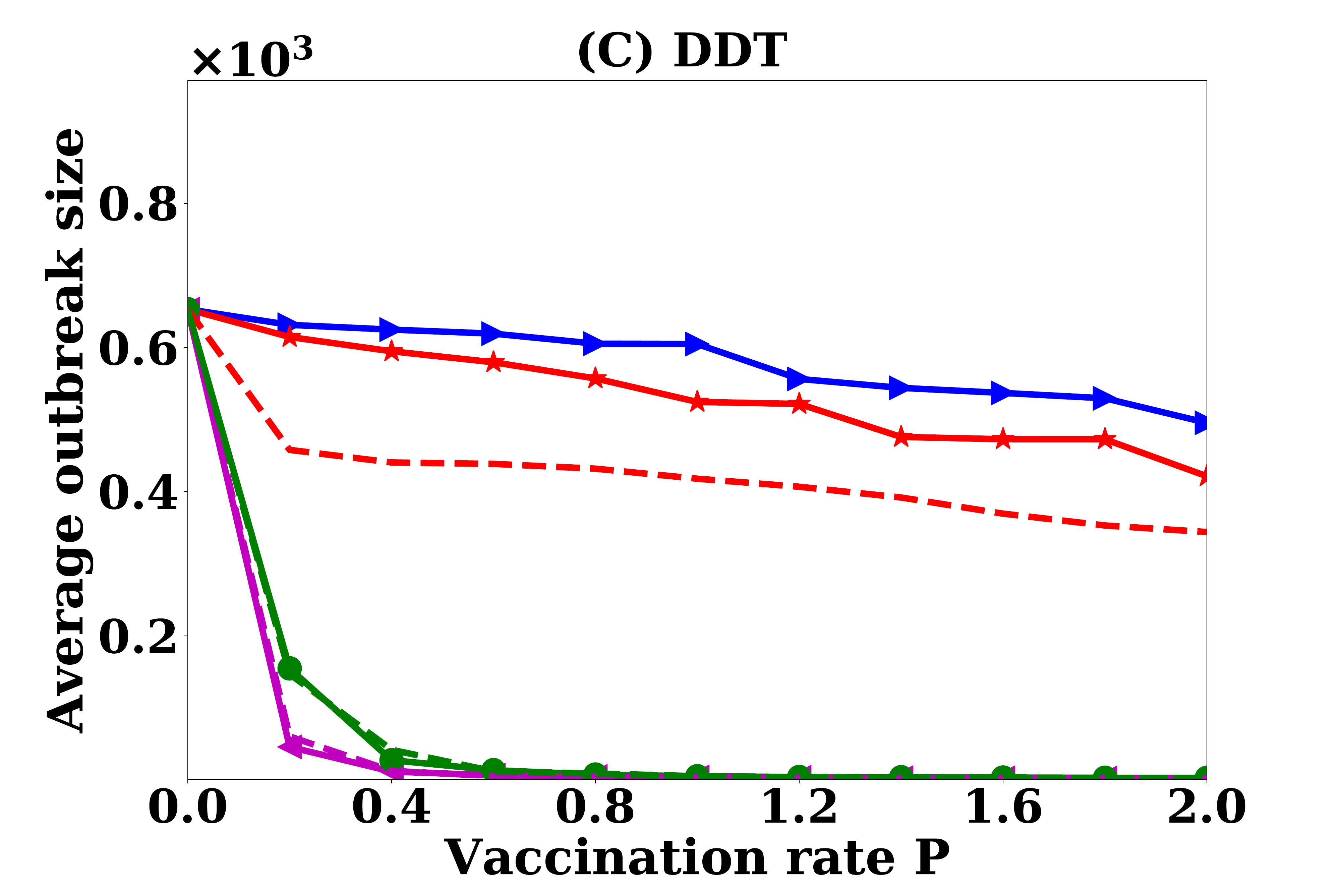}
\includegraphics[width=0.48\linewidth, height=5.5 cm]{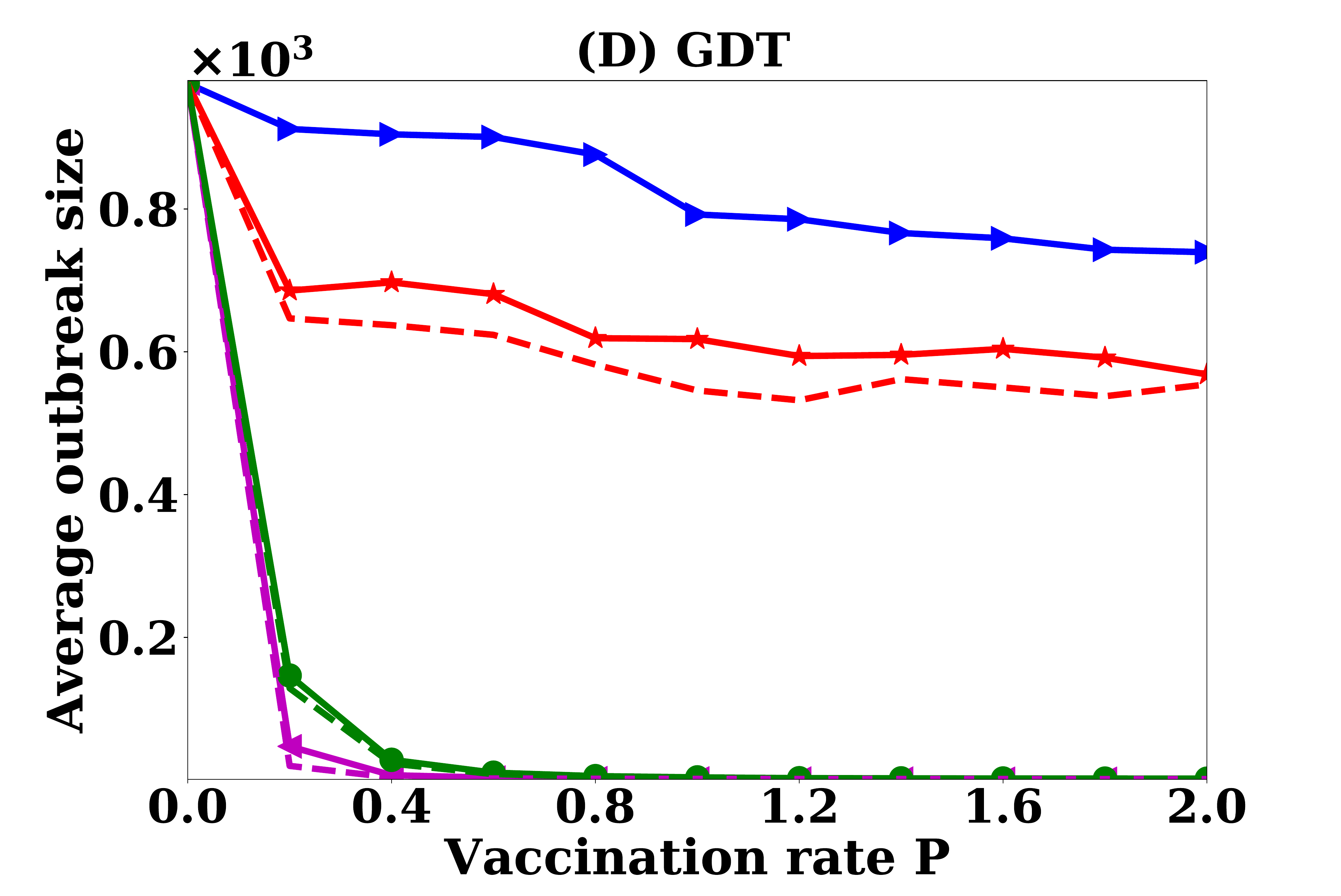}
\caption{Average outbreak sizes at various vaccination rates $P$ (percentage of total nodes) of different strategies: (A, B) nodes are vaccinated with contacts created for direct interactions and (C, D) comparison of outbreak sizes for vaccinating nodes with contacts based on the direct interactions (solid lines) and contact based on any direct or indirect interactions (dashed lines)}
\vspace{-1.5em}
\label{fig:avac}
\end{figure}

\begin{figure}[h!]
\begin{tikzpicture}
    \begin{customlegend}[legend columns=4,legend style={at={(0.12,1.02)},draw=none,column sep=3ex ,line width=2pt,font=\small}, legend entries={RV, AV, IMV, DV}]
    \addlegendimage{solid, color=blue}
    \addlegendimage{color=red}
    \addlegendimage{color=green}
    \addlegendimage{color=magenta}
    \end{customlegend}
 \end{tikzpicture}
 \centering
\includegraphics[width=0.48\linewidth, height=5.5 cm]{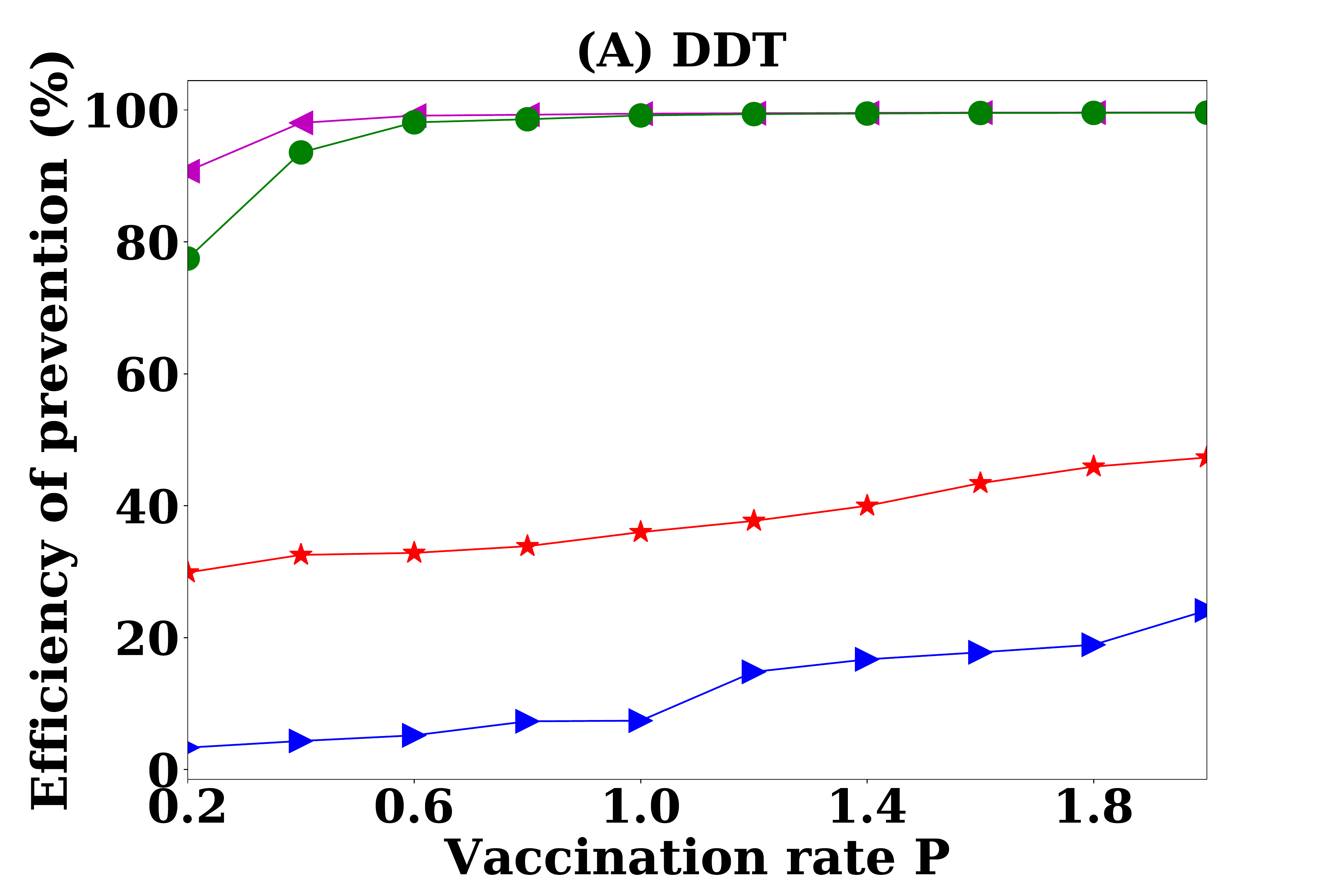}
\includegraphics[width=0.48\linewidth, height=5.5 cm]{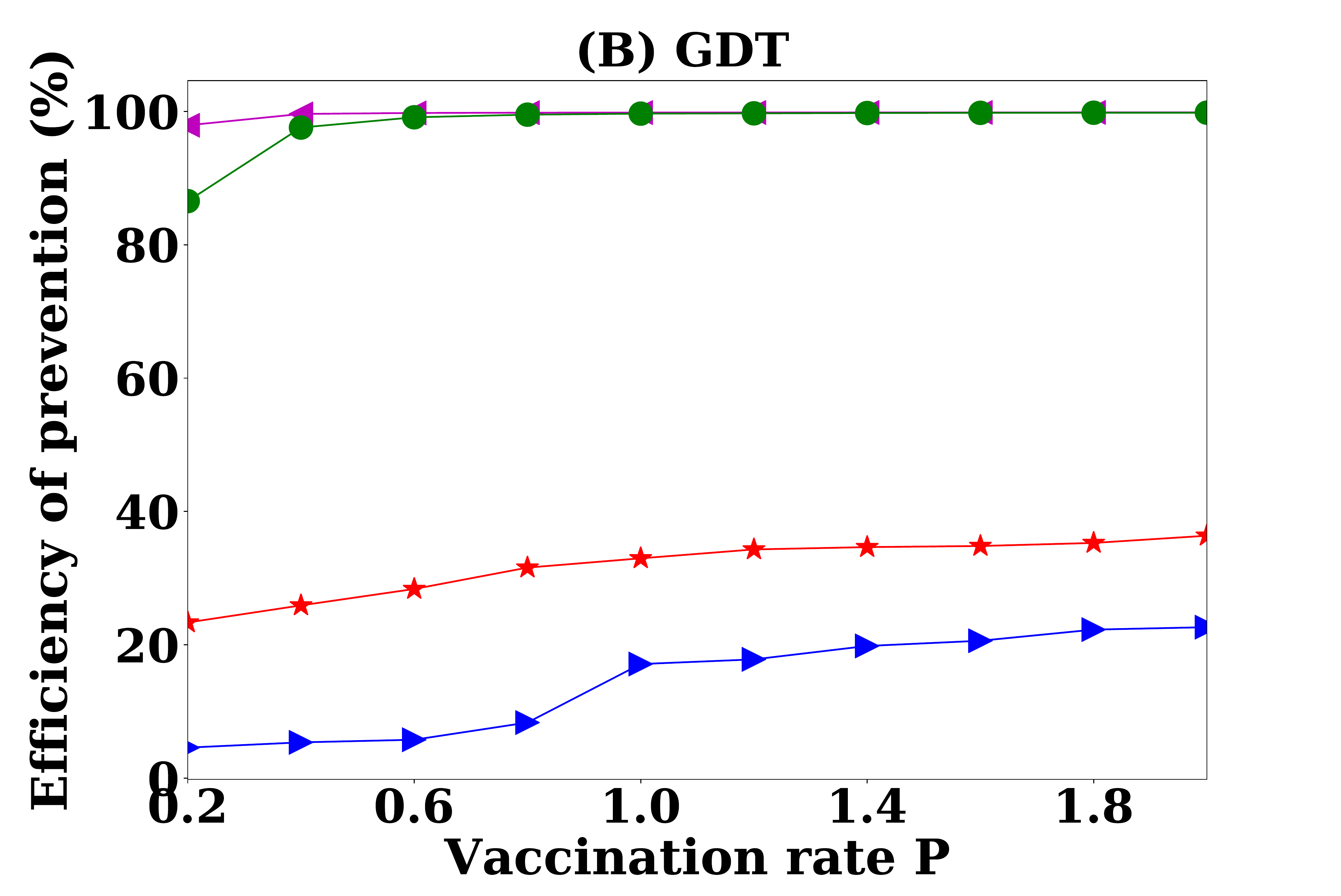}\\

\includegraphics[width=0.48\linewidth, height=5.5 cm]{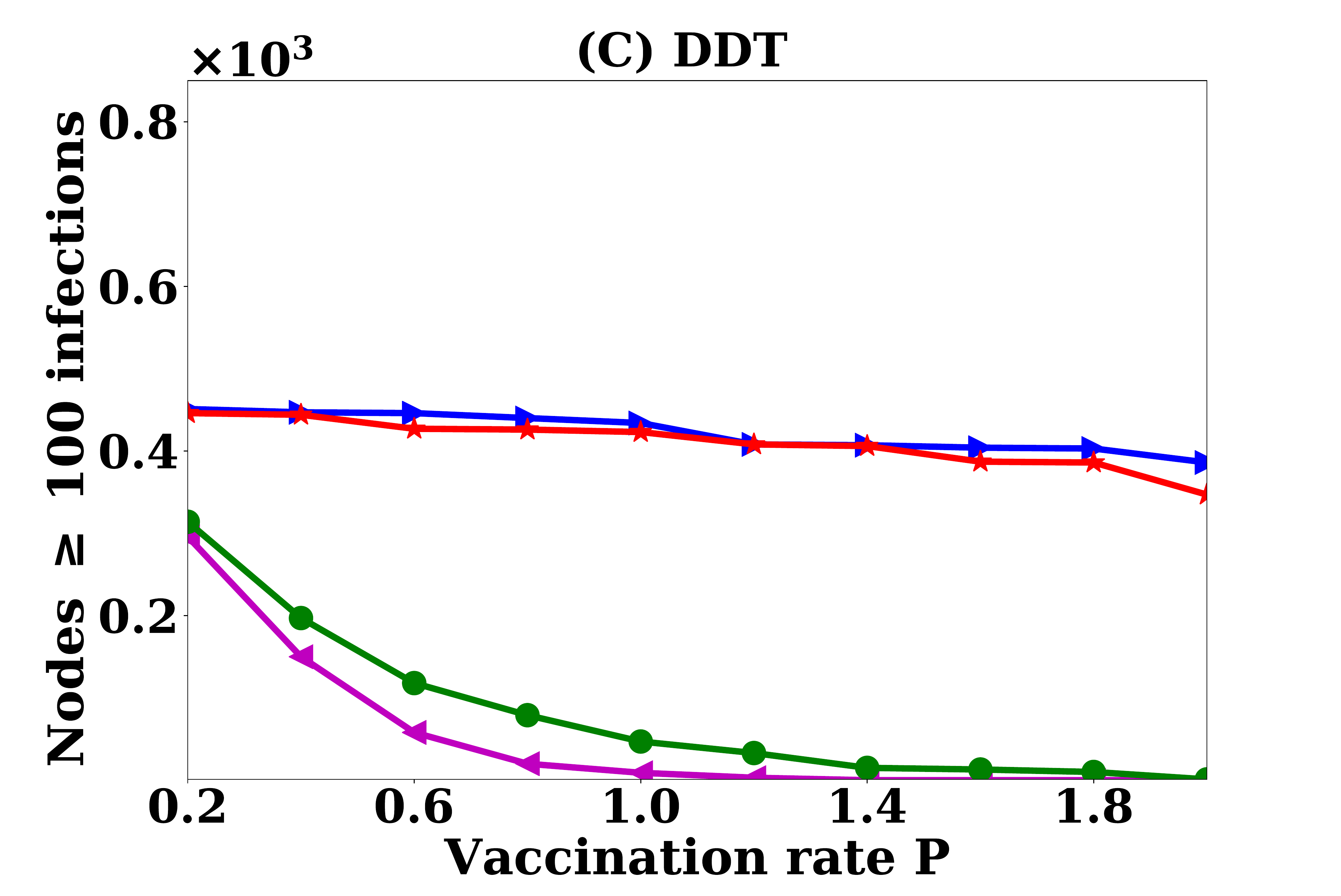}
\includegraphics[width=0.48\linewidth, height=5.5 cm]{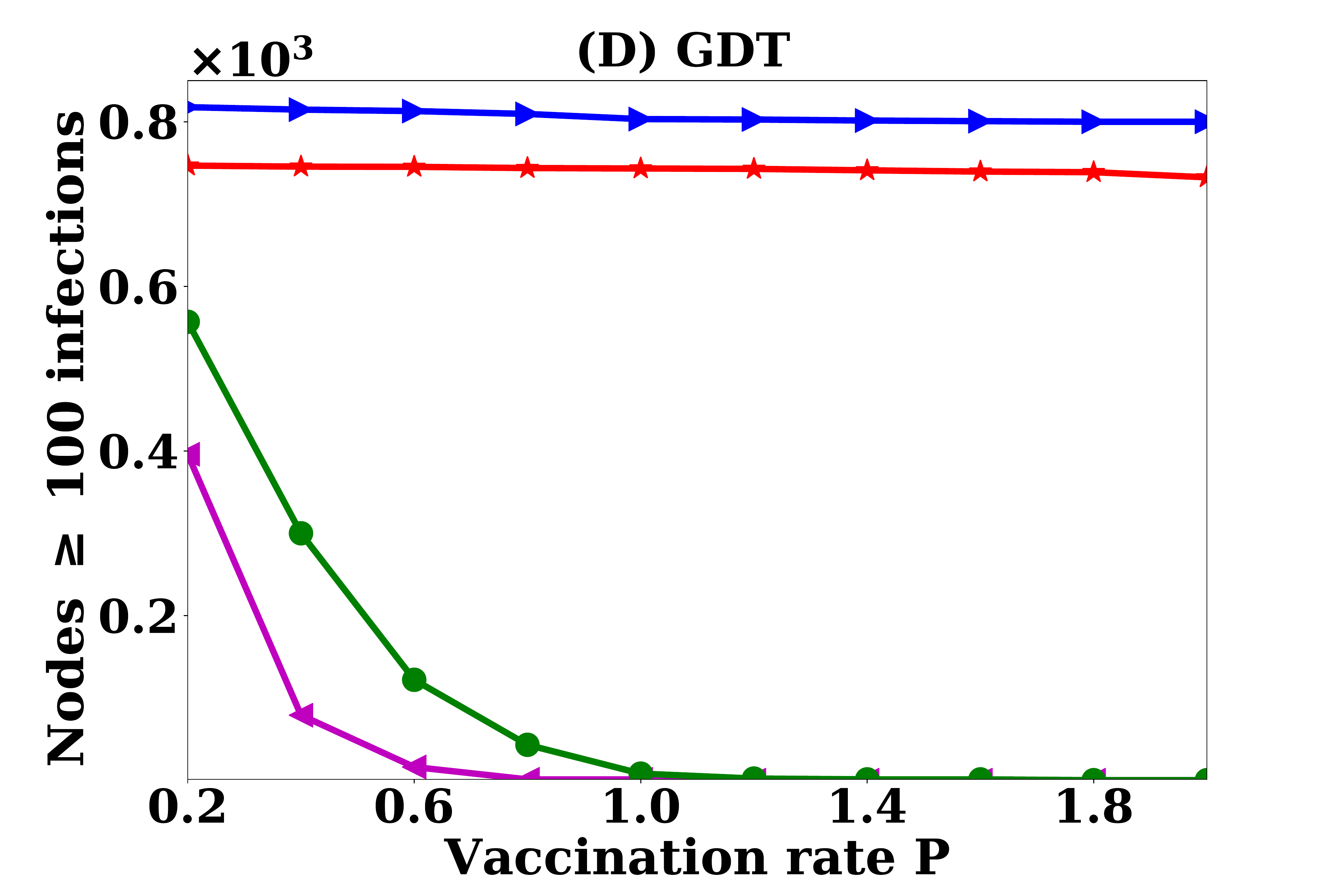}\\

\includegraphics[width=0.48\linewidth, height=5.5 cm]{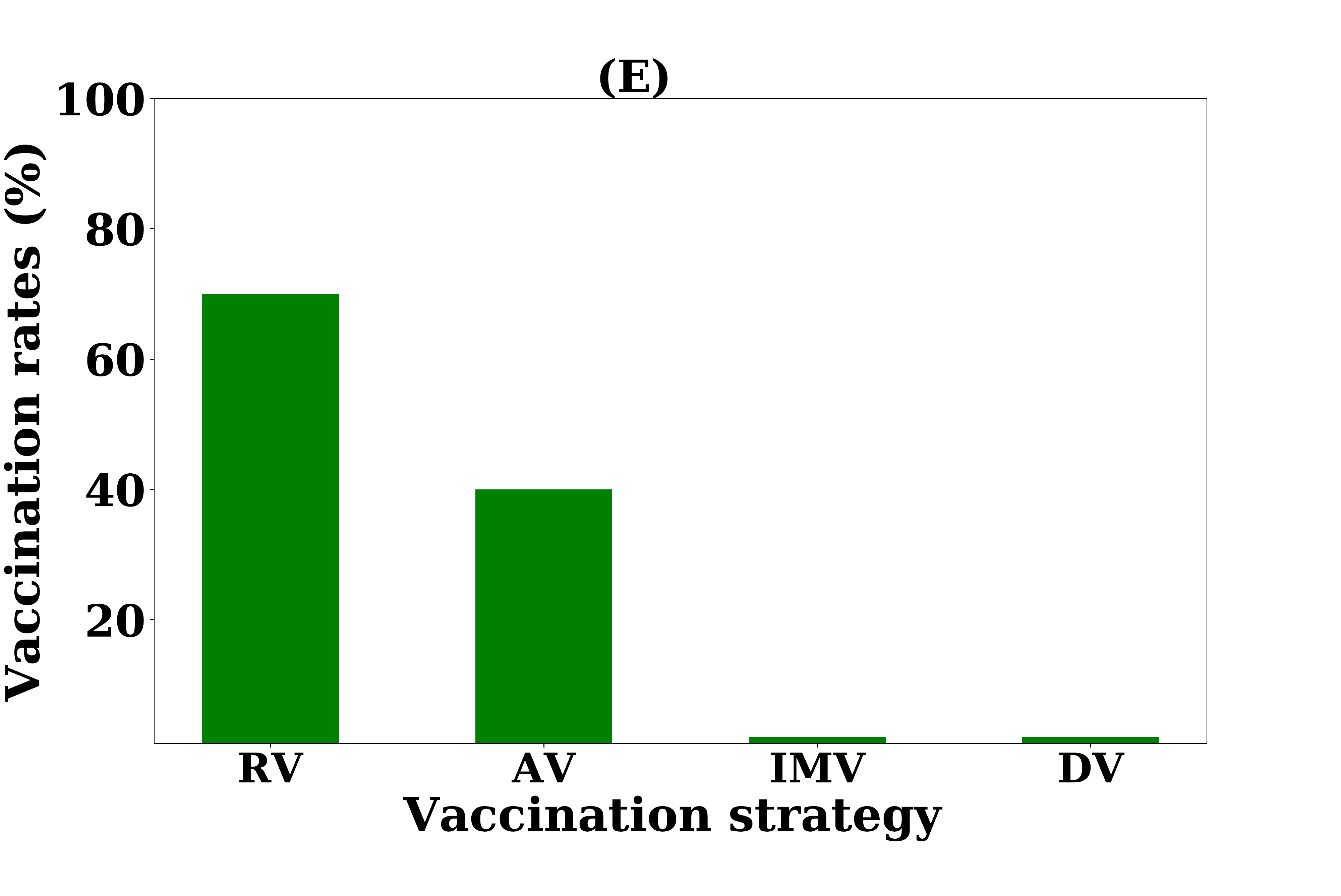}
\includegraphics[width=0.48\linewidth, height=5.5 cm]{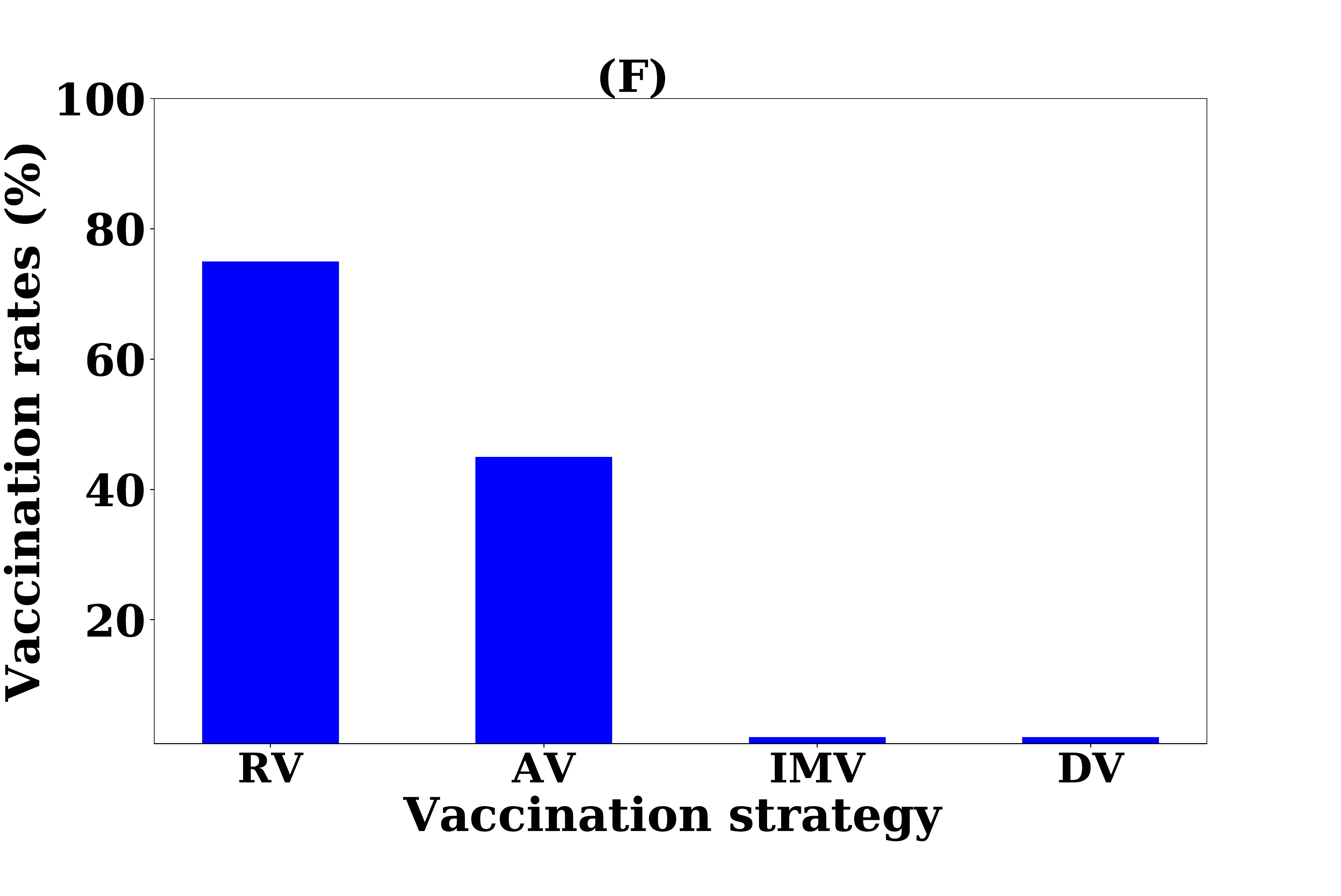}
\vspace{-1.0em}
\caption{Efficiency of vaccination strategies to prevent disease spreading: (A, B) preventive efficiency as a measure of reduction in average outbreak sizes due to vaccination, (C, D) number of seed nodes out of 5000 having outbreak sizes greater than 100 infections, and (E, F) required vaccination rates for strategies to keep outbreak sizes below 100 infections from all seed nodes}
\label{fig:vacef}
\vspace{-1.0em}
\end{figure}

The performances of vaccination strategies are studied with varying vaccination rates $P$ in the range [0.2-2]\% with the step of 0.2\%. Thus, $PN$ number of nodes are chosen based on the vaccination strategies and are vaccinated them assigning their status as recovered. Then, a random node is chosen as a seed node and outbreak size is obtained by running the disease spreading simulation over 35 days from the seed node. This process is iterated through 5000 different seed nodes. Seed nodes are infectious for 5 days and then recover. When individuals are asked about their movements, they usually provide the information based on the number of individuals they have seen in particular locations. They may not notice the number of individuals who have SPDT links through indirect interactions. Thus, the nodes are ranked based on the contacts they have seen, i.e. considering only the direct interactions where both the host and neighbour nodes stay together in a location. Then, the performance is also examined accounting the indirect links along with the direct links. The average of outbreak sizes for vaccination strategies with various $P$ are presented in Figure~\ref{fig:avac}. The average outbreak size without vaccination is obtained by simulating disease from 5000 single seed nodes in both networks. These are 653 infections in the DDT network and 976 infections in the GDT network. Based on these values, the efficiency is measured at each $P$ for all vaccination strategies and shown in Figure~\ref{fig:vacef}. To clearly understand the efficiency of a strategy, the number of seed nodes having outbreak sizes of more than 100 infections are also counted and presented in Figure~\ref{fig:vacef}. This indicates the efficiency of a strategy to hinder super-spreader nodes from spreading disease. If there is no seed nodes of 100 infections in a network, it is assumed that the protection from infection is sufficient for the applied strategy.

The results show that the average outbreak sizes for random vaccination (RV) are higher in both DDT and GDT networks at all values of $P$. Furthermore, increasing $P$ does not reduce the outbreak sizes significantly (see Fig.~\ref{fig:avac}). At $P=0.2$\% (with vaccinating 720 nodes out of total 364K nodes), the average outbreak size for RV strategy is about 624 infections in the DDT network and 912 infections in the GDT network. The average outbreak size reduces to 495 infections in the DDT network and 739 infections at $P=2$\% (with vaccinating 7200 nodes). Thus, the preventive efficiency of RV strategy is about 18\% in the DDT network and 16\% in the GDT network at the vaccination rate $P=2$\% (see Fig.~\ref{fig:vacef}). On average 451 seed nodes out of 5000 have outbreaks of more than 100 infections in the DDT network and 812 nodes in the GDT network at $P=0.2$\%. These number slightly reduces at $P=2$\%. The preventive performance of RV strategy with this range of vaccination rate is very poor. The proposed IMV strategy, however, shows significantly low average outbreak sizes even at $P=0.2$\% with 139 infections in the DDT network and 146 infections in the GDT network. The preventive efficiency of IMV strategy at $P=0.2$\% is 78\% in the DDT network and is 82\% in the GDT network. Unlike RV strategy, the average outbreak sizes quickly decrease with increasing $P$ in IMV strategy. If the vaccination rate is $P=0.6$\% (with vaccinating 2880 nodes), the average outbreak size becomes only 7 infections in DDT network and 10 infections in GDT network and the preventive efficiency is about 98\% in both networks. At $P=0.6$\%, only 71 nodes in the DDT network and 67 nodes in the GDT network have outbreaks of greater than 100 infections whereas they were 314 infections and 572 infections respectively in DDT network and GDT network at $P=0.2$\%. The seed nodes with outbreak sizes of greater than 100 infections become zero at $P=2$\% in DDT network and at $P=1.4$\% in GDT network. Simulations are also run by increasing $P$ for RV strategy until the preventive efficiency reaches the stage where no seed node has outbreak greater than 100 infections. It is found that RV strategy requires to vaccinate 70\% of nodes to achieve such preventive efficiency in both the DDT and GDT networks (see Fig.~\ref{fig:vacef}). Thus, the proposed local information based IMV strategy achieves significantly higher efficiency than the random vaccination strategy.

Acquaintance vaccination (AV) implements targeted vaccination using the local contact information and selecting random acquaintance. It is a modification of the AV strategy to implement targeted vaccination with local contact information. AV strategy improves the efficiency of the RV strategy. The results show that the proposed IMV strategy is still much better than the targeted AV strategy. At $P=0.2$\%, the average outbreak size in the DDT network is 614 infections and 735 infections in GDT network. Thus, the efficiency is 28\% in DDT network and 20\% in GDT network. However, the improvement in efficiency in AV strategy is still very low compared to that of IMV strategy. The IMV strategy shows only one seed node with outbreak size greater than 100 infections at vaccination rate $P=1.4$\% in GDT network while AV strategy still has about 700 seed nodes with outbreak sizes greater than 100 infections in the GDT network.  AV strategy requires about 40\% of nodes to be vaccinated to achieve preventive efficiency where no seed node has outbreaks greater than 100 infections (see Fig.~\ref{fig:vacef}). Thus, the IMV strategy achieves much better efficiency than the local information based targeted AV strategy.  IMV strategy also achieves the efficiency that is very close to that of global information based DV strategy (degree based vaccination).  DV strategy has average outbreak sizes of 47 infections in both the DDT and GDT networks at $P=0.2$\%. The preventive efficiency is about 92\% in the DDT network and 98\% in the GDT network. This is better than the IMV strategy. However, the efficiency of IMV strategy becomes closer to that of DV strategy as $P$ increases. To achieve strong preventive efficiency, both strategies require almost the same rates of vaccination. Thus, the local coarse-grained information based IMV strategy achieves the performance of degree based vaccination strategy. 

The above results are obtained by ranking the nodes based on the contacts created for direct interactions, i.e. neighbour nodes who are connected with direct links are only considered in the ranking process. Figure~\ref{fig:avac} shows (dashed lines) the results for vaccination strategies with indirect interactions, i.e. neighbour nodes connected with indirect links are also considered in calculating the ranking score for a strategy. It is observed that the DV strategy and proposed IMV strategy do not vary the average outbreak sizes largely for any vaccination rates due to including indirect links. As the movement information is the same for creating direct or indirect interactions, the performance of IMV strategy does not decrease. The performance of the other neighbour based strategy AV increases slightly when considering indirect interactions as some nodes may become important with indirect links and this is not captured by the direct interaction based implementation. In addition, some new neighbour nodes appear when indirect links are counted and this changes the score of RV strategy. Thus, the nodes are ranked more precisely in AV strategy when indirect links are counted and the effectiveness of vaccination increases.

\subsection{Sensitivity analysis}
The performance of the strategies is now studied for varying the scale of information availability regarding nodes contact. For IMV strategy, the coarse-grained information of an individual's movement is applied through classifying the location into groups. It is interesting to know how much improvement can be achieved if exact contact information of nodes is applied by ranking process. Accordingly, a new ranking score is defined as
\begin{equation}\label{eq:vtrans}
    w_i=1-(1-\beta)^{d_i}
\end{equation}
where $w_i$ is the probability of transmitting disease to neighbouring nodes for a visit $i$ at a location where infected nodes meet $d_i$ number of other nodes. Thus, the node's rank is given by
\begin{equation}
    W=\sum_{i=1}^{n}w_i
\end{equation}
This equation finds the score considering exact number of neighbouring nodes for each visit during the observation period. IMV strategy is analysed for various vaccination rate $P$ implemented with theses scores. The other important factor for disease spreading through visiting location is the duration of stay. If an infected individual stays longer at a location, they may transmit disease to more susceptible individuals. Thus, the temporal information is integrated with the ranking process as follows. It is assumed that the probability of transmitting disease for visiting a location increases exponentially. Therefore, the transmission probability $\beta$ is defined as 
\begin{equation}
    \beta=1.6 \beta_{0} \left(1-e^{-\frac{t_{i}}{t_0}}\right)
\end{equation}
where $t_i$ is the stay time of the node for a visit $i$ to a location, $t_0$ is the average stay time of users of Momo App, and $\beta_{0}$ is the disease transmission probability if one infected node and one susceptible node stay together for $t_0$ time. This equation captures the potential to transmit disease for a time together. Integrating this time dependency on transmission probability of Equation~\ref{eq:vtrans} can capture the impact of the duration of stay at the visited locations. Therefore, IMV strategy is also studied based the nodes rank with temporal information. Finally, performance of all vaccination strategies is studied for the scenarios where $F$ proportion of nodes provides contact information for ranking process and nodes to be vaccinated are chosen from them. The experiment is conducted for $F=\{0.25, 0.5, 0.75\}$. 

\begin{figure}[h!]
\includegraphics[width=0.48\linewidth, height=5.5 cm]{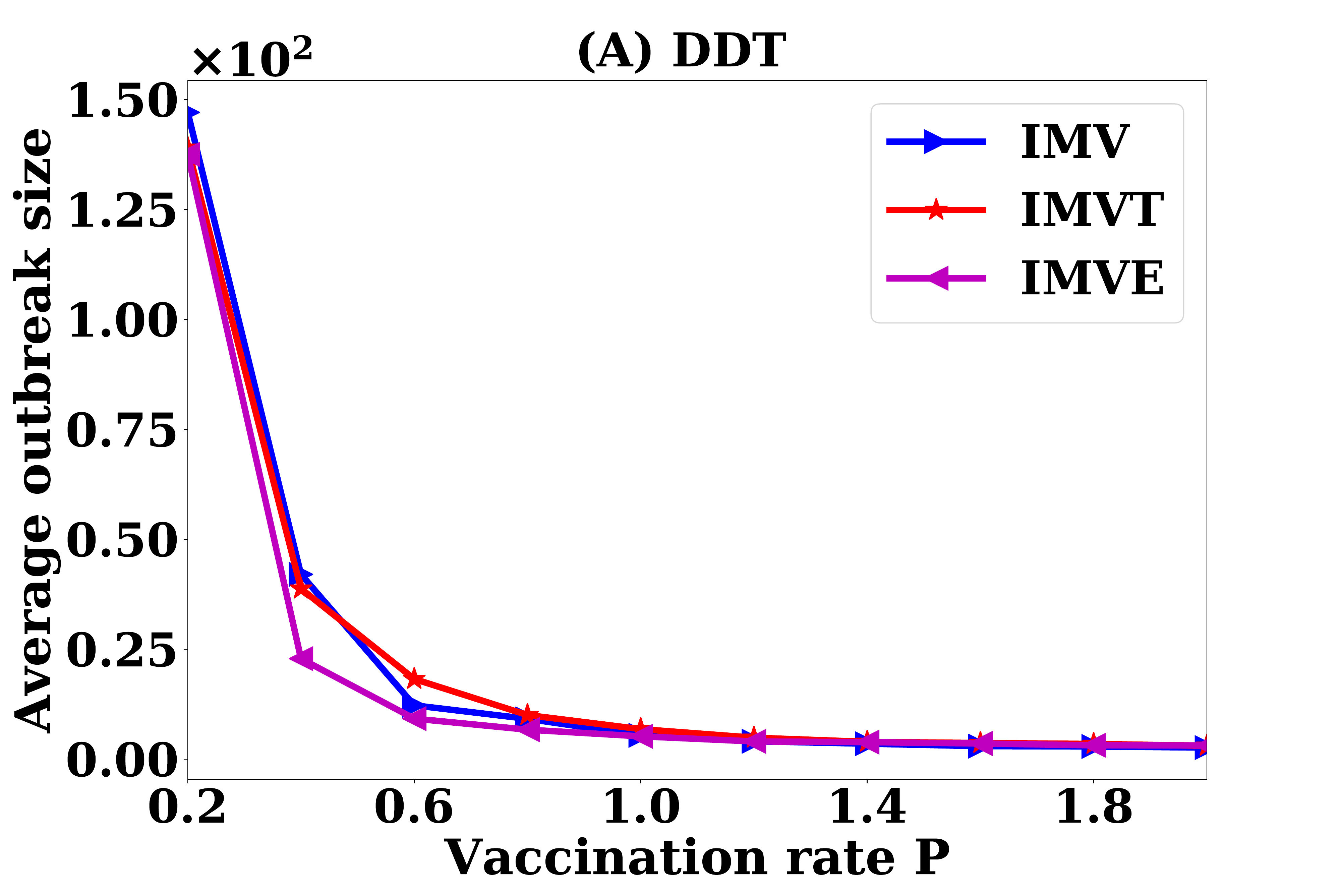}
\includegraphics[width=0.48\linewidth, height=5.5 cm]{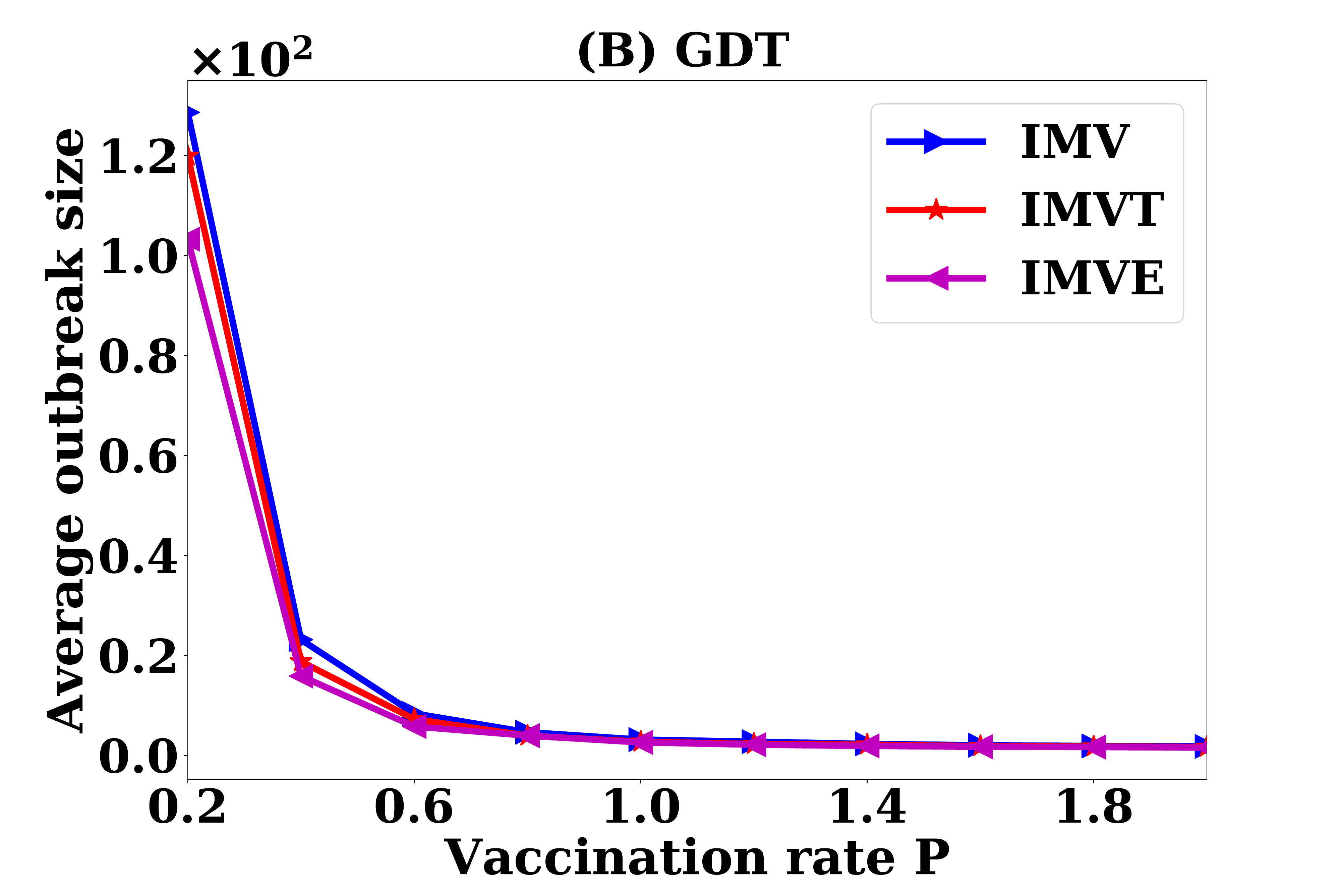}\\
\caption{Variations in average outbreak sizes of IMV strategy for accounting the temporal information and exact contact information}
\label{fig:vacd}
\end{figure} 

Nodes are now ranked with the exact contact information and temporal information for vaccinating with IMV strategy. Theses vaccination strategies of IMV are named as IMVE (exact information based) and IMVT (with temporal information based). Similar to the previous experiments, simulations are conducted from 5000 different single seed nodes and average outbreak sizes are computed for each $P$ in the range [0.2,2]\% with a step of 0.2\%. The results are presented in Figure~\ref{fig:vacd}. For applying exact information, IMVE strategy shows average outbreak size of 119 infections in DDT network and 139 infections in GDT network at $P=0.2$\% while they are 146 and 154 infections in IMV. The inclusion of temporal information (IMVT strategy) also show slight improvement at $P=0.2$\% with 103 infections in DDT network and 137 infections in GDT network. However, if $P$ is increased, these variations compared to the IMV strategy is minimised. At $P=0.8$\%, the average outbreak sizes are about three infections for all versions of IMV strategy. Thus, it is concluded that the inclusion of temporal information (stay duration) does not have a substantial impact when the requirement of preventive efficiency is high and a high vaccination rate is applied. The coarse-grained information is sufficient to capture the individual movement information. 

\begin{figure}[h!]
\includegraphics[width=0.48\linewidth, height=5.5 cm]{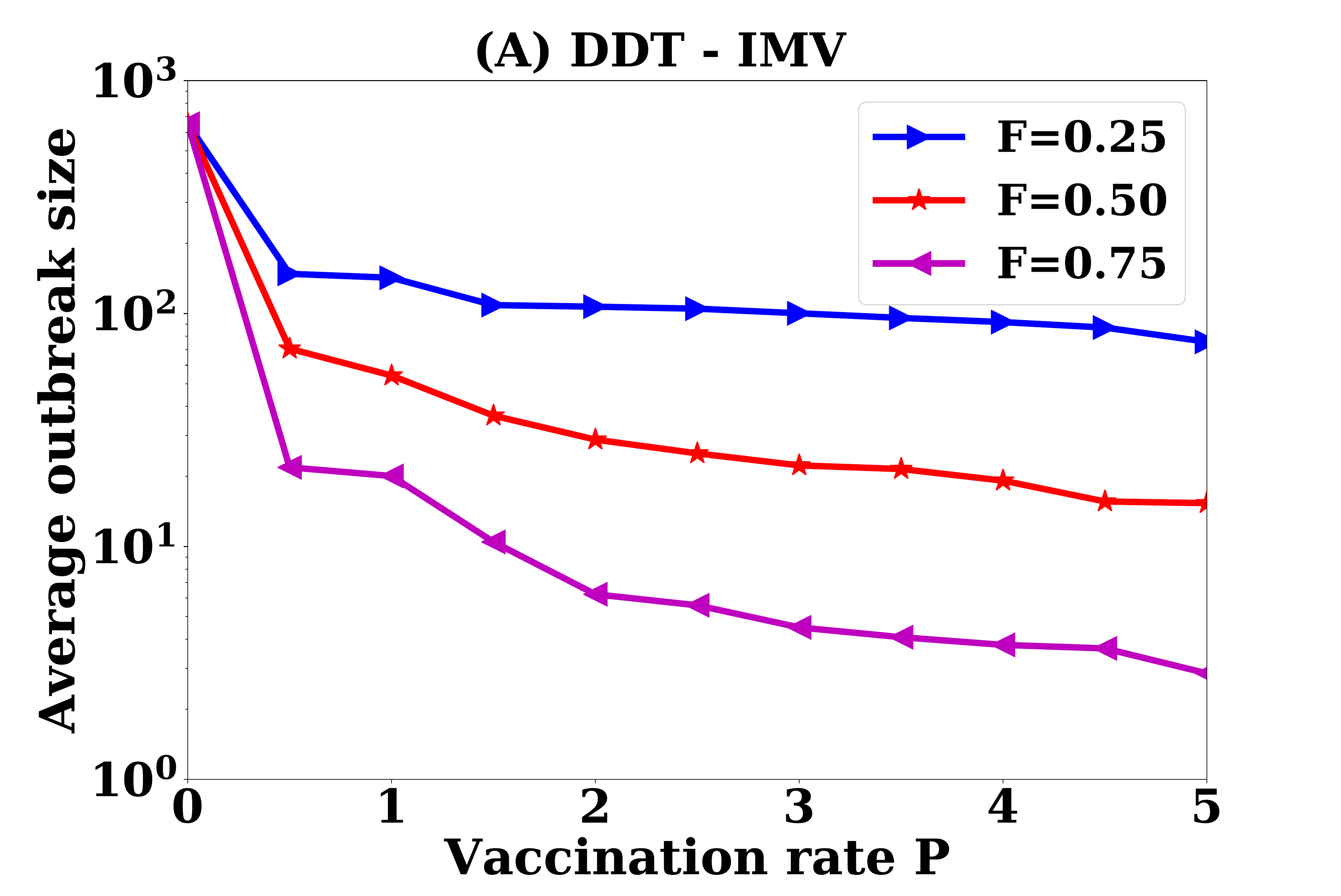}
\includegraphics[width=0.48\linewidth, height=5.5 cm]{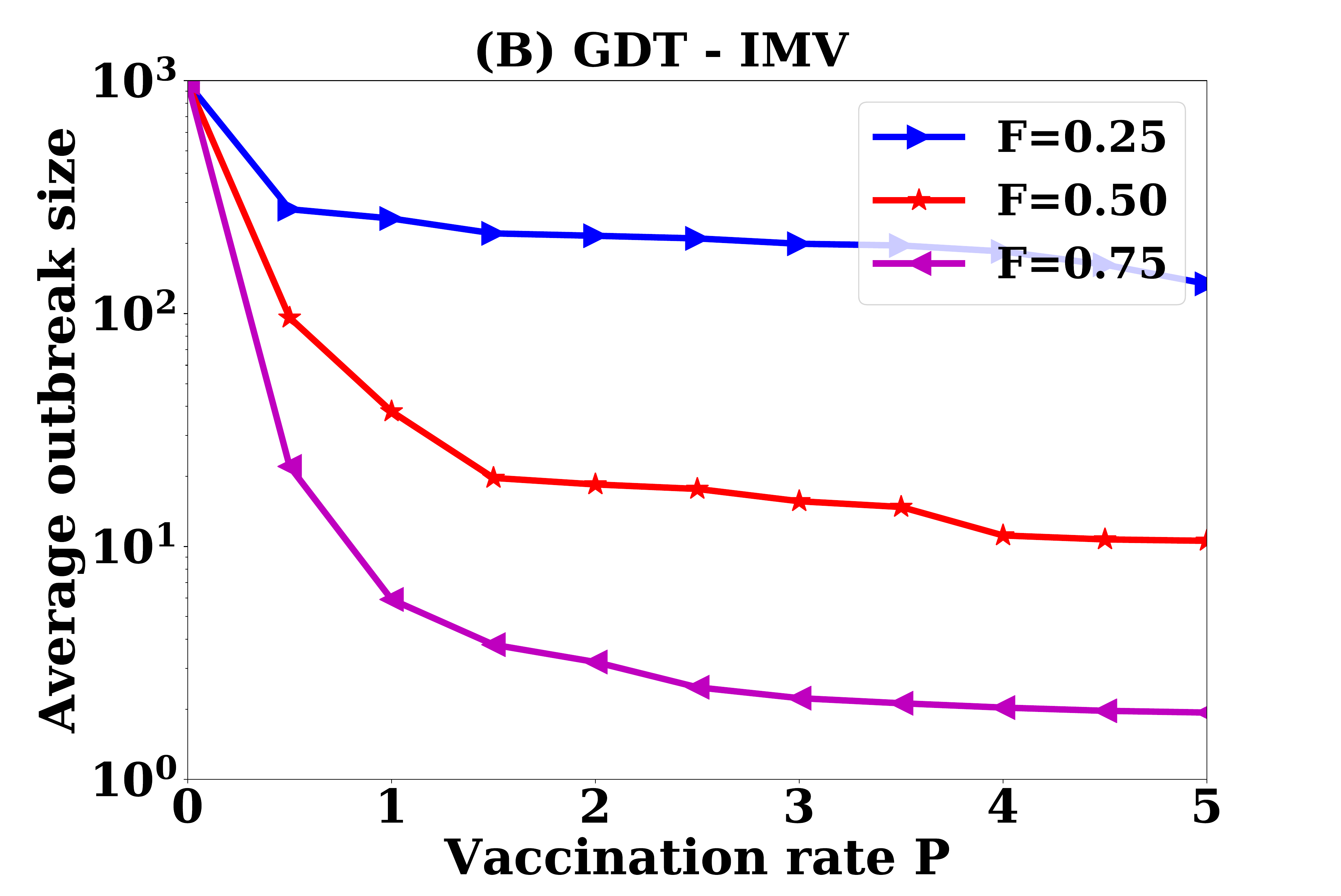}\\
\includegraphics[width=0.48\linewidth, height=5.5 cm]{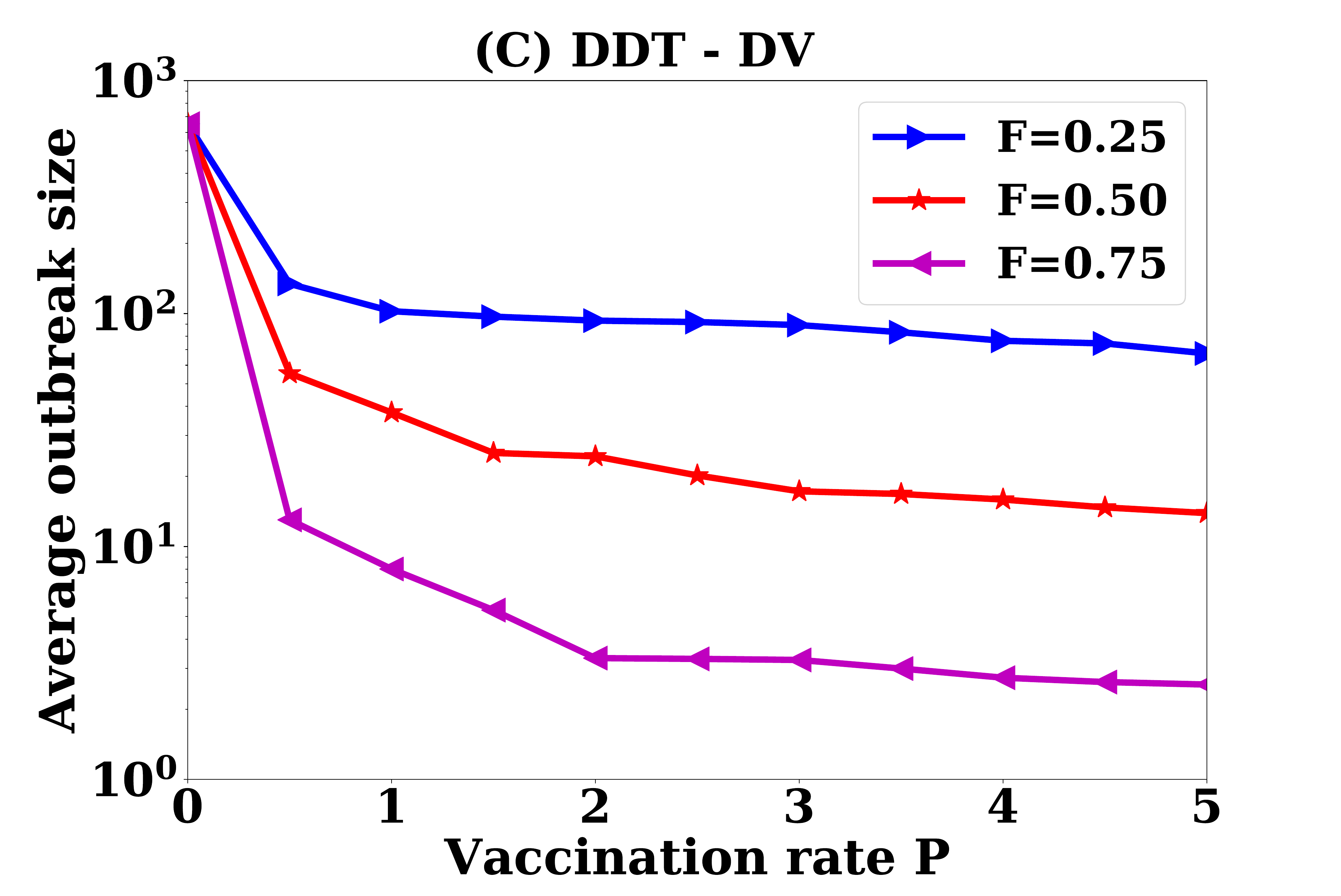}
\includegraphics[width=0.48\linewidth, height=5.5 cm]{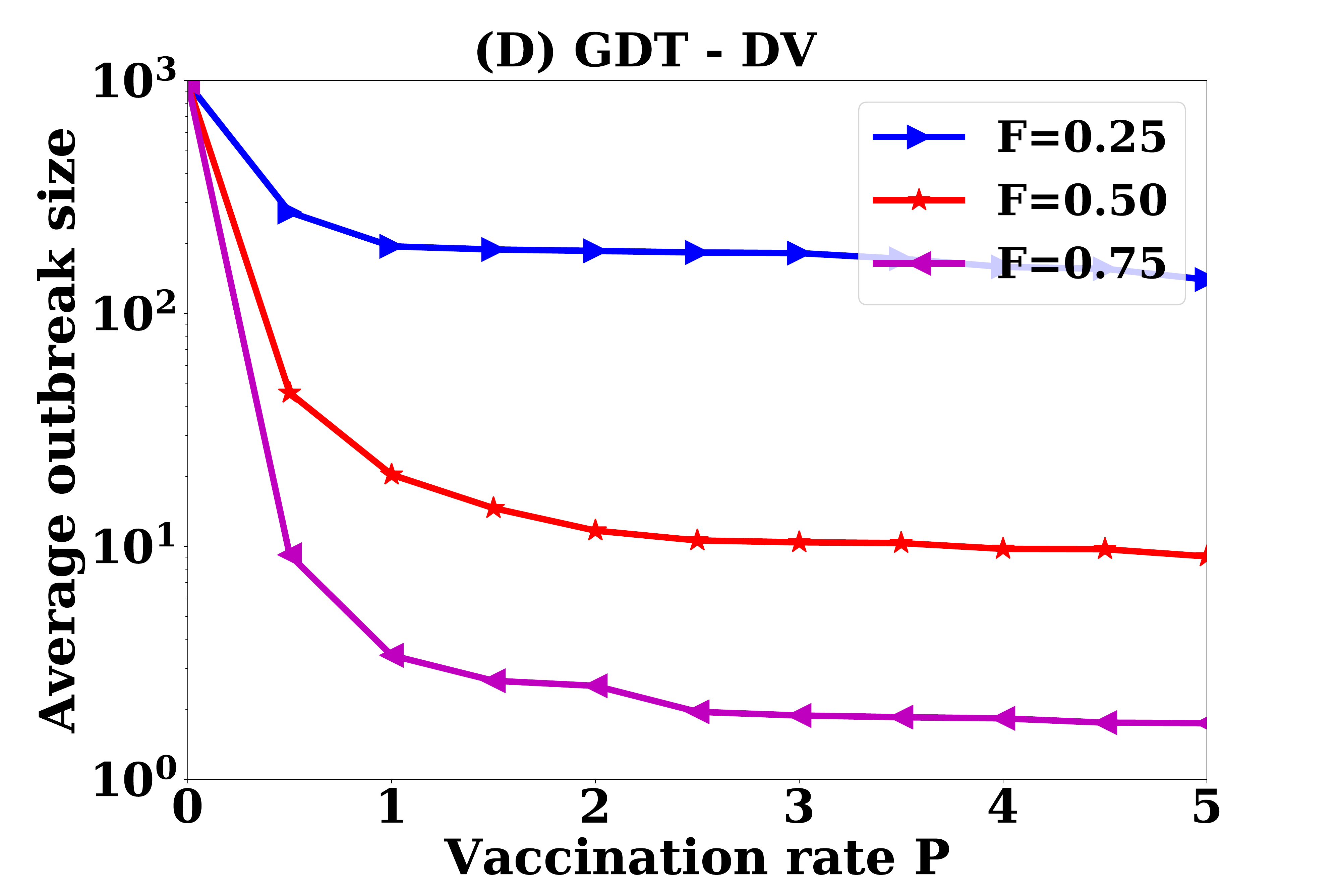}\\
\includegraphics[width=0.48\linewidth, height=5.5 cm]{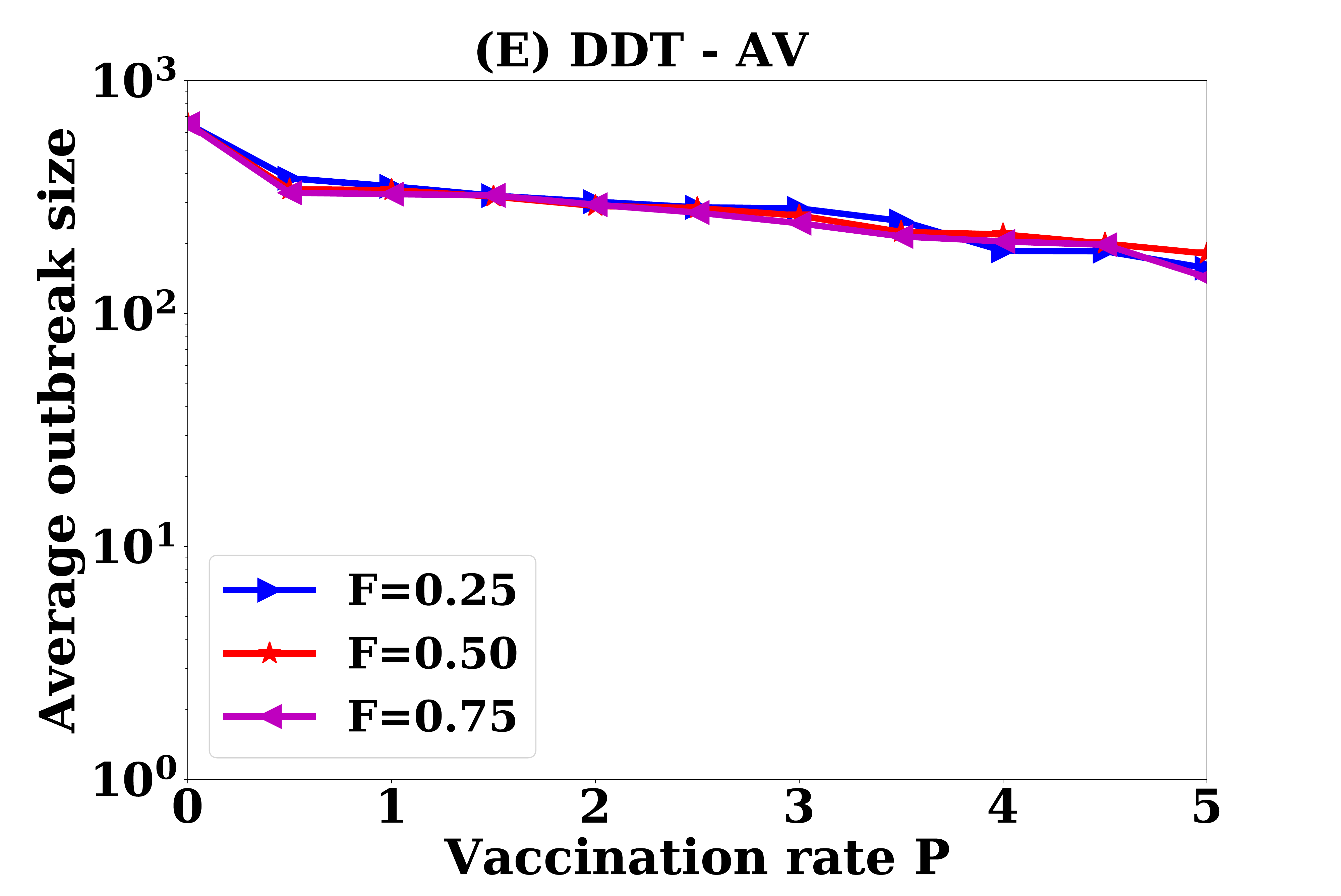}
\includegraphics[width=0.48\linewidth, height=5.5 cm]{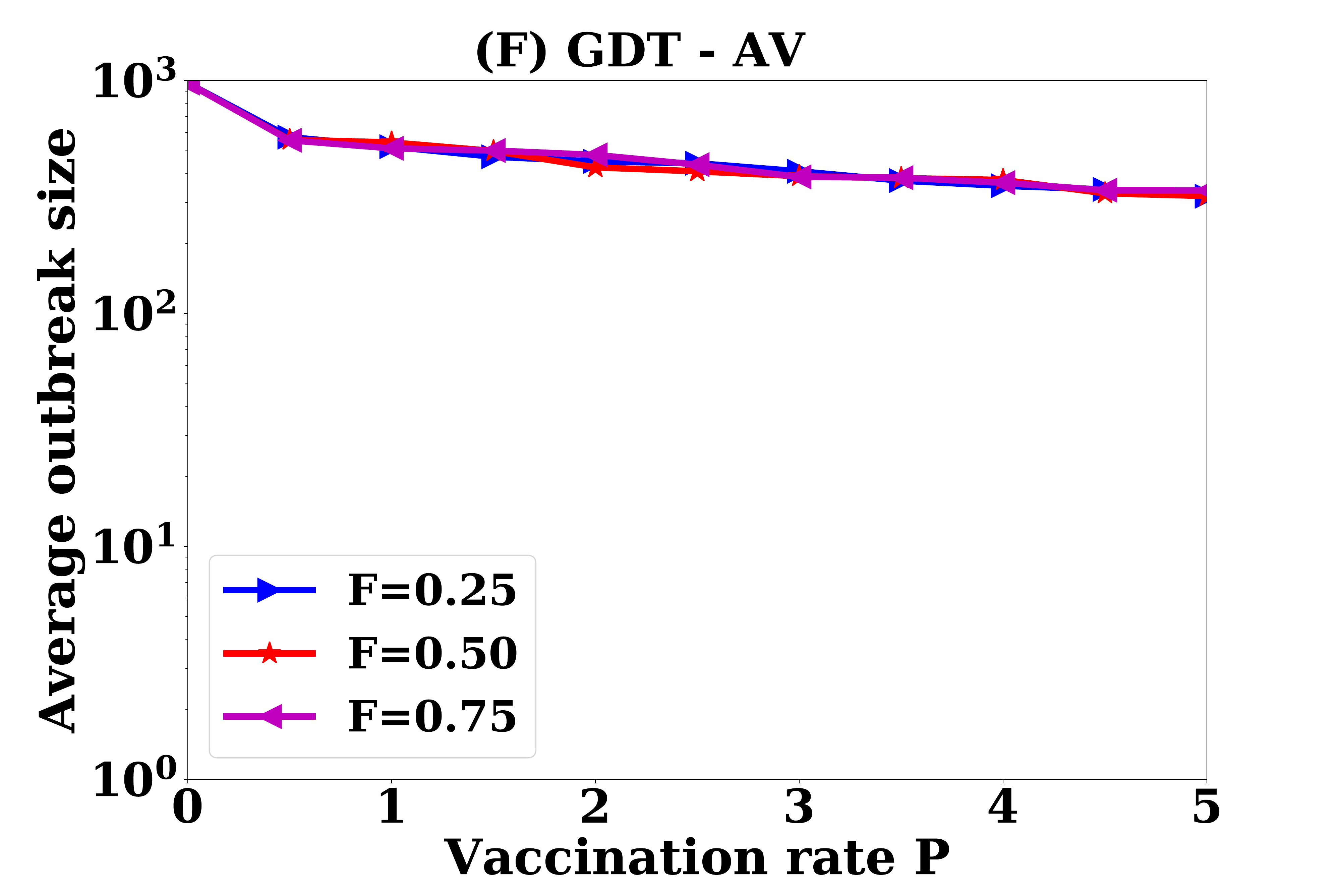}

\caption{Performance of vaccination strategies at various scale of information availability $F$: (A,B) proposed vaccination strategy, (C,D) degree based vaccination strategy and (E,F) acquaintance vaccination strategy}
\label{fig:sens}
\end{figure} 

\begin{figure}[h!]
\includegraphics[width=0.45\linewidth, height=6 cm]{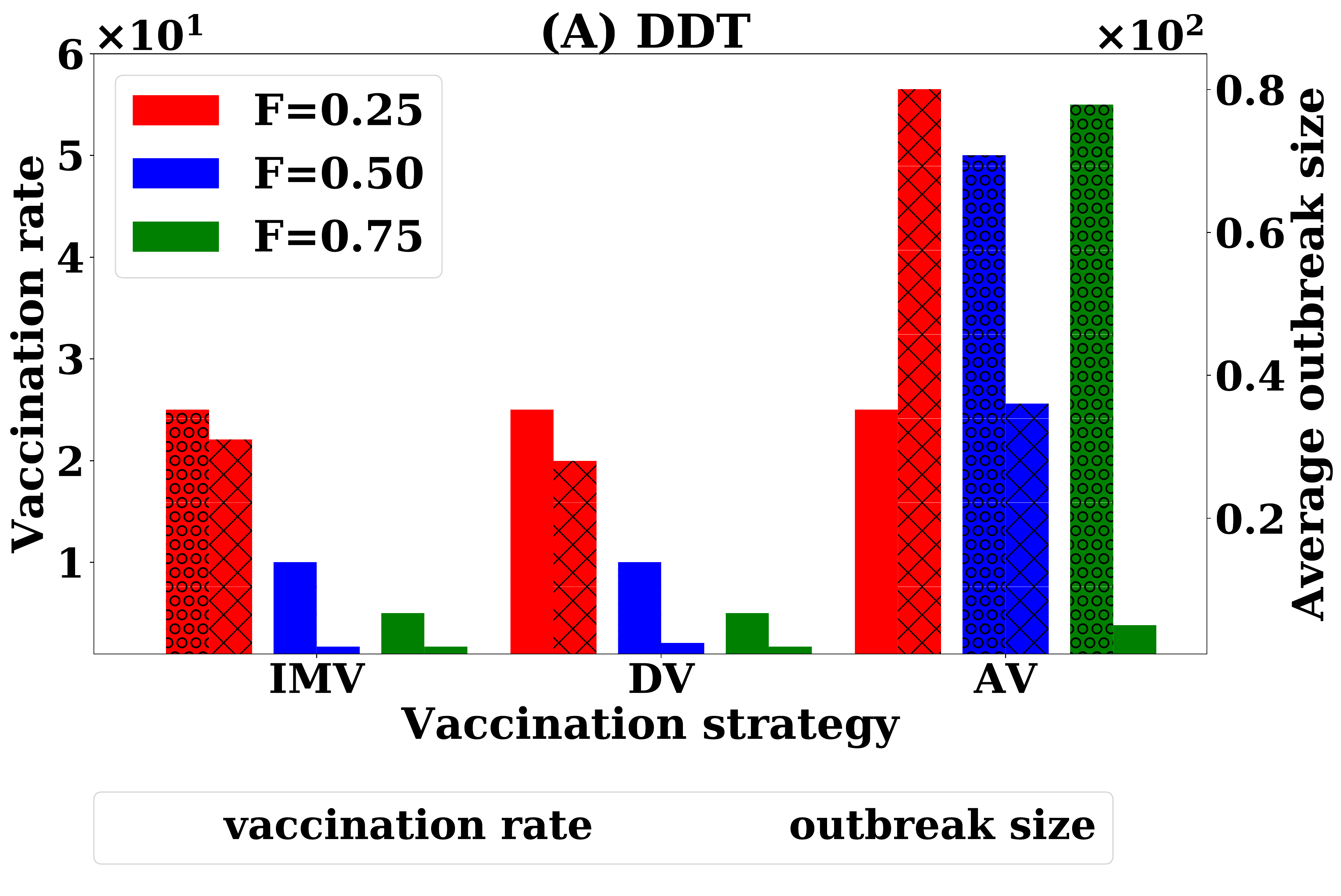} \quad
\includegraphics[width=0.45\linewidth, height=6 cm]{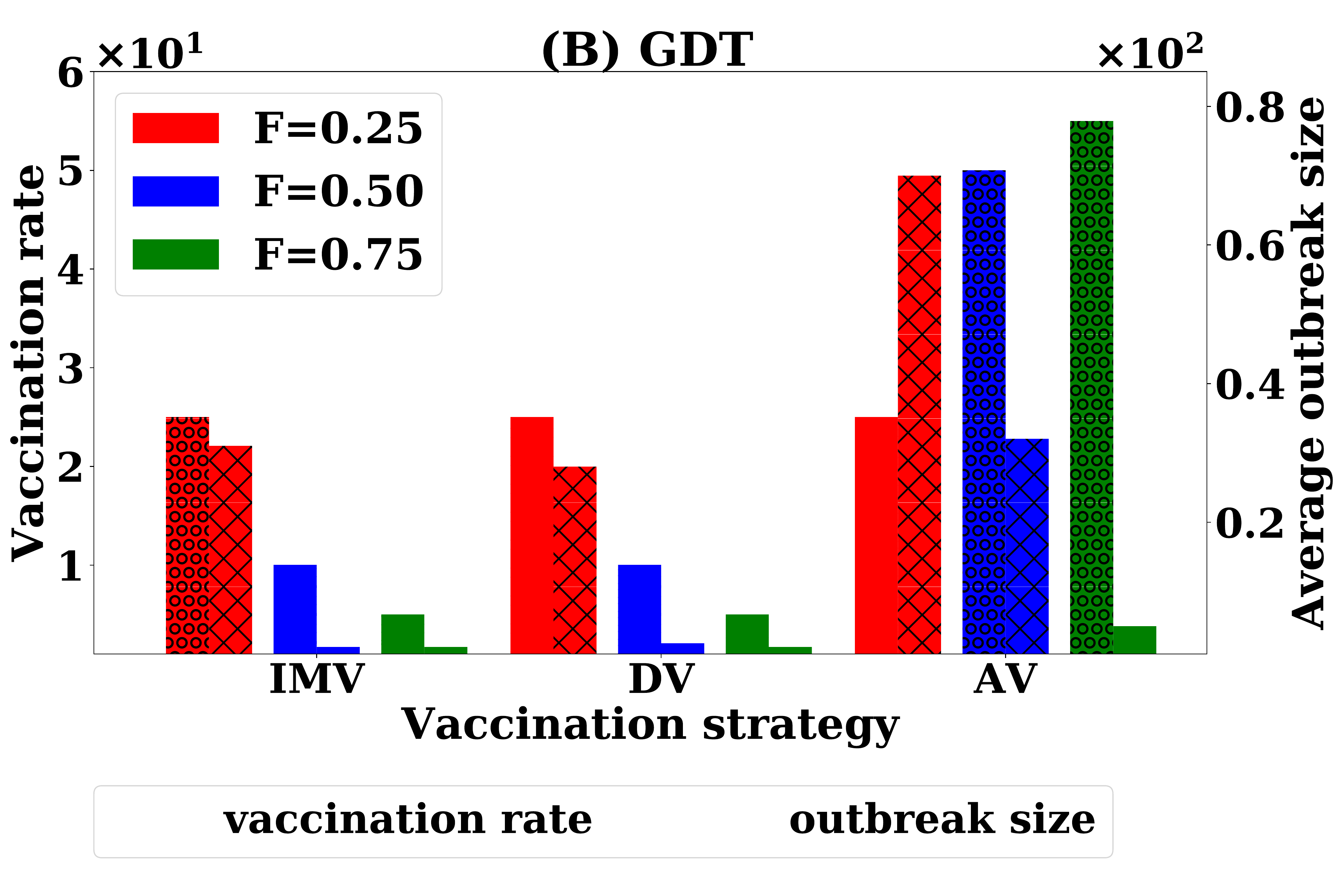}\\ [3ex]
\includegraphics[width=0.45\linewidth, height=6 cm]{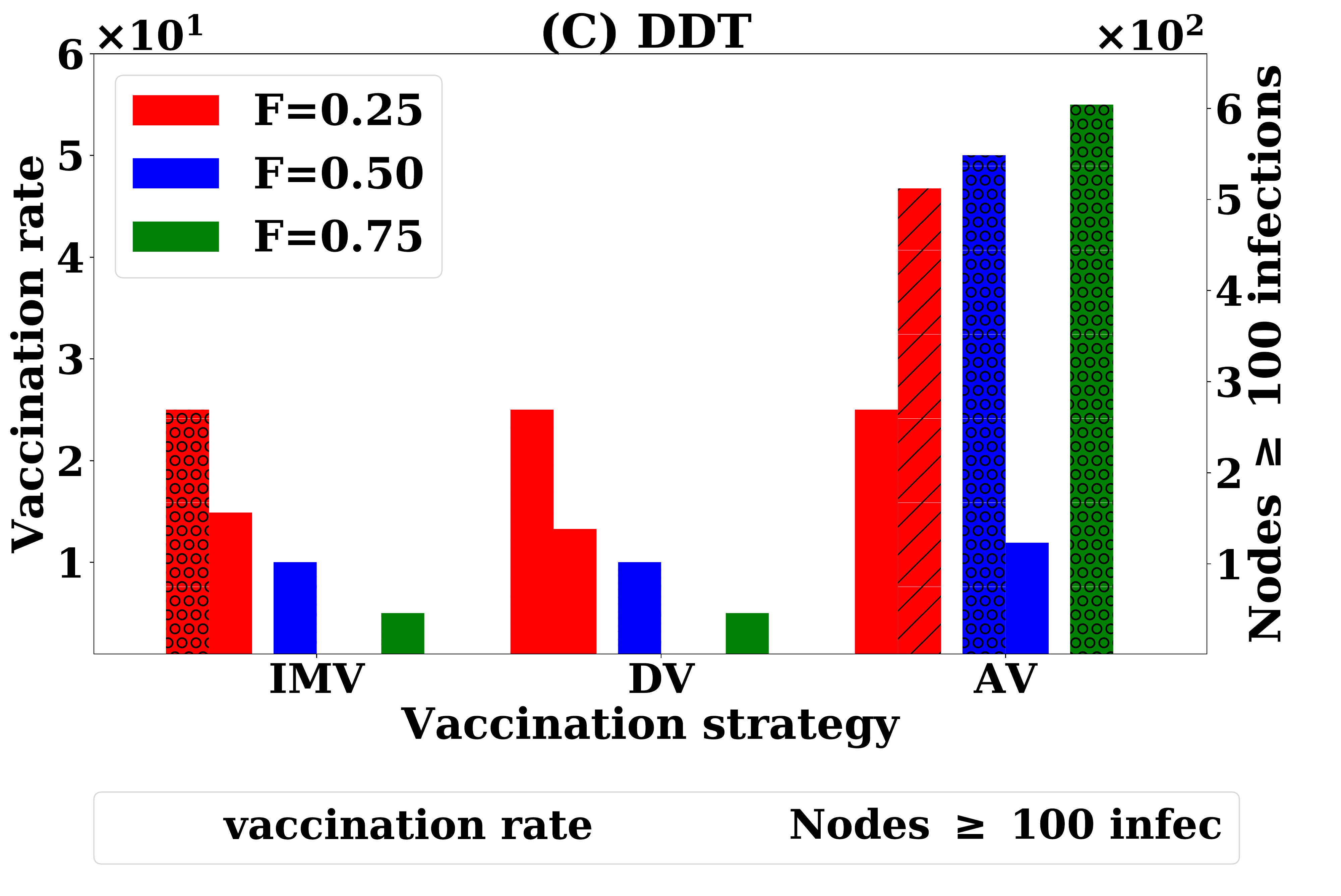}\quad
\includegraphics[width=0.45\linewidth, height=6 cm]{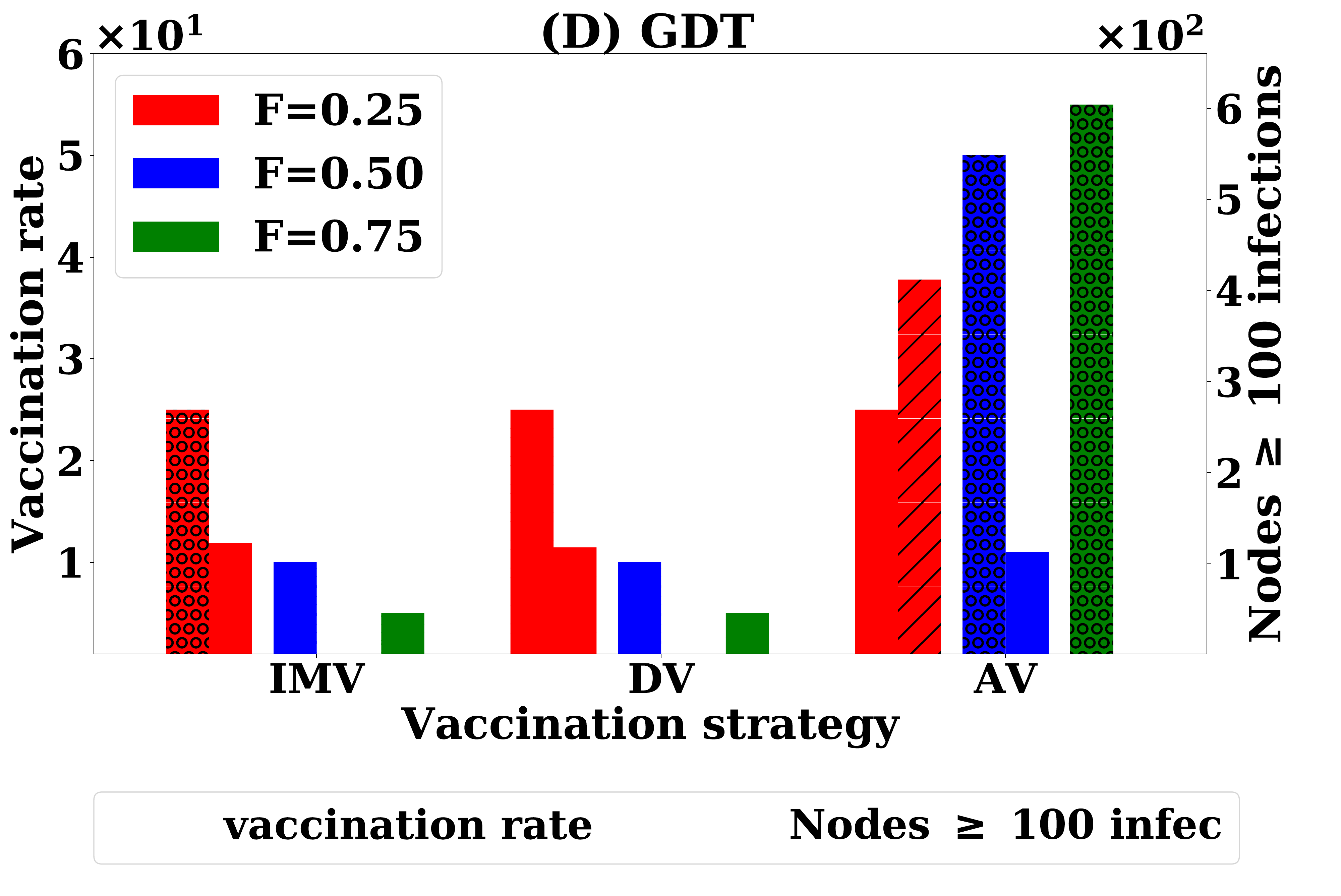}

\caption{Trade-off among information collection cost, vaccination cost and infection cost for various strategy - vaccination rate is on the left y-axis while others are on the right y-axis: (A, B) average outbreak sizes with vaccination rates for different $F$, and (C, D) number of seed nodes having outbreak greater than 100 infections with $P$ and $F$}
\label{fig:sens3}
\end{figure}

In the previous experiments, the upper bound of the efficiency for vaccination strategies was studied assuming that information of all nodes can be collected. However, it is not possible to collect movement information of all individuals in real world scenarios. Thus, it is required to understand the effectiveness of the strategies with the scale of information collected regarding nodes contact. Therefore, this experiment analyses the efficiency of the strategies varying the proportion $F$ of nodes that are picked for collecting their movement information. The required vaccination rates $P$ for a strategy also depend on the value of $F$. In the simulations, a proportion $F$ of nodes are picked up randomly from the first seven days of networks and their ranks are calculated based on the applied vaccination strategy. Then, a vaccination rate $P$ is implemented with the sampled nodes. The impacts of scale of information availability is studied for varying $F$ to 25\%, 50\% and 70\% . The efficiency of strategies is first analysed varying $P$ from 0.5 to 5\% with the step of 0.5\%. The simulations are conducted for IMV, DV and AV strategies on both the DDT and GDT networks. The average outbreak sizes are presented in Figure~\ref{fig:sens}. Then, the simulations are also run for larger value of $P\geq 5$\% until the preventive efficiency is achieved where no seed node has outbreak of more than 100 infections. The results are presented in Figure~\ref{fig:sens3} showing the trade-off between the performance and vaccination rates with information availability scale $F$.

The results show that there is trade-off among vaccination rate, infection cost and information collection cost for implementing a vaccination strategy. If contact information of $F=25$\% nodes is applied to select the nodes to be vaccinated, the average outbreak sizes are about 100 infections for IMV and DV strategy in the DDT network and 126 infections in the GDT network even at the vaccination rate $P=5$\% (Figure~\ref{fig:sens}). Then, the vaccination rate is increased to $P=25$\% for understanding maximum efficiency at $F=25$\%. There is still substantially high average outbreak sizes in both DDT and GDT networks. Besides, the number of seed nodes having more than 100 infections is 167 nodes in DDT network and 131 nodes in GDT network. With the information collection cost $F=25\%$, therefore, vaccination cost is high along with less protection of infection. If $F$ is increased to $F=0.50$, the outbreak sizes reduce to about 21 infections at $P=5$\% in GDT network for IMV strategy while it is still 25 infections in DDT network. However, the degree based vaccination (DV) shows low average outbreak sizes compared to that of IMV strategy with lower $P$. But, the average outbreak sizes are not below 20 infections within the vaccination rate of $P=5$\%. However, the preventive efficiency with no seed nodes having outbreak sizes greater than 100 infections is achieved vaccinating 10\% of nodes in both networks. At $F=0.5$, the strong protection from infection can be achieved with the vaccination cost of 10\%. Further increase of $F$ to 0.75 makes possible to reduce outbreak sizes below 10 infections in both networks for DV strategy at $P=2$\%. The number of nodes having outbreak sizes greater than 100 infections become zero for DV strategy. On the other hand, the proposed IMV model reduces outbreak sizes below 10 infections at $P=2$\% in GDT network and at $P=3$\% in DDT network. The number of nodes having outbreak sizes greater than 100 infections become 3 nodes at $P=3$\%. With $F=0.75$, the vaccination cost is about 3\% in IMV strategy while it was 1\% for $F=1$. For AV strategy, average outbreak sizes is very high up to $F=0.5$ and there is a large number of seed nodes having outbreak sizes greater than 100 infections. However, the strong preventive efficiency is achievable at $F=0.75$ with vaccinating of 55\% nodes where no seed node has outbreak greater than 100 infections. With the low scale of information, the IMV strategy performs better than AV strategy and close to DV strategy.

%% file: 6.5_chap_postout.tex
\section{Post-outbreak vaccination}
The previous section has investigated the performance of the proposed method for preventive scenarios. Even if the preventive vaccination is implemented within a population, the outbreaks can happen with the new strains of disease. Therefore, it is crucial to develop proper vaccination for post-outbreak scenarios as well. The proposed vaccination strategy is now studied for the post-outbreak scenarios which can be implemented in two ways. The first way is similar to the mass vaccination and is called population-level vaccination where a proportion $P$ of the population is randomly chosen and is vaccinated. In the simplest version, behaviours of the selected individuals are not considered, randomly picked up to vaccinate, and hence information collection cost is minimal. However, the resource cost is often very high as it requires to vaccinate a large number of individuals. This approach is improved by vaccinating individuals who have specific charecteristics relevant to disease spreading. For example, if an individual has interactions with many other individuals, he is chosen to be vaccinated (degree based vaccination). In this section, the applied three ranking methods (IMV, DV and AV) of nodes from the previous sections are used to select the nodes to be vaccinated. In addition, random vaccination (RV) method is also examined along with these three ranking based methods. 

In the second way of post-outbreak vaccination implementation, the infected individuals are at the focus point where susceptible individuals who have contact with infected individuals are vaccinated to hinder further spreading of disease. Ring vaccination is one of the strategies from this class. This approach of implementing post-outbreak vaccination is named as node level vaccination. In the ring vaccination, a proportion $P$ of neighbour nodes are vaccinated and the neighbour nodes can be chosen randomly or based on specific criterion. The two ranking methods (IMV, and DV) are applied to select neighbours in the ring vaccination and the performance of node level vaccination is studied. The performance are compared with the strategy of random neighbour selection (RV). Both population level and node level vaccination are investigated in this section on both DDT and GDT contact networks. The population level vaccination is also examined with the scale $F$ of information availability on node's contact. In case of node level vaccination, all infected node might not be identified. Thus, the performance is analysed if a proportion $F$ of infected nodes are identified only.

\subsection{Population level vaccination}
\begin{figure}[h!]
\includegraphics[width=0.48\linewidth, height=5.5 cm]{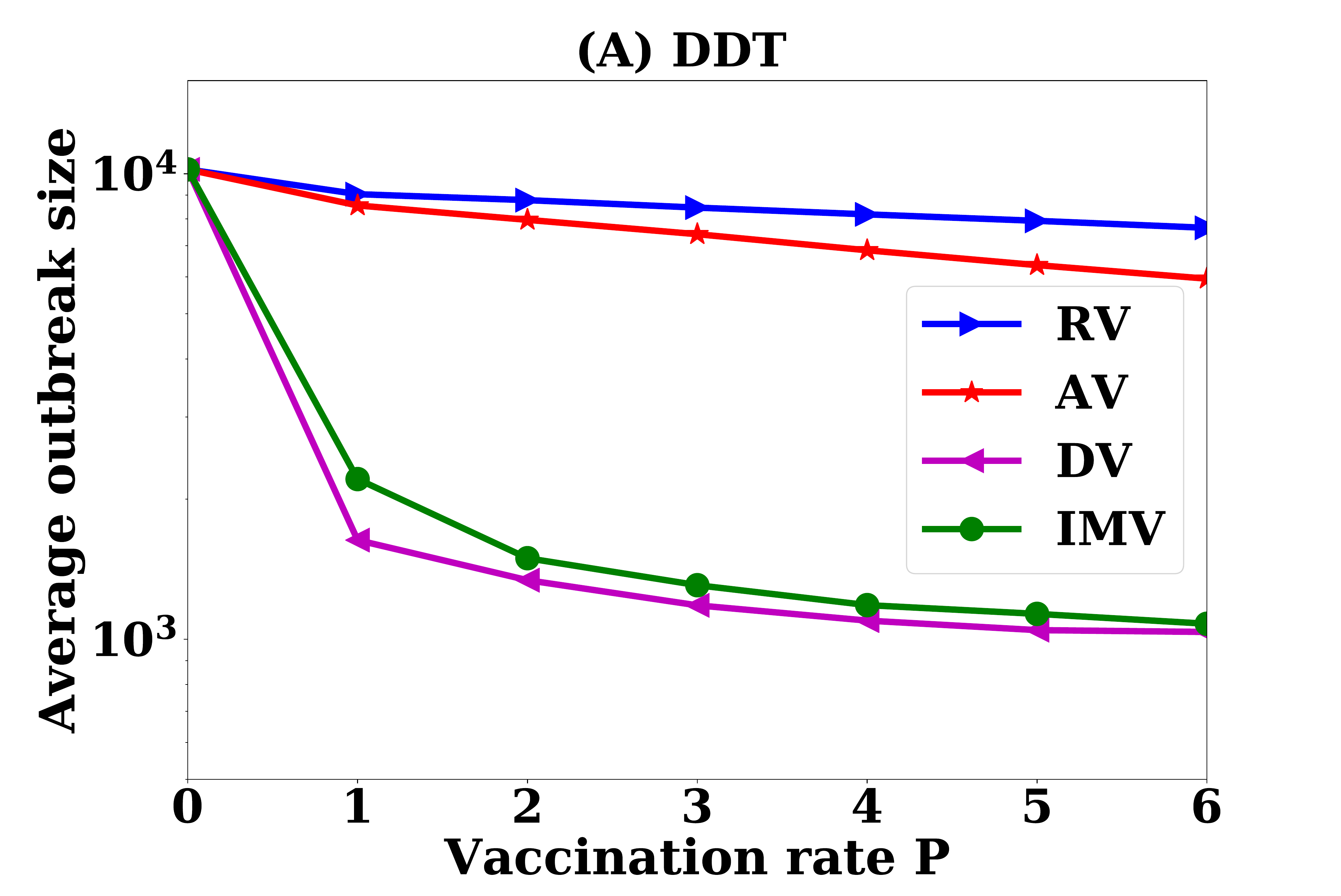}
\includegraphics[width=0.48\linewidth, height=5.5 cm]{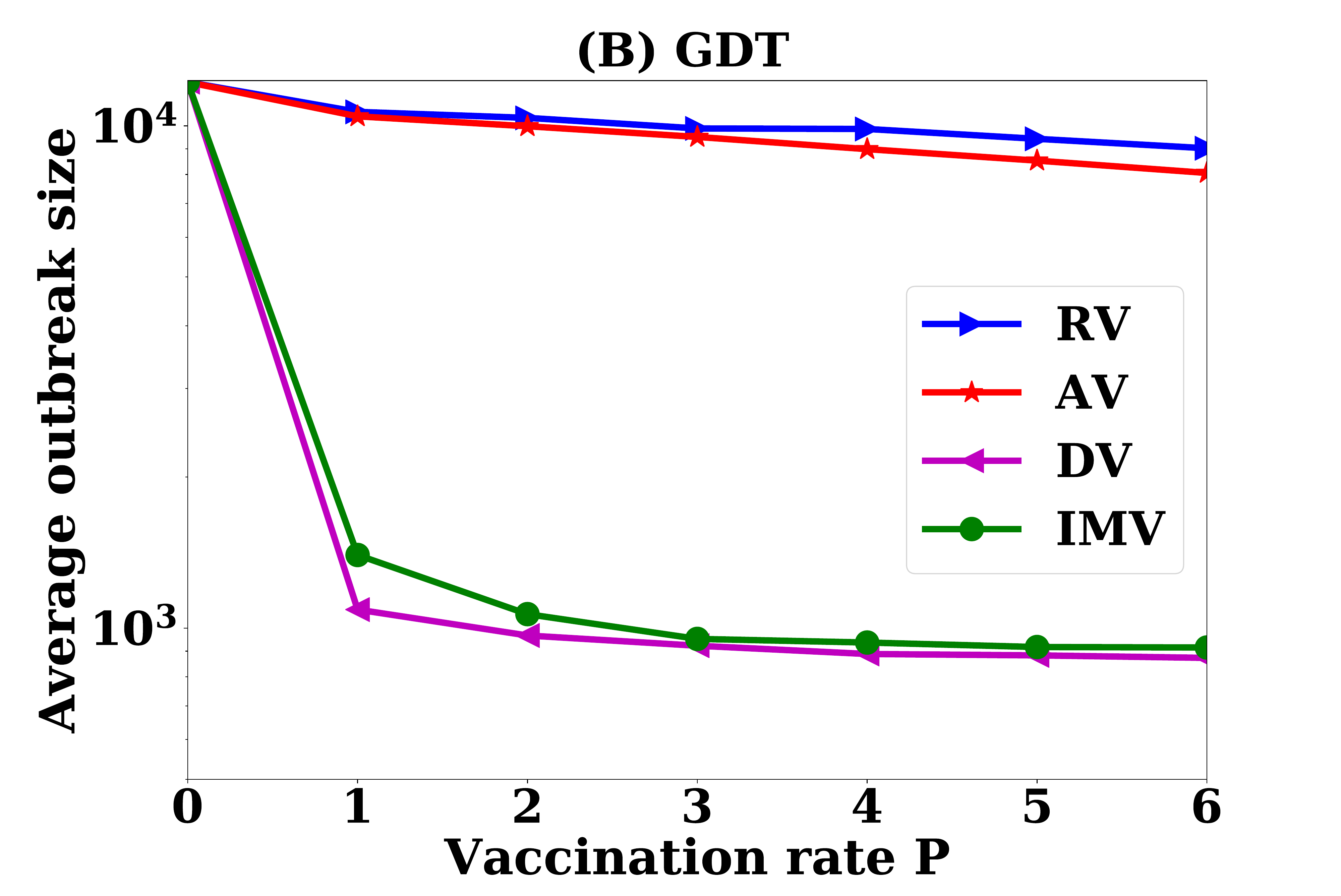}\\
\caption{Average outbreak sizes for different vaccination strategies at various vaccination rates $P$ in post-outbreak vaccination}
\label{fig:vmpf}
\end{figure}

In these simulations, disease starts with 500 seed nodes and continues for 42 days. Nodes are infectious for $\tau$ days randomly chosen in the range [3-5] days. The vaccination is implemented at the 7th day of simulation assuming that these days are required to notice the emergence of disease and to collect the contact information. The node's rank, based on the applied vaccination strategy, is calculated from the contact data of the first seven days. Then, a proportion $P$ of nodes are vaccinated (assigned recovered status) and final outbreak sizes are calculated for 42 days of simulations. For each vaccination strategy, simulations are conducted for different $P$ in the range [1,6]\% with a step of 1\%. At each value of $P$, the simulations are run for 1000 times and the average outbreak sizes are presented in Figure~\ref{fig:vmpf}.  

The simulation of disease spreading without vaccination makes outbreak of on average 10K infections in the DDT network and 12K infections in the GDT network. This average outbreak size is used as the reference to understand the efficiency of the applied vaccination strategy. Random vaccination (RV) strategy does not reduce the average outbreak sizes significantly in both networks (Fig.~\ref{fig:vmpf}). At the vaccination rate $P=1$\% (vaccinating 3600 nodes), the average outbreak size is 9K infections in the DDT network and 10K infections in the GDT network. On changing $P$ from 1\% to 6\% (vaccinating 21600 nodes), there is a 1K infections reduction in the average outbreak size for RV strategy. If the ranking score of acquaintance vaccination (AV) is applied to vaccinate nodes, the reduction in outbreak sizes slightly increases. At $P=1$\%, the average outbreak size is similar to that of RV strategy. If $P$ is, however, increased to 6\%, the average outbreak size is 8K infections in the DDT network and 10K infections in the GDT network. Within the range of vaccination rates, RV and AV strategies fail to contain the disease spreading. Increasing vaccination rates further for RV and AV strategies show that the RV strategy requires 75\% of nodes to be vaccinated in both networks and the AV strategy requires vaccination of 40\% nodes in DDT network and 35\% in GDT network to contain disease spreading within the outbreaks of 1K infections.

The proposed vaccination strategy (IMV) shows the significant improvements in reducing the final average outbreak sizes compared to the AV and RV strategies (Fig.~\ref{fig:vmpf}). With vaccinating 1\% of nodes (vaccinating 3600 nodes), the average outbreak size reduces to 2K infections in both networks. If the vaccination rate $P$ is increased to 4\% in the DDT network, the outbreak size reduces to about 1K infections but a further increase in $P$ does not reduce outbreak sizes substantially. In the GDT network, the average outbreak sizes are about 1K infections with 3\% (vaccinating 10800 nodes) vaccination. The degree based vaccination shows similar performance to that of IMV strategy. Similar to the preventive vaccination, the DV strategy initially shows better performance and then becomes similar to the IMV strategy at the higher vaccination rates. The coarse-grained information based IMV strategy achieves the same performance of DV strategy for the post-outbreak vaccination scenarios as well.   

\begin{figure}[h!]
\includegraphics[width=0.48\linewidth, height=5.5 cm]{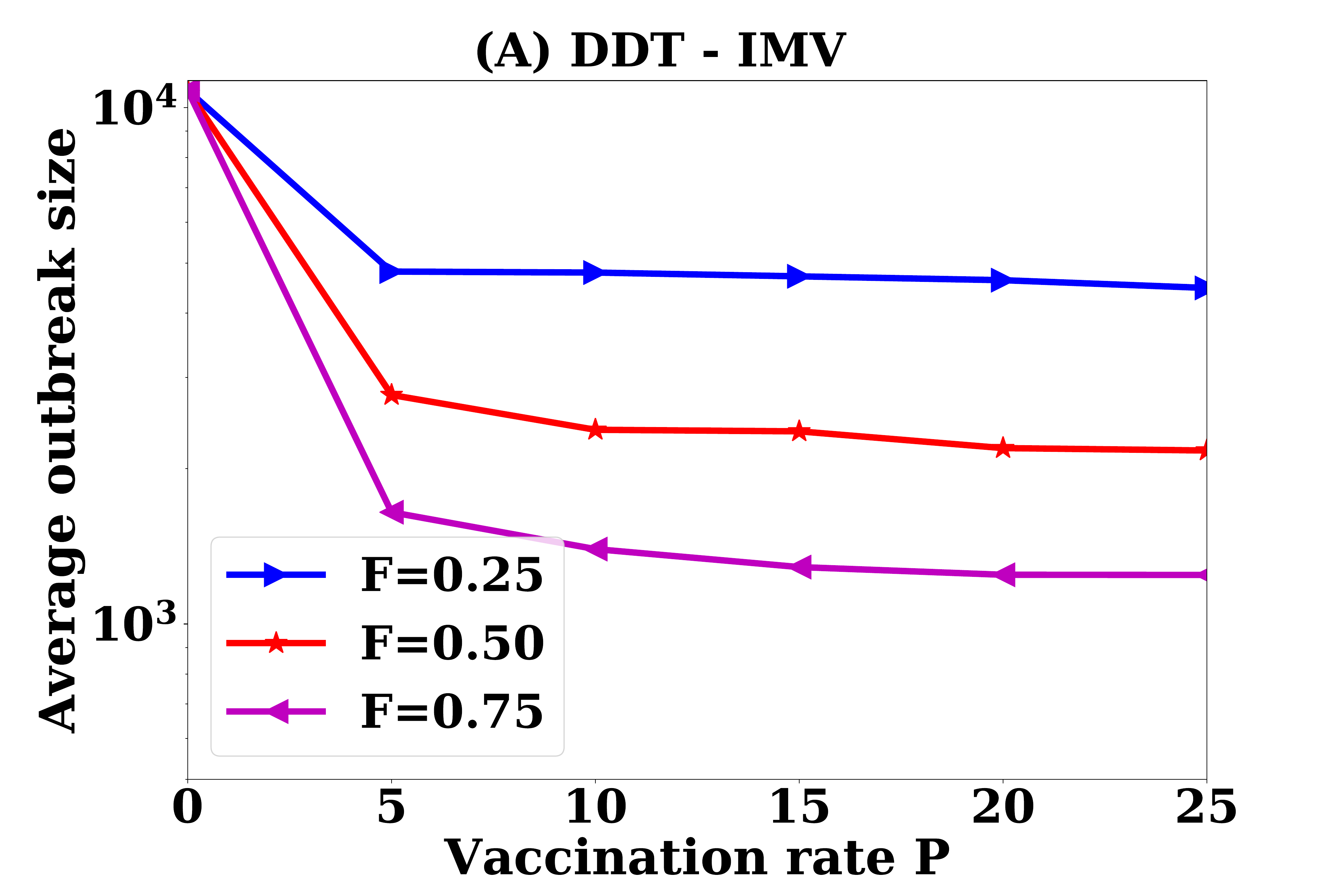}
\includegraphics[width=0.48\linewidth, height=5.5 cm]{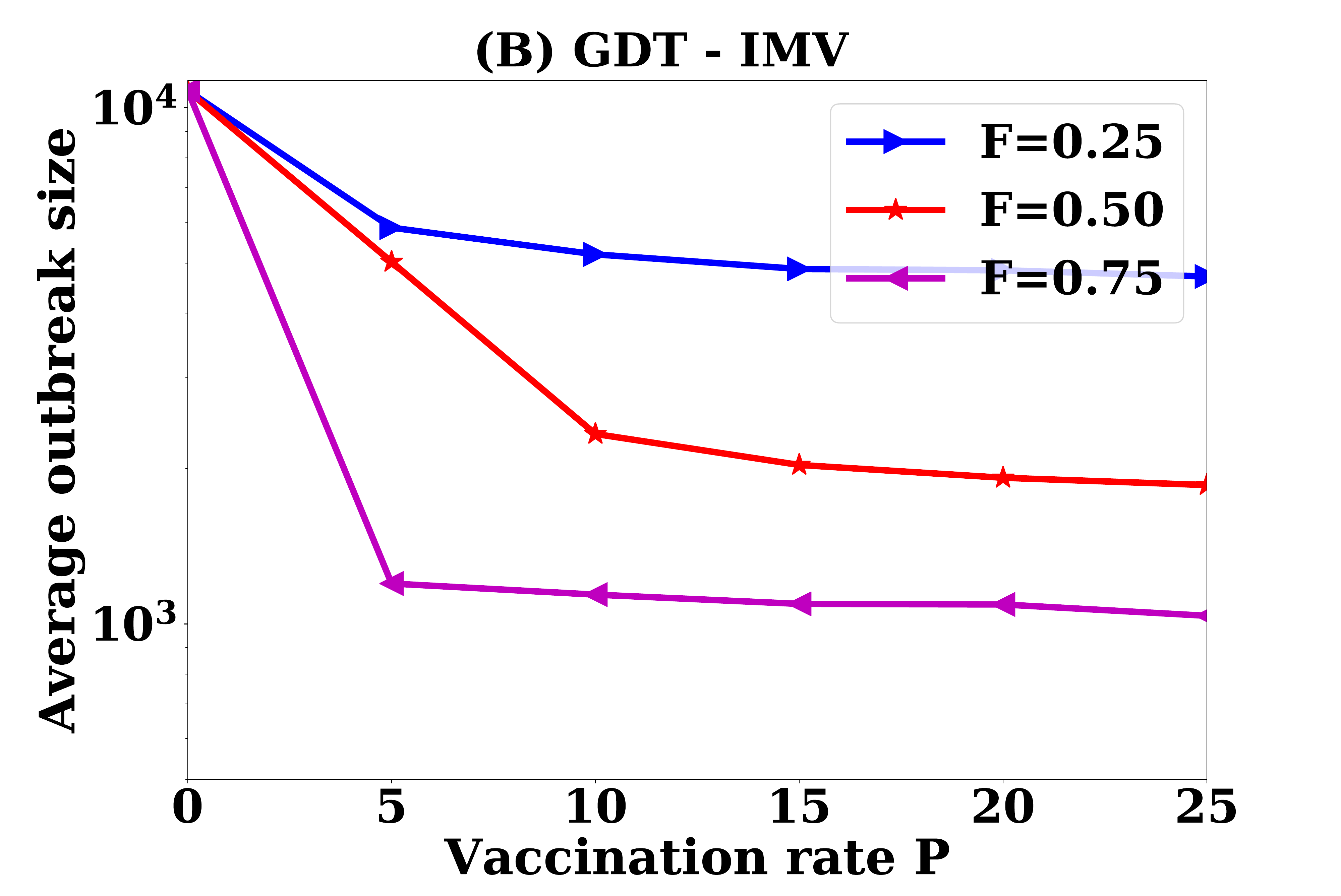}\\
\includegraphics[width=0.48\linewidth, height=5.5 cm]{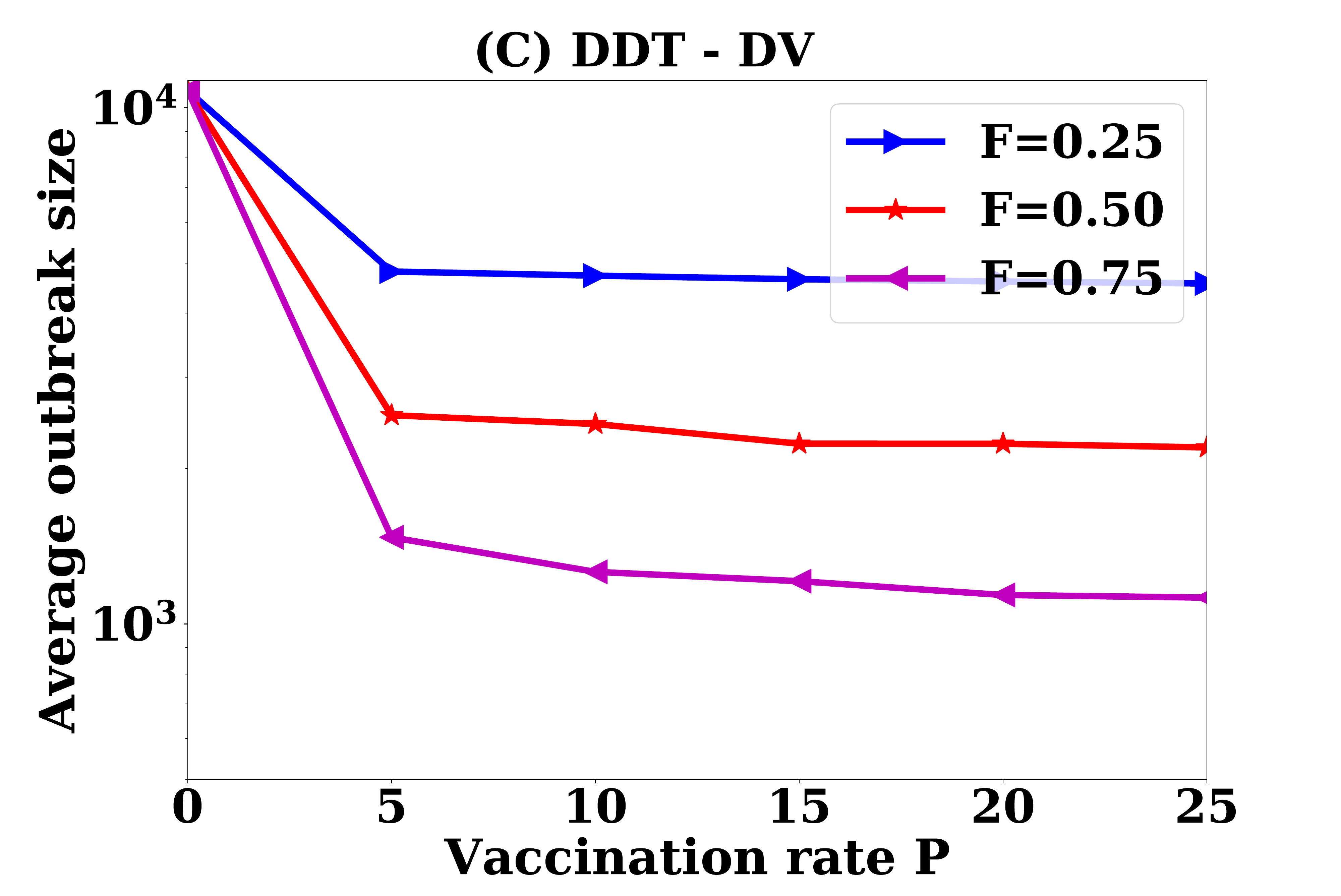}
\includegraphics[width=0.48\linewidth, height=5.5 cm]{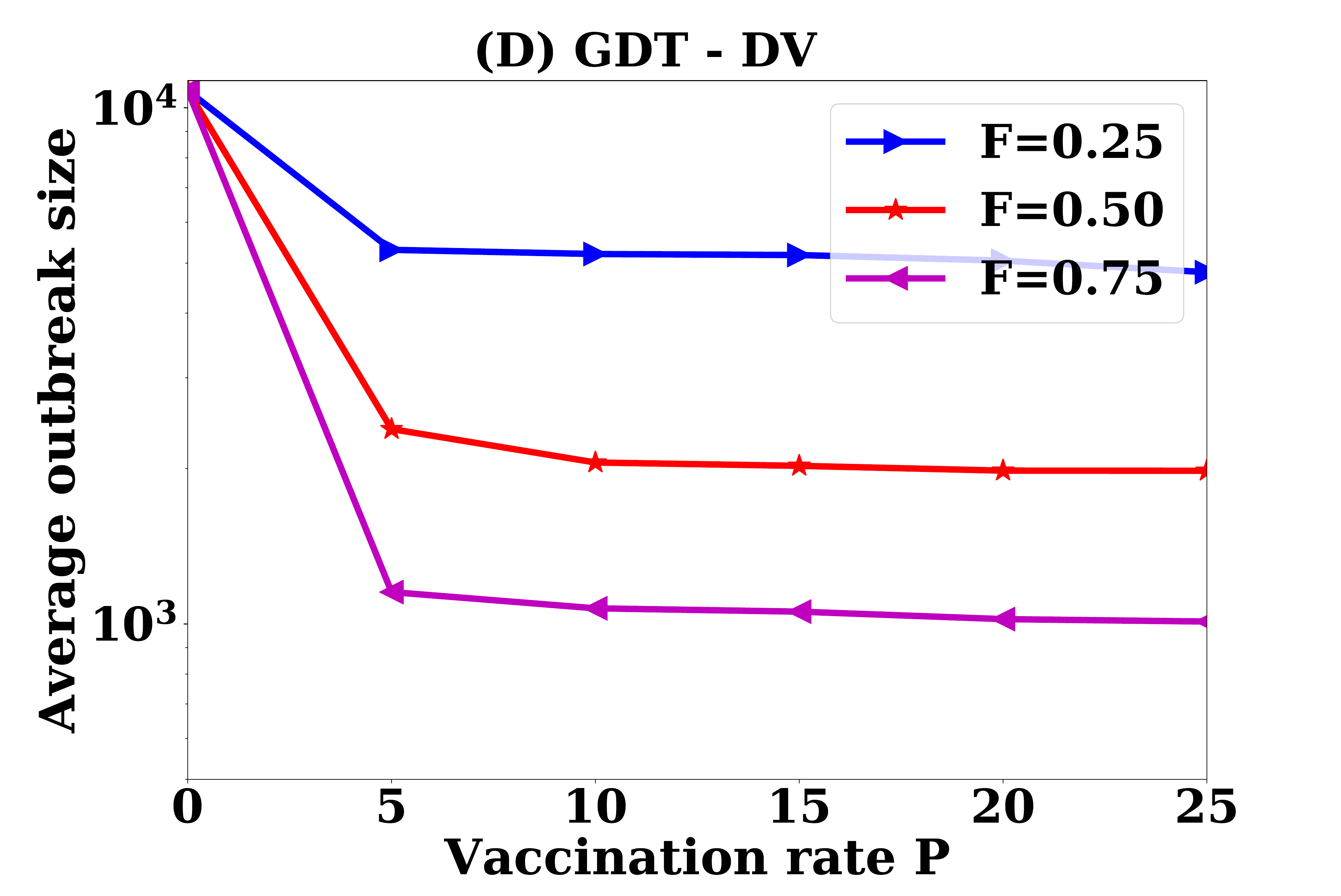}\\
\includegraphics[width=0.48\linewidth, height=5.5 cm]{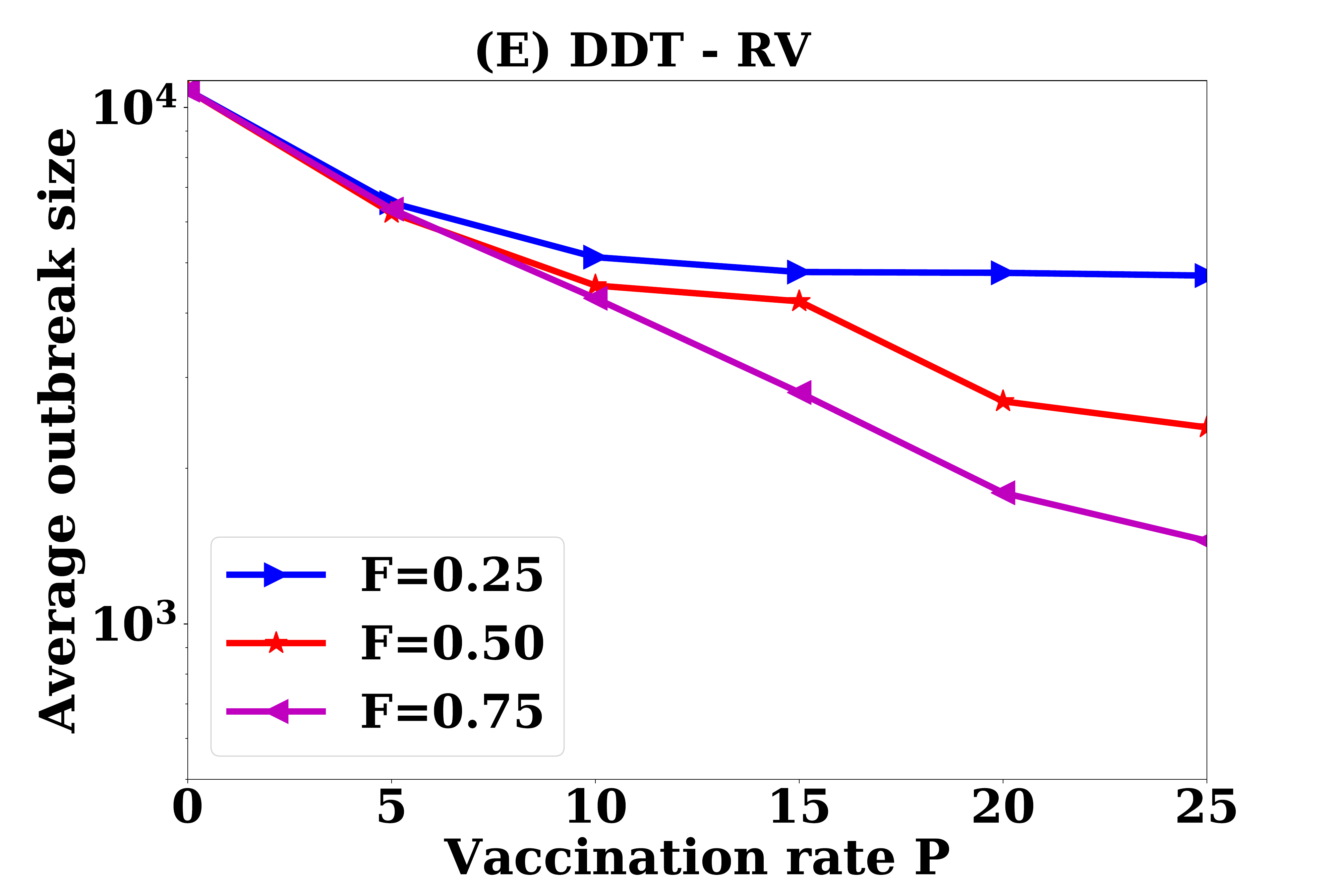}
\includegraphics[width=0.48\linewidth, height=5.5 cm]{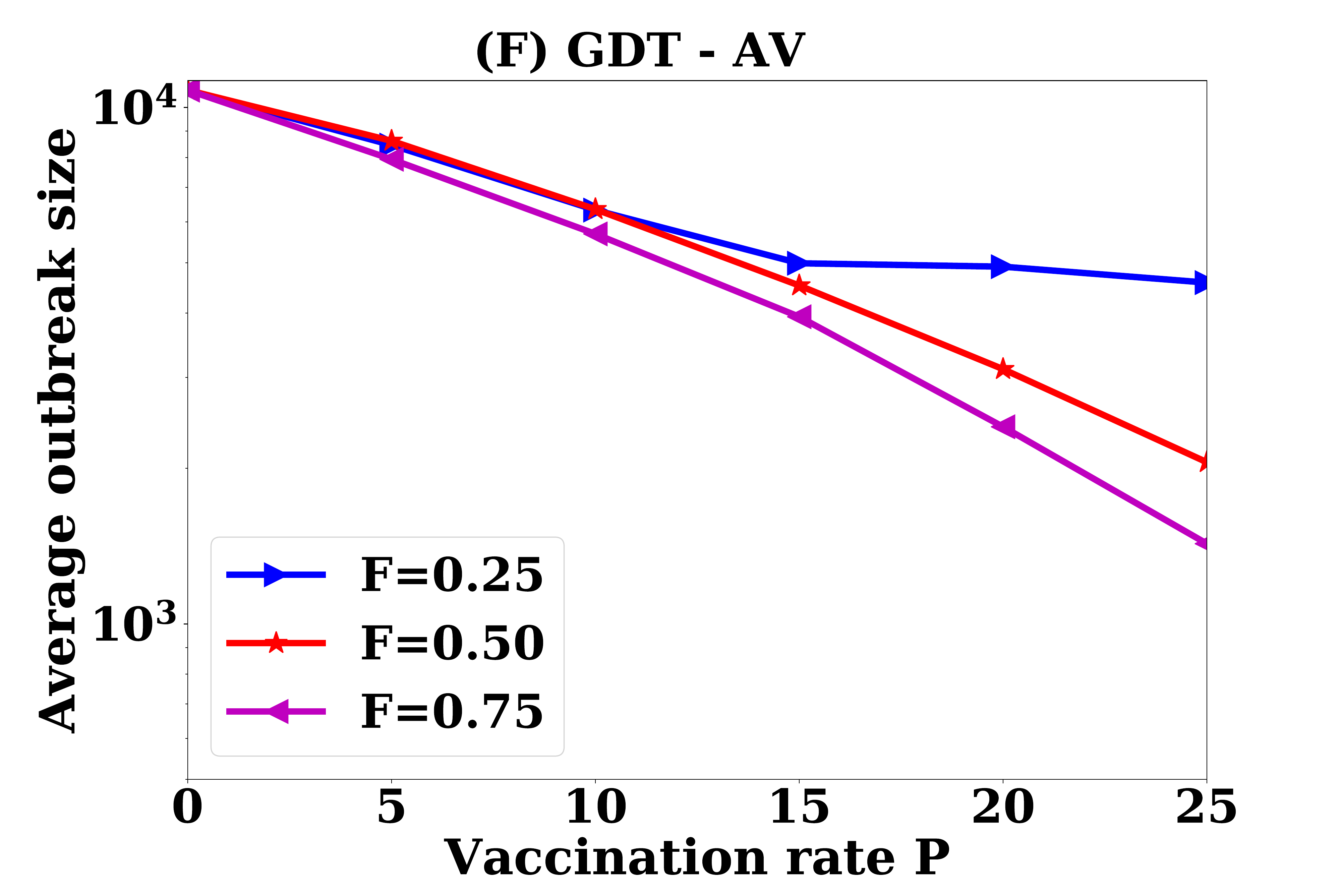}

\caption{Variation in post-outbreak performances with scale of information availability of node contacts: (A,B) proposed vaccination strategy, (C,D) degree based vaccination, and (E,F) acquaintance based vaccination}
\label{fig:fpvac}
\end{figure} 

\begin{figure}[h!]
\begin{tikzpicture}
    \begin{customlegend}[legend columns=2,legend style={at={(0.32,1.00)},draw=none,column sep=3ex ,line width=10pt,font=\small}, legend entries={vaccination rate, outbreak size}]
    \addlegendimage{solid, color=red}
    \addlegendimage{color=blue}
    \end{customlegend}
 \end{tikzpicture}
\centering
\includegraphics[width=0.48\linewidth, height=5.0 cm]{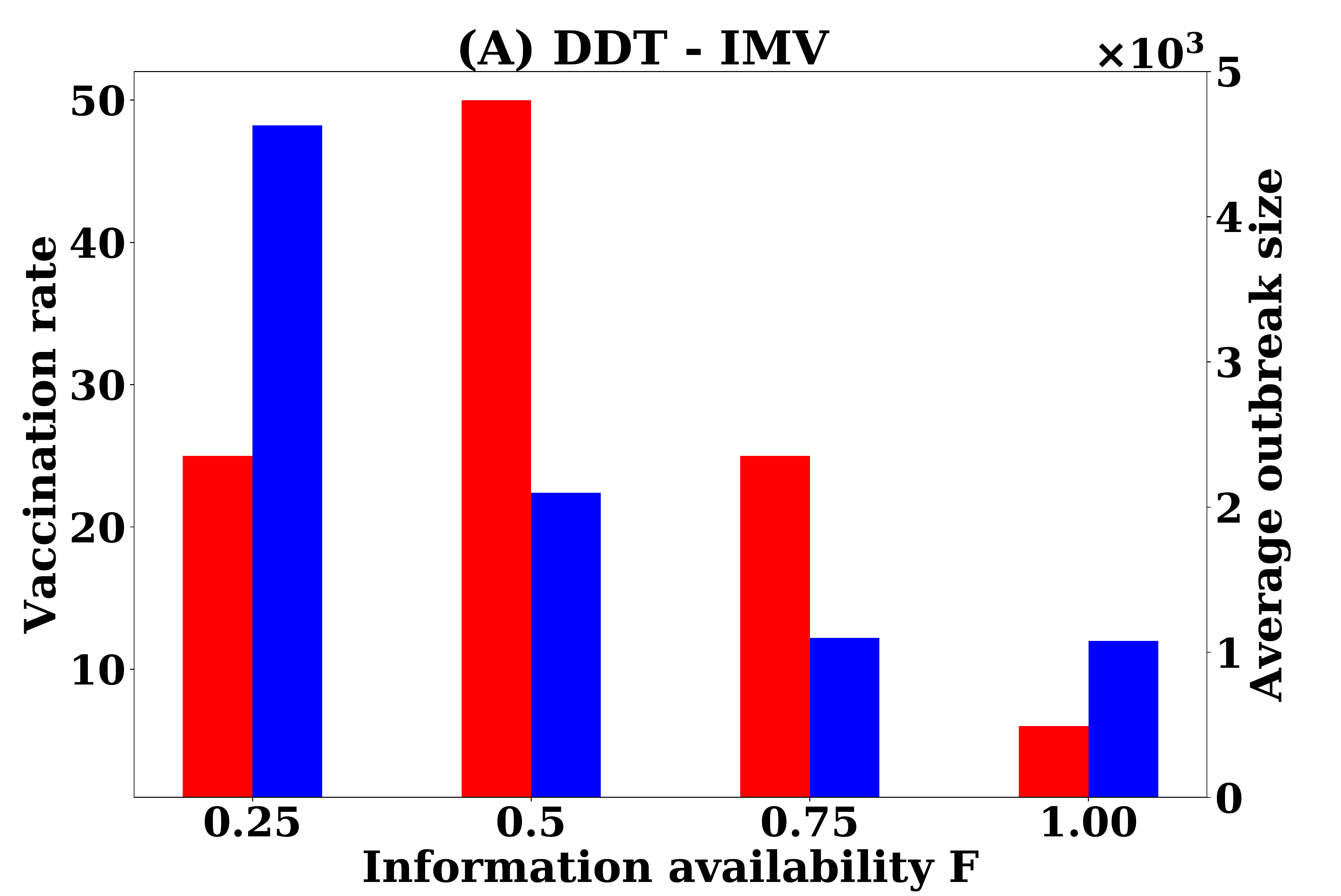}
\includegraphics[width=0.48\linewidth, height=5.0 cm]{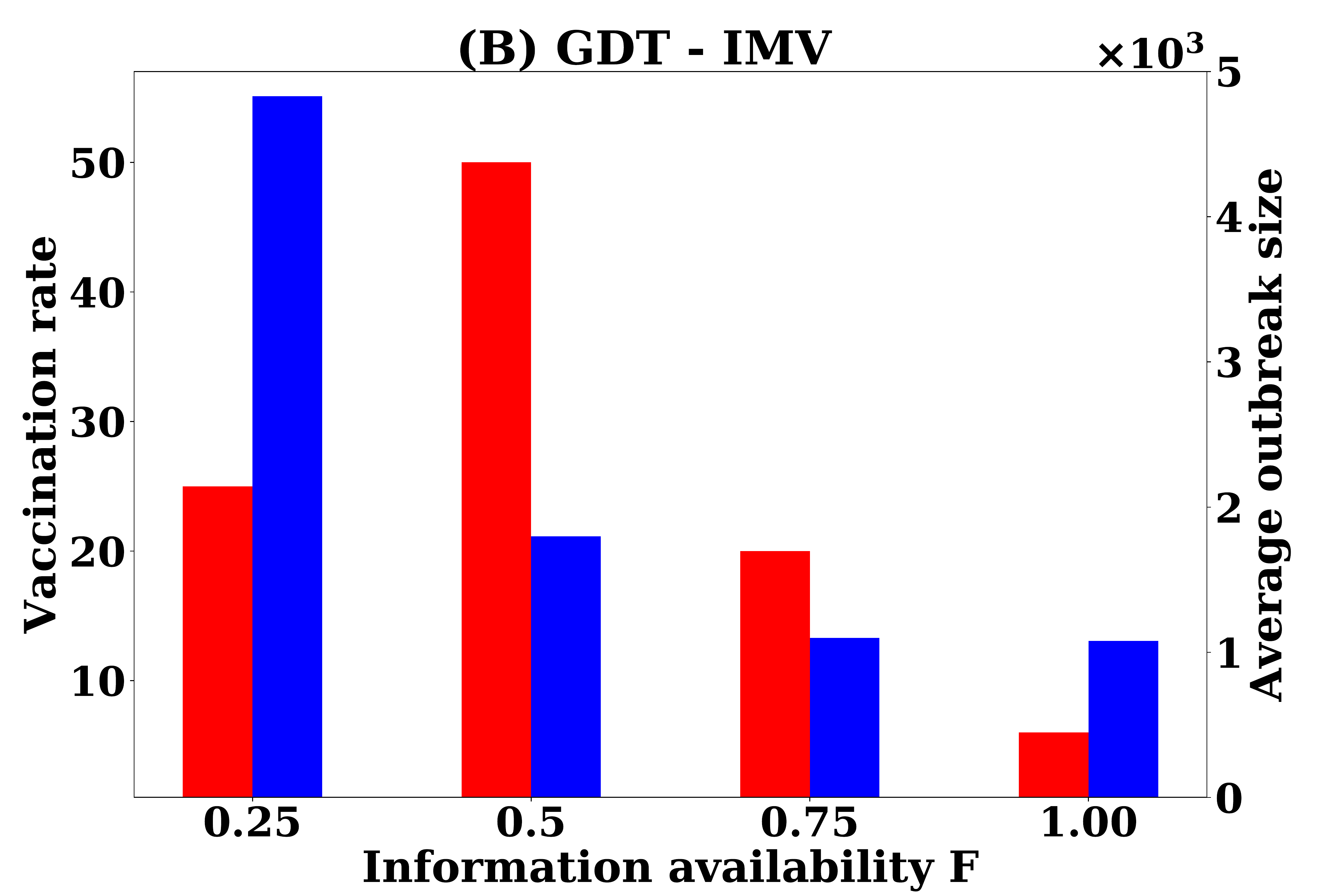}\\ 
\includegraphics[width=0.48\linewidth, height=5.0 cm]{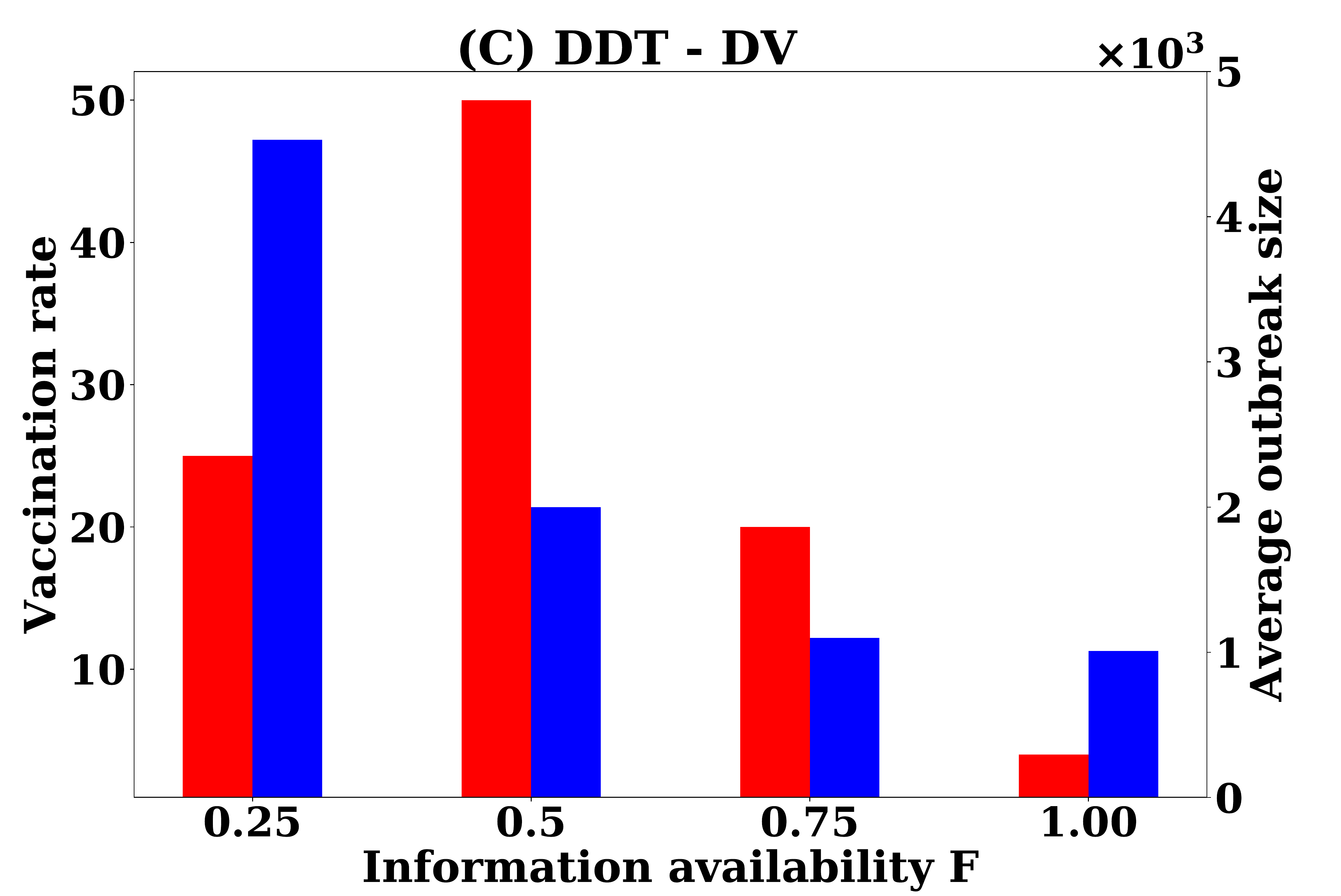}
\includegraphics[width=0.48\linewidth, height=5.0 cm]{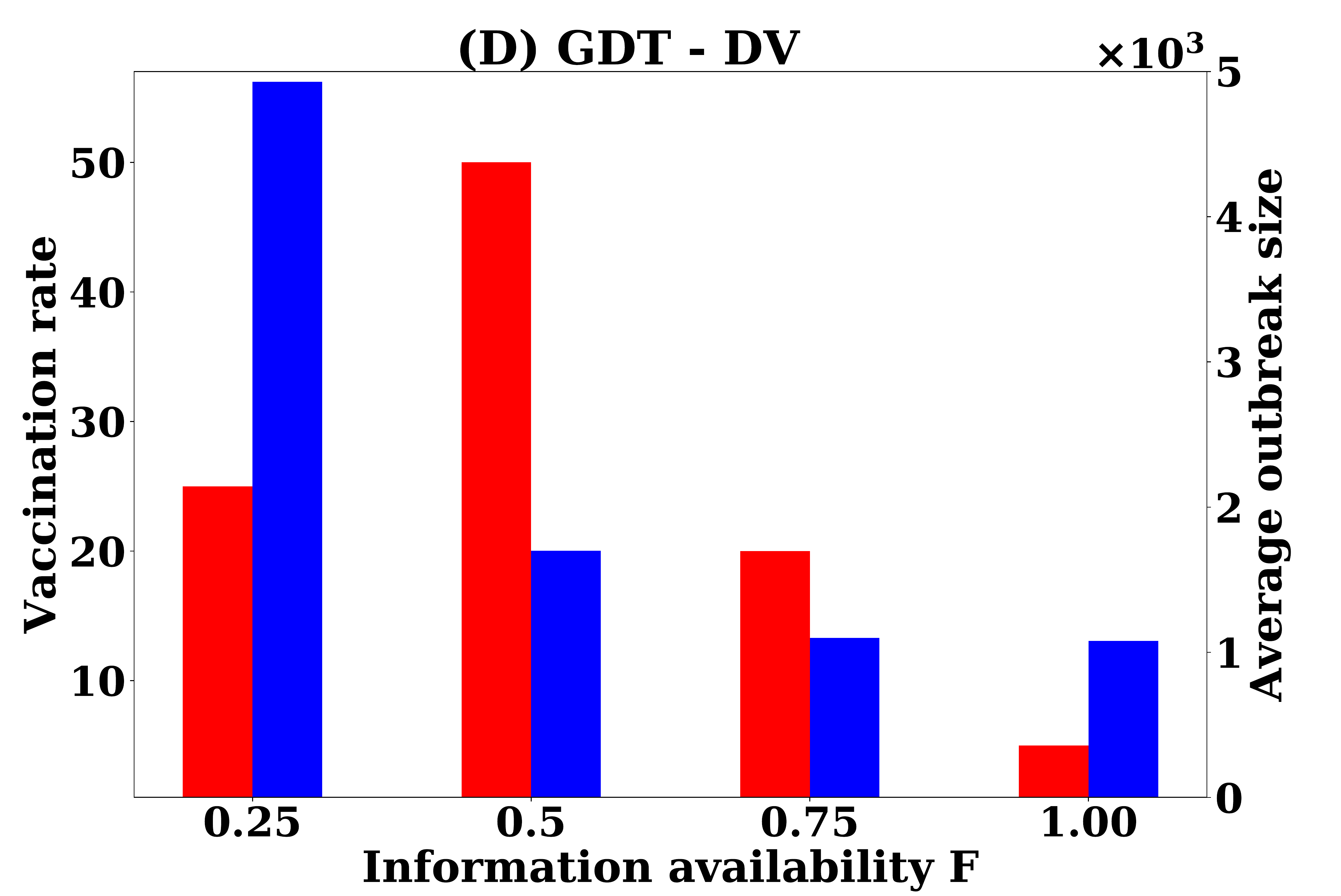}\\ \includegraphics[width=0.48\linewidth, height=5.0 cm]{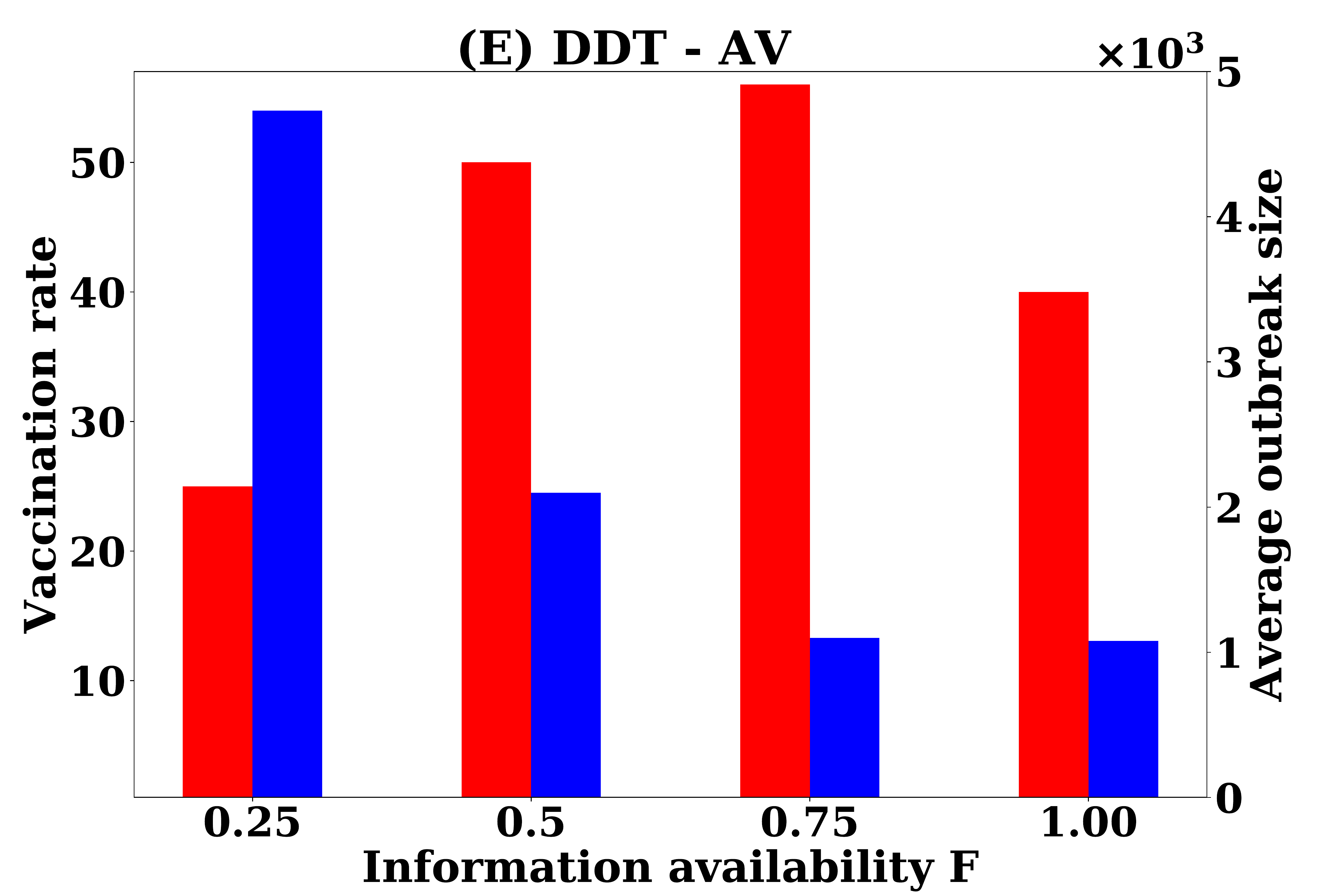}
\includegraphics[width=0.48\linewidth, height=5.0 cm]{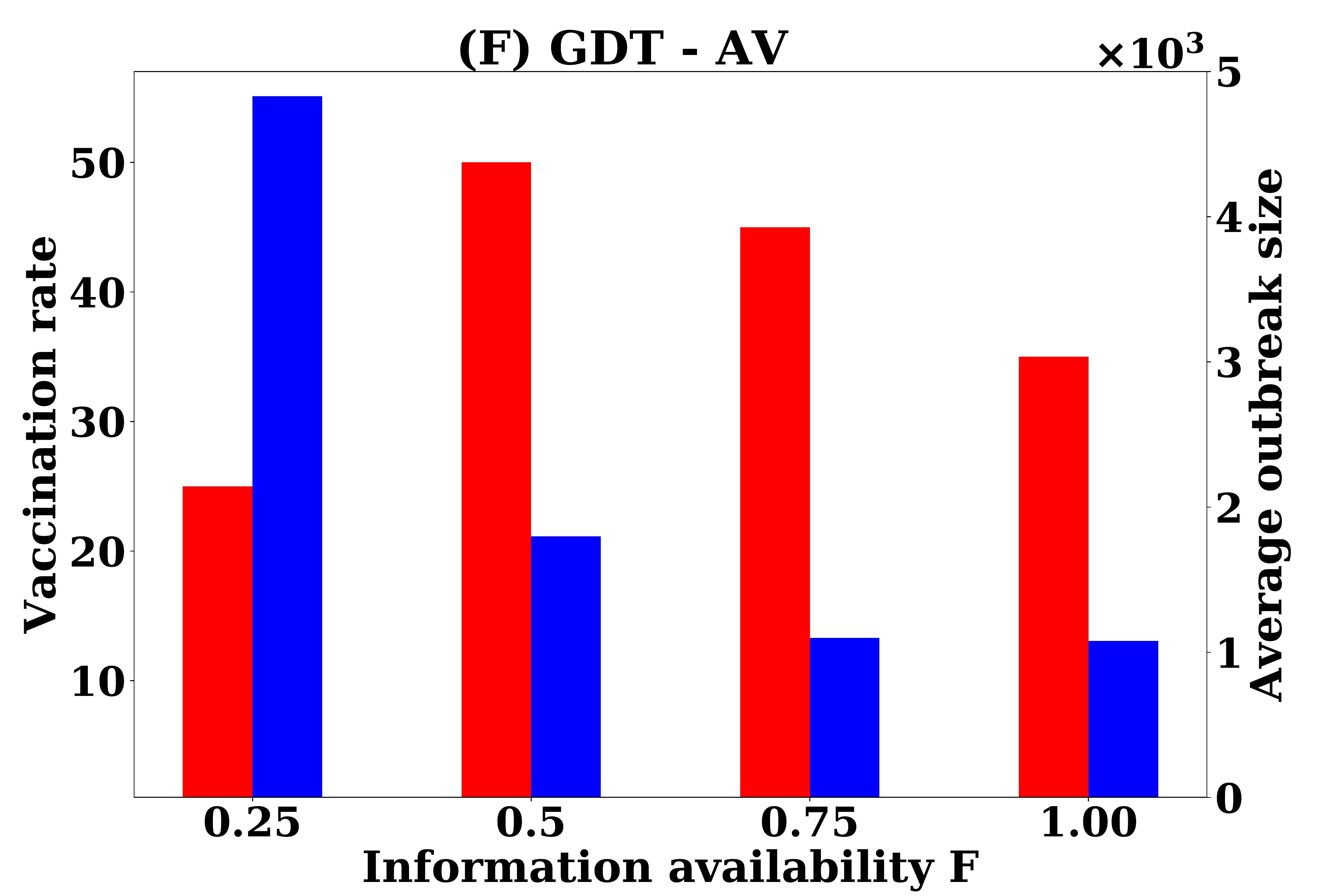}

\caption{Trade-off among information collection cost, vaccination cost and infection cost: (A, B) proposed IMV strategy, (C, D) degree based strategy (DV), and (E,F) acquaintance vaccination (AV) strategy }
\label{fig:fpvacf}
\end{figure}

Simulations are now conducted to understand the effectiveness of vaccination strategies with the scale of information availability regarding node's contact for post-outbreak scenarios. Simulations are carried out for the scenarios where the contact information of $F=\{0.25, 0.5, 0.75\}$ proportion of nodes are available for ranking procedure. At each value of $F$, the performance of the vaccination strategies are analysed for vaccination rates $P$ varying in the range [0,25]\% with a step of 5\%. Similar to the previous experiment, the disease starts with 500 seed nodes and continues for 42 days. The vaccination is implemented at the 7th day of simulations. The final outbreak sizes are obtained on both the DDT and GDT networks. The average outbreak sizes are presented in Figure~\ref{fig:fpvac}. The simulations are also run for all strategies until the average outbreak sizes fall below 1K infections at a certain value of $P$. The obtained results show the trade-off among information collection cost, vaccination cost and infection cost for strategies and are presented in Figure~\ref{fig:fpvacf}.

The performance of the applied strategy is strongly affected by the scale of information availability regarding node contact. The AV strategy slowly reduces the average outbreak sizes within the studied vaccination rates in [0,25]\%. They do not reduce outbreak sizes below 6K infections at $F=0.25$ in both the DDT and GDT networks for any value of $P$.
Therefore, it is not possible to contain the outbreak size by vaccinating nodes if only 25\% of nodes contact information is available. By increasing $F$ to 50\%, both IMV and DV strategies are capable of reducing the outbreak sizes to about 2K infections with vaccinating 50\% of nodes in both networks, i.e. all nodes picked up for ranking procedure are vaccinated. The AV strategy has the same average outbreak size at $F=0.50$ to that of IMV and DV strategy (Fig.~\ref{fig:fpvacf}). All these outbreak sizes are the similar to that of random vaccination with 50\% vaccination rate. The differences at $F=0.5$ are that IMV and DV strategies have more efficiency at lower $P$ comparing to AV (Fig.~\ref{fig:fpvac}). Increasing $F$ to 75\% reduces the requirement of vaccination rate significantly in both IMV and DV. Now, IMV strategy requires 25\% of the nodes to be vaccinated and DV strategy requires 20\% in both networks. At this value of $F$, acquaintance vaccination (AV) shows the ability to contain outbreak sizes below 1K infections with 55\% vaccination in the DDT network and 45\% vaccination in the GDT network. Up to $F=0.5$, the maximum performance can be achieved by the strategies that is same to random vaccination with 50\%. However, IMV and DV strategies for $F> 0.5$ shows that the strong protection is achievable with lower value of $P$. There is a strong trade-off between information collection cost, vaccination cost and infection cost in applying a vaccination strategy (Fig.~\ref{fig:fpvacf}). The proposed IMV strategy still achieves better efficiency than AV strategy with the constrained of information availability. 

\subsection{Node level vaccination}
\begin{figure}[h!]
\centering
\begin{tikzpicture}
    \begin{customlegend}[legend columns=2,legend style={at={(0.42,1.00)},draw=none,column sep=3ex ,line width=10pt,font=\small}, legend entries={ outbreak size, nodes vaccinated}]
    \addlegendimage{solid, color=red}
    \addlegendimage{color=blue}
    \end{customlegend}
 \end{tikzpicture}\\  \vspace{2ex}
\includegraphics[width=0.48\linewidth, height=5.0 cm]{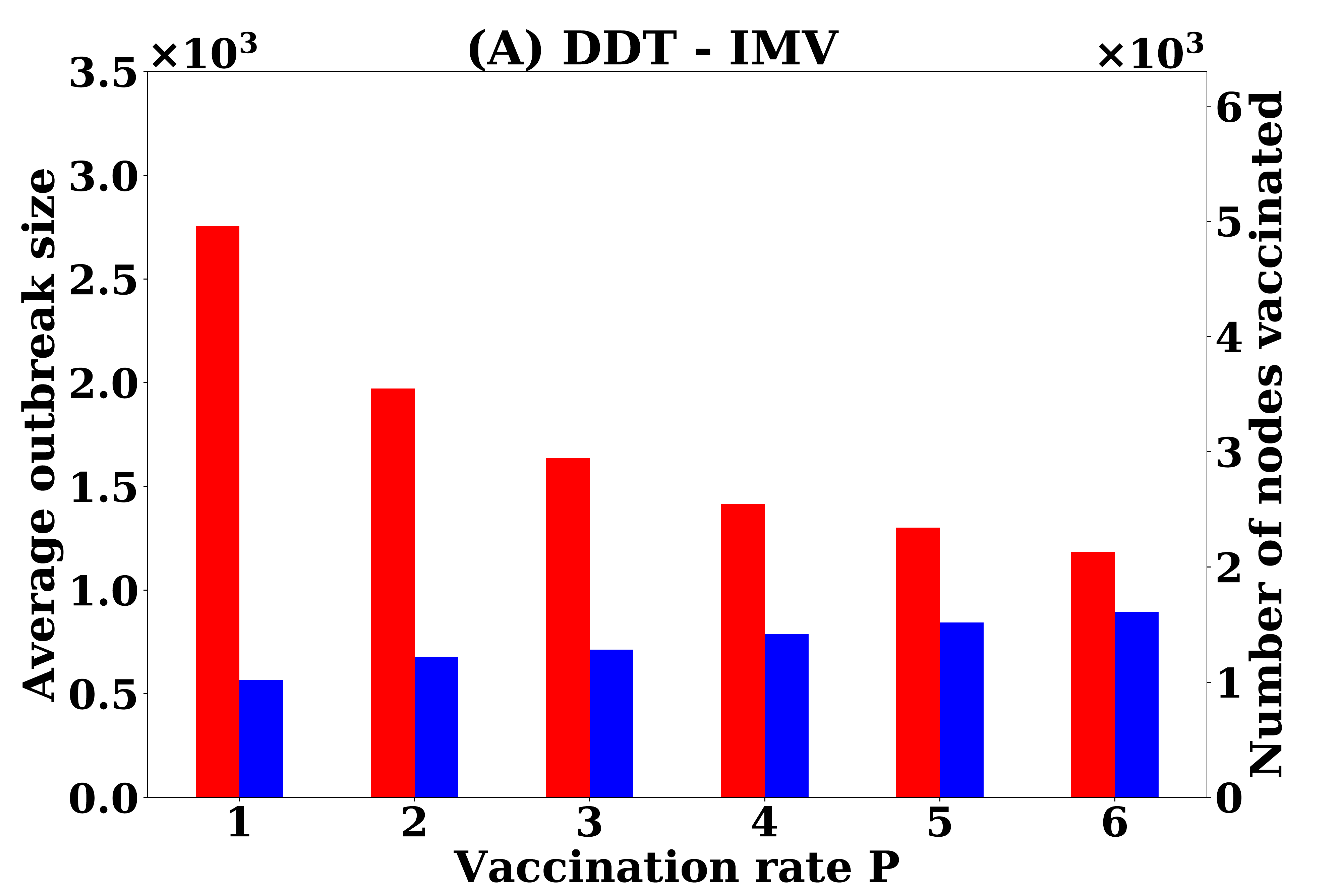}
\includegraphics[width=0.48\linewidth, height=5.0 cm]{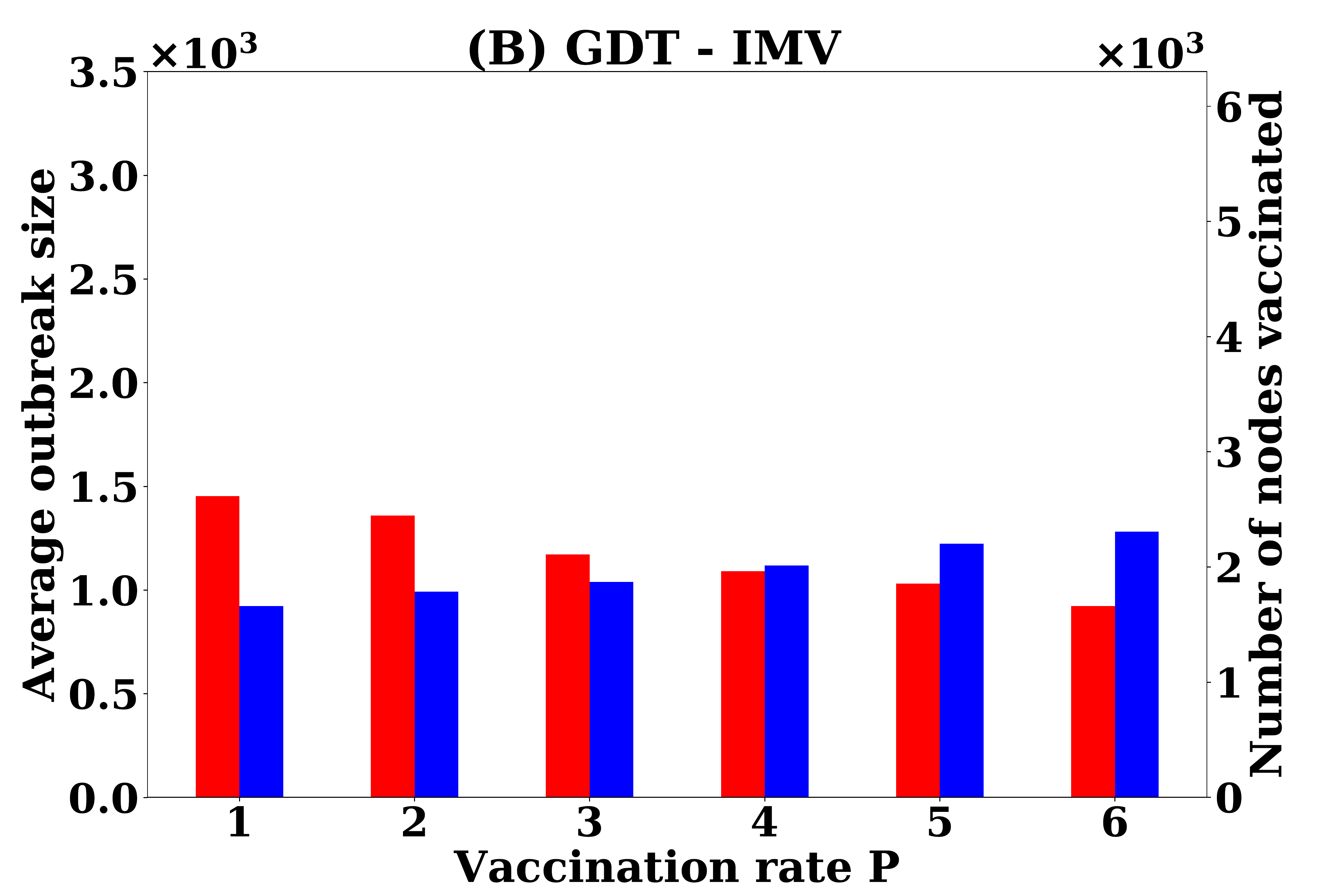}\\ [2ex]
\includegraphics[width=0.48\linewidth, height=5.0 cm]{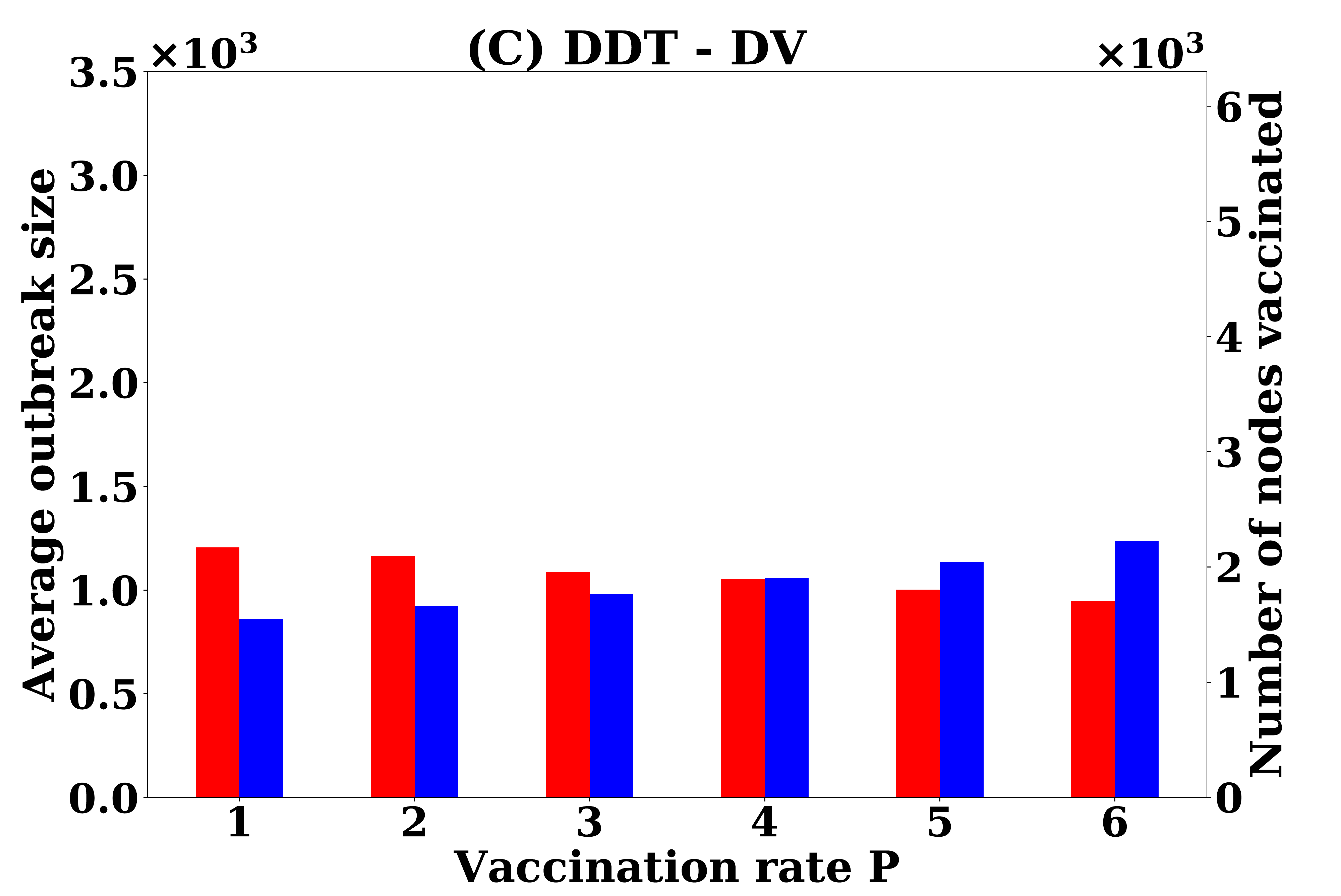}
\includegraphics[width=0.48\linewidth, height=5.0 cm]{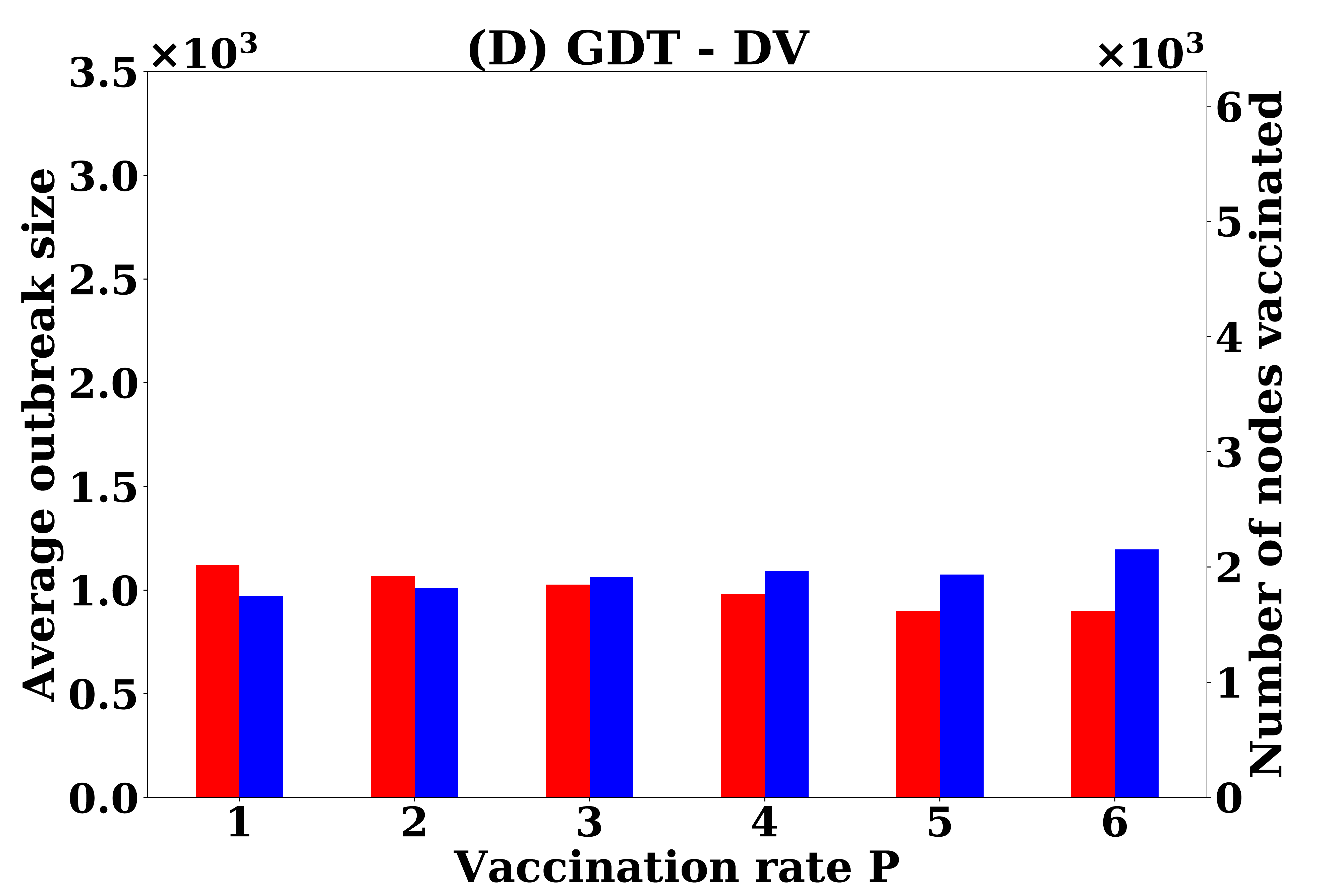}\\ [2ex]
\includegraphics[width=0.48\linewidth, height=5.0 cm]{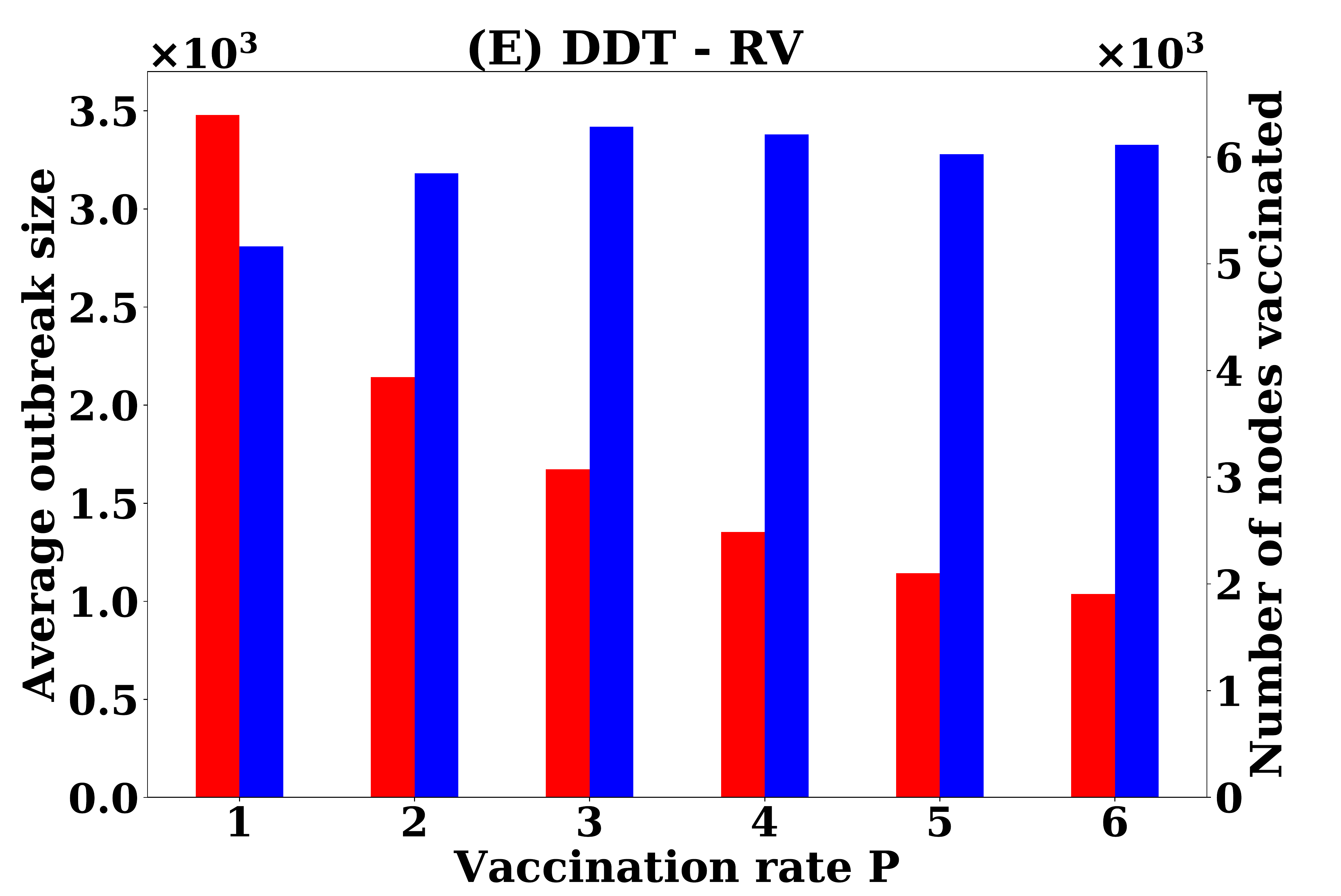}
\includegraphics[width=0.48\linewidth, height=5.0 cm]{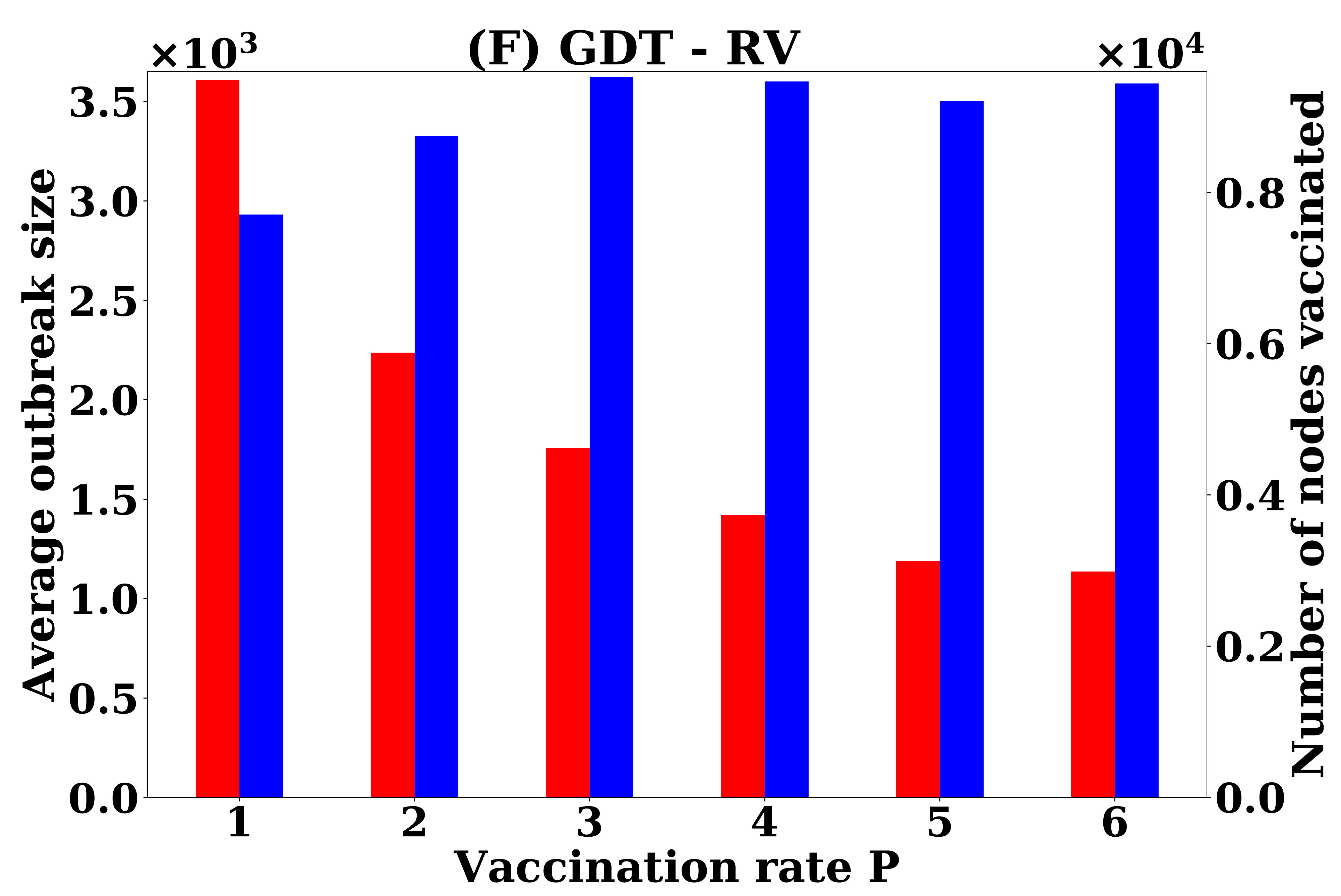}

\caption{Performance of node level vaccination with various strategies: (A, B) proposed IMV strategy, (C, D) DV strategy, and (E,F) RV strategy}
\label{fig:nvaca}
\vspace{-1.0em}
\end{figure}

Population level vaccination requires a high vaccination rate to contain the disease spreading and requirement increases with the constrained of information collection. Alternatively, node level vaccination is often applied through ring vaccination. In this approach, a proportion of infected node's neighbour is vaccinated. In this experiment, neighbour nodes to be vaccinated are selected using three methods, namely random process, degree-based ranks and IMV based ranks. Similar to the previous experiments, the disease starts with 500 seed nodes and continues for 42 days. Initially, nodes are vaccinated at the 7th day of simulation and after that neighbour nodes are vaccinated when a new node is infected. It is assumed in the first experiment that all infected nodes are identified and their neighbour nodes are vaccinated based on applied strategy. Then, a proportion $F$ of all infected nodes is identified and their neighbour nodes are vaccinated based on the applied strategy. In these simulations, a threshold value of ranking score is set to select neighbour nodes following DV and IMV strategies. If a neighbour node has scored more than a threshold, then it will be vaccinated. The threshold value is a ranking score above which a proportion $P$ of total nodes have scored. This ensures that a proportion $P$ of neighbour nodes will be vaccinated through a threshold value. In random vaccination (RV), a proportion $P$ of neighbour nodes are randomly chosen for vaccination. The simulations are run for $P$ in the range [1,6]\% for each strategy and are repeated 100 times for each value of $P$. The main focus in this experiment is to understand the reduction in outbreak sizes against the number of nodes of vaccinated through the above neighbour selection process. The results are presented in Figure~\ref{fig:nvaca}. 

Node level vaccination is more efficient than the population level vaccination. In IMV strategy, average outbreak size at $P=1$\% is about 2.6K infections with the vaccination of 1100 nodes in the DDT network and these are slightly higher in the GDT network. If $P$ are increased to 6\%, the outbreak sizes reduce to about 1K infection in both networks with the vaccination of 2000 nodes. To achieve this performance in population level vaccination, IMV strategy requires vaccination of about 4\% (14,400) nodes. The DV strategy in DDT network shows the average outbreak of 1.2k infections with vaccination of 1700 nodes at $P=1$\% and average outbreak size reduces to about 1k infections with 2000 nodes at $P=4$\%. The GDT network with DV strategy shows slightly higher outbreak size at $P=1$\% but is reduced to 1K infections with increasing $P=6$\% and vaccination of 2600 nodes. However, RV strategy shows very poor performance at a lower value of $P$ with the vaccination of about 5000 nodes in both networks. Increasing of $P$ reduces outbreak sizes quickly with the increasing of the number of nodes to be vaccinated. It is found that the number of nodes vaccinated is stabilised at about 6000 nodes in the DDT network and 10000 nodes in the GDT network regardless of $P$ in RV strategy. Further, increasing of $P$ shows that RV strategy reduces outbreak sizes to about 1K infections at $P=6$\%. For node level vaccination, coarse-grained information based IMV strategy achieves the performance of DV strategy and better than RV strategy.

\begin{figure}[h!]
\centering
\begin{tikzpicture}
    \begin{customlegend}[legend columns=3,legend style={at={(0.32,1.00)},draw=none,column sep=3ex ,line width=10pt,font=\small}, legend entries={ F=0.25,F=0.5, F=0.75}]
    \addlegendimage{solid, color=red}
    \addlegendimage{color=blue}
        \addlegendimage{color=green}
    \end{customlegend}
 \end{tikzpicture}\\  \vspace{2ex}

\includegraphics[width=0.48\linewidth, height=5.5 cm]{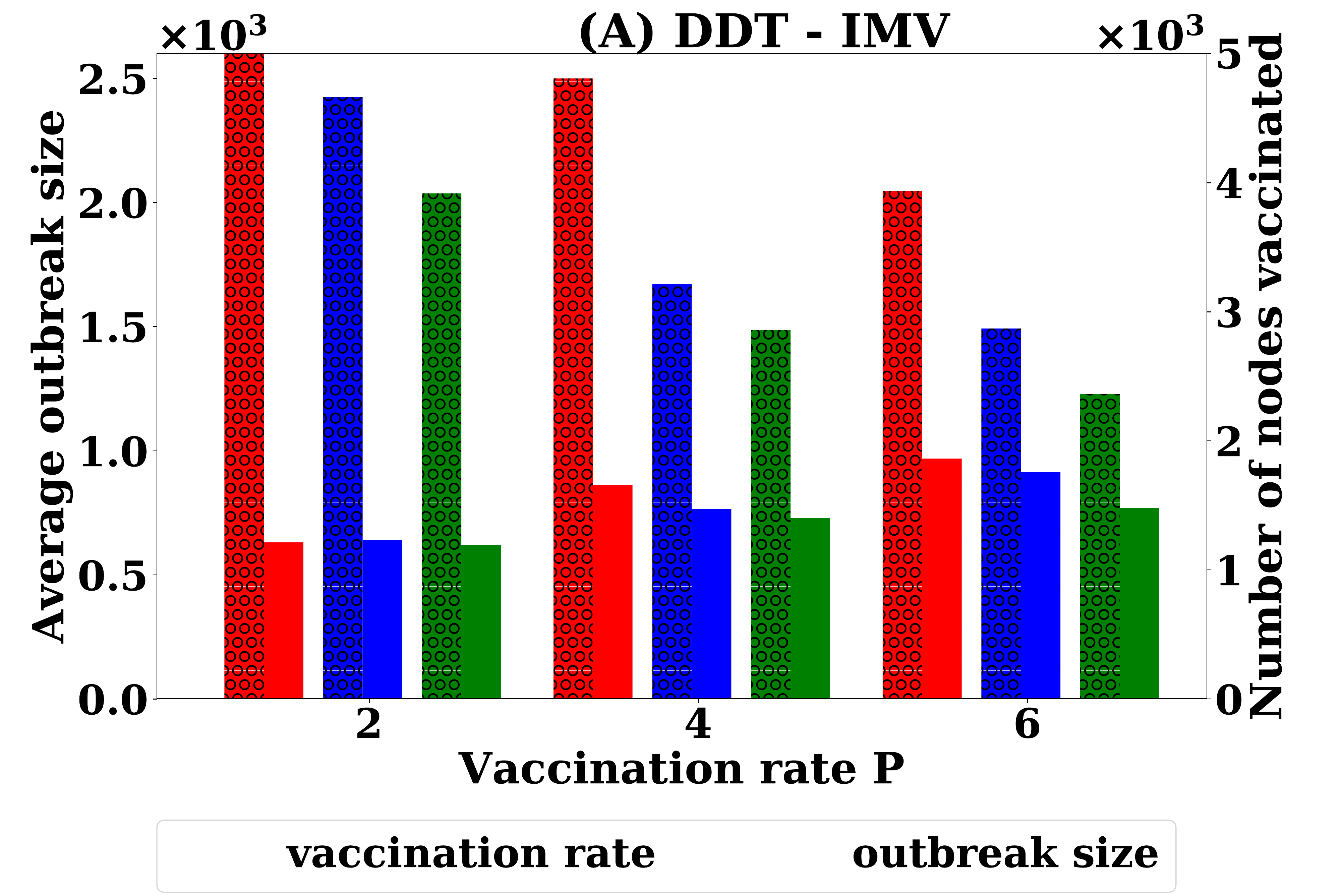}
\includegraphics[width=0.48\linewidth, height=5.5 cm]{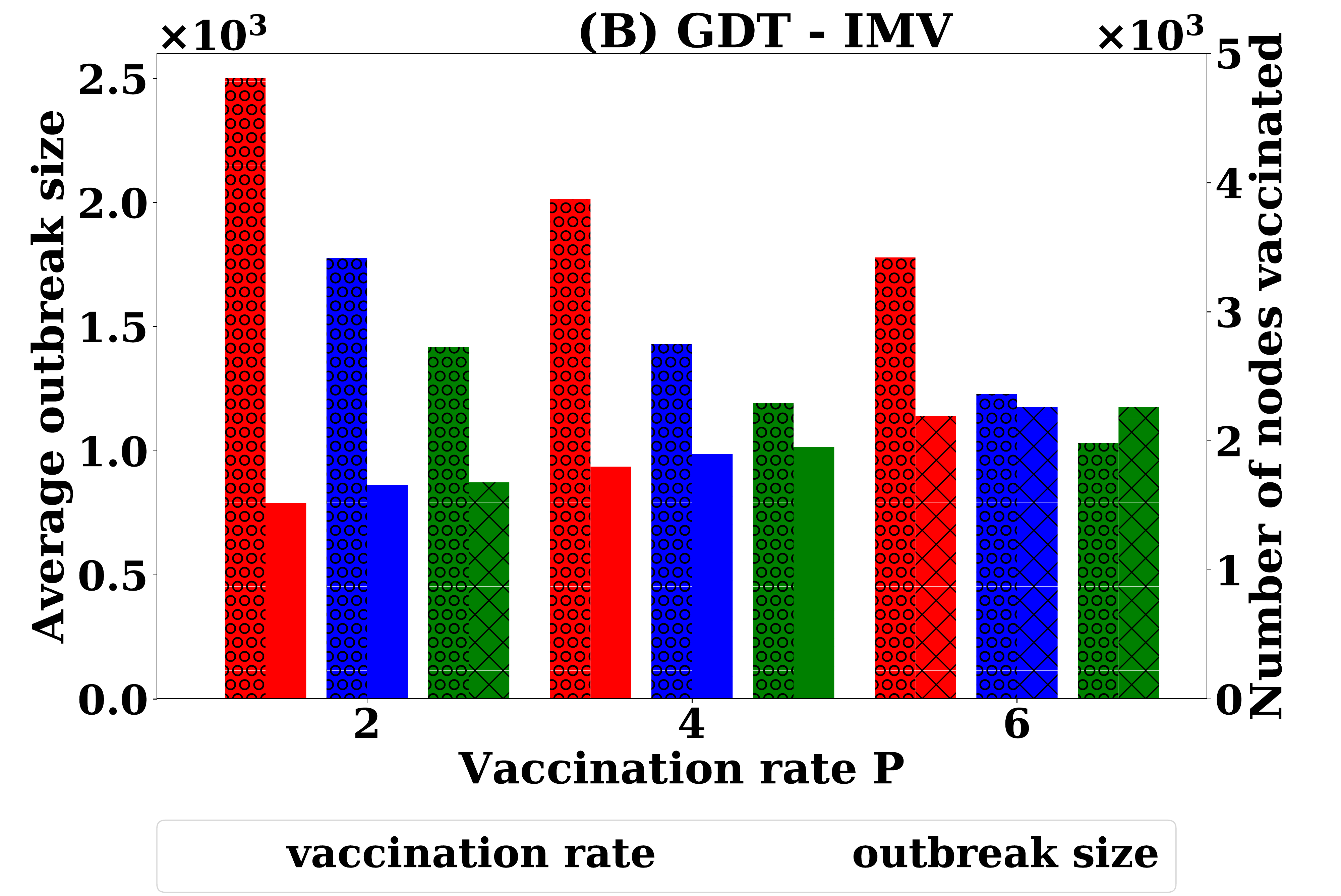}\\ [3ex]
\includegraphics[width=0.48\linewidth, height=5.5 cm]{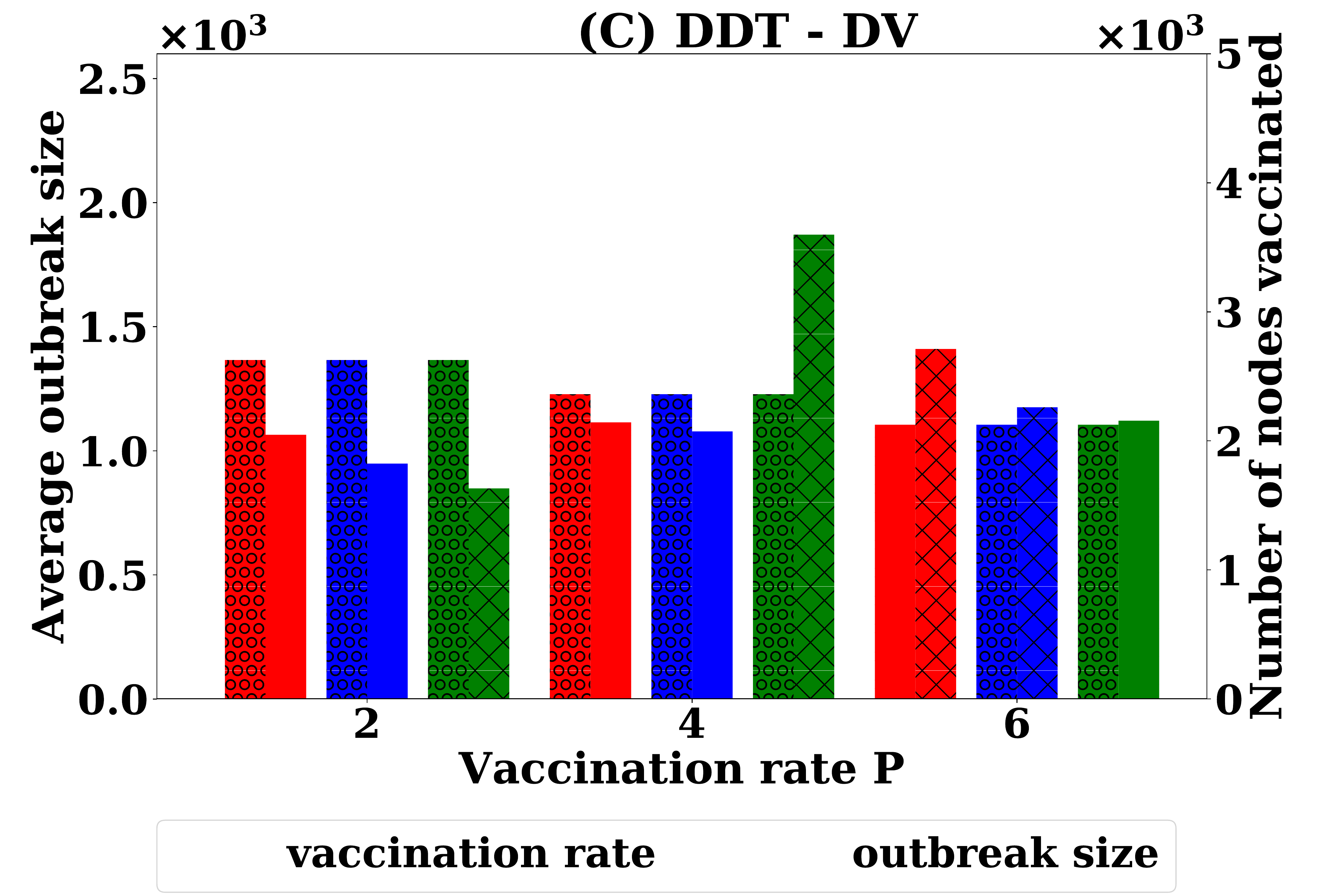}
\includegraphics[width=0.48\linewidth, height=5.5 cm]{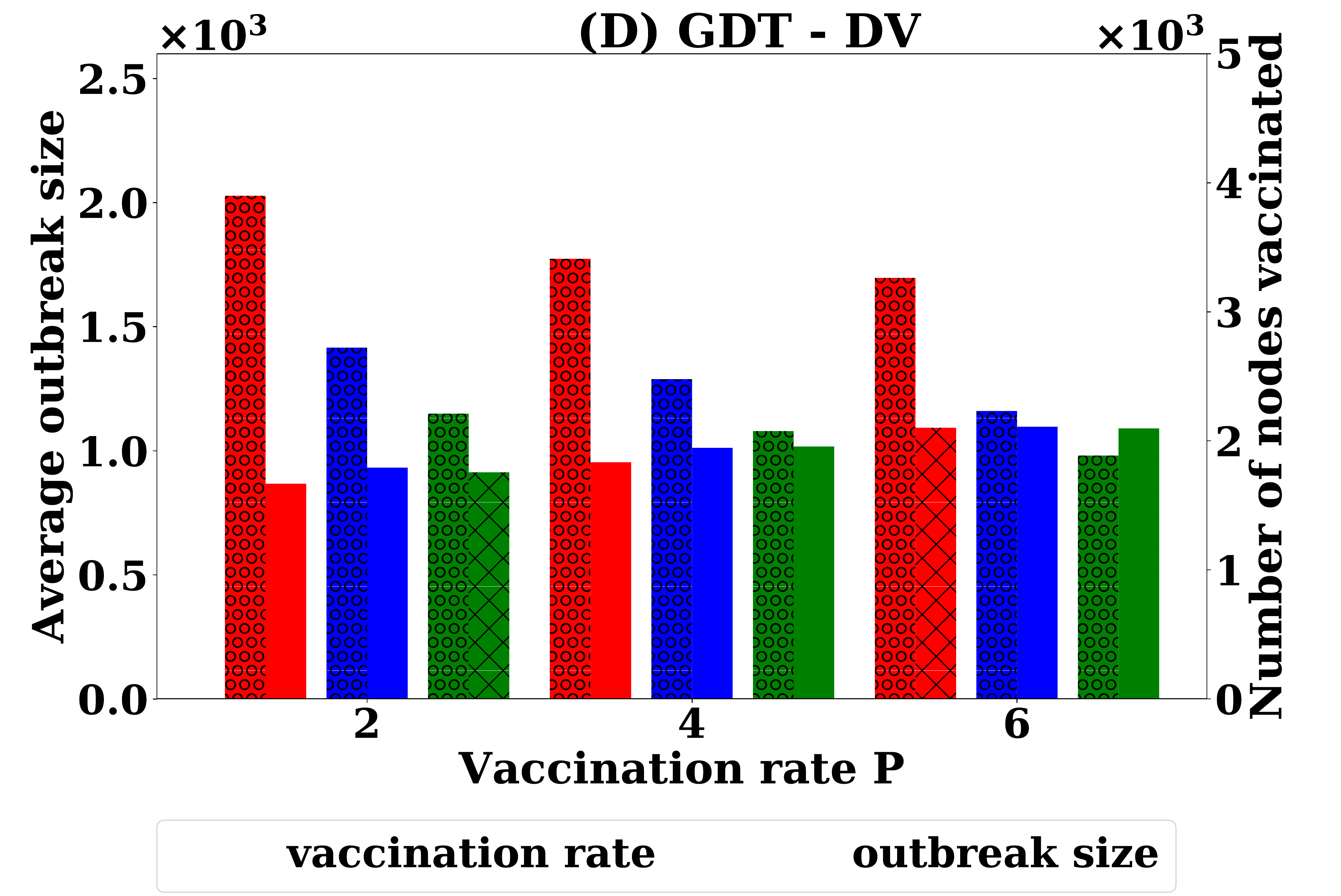}\\ [3ex]
\includegraphics[width=0.48\linewidth, height=5.5 cm]{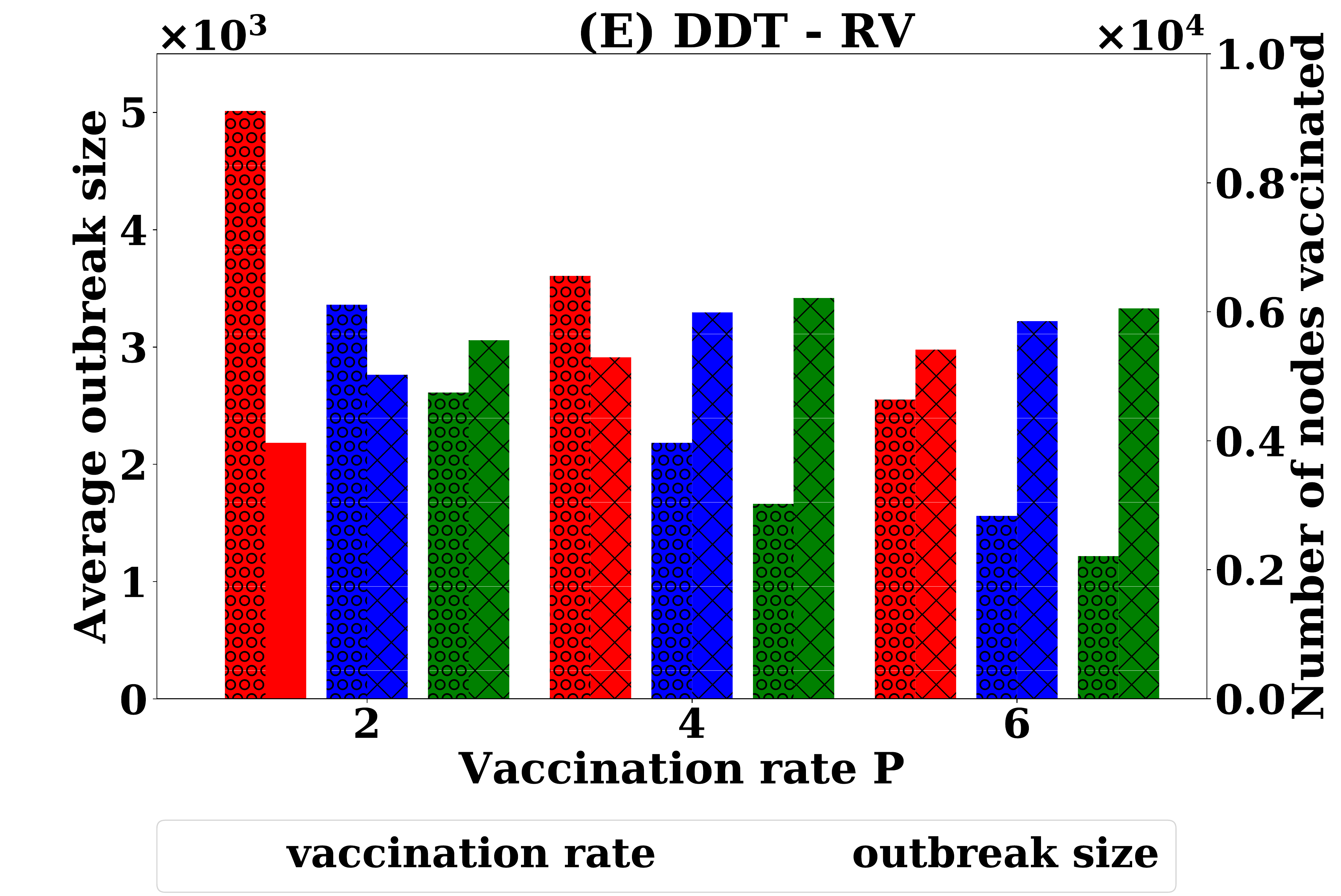}
\includegraphics[width=0.48\linewidth, height=5.5 cm]{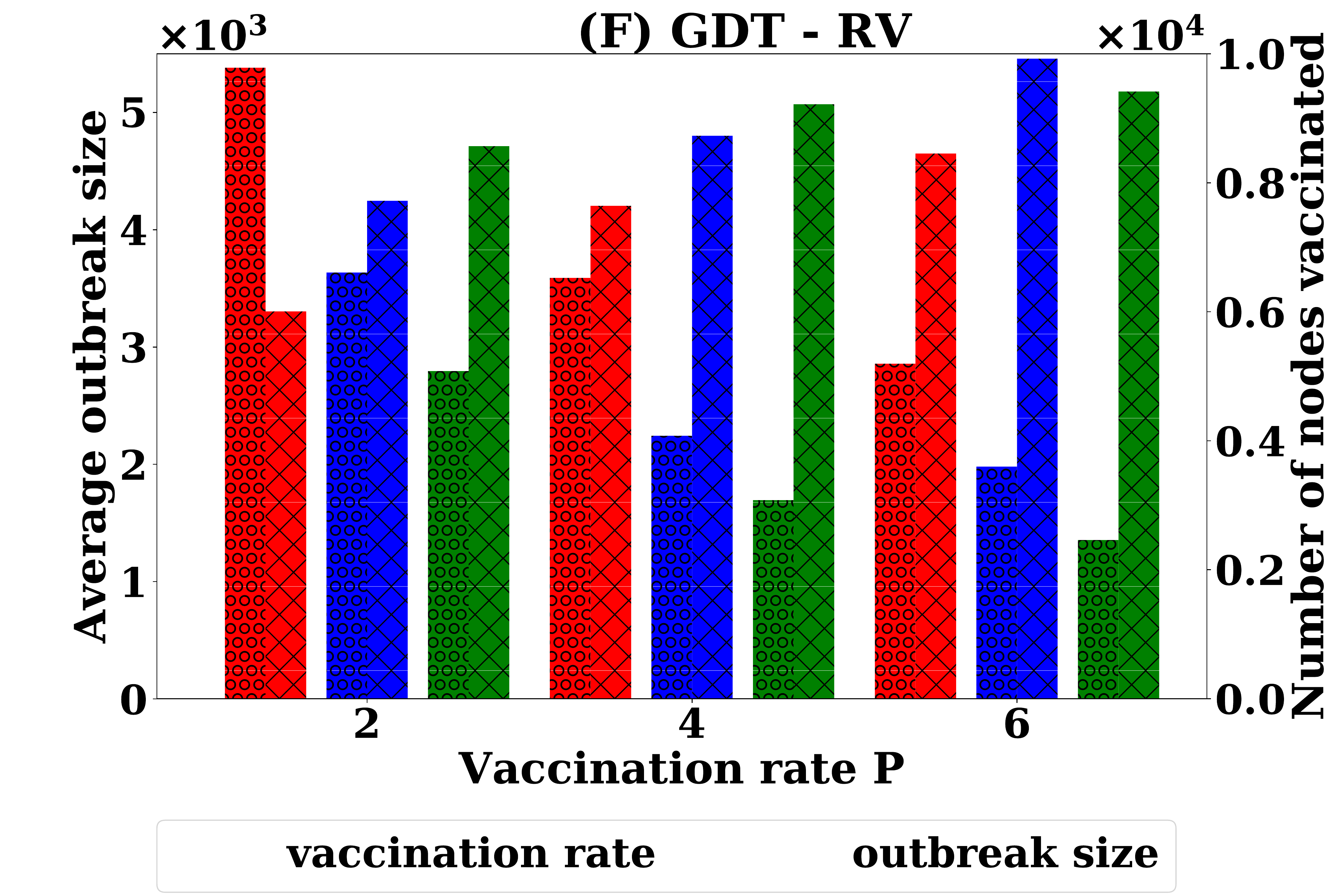}

\caption{Performances of node level vaccination with F proportion of infected nodes identified: (A, B) proposed vaccination strategy, (C, D) degree based vaccination, and (E, F) acquaintance based vaccination}
\label{fig:nvacb}
\vspace{-1.0em}
\end{figure} 

\begin{figure}[h!]
\centering
\begin{tikzpicture}
    \begin{customlegend}[legend columns=3,legend style={at={(0.32,1.00)},draw=none,column sep=3ex ,line width=10pt,font=\small}, legend entries={ F=0.25,F=0.5, F=0.75}]
    \addlegendimage{solid, color=red}
    \addlegendimage{color=blue}
        \addlegendimage{color=green}
    \end{customlegend}
 \end{tikzpicture}\\  \vspace{2ex}
\includegraphics[width=0.46\linewidth, height=5.0 cm]{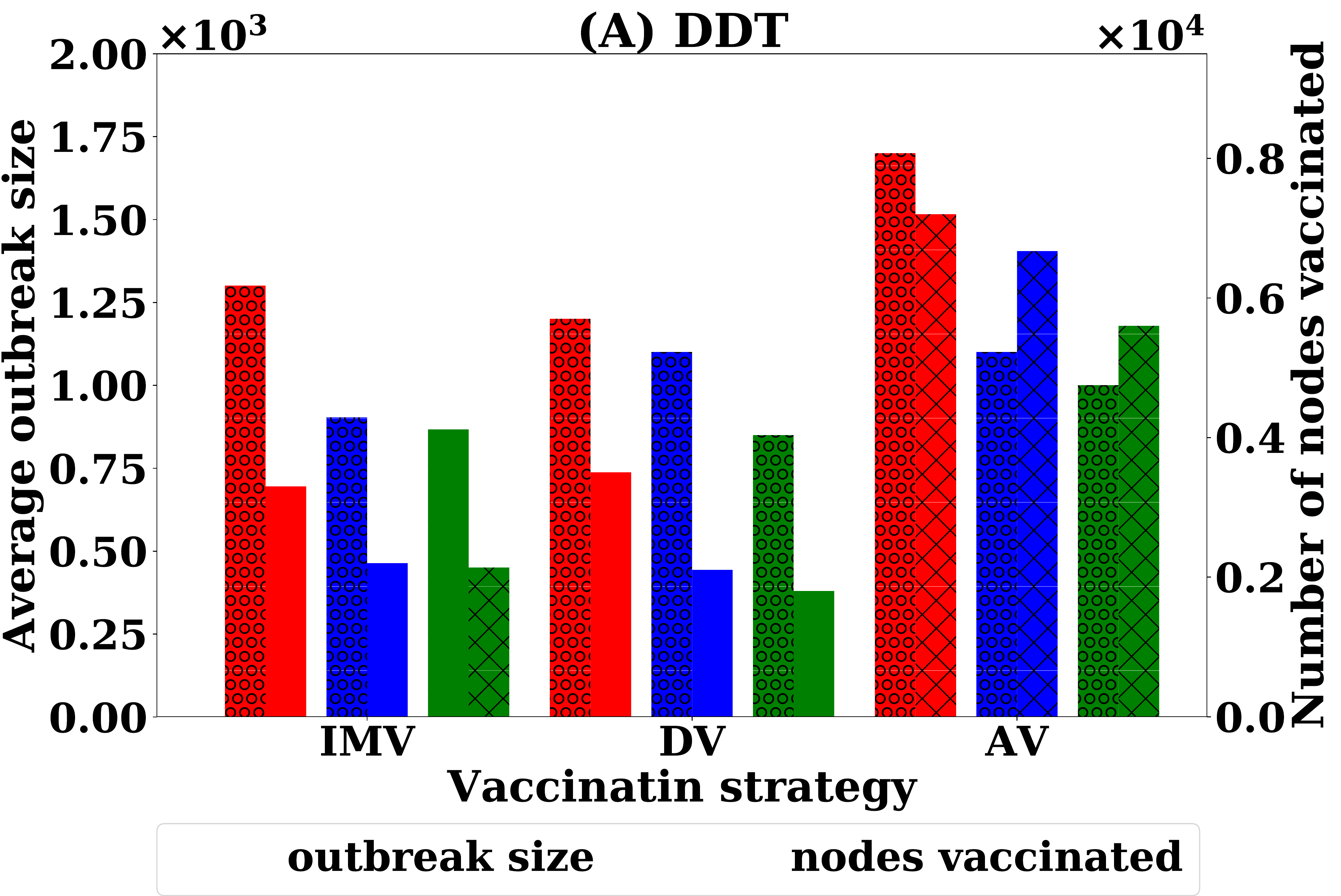}\quad
\includegraphics[width=0.46\linewidth, height=5.0 cm]{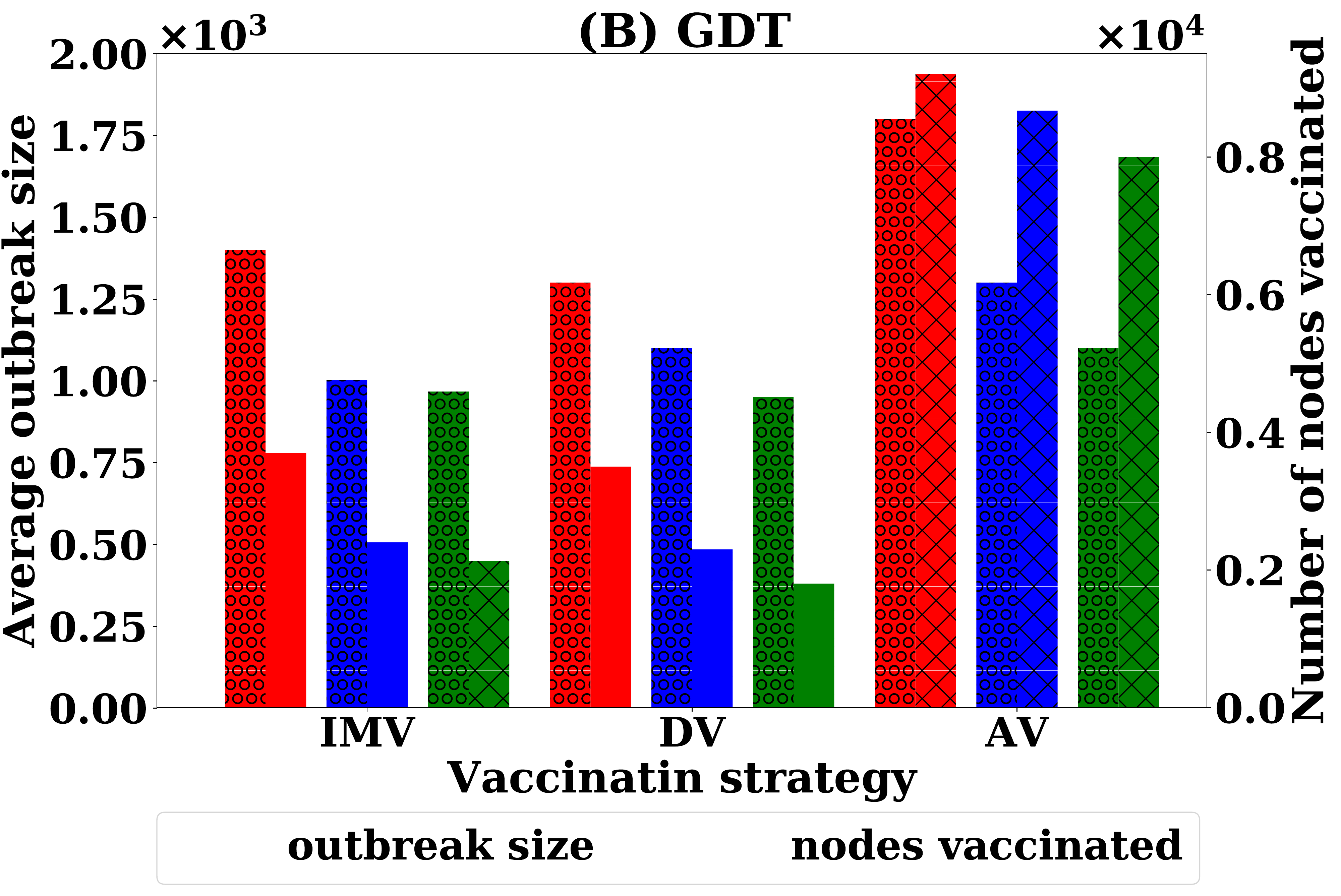}

\caption{Required number nodes to be vaccinated in node level vaccination for achieving outbreak sizes below 1K infections with F proportion of infected nodes identified with different strategies}
\label{fig:nvacc}
\end{figure}

The performance is now studied varying $F$, the proportion of infected nodes can be identified, for each vaccination strategy. In these simulations, a proportion $F$ of infected nodes are selected and then the above threshold based procedures of vaccinating nodes are used according to the applied vaccination strategy. Similarly, disease spreading is simulated for 42 days with the initial set of 500 seed nodes. The simulations are conducted for $P$ in the range [2,6]\% for each strategy and are repeated 1000 times for each value of $P$. The average outbreak sizes and the corresponding number of nodes vaccinated at different $F$ for each vaccination strategy are presented in Figure~\ref{fig:nvacb}. Then, simulations are run for each strategy at different $F$ until the average outbreak sizes become below 1K infections. The results are presented in Figure~\ref{fig:nvacc}. In the IMV strategy, the outbreak sizes is about 2k infections at F=0.25 in both networks at the highest $P=0.6$ with vaccinating about 2K nodes (Fig.~\ref{fig:nvacb}A and Fig.~\ref{fig:nvacb}B). For other values of F, the outbreak sizes reduce and the number of nodes vaccinated reduces in the DDT network at $P=0.6$ while the number of nodes vaccinated is same to 2K nodes in GDT network with reducing outbreak sizes. For IMV strategy, it is not possible to reduce outbreak sizes below 1K infections at F=0.25 in both networks while for other F it is possible by vaccinating 2K nodes. For $F>0.25$, the required number of vaccination to achieve outbreak sizes about 1K infections are same as the 2K nodes (Fig~\ref{fig:nvacc}). The similar trend is found for the DV strategy. However, the RV strategy requires a large number of nodes to be vaccinated for reducing outbreak size below 1K infections. Similar to the other strategy, the outbreak sizes do not reduce to 1K infections in RV strategy even vaccinating 11K nodes. Under the constrained of information collection, IMV strategy still performs well compared to RV strategy.

%% file: 7_chap_conclusion.tex

\chapter{Conclusion}
This thesis has investigated how the inclusion of indirect transmissions influence diffusion phenomena on dynamic contact networks. In particular, this thesis has developed a diffusion model called the same place different time transmission based diffusion model (SPDT diffusion) and a graph model called SPDT graph for investigating diffusion phenomena with indirect transmission links. SPDT diffusion behaviours and how to control them were explored on dynamic contact networks. This chapter describes the main findings, the limitations of conducted research and the future research directions of SPDT diffusion.

\subsubsection{Research outcomes and analysis}
Current diffusion models cannot capture indirect transmission opportunities for dynamic diffusion. To fill this gap, a novel diffusion model called the same palace different time transmission based diffusion (SPDT diffusion) has been developed in Chapter 3. An infection risk assessment method has been integrated in the SPDT diffusion model to account for indirect transmissions. Therefore, the SPDT diffusion model can determine transmission probability accounting for the arrival and departure times of both susceptible and infected individuals at the interaction locations. The risk assessment method has also defined a particle decay rate which reflects the removal of contagious items from the interaction location against a particle generation rate. Thereby, the SPDT model captures dynamic transmission probabilities of links through random particle decay rates and different arrival and departure times of both susceptible and infected individuals. The SPDT diffusion model captures the contact properties at a granular level.

The inclusion of indirect transmissions with the SPDT diffusion model introduces a new dimension in network dynamics where the underlying contact dynamics vary with particle decay rates. This is because the length of time infectious particles are present at a location varies with the particle decay rates and thereby changes the number of individuals to be exposed to the infectious particles. Therefore, the underlying network connectivity in the SPDT model changes according to the decay rates and hence the speed of diffusion. Investigations in Chapter 3 has shown that there is a threshold value of decay rates above which the disease prevalence increases in the networks. Otherwise, the disease dies out quickly. Disease parameters such as infectiousness of disease and recovery rates also vary the influence of decay rate. The interesting finding is that the impact of SPDT links is not reproducible with the direct SPST links as the decay properties of indirect links are not captured with direct links.

A new graph model called SPDT graph has been developed in Chapter 4 for understanding SPDT diffusion behaviours in the synthetic contact networks and to provide a tool for further exploration of SPDT diffusion in if-else scenarios. The principle of activity driven time-varying network modelling (ADN) has been adopted for this model. Then, the graph generation methods have been developed using simple statistical distributions and fitted with real contact graphs of Momo users. Analysis of the model has shown that the generated graph assumes the same network properties of real SPDT graphs. The SPDT graph is also capable of simulating SPDT diffusion processes and responds to the diffusion model parameters substantially. Comparing to the traditional SPST models and ADN models, the SPDT graph model captures SPDT diffusion dynamics of real contact networks up to 30\% more accurately.

The SPDT model has introduced a new property in the contact networks through \textit{hidden spreader} nodes: they do not have direct transmission links during their infectious periods. In Chapter 5, the spreading potential of indirect links has been studied through these hidden spreaders as they do not transmit disease through direct links. The potential of the indirect transmission links has thus been characterised by the spreading potential of hidden spreaders. The analysis has shown that the indirect links have at least the same potential of direct transmission links. The interesting finding is that hidden spreader nodes can be super-spreaders whilst by definition they have zero spreading potential in SPST networks. This chapter has also revealed that the networks with only hidden spreaders can lead to outbreaks of significant sizes. This indicates that if face-to-face meetings (SPST scenarios) are limited during outbreaks, there might be still outbreaks through co-location indirect interactions. The other finding is that the number of super-spreaders doubles in SPDT models due to including indirect transmission links. Therefore, the likelihood of emerging infectious disease increases.

The control strategies based on the direct transmission links fail to control diffusion in the SPDT contact networks (Chapter 5). Thus, an individual movement based vaccination (IMV) strategies have been introduced in Chapter 6 where individuals (nodes) are ranked based on their movement behaviours. This strategy depends on locally obtainable contact information and approximate contact information based on Points-of-Interests instead of exact contact information. The proposed IMV strategy requires up to 67\%  fewer vaccinations than the random vaccination (RV) strategy and up to 37\% fewer than the acquaintance vaccination (AV) strategy. In addition, the required rate of vaccination reduces to 0.5\% of the nodes for IMV strategy with ring vaccination approach. No vaccination strategy can contain disease outbreaks effectively with the contact information availability of 25\% of the nodes. However, IMV strategy can work for the contact information availability of 50\% of the nodes while RV and AV strategy fails. With limited contact information of nodes, IMV strategy with ring vaccination approach outperforms other strategies.


\subsubsection{Limitations of the conducted research}
This research has three types of limitations: i) simplified disease propagation, ii) sparsity of data and, iii) simplified methodology. 
\begin{quote}
 i) Simplified disease propagation: The decay rates combine all removal processes of contagious items (described in Chapter 3). However, the removal of particles can occur for several factors such as weather conditions or properties of contagious items and may have different impacts. In the simulations, the contact information is aggregated over one day to calculate the infection probability assuming that this will produce similar results of the separate contact information.  
\end{quote}

\begin{quote}
ii) Sparsity of data: The applied empirical contact networks constructed by Momo users provide the partial contact information of real-world scenarios. Applying complete information may change diffusion dynamics. In the investigation of vaccination strategies, the classification of locations has been done using common sense. The real classification may vary and the obtained results can change. The obtained diffusion dynamics in the experiments have not been verified with real-world disease spreading cases.
\end{quote}

\begin{quote}
iii) Simplified methodology: In developing the graph model (Chapter 4), the spatial diversity of nodes has been abstracted. This could also vary the diffusion dynamics. The stability of the graph model has not been studied to define the boundary conditions. This thesis has not studied the enhancement in a specific network property to correlate it with the resultant diffusion dynamics. During the outbreak, individuals may use the self-protection such as a mask. The diffusion controlling strategy did not consider the impact of such self-protection. The performance of IMV strategy is not also compared with any strategy that uses temporal information. 

\end{quote}

\subsubsection{Future research directions}
The SPDT diffusion model introduces interesting and innovative research problems in the field of diffusion study. The exploration of individual-level movement data allows researchers to develop data-driven diffusion models. However, the obtained results are often validated with an analytical model. Thus, it is important to develop a mathematical model for SPDT diffusion as well. In this study, the realistic disease parameters from the literature are used to simulate the disease spreading. It would be an interesting research direction to collect real disease incidence and compare the simulation results with them. The model developed can be studied in other scenarios such as message dissemination in online social networks.

There are several interesting research directions for future development of the graph model developed in Chapter 4. For simplicity, it is assumed that the maximum indirect link creation delay is constant. Thus, the model might miss some of the contacts that are created after that period and their impact on diffusion. The network generation process can be improved by including the unlimited link creation delay which is limited in this thesis to 3 hours. The model currently aggregates the node states of the basic activity driven network model. It will be important in the future to quantify the effect of this aggregation simplification along with other simple assumptions that may affect the model's resolution and representativeness. The critical conditions for network stability should also be studied further. 

The creation of indirect links in SPDT model for visiting the same locations provide opportunities to connect one individual to many other individuals comparing to that in the SPST model. Thus, the community structure may change in the SPDT model. It is widely discussed in the literature that the community structure has a strong influence on diffusion dynamics. It would be interesting to know how the community structure changes in the SPDT model and impacts diffusion dynamics. The indirect links creation opportunities also change the temporal paths in SPDT model compared to the SPST model. Thus, the other interesting direction of research is to study the temporal properties and their impacts on diffusion dynamic in the SPDT model. One interesting research direction is to investigate the impacts of hidden spreaders on the efficiency of current models of tracing the source of infection. It is often hard to obtain all contact data from outbreak scenarios. Thus, SPDT model can be used to estimate the extent of errors based on the available data where it can be assumed that direct contacts represent the available data while indirect contacts represent the missing data.

The implementation methods of vaccination can determine the effectiveness of the strategies. For example, a strategy provides better performance if vaccination is implemented based on community or spatial regions. Instead of vaccinating many individuals from the same community, it may be more efficient to vaccinate individuals from different communities or regions. This can be a potential research direction for studying SPDT diffusion control. The proposed strategy is not compared to strategies integrating temporal contact information. The performance of the control strategy can be studied with temporal information based methods. For the post-outbreak scenarios, closures methods (school close, popular location closes) are studied for post-outbreak vaccinations~\cite{costill2016targeted,ciavarella2016school}. Thus, the performance of IMV strategy can be compared with these closure methods as well.

\subsubsection{Significance and applicability of outcomes}
The SPDT diffusion model captures a realistic disease reproduction and shows its applicability to be applied in real-world applications. The model can be applied for vector-borne diseases such as malaria~\cite{tatem2014integrating,adams2009man}. The variations in SPDT diffusion dynamics according to decay rates can be applied to model the seasonal variations in infectious disease spreading~\cite{yan2018infectious,biggerstaff2014estimates}. The SPDT model can also be used to study the spreading of information in online social networks (OSN). For example, the popularity of a message posted on the Facebook wall of a user decays with time~\cite{nasrinpour2016agent}. Therefore, this link for transmitting messages has decay properties that are similar to indirect links in the SPDT model. The SPDT model can also be used for directed weighted networks as it is directed. Heterogeneous particle generation rates and decay rates can be applied to model weighted links.  


Contact networks generated by the developed graph model can simulate the SPDT diffusion and reproduce diffusion dynamics of real contact networks. The graph model can also generate SPST contact networks for co-location direct interactions. Thus, the SPDT model allows one to study if-else scenarios of disease spreading in both SPDT and SPST models. The co-location based applications can be studied with the generated networks~\cite{meyners2017role,molitor2016location,lu2016towards}. The other interesting application of SPDT graphs is to apply it for modelling diffusion on multiple social network platforms~\cite{jain2013seek}. A user can be active in multiple social networks and interact with friends which can be captured by the concept of active copy introduced in the graph model to represent activities at different social networks. Similar to the SPDT diffusion model, the SPDT graph model can be used to generate directed networks with weighted links as the model parameters active period, link creation delay, and link duration assume heterogeneous values for each link. 

The developed vaccination strategy effectively protect networks from infection. It is also found that the proposed strategy performs well under the constraint of information availability as well. These finding will help health authority to plan vaccination of infectious diseases. This study was conducted for co-location interactions contact networks. However, there are many infectious diseases that do not require co-location interaction to be transmitted. The proposed vaccination strategy IMV can be applied for other diseases as well with some reconfiguration. For example, the IMV strategy can be applied for SIT disease where visiting a partner means activating at a location and where contact degree is one~\cite{ghani2000risks}. The IMV strategy can be applied to maximise the spreading of products through co-location interactions~\cite{yuan2015hotspot}. 

%% file: listofsymbols.tex
\chapter{List of Symbols}

\textbf{List of abbreviations}

\begin{longtable}{c|c} 

Acronym & Descriptions\\ \hline

ADN & Activity driven network modelling \\
AV & Acquaintance vaccination \\

APV & Absolute percentage variation \\

CIP & co-location interaction parameters \\
CN & Common neighbours \\
DST network & Dense SPST network \\
DDT network & Dense SPDT network \\
DDT1 & Vaccinating neighbours in DDT network with direct links\\
DDT2 & Vaccinating neighbours in DDT network with any links\\

GDT & Generated SPDT network with 364K nodes \\
GST & Generated SPST network with 364K nodes \\

IMV & Individual movement based vaccination strategy \\
 
IMVE & Individual movement based vaccination strategy with exact information \\

IMVT & Individual movement based vaccination strategy with temporal information \\

LST & SPDT network with the same number of links that of DDT network \\
LST & SPST network with the same number of links that of DST network \\

MLE & Maximum likelihood estimator \\

OSN & Online social network \\
PFU & Plaque-forming unit \\

RV & Random vaccination \\
DV & Degree vaccination \\

RSE & Root squared error \\
SPST & Same place same time transmission \\
SPDT & Same place different time transmission \\
SIR  & Susceptible-infected-recovered \\
SDT network & Sparse SPDT network of links having direct and indirect components \\
SST network & Sparse SPST network of links having indirect component only\\
SPDT graph & graph based on SPDT diffusion \\ \hline

\end{longtable}

\noindent
\textbf{List of symbols}
\begin{longtable}{c|c} 

Symbol & Descriptions\\ \hline

A & Set of active copies of nodes in SPDT graph \\

b & active particle decay rates from an area of interaction \\

$a_i$ & activity potential of node \\   

$a_1, a_2, a_3,$ & active periods of a node \\   

C & Particle concentration in interaction area \\   

C$_1$, C$_2$,C$_3$  & scenario 1, 2 and 3 \\    

d & Activation degree - number of SPDT links created during an activation \\   

E & Intake dose or exposure of infectious particles\\   

$\bar f$ & Average volume fraction of room air introduced by exhaled breath\\   

f & distribution function \\   
 
F & Disease spreading force in the network at the current day of simulation \\   

$F_a$ & Average disease spreading force in the network \\   

$G$  & Graph  \\   

$G_t$  & Dynamic graph  \\   

g & Particle generation rate by an infected individual \\   

h & Activation frequency \\   

I & Number of infected individuals in the system \\   

$I_n$ & Number of infected individuals at a simulation day \\   

$I_p$ & Number of infected individuals in the system at the current time that disease prevalence \\   

$I_a$ & Number of infected individuals up to a simulation day \\   
L & links set \\   

N & Total number of individuals, nodes, users \\   

p & Pulmonary rate of susceptible individual \\   

$P_I$ & Infection probability for an intake dose\\   

$p_c$ & Probability of creating a link during an activation \\   

$p_b$ & Probability of breaking a created link \\   

Q & air exchange rate from an area \\   

q & Transition probability for changing inactive to active state \\   

R  & disease reproduction ability of an infected individual \\   

r & Particle removal rate from interaction area \\   

r$_t$ & Median of particles removal rates \\   

S & Number of susceptible individual \\   

T & Simulation period or disease observation period \\   

$t_a$ & Activation period or period host user or node stays at the interacted location \\   

$t_c$ & Link creation delay or delay neighbour user or node arrives at the interacted location \\   

$t_d$ & Stay duration of user or node stays at the interacted location \\   

V & Air volume of interaction area \\   

$w_1$,$w_2$, $w_3$  & waiting periods of a node \\   

Y & Labelling sets in graph \\   

X & updates \\   

z & Number of time step \\   

Z & set of nodes in the SPDT graph \\   

 $\alpha$ & power law exponent \\   
 
$\beta$ & Infection rate at the current day of simulation in the network for an infected individual\\   

$\delta$ & Indirect transmission period \\   

$\eta$ & central tendency \\   

$\lambda $ & Scaling parameter of activation degree distribution \\   

$\mu$ & neighbour proportion \\   

$\omega$  & average volume fraction of room air that is exhaled by an susceptible individual \\   

$\phi$ & links presence function \\   

$\pi$ & state probability \\   

$\Psi$ & nodes presence function \\   

$\rho$ & switching probability form active to inactive states \\   

$\sigma$ & Infectiousness of infection particles \\   

$\theta$ & Fraction of dose or exposure reaches to the target infection site \\   

$\tau$ & Duration that virus is generated or infectious period of infected individual\\   

$\tau_i$ & inter-event time for node $i$ in activity driven networks \\   

$\xi$ & lower limit of active degree distribution \\   

$\vartheta$ & activation potential in ADN networks \\   

$\varphi$  & particle accumulation rate \\   
$\rho$ & Transition probability for changing active to inactive state \\   \hline


\end{longtable}


%% file: chap_appendix.tex
\chapter{Appendix}
\begin{appendix}
\renewcommand{\thesection}{A\arabic{section}}
\section{Fitting Graph Model Parameters}
In this section, we describe the formulations for fitting model parameters.

\subsection{Formulating Maximum Likelihood Estimation for Node Activation}
The distribution function for the frequency of node activation given $z$ number of time step in the system, $\rho$ scaling parameter for active periods $t_a$ and the waiting time parameter $q$, we can write
\begin{equation*}
Pr(h\mid q)=\frac{\left(\frac{z \rho q}{q+\rho}\right)^ {h}e^{-\frac{z \rho q}{q+\rho}}}{h!}
\end{equation*}

Thus, the Likelihood function for a given set $h=\{h_1,h_2\ldots,h_m\}$ is 
\begin{equation*}
L(q\mid h)=\prod_{i=1}^{m}\frac{\left(\frac{z \rho q}{q+\rho}\right)^ {h_1}e^{-\frac{z \rho q}{q+\rho}}}{h_1!}
\end{equation*}

Taking log on the both sides as it increases monotonically, we obtain
\begin{equation*}
\ln\left( L\left(q\mid h\right)\right)=\ln \left(\frac{zq\rho}{\rho +q}\right)\sum_{i=1}^{m}h_i-\frac{mzq\rho}{q+\rho}-\sum_{i=1}^{m}\ln h_i
\end{equation*}

Differentiating partial derivatives with respect to $q$, we get
\begin{equation*}
\frac{\partial \ln \left(L\right)}{\partial q}=\frac{\rho}{q(\rho+q)}\sum_{i=1}^{m}h_i-\frac{mz\rho^2}{(q+\rho)^2}
\end{equation*}

At the maximum, $\frac{\partial \ln \left(L\right)}{\partial q}=0$ and we can write

\begin{equation*}
0=\frac{1}{q}\sum_{i=1}^{m}h_i-\frac{mz\rho}{q+\rho}
\end{equation*}

Thus, the maximum likelihood estimation condition will be
\begin{equation}
\frac{qz\rho}{q+\rho}=\frac{1}{m}\sum_{i=1}^{m}h_i
\end{equation}

\subsection{Link creation delay distribution}

The probability function for link creation delay $t$ is given by the following equation
\begin{equation*}
f\left ( t_{c}\mid p_{c} ,t_{a}  \right )=\frac{p_{c} (1-p_{c})^{t_{c}-1} }{1-(1-p_{c})^{t_{a}+\delta}}
\end{equation*}
The probability of $t_c$ given $p_c$ can be written according to law of total probability as
\begin{equation*}
f\left(t_c\mid p_c\right)=\int f\left(t_c\mid p_c,t_a\right) f\left(t_a\right) dt_a
\end{equation*}
Approximating the above equation for m values of $t_a$ applying quadrature, we get
\begin{equation*}
f\left(t_c\mid p_c\right)\approx \frac{1}{m} \sum_{l=1}^{m} f\left(t_c\mid p_c,t^{l}_{a}\right)
\end{equation*}
\begin{equation*}
\approx \frac{1}{m} \sum_{l=1}^{m} \frac{p_{c} (1-p_{c})^{t_{c}-1} }{1-(1-p_{c})^{t^{l}_{a}+\delta}}
\end{equation*}

Now we define a likelihood function $L(p_c\mid t_c)$ for the probability distribution function $f(t_c\mid p_c)$. Function $L(p_c\mid t_c)$ will be maximizes for the observations $t^{1}_{c},t^{2}_{c},\ldots,t^{n}_{c}$. The approximated likelihood function can be written as
\begin{equation*}
L(p_c\mid t_c)\approx \prod_{k=1}^{n}\frac{1}{m} \sum_{l=1}^{m} \frac{p_{c} (1-p_{c})^{t^{k}_{c}-1} }{1-(1-p_{c})^{t^{l}_{a}+\delta}}
\end{equation*}
Since the log function is monotonically increasing, we can maximize $\ln L(p_c\mid t_c)$ instead $L(p_c\mid t_c)$. Therefore, we obtain
\begin{equation*}
\ln L(p_c\mid t_c)= n\ln(p_c)-n\ln(m) +\sum_{k=1}^{n} \ln \left(\sum_{l=1}^{m} \frac{(1-p_{c})^{t^{k}_{c}-1} }{1-(1-p_{c})^{t^{l}_{a}+\delta}}\right)
\end{equation*}
Differentiating the above equation, we get
\begin{equation*}
\frac{\partial}{\partial p_c}\left(\ln L(p_c\mid t_c)\right)=\frac{n}{p_c}-\sum_{k=1}^{n} \frac{\sum_{l=1}^{m}\frac{(t_{c}^{k}-1)(1-p_c)^{t_{c}^{k}-2}(1-(1-p_c)^{t_a^{l}+\delta})+(t_{a}^{l}+\delta)(1-p_c)^{t_c^k +t_a^l +\delta -2}}{(1-(1-p_c)^{t^{l}_{t_a}+\delta})^2}}{\sum_{l=1}^{m}\frac{(1-p_{c})^{t^{k}_{c}-1} }{1-(1-p_{c})^{t^{l}_{a}+\delta}}}
\end{equation*}
For maximizing, $\hat p_c$ we set $\frac{\partial}{\partial p_c}\left(\ln L(p_c\mid t_c)\right)=0$ and get
\begin{equation}
0=\frac{n}{p_c}-\sum_{k=1}^{n} \frac{\sum_{l=1}^{m}\frac{(t_{c}^{k}-1)(1-(1-p_c)^{t_a^{l}+\delta})+(t_{a}^{l}+\delta)(1-p_c)^{t_a^l +\delta}}{(1-p_c)(1-(1-p_c)^{t^{l}_{t_a}+\delta})^2}}{\sum_{l=1}^{m} ((1-(1-p_{c})^{t^{l}_{a}+\delta})^{-1}}
\end{equation}
We can solve this equation numerically to find $\hat p_c$.

\subsection{Activation degree distribution}

The probability density function for activation degree $d$ given $\lambda$ is,
\begin{equation*}
f(d\mid \lambda)=(1-\lambda)\lambda^{d-1}
\end{equation*}
and $\lambda$ is drawn from the following power law distribution
\begin{equation*}
f\left( \lambda \right)= \frac{\beta \lambda ^{-(\beta+1)}}{\xi^{-\beta}-\psi^{-\beta}}
\end{equation*}
Therefore, the probability distribution of $d$ for any $\lambda$ can be found according to Law of Total Probability as follows
\begin{equation*}
f(d)=\frac{\beta}{\xi^{-\beta} - \psi^{-\beta}}\int_{\xi}^{\psi}(1-\lambda)\lambda^{d-1}\lambda^{-(\beta+1)} d\lambda=\frac{\beta }{\xi^{\beta}-\psi^{\beta}}\int_{\xi}^{\psi}( \lambda^{d-\beta-2} -\lambda^{d-\beta-1}) d\lambda
\end{equation*}
\begin{equation*}
=\frac{\beta}{\xi^{-\beta} - \psi^{-\beta}}\left(\frac{\psi^{d-\beta-1}-\xi^{d-\beta-1}}{d-\beta-1}-\frac{\psi^{d-\beta}-\xi^{d-\beta}}{d-\beta}\right)
\end{equation*}

Now we would like to derive the maximum likelihood estimator (MLE) conditions to find the parameters $\beta$, $\xi$ and $\psi$ given $d$ samples from real world network. The likelihood function $L$ for the random values of $d=\{d_1,d_2,\ldots d_n\}$ can be written as 
\begin{equation*}
L\left(\beta,\xi,\psi \mid d\right)=\prod_{k=1}^{n}\frac{\beta}{\xi^{-\beta} - \psi^{-\beta}}\left(\frac{\psi^{d_k-\beta-1}-\xi^{d_k-\beta-1}}{d_k-\beta-1}-\frac{\psi^{d_k-\beta}-\xi^{d_k-\beta}}{d_k-\beta}\right)
\end{equation*}
Taking log on both sides, we get
\begin{equation}\label{actd}
\ln L\left(\beta,\xi,\psi \mid d\right)=n\ln\beta-n\ln(\xi^{-\beta} - \psi^{-\beta})+\sum_{k=1}^{n}\ln \left(\frac{\psi^{d_k-\beta-1}-\xi^{d_k-\beta-1}}{d_k-\beta-1}-\frac{\psi^{d_k-\beta}-\xi^{d_k-\beta}}{d_k-\beta}\right)
\end{equation}
To find the maximum of likelihood function $L(\beta,\xi,\psi)$ for the value of $\beta$, we take the partial derivative of \ref{actd} with respect to $\beta$ and set zero to result for the maximum of $\hat \beta$
\begin{equation*}
0=\frac{n}{\beta}-n\frac{\psi^{-\beta}\ln\psi-\xi^{-\beta}\ln\xi}{\psi^{-\beta}-\psi^{-\beta}}+\sum_{k=1}^{n}\frac{\frac{\xi^{d_k-\beta -1} \ln\xi -\psi^{d_k-\beta -1}\ln\psi}{d_k-\beta-1}-\frac{\psi^{d_k-\beta -1}-\xi^{d_k-\beta -1}}{(d_k-\beta -1)^2}-\frac{\xi^{d_k-\beta}\ln\xi-\psi^{d_k-\beta} \ln\psi}{d_k-\beta}-\frac{\psi^{d_k-\beta}-\xi^{d_k-\beta}}{(d_k-\beta)^2}}{\frac{\psi^{d_k-\beta-1}-\xi^{d_k-\beta-1}}{d_k-\beta-1}-\frac{\psi^{d_k-\beta}-\xi^{d_k-\beta}}{d_k-\beta}}
\end{equation*}
Since $\psi \approx 1 $, we assume $\ln \psi = 0$ and $\psi^{x}=1$ for any $x$. We can simplify as 
\begin{equation}
0=\frac{n}{\beta}-\frac{n\xi^\beta \ln\xi}{\xi^\beta-\psi^{-\beta}}+\sum_{k=1}^{n}\frac{\frac{\xi^{d_k-\beta -1} \ln\xi}{d_k-\beta-1}-\frac{1-\xi^{d_k-\beta -1}}{(d_k-\beta -1)^2}-\frac{\xi^{d_k-\beta} \ln\xi}{d_k-\beta}-\frac{1-\xi^{d_k-\beta}}{(d_k-\beta)^2}}{\frac{1-\xi^{d_k-\beta-1}}{d_k-\beta-1}-\frac{1-\xi^{d_k-\beta}}{d_k-\beta}}
\end{equation}\\
If we set $y=$
In the similar way, we can find $\hat \xi$ derivating $L(\beta,\xi,\psi)$ ( Eq.\ref{actd}) with respect to $\xi$ and setting zero for maximal of $\hat \xi$

\begin{equation*}
0=\frac{n \beta\xi^{-\beta -1}}{\xi^{-\beta}-\psi^{-\beta}}+\sum_{k=1}^{n}\frac{\frac{(d_k-\beta)\xi^{d_k-\beta -1}}{d_k-\beta}-\frac{(d_k-\beta-1)\xi^{d_k-\beta -2}}{d_k-\beta-1}}{\frac{\psi^{d_k-\beta-1}-\xi^{d_k-\beta-1}}{d_k-\beta-1}-\frac{\psi^{d_k-\beta}-\xi^{d_k-\beta}}{d_k-\beta}}
\end{equation*}

With simplification we get
\begin{equation}
0=\frac{n(\beta +1)}{\xi^{-\beta}-\psi^{-\beta}}+\sum_{k=1}^{n}\frac{\xi^{d_k}-\xi^{d_k-1}}{(\psi^{d_k-\beta-1}-\xi^{d_k-\beta-1})(d_k-\beta)^{-1}-(d_k-\beta-1)^{-1}(\psi^{d_k-\beta}-\xi^{d_k-\beta})}
\end{equation}

To estimate $\hat \psi$, we derivate $L(\beta,\xi,\psi)$ ( Eq.\ref{actd}) with respect to $\psi$ and setting zero for maximal of $\hat \psi$
\begin{equation*}
0=-\frac{n\beta \psi^{-\beta -1}}{\xi^{-\beta}-\psi^{-\beta}}+\sum_{k=1}^{n}\frac{\frac{(d_k-\beta -1)\psi^{d_k-\beta -2}}{d_k-\beta-1}-\frac{(d_k-\beta)\psi^{d_k-\beta -1}}{d_k-\beta}}{\frac{\psi^{d_k-\beta-1}-\xi^{d_k-\beta-1}}{d_k-\beta-1}-\frac{\psi^{d_k-\beta}-\xi^{d_k-\beta}}{d_k-\beta}}
\end{equation*}
\begin{equation}
0=-\frac{n(\beta +1)}{\xi^{-\beta}-\psi^{-\beta}}+\sum_{k=1}^{n}\frac{\psi^{d_k-1}-\psi^{d_k}}{(\psi^{d_k-\beta-1}-\xi^{d_k-\beta-1})(d_k-\beta)^{-1}-(d_k-\beta-1)^{-1}(\psi^{d_k-\beta}-\xi^{d_k-\beta})}
\end{equation}
\end{appendix}